\documentclass{aa}  
\usepackage{graphicx}
\usepackage{lscape}
\usepackage{txfonts}
\usepackage{float}
\usepackage{hyperref}
\usepackage{soul}

\begin{document}

   \title{LOFAR 58~MHz Legacy Survey of the 3CRR Catalog}


   \author{J.~M.~Boxelaar\inst{1}\fnmsep\inst{2}\fnmsep\thanks{\email{j.boxelaar@ira.inaf.it}}
          \and
          F.~De~Gasperin\inst{1}
          \and
          M.~J.~Hardcastle\inst{3}
          \and 
          J. H. Croston\inst{4}
          \and
          L. K. Morabito\inst{5}\fnmsep\inst{6}
          \and
          R.~J.~van~Weeren\inst{7}
          \and
          H. Edler\inst{8}
          }

   \institute{INAF–Istituto di Radioastronomia, Via P. Gobetti 101, 40129 Bologna, Italy
         \and
             Dipartimento di Fisica e Astronomia, Università di Bologna, via P. Gobetti 93/2, 40129 Bologna, Italy
        \and
            Centre for Astrophysics Research, University of Hertfordshire, College Lane, Hatfield AL10 9AB, UK
        \and
            School of Physical Sciences, Open University, Walton Hall, MK7 6AA, UK
        \and
            Centre for Extragalactic Astronomy, Department of Physics, Durham University, Durham DH1 3LE, UK
        \and
            Institute for Computational Cosmology, Department of Physics, Durham University, Durham DH1 3LE, UK
        \and
            Leiden Observatory, Leiden University, PO Box 9513, 2300 RA Leiden, The Netherlands
        \and 
            ASTRON, Netherlands Institute for Radio Astronomy, Oude Hoogeveensedijk 4, 7991 PD Dwingeloo, The Netherlands
             }

   \date{Received \today; accepted NA}


\abstract
{The Low Frequency Array (LOFAR) is uniquely able to perform deep, 15\arcsec\ resolutions imaging at frequencies below 100~MHz. Observations in this regime, using the Low Band Antenna (LBA) system, are significantly affected by instrumental and ionospheric distortions. Recent developments in calibration techniques have enabled routine production of high-fidelity images at these challenging frequencies.}
{The aim of this paper is to obtain images of the radio sources included in the Third Cambridge catalog, second revised version (3CRR), at an observing frequency of 58~MHz, with an angular resolution of 15\arcsec\ and sensitivity to both compact and diffuse radio emission. This work also aims to produce accurate flux measurements for all sources. This dataset is designed to serve as a reference for low-frequency radio galaxy studies and future spectral aging analyses.}
{We present the data reduction and calibration procedures developed for narrow-band observations of bright sources with the LOFAR LBA. These include tailored direction-independent calibration strategies optimized for mitigating ionospheric phase corruptions and instrumental effects at 58~MHz. Imaging techniques were refined to reliably recover both small- and large-scale radio structures reliably.}
{We deliver 58~MHz radio images for the complete 3CRR sample including flux density measurements. We determined that the LBA has an accurate flux density scale with an average flux uncertainty of 10\%. This is an important confirmation for any future works using the LOFAR LBA system. 
With these results we characterize the bright radio galaxy population with new high-resolution low-frequency images. We also provide high-resolution models of these sources which will be useful for calibrating future surveys. }
{This legacy survey significantly expands the available high-resolution data at low frequencies and is the first fully imaged high-resolution sample at ultra low frequencies ($< 100$~MHz). It lays the foundation for future studies of radio galaxy physics, low-energy cosmic-ray populations, and the interplay between radio jets and their environments.}


\maketitle

%

\section{Introduction}

The radio sky looks vastly different from the sky that is observed at other wavelength regimes. But even within the broad radio band (spanning from millimeter to decameter wavelengths) there can be large differences in what can be observed. It is therefore essential to explore the radio sky down to the lowest possible frequencies to  better understand the objects we observe. At gigahertz frequencies there exist plenty of deep, high-resolution data. This is not yet the case for observations at ultra-low frequencies ($< 100$ MHz\footnote{The term ultra-low is chosen to discriminate between the low frequencies observed with LOFAR at 150 MHz and the frequency range used in this paper.}). 

It has proven difficult to obtain deep, high-resolution images at ultra-low frequencies because of two main challenges. Firstly, as resolution $\propto B/\lambda$ (with $B$ the maximum baseline length and $\lambda$ the wavelength), observations at low frequency require longer baselines to achieve the same resolution as higher frequency interferometers, which poses practical challenges in the design of the interferometer, e.g. regarding data transfer and clock synchronization. Second, atmospheric turbulence in the ionosphere has a great influence on radio signals at these low frequencies and corrupts data in an unpredictable way \citep[e.g.][]{1967Slee, 1986Thompson, 2009Intema, 2018Gasperin}. For interferometers, the turbulent and inhomogeneous ionosphere causes phase shifts that are time and frequency dependent. For high frequencies, ionospheric corruption is less prominent as the dominant scalar phase delay scales approximately as $\Phi_{\text{iono}}\propto\nu^{-1}$. For ultra low frequencies, the second order differential effect called Faraday rotation ($\propto\nu^{-2}$) becomes important. Faraday rotation causes a rotation in polarization angle of the incoming radiation and needs to be corrected for using a rotation matrix. Higher-order effects up to $\nu^{-3}$ become non-negligible below 30~MHz \citep{Datta-Barua2008, 2018Gasperin}. Furthermore, varying ionospheric conditions across the interferometer make these effects direction dependent and thus complicate observations at low frequencies even further.

Given these difficulties, the LOw Frequency ARray \citep[LOFAR,][]{vanHaarlem2013} is currently the most capable instrument for performing high-resolution observations at low frequencies. Observations at ultra-low frequencies are done by the Low Band Antenna (LBA), operating in the range $10-90$~MHz. 
In the past years, ongoing algorithmic improvements made it possible to routinely correct for the severe ionospheric effects in the LBA \citep{2019Gasperin, Gasperin2020, Edler2022, Gasperin2023}.
Recently, this has even also allowed observations to be carried out at arcsecond resolution (with the International LOFAR Telescope, ILT) down to 30 MHz \citep{2016MNRAS.461.2676M, Gasperin2020, Groeneveld2022}. 

In this paper, we apply these recently developed advanced calibration techniques to obtain 15\arcsec\ resolution images of the entire second revision of the Third Cambridge Catalog of Radio Sources (3CRR, \citealt{Laing1983}), consisting of the brightest radio sources in the northern hemisphere 
at 178~MHz. The vast majority of the sources show emission originating from active Galactic nuclei (AGN), which are powered by super-massive accreting black holes at the center of their host galaxies. These black holes accrete gas and dust and eject plasma as powerful bipolar jets. Electrons in the jets are accelerated to highly relativistic energies, resulting in observable synchrotron emission. AGN jets can reach sizes of hundreds of kiloparsec \citep[e.g.][]{Birkinshaw1981MNRAS.197..253B,Alexander1987MNRAS.225...27A} up to several megaparsec \citep{Oei2024Natur.633..537O}.

Although the sources in this catalog have been studied in detail at higher frequencies both in surveys and in individual targeted studies, this is the first high-resolution survey of the full 3C catalog at ultra-low frequencies. Surveys like the 38~MHz 8th Cambridge survey (8C; \citet{1990MNRAS.244..233R, 1995MNRAS.274..447H}) and the 78~MHz VLA Low-Frequency Sky Survey (VLSSr; \citet{2007AJ....134.1245C,2014MNRAS.440..327L}) previously observed below 100~MHz.  However, 8C only observed above $\delta > 60^{\circ}$ with multiple arc-minute resolution, while VLSSr observed the entire northern sky (down to $-30$ Dec) at 78~MHz with a resolution of 75\arcsec, about 5 times lower than the average resolution of the maps presented here.

The synchrotron emission of radio galaxies is usually well described by a synchrotron spectrum across a wide frequency range. At low frequencies, spectra will deviate from a pure synchrotron spectrum. Thus, studying AGN at the low frequency end of the electromagnetic spectrum is important for our understanding of the absorption mechanisms at play. At low frequencies, synchrotron sources will eventually turn over due to e.g. synchrotron self-absorption, as at sufficiently low frequencies, electrons can absorb radiation emitted by electrons of the same plasma \citep{Pacholczyk1970, 2016era..book.....C}. Other effects may cause a low frequency turn over, such as free-free absorption and a low-energy cut-off in the electron energy distribution \citep{Kellerman1966, Bicknell1997, Tingay2003}. Low-frequency observations help distinguish between these processes.


Synchrotron spectral aging is most often observed as exponential cut-offs in the GHz regime of the spectrum \citep{Longair1973} but standard models for spectral aging (e.g. the JP \citep{Jaffe1973} and KP \citep{Kardashev1962, Pacholczyk1970} models) assume an injection index\footnote{We define the spectral index such that $S_{\nu}\propto\nu^{-\alpha}$} $\alpha \geq 0.5$ for the aging spectrum that has to be constrained by low frequency data. This is because the low frequency part of the spectrum ages most slowly. Such analysis can only be properly done for spatially resolved sources. This emphasizes the need for high resolution maps at ultra-low frequencies that can significantly help to constrain the injection index for these aging models.
Good constraints on the electron energy index at the lowest frequencies are important for reliable estimates of the total energy stored in the synchrotron-emitting plasma, because the internal energy is dominated by the lowest energy particles. This is important for investigations of the energetic impact of jets, and for magnetic field strength estimates.
Furthermore, due to the suppressed spectrum of the aged emission at higher frequencies, old emission at the end of jets and lobes is possibly only observed at low frequencies. For sources with extended, aged radio emission, combined LOFAR LBA and HBA ($110-240$~MHz) observations provide a unique possibility to determine spectral indices.  

In this paper we present a catalog of narrow-band (3 MHz bandwidth) observations and high-resolution imaging of all sources in the 3CRR catalog with LOFAR at an ultra-low central frequency of 58~MHz. We begin by describing the calibration pipeline that allowed us to obtain images at a resolution of 15\arcsec\ and rms noise of $\sim$10~mJy~beam$^{-1}$. In addition, we present flux densities for all sources and complement this with flux densities from HBA to estimate spatially integrated synchrotron spectral indices (using the convention $S_{\nu}\propto\nu^{-\alpha}$). The main results of this survey include independent flux density measurements at 58 MHz of all 3CRR sources. Because this catalog has been widely studied at all radio frequencies up to $\sim10$~GHz, we are also able to determine the flux scale of the LBA system. 
We also present the first high resolution ($\sim15$\arcsec) catalog of total intensity maps for all 3C sources at ultra-low frequencies. The radio maps have a typical rms noise between $\sim 8-15$~mJy~beam$^{-1}$. Along with the maps we publish high-resolution models in the \texttt{Library for Low Frequencies (LiLF\footnote{\url{https://github.com/revoltek/LiLF}})}. These models are useful for LBA calibrations of future survey projects. These wide-field studies need to observe faint sources and thus require that 3C sources are subtracted from the side-lobes during self-calibration.

In this work, we will discuss in more detail one of the uses of this survey in Section~\ref{sec:ssa}, where we use peaked spectrum sources to estimate the magnetic field in the source. This is a unique possibility that low-frequency observations provide as measurements in this regime are essential to constrain the turnover. In a second work (referred to as Paper~II) on the data presented here, we will select sources with large angular size to produce low-frequency spectral index maps and test the validity of conventional aging models (such as JP and KP) in this regime.   

The outline of the paper is as follows: In Sect.~\ref{sec:observation} we further describe the sample and the observations. Data reduction and calibration are described in Sect.~\ref{sec:data_reduction}. In Sect.~\ref{sec:result} we describe the flux density measurements and validation, followed by a discussion of the obtained spectral energy diagrams and the prospects of ultra-low frequency observations and final conclusions in Sect.~\ref{sec:discussion}~\&~\ref{sec:conclusion}.

Throughout this work, we assume a flat $\Lambda$CDM cosmology with $\Omega_{\Lambda} = 0.7$, $\Omega_m = 0.3$ and $H_0 = 70$~km~s$^{-1}$~Mpc$^{-1}$.

\section{Sample and observations}
\label{sec:observation}

The Third Cambridge Catalog of Radio Sources (3CRR catalog) by \citet{Laing1983} is the second revision, hence the double R, of the original 3C catalog by \citet{Edge1959}. The 3CRR catalog consists of the brightest extragalactic radio sources with flux densities $S_{\nu}> 10$~Jy\footnote{In the original paper. $S>10.9$~Jy for the flux scale adopted in this work} at 178~MHz. Sources are constrained by having Dec~$>10\degr$ and to be $10\degr$ below or above the Galactic plane. This sample contains 173 sources, ordered by right ascension. Sources discovered after the original paper therefore have a numeric suffix (e.g., 3C~314.1) to avoid renumbering of all subsequent sources. Other sources have been discovered to consist of two distinct emission regions. To differentiate these distinct regions, the sources have an alphabetic suffix (e.g., 3C~225B). Although technically the observed sample is 3CRR, we will sometimes refer to it as the 3C~catalog for the rest of this work.

\begin{figure*}[th]
	\centering
	\includegraphics[width=1.2\columnwidth, trim={2.cm 2.cm 1.5cm 2.cm},clip]{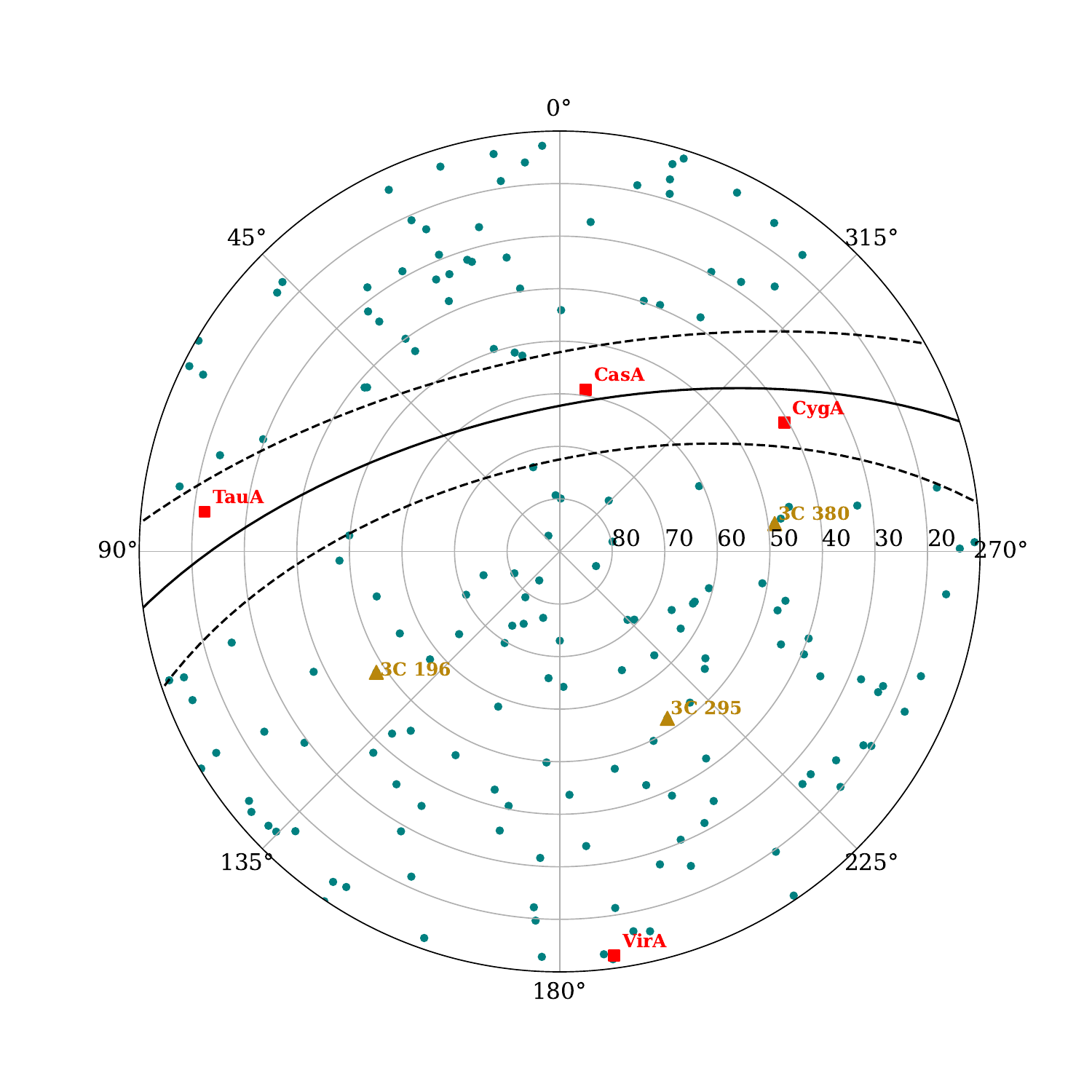}
	\caption{RA/Dec radial plot of the northern hemisphere containing all sources in the catalog. The four brightest radio sources are shown as red squares and the calibrators are depicted by the triangles. The solid black line shows the galactic plane and the dashed lines around it the $\pm10$\degr offset from the plane. The figure extends to 10\degr~Dec only because the sample does not contain sources below that threshold.}
	\label{fig:all_sky_map}
\end{figure*}

The FR~II type radio galaxies \citep{Fanaroff1974}, showing the highest surface brightness at the outer lobe edges, make up about 76\% of the sample. Ten percent are identified as FR~I type radio galaxies, where the highest surface brightness is located around the central host galaxy. 
About 10 of the galaxies are giant radio galaxies (GRG), with size $\gtrsim 1$~Mpc \citep{Orru2010}. As a result of the selection criteria, all galaxies lie within $z=2$. Furthermore, one of the sources' emission does not originate from AGN outflows but from Galactic emission related to star formation. This source is 3C~231, better known as the starburst galaxy M82. 
All sources, including the results obtained in this work, are published separately as a catalog. part of the catalog is listed in the Table in Appendix~\ref{app:all_sources}. Figure~\ref{fig:all_sky_map} displays the 3C sources on the northern hemisphere, highlighting the location of the three calibrators as well as the location of the four brightest sources and the Galactic plane $\pm 10\degr$.

Observations were carried out with the LOFAR telescope using the low-band antenna (LBA) system under project LC9\_17 \& LC10\_020 (PI: de~Gasperin) over a time period between 29-11-2017 and 27-09-2018, totaling 48 hrs of observing time. 
Observations were performed in LBA\_OUTER mode, in which only the outer half of the LBA dipoles in a station are used. Observations are carried with the full Dutch array (25 Core and 14 Remote stations) with the exception of CS013 and CS031 which were malfunctioning and therefore flagged in almost all of the data. For 8 observing hours, RS310 was also flagged, which yields almost no data for that station for about 30 sources. This is particularly problematic for the \emph{uv}-coverage in the east-west direction since RS310 is the westernmost station in the otherwise north-south oriented Dutch array. Therefore, sources observed without RS310 have a rather elongated beam in the east-west direction. These sources are listed in Section~\ref{sec:result}.

For this survey, the observing strategy is somewhat unorthodox as it utilizes LOFAR's unique multi-beam capabilities. 
We aimed to observe the entire set of 173 sources of the 3CRR catalogue with a long exposure (8~hrs) to increase the \emph{uv}-coverage. Since observing the targets one after the other would have been prohibitively expensive in terms of telescope time, we reduced the bandwidth by increasing the number of simultaneous beams. We therefore observed with 30 beams directed towards 29 3C sources and one calibrator. The phased array design of LOFAR makes this unique set-up possible. The calibrators used here are 3C~196, 3C~380 and 3C~295, which are themselves part of the catalog. They are chosen because they are approximately equally spaced in RA and are compact bright sources. Every observing hour, a calibrator is chosen based on the maximum elevation at that time. These sources are well-known standard calibrators with defined flux densities on the Scaife \& Heald (SH) flux scale \citep{Scaife2012}. For the observing time per target, see Table~\ref{tab:calibrators}.

\begin{table}
	\centering
	
	\caption{ Summary of observations. Fluxes are inferred from the SH scale.}
    \label{tab:calibrators}
	\begin{tabular}{lcccc}
		\hline\hline
		Object & Time & Freq. & Bandwidth & $S_{58\text{ MHz}}$\tablefootmark{a}\\
		& [h] & [MHz] & [MHz] & [Jy]\\
		\hline
		3C 196 & 22 & 57.7 & 3 & 155.10\\
		3C 380 & 17 & 57.7 & 3 & 134.26 \\
		3C 295 & 24 & 57.7 & 3 & 160.96 \\
		\hline
		Targets & 8 & 57.7 & 3 &\\
		\hline
	\end{tabular}
	
	\tablebib{(a) flux scale from \citet{Scaife2012}.}
\end{table}

By using 30 simultaneous beams, every direction has 15 of the total 480
sub-bands assigned to it, resulting in a total bandwidth of 3.125~MHz per direction centered around 57.7~MHz where the LBA is most sensitive. In this way, the total observing time is cut by approximately a factor of 10 compared to the time it would have taken with typical wide band (30 MHz) observations for all sources. The total observing time for the entire catalog is just 46 hours. Clearly the drawback is the limited bandwidth, but because we are targeting the brightest sources in the northern sky, it is possible to achieve an adequate signal to noise even with the limited bandwidth. Observations were taken between November 2017 and September 2018 at irregular intervals between observations, resulting in a variety of LSTs and thus maximizing the \emph{uv}-coverage. 

During pre-processing, the data were flagged for RFI using \texttt{AOflagger} \citep{Offringa2010,Offringa2012} and subsequently CasA and CygA were demixed using the demixing procedure \citep{vdtol2007,Gasperin2020b}. During demixing, sidelobe contributions from the brightest sources in the sky (A-team sources) are subtracted from the data. The models for the A-team sources VirA and TauA were taken from \citet{Gasperin2020b}. Data is furthermore compressed using the dysco software \citep{Offringa2016} to reduce the data to a more manageable size. After that, data were averaged to 4~s time intervals and 4~channels per subband. The data were then calibrated and imaged using the Default Pre-Processing Pipeline (DP3; \citet{Diepen2018}) and WSClean \citep{Offringa2014, Offringa2017}. The process will be explained in more detail in the following section.

\section{Data reduction}
\label{sec:data_reduction}

The calibration strategy for the calibrator data is based on previous work described by \citet{2019Gasperin}, available in LiLF. Since the calibration strategy of LBA data is still evolving, the strategy used here differs in some aspects from what is described there. 

Before solving and applying any corrections, we use existing lists of clean components for the calibrators as initial models. These models have total fluxes equal to that from the SH flux density scale models and the calibrators will thus have the same flux density by definition (see Table~\ref{tab:calibrators}). Before solving for any systematic effects, the visibilities are smoothed in frequency and time with a Gaussian kernel whose size is inversely proportional to the baseline length. This baseline dependent smoothing is done to reduce the noise. The resulting reduction in  field of view (FoV) has no significant influence as our sources are small compared to the FoV and at the phase center. Additionally, since the smoothing is baseline dependent, the least amount of smoothing happens on long baselines.
As almost all sources are expected to appear unpolarized at LBA frequencies, the off-diagonal elements of a full Jones calibration matrix are expected to have incoherent, quickly varying phases and negligible amplitudes. Therefore smoothing has no significant effect and to reduce computation time we only smoothed the diagonal polarizations.

\begin{figure*}[ht]
	\centering
	\includegraphics[width=.75\linewidth, trim={0.cm 0.cm 0.cm 0.cm},clip]{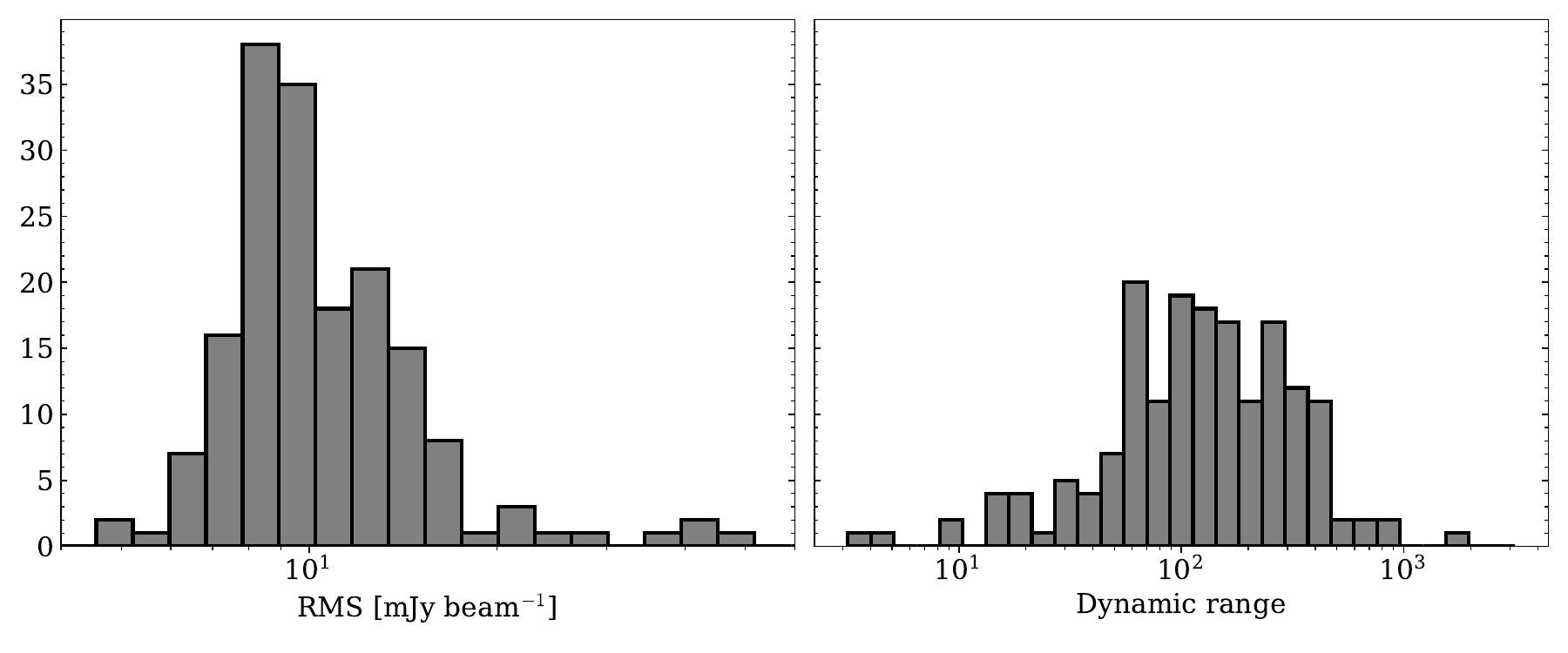} 
	\caption{Noise and dynamic range distribution of the produced radio maps.}
	\label{fig:image_statistics}
\end{figure*}

Calibrator data are corrected following the signal path starting at the receiver. The first effect that is corrected for is polarization alignment. This corrects for differential delays between the X and Y dipoles. Subsequently, the dipole beam correction is applied. This has to be done before the Faraday rotation correction, as the two matrices describing the correction do not commute. The Faraday rotation effect is solved for and corrected in a circular polarization basis, so that the effect is described by a diagonal matrix of phase terms \citep{Smirnov2011}. Next, the data are corrected for the time-averaged bandpass response. Finally, we solve for scalar phases, capturing the underlying ionospheric and station clock delays. 

For the target field, we also start by flagging and demixing the data. After this, the calibrator solutions of the corresponding calibrator at the time of observation are applied. Because of the observing strategy, targets may have different calibrators depending on the observing time. Solutions transferred from the calibrator are: polarization alignment, amplitude bandpass, clock delay, and first-order ionospheric effects. Beam effects are estimated from models and also applied to the targets. We do not apply solutions from Faraday rotation on the target because the frequency-dependent effect is negligible and difficult to fit to since our total bandwidth is only 3~MHz. Any small contributions should be corrected for during self-calibration. After these corrections are transferred, the data are backed up and a copy of the measurement set is created containing only the core stations. 

\begin{figure*}[th]
	\centering
	\includegraphics[width=.325\linewidth, trim={.cm .cm .cm .cm},clip]{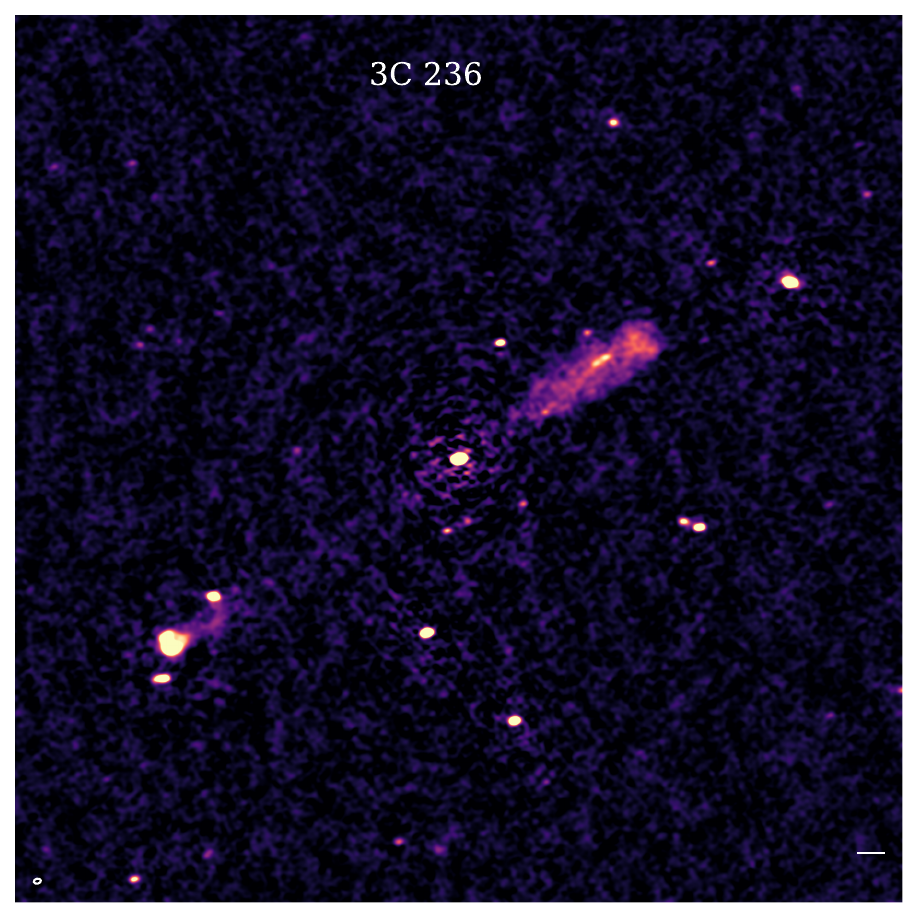}
	\includegraphics[width=.33\linewidth, trim={.cm .cm .cm .cm},clip]{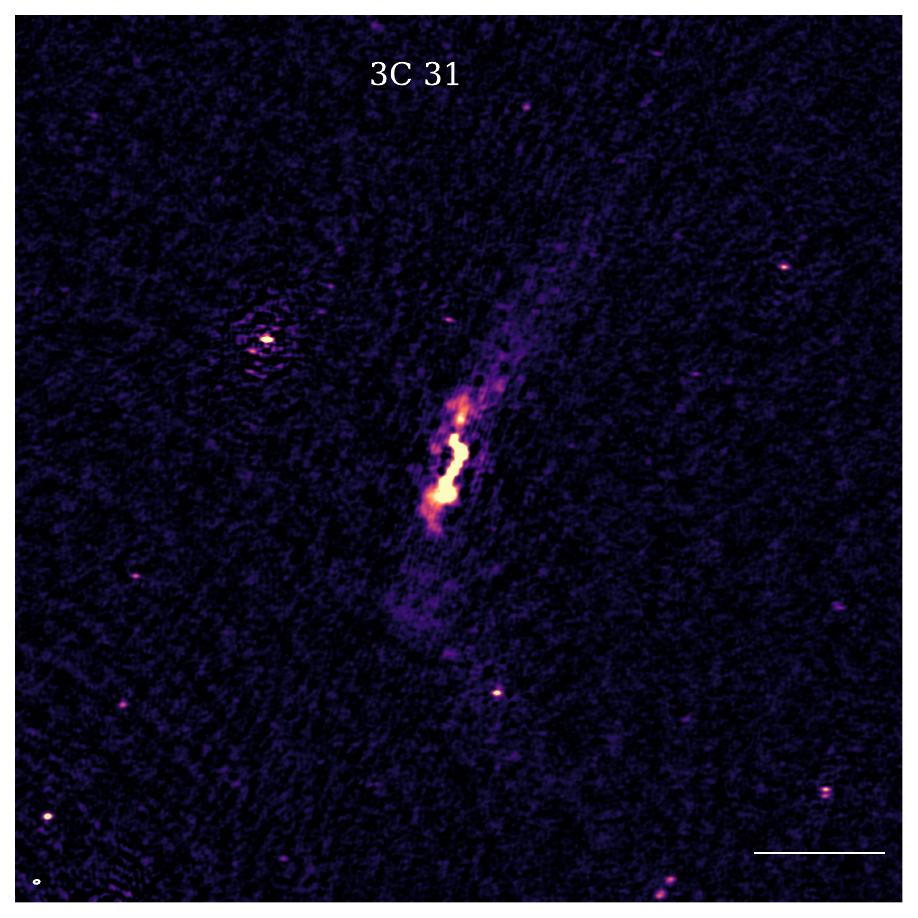}
	\includegraphics[width=.335\linewidth, trim={.cm .cm .cm .cm},clip]{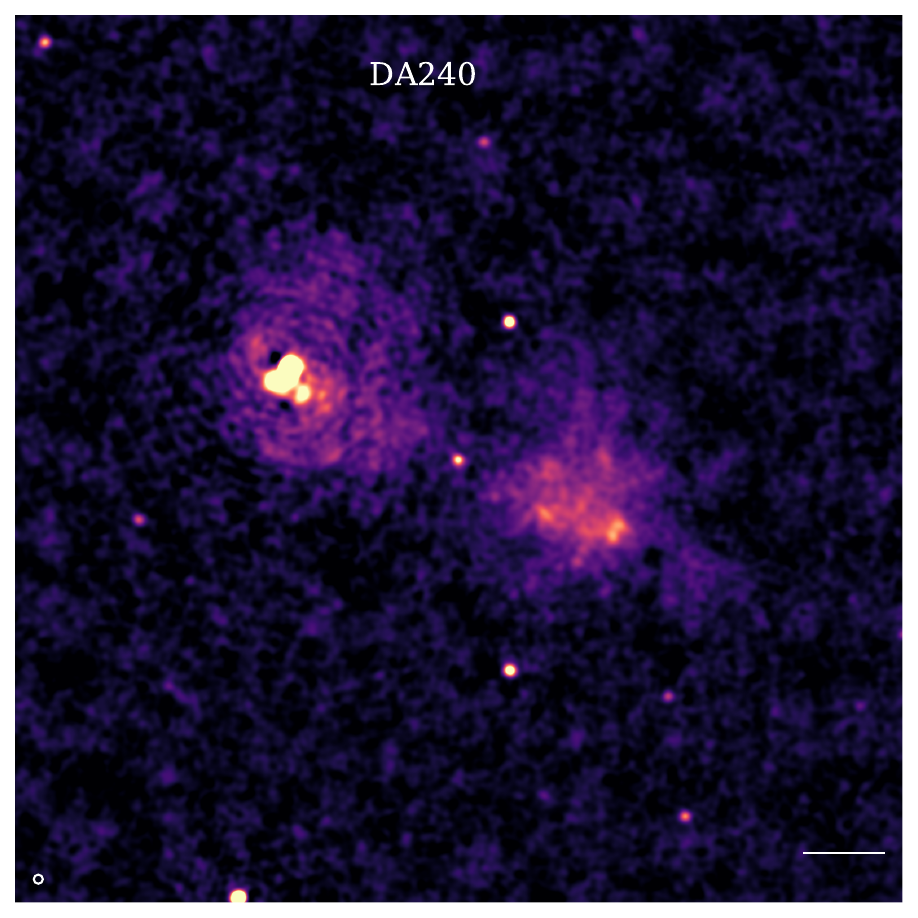}
	\includegraphics[width=.33\linewidth, trim={.cm .cm .cm .cm},clip]{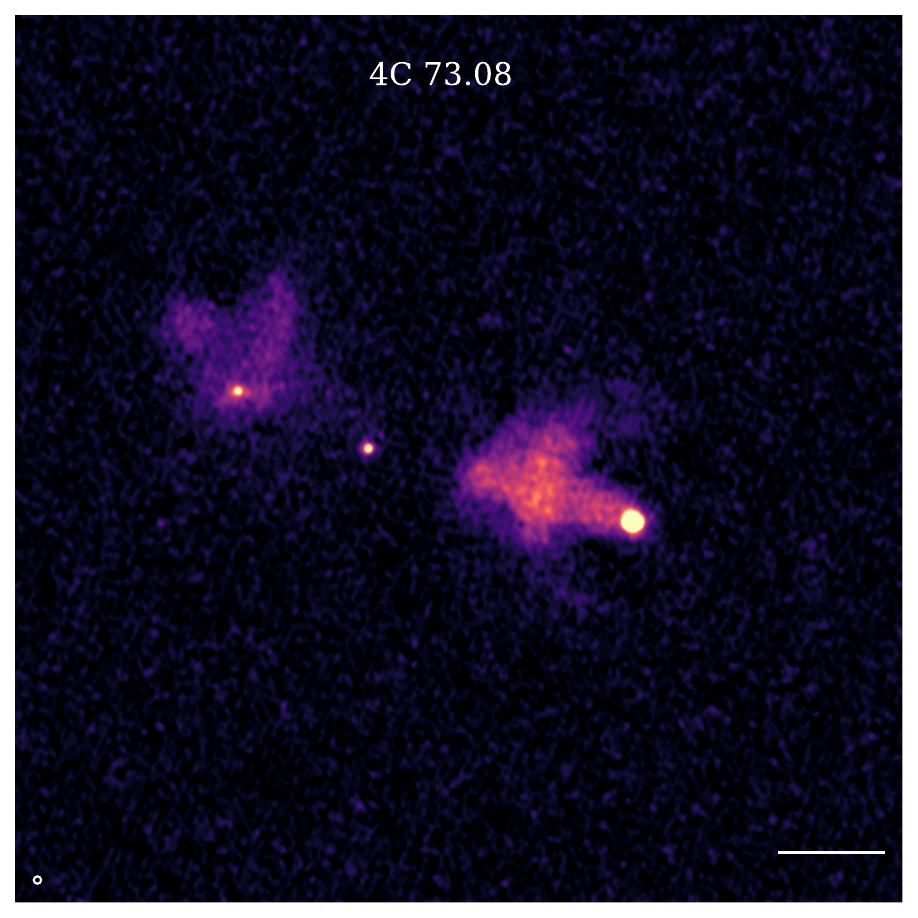}
	\includegraphics[width=.33\linewidth, trim={.cm .cm .cm .cm},clip]{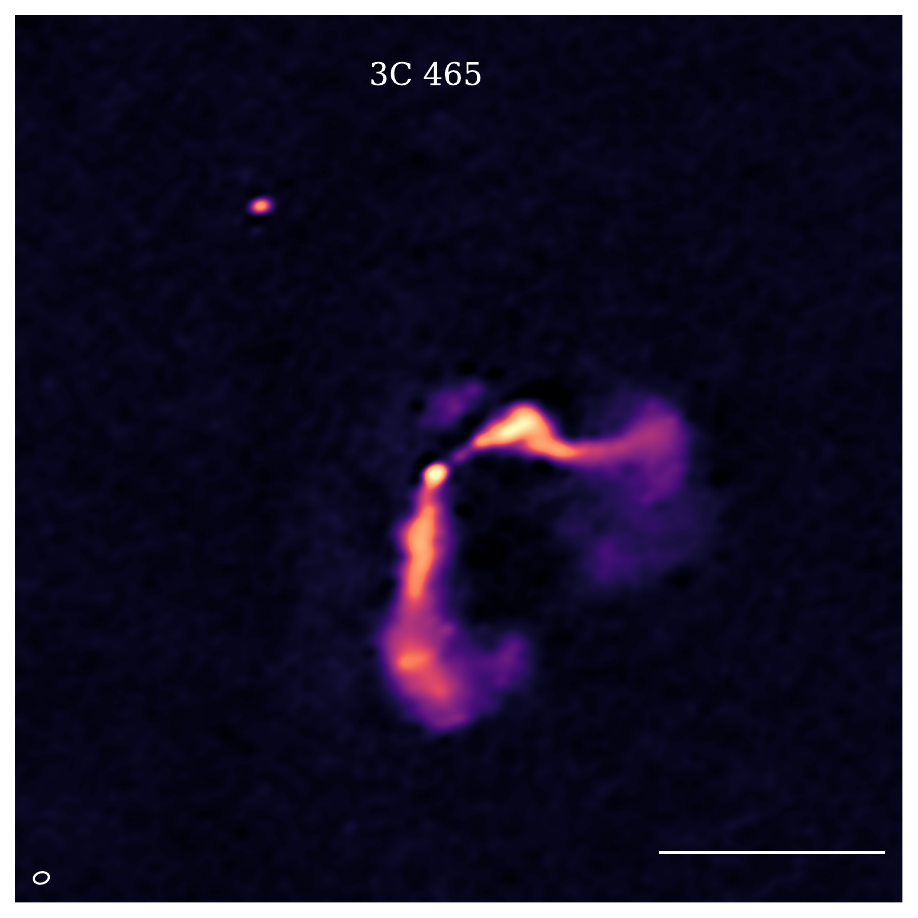}
	\includegraphics[width=.33\linewidth, trim={.cm .cm .cm .cm},clip]{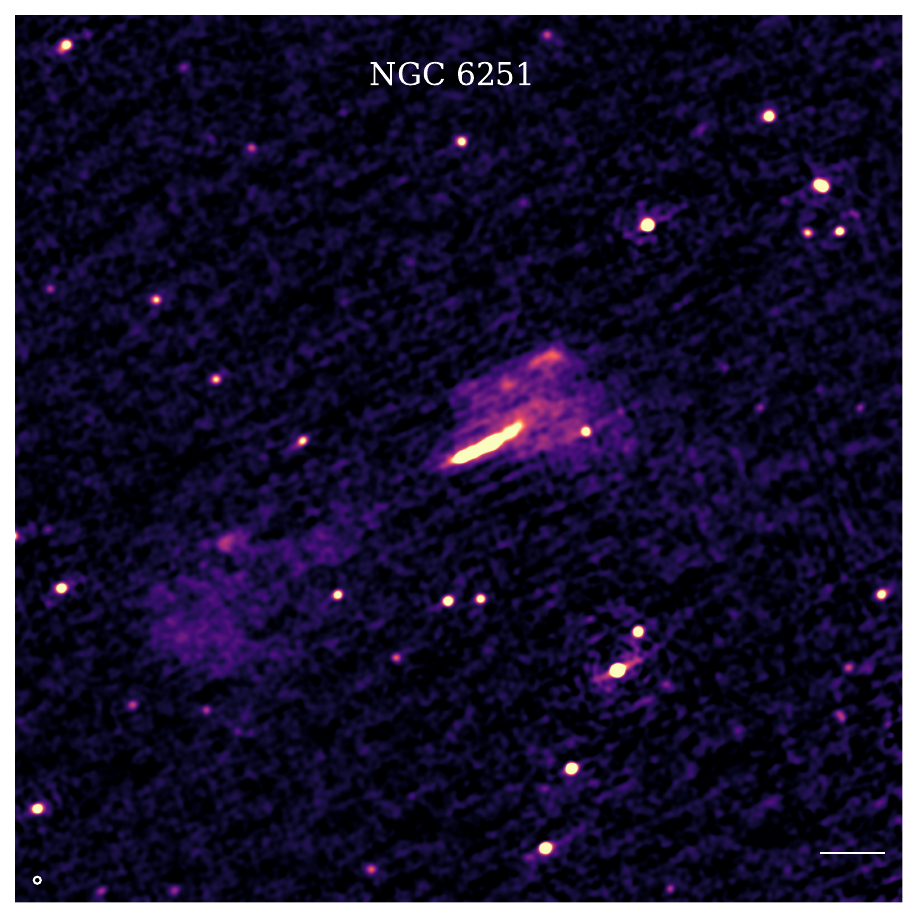}
	\caption{Some of the most extended sources in the sample, some with low signal to noise emission. The beam is indicated in the bottom left of every panel. The scale bar in the bottom right are 200~kpc for each panel. The color scales are different for every panel to highlight the diffuse emission.}
	\label{fig:faint_sources}
\end{figure*}

The first few cycles of self-calibration are done on core stations only to establish an excellent solution on the short baselines; this process reduces the complexity and degrees of freedom for the solver when adding back the longer baselines. The first cycle of self-calibration is a scalar phase solve at the highest time resolution, done against models taken from TGSS. Solving with a time resolution equal to the data resolution is possible here because of the high brightness of the sources and the high signal to noise on the core baselines. This first imaging cycle sets the flux scale for the rest of the calibration process and we therefore normalize all subsequent amplitude solutions to average to unity to preserve the flux scale. 
The following self-calibration cycles solve for scalar phase and full-Jones matrix respectively. The full-Jones matrix should correct for any small, higher order effects that were not accounted for by the calibrator solutions or the phase solve. The full-Jones matrix is solved for on a much larger time scale of about 13 minutes to reduce the degrees of freedom and prevent noise from dominating in the solutions. The diagonal amplitude part of the full Jones matrix is scaled to unity to preserve the flux scale set in the first cycle. The correct method would be to avoid the scaling and re-calibrate the diagonal elements to compensate for any deviation introduced by the full-Jones calibration step. However, we determined it was not necessary because of the small off-diagonal polarization contribution. Moreover, the final flux is checked (see Section~\ref{sec:flux_scale}) and found to be correct.

After every calibration cycle, the corrected data are imaged and the model is updated accordingly. For the core stations, imaging of the full 6.5~degree FoV was performed at a pixel scale of 50~\arcsec\ using a Briggs weighting with robustness parameter $-0.8$ and a \emph{uv}-cut of $30$~k$\lambda$. This results in a resolution of about 4.5\arcmin-6\arcmin. The low-resolution images allow for measuring the flux density of even the most diffuse emission from the 3C sources despite the high noise level of the produced images of $\sim35 - 100$~mJy~beam$^{-1}$.

Once the core stations are mutually well calibrated, we phase up the core stations to form a virtual super station, which has become common practice for LOFAR VLBI observations using the high-band antennas \citep{Morabito2022} as well as for some prior LBA observations \citep[e.g.][]{2016MNRAS.461.2676M, Groeneveld2022}. The main advantage of this is that it reduces the degrees of freedom and results in a much higher signal to noise on baselines connecting remote stations to the virtual super station. Additionally, it drastically decreases the field of view for the CS-RS baselines, thereby suppressing the impact of other sources in the field. If all 23 core stations are phased up, the new station would have an effective diameter of 4231~m, reducing the field of view of this station to 5.46\arcmin\ (using $\text{FOV}=1.3\lambda/D$). To prevent the loss of flux, we decided that sources must have an angular diameter smaller than 1.49\arcmin\ to phase up the core in the measurement set. Since many sources are larger than that, we dynamically phase up core stations only up to the point where no flux is lost on shorter baselines. For sources with large angular sizes e.g. 3C~31 or NGC~6251, this means no phase-up is possible, while for many point-like sources there is no issue. 

Calibrating the entire Dutch array is done by iteratively solving for scalar phase corrections and imaging to the point where there is no significant improvement in both the rms noise and dynamic range. Phase solutions are constrained to be smooth on 1~MHz scales and solved for at 4~s time resolutions. Once there are no improvements to the image quality in the phase calibration iterations, we add a subsequent full-Jones matrix solve to attempt to capture residual corruptions of the data. Full-Jones calibration starts at a time resolution of 13.3 minutes and this interval is gradually decreased to the point where the solutions become noisy and worsen the image quality. Brightness maps are produced with a FoV of 1.4 degrees radius and a resolution of about 12~\arcsec\ using a robustness parameter $-0.8$.

\begin{figure*}
	\centering
	\includegraphics[width=.334\linewidth , trim={.1cm .cm 1.6cm 1.cm},clip]{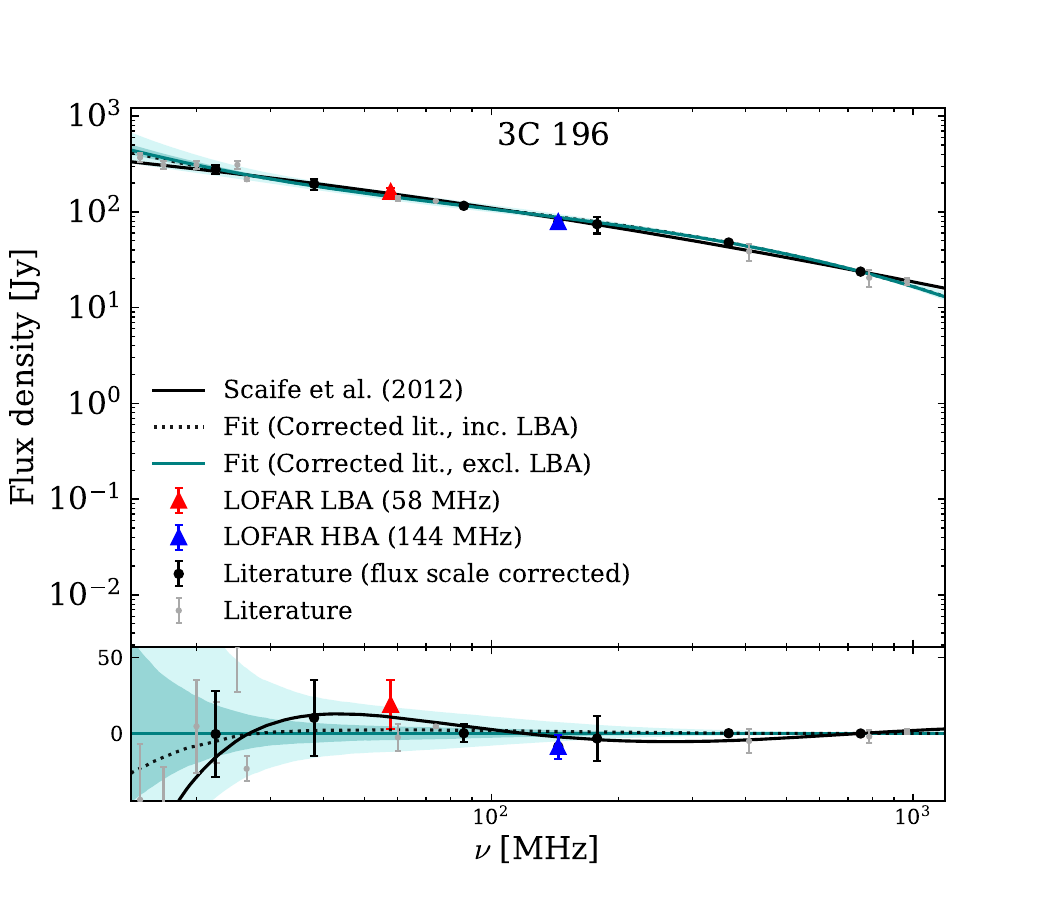}
	\includegraphics[width=.32\linewidth, trim={.7cm 0.cm 1.6cm 1.cm},clip]{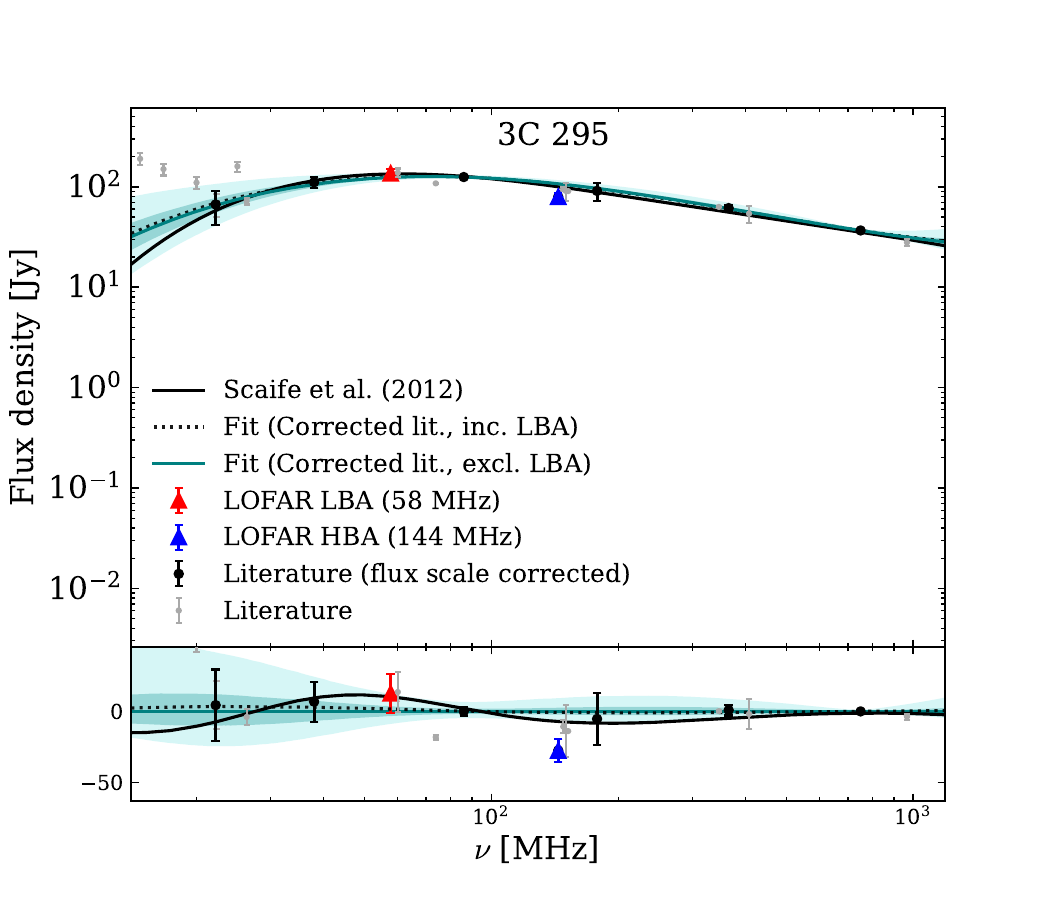}
	\includegraphics[width=.31\linewidth, trim={1.1cm 0.cm 1.6cm 1.cm},clip]{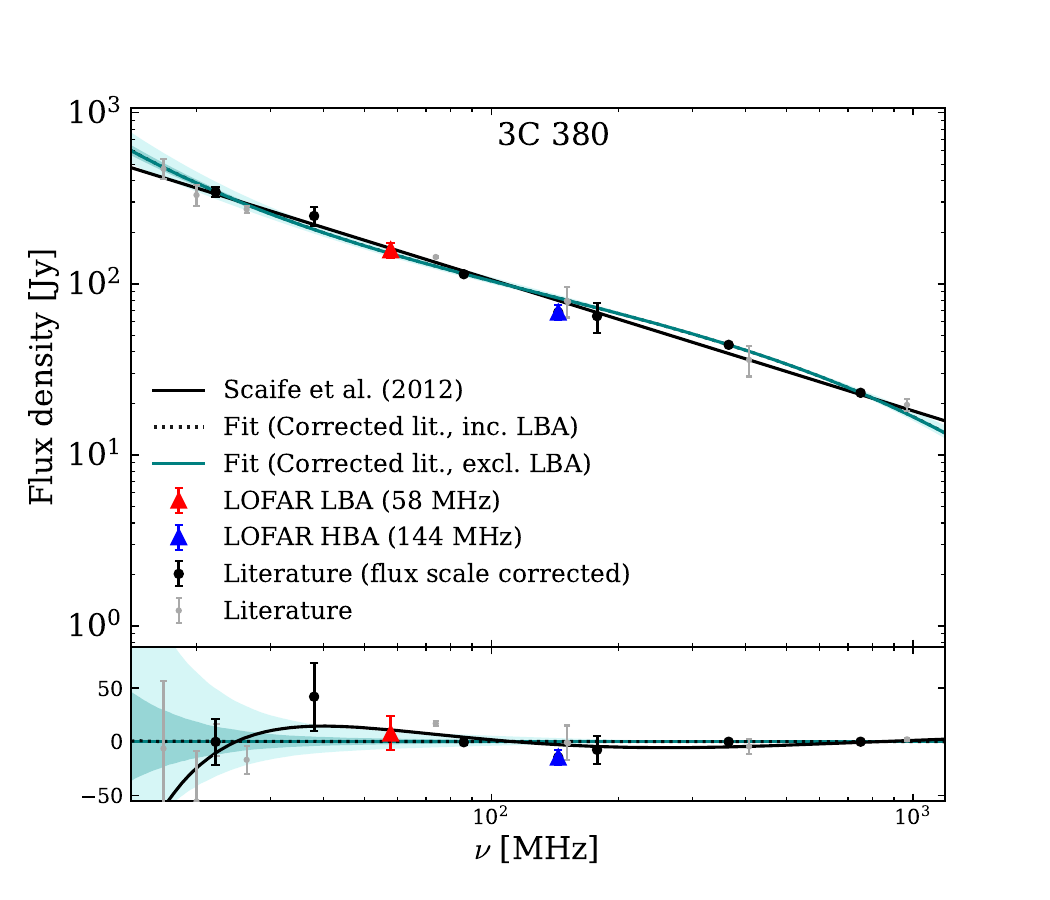}
	\caption{\emph{Top panels:} Synchrotron spectrum of the calibrators. The red triangle shows indicates our LOFAR LBA measurement with an added illustrative uncertainty of 10\% and the circle points the entries queried from NED. Only the black dots were scaled to the SH flux scale, gray points are not. The blue line shows the MCMC fitted spectrum through the black points, the black dotted line shows the fitted spectrum using all scaled literature values including LBA. The blue line shows our fit with 1$\sigma$ and $2\sigma$ model uncertainties. The fit is performed on the literature values with flux scale corrected fluxes and fluxes} at frequencies $> 1$~GHz. \emph{Bottom panels:} residual of the three fits in linear space.
	\label{fig:calibrator_sed}
\end{figure*}
 
\section{Results}
\label{sec:result}

\subsection{Radio maps}


One of the aims of this work is to produce LOFAR LBA radio maps of the entire 3C catalog. The calibration strategy to arrive at these images was discussed in the previous section. This section will describe the maps in more detail and highlight some special sources that required additional steps. The collection of all maps can be found in Figure~\ref{fig:all-maps} in Appendix~\ref{app:images}. The maps' color bars in the appendix are scaled based on the image properties and thus different for every image. The scale indications use the redshift taken from the NASA/IPAC Extragalactic Database (NED). The cutouts are $10\arcmin\times 10\arcmin$ by default, $40\arcmin\times 40\arcmin$ for extended sources and $50\arcmin\times 50\arcmin$ for very extended sources. The $\sigma_{\text{rms}}$ varies roughly between 6~and~16~mJy~beam$^{-1}$ with a dozen outliers above 20~mJy~beam$^{-1}$. See the accompanying catalog for all values and Figure~\ref{fig:image_statistics} for the statistics. The dynamic range is roughly between 20 and 800, depending on the source.

During imaging, we distinguish a few sources as extended or very extended. A selection of those is highlighted in Figure~\ref{fig:faint_sources}. For extended targets, multi-scale cleaning is performed for scales up to 108\arcsec\ and robustness parameter $-0.7$. For very extended targets, multi-scale cleaning is performed for scales up to 240\arcsec\ and robustness parameter $-0.6$. 
Eight sources have extended emission that is too faint to be detected at the default resolution and are therefore cleaned with a Gaussian taper. These sources and their taper are 3C~236 at 25\arcsec\ and  3C~274, 3C~31, 3C~449, 3C~465, NGC~6109 and NGC~6251 at 20\arcsec. However, the high rms noise of our observations, due to the small bandwidth, inevitably results in the inability to detect diffuse, low surface brightness emission at $\sim12\arcsec$ resolution. For many of the extended sources $> 2.5\arcmin$, we did not detect all diffuse emission with a signal to noise above $3\sigma_{\text{rms}}$. Of these sources, 3C~31, 3C~264, 3C~449 and NGC~7385 exhibit diffuse emission detected in (unreleased) LoTSS \citep{Shimwell2019} data that remains undetected in our survey unless we go to arc-minute resolutions. All of the listed sources are extended FR-I type galaxies of which the most diffuse, low brightness parts at the end of the lobes are invisible. 

The final radio maps were not corrected for astrometric offsets. From the spectral analysis (performed in an upcoming work), where LBA maps were aligned with other surveys, the astrometric error is on the order of $2\arcsec-5\arcsec$.



\subsubsection{Challenging sources}

Catalog sources that are spatially close together are particularly difficult to calibrate, as the neighboring 3C source is essentially a bright off-axis source contaminating the target. In a few cases, it was necessary to subtract these off-axis sources using a peeling procedure \citep{Smirnov2011} before it became possible to calibrate them properly. This additional step was performed for 3C~272.1, where 3C~274 (M87) was peeled, and 3C~381, where 3C~380 was peeled. We checked the flux of the sources with and without peeling of the neighboring source and there was no significant flux suppression due to the peeling.
The pairs A1552 \& M87 and 3C~84 \& 3C~83.1B are so close together and bright that peeling would introduce artifacts around the companion. For these two pairs the calibration and imaging was done jointly. In order to image A1552, which is only 41.5\arcmin\ away from M87, we phase shifted the data to M87 and calibrated the two sources jointly. The figure of A1552 in Appendix~\ref{app:images} still shows some contamination from M87 on the left side of the image, but it is a great improvement over the default pipeline we utilized.  

For self-calibration are 3C~84 (NGC~1275) and 3C~83.1B (NGC~1265), the two most luminous radio galaxies in the Perseus cluster, separated by only 27\arcmin. The challenging component of the cluster is 3C~84, which has a tens of Jy bright, compact core embedded in diffuse emission. Some of the diffuse emission has been observed by \citet{Gendron-Marsolais2020} with the VLA at 230–470~MHz, but recent observations with the LOFAR HBA by \citet{Weeren2024} suggest that the radio galaxy is at the centre of a giant radio halo. The superposition of bright compact emission and diffuse faint emission requires additional data reduction. Other broadband observations of the cluster were recently done by Groeneveld et al. (in prep.) and show more of the diffuse emission in the halo. Our observations primarily detect the bright core.

\begin{figure*}[th]
	\centering
	\includegraphics[width=1.\columnwidth, trim={.5cm 0.cm 1.6cm 1.cm},clip]{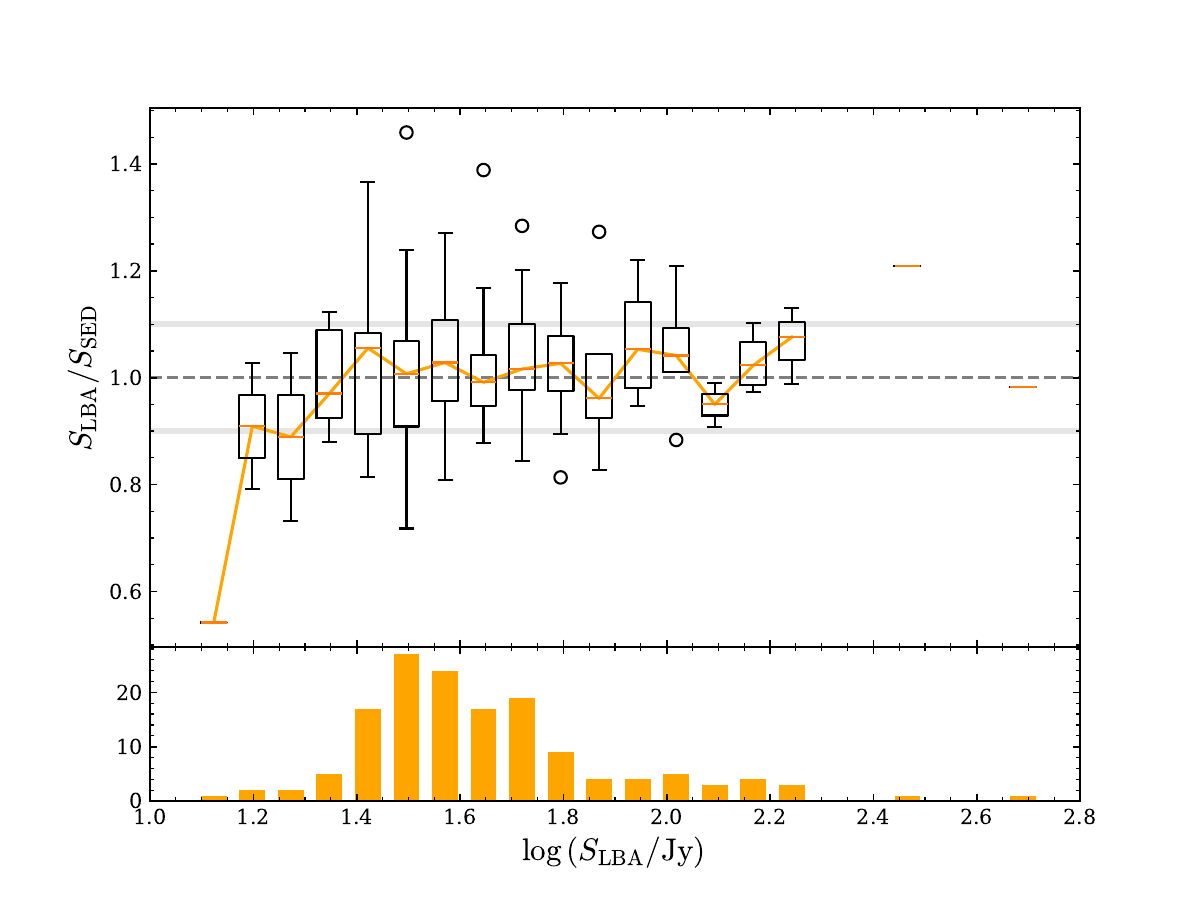 }
	\includegraphics[width=1.\columnwidth, trim={.5cm 0.cm 1.6cm 1.cm},clip]{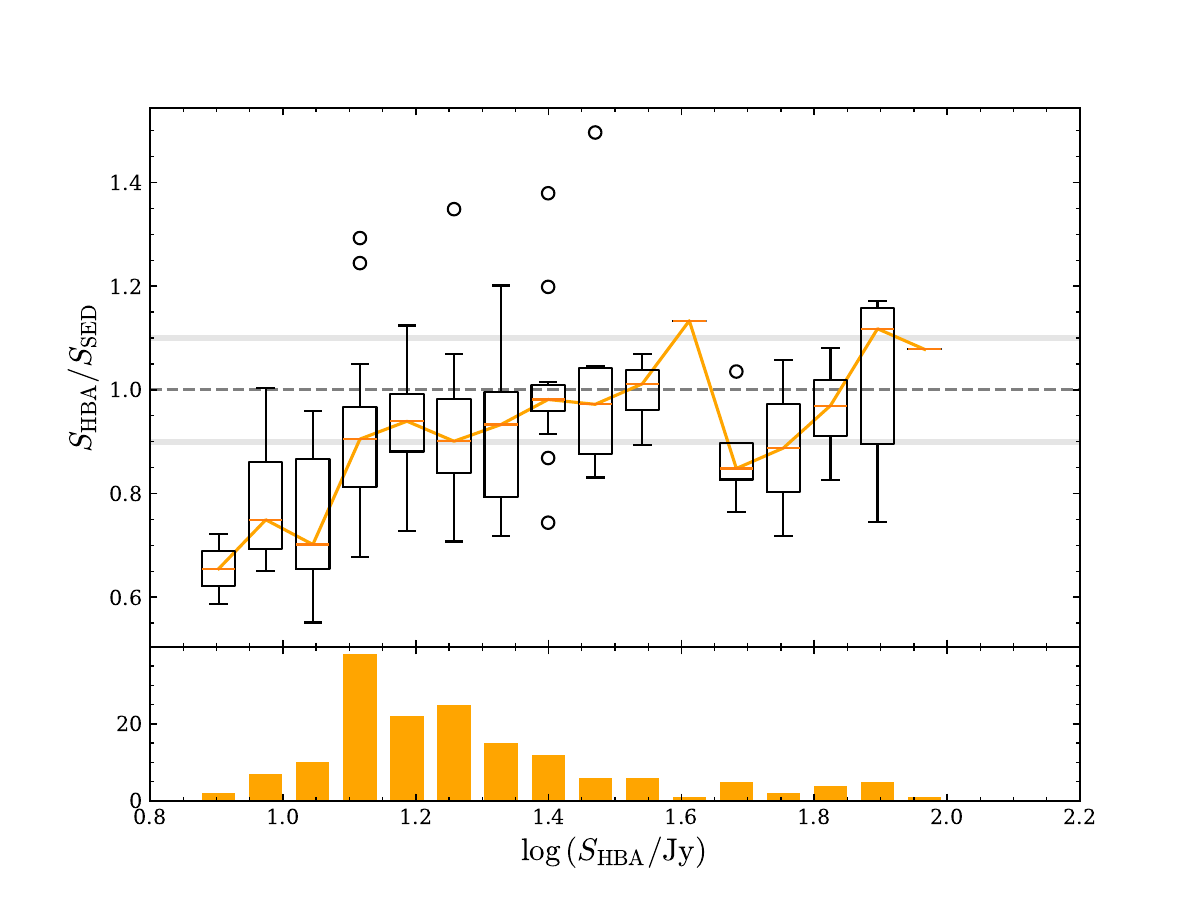 }
	\caption{\emph{Left:} Statistics of flux density accuracy of sources in the 3C sample. The box plot in the upper panel shows the distribution with deviations from the expectation according to the SEDs over a range of fluxes. The `box' in the box plot denotes that the middle 50\% of fluxes fall within this range, while the lower and upper 25\% of fluxes fall within the error bars extending from it. The circles outside the boxes are outliers. The orange lines denote the weighted average of all the fluxes in that flux bin. \emph{Right:} Same as the left panel but for the measurements from LoTSS. Note how the average flux seems to be systematically underestimated by about 5\% where the statistics are best. }
	\label{fig:flux_accuracy}
\end{figure*}

\subsection{LOFAR LBA flux density scale}
\label{sec:flux_scale}
The source flux densities were independently estimated, adopting the flux density scale of \citet{Scaife2012} for the calibrators. For any other catalog of extragalactic sources, it would be challenging to assess the accuracy and precision of flux measurements because of the lack of reliable measurements for a large range of frequencies, especially low frequencies. However, this sample contains some of the most studied and brightest radio sources with measurements dating back several decades and extending down to decameter frequencies \citep{Braude1969}, making this an excellent sample to test the LBA flux scale. To do this, we compare a spectral energy distribution (SED) composed of literature measurements with our measurement at 58~MHz. The deviation from the `expected' value from a polynomial fit is a measure of the accuracy (offset from the true flux) and precision (standard deviation from the true flux) of the LBA flux density measurement. A similar analysis has been performed by \citet{Gasperin2021} who concluded that an estimate of the systematic calibration error of 10\% was sufficient, although this analysis used mainly two higher frequency measurements and a linear extrapolation method. \citet{Gasperin2023} repeated the analysis by performing quadratic fitting, although mostly using extrapolation from two higher frequency bands. In the present work, we characterize the flux density scale more rigorously by taking a larger sample of measurements to construct the SED to which we fit, as well as requiring multiple observations below 58~MHz to properly constrain both sides of the measurement. 
 
 Where available, we collected 20\arcsec\ resolution cutouts from LoTSS at 144~MHz from yet unreleased DR3 data. The 144~MHz data points help to constrain the SED fit as well as determine the low frequency spectral index for nearly the entire catalog. The few 3C sources not covered by LoTSS are either close to M87 (3C~274, which is excluded from LoTSS due to the difficulty of self-calibrating close to this kJy source)\footnote{M87 has been observed separately in the LOFAR HBA Virgo Cluster Survey \citep{Edler2023}.} or have a separation from the Galactic plane $\lesssim 23\degr$.
 
 Flux densities were measured using manually drawn regions for every source roughly following the 20\arcsec\ HBA 2$\sigma$ outline of the source. This was done so that the same regions could be used for both HBA and LBA to calculate the total flux. All flux measurements, LoTSS fluxes as well as $58-144$~MHz integrated spectral indices are published in an accompanying catalog and are, for illustrative purposes, listed in the Table in Appendix~\ref{app:all_sources}. Uncertainties for all spectral indices are 0.155 assuming the uncertainties for both LBA and HBA are 10\% of the measured flux.  
 
Integrated flux density measurements and uncertainties were collected for all sources using the NASA Extragalactic Database. These can be measurements appearing in a catalog or in a single source study at frequencies ranging from 20~MHz up to 10~GHz. The measurements from the literature also vary widely in publication date, going back to the 1960s. Historically, there are a handful of different flux density scales on which flux measurements are defined. Depending on the frequency and the scale used, the discrepancy can be up to 20\%. Widely used flux scales include the Roger, Bridle \& Costain scale \citep{Roger1973}, the \citet{Baars1977} scale and the Kellerman, Pauliny-Toth \& Williams \citep{Kellerman1969} scale. Other works define their own flux scale and mention how it relates to one of the more commonly used scales.

To prevent going through the cumbersome work of checking the flux scale of every entry for every source, we convert only a few literature values from their scale to the SH scale. For this, we use publications containing a large portion of the sources in our sample, greatly reducing the amount of manual work. Table~2 in \citet{Scaife2012} summarizes the respective works, their conversions to the SH scale and frequency of the measurement. This work adds some additional references with frequent occurrences in the NED queries, which are listed in Tab.~\ref{tab:corrections}.

  \begin{table}
	\caption{\label{tab:corrections} Scaling factors used for broadband SED}
	\centering
	\begin{tabular}{llc}
		\hline\hline
		Freq. & Reference & factor\\
		\hline
10~MHz & \citet{1968AJ.....73..717B} & 1.20 \\
 & \citet{Roger1973} & - \\
22.25~MHz & \citet{Roger1973} & - \\
 & \citet{1969AJ.....74..366R} & 1.15 \\
 & \citet{1980MNRAS.190..903L} & - \\
38~MHz & \citet{Kellerman1969} & 1.18 \\
 & \citet{1990MNRAS.243..637R,1990MNRAS.244..233R} & - \\
81.5~MHz & \citet{1971MNRAS.154P..19S} & 0.90 \\
86~MHz & \citet{1980MNRAS.190..903L} & 0.94 \\
 & \citet{1969SvA....12..567A} & 0.94 \\
151~MHz & \citet{1985MNRAS.217..717B} & - \\
 & \citet{1993MNRAS.263...25H} & - \\
178~MHz & \citet{Kellerman1969} & 1.09 \\
325~MHz & \citet{1997AAS..124..259R} & 0.90 \\
365~MHz & \citet{1981AAS...45..367K} & 0.94 \\
 & \citet{1996AJ....111.1945D} & 0.98 \\
408~MHz & \citet{1970AAS....1..281C,1972AAS....7....1C,1973AAS...11..291C} & - \\
 & \citet{1974AAS...18..147F} & - \\
 750~MHz & \citet{Kellerman1969} & - \\
 & \citet{1980MNRAS.190..903L} & - \\
960~MHz & \citet{1997BSAO...44...50K} & 0.96 \\
		\hline
	\end{tabular}
\end{table}

Below 1~GHz, only flux densities rescaled to SH are used to fit a polynomial synchrotron spectrum of the form 
\begin{align}
    &&S = \prod_{i=0}^410^{A_i\log^i\left(\frac{\nu}{150\text{ MHz}}\right)}
\end{align}
We fitted this spectrum with a Markov Chain Monte Carlo implementation up to the $A_4$ term, or up to term $N$ where we have $N-1$ data points available. The fitted model at 58~MHz is then used as a proxy of the true flux. The LOFAR LBA flux densities are plotted against the literature SED together with the fitted synchrotron spectrum
\footnote{We were not able to make an SED using this method for A~1552 as it lacks flux density entries in NED.}. The spectra for the calibrators are highlighted in Figure~\ref{fig:calibrator_sed}, where the black solid line indicates the fit from SH. The figure shows that our measured fluxes are in perfect agreement with SH. For the calibrators, the flux is measured to be $156.7$~Jy for 3C~196, $136.5$~Jy for 3C~295 and $163.05$~Jy for 3C~380. The collection of all SEDs can be found in Appendix~\ref{app:seds}.

Comparing the measured fluxes to the expected flux based on the spectrum fit, we can derive an estimate of the flux density accuracy and precision of the LOFAR LBA for high-SNR sources, visualized in Figure~\ref{fig:flux_accuracy}. The figure shows an overall accurate flux scale with some clear outliers and larger scatter around unity at low brightness and high noise. Since synchrotron spectra tend to turn over due to self-absorption at ultra-low frequencies, we only included sources with literature measurements below 58~MHz, eliminating the possibility that our measurement has a poor fit due to possible self-absorption that was not previously observed. Thus, sources like 3C~225B or 4C~14.11 are excluded from the statistics of Figure~\ref{fig:flux_accuracy}. The left panel of Figure~\ref{fig:flux_accuracy} shows that the flux measurement becomes excessively unreliable for the few sources with fluxes below 25~Jy. The majority of these few sources (specifically 3C~287, 3C~343.1, 3C~345 and 3C~454.3) driving down $S_{\text{LBA}}/S_{\text{SED}}$ are compact quasars with potentially high variability \citep[e.g.][]{2001ASPC..224..265F,2005ApJ...618..108B,2024ApJ...977..166T}. This idea is confirmed by their inconsistent low-frequency SEDs in Figure~\ref{app:seds}. 
Excluding these sources from the statistic yields a flux accuracy of 
\begin{align*}
&&S_{\text{LBA}}/S_{\text{SED}}	= 1.00 \pm 0.09
\end{align*}
This shows that the LOFAR LBA flux scale is accurate with roughly 9\% calibration error. We will round this up to 10\% and adopt that percentage as our uncertainty estimate on the measured flux densities for the remainder of this work. The right panel of Figure~\ref{fig:flux_accuracy} shows the same as the left panel but for LoTSS (HBA), showing the same inaccuracy at the low flux end. More importantly, it shows a slight systematic underestimation of flux as to what is expected from the literature. For the HBA fluxes, the ratio is $0.95 \pm 0.11$, a slight systematic underestimation of about 5\% (We discuss this result in more detail in Section~\ref{sec:discussion}). The results presented here regarding the LBA flux scale show it to be accurate in the field centre. This is in contrast with what found in \citet{Gasperin2023}, who noticed a systematic underestimation of around 10\% using the extrapolation method, although their work used sources out to the edges of the field of view. Our result is in good agreement with their less significant estimate that includes 8C measurements at 38~MHz. This suggests that a significant portion of sources start turning over due to self-absorption at 57~MHz, giving rise to the underestimation from extrapolation. 

\begin{figure}[th]
	\centering
	\includegraphics[width=1.\columnwidth, trim={.5cm .cm 1.5cm 1.5cm},clip]{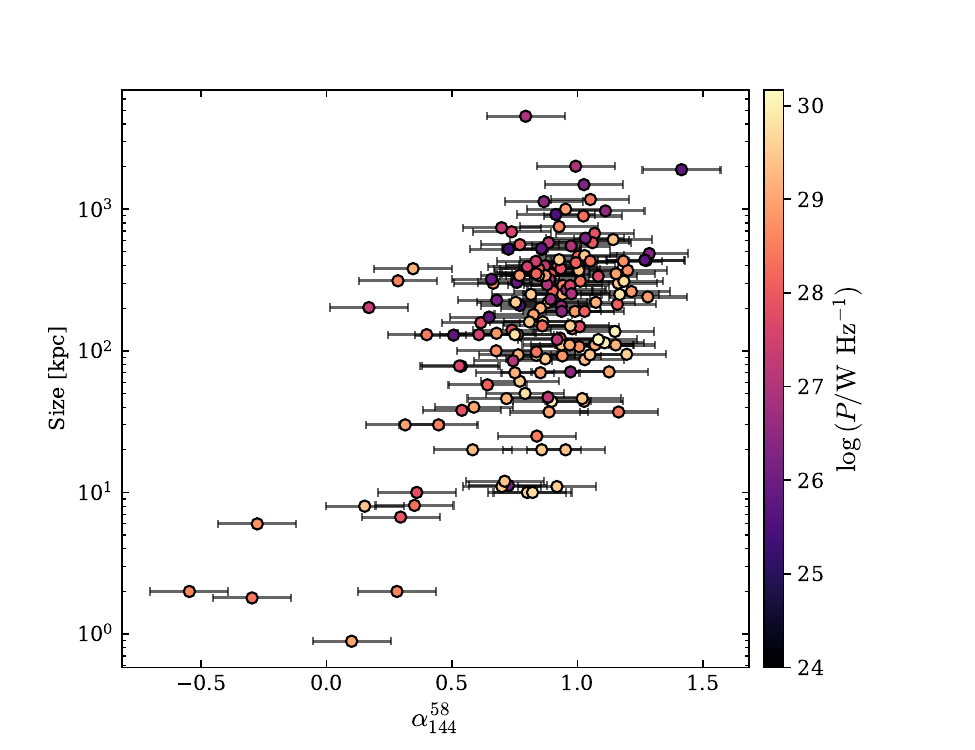 } 
	\caption{Spectral index from LBA and HBA fluxes. The radio power is calculated from the LBA fluxes using the redshift mentioned in the catalog.}
	\label{fig:flux_distribution}
\end{figure}

Figure~\ref{fig:flux_distribution} shows the distribution of the low-frequency spectral index $\alpha^{58}_{144}$ with respect to the physical size\footnote{Sizes were taken from \url{https://3crr.extragalactic.info} and are measured from various high-resolution sources.} and radio power at 58~MHz. The uncertainties were propagated using 10\% uncertainty for both LBA and HBA. Low-frequency spectral indices are distributed over a large range up to 1.5 and a median spectral index of 0.88. Three sources (3C~147, 3C~343.1 and 3C~343) have an LBA flux lower than their respective HBA flux. The smallest sources indicate that there is a proportionality between the source's physical size and the low-frequency spectral index.This is possibly related to the relation between peaked spectrum sources and source size discussed in Section~\ref{sec:peak_spectrum_sources}.

\begin{figure}[th]
	\centering
	\includegraphics[width=1.\columnwidth, trim={0.cm 0.cm 0.cm 0.cm},clip]{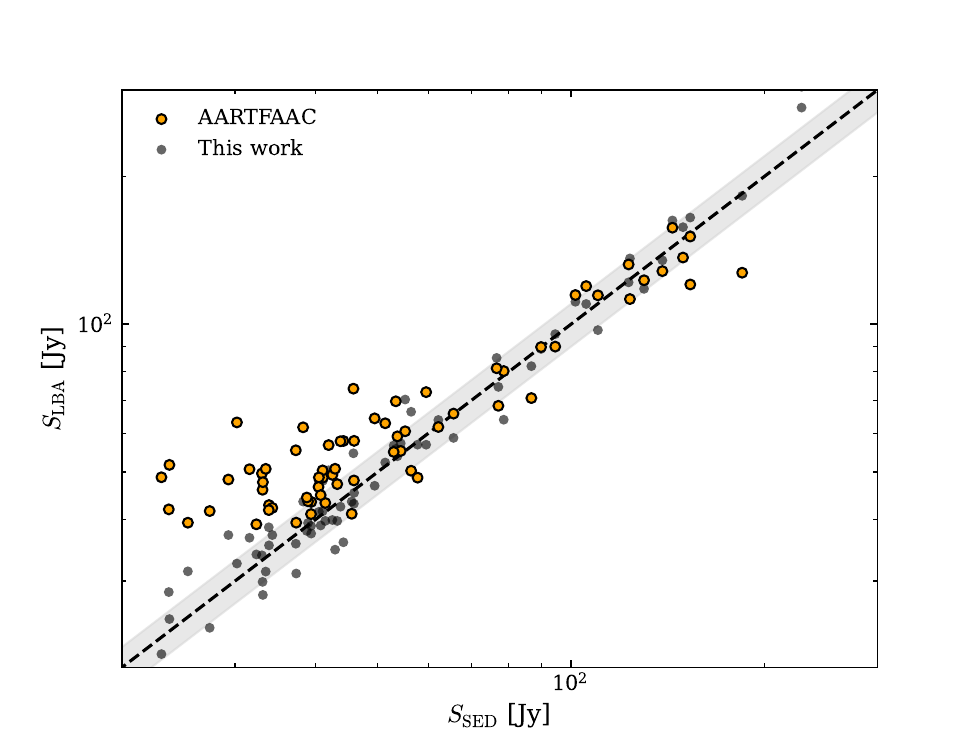 } 
	\caption{LBA flux for 70 sources as measured by AARTFAAC \citep{Kuiack2019} in orange and as measured in this work in grey. Fluxes are compared to the expected flux from the SEDs. The dashed line denotes $S_{\text{LBA}}=S_{\text{SED}}$. The shaded region denotes $\pm10\%$ from this expectation.  
	} 
	\label{fig:aartfaac_comp}
\end{figure}

We compared the flux density values from this work with the ones available available from the Amsterdam ASTRON Radio Transient Facility And Analysis Centre (AARTFAAC, \citealt{Kuiack2019}). Their catalog of sources at 60~MHz, observed with the 12 innermost core stations in an all-sky map, reported flux densities for 70 sources that are in the 3CRR catalog. We compared their measurements to the ones in this work in reference to the SED fit. Figure~\ref{fig:aartfaac_comp} shows the AARTFAAC fluxes deviate significantly below 50~Jy, while above that the flux scale seems consistent with what we find here.  This indicates the flux densities in AARTFAAC are not accurate below 50~Jy. 

Besides AARTFAAC, LBA flux density measurements exist for only a few non-calibrator 3C sources. For 3C~225B, our measurement of $55.6\pm5.5$~Jy is higher though in accordance with the $45\pm20$~Jy obtained by \citet{Groeneveld2022}, considering the large uncertainty on their measurement. For 3C~223, \citet{2016MNRAS.458.4443H} obtained 37.4~Jy at 51.6~MHz, again consistent with our measurement of $41\pm4.1$~Jy. For 3C~31 and 3C34, \citet{Heesen2018} reports of $47.7\pm1.5$~Jy, respectively $44.3\pm1.3$~Jy, where they assumed a fairly small calibration error of 5\%. It is therefore just in agreement with the $43.7\pm4.4$~Jy and $42.8\pm4.3$ presented in this work. 3C~274 (VirA or M87) has been observed by \citet{Gasperin2020b} who reported a total flux density of 2635~Jy, again in agreement with our $2443\pm244$~Jy. 

\subsection{Peaked spectrum sources}
\label{sec:peak_spectrum_sources}
The collection of SEDs extending to ultra low-frequencies enables us to study peaked spectrum sources and synchrotron self-absorption. At sufficiently low frequencies, synchrotron sources eventually become opaque and turn over. As we will describe later in this section, this spectral characteristic can be used to estimate magnetic field strengths within compact sources. Collective analysis of the SED fits that include the LBA measurement reveals that there are 55 sources that are likely to have a peaked spectrum with peak between $10-500$~MHz. This is 32\% of the catalog. This is determined based on the SED fit that includes the LBA measurement. If this fit has a peak between $10-500$~MHz we consider it peaked. We determine the peak frequency $\nu_p$ from the SED fit that includes the LBA measurement. If we were to exclude the LBA data point, the number of peaked-spectrum sources from the fits is 45. Of these, 3C~236 appears to be peaked at $\nu\sim 100$~MHz but this is ruled out by the LBA measurement. The best fits, including LBA measurements, thus suggest a peaked spectrum for 11 new sources. Without the LOFAR LBA data it would not have been possible to reliably determine that the spectrum was peaked. 

\begin{figure}[th]
	\centering
	\includegraphics[width=1.\columnwidth, trim={.8cm 0.25cm 1.8cm 1.2cm},clip]{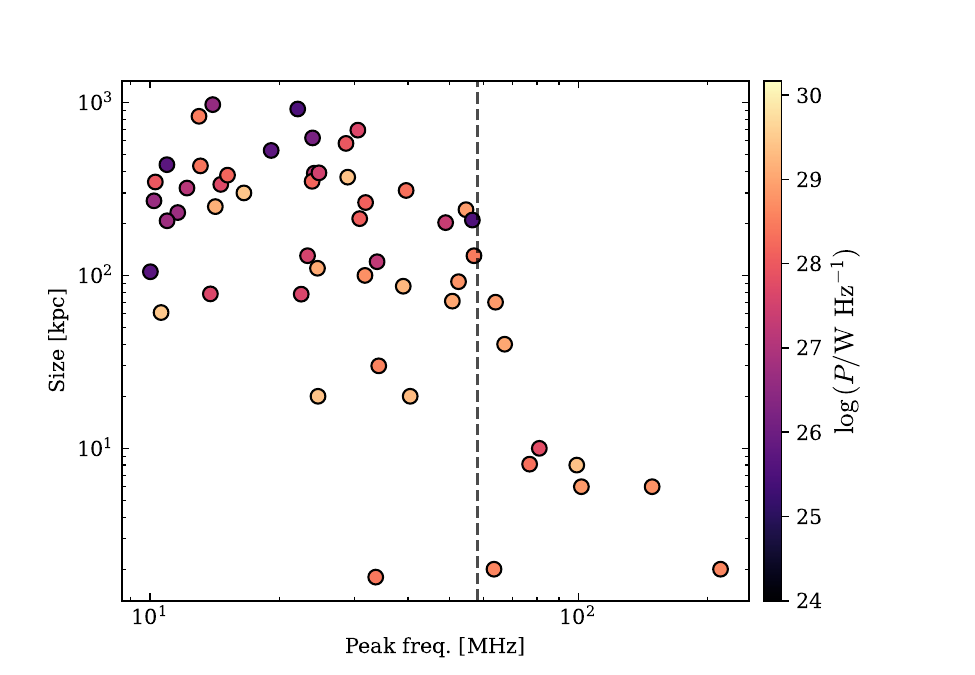 } 
	\caption{Peak frequency as a function of the physical size of AGN. The colormap depicts the power $P$ of the source. The dashed line depicts the LBA frequency (58~MHz). 
	} 
	\label{fig:peak_frequency}
\end{figure}

Despite the low numbers of likely peaked sources for this sub-sample, Figure~\ref{fig:peak_frequency} shows a clear relation between the peak frequency $\nu_p$ at which the source becomes optically thick, the physical size and the total power $P$ of the source. Essentially all sources where $\nu_p>58\text{ MHz}$ are smaller than 10~kpc.

There also seems to be a relation regarding the power of the radio source. Low-power radio galaxies tend to be larger and if they are peaked with $\nu_p>10$~MHz, $\nu_p$ is consistently lower compared to the smaller, luminous sources. For the more compact and luminous sources, it is possible to use the fact that they are peaked and thus self-absorbing to estimate the magnetic field strength of the source. At sufficiently low frequencies, electrons can effectively absorb synchrotron emission emitted within the same medium and the source becomes optically thick. The fact that this happens most clearly in compact luminous sources is to be expected, as the combination of high emitting power and compact structure constitutes a higher density of high-energy electrons which will have a higher chance of absorbing radiation. 

\section{Discussion}
\label{sec:discussion}

This work has presented the first the 3CRR catalog imaged at ultra-low frequencies. Despite the novelty, it is worthwhile to reiterate some clear limitations of the quality of the data presented. Due to the observing strategy, which was set up to observe 30 directions simultaneously, there is only 3~MHz of bandwidth allocated for each target. 
Despite the rather long (given the brightness of the sources) observation time of 8~hours, this caused a suboptimal \emph{uv}-coverage and resulted in high noise levels for some sources. For example, 3C~314.1 (39.9~mJy~beam$^{-1}$) and 3C~315 (51.6~mJy~beam$^{-1}$) have especially low image quality. The average image noise lies around 10~mJy~beam$^{-1}$ as can be seen in Figure~\ref{fig:image_statistics}. The poor image quality is also apparent in the form of a particularly elongated beam in roughly the east-west direction. The sources where this is especially apparent suffered from a totally flagged LOFAR station RS310, an important station for east-west baseline coverage. Another source with missing emission is the core of the Perseus cluster (3C~84). \citet{Weeren2024} observed a large extended halo around the bright core. Due to the sparse amount of data, we were unable to properly clean the image around the 270~Jy bright core, leaving the surrounding diffuse emission lost in calibration artifacts. 

A second effect caused by the narrow-band observations is the relatively low sensitivity in the images that, combined with limitations in the dynamic range, resulted in undetected diffuse emission for some sources. NGC~7385, NGC~6109, 3C~449 and 3C~83.1B have previously observed diffuse emission at 144~MHz in LoTSS that is completely undetected in our data, while for a larger subset of extended sources the emission is detected but with a low signal to noise ratio. It seems that the missing diffuse emission does not have a significant influence on the final measured flux (well within the 10\% calibration error), but it does have an effect on the usefulness of such data in spectral analyses, which requires resolved maps and good fidelity. This limitation will become clear in Paper~II, where we use the most resolved maps for spectral analysis. Some of these extended sources will need new, deeper broadband observations to achieve a more competitive image quality.  

\subsection{Low-frequency SEDs}
This work also provides the best characterization of the LBA flux density scale in the high signal to noise regime, thanks to the great abundance of literature measurements of the 3CRR fluxes at a wide range of frequencies between $0.01-10$~GHz. In particular, the many literature values below 58~MHz allowed us to constrain the spectrum for our measurements. Despite the overall reassuring outcome that the flux scale is accurate with about 10\% error, there are a handful of sources that seriously deviate from the expected SED. 3C~314.1 has the largest overestimation of about 30\%, and the same source has a map with a high rms noise. 3C~231 (M82) has the largest underestimation of 35\%, although in that case, the model does not seem to fit the SED well. The fit relies heavily on the 365~MHz Texas Interferometer Survey \citep{1996AJ....111.1945D} data point. This survey is known to underestimate flux densities for sources larger than 30\arcsec. This data point was therefore not considered for this source and any other source larger than 30\arcsec. Fitting again to the remaining data results in an underestimation of 12\%. 

A possible hidden bias in these SEDs is the error on the individual entries. We did not check the reliability of the errors and whether they take into account just statistical uncertainties or also systematic ones. If a reported uncertainty in NED is lower than the actual total uncertainty, the fit will be statistically more dependent on that entry than others that have a higher uncertainty. Since for this analysis we are more interested in the statistical accuracy of the flux scale, we do not deem in necessary to take into account these small biases.

Repeating the same analysis on the flux scale with 20\arcsec\ LoTSS DR3 images showed a slight systematic underestimation of the flux in DR3 of on average 5\%, fairly small compared to the average 10\% calibration error. Since many of the sources are part of the upcoming LoTSS DR3, the systematic deviation on the flux is not alarming, as the study of the flux density scale of the survey is still ongoing. 
As of this moment, the flux scales of the survey are still under investigation. The HBA images will be used in Paper~II to construct low-frequency spectral index maps. 

\subsection{Synchrotron self-absorption}
\label{sec:ssa}
Synchrotron self-absorption is a possible physical explanation for the low-frequency turnover we observe in some of the sources. Assuming this is the mechanism, we are able to make an estimate of the magnetic field in the source \citep{1969ApJ...155L..71K}. The brightness $I_{\nu}$ at some frequency $\nu$ is related to the brightness temperature $T_b$ in the Raleigh-Jeans limit
\begin{align}
	&&T_b=\frac{c^2}{2k\nu^2}I_{\nu},
\end{align}
At the same time, the energy of emitting electrons $\gamma m_e c^2$ is expressed through the effective temperature $T_e$ through $3T_ek = \gamma m_e c^2$, where the factor 3 is valid for ultra-relativistic plasma. Electrons emit most synchrotron energy at a certain frequency $\nu\sim{\gamma^2eB}/{2\pi m_ec}$ and we can use this to obtain an equation that relates the effective temperature of electrons emitting mostly at $\nu$ in terms of the magnetic field $B$ and $\nu$ by substituting $\gamma$, yielding
\begin{align}
&&T_e\approx\left(\frac{2\pi m_ec\nu}{eB}\right)^{1/2}\frac{m_ec^2}{3k}.
\end{align}
Since the energy of the emitted radiation cannot exceed the electron energy, we require $T_b \leq T_e$. Eventually, at low frequencies, $T_b$ approaches $T_e$ and this is the point where the source becomes optically thick. Setting $T_b\approx T_e$ yields an equation for the optically thick part of the spectrum that is only dependent on the magnetic field such that $I_{\nu}\propto\nu^{5/2}B^{-1/2}$. When taking a single brightness temperature for the entire source, we assume it to be homogeneous. For more realistic sources, the final self-absorbed spectrum will have a broader peak.  

\citet{Pacholczyk1970} more precisely calculated the $I_{\nu}\propto\nu^{5/2}B^{-1/2}$ relation in CGS units replacing $I_{\nu}$ by the total flux over the source solid angle $\Omega_s$
\begin{align}
\label{eq:magnetic_field_analytical}
	&& \frac{S_{\nu}}{\Omega_s}=c^{-1}_{14}(B\sin{\theta})^{-1/2}\nu^{5/2}\left(\frac{\nu}{\nu_p}\right)^{-\beta}\left[1-\exp\left(-\left(\nu/\nu_p\right)^{-(\gamma+4)/2}\right)\right],
\end{align}
which reduces to $S_{\nu}\propto\nu^{5/2-\beta}$ for $\nu\ll\nu_p$ and to the classic power-law $S_{\nu}\propto\nu^{-\alpha}$ for $\nu\gg\nu_p$ with $\alpha = (\gamma-1 + 2\beta)/2$. Here, $\beta=0$ for homogeneous sources while for inhomogeneous sources it takes a value between 0 and 2.5 and flattens the inverse slope \citep[see e.g.][]{2016era..book.....C,2024A&A...682L...3P}. Introducing $\beta$ as a free parameter to determine the inverse slope is not always possible due to insufficient coverage at those frequencies. In practice, increasing $\beta$ strongly increases the magnetic field. Therefore, taking $\beta=0$ gives a minimum magnetic field strength estimate.  

The pseudo-constant $c_{14}$ is tabulated in \citet{Pacholczyk1970} and has a value of $4.65\times10^{40}$ for a typical spectral index $\alpha=0.75$. For a source of solid angle $\Omega_s$, the synchrotron self-absorption (SSA) magnetic field (in Gauss) can be estimated to be 
\begin{align}
	\label{eq:magnetic_field}
	&&B_{ssa}\sim B\sin\theta \approx 2.555\times10^{-7} \left(\frac{\Omega_s}{\text{arcsec}^2}\right)^2 \left(\frac{S_{\nu}}{\text{Jy}}\right)^{-2}\\&&\times\left\{ 
	\begin{aligned} 
  		&\left(\frac{\nu}{\text{MHz}}\right)^5\left(\nu/\nu_p\right)^{-2\beta},&\nu\ll\nu_p \nonumber \\
  		&\left(\frac{\nu_p}{\text{MHz}}\right)^5\left(\nu/\nu_p\right)^{-2\alpha},&\nu\gg\nu_p
	\end{aligned} \right.
\end{align} 
The main practical issue for such magnetic field estimation is that the size of the synchrotron source is required to be known, which can be difficult since the most apparent self-absorption happens for the smallest sources. Fortunately, very high-resolution images are available for 3CRR sources. We can apply this method to the sources in the sample with the highest $\nu_p$ and obtain a very rough estimate of their magnetic fields by approximating their solid angle using the maximum angular extent and a length to width ratio of 1:4. Eq.~\ref{eq:magnetic_field_analytical} was fitted to the SED data (including LBA measurement) to find new $\nu_p$ and $\alpha$ that best fitted this function. Although the LBA measurement itself cannot be used to evaluate Eq.~\ref{eq:magnetic_field} (since 58~MHz~$\sim\nu_p$), it is important in constraining the peak frequency parameter.  Magnetic field estimates are summarized in Table~\ref{tab:magnetic_field}.

We limited the analysis to sources with a turnover at $\nu_p> 58$~MHz to better constrain the absorbing part of the spectrum. $\beta$ was only allowed as a free parameter for SEDs where there is a measurement below 40~MHz. For other sources $\beta=0$ as the inverse slope cannot be determined with the available data. The $\nu_p$ values in Table~\ref{tab:magnetic_field} are the resulting values from fitting the data to Eq.~\ref{eq:magnetic_field}.

\begin{table}
\caption{\label{tab:magnetic_field} Magnetic field estimations $B_{ssa}$ assuming synchrotron self-absorption. The table also summarizes other parameters for Eq.~\ref{eq:magnetic_field}. LAS: Largest angular size.} 
\centering
\begin{tabular}{lccccc}
	\hline\hline
	Object & $B_{ssa}$ & $\beta$ & $\nu_p$ & $\alpha$ & LAS\\
    & [mG] && [MHz] & &\\
	\hline
3C 138 & 3.4 & 0.39 & $63.53\pm4.42$ & 0.54 & 0.81\arcsec \\
3C 147 & 25.1 & 1.16 & $118.39\pm14.82$ & 0.65 & 0.94\arcsec \\
3C 241 & 34.0 & & $74.48\pm3.70$ & 0.99 & 0.94\arcsec \\
3C 295\tablefootmark{a} & 709 & 1.41 & $60.97\pm14.45$ & 0.72 & 6.85\arcsec \\
3C 343 & 0.6 & & $90.73\pm6.79$ & 0.61 & 0.25\arcsec \\
3C 48\tablefootmark{b} & 0.2 & 1.96 & $93.67\pm10.44$ & 0.75 & 0.39\arcsec \\
3C 49\tablefootmark{b} & 67.0 & & $62.61\pm6.98$ & 0.68 & 1.19\arcsec \\
\hline

\end{tabular}
\tablefoottext{a}{$\sim 1$~G field strength, likely free-free absorption.}
\tablefoottext{b}{Turnover due to low energy cut-off not ruled out.}
\end{table}

Besides self-absorption, two other explanations exist for turnovers. One is due to the lack of electrons emitting at these frequencies due to a low-energy cutoff in the electron population $N(E)\propto E^{-\delta}$, while the other is free-free absorption. In an idealized case, low frequency spectral turnovers due to a low energy cut-off have a slope of $\alpha=-1/3$ \citep{Pacholczyk1970}, while SSA has a slope of $\alpha=-5/2$. All sources except 3C~48, 3C~49 and 3C~241 have significantly steeper inverted slopes between the lowest available data points, therefore weakening the case for the low energy cut-off scenario.
Ruling out free-free absorption is difficult without additional data from X-ray or optical telescopes. It is possible to determine the feasibility of SSA by comparing the magnetic field to measurements from e.g. equipartition. \citet{Taylor1992} argued that 3C~295 was more likely turning over due to free-free absorption than SSA based on X-ray and optical observations as well as the incompatibility of the obtained $B_{ssa}$ with other methods. They claim 3C~295 to have $B_{ssa}=1$~G, which is consistent with our 0.7~G, while the equipartition magnetic field is around 1~mG, which is indeed a discrepancy of three orders of magnitude. 

However, our estimates for e.g. 3C~138 and 3C~147 are in fair agreement with equipartition estimates of respectively 2~mG \citep{Geldzahler1984} and $\sim15$~mG \citep{Zhang1991}. The low frequency turnover of compact sources can be seen as a problem; as they become faint they lose their value as in-field calibrators for ultra low-frequency VLBI data reduction. The exercise that we have carried out here shows the usefulness of such sources and demonstrates that low frequency observations can be an important tool to estimate magnetic field strengths in sources with a spectral turnover. Ultra low-frequency observations can also constrain the physical cause of the observed turnover fairly well. As these spectrum peaks occur mostly at ultra-low frequencies, performing more observations from e.g. LBA could lead to an increased use of this method to estimate magnetic fields, both in resolved and unresolved sources. In particular, decameter observations with LOFAR (between $10-30$~MHz, e.g.~\citealt{Groeneveld2024}) could help in constraining the turnover, considering that the lower one goes in frequency, the more sources will show a turnover.

\section{Conclusions}
\label{sec:conclusion}

We have presented the first survey of the 3CRR catalog \citep{Laing1983} at ultra low-frequencies with data obtained using the LOFAR Low Band Antenna. Only a handful of these sources have already been observed with LOFAR at these frequencies. This survey at 58~MHz provides total intensity maps for all 173 sources at a median resolution of $15\arcsec\times15\arcsec$ and a typical RMS noise of 10~mJy beam$^{-1}$. Besides the radio maps, we have also made independent flux measurements for all sources in the catalog and have validated them against a carefully constructed SED made up from selected literature sources. With this validation, we were able to determine the flux density scale of the LBA to have no systematic offset was observed and the average calibration error on flux measurements is 9\%, which we rounded up to 10\% in the final results. Fluxes in the catalog range between 8.6~Jy for 3C~343.1 up to 2443~Jy for 3C~274 (M87). 

All flux measurements and image statistics are published in an accompanying catalog and a shortened version is summarized in the Table in Appendix~\ref{app:all_sources}. This table also notes the flux density for almost the entire catalog using data from yet to be released LoTSS DR3 data. The systematic offset on the HBA fluxes was non-negligible with about 5\% underestimation of total flux. Finally, the table also notes the spectral index between the two LOFAR frequency bands 58~MHz and 144~MHz which is on average 0.86. 

With the addition of the LBA measurements to the SEDs, we identified 55 sources that show hints of having a peaked spectrum at low frequencies based on the SED fits. This is 33\% of the sample. For the sources with the most evident turnover happening at frequencies above 58~MHz, we were able to estimate the magnetic field in the emitting regions, assuming that the turnover is due to synchrotron self-absorption. We did not investigate the possibility of the occurrence of free-free absorption instead of synchrotron self-absorption; however, we were able to exclude the existence of a simple low-energy cut-off in the plasma.  The analysis adopted here could be used for any of the other sources for which the fitted SED has a peak frequency, but since there is not enough data below 58~MHz it is impossible to make any sensible estimate of a magnetic field. More observations at 58~MHz, as well as down to decameter wavelengths (10~MHz), are necessary to increase the sample of sources for which such analysis is possible. Given the low resolution of the LBA with Dutch baselines and that the source size needs to be known to estimate a magnetic field, we would need VLBI observations with LOFAR LBA international stations so that we do not have to rely on high-frequency data for the analysis. 

This survey also provides a selection of extended sources that have a lot of scientific potential in their own right. In an upcoming work, we will use the maps of a sub-sample of the 25 most extended sources complemented by LoTSS \citep{Shimwell2019} and three RACS bands \citep{RACS2020, RACS-mid2023} to study their resolved spectral properties. We will construct resolved spectral index maps of these extended sources down to 58~MHz, something which has only been done for four sources in the catalog to date. These resolved maps will also provide an opportunity to study whether conventional spectral aging models, which have been tested extensively at higher frequencies, also hold up at the low-frequency end of the spectrum. The full survey of the 3CRR catalog at 58~MHz will be made available via the LOFAR Surveys website \url{https://lofar-surveys.org}.

\begin{acknowledgements}
JMB and FdG acknowledge the support of the ERC Consolidator Grant ULU 101086378. MJH thanks the UK STFC for support [ST/Y001249/1].
LKM is grateful for support from UKRI [MR/Y020405/1] and STFC [ST/V002406/1].

The Low Frequency Array, designed and constructed by ASTRON, has facilities in several countries, which are owned by various parties (each with their own funding sources), and which are collectively operated by the International LOFAR Telescope (ILT) foundation under a joint scientific policy. 

The ILT resources have benefited from the following recent major funding sources: CNRS-INSU, Observatoire de Paris and Université d’Orléans, France; BMBF, MIWF-NRW, MPG, Germany; Science Foundation Ireland (SFI), Department of Business, Enterprise and Innovation (DBEI), Ireland; NWO, The Netherlands; The Science and Technology Facilities Council, UK; Ministry of Science and Higher Education, Poland; Istituto Nazionale di Astrofisica (INAF), Italy.

This research made use of the Dutch national e-infrastructure with support of the SURF
Cooperative (e-infra 180169) and the LOFAR e-infra group. The Jülich LOFAR Long Term
Archive and the German LOFAR network are both coordinated and operated by the Jülich
Supercomputing Centre (JSC), and computing resources on the supercomputer JUWELS at JSC
were provided by the Gauss Centre for Supercomputing e.V. (grant CHTB00) through the John
von Neumann Institute for Computing (NIC).
This research made use of the University of Hertfordshire high-performance computing facility
and the LOFAR-UK computing facility located at the University of Hertfordshire and supported
by STFC [ST/P000096/1], and of the Italian LOFAR IT computing infrastructure, including the resources within the PLEIADI special "LOFAR" project by USC-C of INAF, supported and
operated by INAF, and by the Physics Department of Turin university (under an agreement with
Consorzio Interuniversitario per la Fisica Spaziale) at the C3S Supercomputing Centre, Italy.

This research has made use of the NASA/IPAC Extragalactic Database (NED), which is operated by the Jet Propulsion Laboratory, California Institute of Technology, under contract with the National Aeronautics and Space Administration. 
This research has made use of the VizieR catalogue access tool, CDS,
Strasbourg, France (DOI : 10.26093/cds/vizier). The original description 
of the VizieR service was published in \citet{vizier2000}.
This research made use of Astropy,\footnote{\url{http://www.astropy.org}} a community-developed core Python package for Astronomy \citep{2013A&A...558A..33A, 2018AJ....156..123A}
\end{acknowledgements}

%
%

\bibliography{main}

\newpage
\onecolumn
\begin{appendix} 

\section{Final maps of the 3CRR catalog}
\label{app:images}

\begin{figure}[H]
\centering
\includegraphics[width=0.162\linewidth, trim={0.cm 0.cm 0.cm 0.cm},clip]{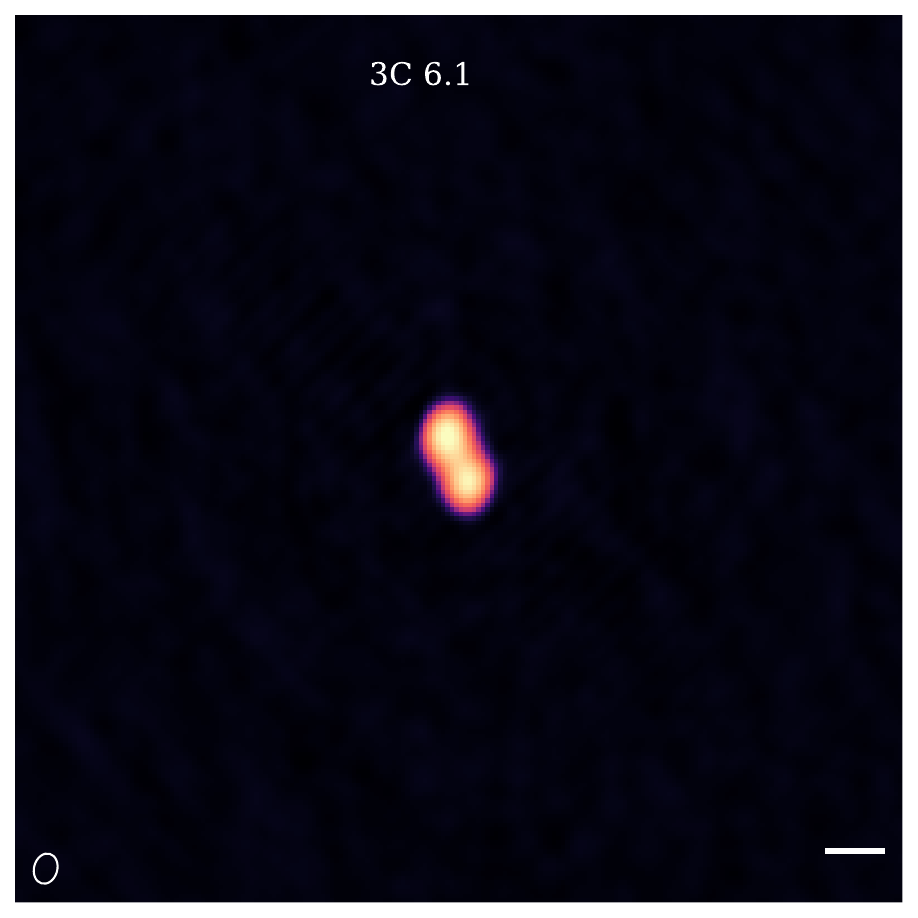}
\includegraphics[width=0.162\linewidth, trim={0.cm 0.cm 0.cm 0.cm},clip]{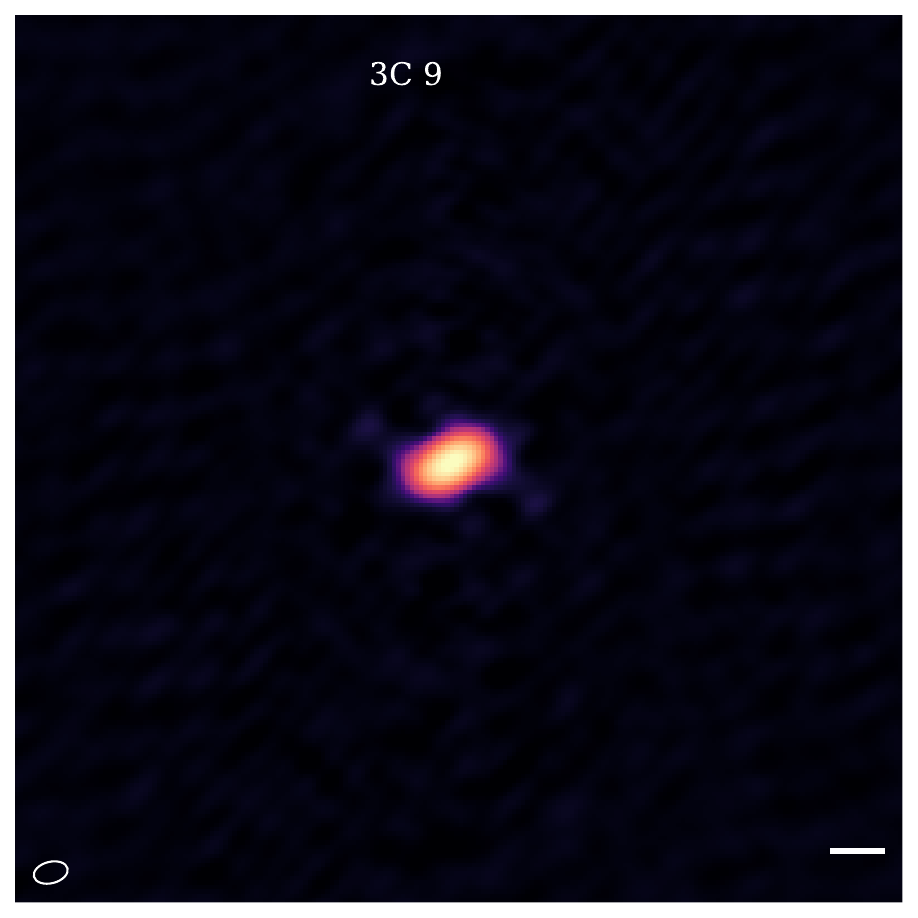}
\includegraphics[width=0.162\linewidth, trim={0.cm 0.cm 0.cm 0.cm},clip]{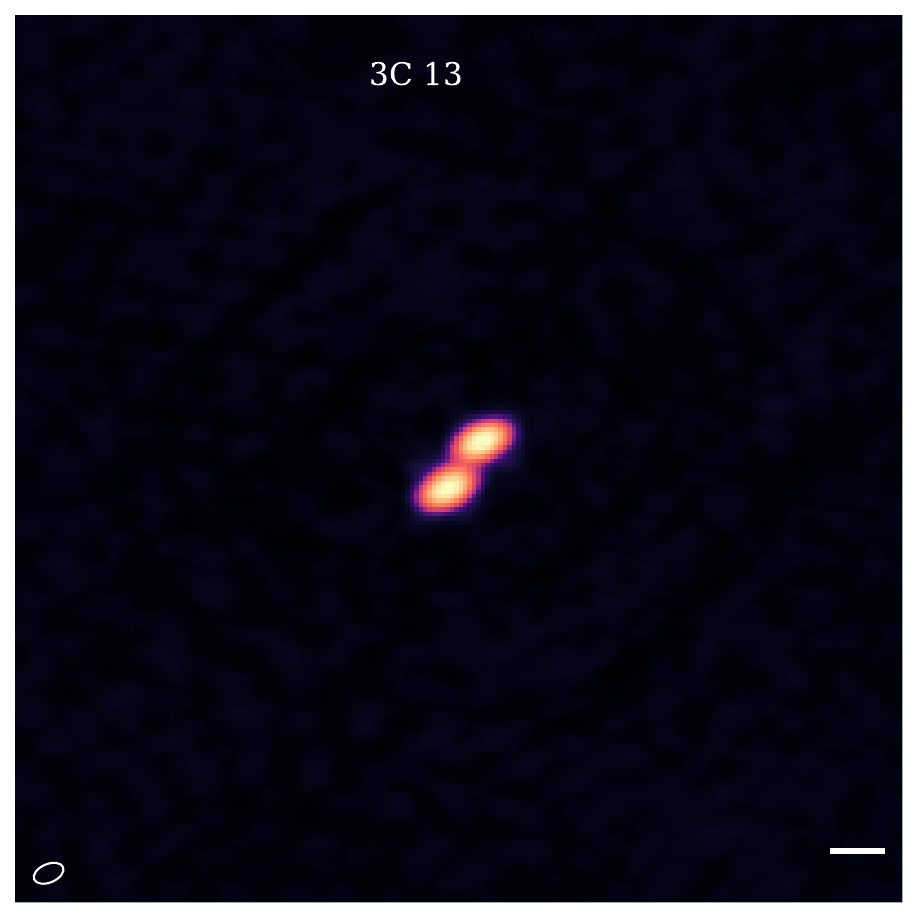}
\includegraphics[width=0.162\linewidth, trim={0.cm 0.cm 0.cm 0.cm},clip]{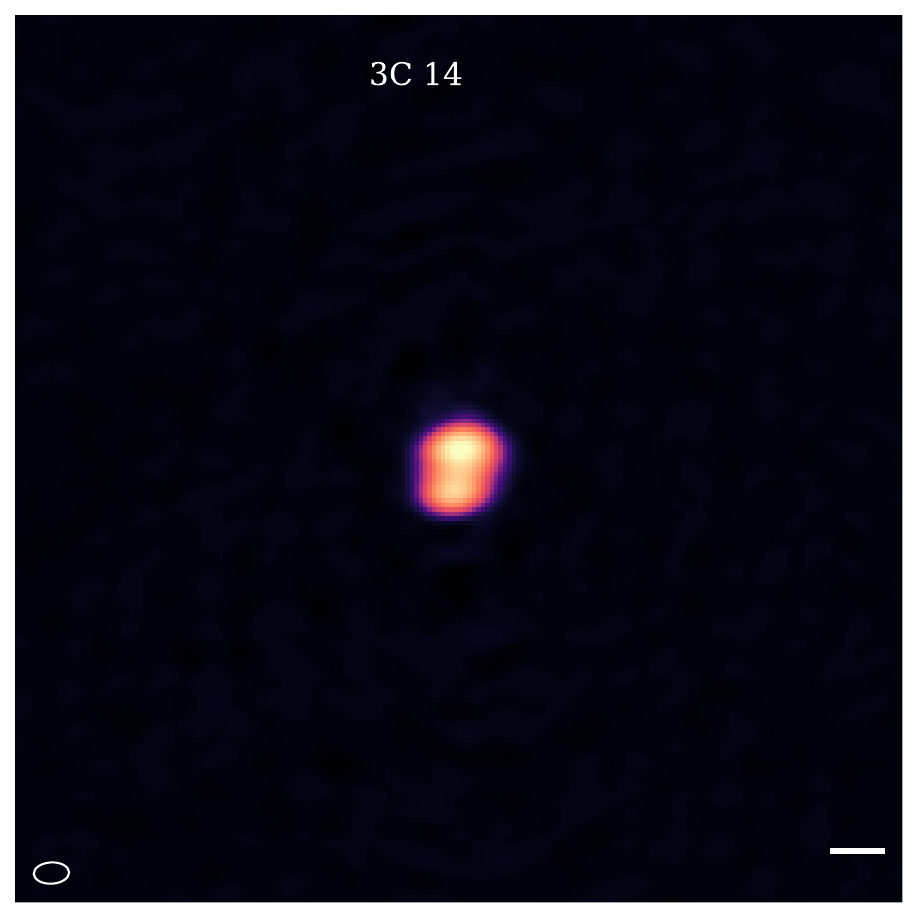}
\includegraphics[width=0.162\linewidth, trim={0.cm 0.cm 0.cm 0.cm},clip]{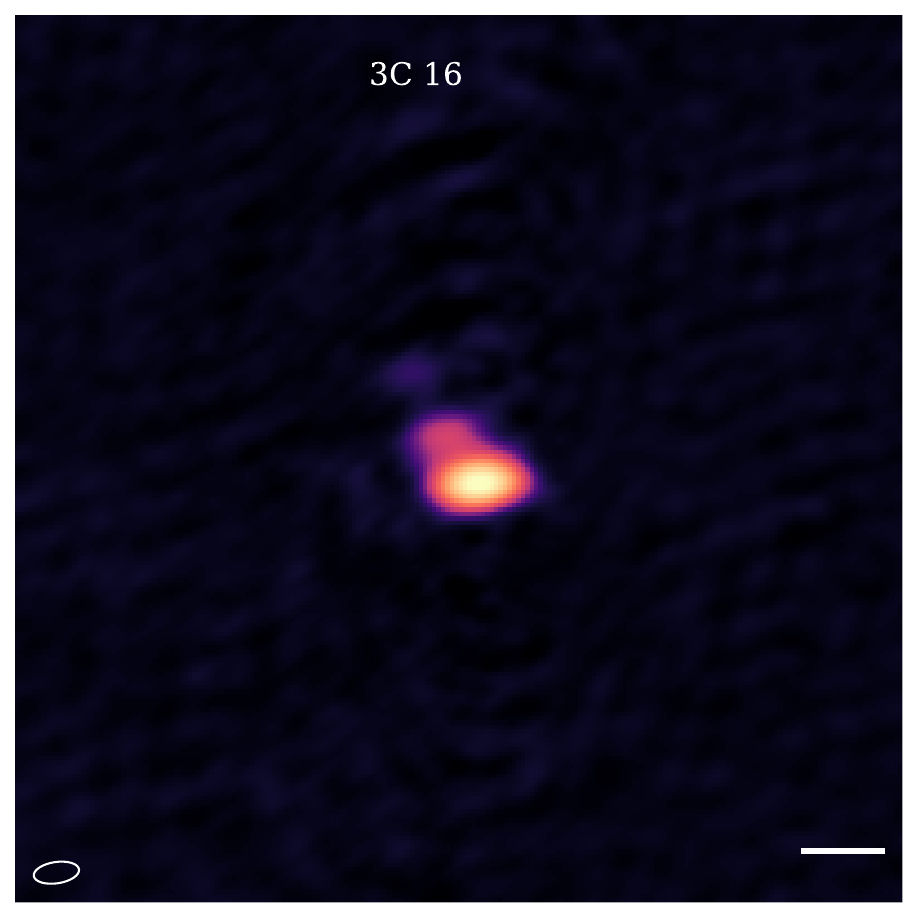}
\includegraphics[width=0.162\linewidth, trim={0.cm 0.cm 0.cm 0.cm},clip]{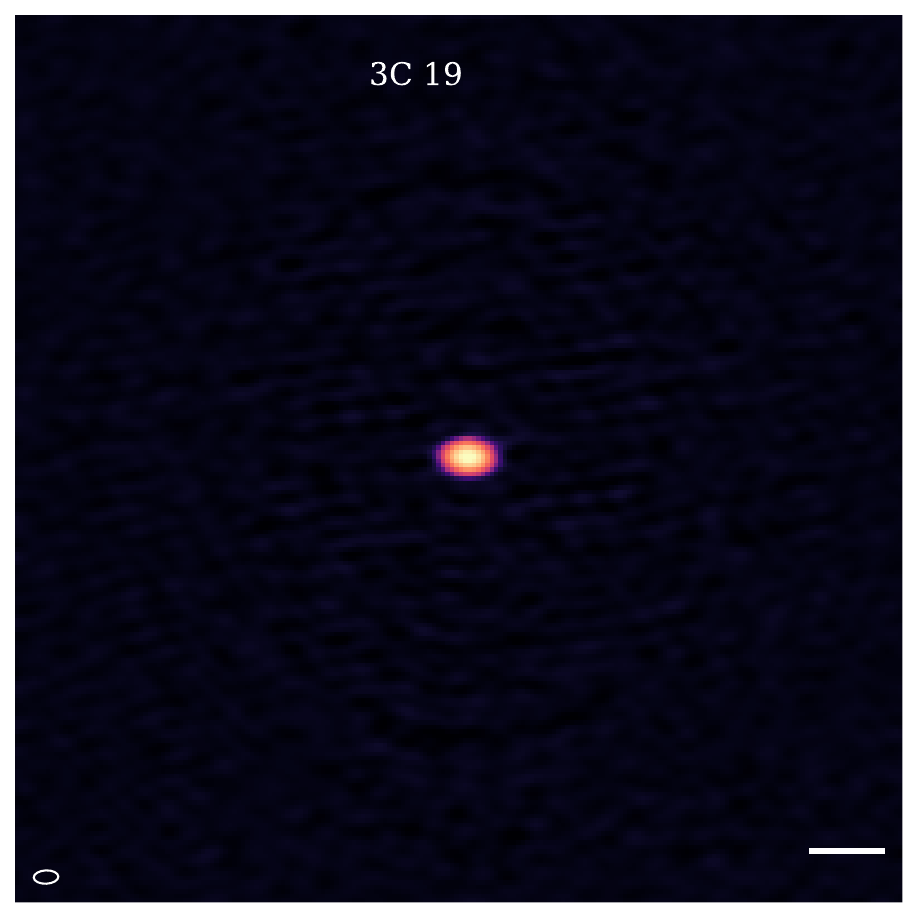}
\includegraphics[width=0.162\linewidth, trim={0.cm 0.cm 0.cm 0.cm},clip]{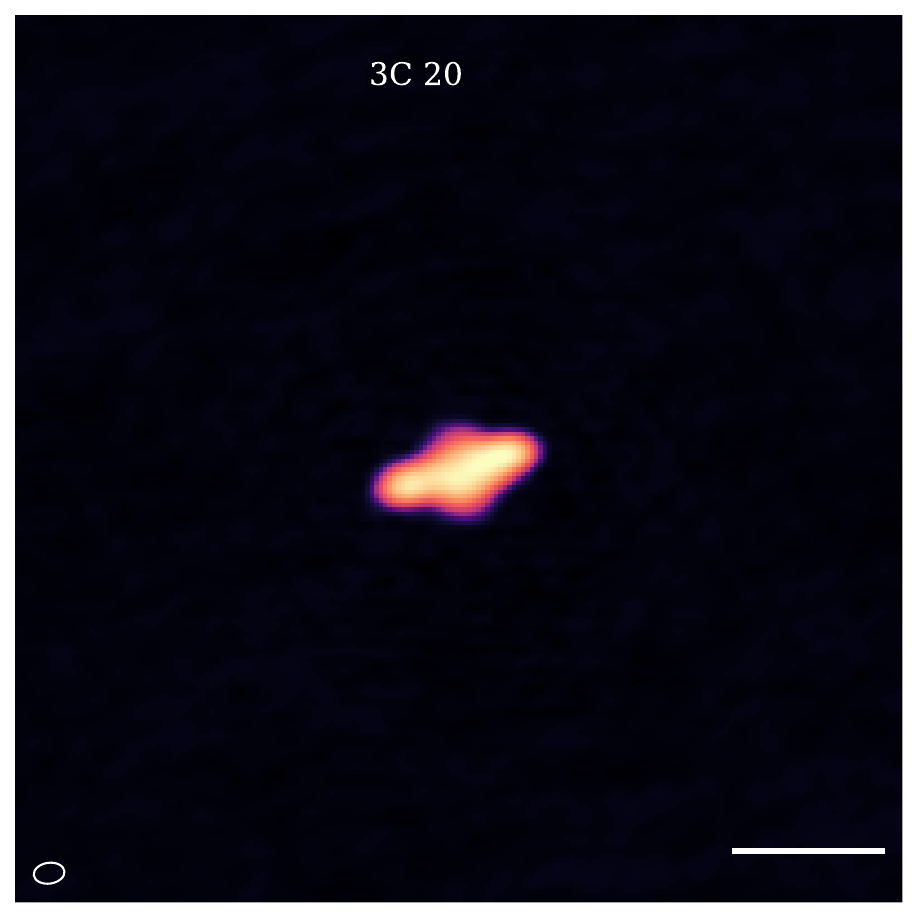}
\includegraphics[width=0.162\linewidth, trim={0.cm 0.cm 0.cm 0.cm},clip]{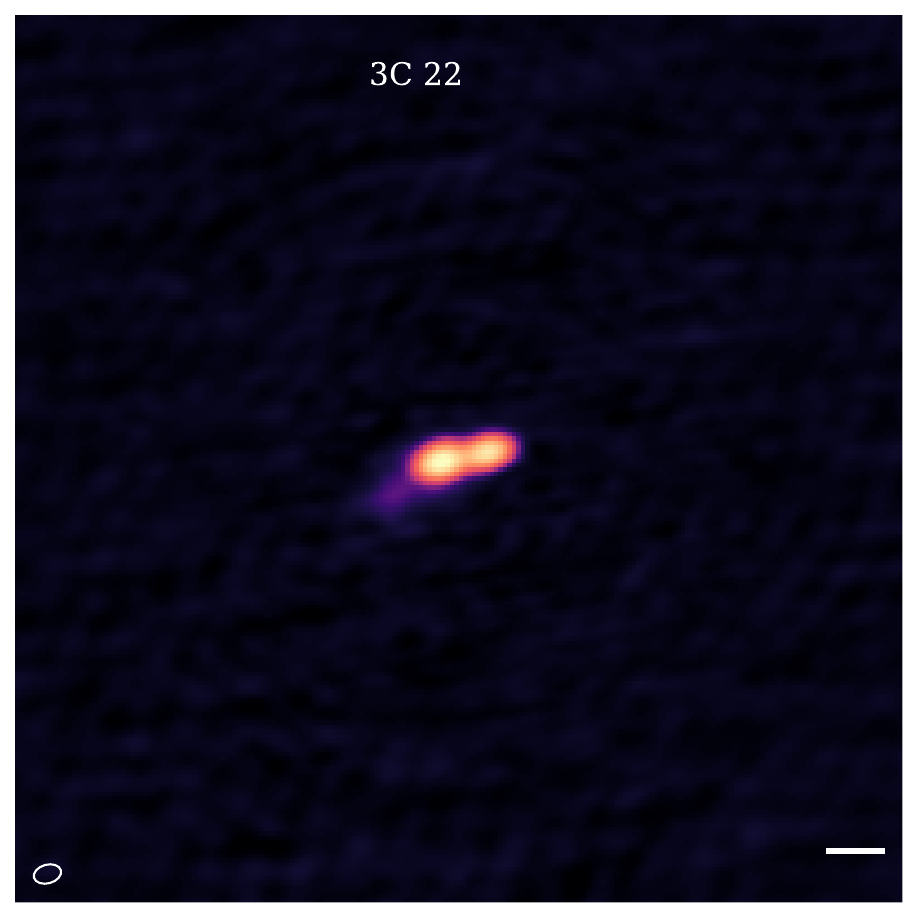}
\includegraphics[width=0.162\linewidth, trim={0.cm 0.cm 0.cm 0.cm},clip]{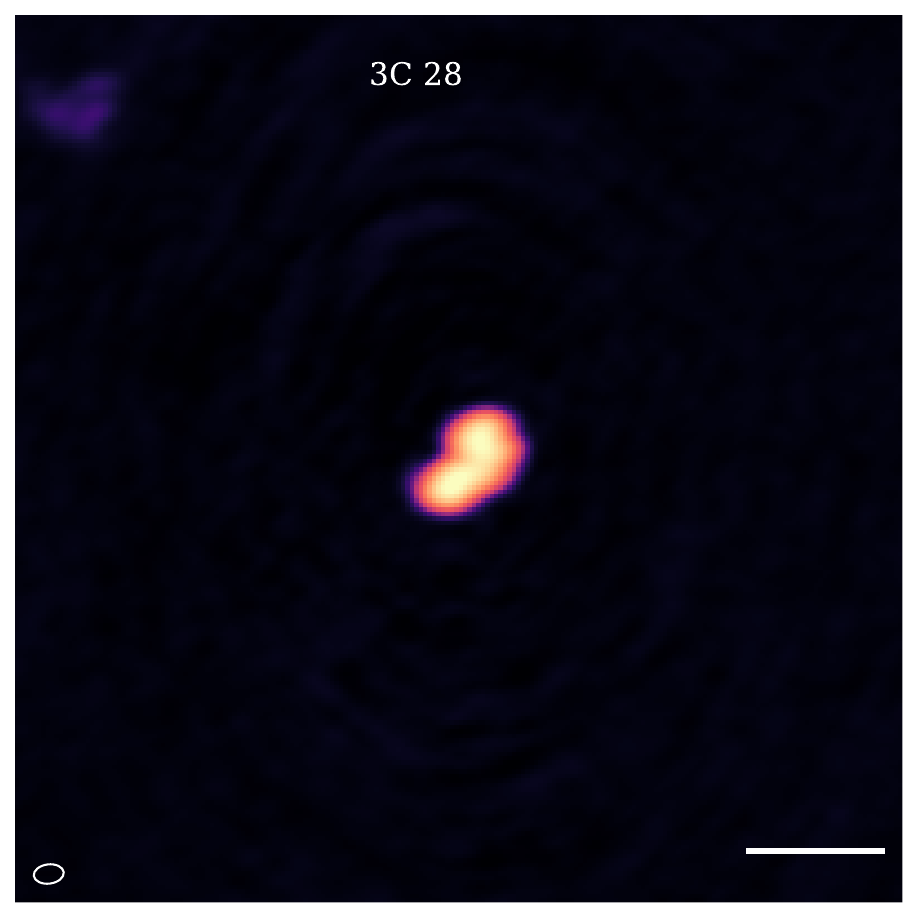}
\includegraphics[width=0.162\linewidth, trim={0.cm 0.cm 0.cm 0.cm},clip]{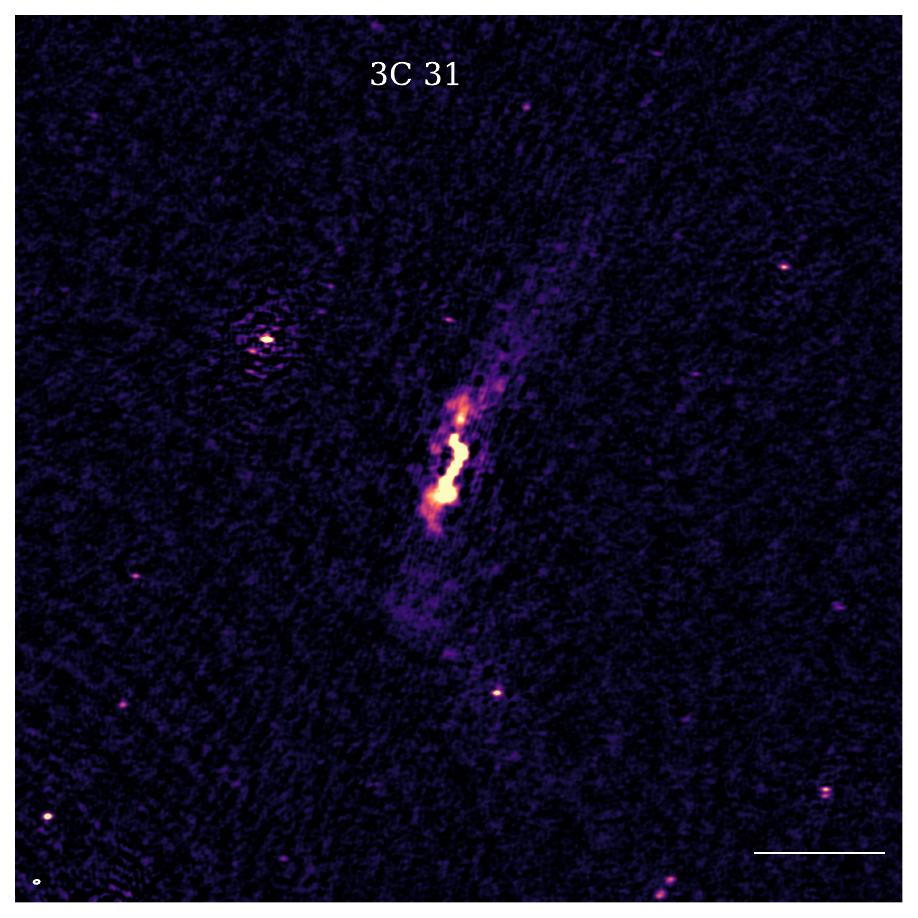}
\includegraphics[width=0.162\linewidth, trim={0.cm 0.cm 0.cm 0.cm},clip]{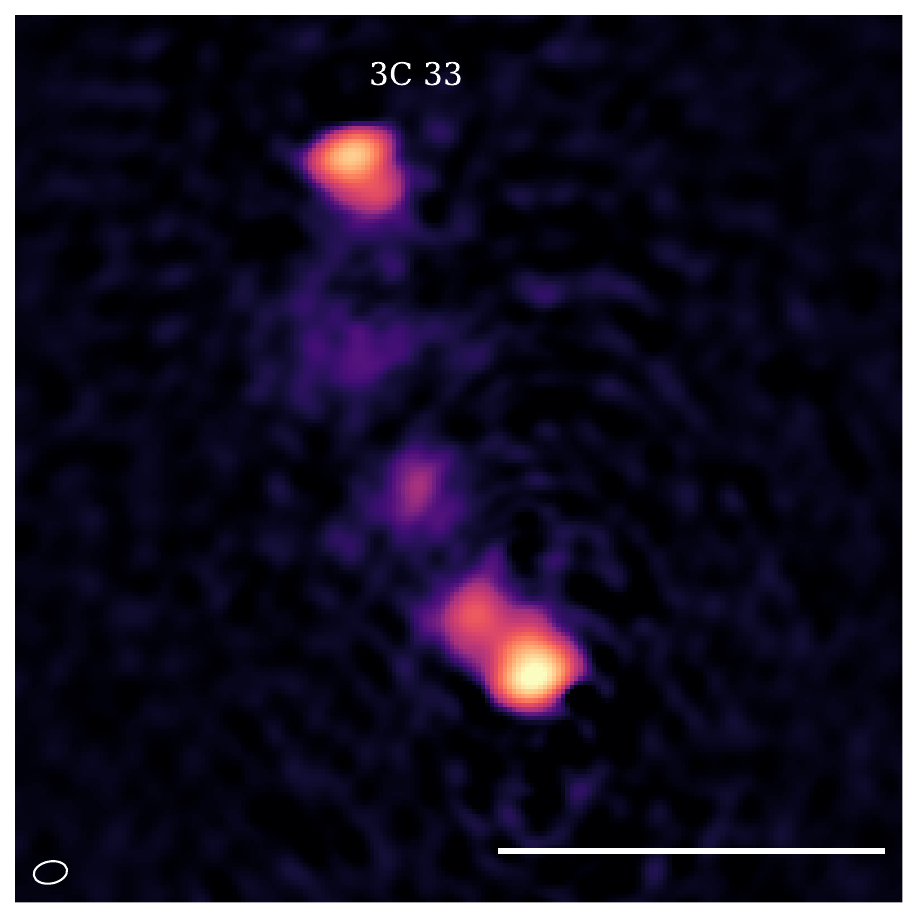}
\includegraphics[width=0.162\linewidth, trim={0.cm 0.cm 0.cm 0.cm},clip]{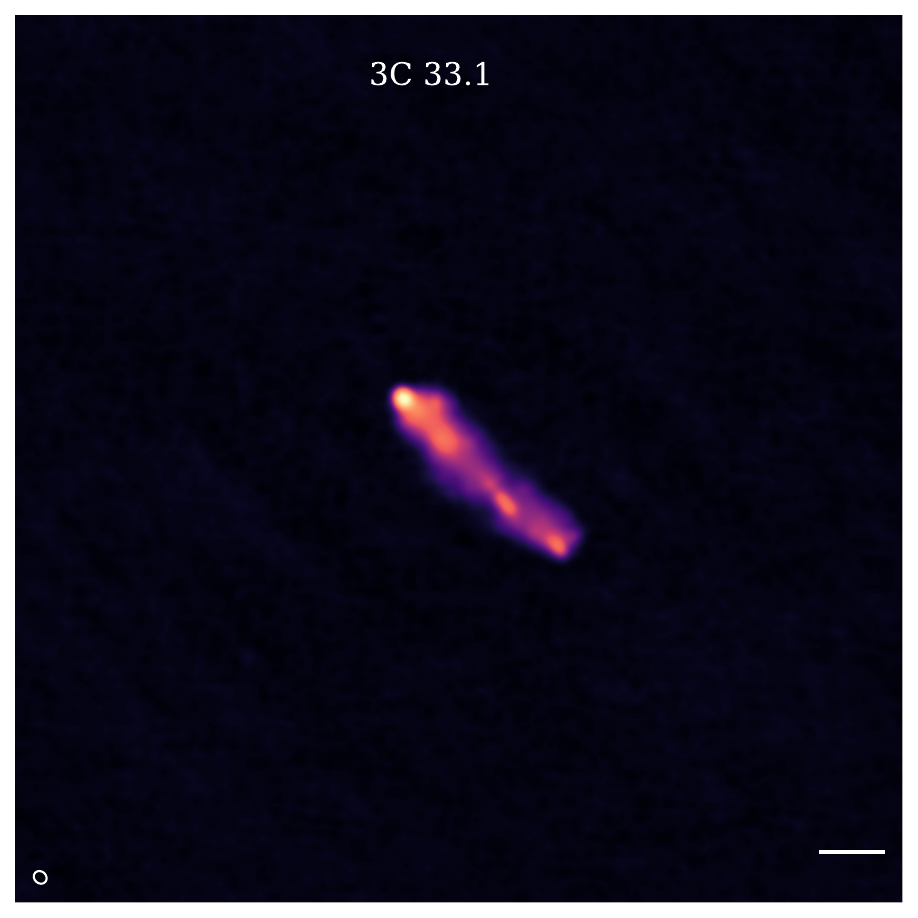}
\includegraphics[width=0.162\linewidth, trim={0.cm 0.cm 0.cm 0.cm},clip]{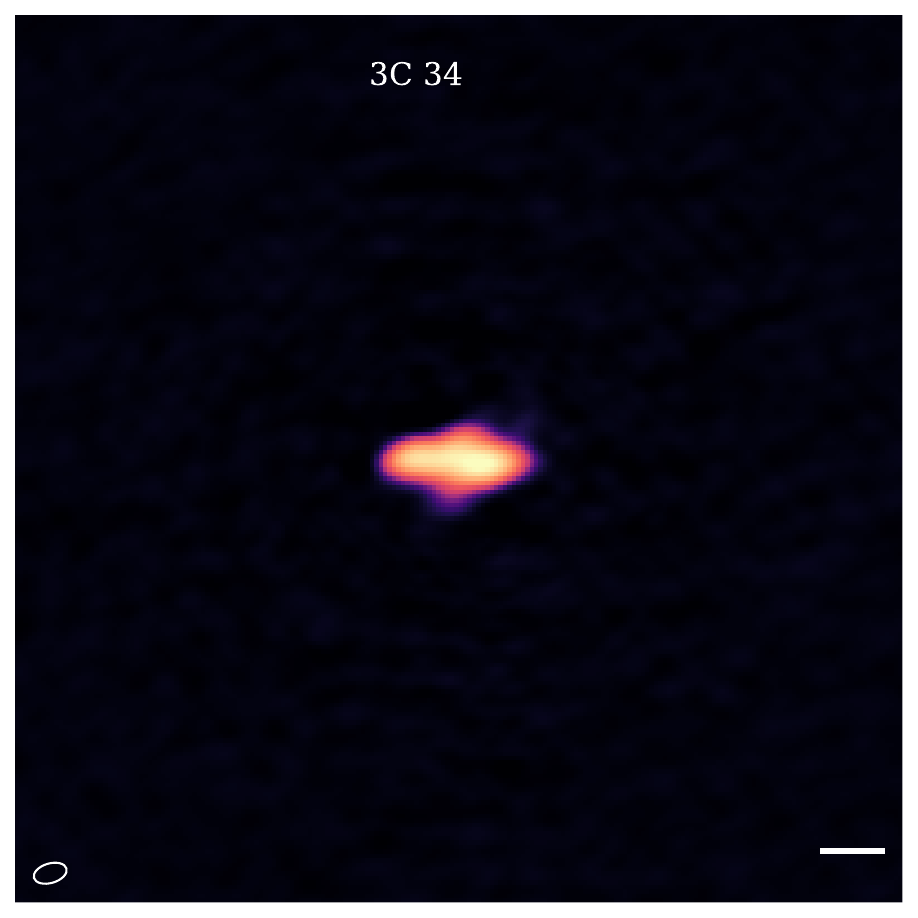}
\includegraphics[width=0.162\linewidth, trim={0.cm 0.cm 0.cm 0.cm},clip]{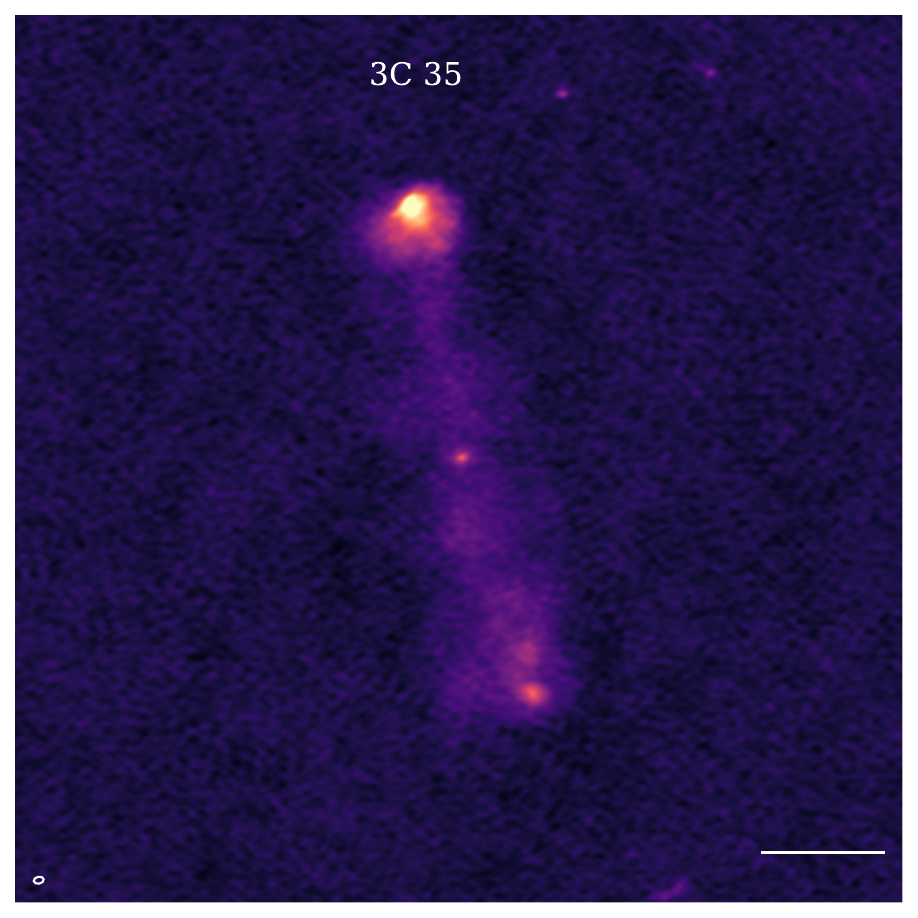}
\includegraphics[width=0.162\linewidth, trim={0.cm 0.cm 0.cm 0.cm},clip]{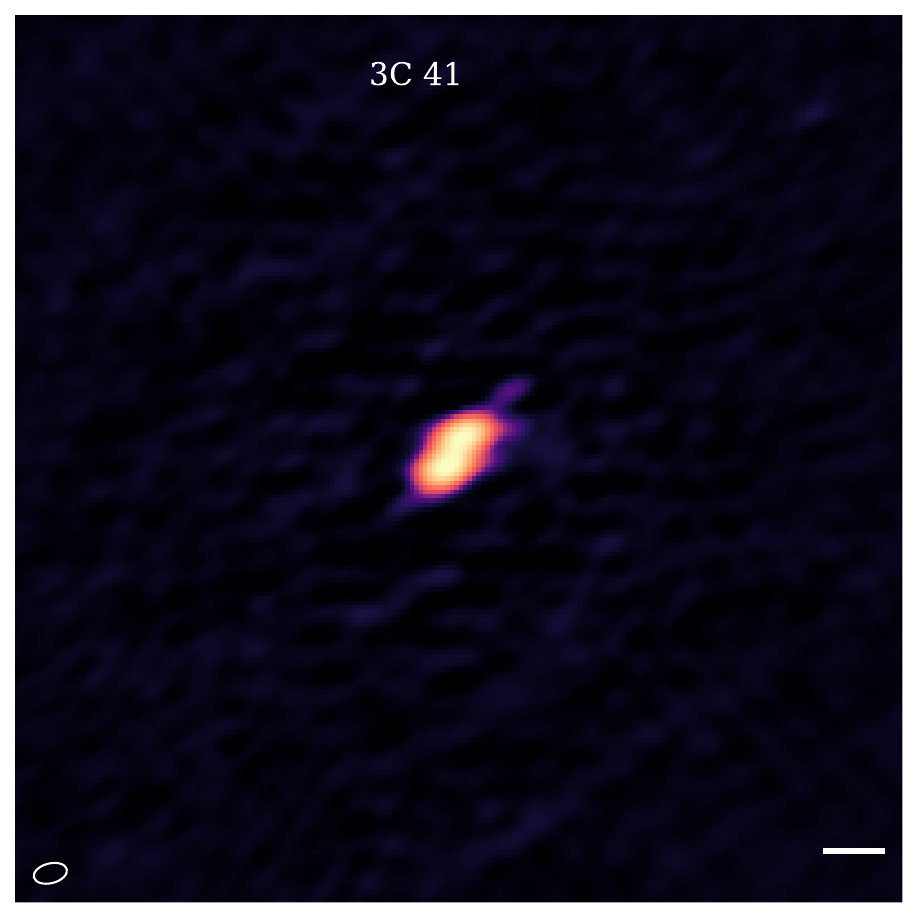}
\includegraphics[width=0.162\linewidth, trim={0.cm 0.cm 0.cm 0.cm},clip]{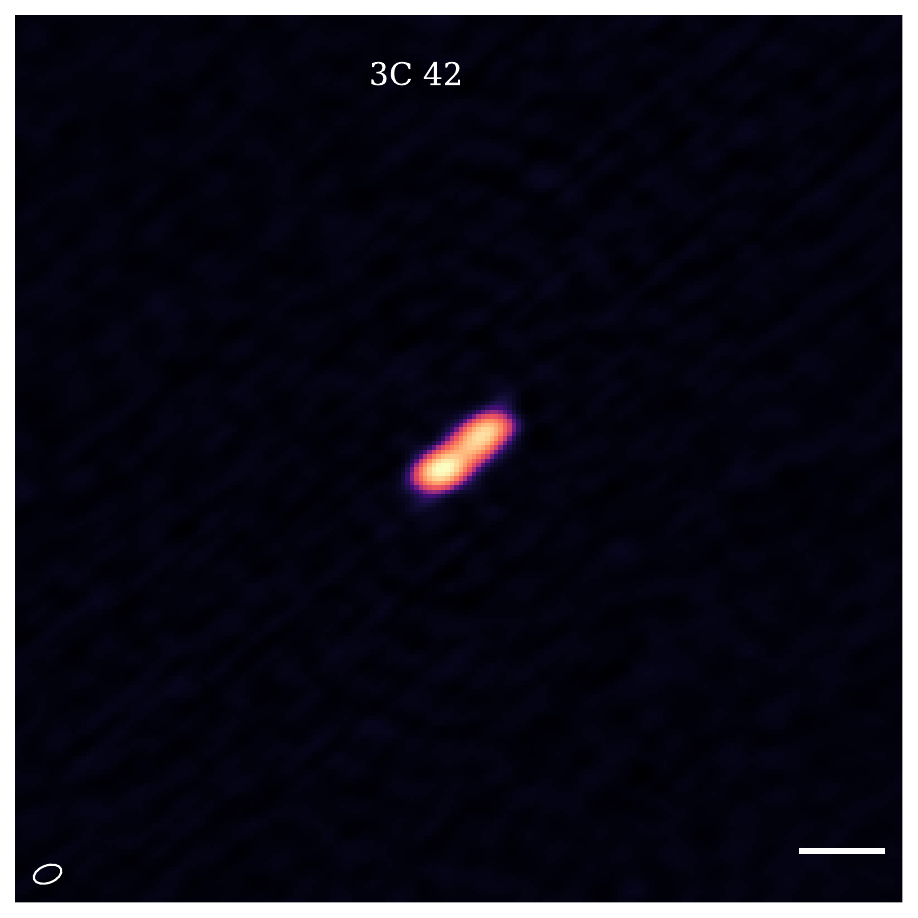}
\includegraphics[width=0.162\linewidth, trim={0.cm 0.cm 0.cm 0.cm},clip]{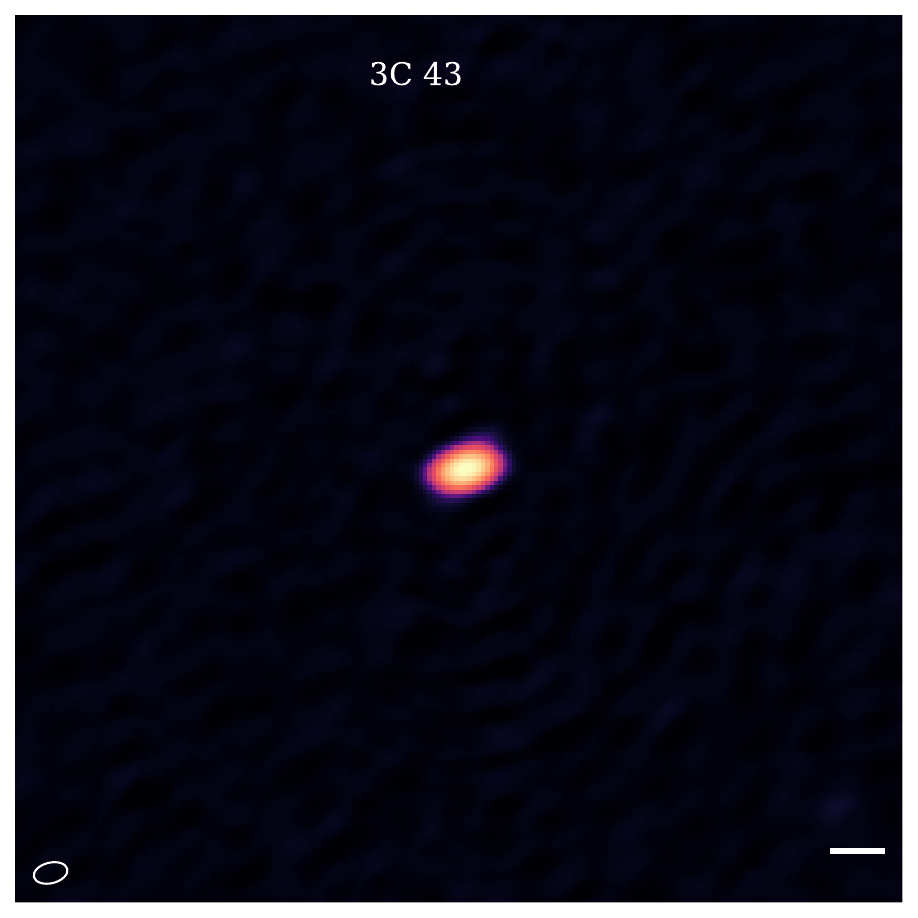}
\includegraphics[width=0.162\linewidth, trim={0.cm 0.cm 0.cm 0.cm},clip]{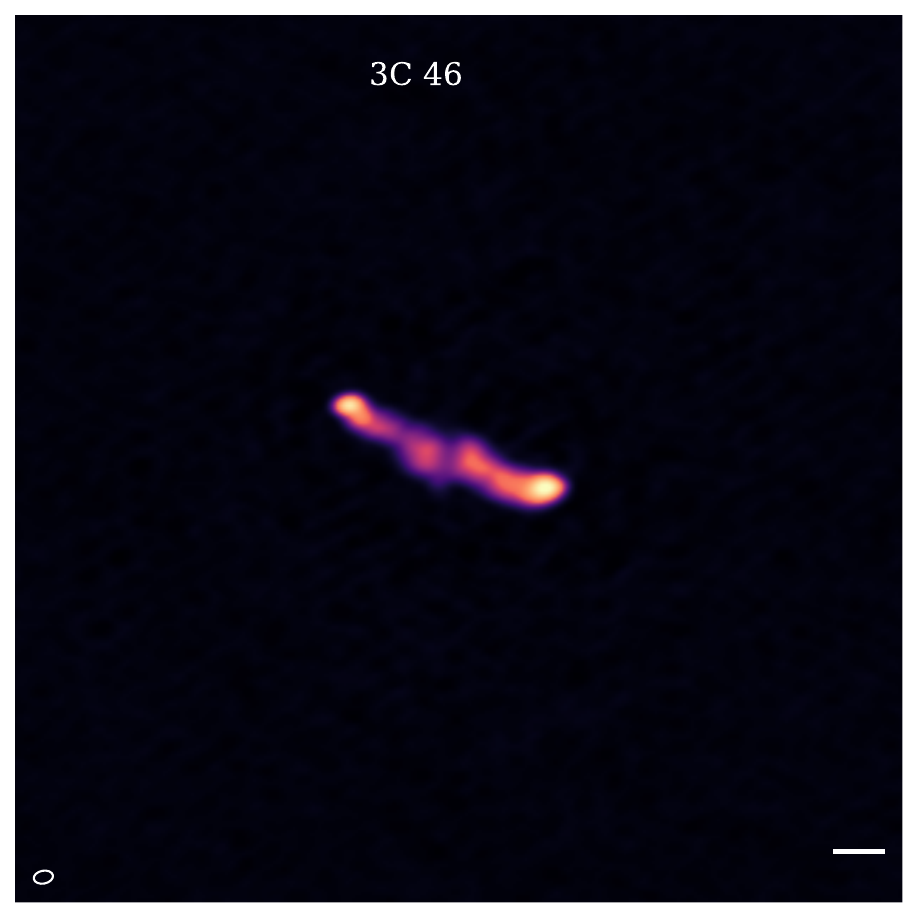}
\includegraphics[width=0.162\linewidth, trim={0.cm 0.cm 0.cm 0.cm},clip]{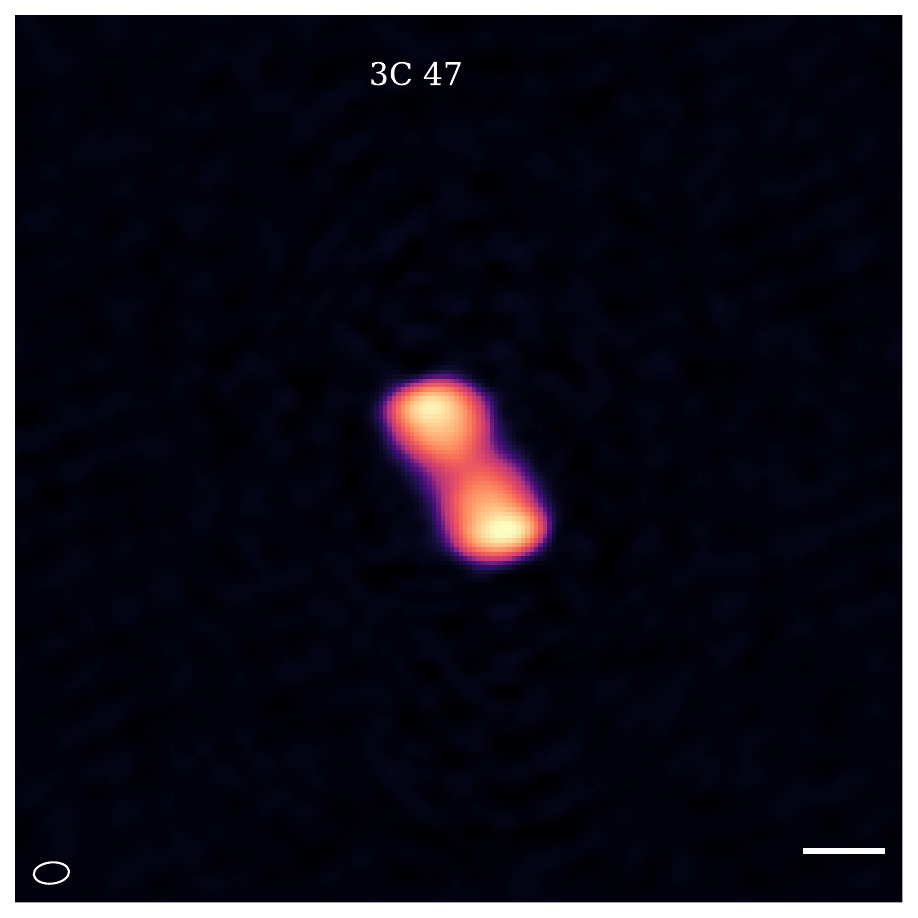}
\includegraphics[width=0.162\linewidth, trim={0.cm 0.cm 0.cm 0.cm},clip]{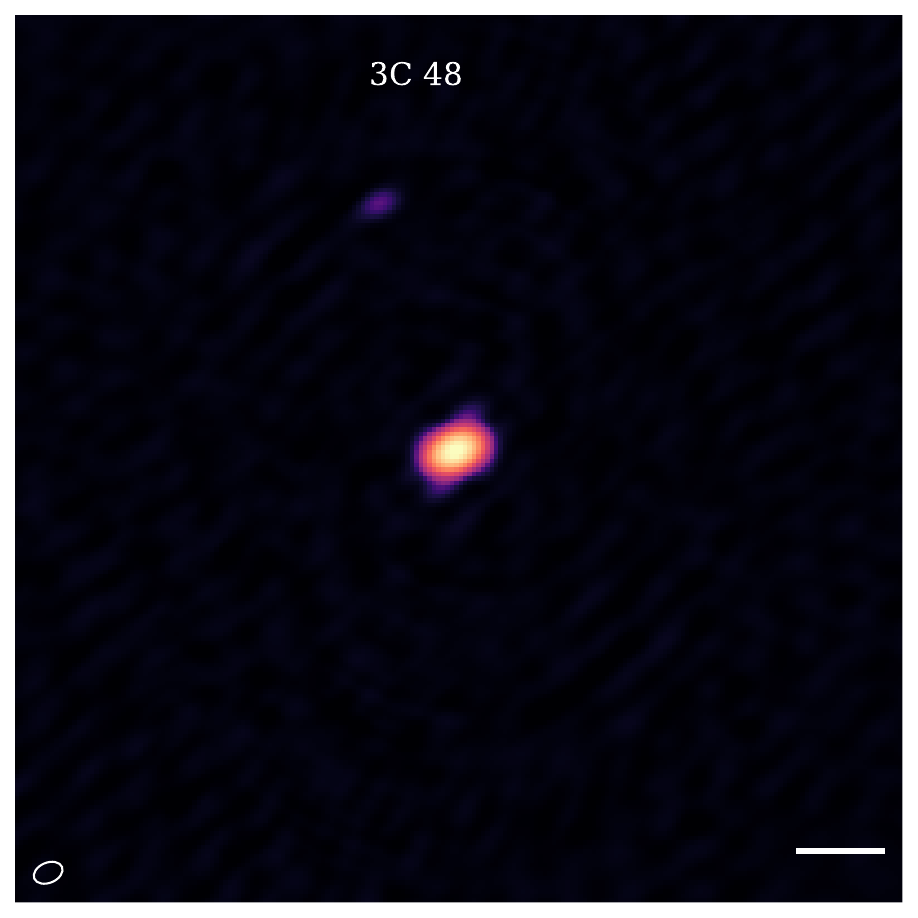}
\includegraphics[width=0.162\linewidth, trim={0.cm 0.cm 0.cm 0.cm},clip]{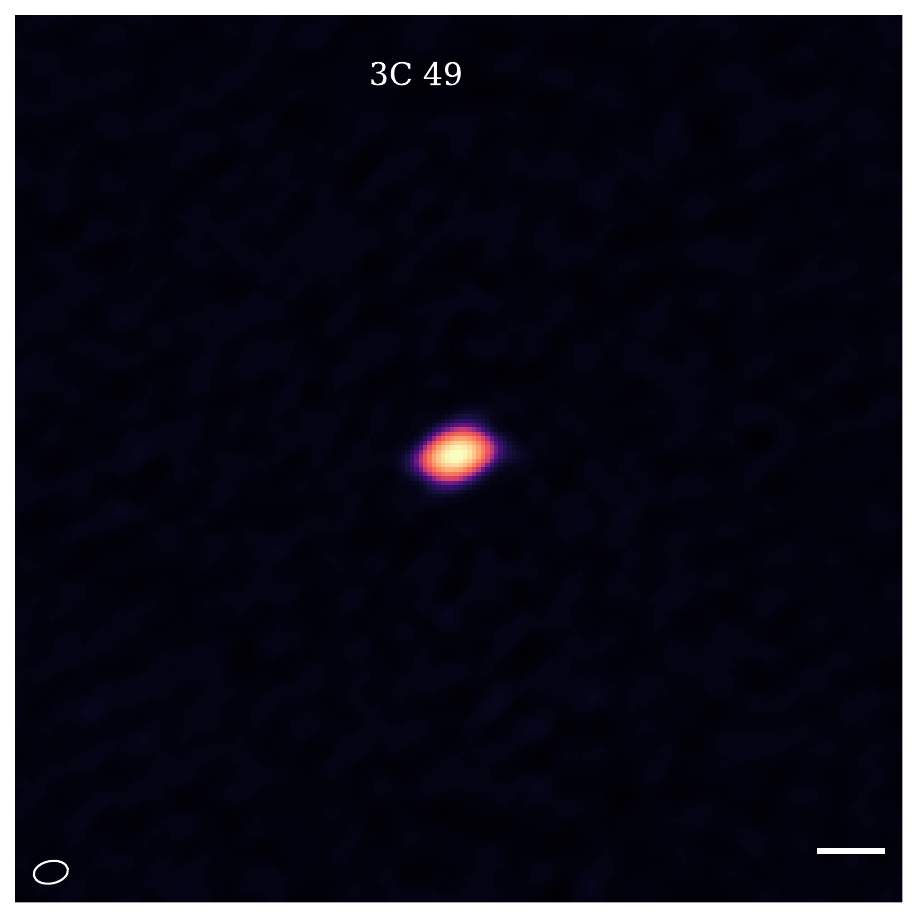}
\includegraphics[width=0.162\linewidth, trim={0.cm 0.cm 0.cm 0.cm},clip]{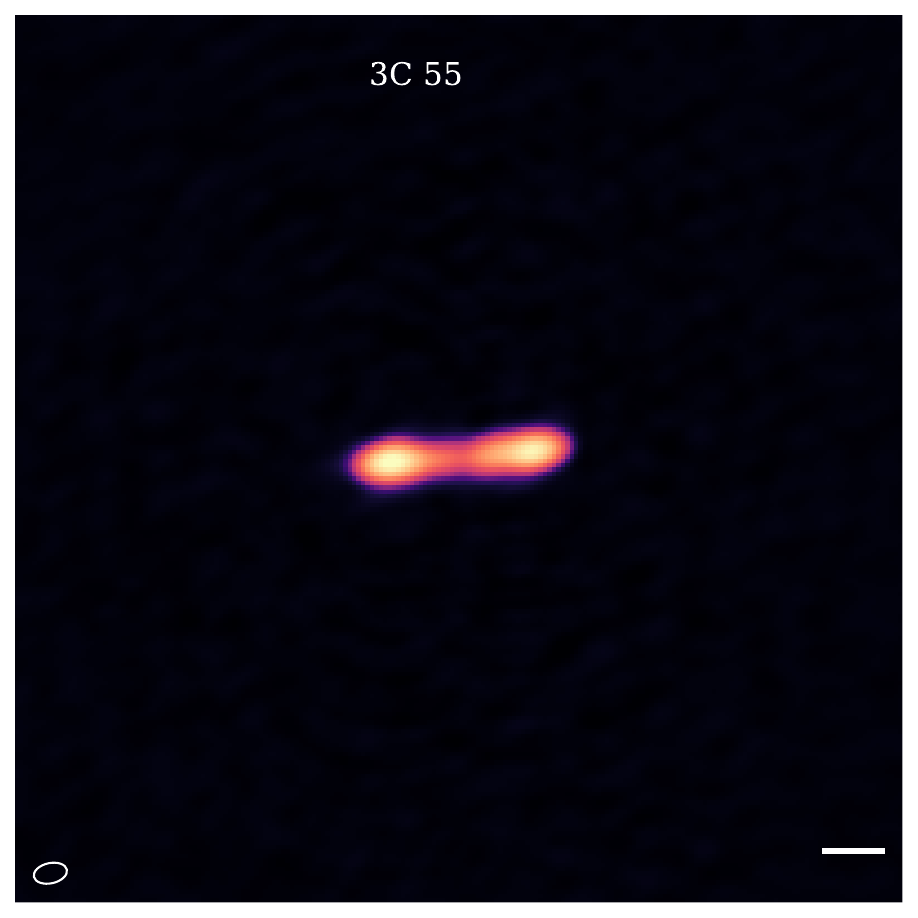}
\includegraphics[width=0.162\linewidth, trim={0.cm 0.cm 0.cm 0.cm},clip]{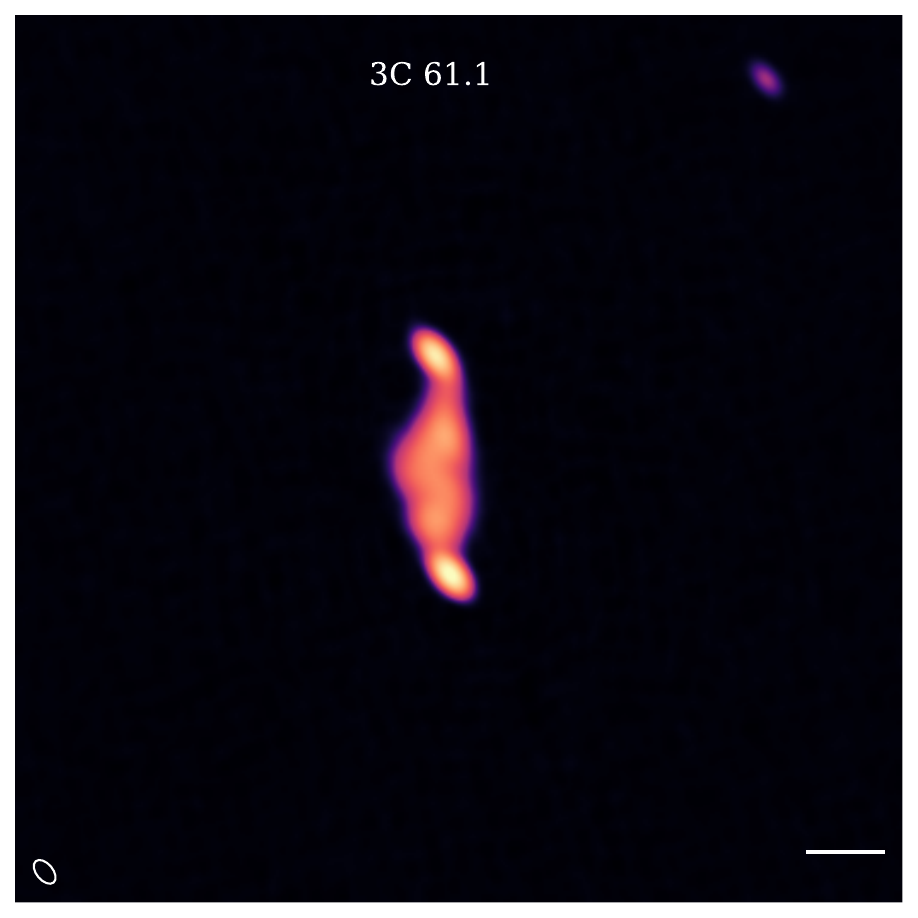}
\includegraphics[width=0.162\linewidth, trim={0.cm 0.cm 0.cm 0.cm},clip]{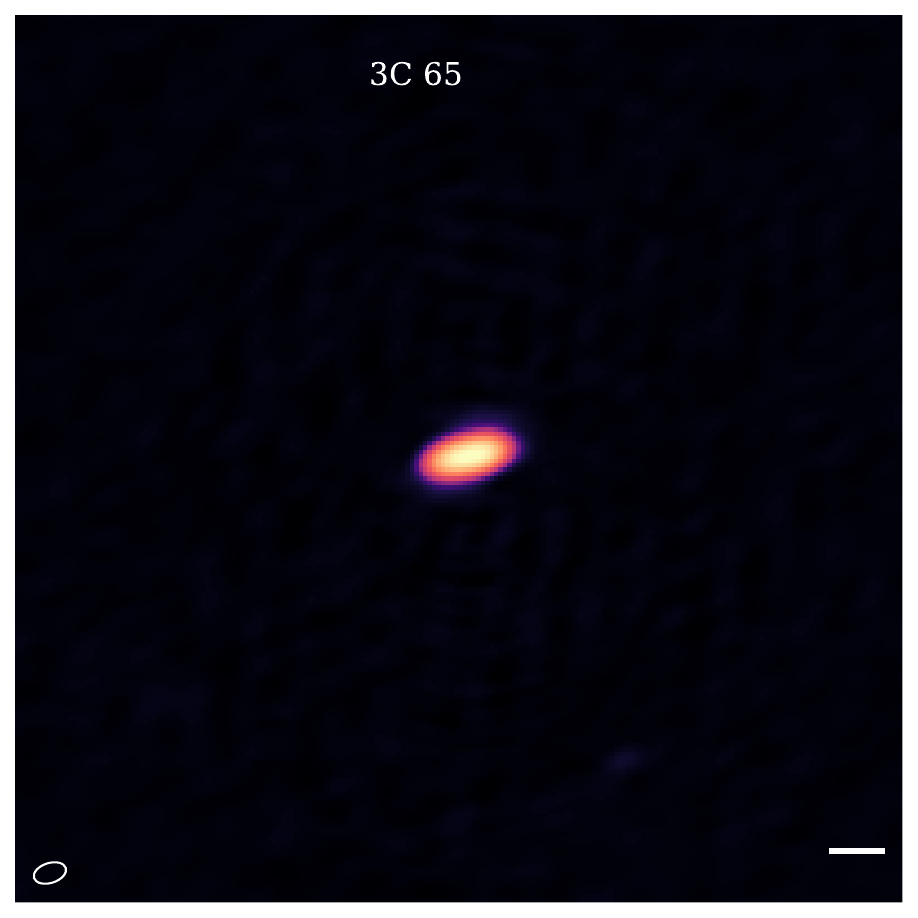}
\includegraphics[width=0.162\linewidth, trim={0.cm 0.cm 0.cm 0.cm},clip]{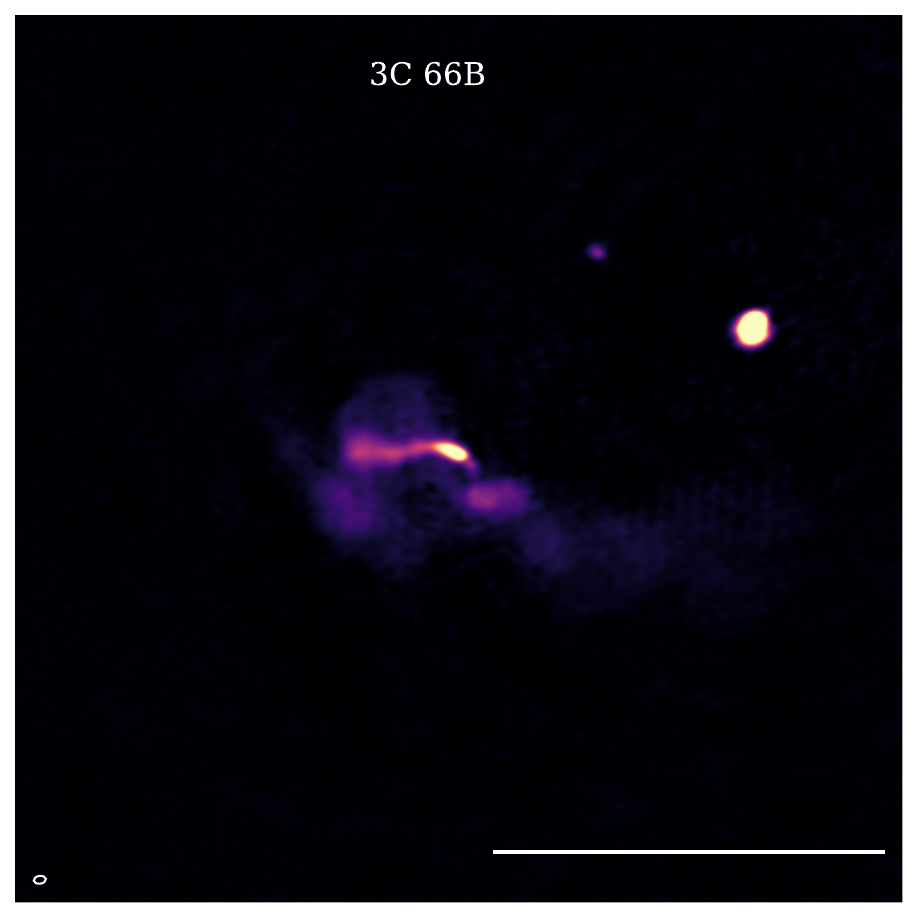}
\includegraphics[width=0.162\linewidth, trim={0.cm 0.cm 0.cm 0.cm},clip]{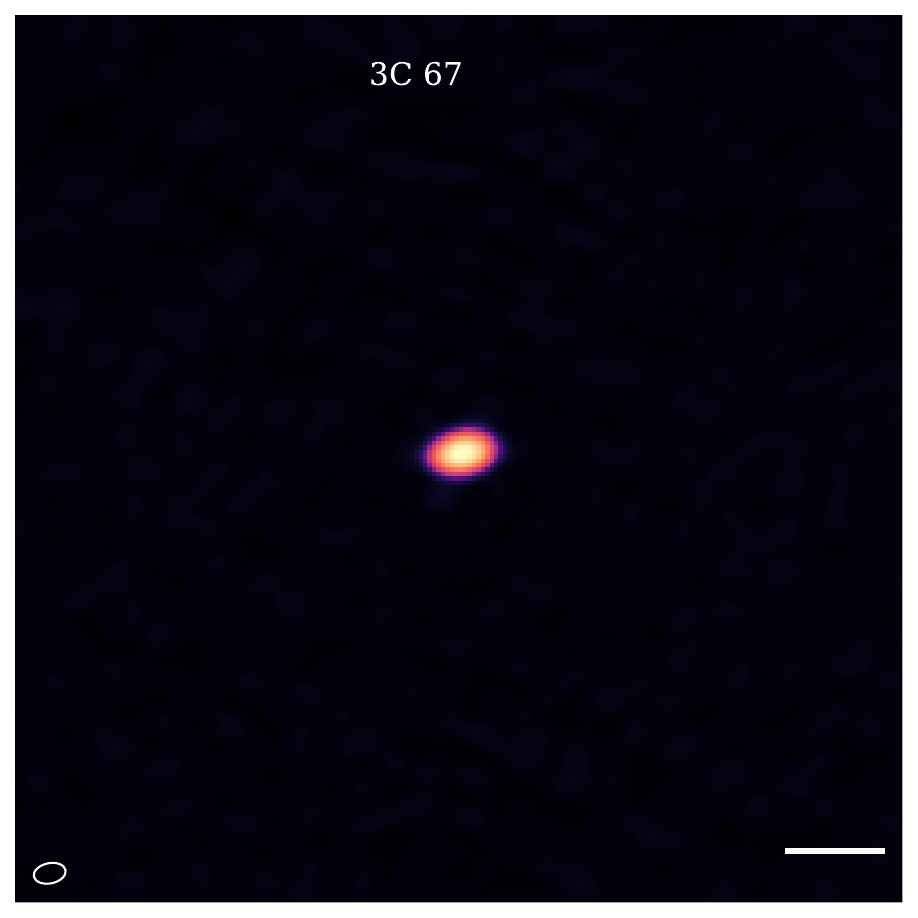}
\includegraphics[width=0.162\linewidth, trim={0.cm 0.cm 0.cm 0.cm},clip]{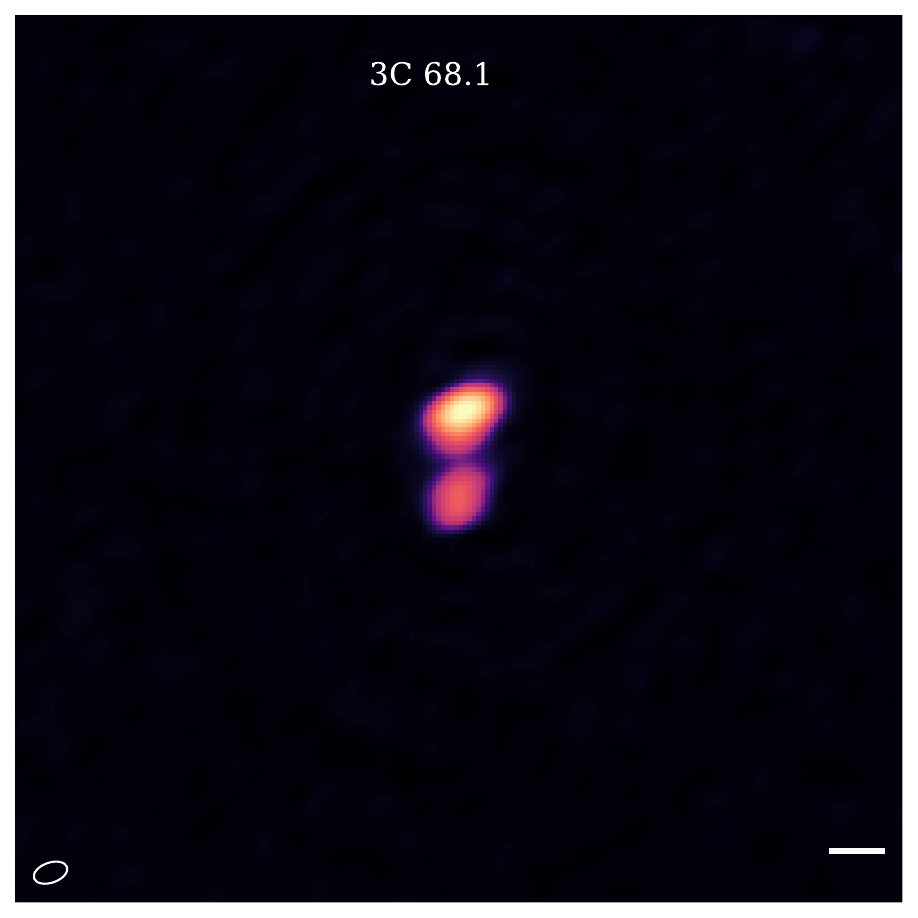}
\includegraphics[width=0.162\linewidth, trim={0.cm 0.cm 0.cm 0.cm},clip]{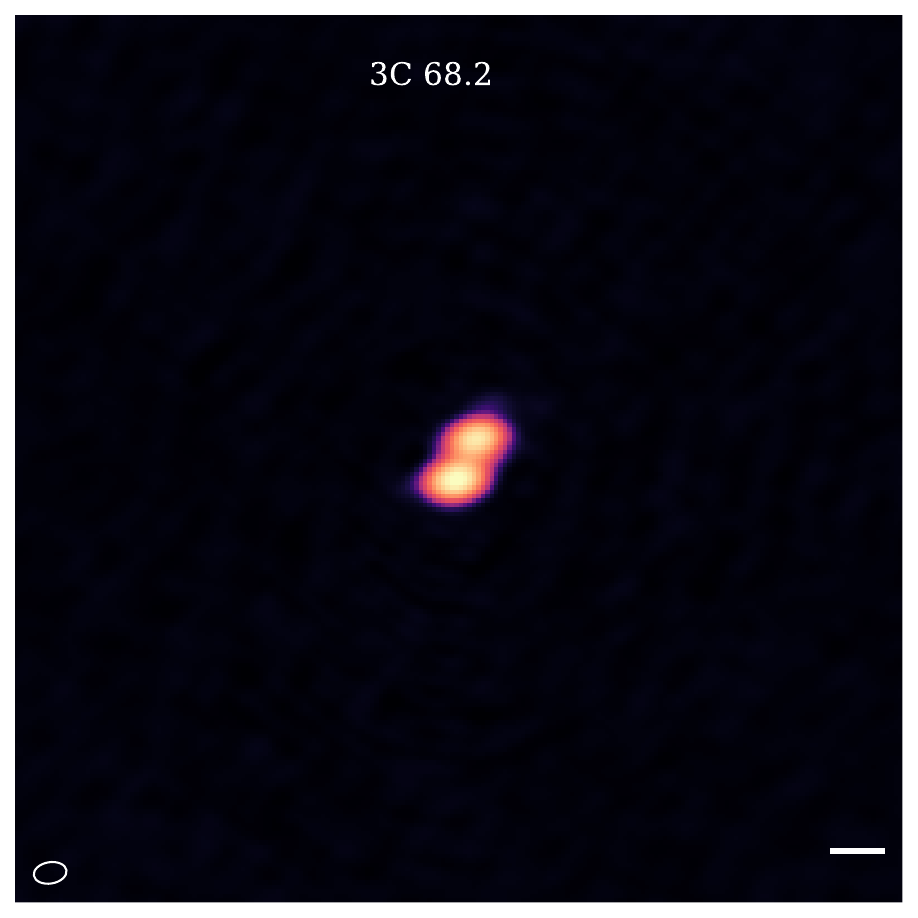}
\includegraphics[width=0.162\linewidth, trim={0.cm 0.cm 0.cm 0.cm},clip]{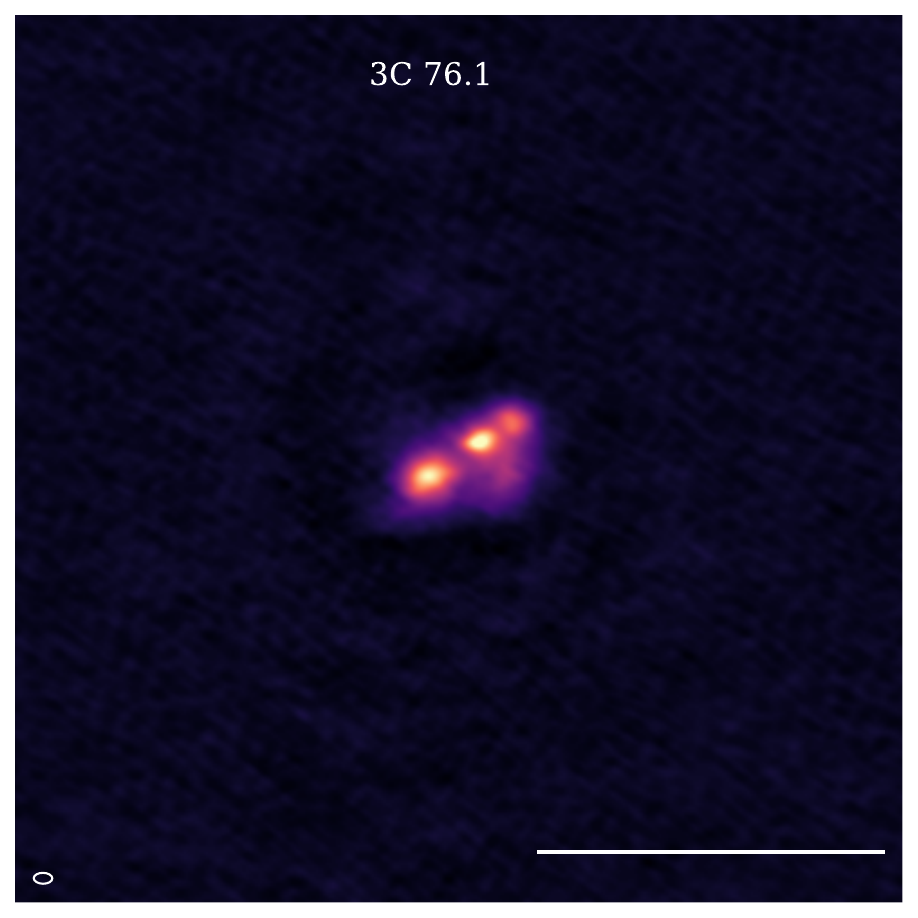}
\includegraphics[width=0.162\linewidth, trim={0.cm 0.cm 0.cm 0.cm},clip]{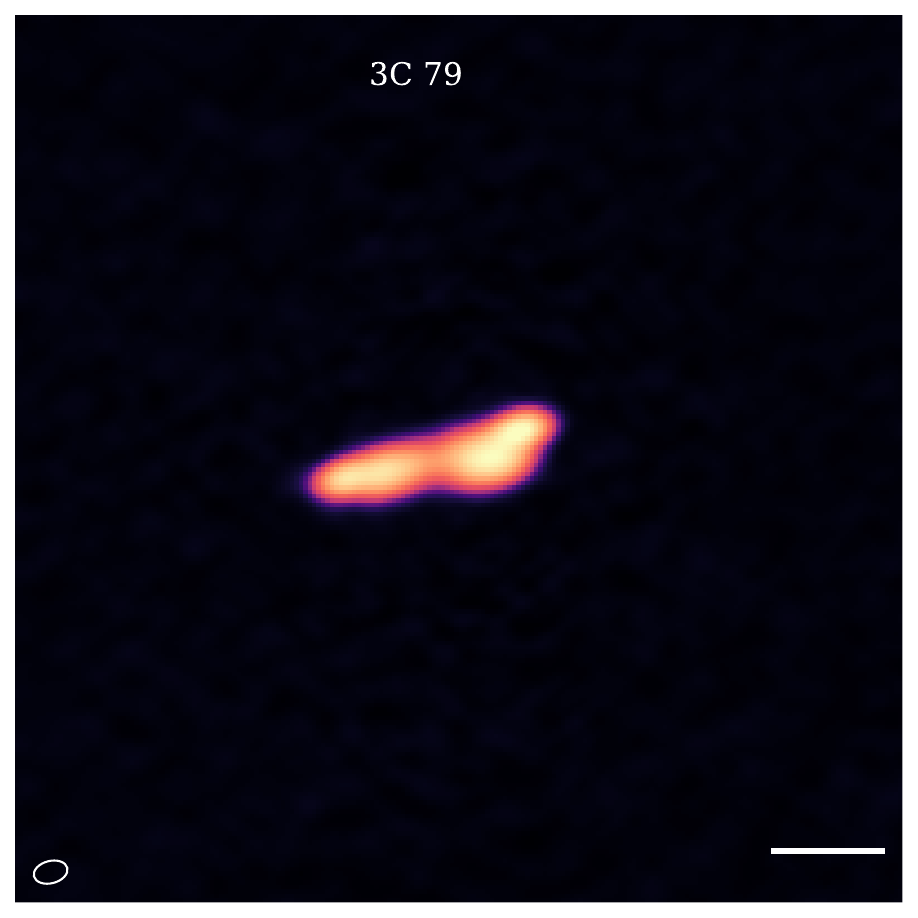}
\includegraphics[width=0.162\linewidth, trim={0.cm 0.cm 0.cm 0.cm},clip]{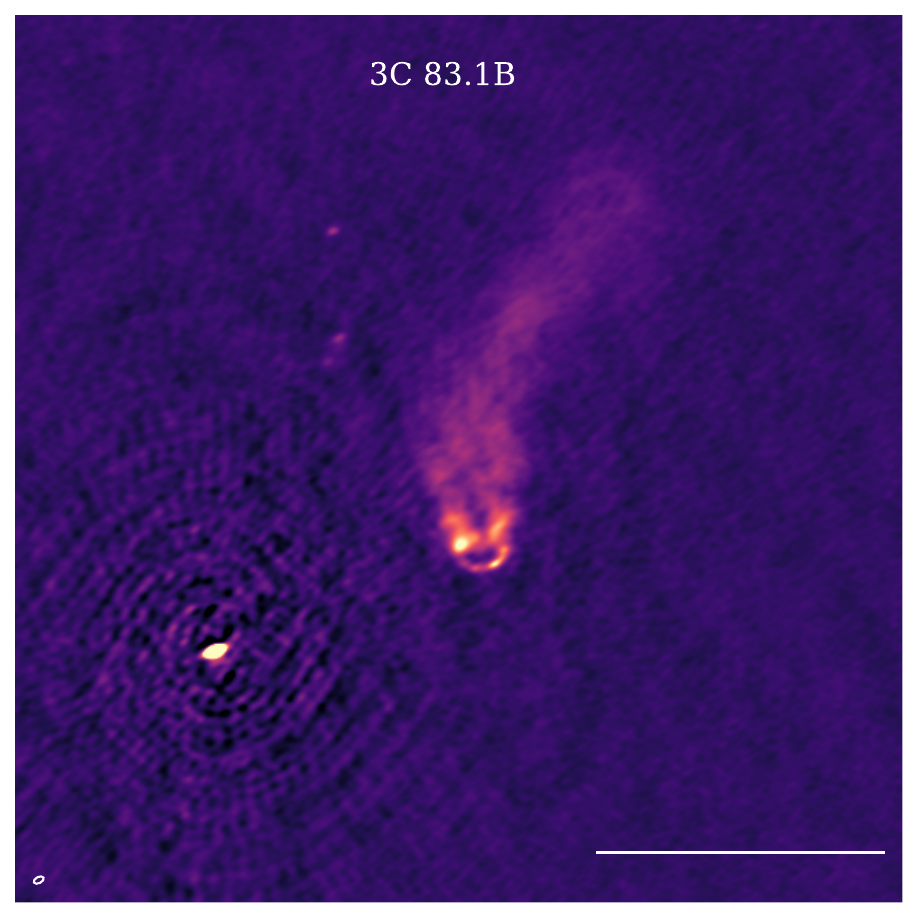}
\includegraphics[width=0.162\linewidth, trim={0.cm 0.cm 0.cm 0.cm},clip]{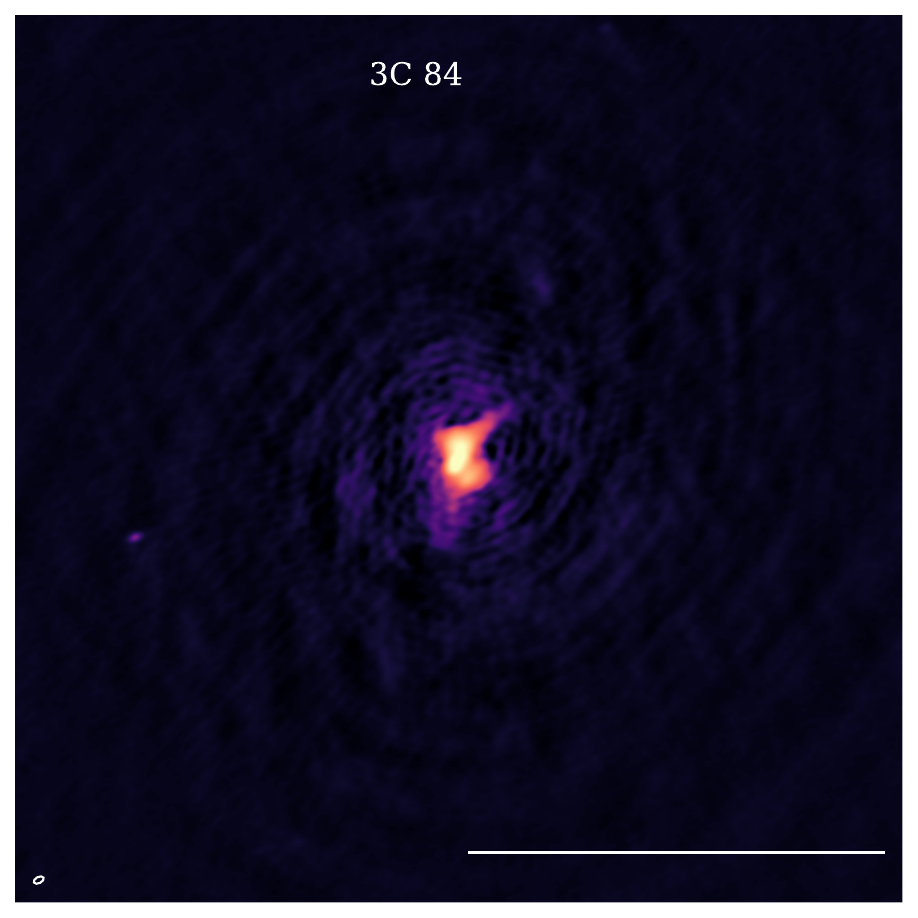}
\includegraphics[width=0.162\linewidth, trim={0.cm 0.cm 0.cm 0.cm},clip]{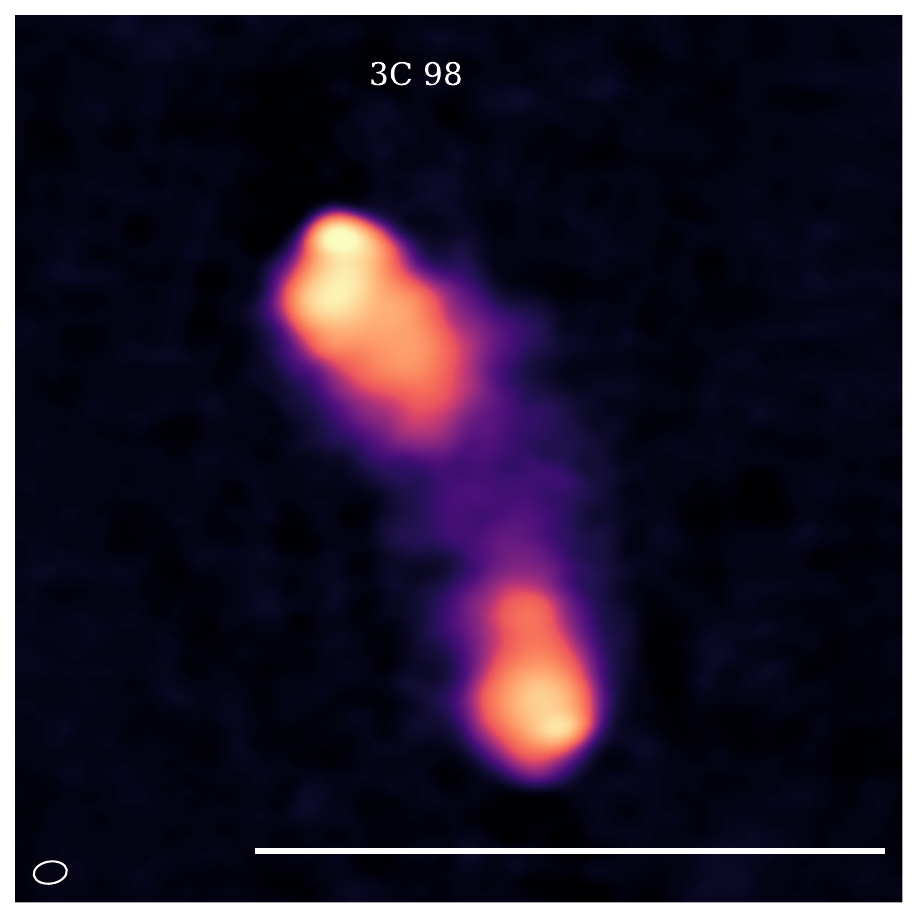}
\includegraphics[width=0.162\linewidth, trim={0.cm 0.cm 0.cm 0.cm},clip]{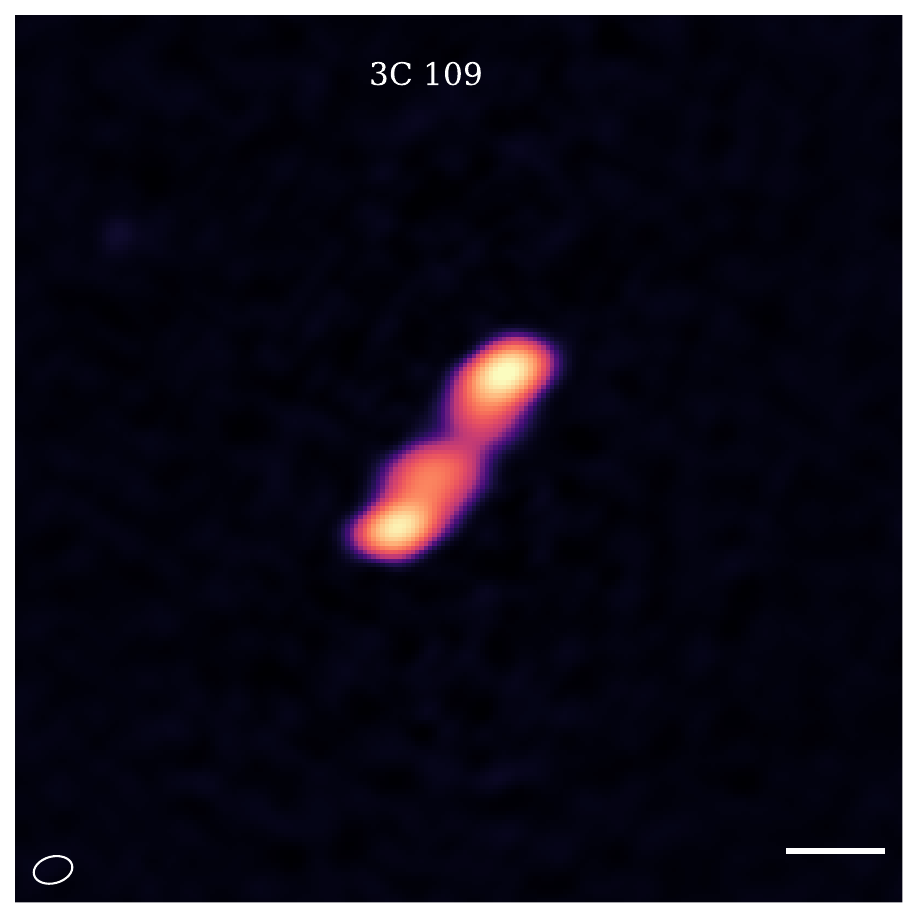}
\includegraphics[width=0.162\linewidth, trim={0.cm 0.cm 0.cm 0.cm},clip]{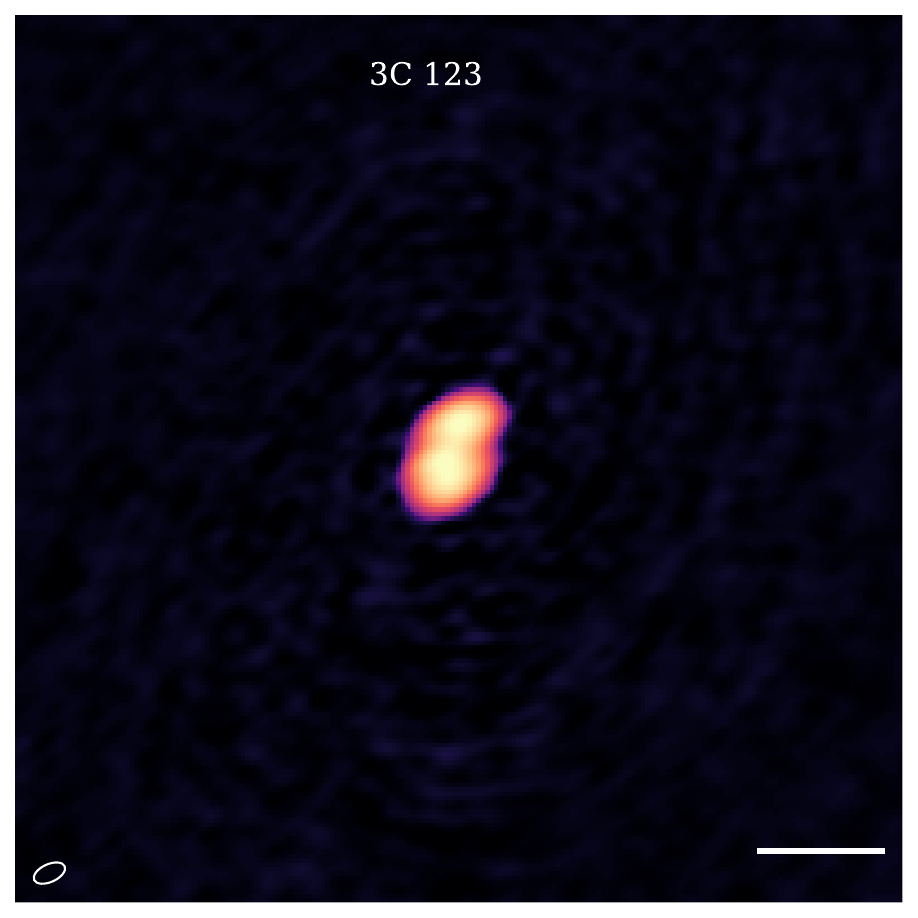}
\includegraphics[width=0.162\linewidth, trim={0.cm 0.cm 0.cm 0.cm},clip]{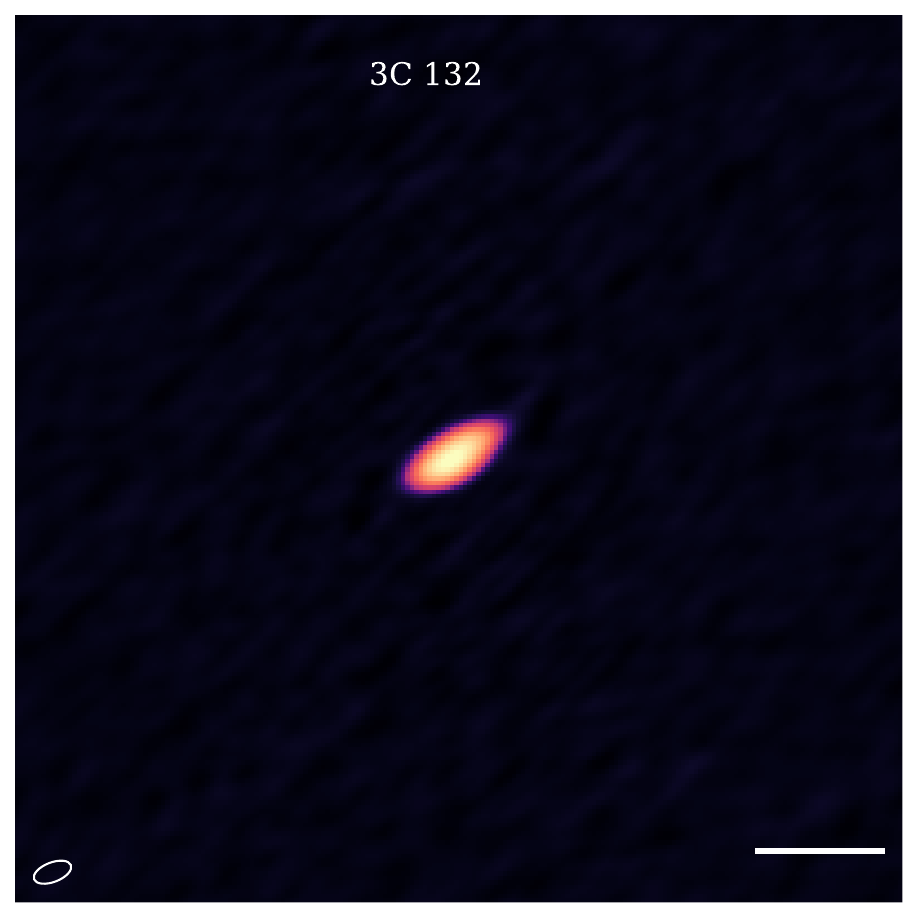}
\includegraphics[width=0.162\linewidth, trim={0.cm 0.cm 0.cm 0.cm},clip]{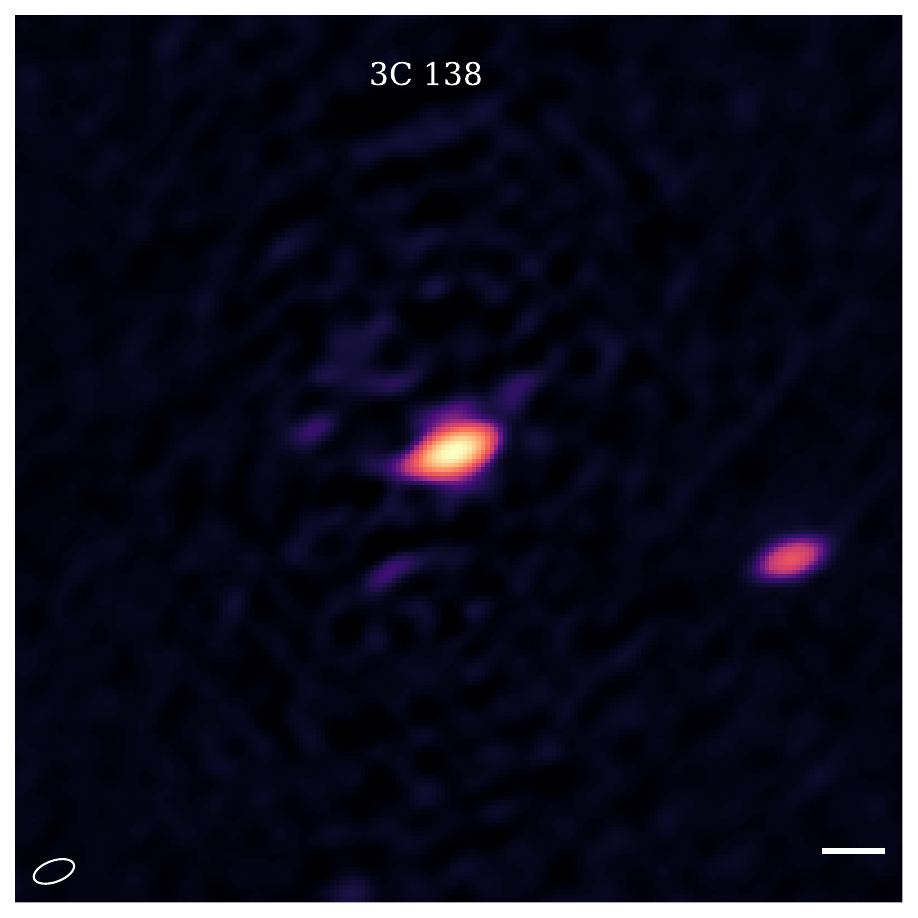}
\includegraphics[width=0.162\linewidth, trim={0.cm 0.cm 0.cm 0.cm},clip]{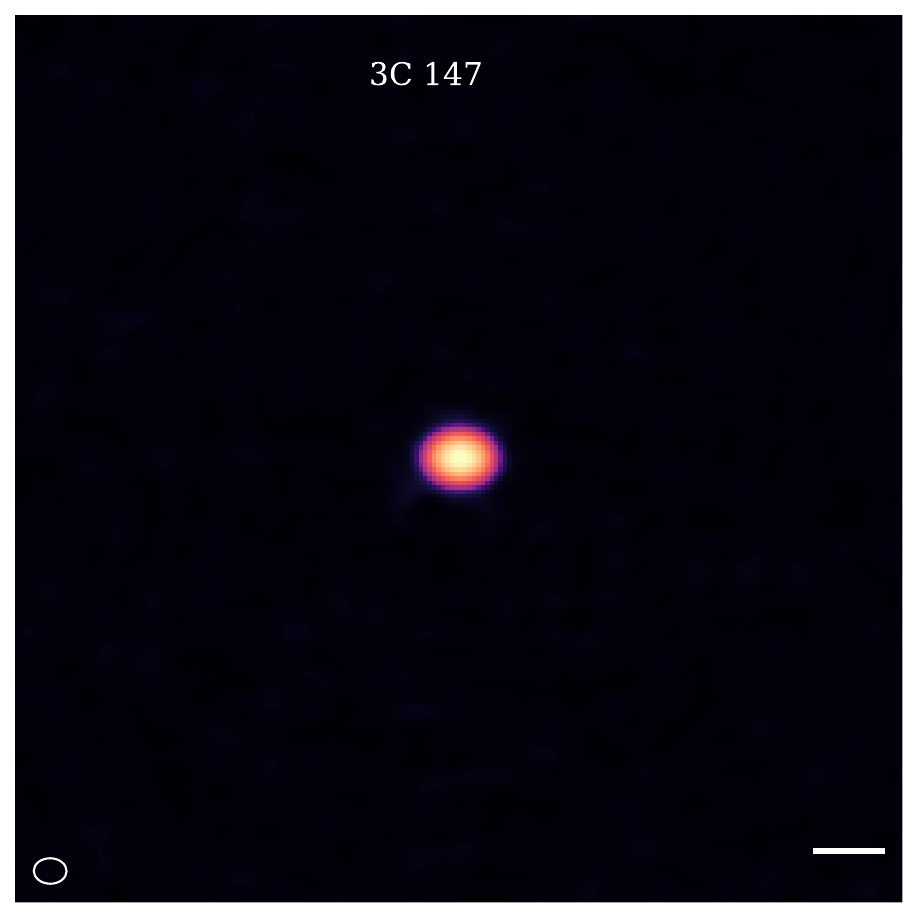}
\includegraphics[width=0.162\linewidth, trim={0.cm 0.cm 0.cm 0.cm},clip]{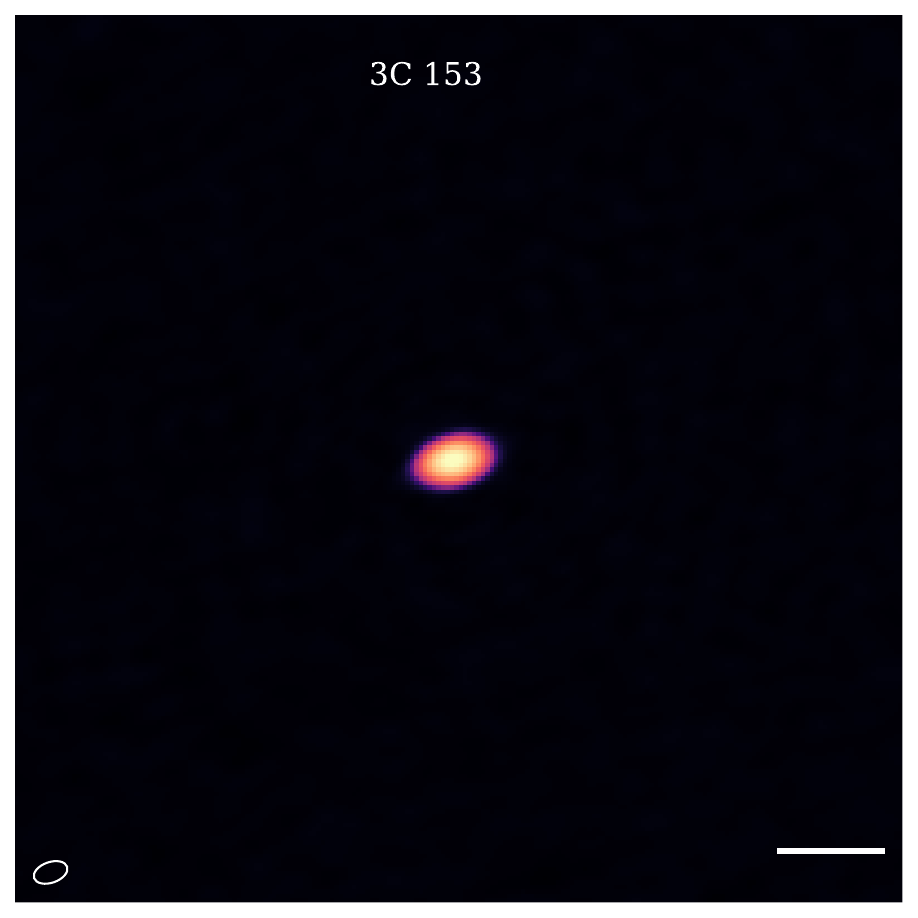}
\includegraphics[width=0.162\linewidth, trim={0.cm 0.cm 0.cm 0.cm},clip]{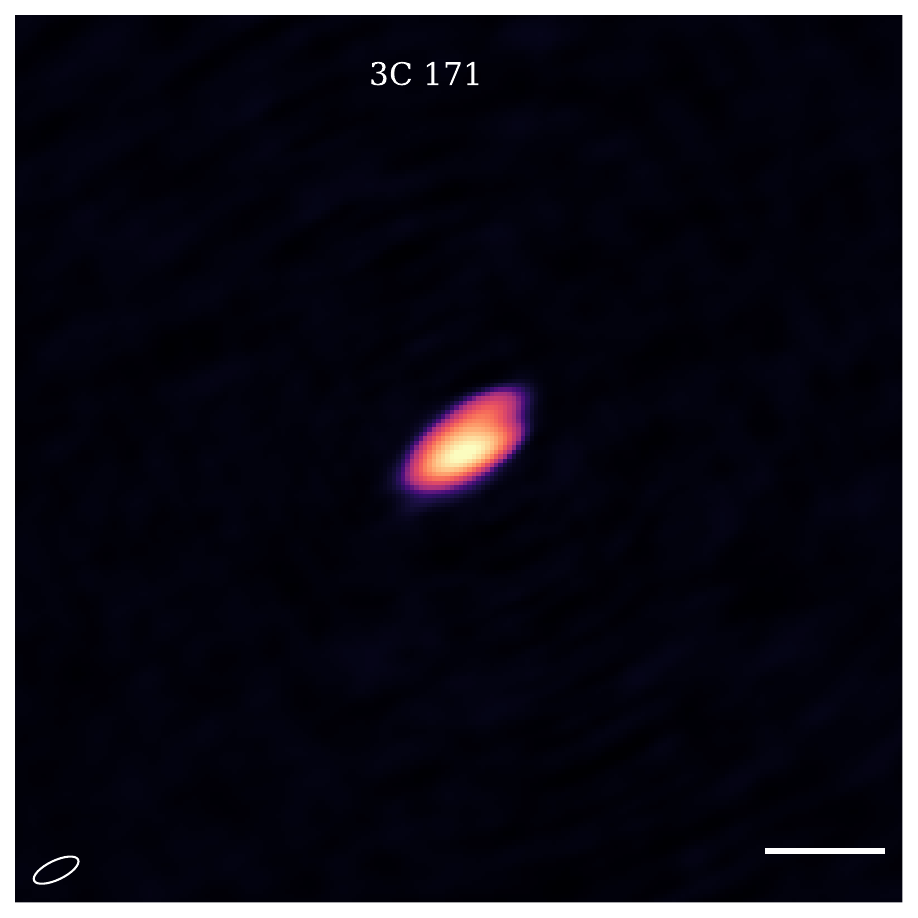}
\includegraphics[width=0.162\linewidth, trim={0.cm 0.cm 0.cm 0.cm},clip]{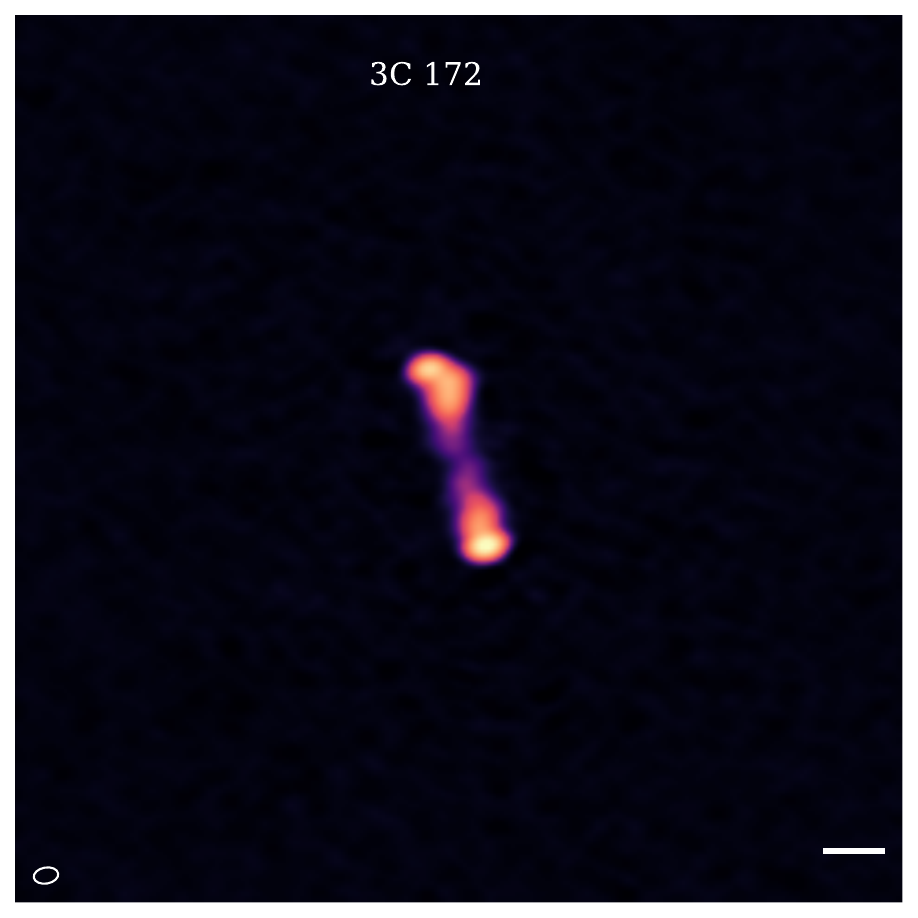}
\includegraphics[width=0.162\linewidth, trim={0.cm 0.cm 0.cm 0.cm},clip]{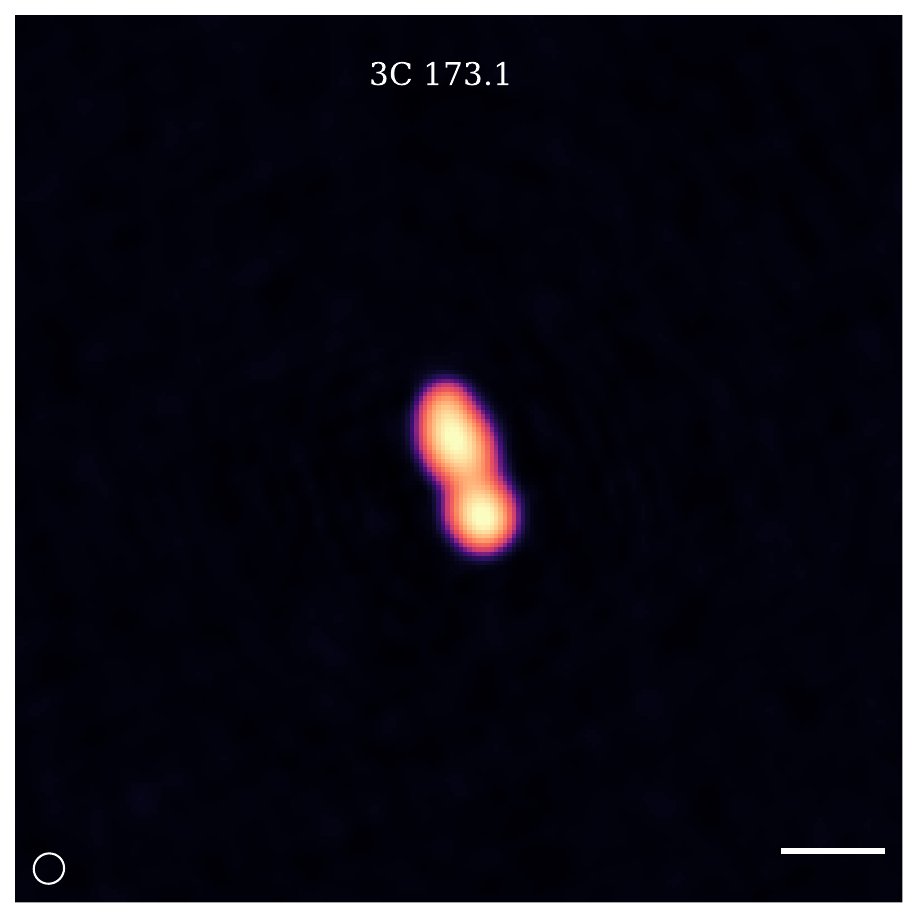}
\caption{Radio maps of 3C sources. The color map is not indicated and varies depending on the surface brightness of the source. Every map  indicates the beam for that map. the physical scale bar denotes 200~kpc with the exception of 3C~231 and 3C~272.1 (10~kpc), and 3C~274 and 3C~386 (100~kpc).}
\end{figure}
\setcounter{figure}{0} 

\begin{figure}[H]
\centering

\includegraphics[width=0.162\linewidth, trim={0.cm 0.cm 0.cm 0.cm},clip]{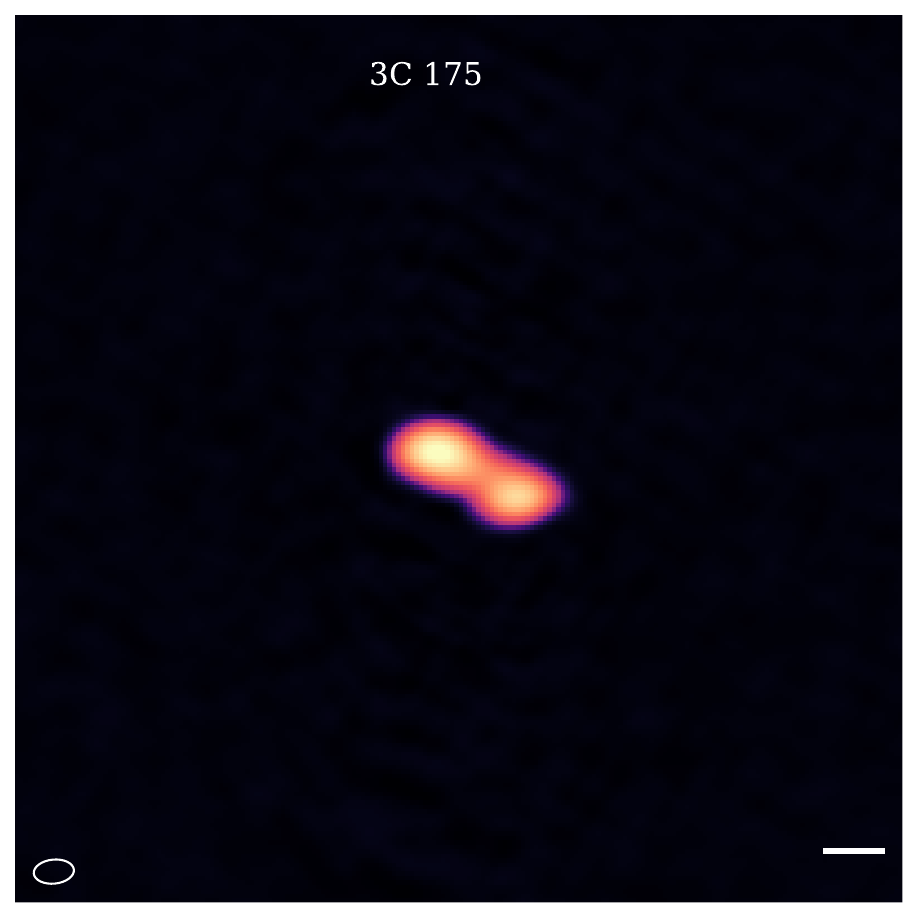}
\includegraphics[width=0.162\linewidth, trim={0.cm 0.cm 0.cm 0.cm},clip]{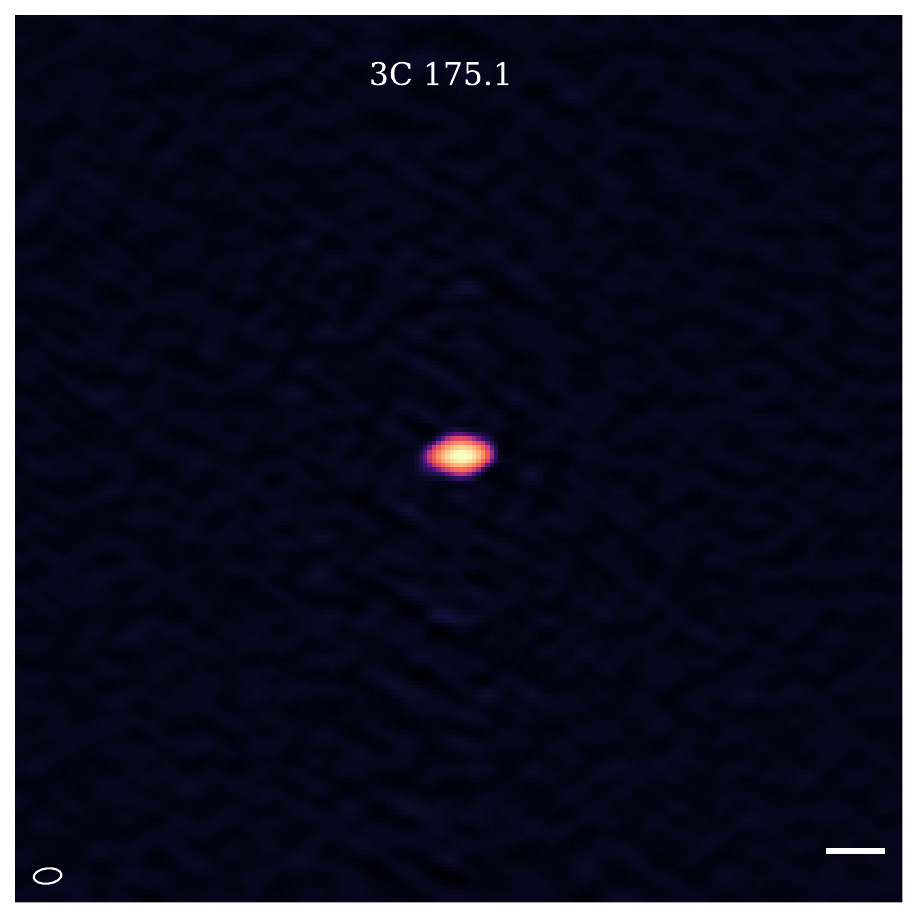}
\includegraphics[width=0.162\linewidth, trim={0.cm 0.cm 0.cm 0.cm},clip]{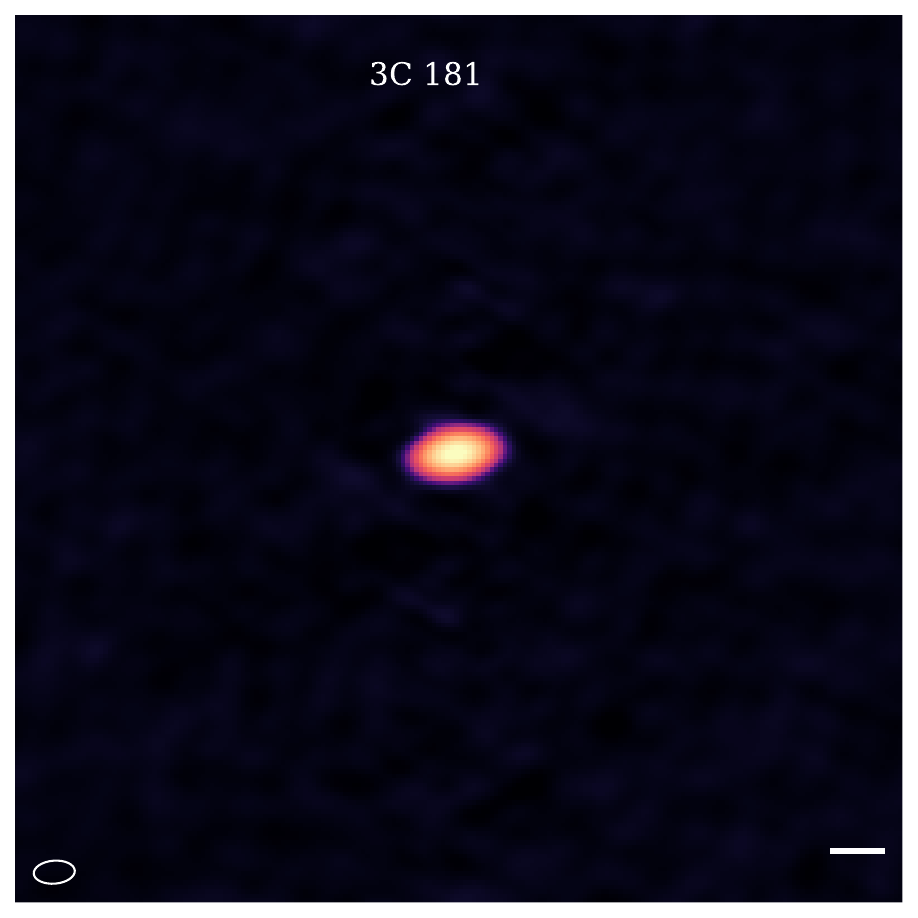}
\includegraphics[width=0.162\linewidth, trim={0.cm 0.cm 0.cm 0.cm},clip]{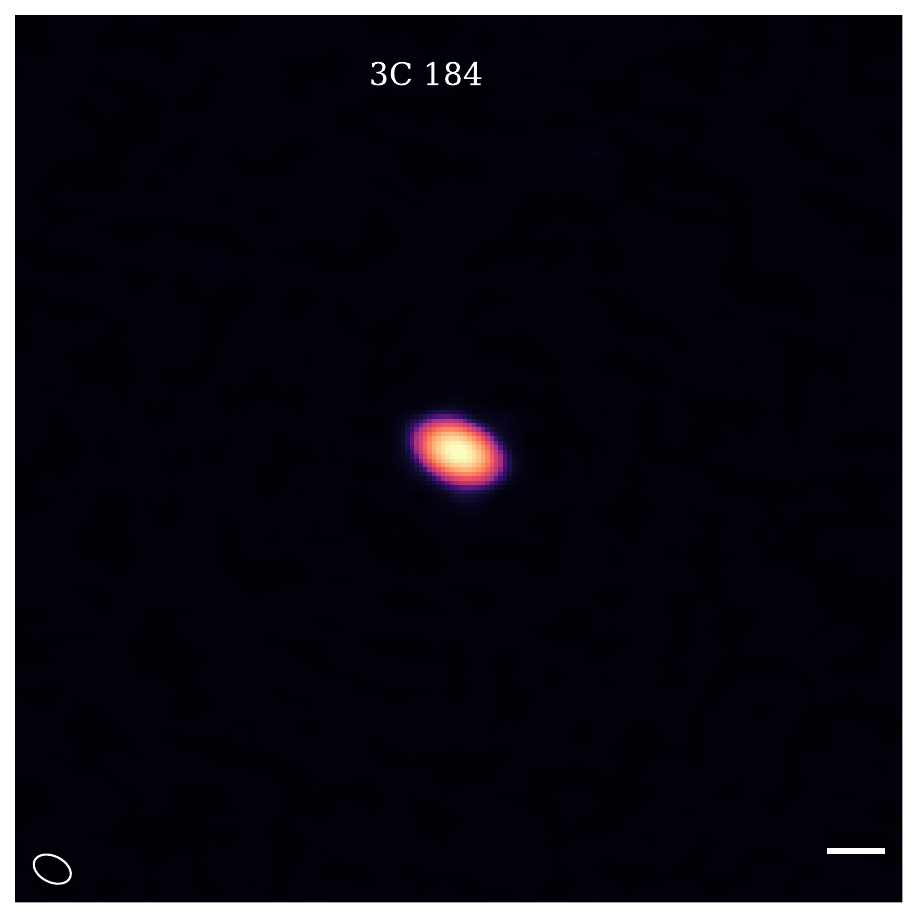}
\includegraphics[width=0.162\linewidth, trim={0.cm 0.cm 0.cm 0.cm},clip]{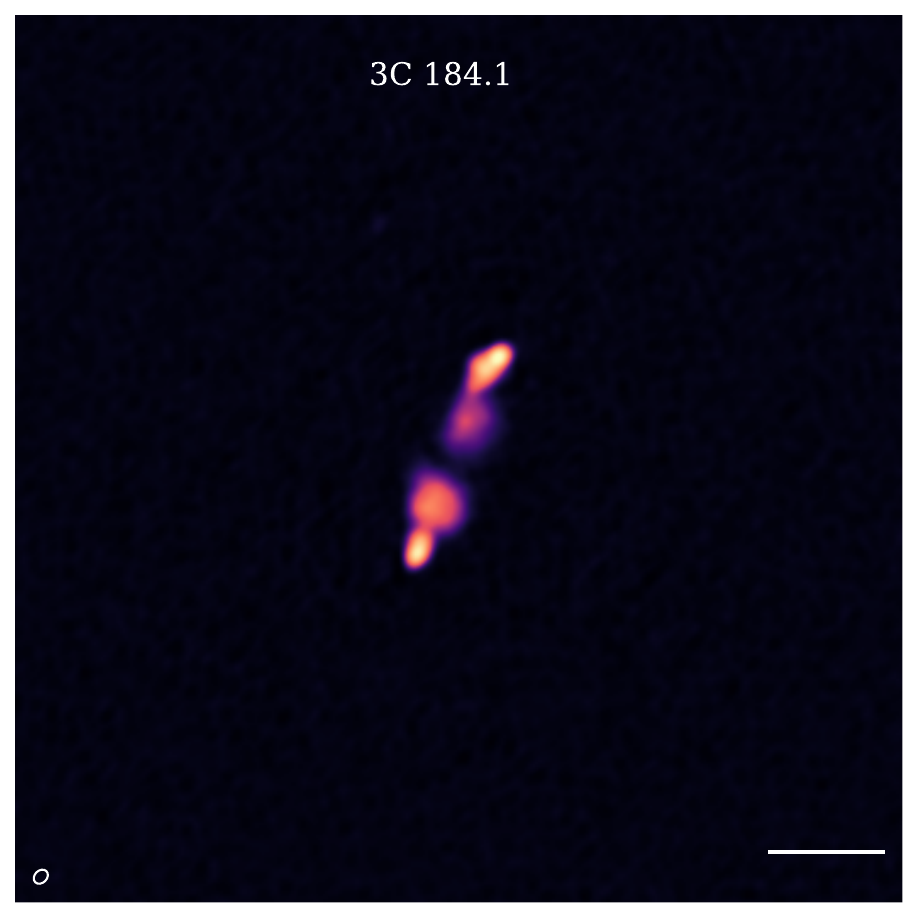}
\includegraphics[width=0.162\linewidth, trim={0.cm 0.cm 0.cm 0.cm},clip]{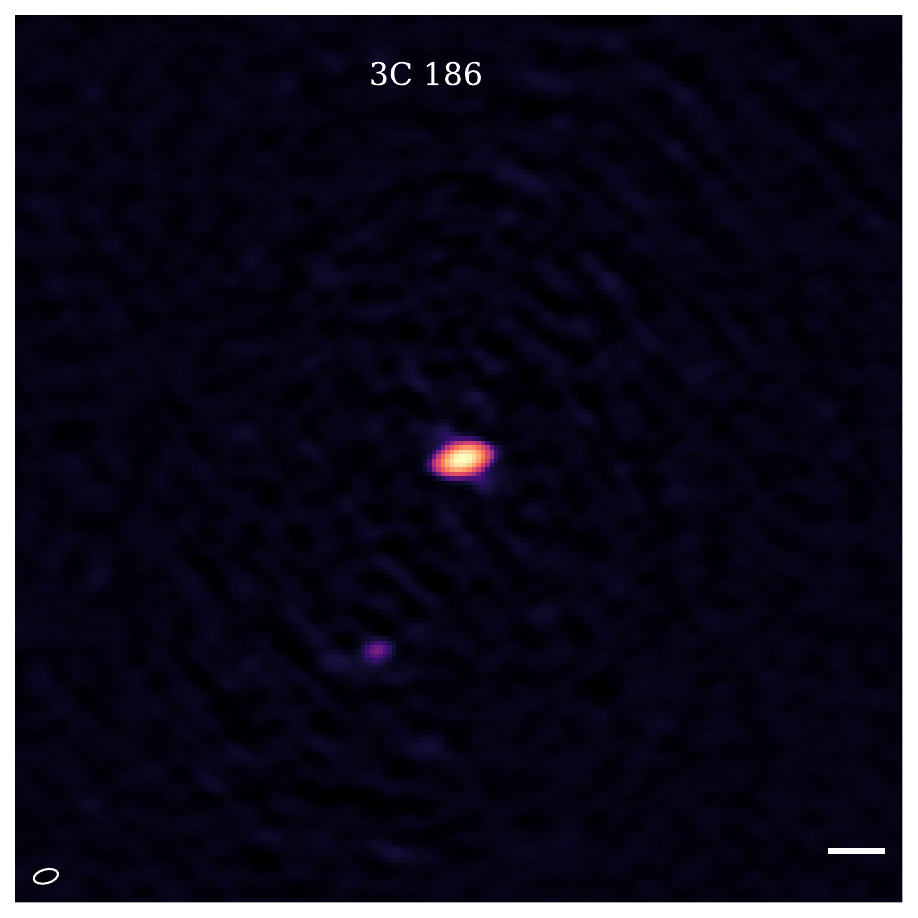}
\includegraphics[width=0.162\linewidth, trim={0.cm 0.cm 0.cm 0.cm},clip]{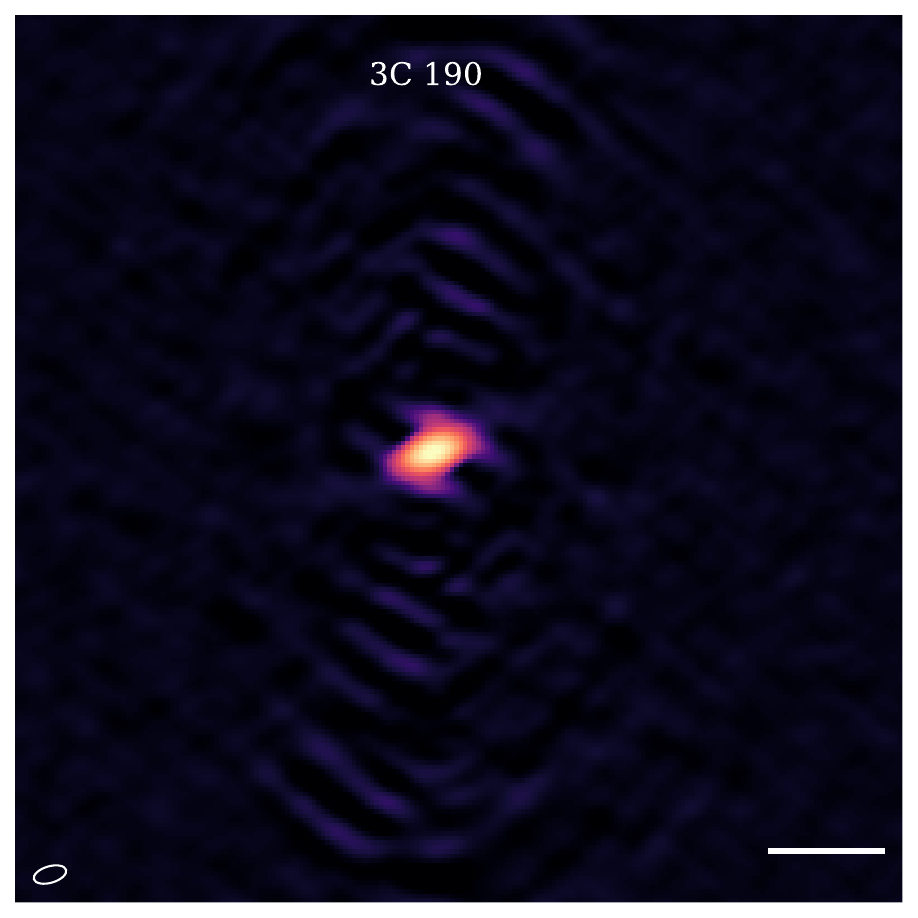}
\includegraphics[width=0.162\linewidth, trim={0.cm 0.cm 0.cm 0.cm},clip]{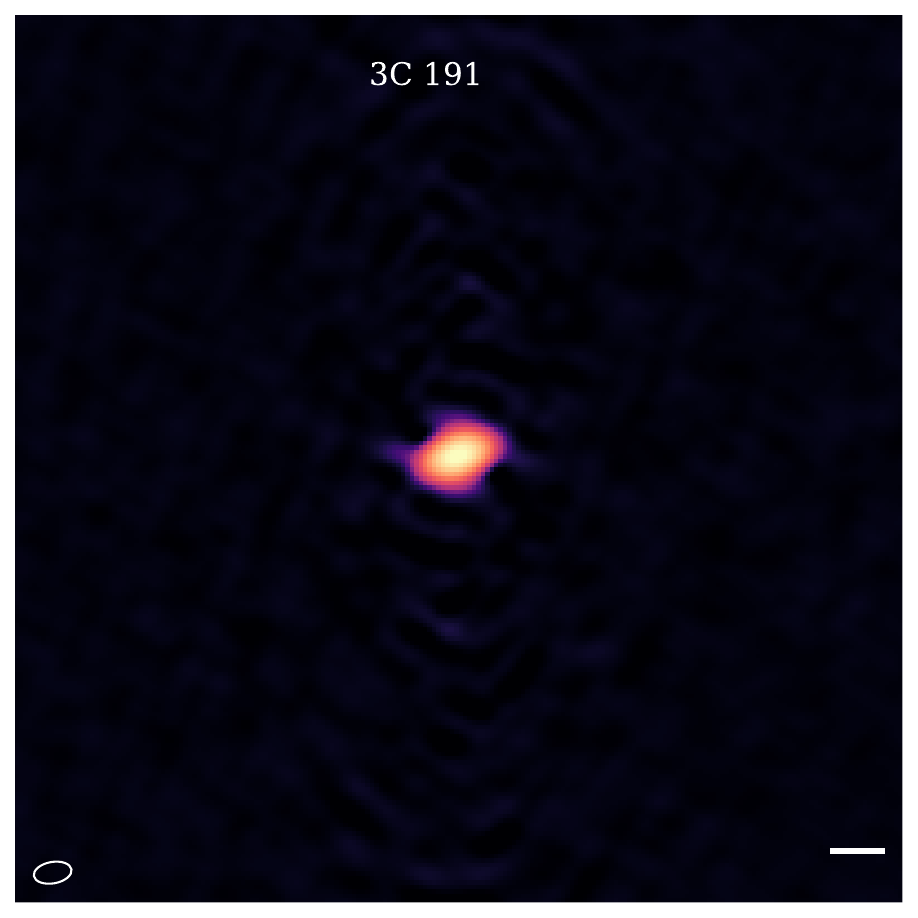}
\includegraphics[width=0.162\linewidth, trim={0.cm 0.cm 0.cm 0.cm},clip]{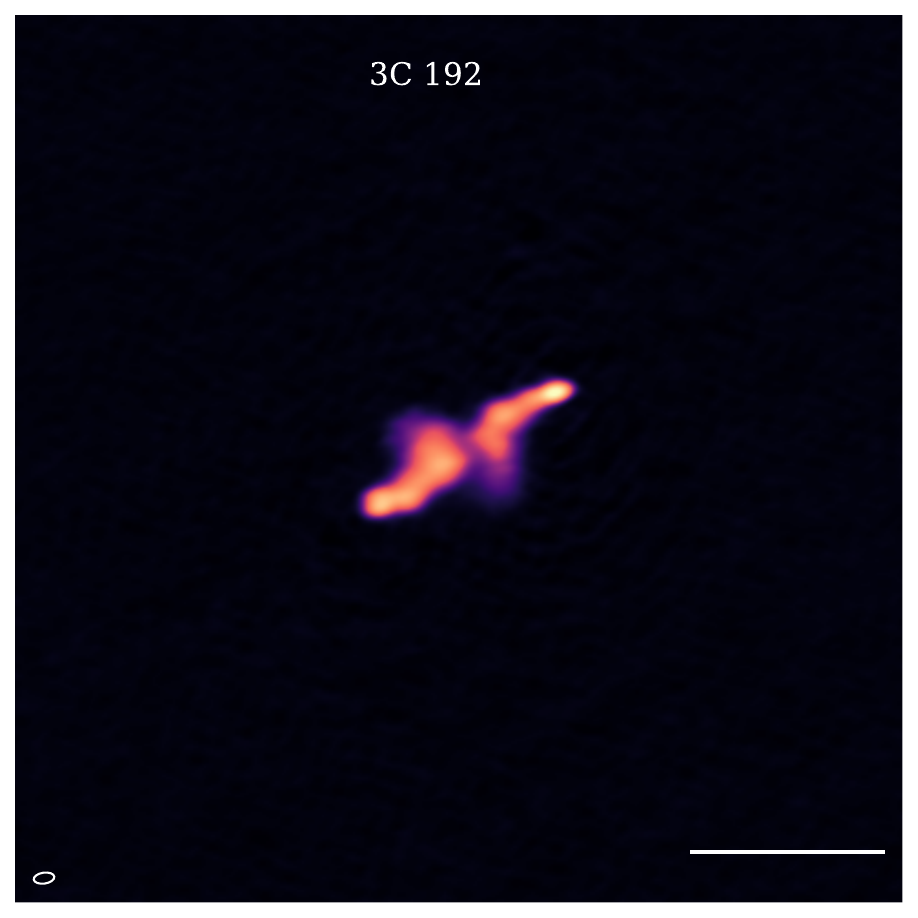}
\includegraphics[width=0.162\linewidth, trim={0.cm 0.cm 0.cm 0.cm},clip]{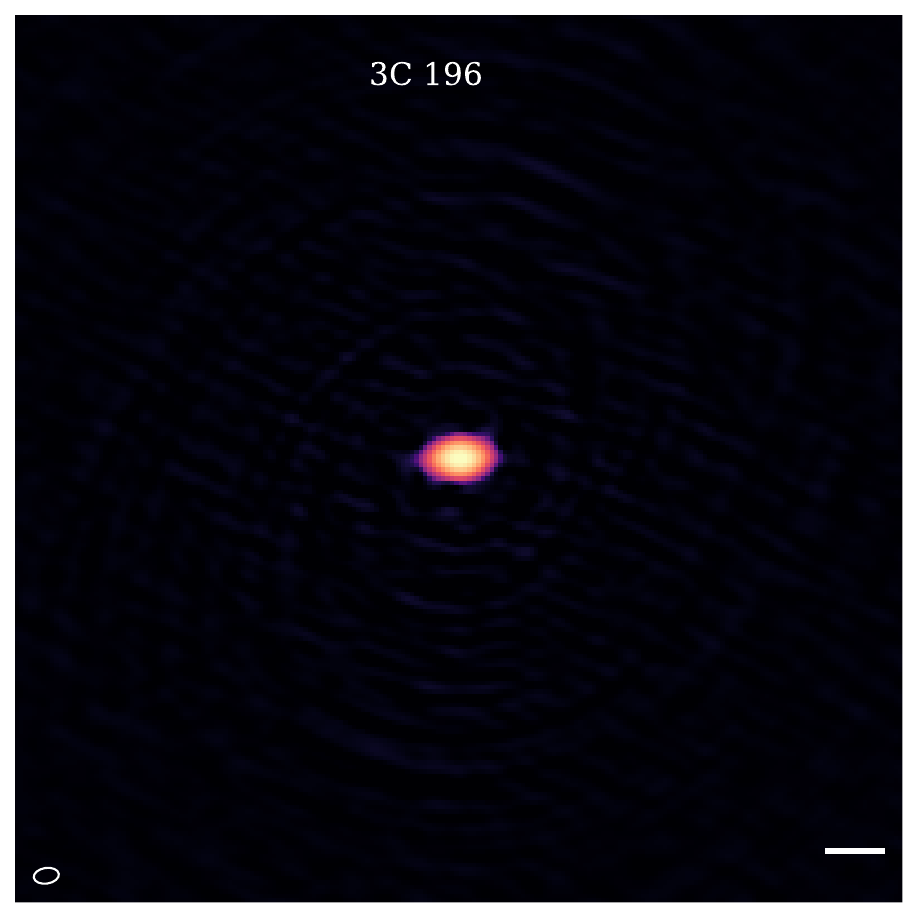}
\includegraphics[width=0.162\linewidth, trim={0.cm 0.cm 0.cm 0.cm},clip]{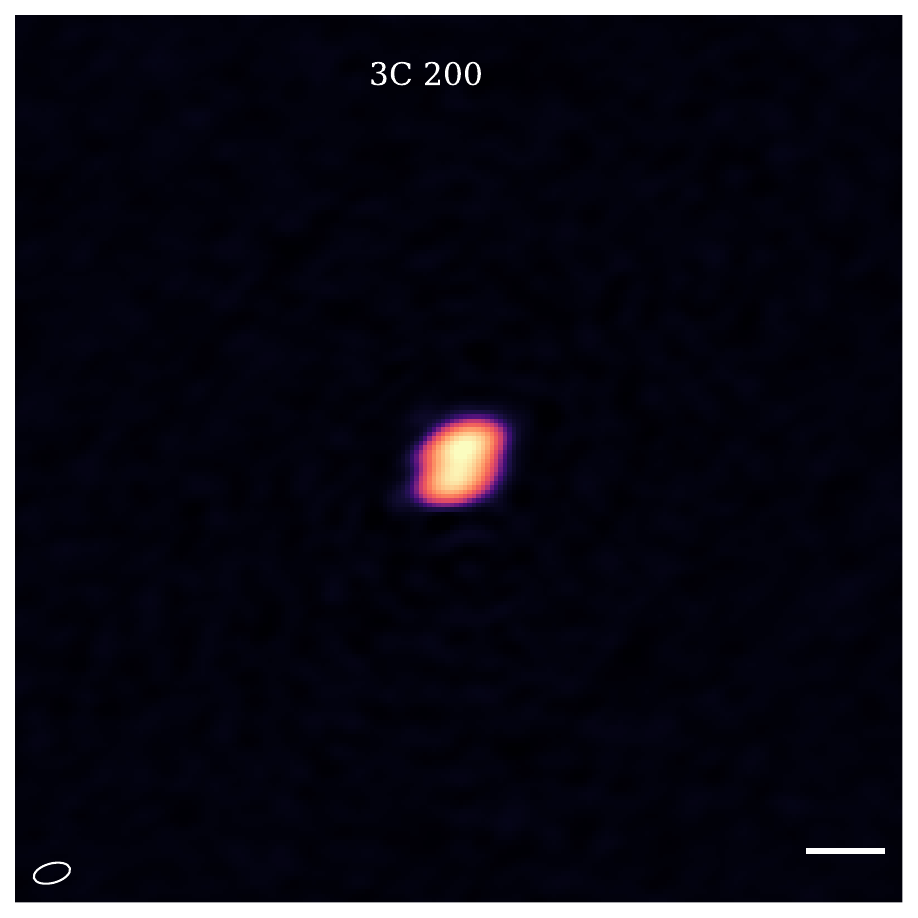}
\includegraphics[width=0.162\linewidth, trim={0.cm 0.cm 0.cm 0.cm},clip]{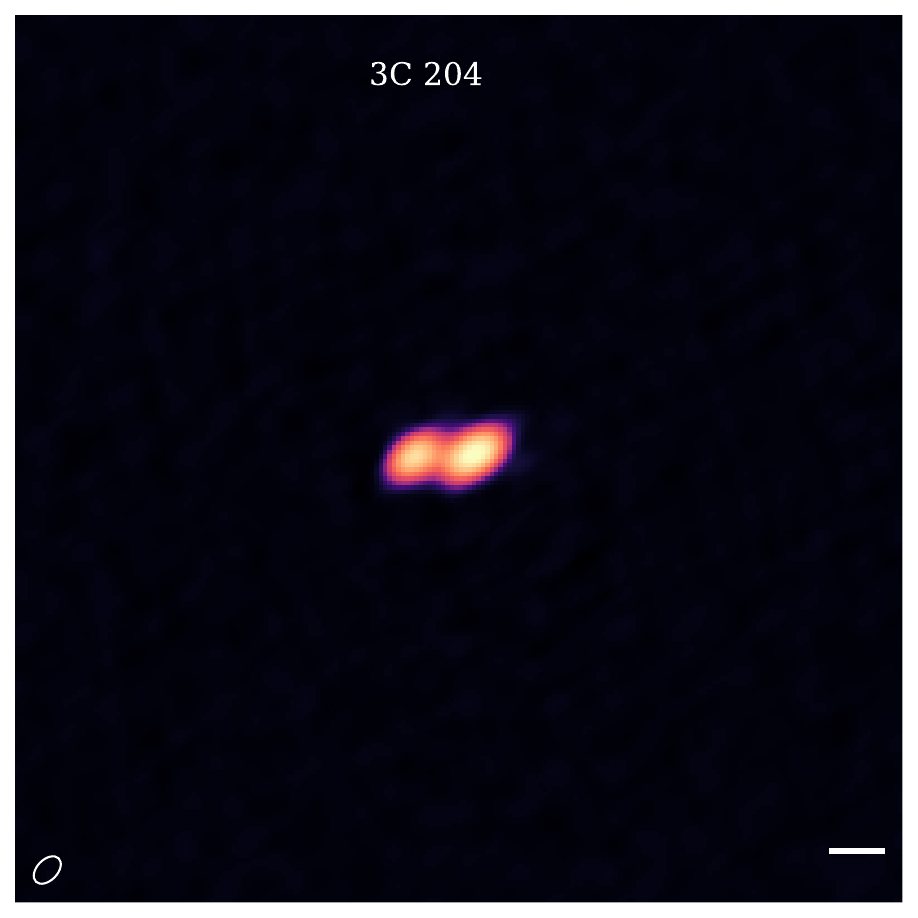}
\includegraphics[width=0.162\linewidth, trim={0.cm 0.cm 0.cm 0.cm},clip]{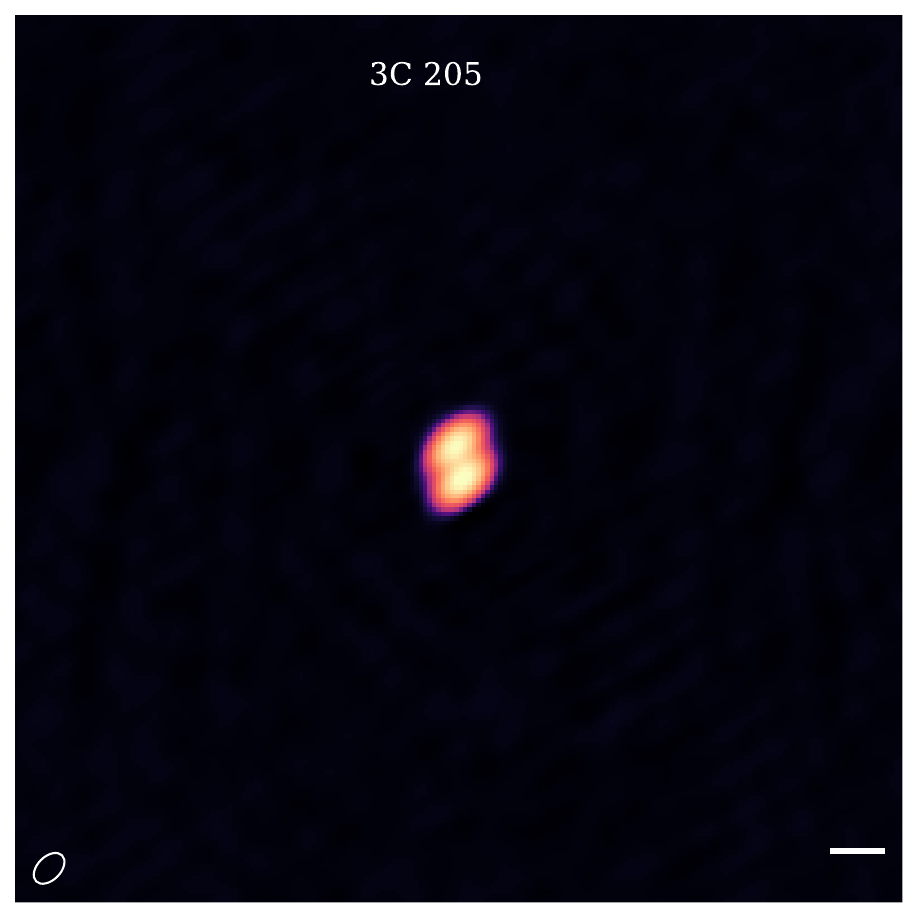}
\includegraphics[width=0.162\linewidth, trim={0.cm 0.cm 0.cm 0.cm},clip]{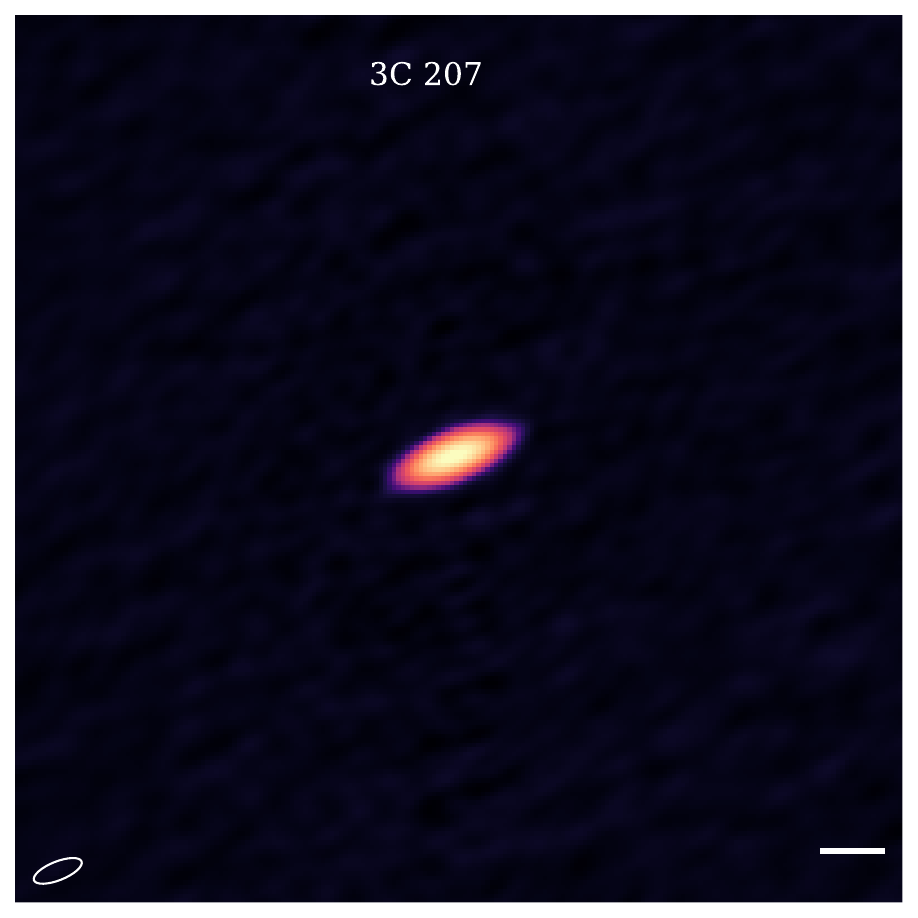}
\includegraphics[width=0.162\linewidth, trim={0.cm 0.cm 0.cm 0.cm},clip]{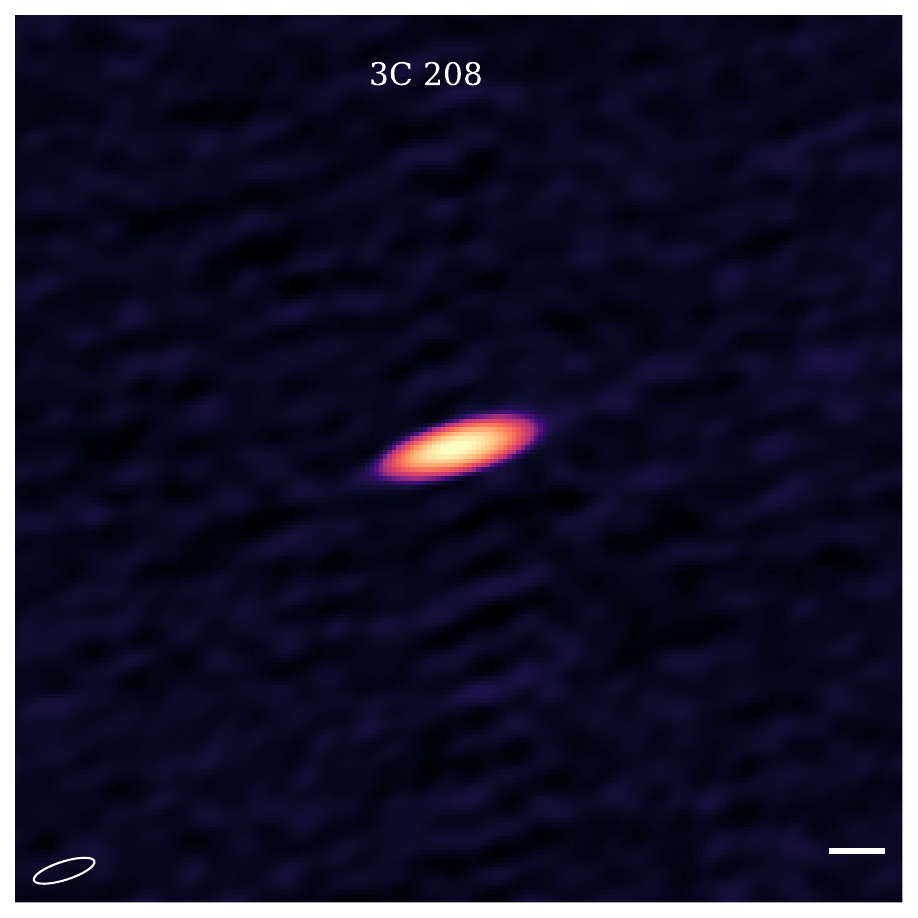}
\includegraphics[width=0.162\linewidth, trim={0.cm 0.cm 0.cm 0.cm},clip]{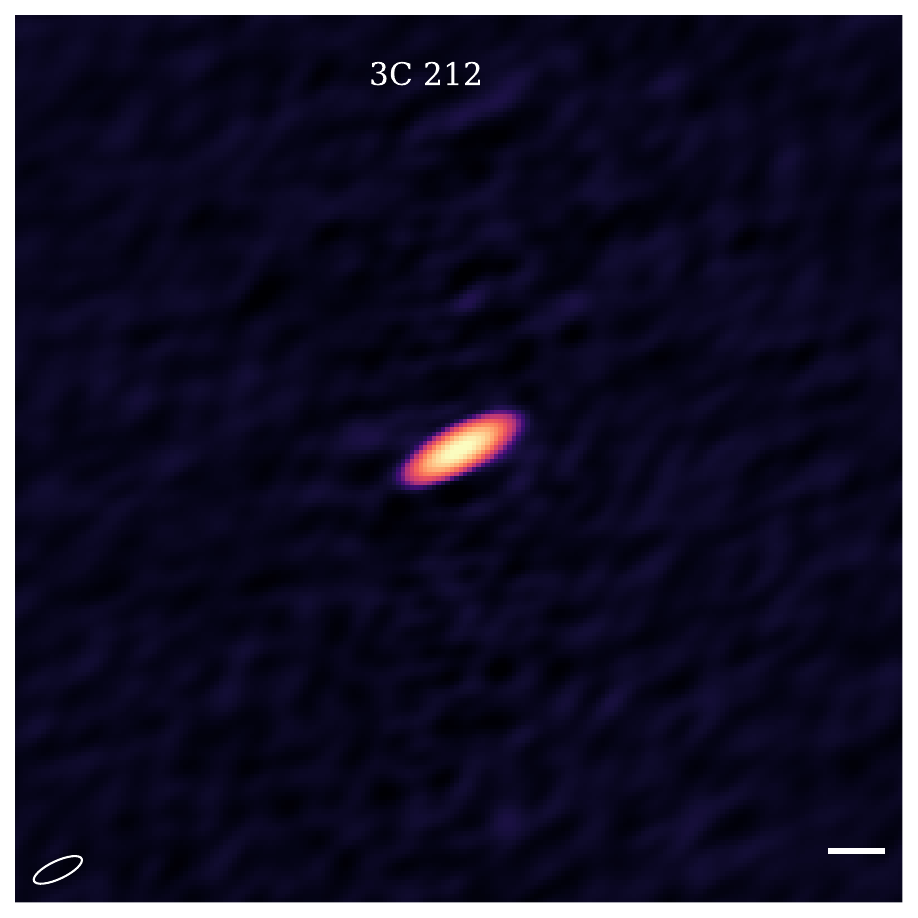}
\includegraphics[width=0.162\linewidth, trim={0.cm 0.cm 0.cm 0.cm},clip]{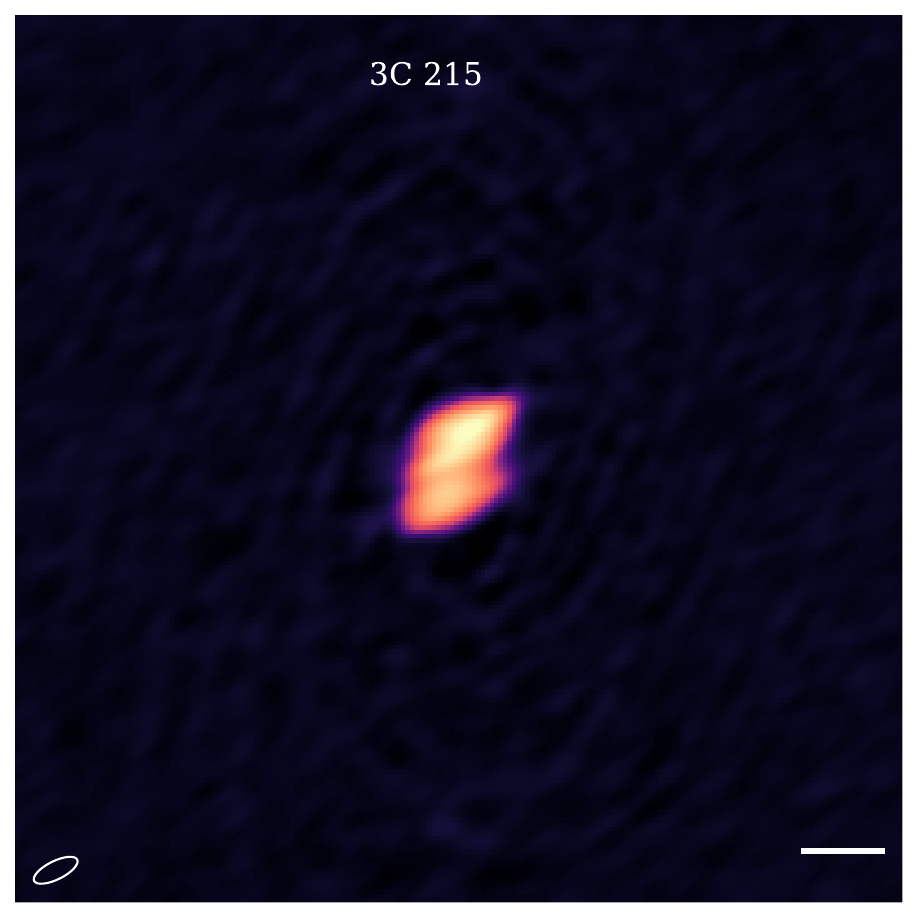}
\includegraphics[width=0.162\linewidth, trim={0.cm 0.cm 0.cm 0.cm},clip]{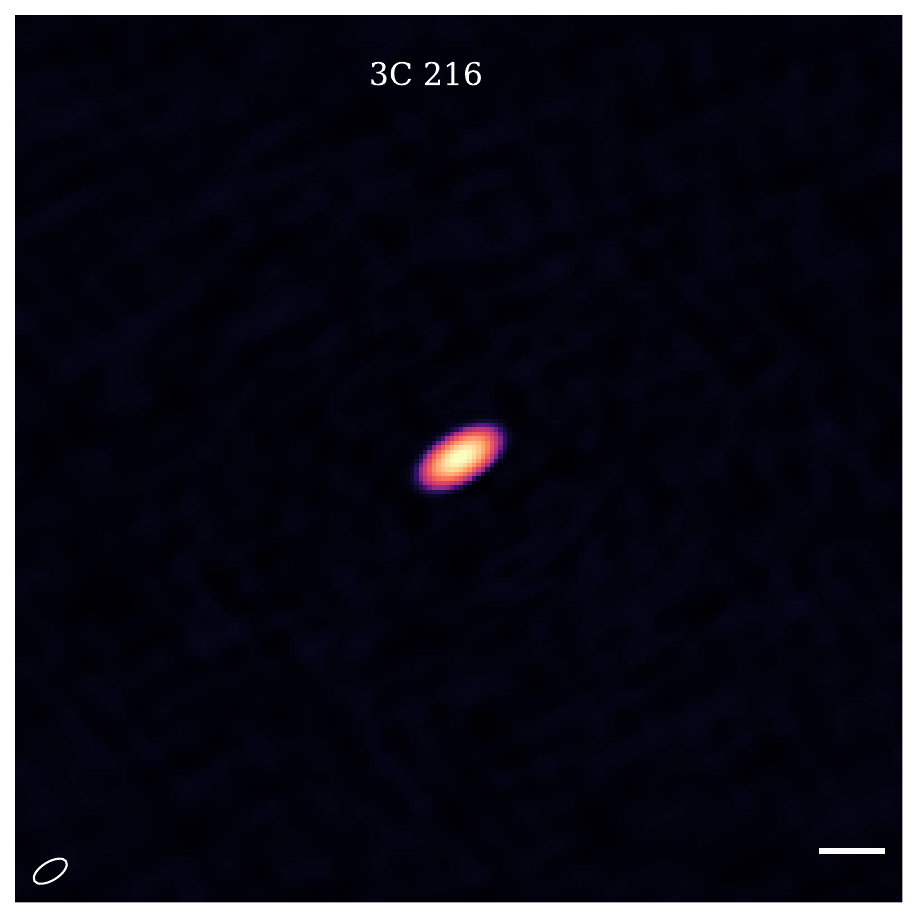}
\includegraphics[width=0.162\linewidth, trim={0.cm 0.cm 0.cm 0.cm},clip]{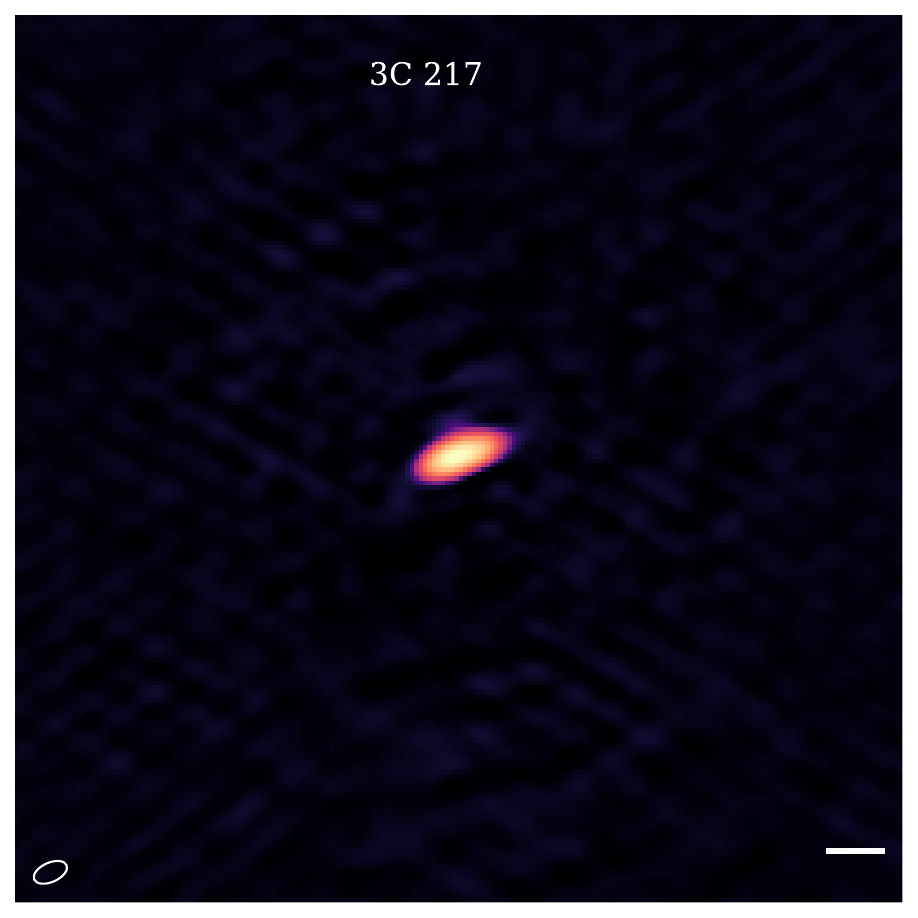}
\includegraphics[width=0.162\linewidth, trim={0.cm 0.cm 0.cm 0.cm},clip]{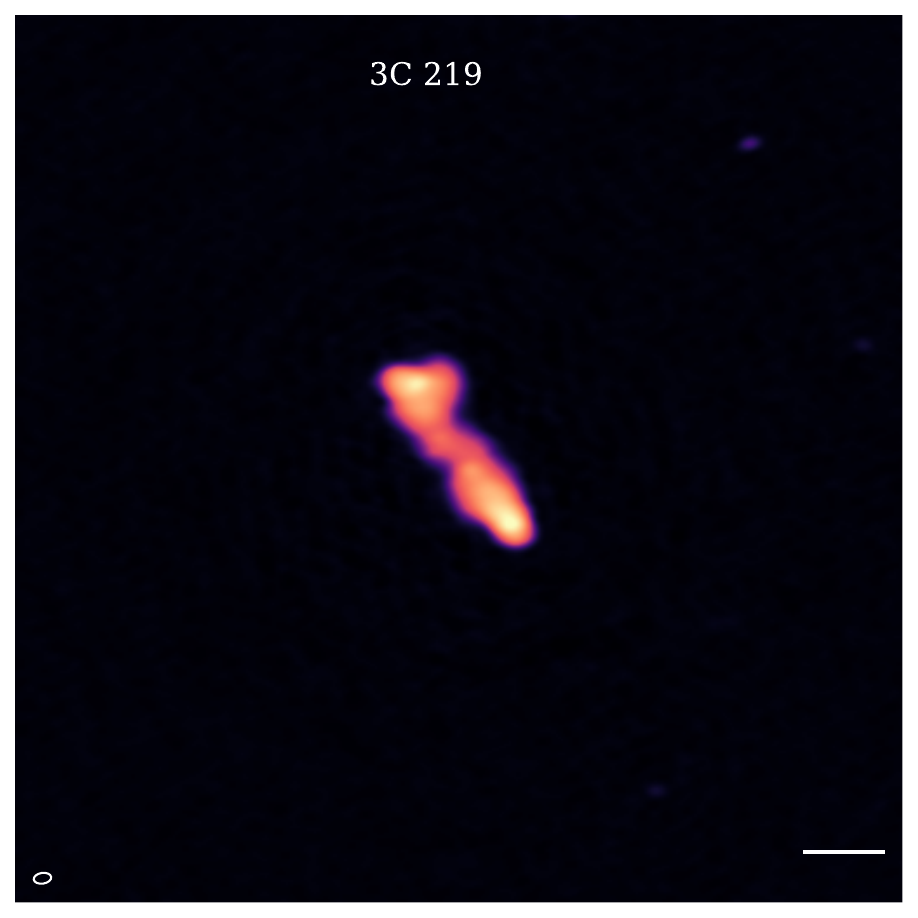}
\includegraphics[width=0.162\linewidth, trim={0.cm 0.cm 0.cm 0.cm},clip]{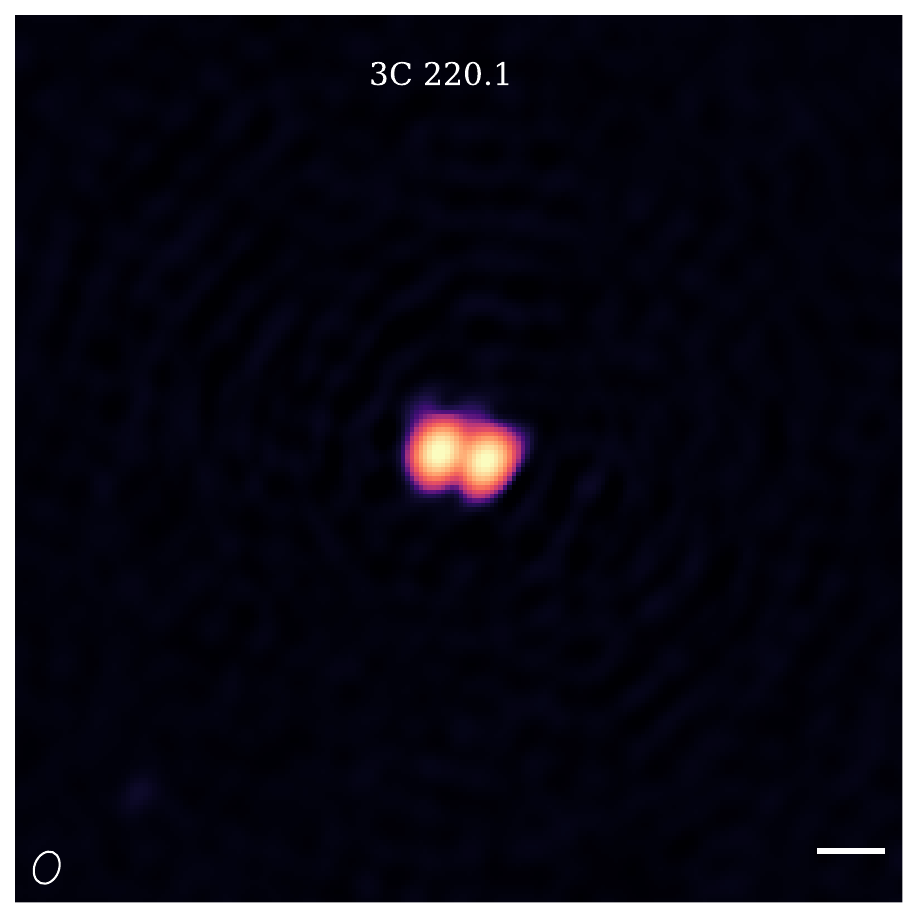}
\includegraphics[width=0.162\linewidth, trim={0.cm 0.cm 0.cm 0.cm},clip]{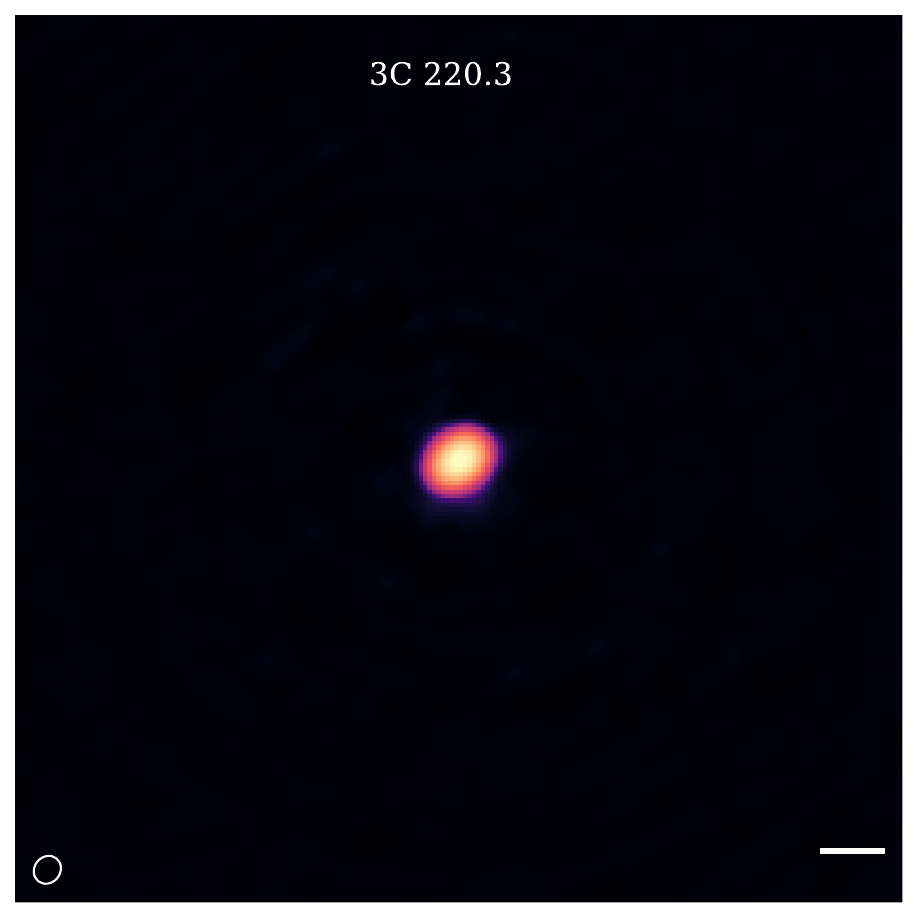}
\includegraphics[width=0.162\linewidth, trim={0.cm 0.cm 0.cm 0.cm},clip]{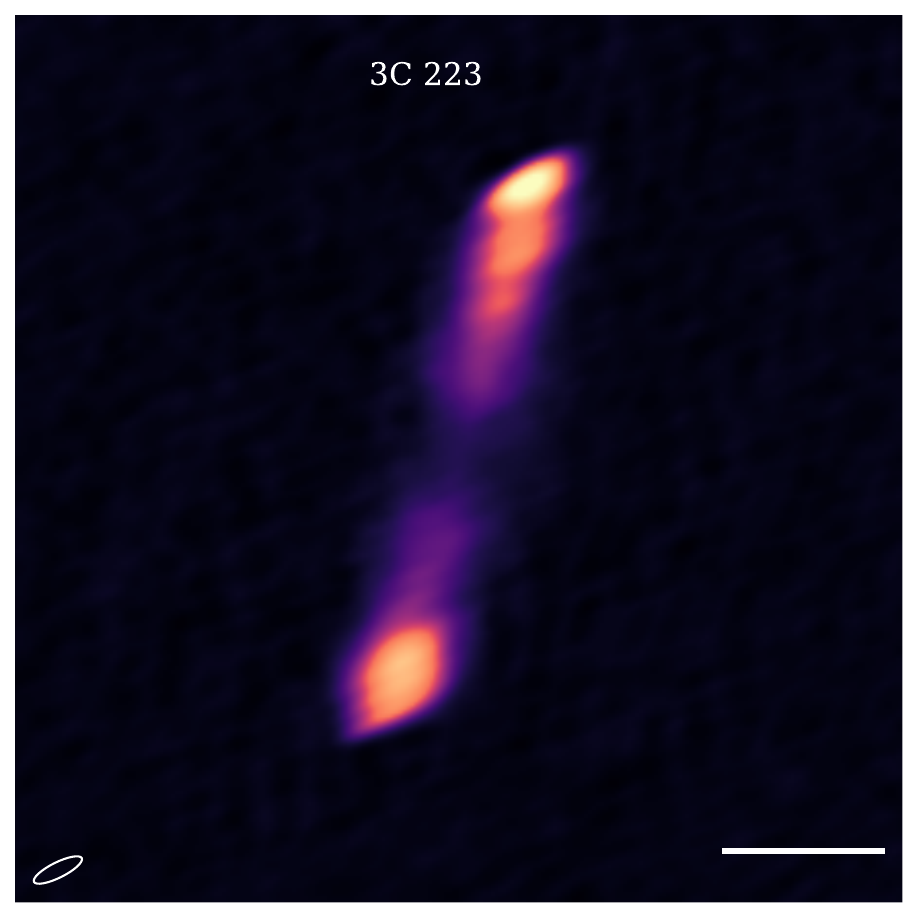}
\includegraphics[width=0.162\linewidth, trim={0.cm 0.cm 0.cm 0.cm},clip]{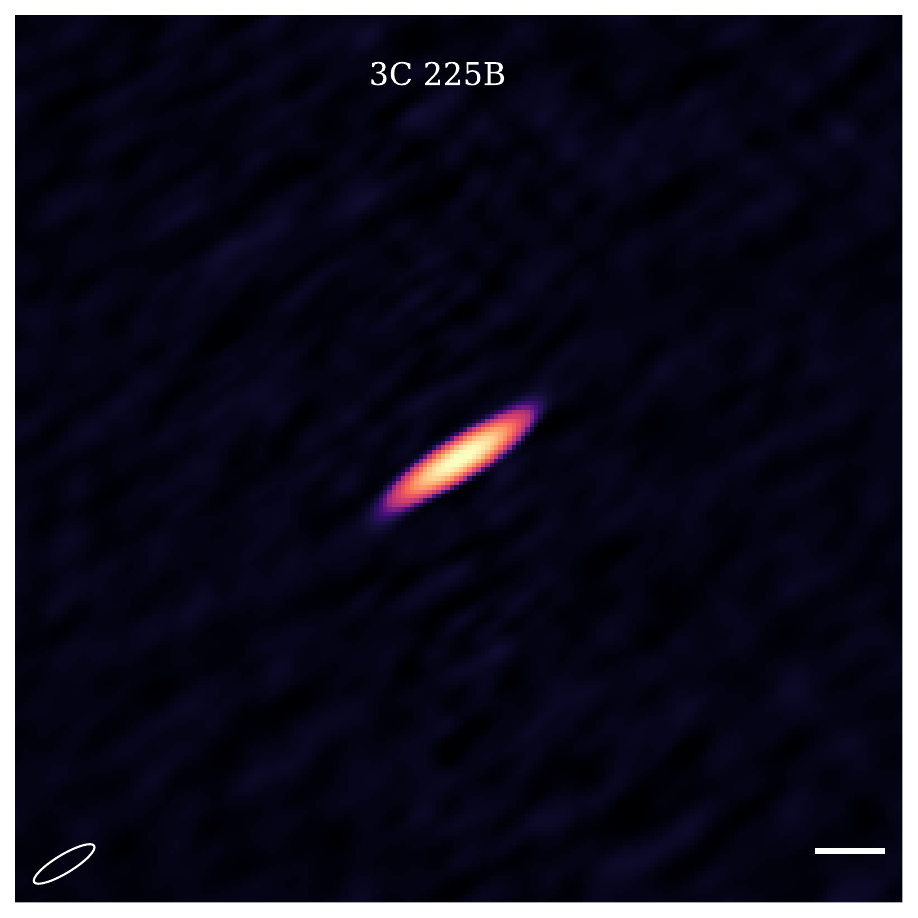}
\includegraphics[width=0.162\linewidth, trim={0.cm 0.cm 0.cm 0.cm},clip]{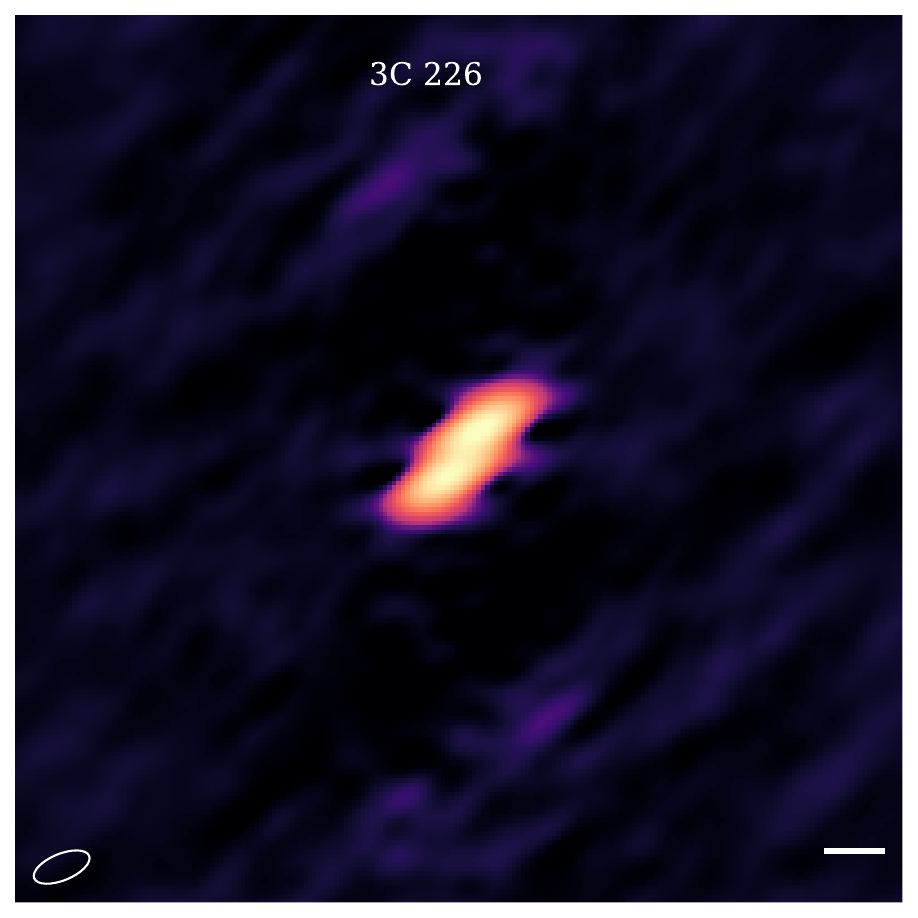}
\includegraphics[width=0.162\linewidth, trim={0.cm 0.cm 0.cm 0.cm},clip]{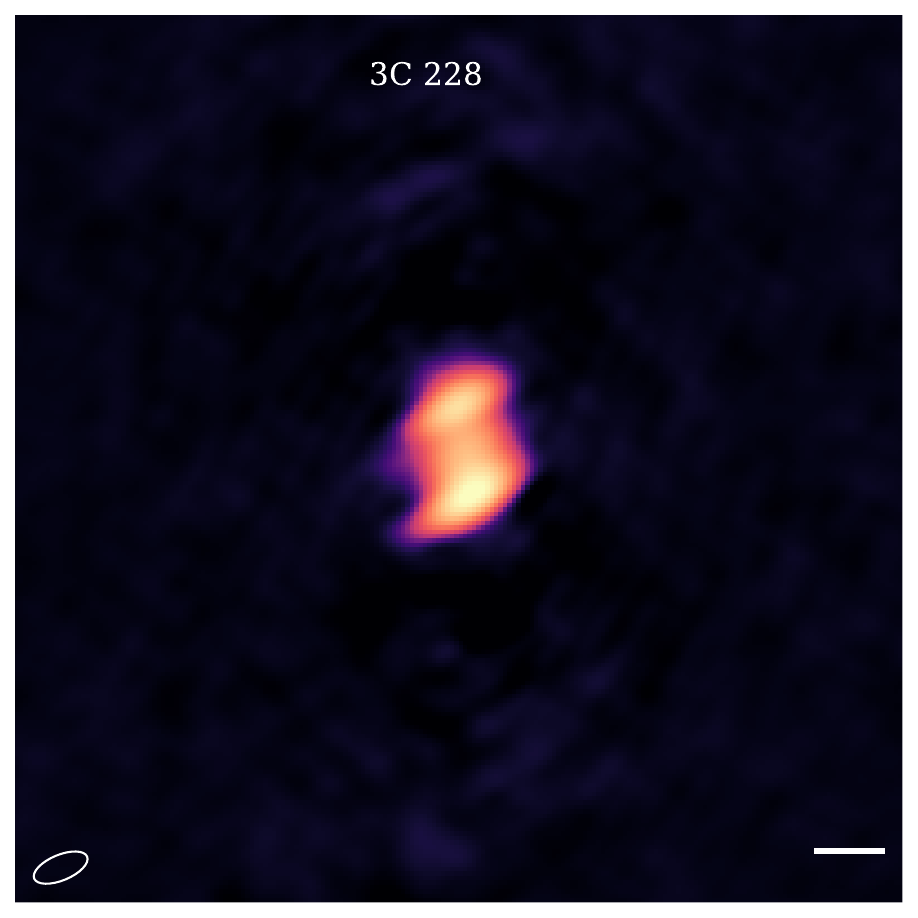}
\includegraphics[width=0.162\linewidth, trim={0.cm 0.cm 0.cm 0.cm},clip]{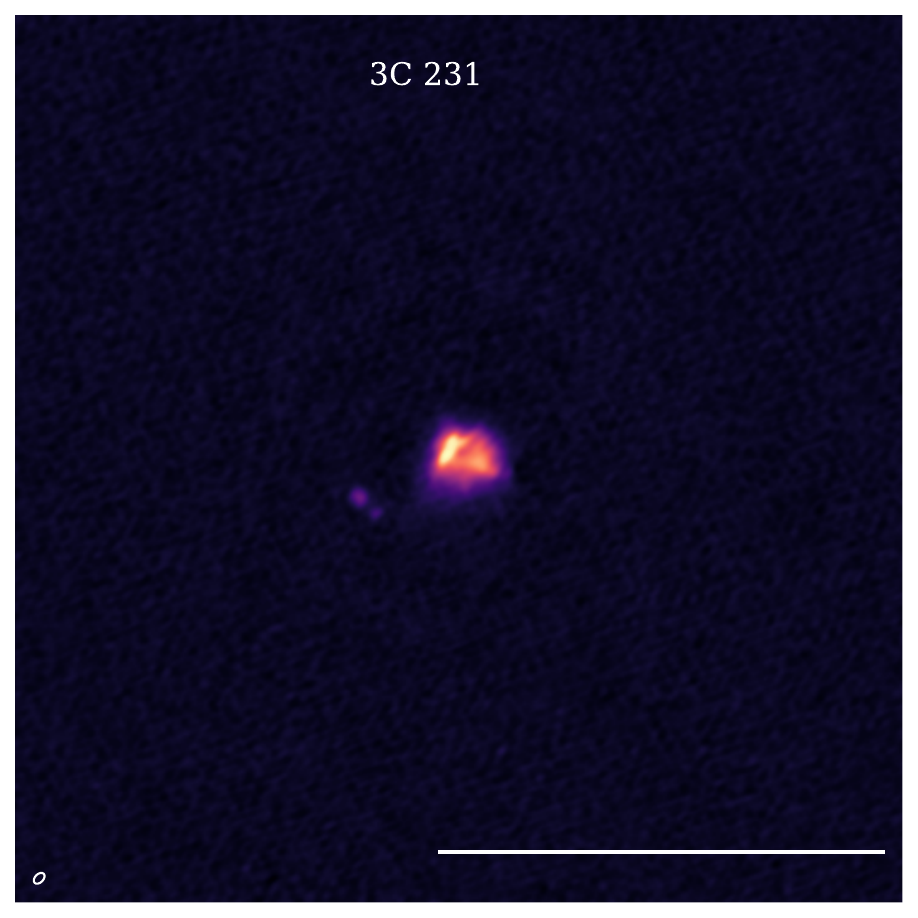}
\includegraphics[width=0.162\linewidth, trim={0.cm 0.cm 0.cm 0.cm},clip]{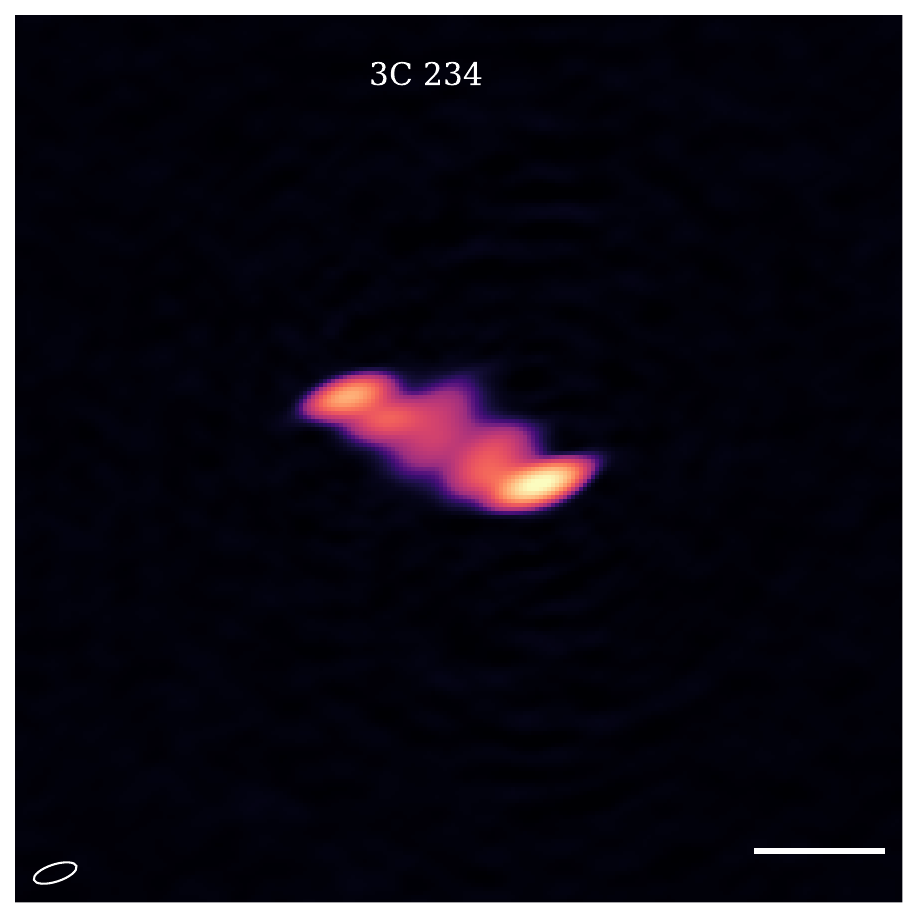}
\includegraphics[width=0.162\linewidth, trim={0.cm 0.cm 0.cm 0.cm},clip]{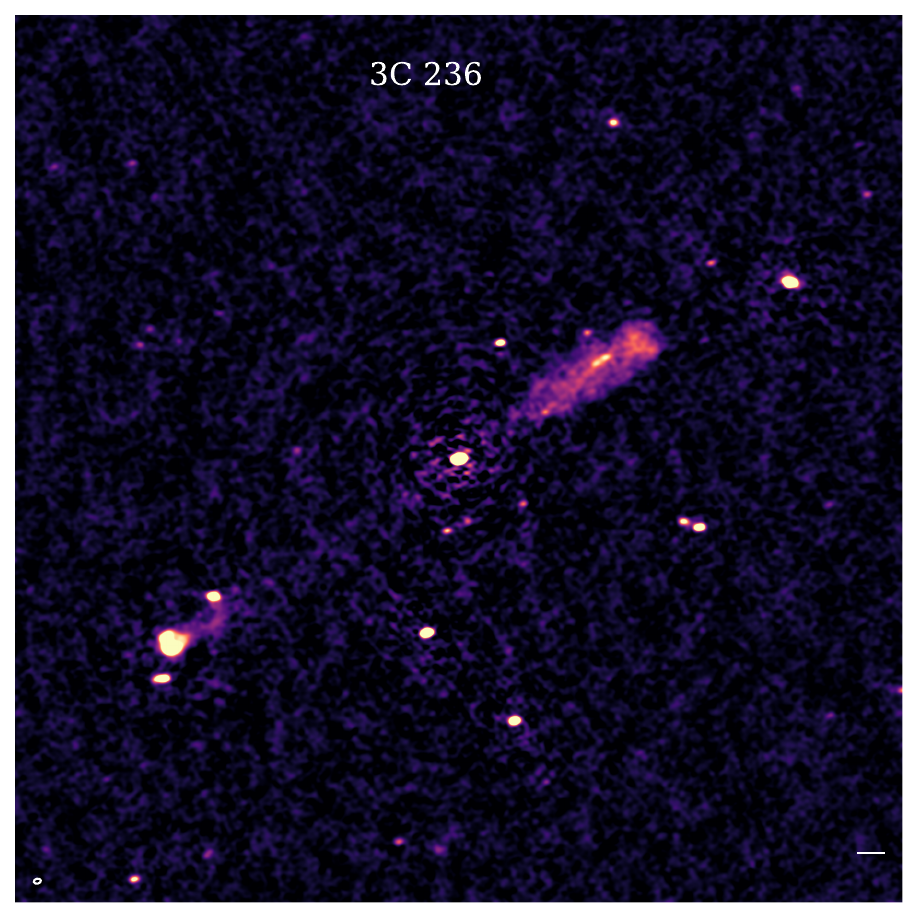}
\includegraphics[width=0.162\linewidth, trim={0.cm 0.cm 0.cm 0.cm},clip]{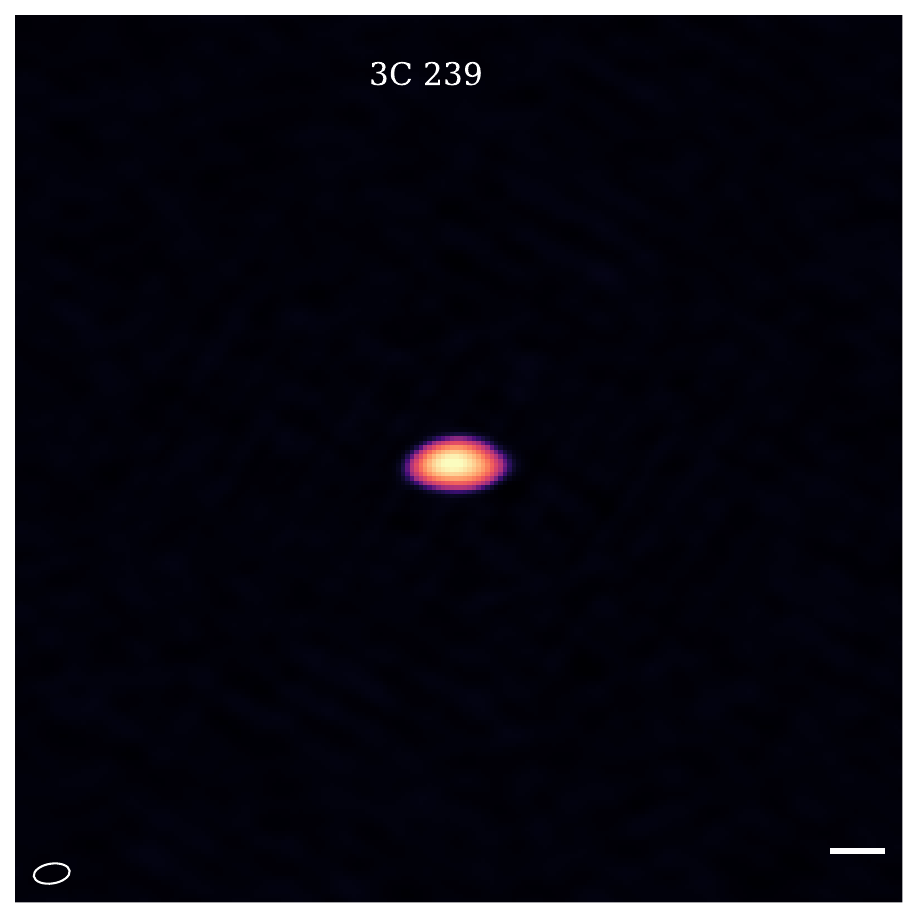}
\includegraphics[width=0.162\linewidth, trim={0.cm 0.cm 0.cm 0.cm},clip]{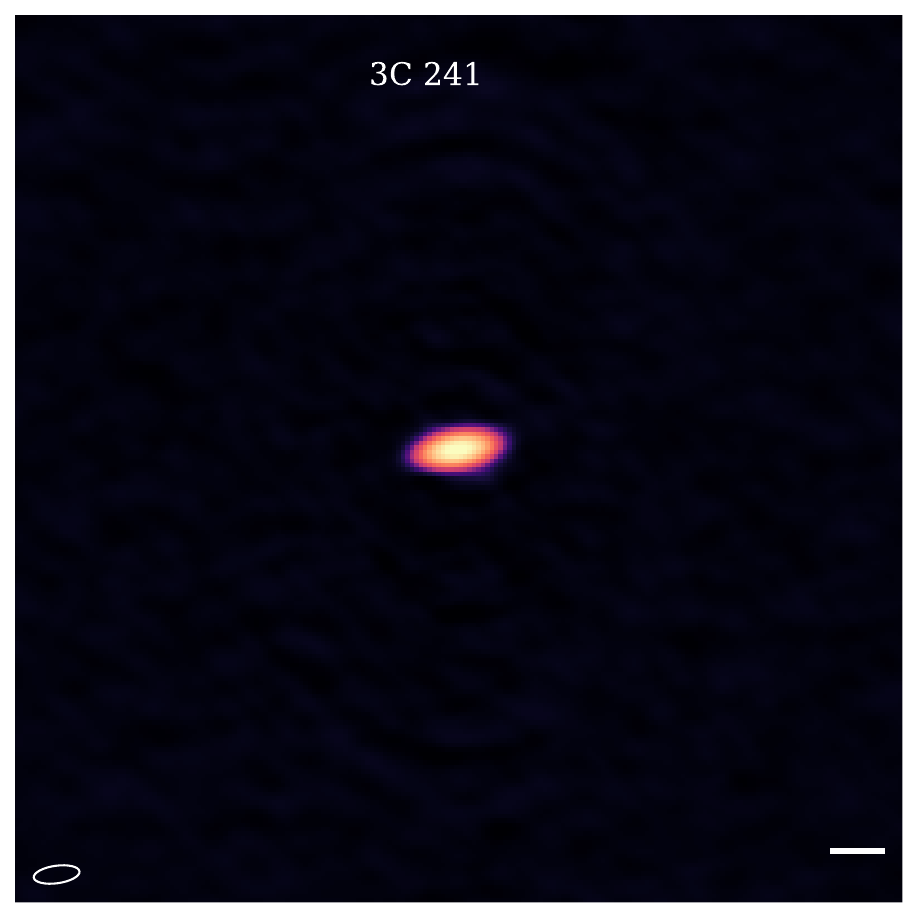}
\includegraphics[width=0.162\linewidth, trim={0.cm 0.cm 0.cm 0.cm},clip]{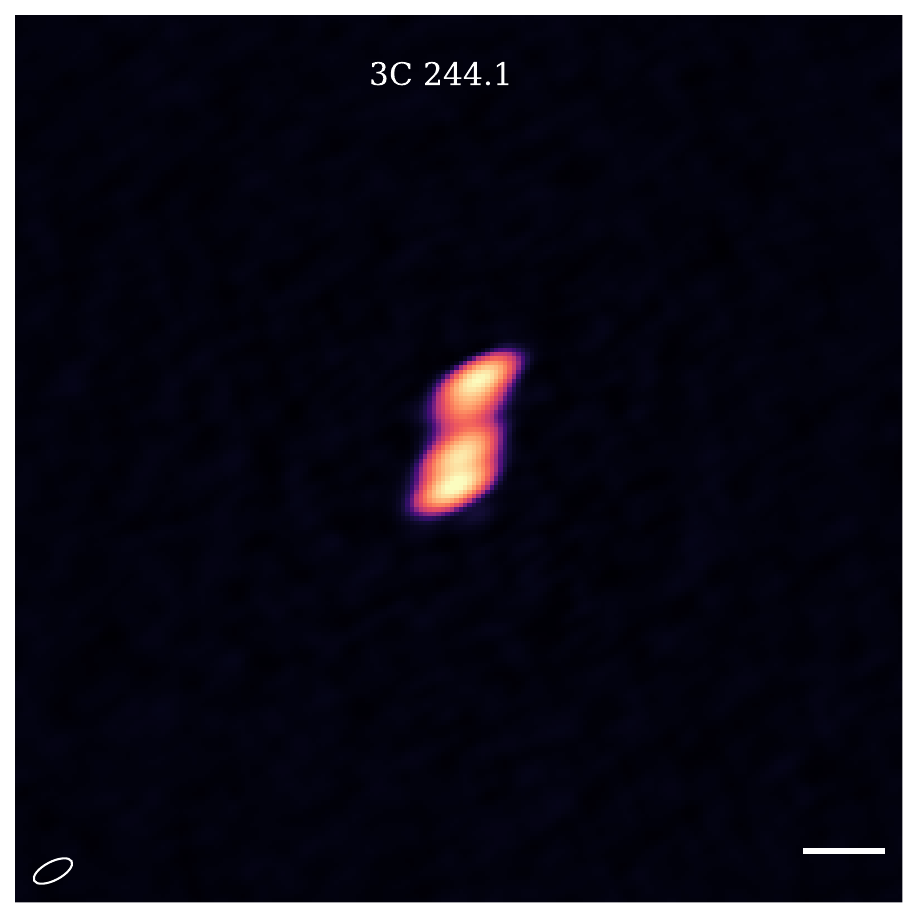}
\includegraphics[width=0.162\linewidth, trim={0.cm 0.cm 0.cm 0.cm},clip]{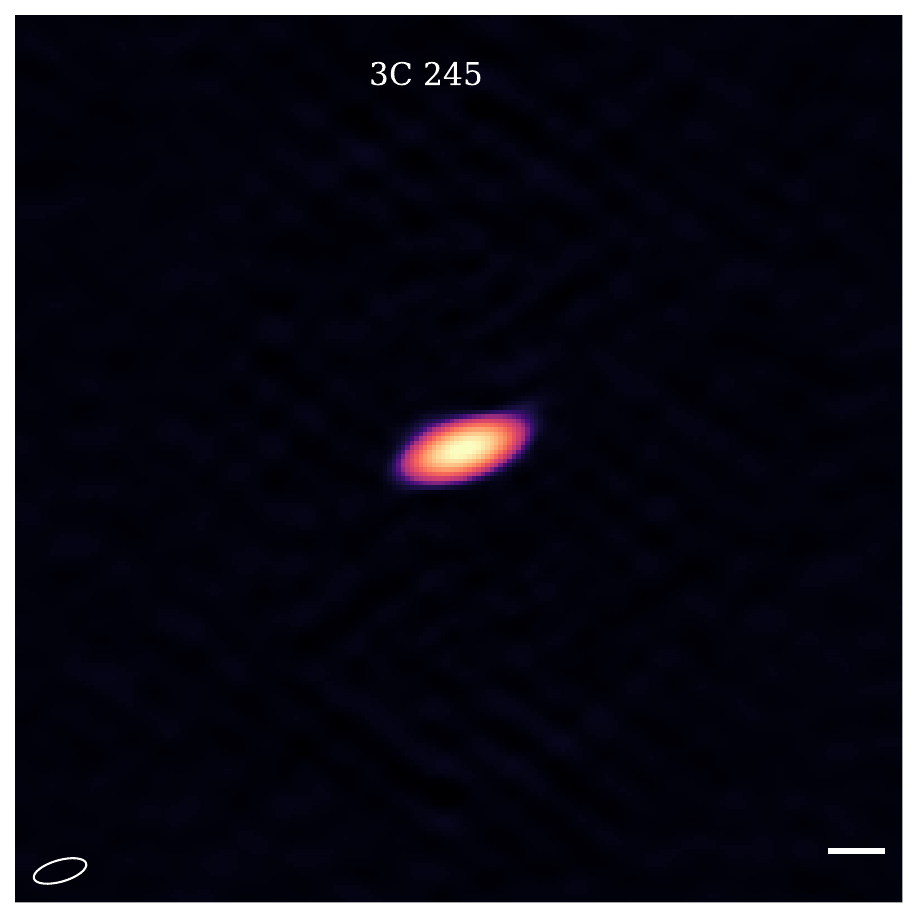}
\includegraphics[width=0.162\linewidth, trim={0.cm 0.cm 0.cm 0.cm},clip]{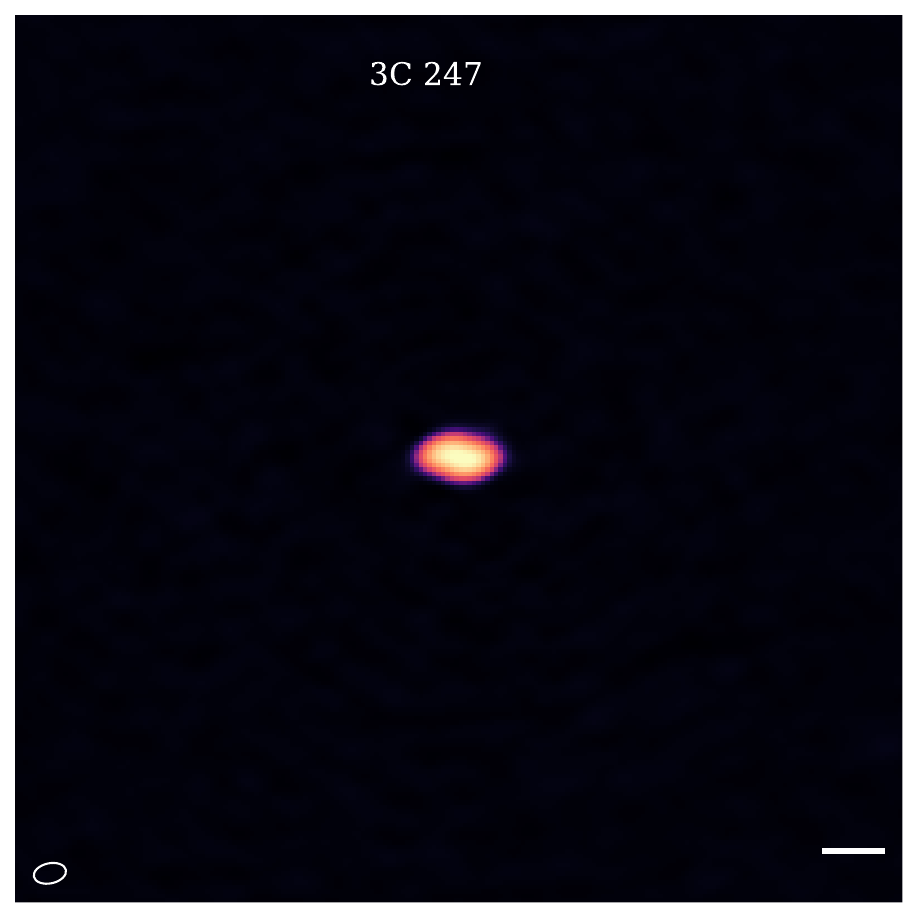}
\includegraphics[width=0.162\linewidth, trim={0.cm 0.cm 0.cm 0.cm},clip]{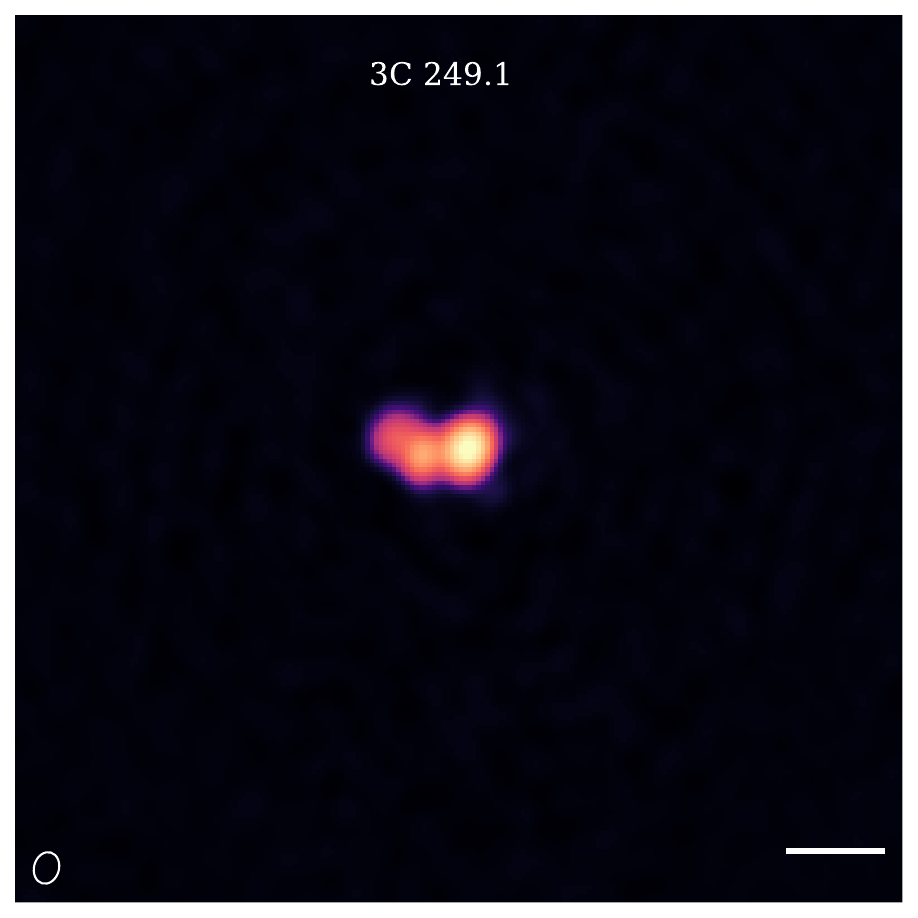}
\includegraphics[width=0.162\linewidth, trim={0.cm 0.cm 0.cm 0.cm},clip]{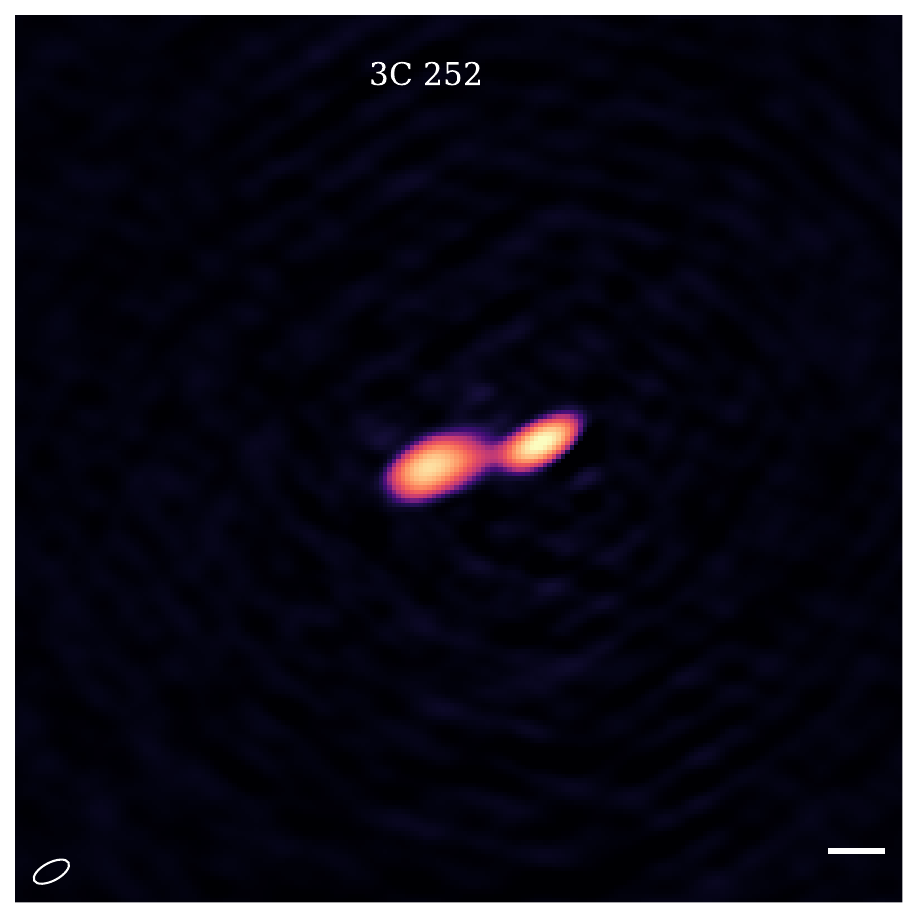}
\includegraphics[width=0.162\linewidth, trim={0.cm 0.cm 0.cm 0.cm},clip]{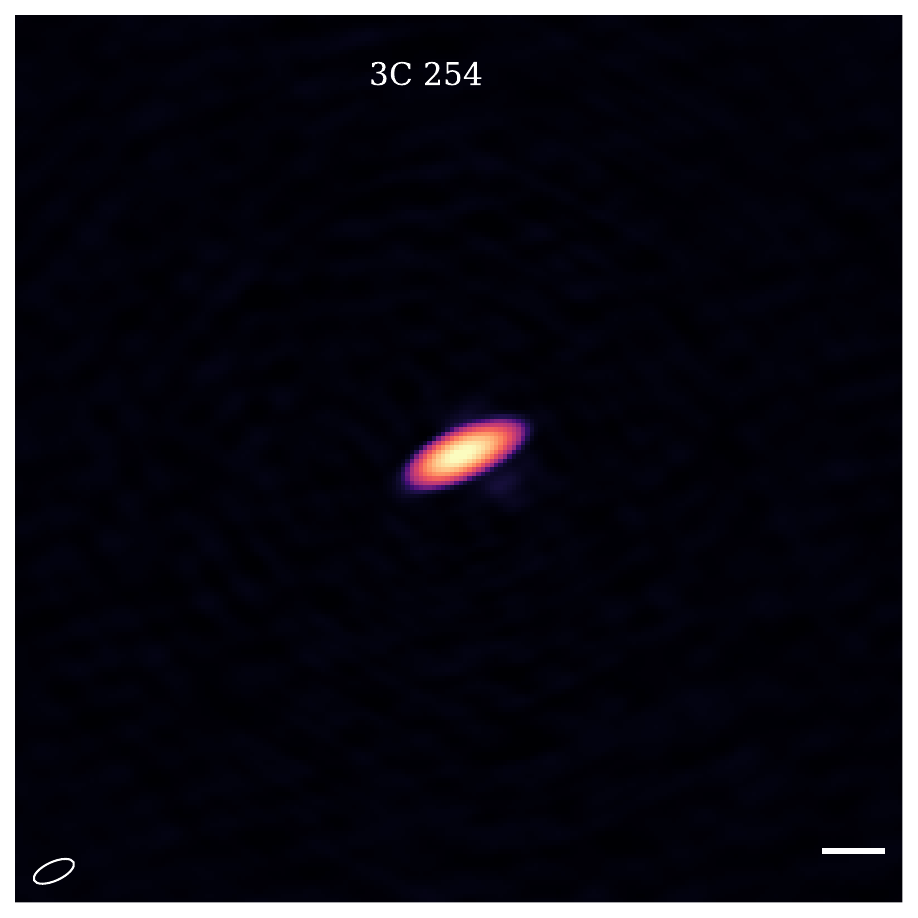}
\includegraphics[width=0.162\linewidth, trim={0.cm 0.cm 0.cm 0.cm},clip]{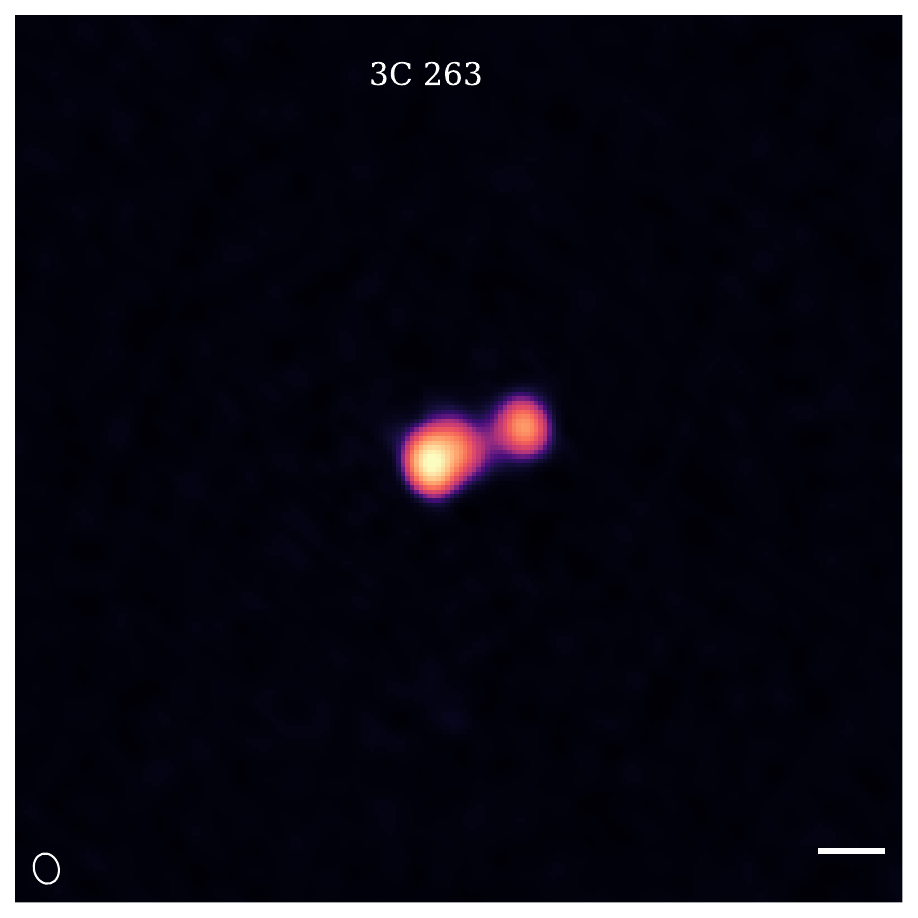}
\includegraphics[width=0.162\linewidth, trim={0.cm 0.cm 0.cm 0.cm},clip]{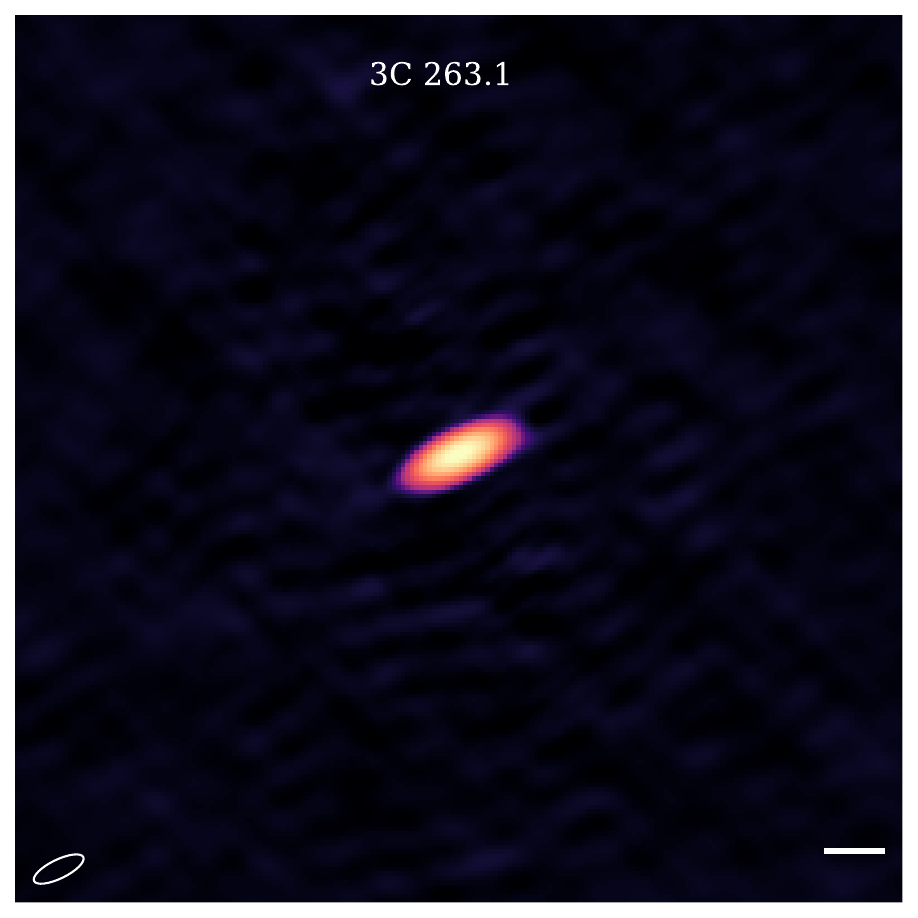}
\includegraphics[width=0.162\linewidth, trim={0.cm 0.cm 0.cm 0.cm},clip]{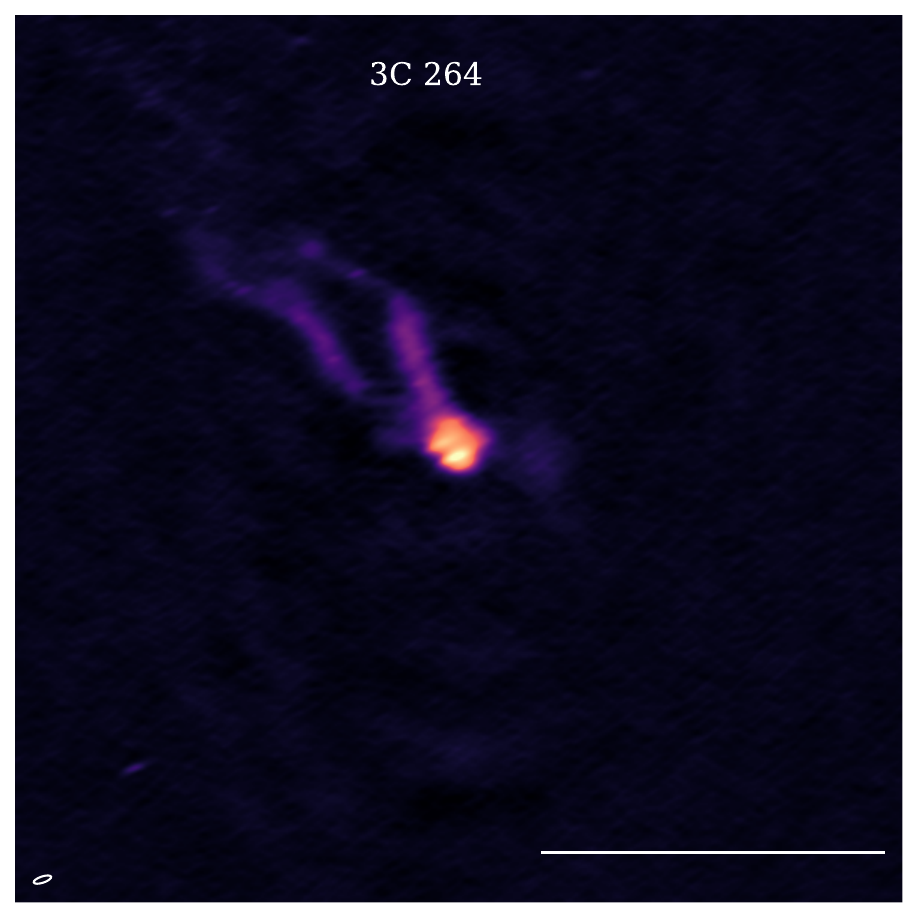}
\includegraphics[width=0.162\linewidth, trim={0.cm 0.cm 0.cm 0.cm},clip]{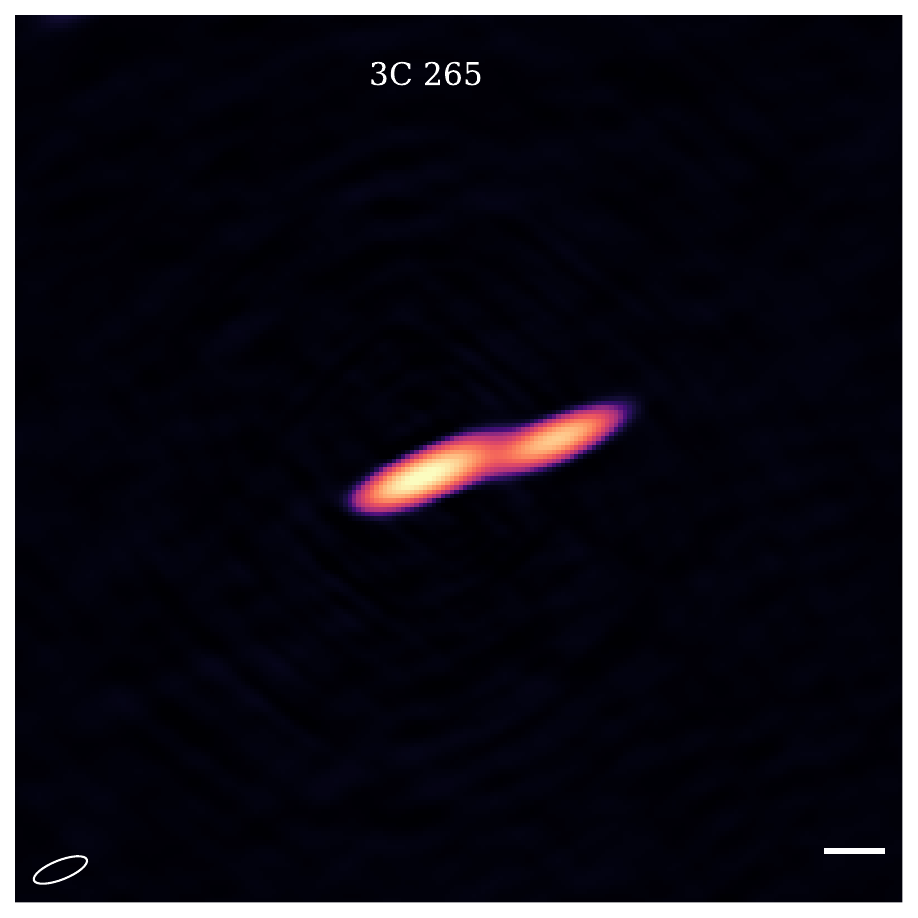}
\includegraphics[width=0.162\linewidth, trim={0.cm 0.cm 0.cm 0.cm},clip]{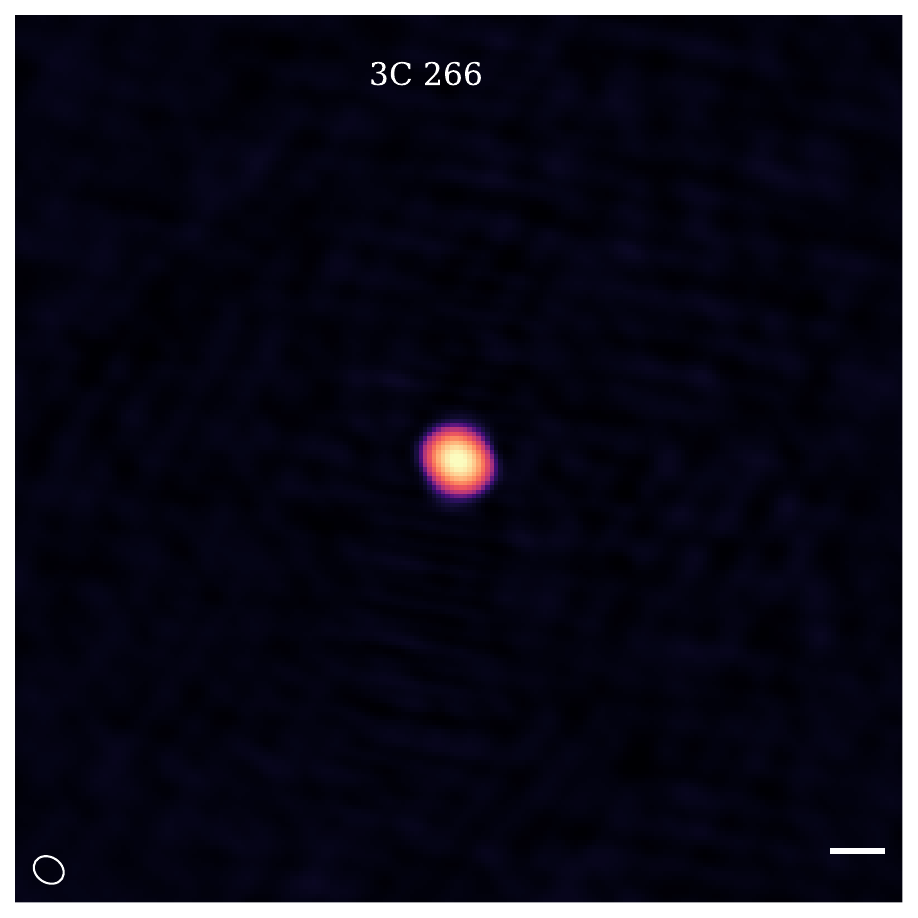}
\includegraphics[width=0.162\linewidth, trim={0.cm 0.cm 0.cm 0.cm},clip]{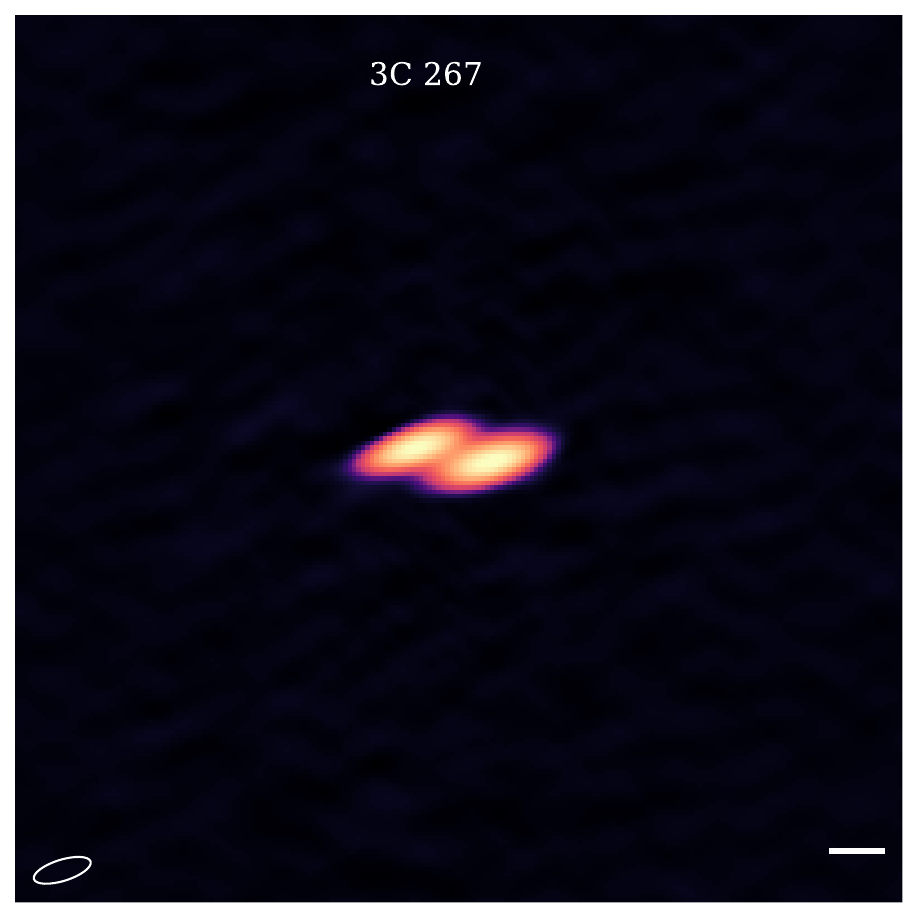}
\includegraphics[width=0.162\linewidth, trim={0.cm 0.cm 0.cm 0.cm},clip]{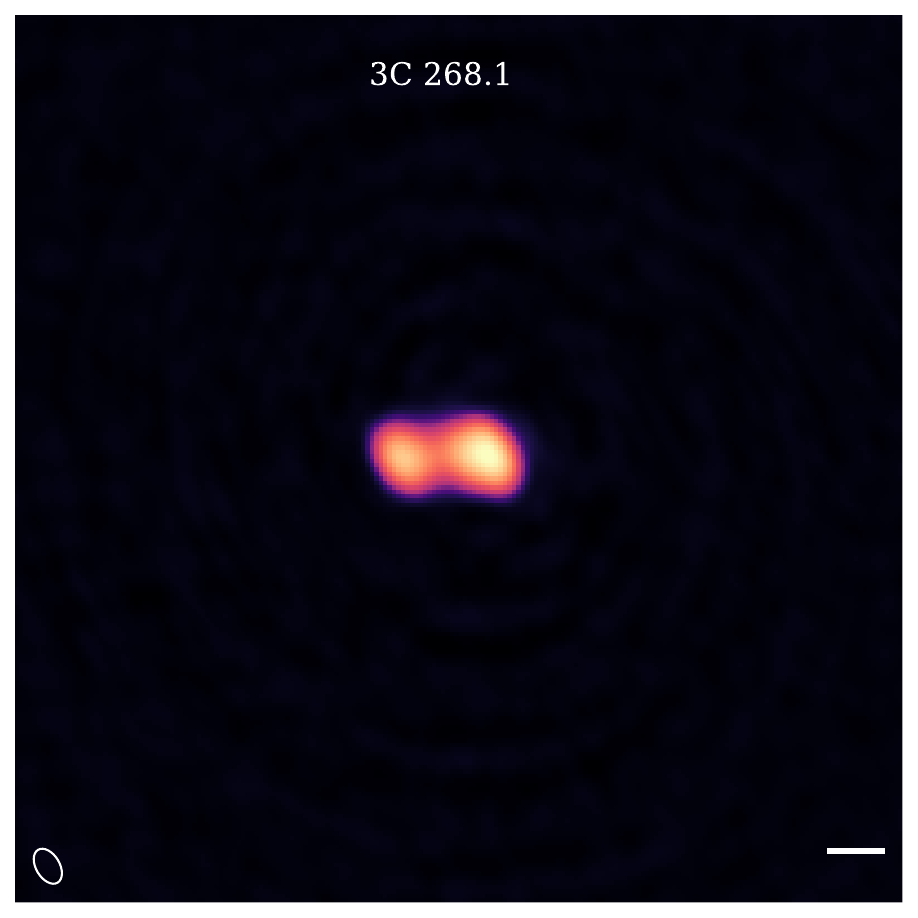}
\includegraphics[width=0.162\linewidth, trim={0.cm 0.cm 0.cm 0.cm},clip]{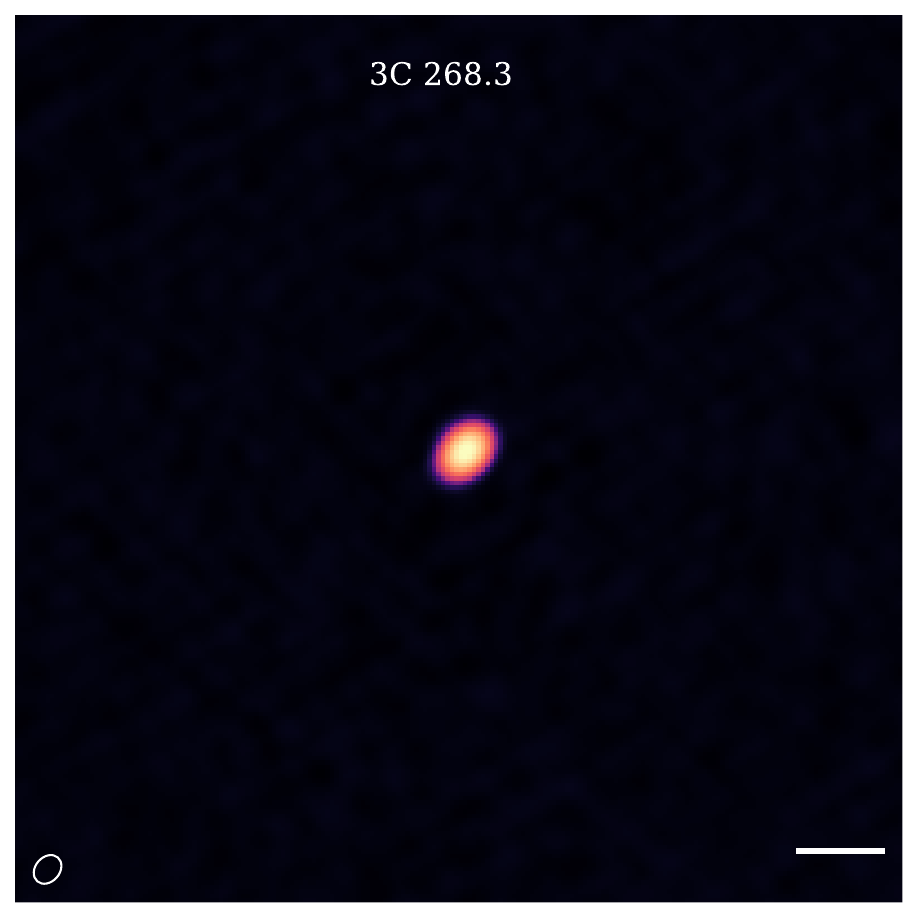}
\includegraphics[width=0.162\linewidth, trim={0.cm 0.cm 0.cm 0.cm},clip]{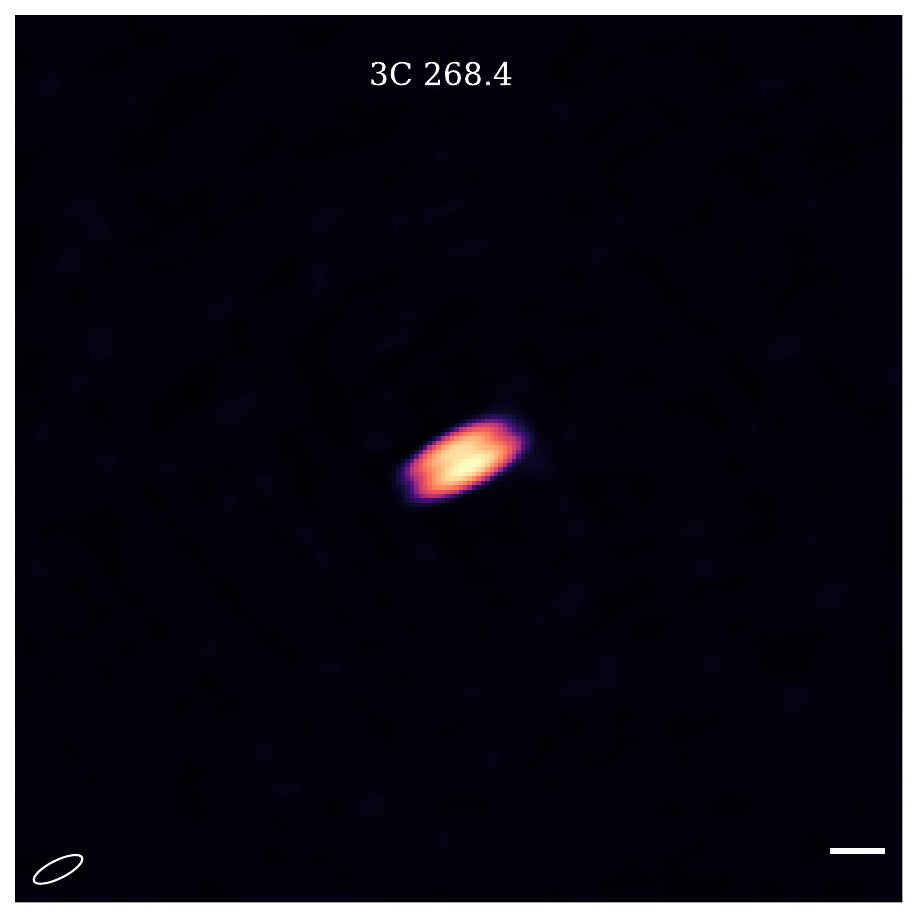}
\includegraphics[width=0.162\linewidth, trim={0.cm 0.cm 0.cm 0.cm},clip]{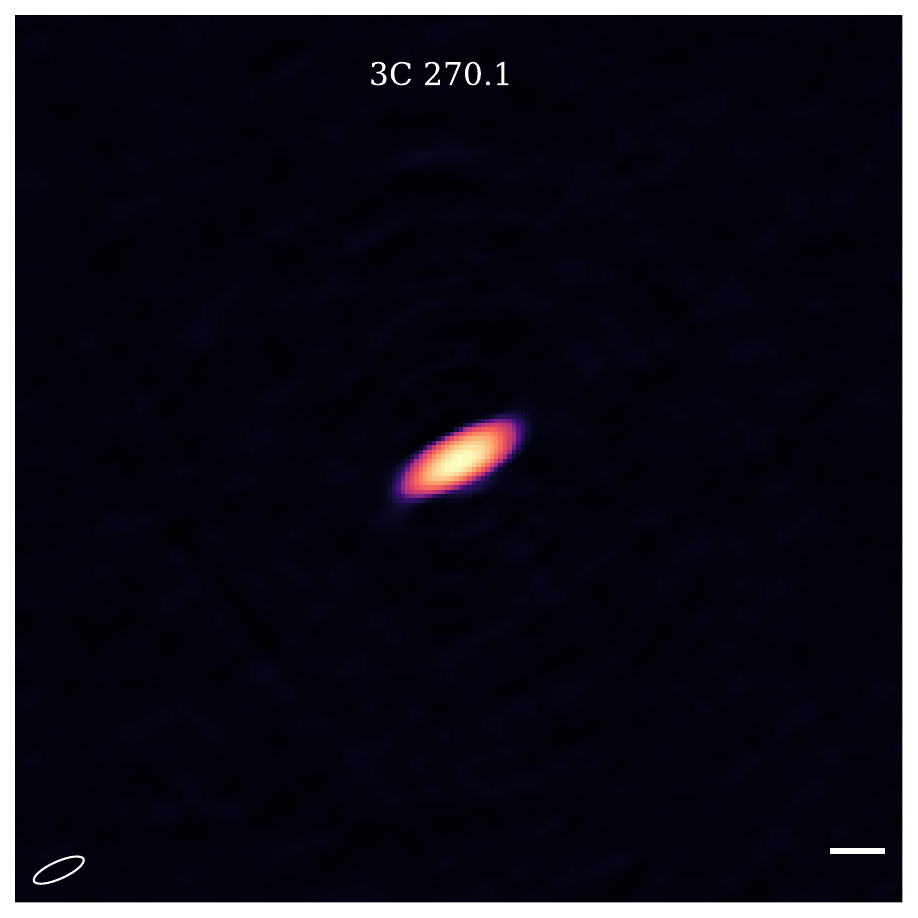}
\includegraphics[width=0.162\linewidth, trim={0.cm 0.cm 0.cm 0.cm},clip]{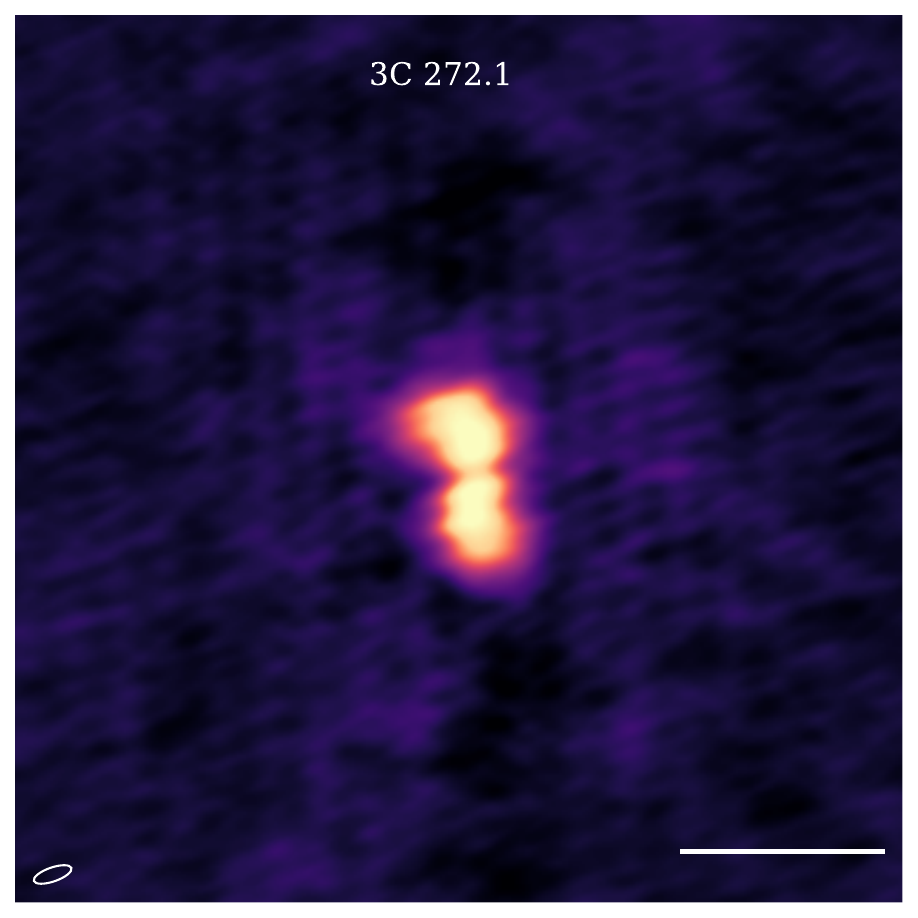}
\caption{continued.}
\end{figure}
\setcounter{figure}{0}

\begin{figure}[H]
\centering
\includegraphics[width=0.162\linewidth, trim={0.cm 0.cm 0.cm 0.cm},clip]{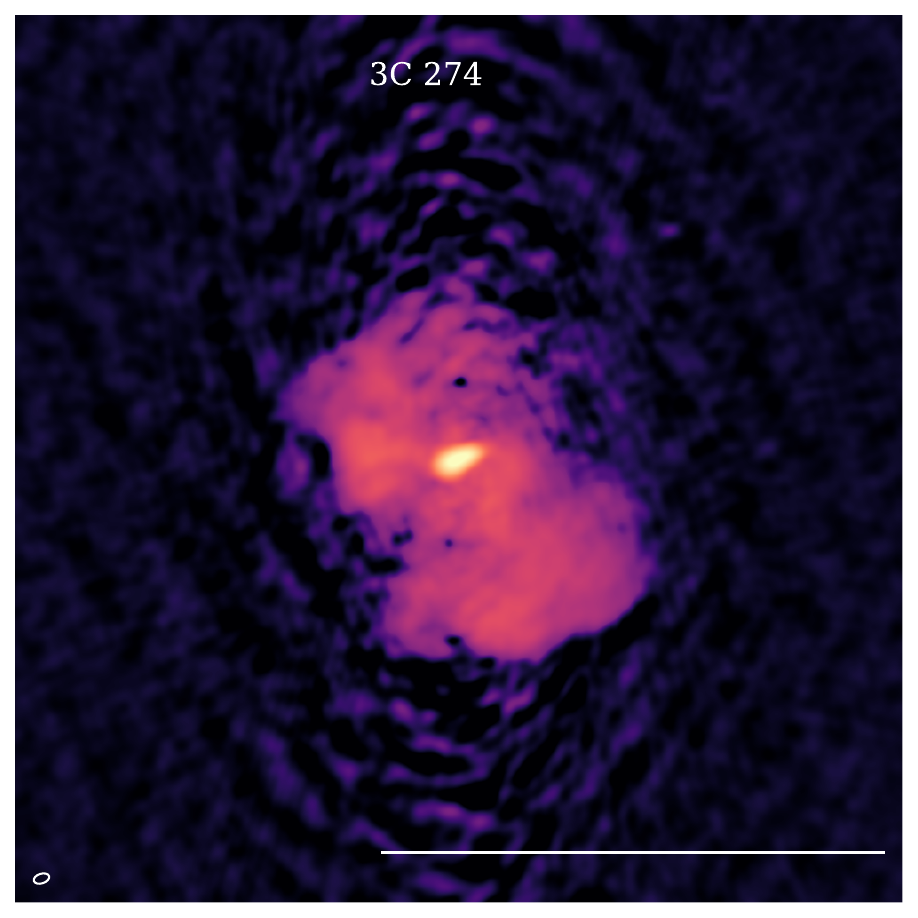}
\includegraphics[width=0.162\linewidth, trim={0.cm 0.cm 0.cm 0.cm},clip]{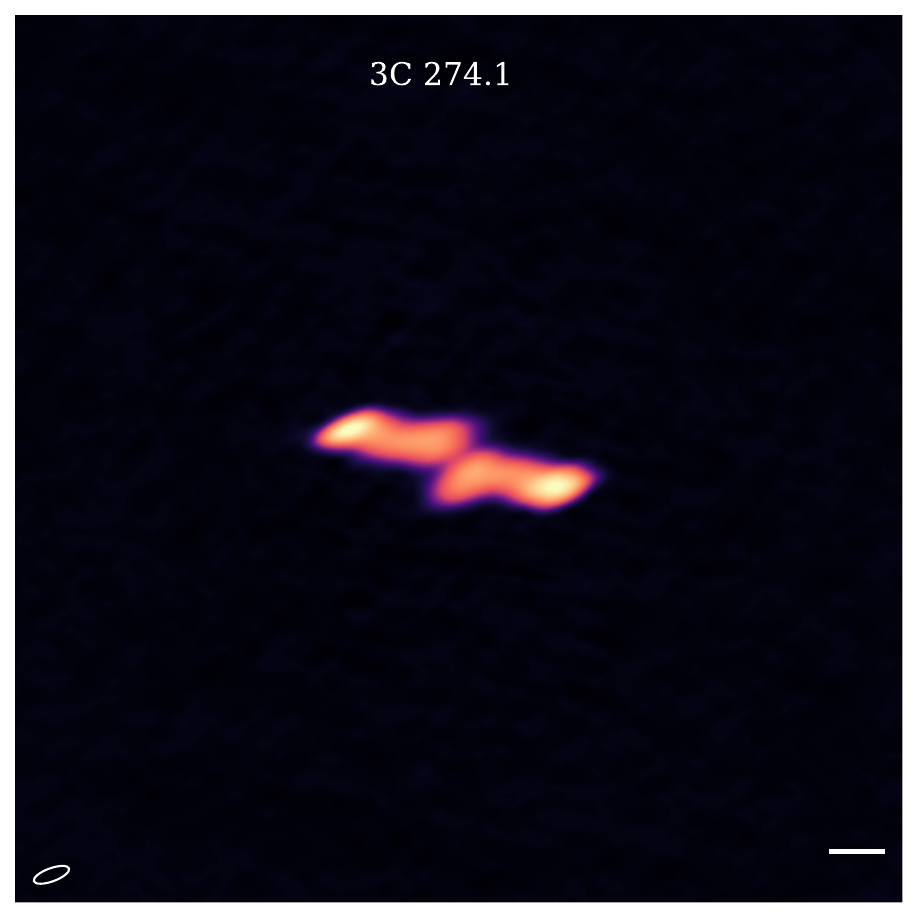}
\includegraphics[width=0.162\linewidth, trim={0.cm 0.cm 0.cm 0.cm},clip]{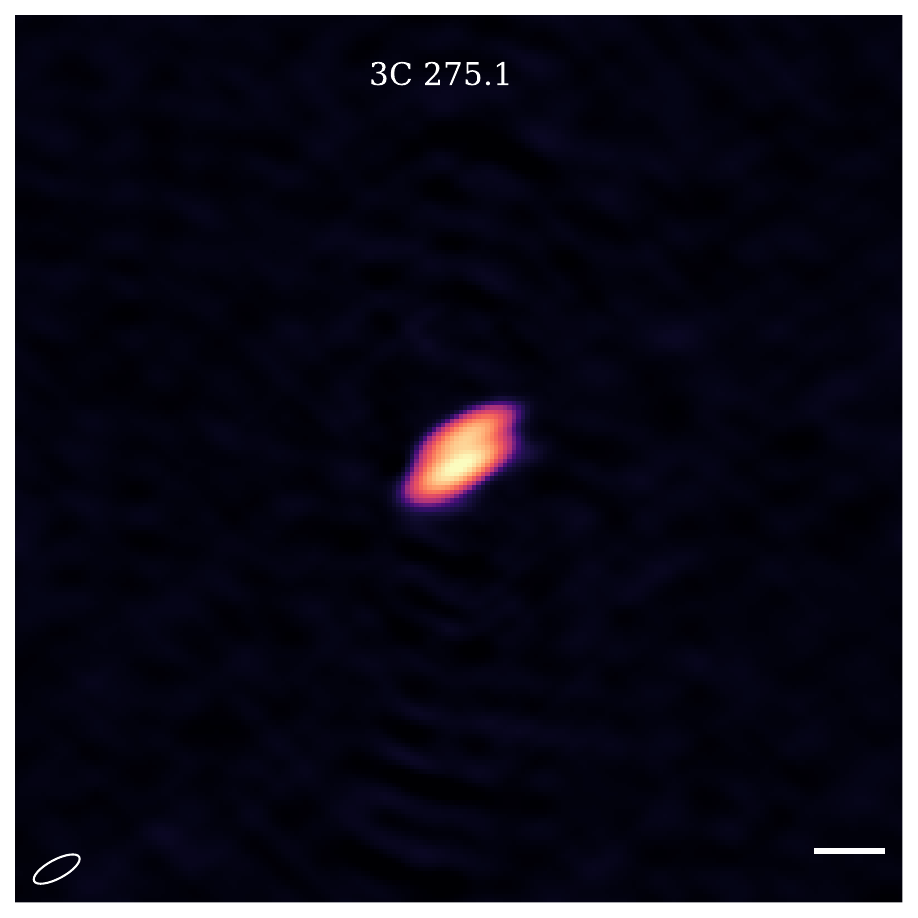}
\includegraphics[width=0.162\linewidth, trim={0.cm 0.cm 0.cm 0.cm},clip]{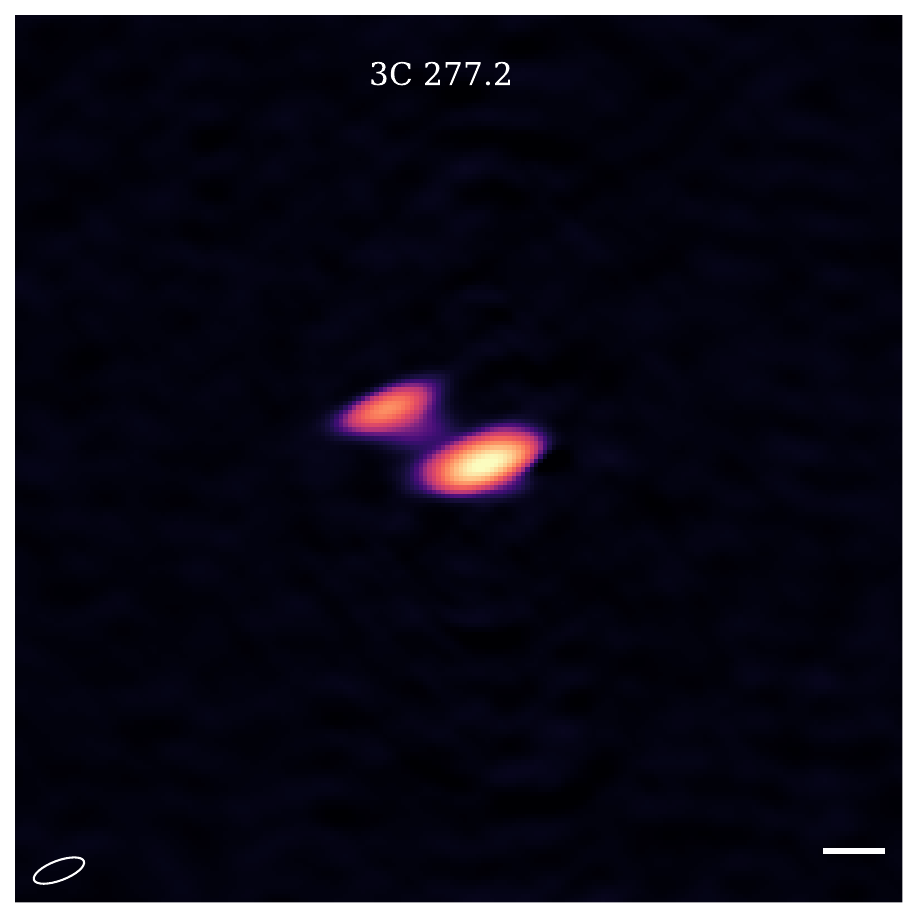}
\includegraphics[width=0.162\linewidth, trim={0.cm 0.cm 0.cm 0.cm},clip]{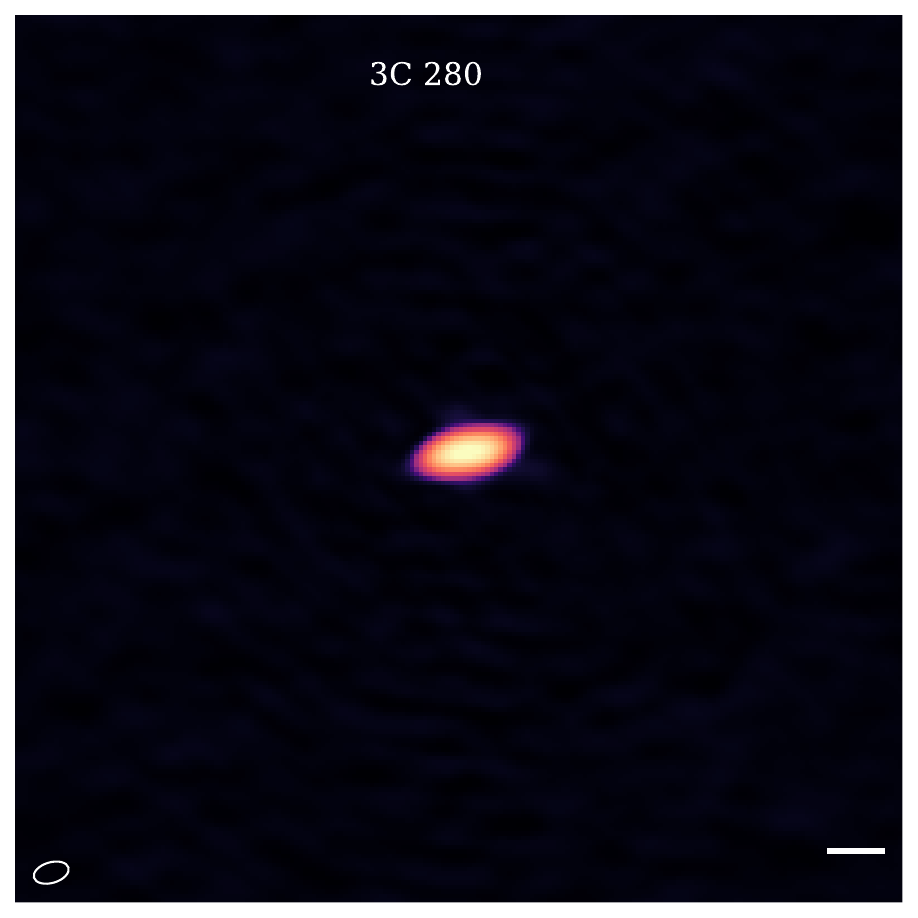}
\includegraphics[width=0.162\linewidth, trim={0.cm 0.cm 0.cm 0.cm},clip]{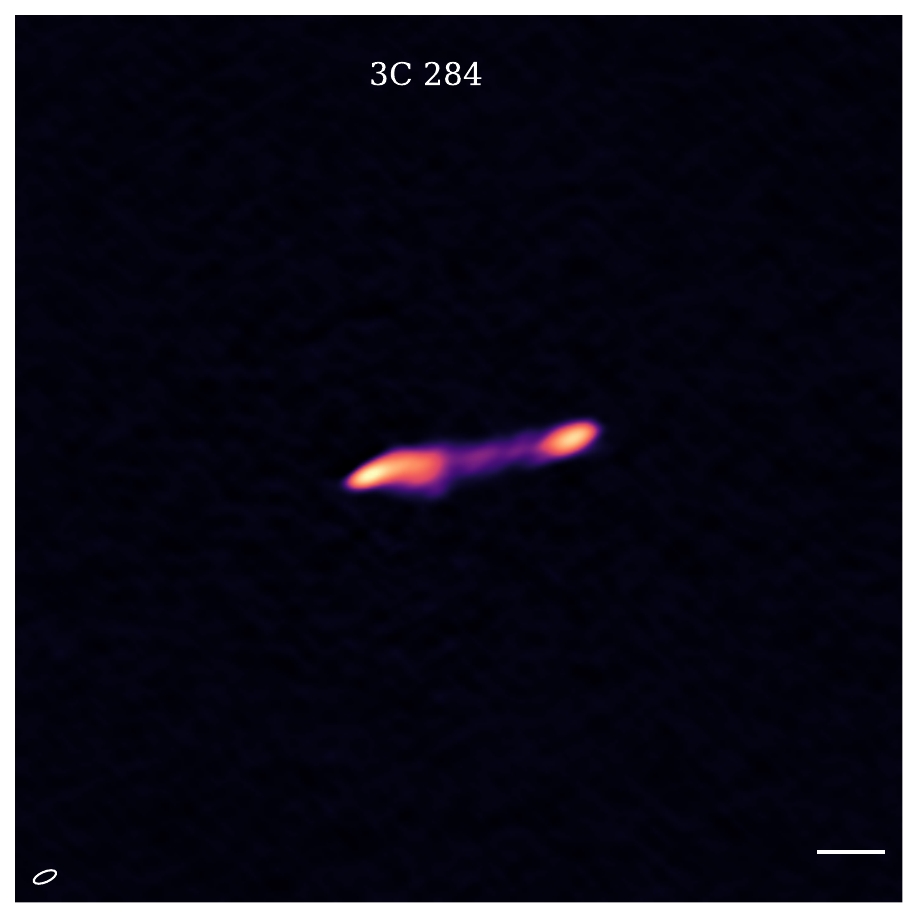}
\includegraphics[width=0.162\linewidth, trim={0.cm 0.cm 0.cm 0.cm},clip]{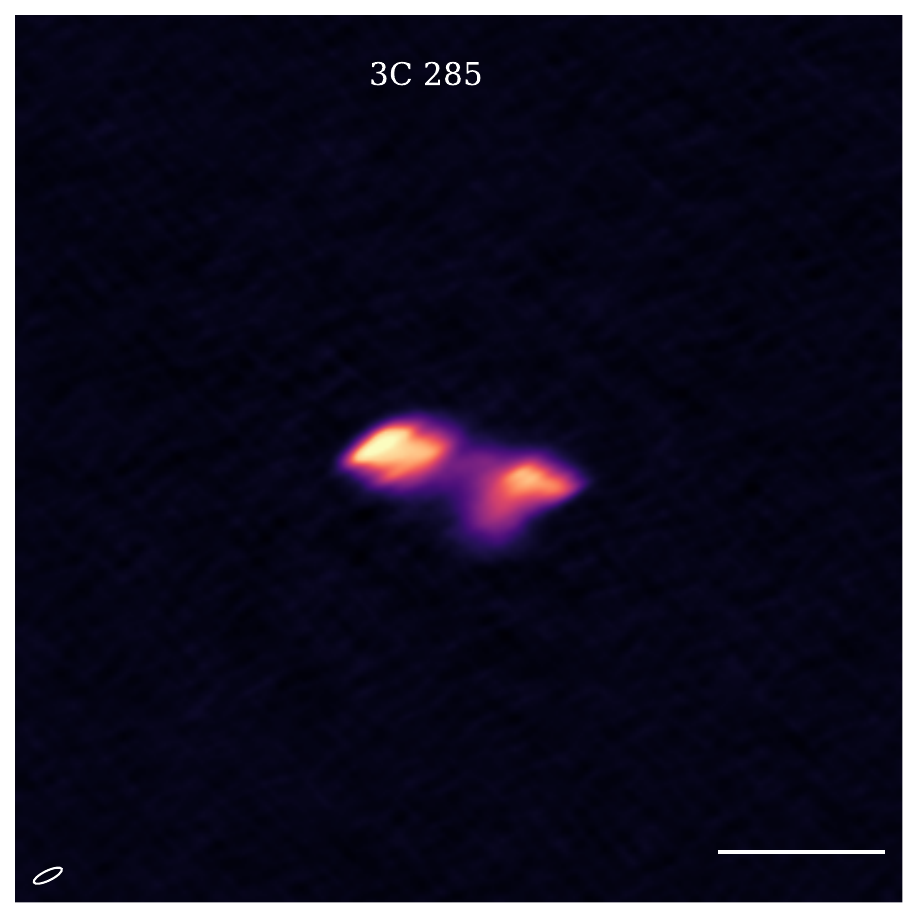}
\includegraphics[width=0.162\linewidth, trim={0.cm 0.cm 0.cm 0.cm},clip]{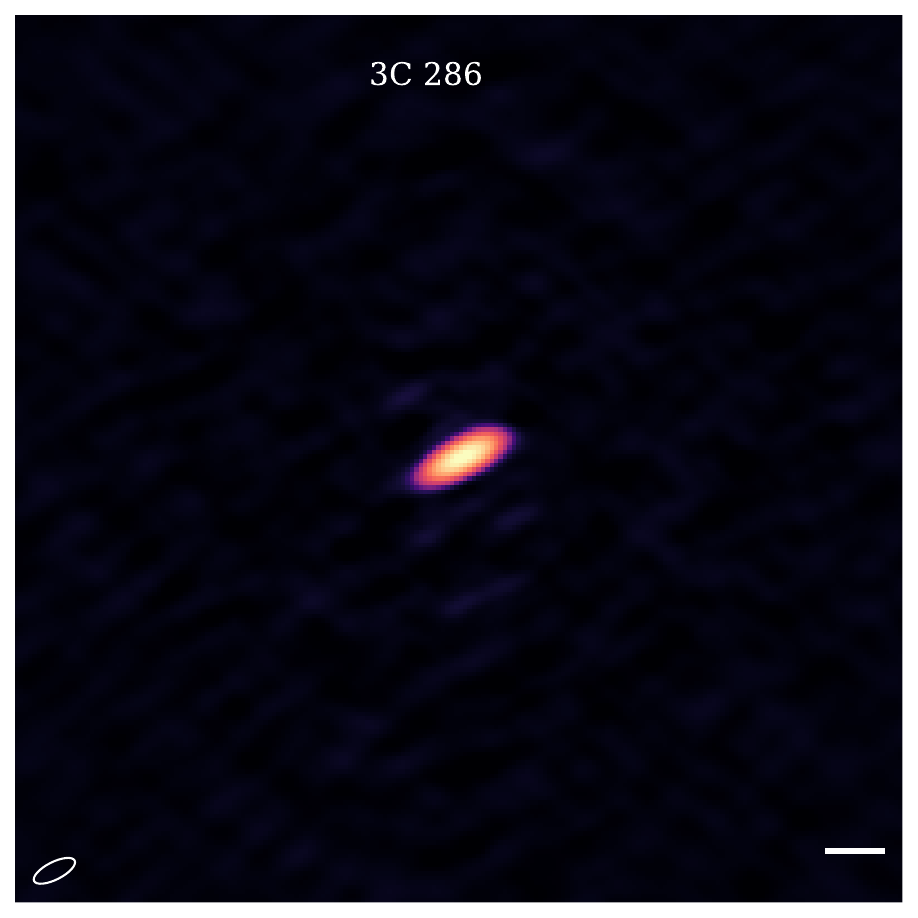}
\includegraphics[width=0.162\linewidth, trim={0.cm 0.cm 0.cm 0.cm},clip]{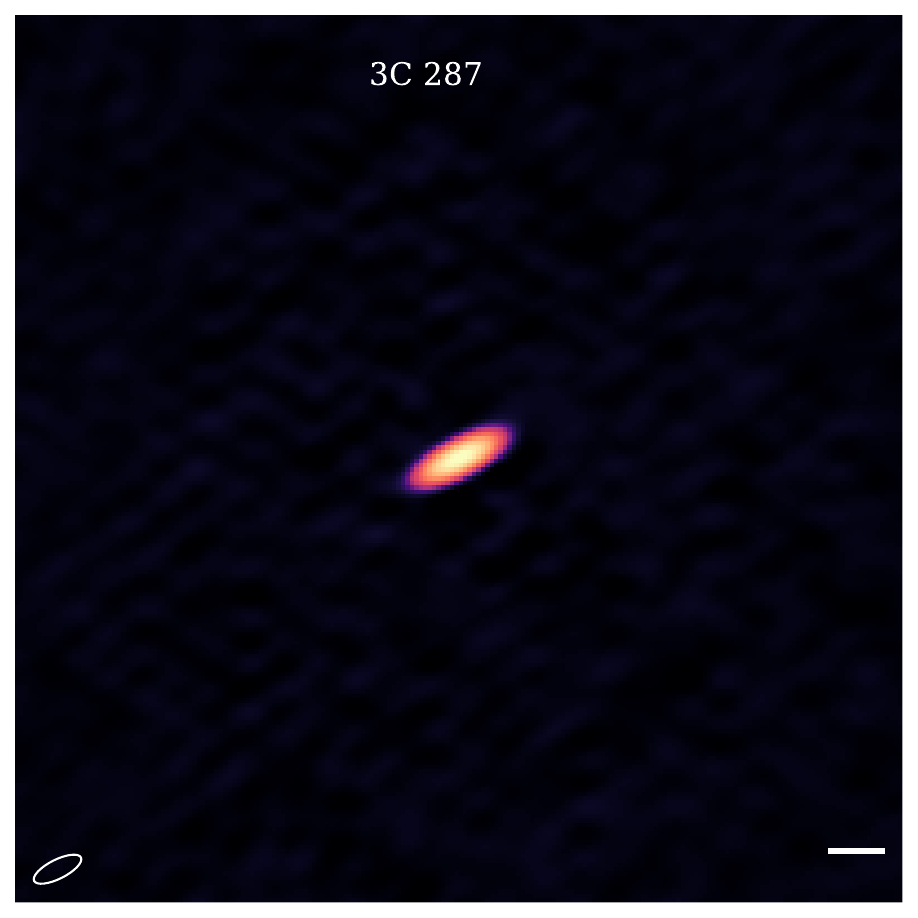}
\includegraphics[width=0.162\linewidth, trim={0.cm 0.cm 0.cm 0.cm},clip]{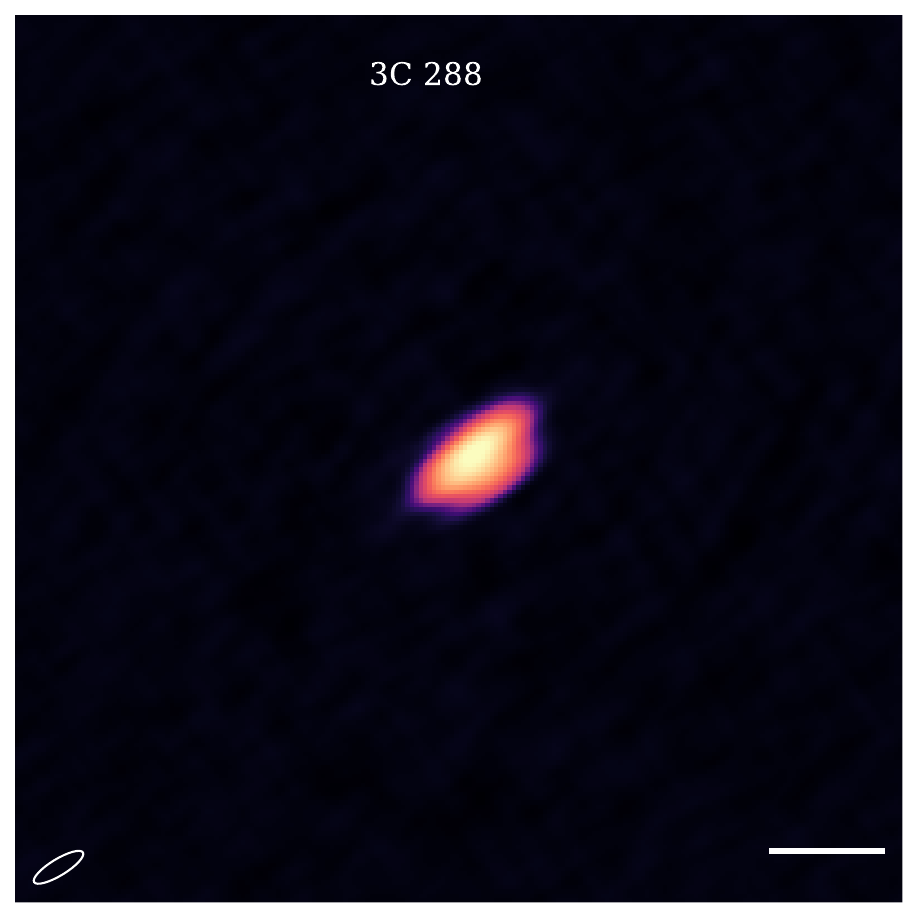}
\includegraphics[width=0.162\linewidth, trim={0.cm 0.cm 0.cm 0.cm},clip]{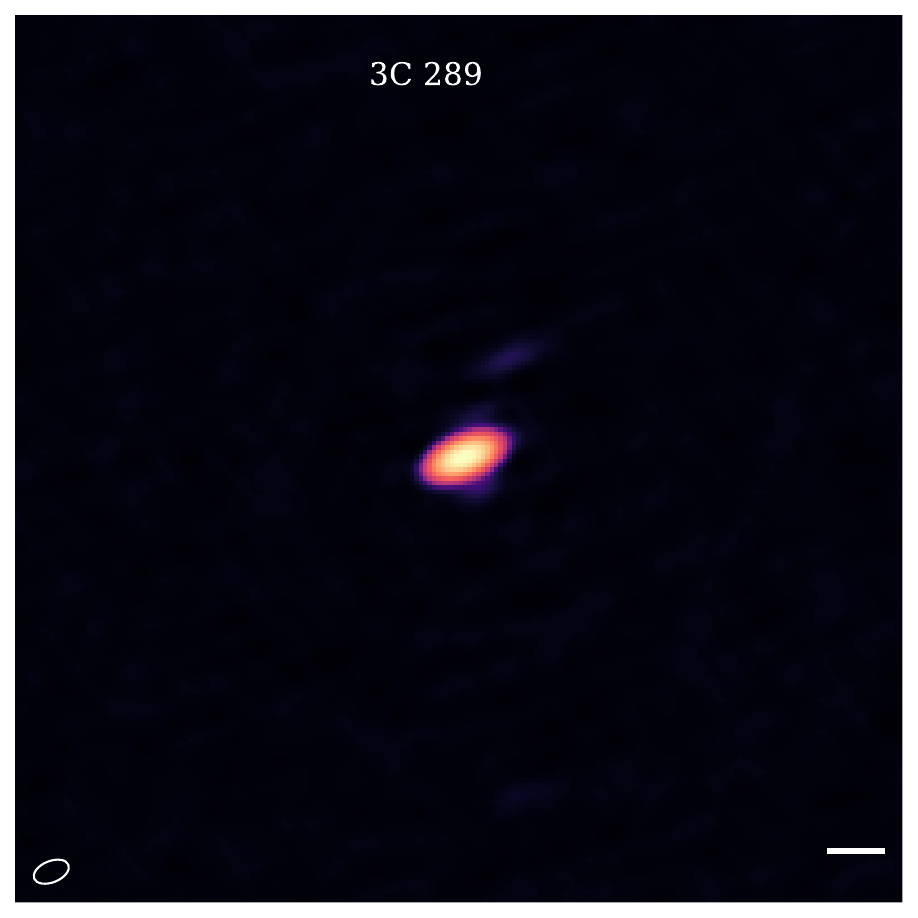}
\includegraphics[width=0.162\linewidth, trim={0.cm 0.cm 0.cm 0.cm},clip]{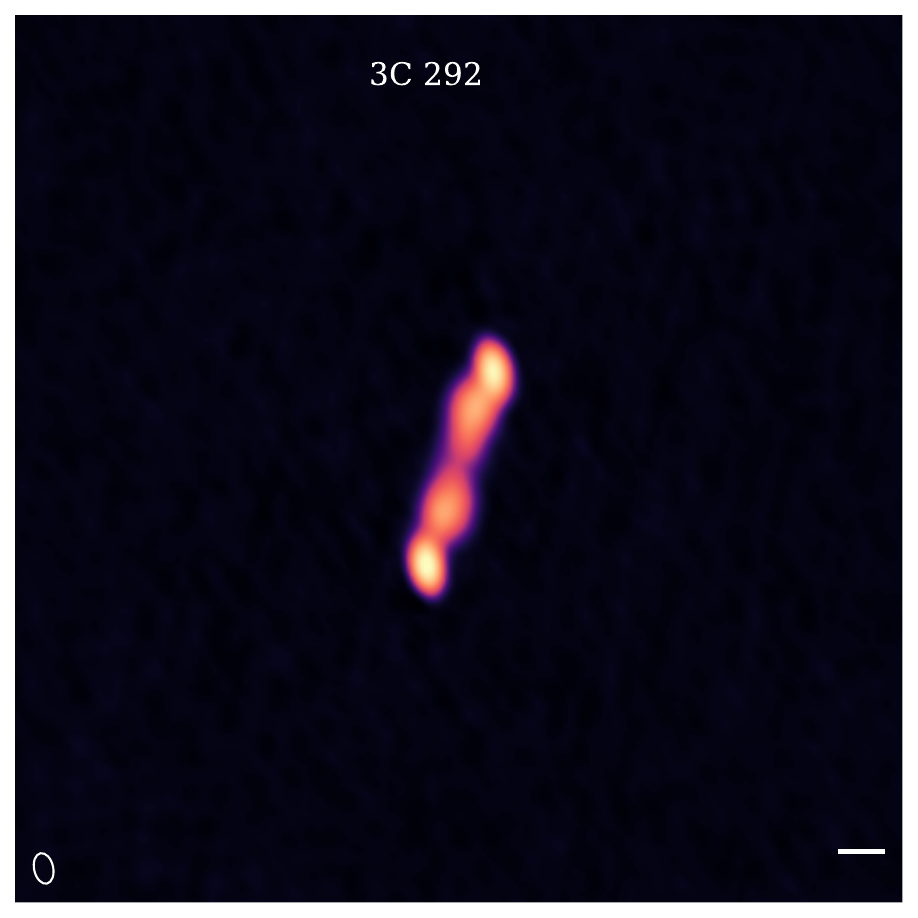}
\includegraphics[width=0.162\linewidth, trim={0.cm 0.cm 0.cm 0.cm},clip]{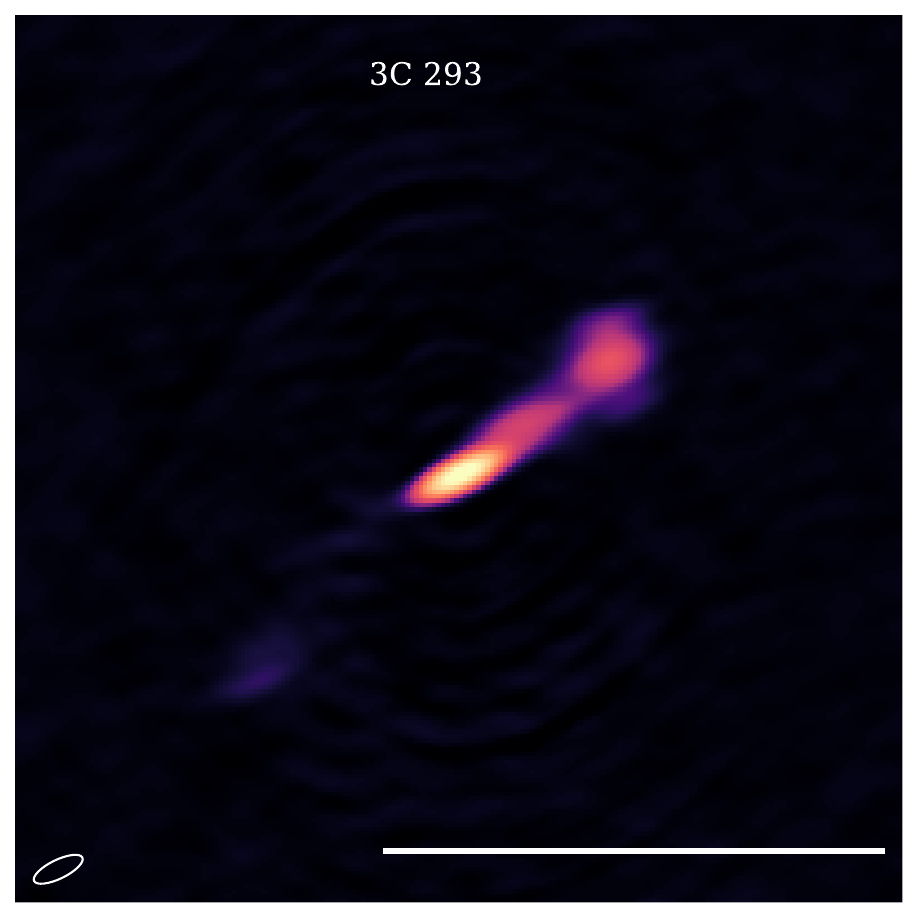}
\includegraphics[width=0.162\linewidth, trim={0.cm 0.cm 0.cm 0.cm},clip]{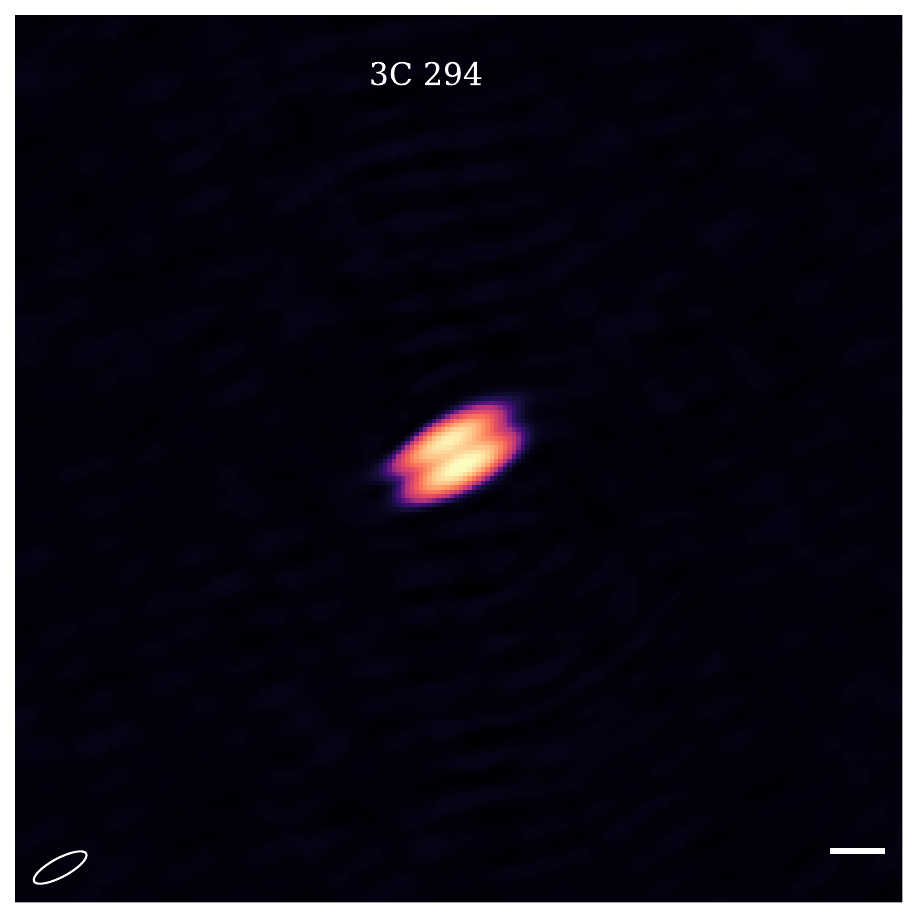}
\includegraphics[width=0.162\linewidth, trim={0.cm 0.cm 0.cm 0.cm},clip]{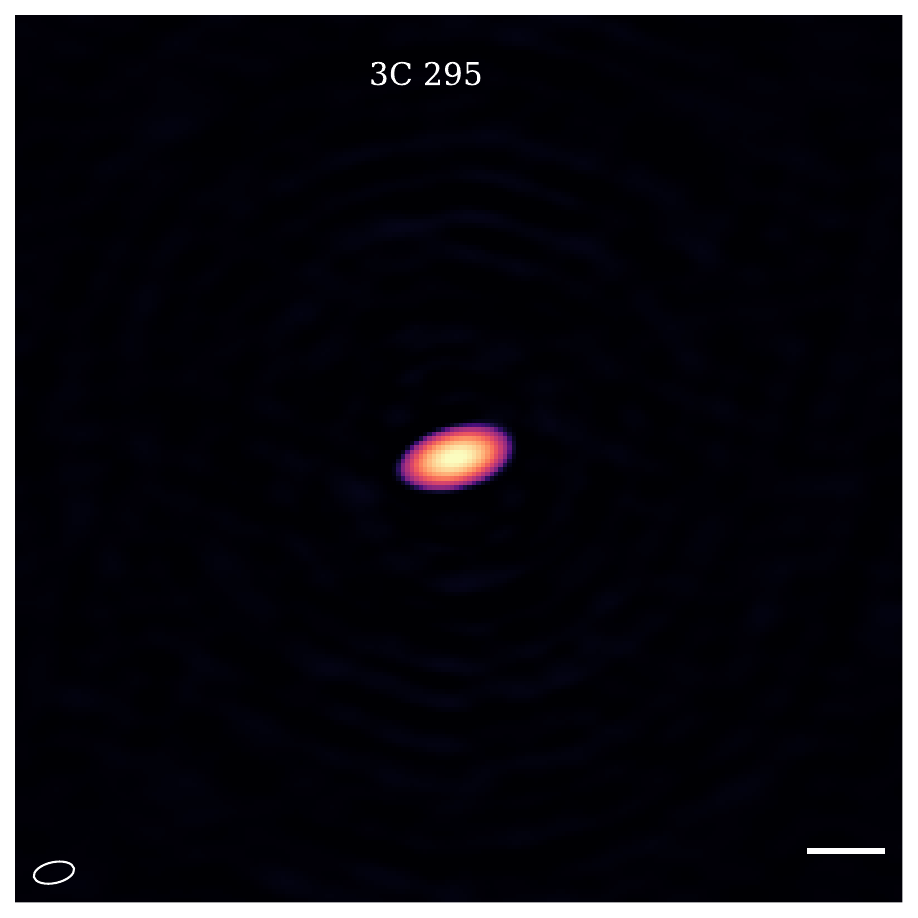}
\includegraphics[width=0.162\linewidth, trim={0.cm 0.cm 0.cm 0.cm},clip]{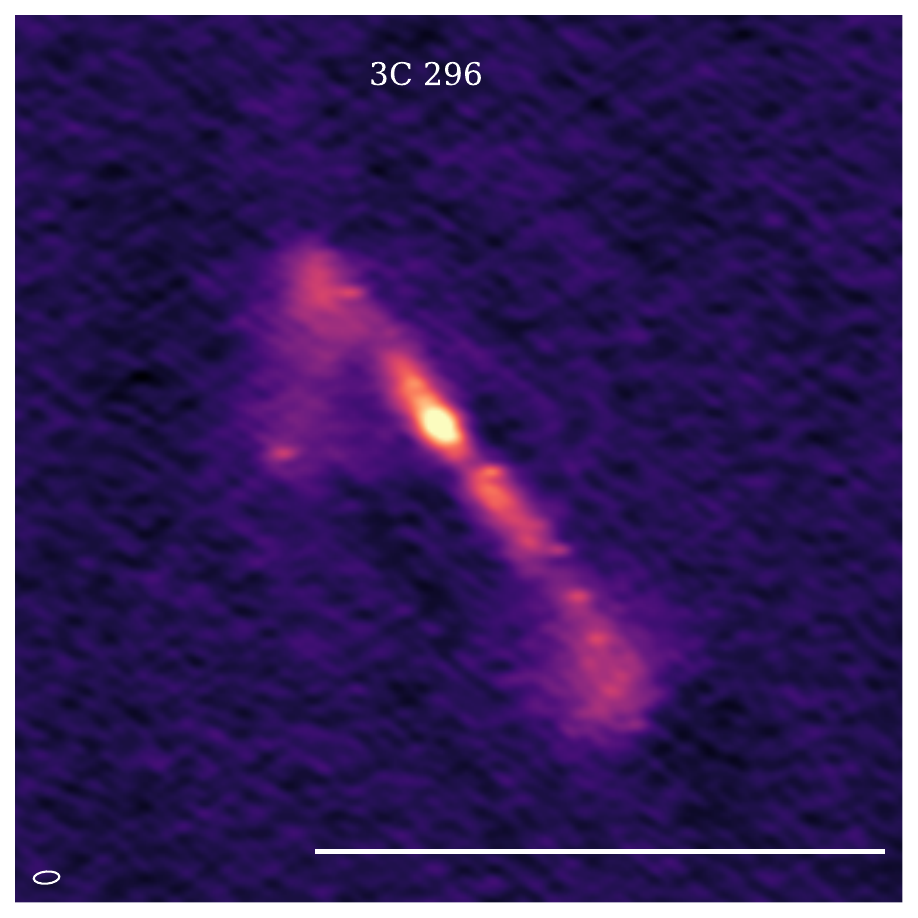}
\includegraphics[width=0.162\linewidth, trim={0.cm 0.cm 0.cm 0.cm},clip]{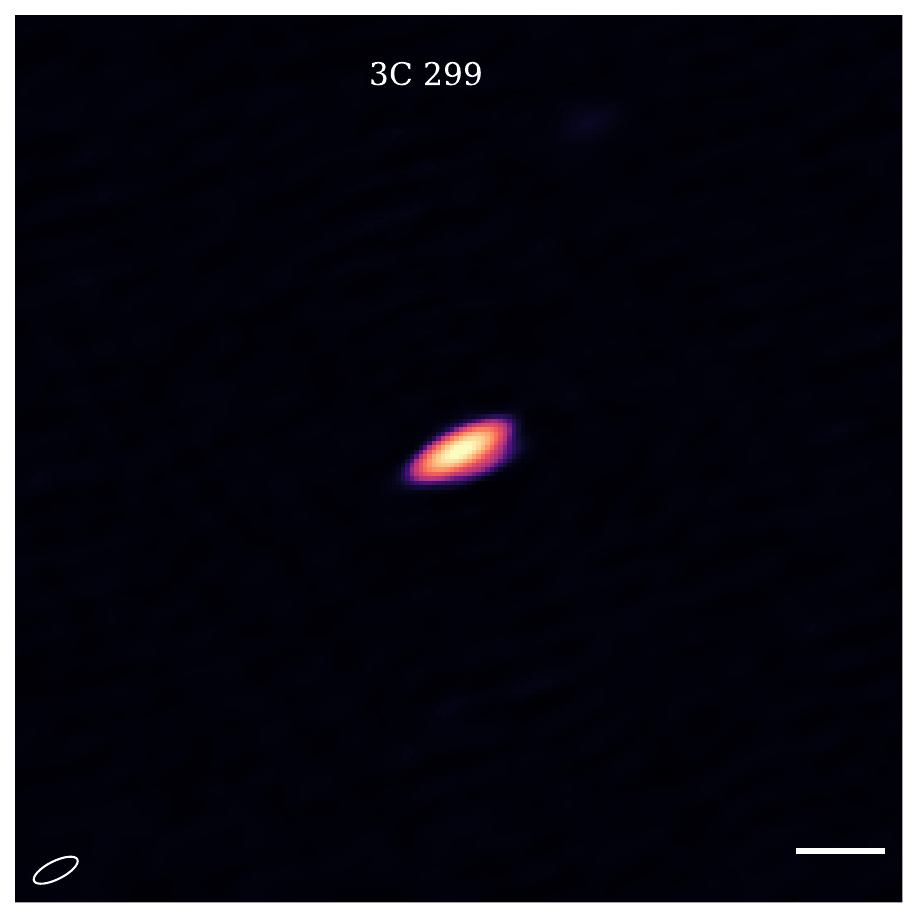}
\includegraphics[width=0.162\linewidth, trim={0.cm 0.cm 0.cm 0.cm},clip]{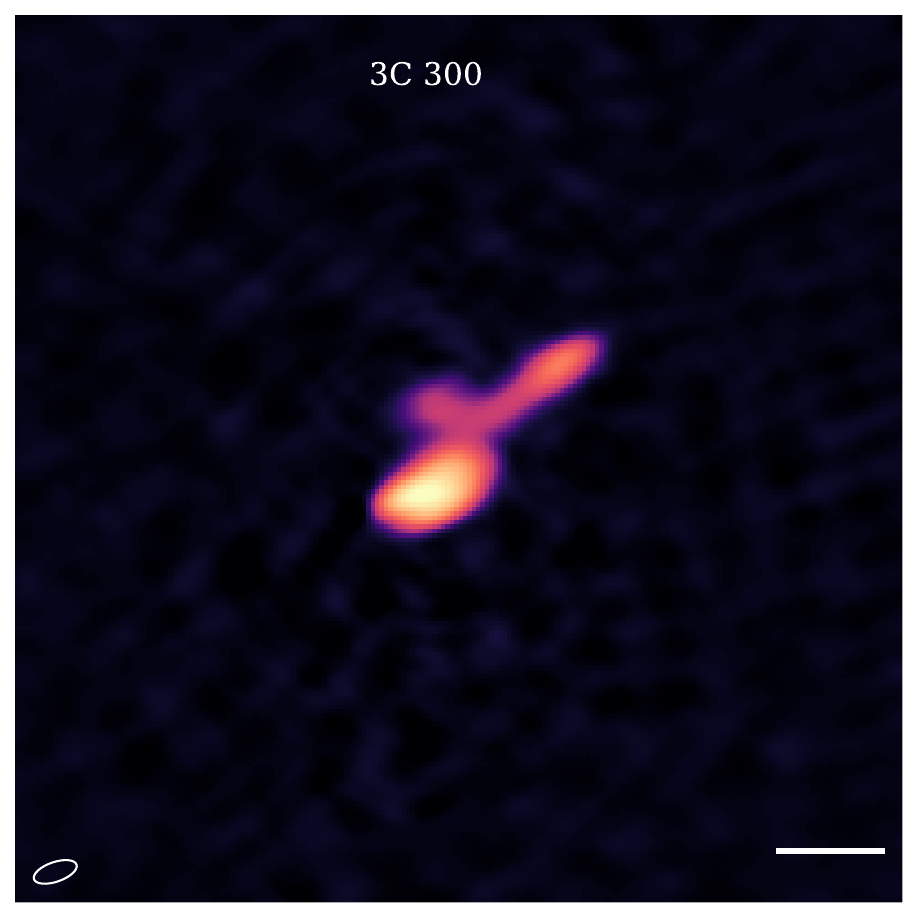}
\includegraphics[width=0.162\linewidth, trim={0.cm 0.cm 0.cm 0.cm},clip]{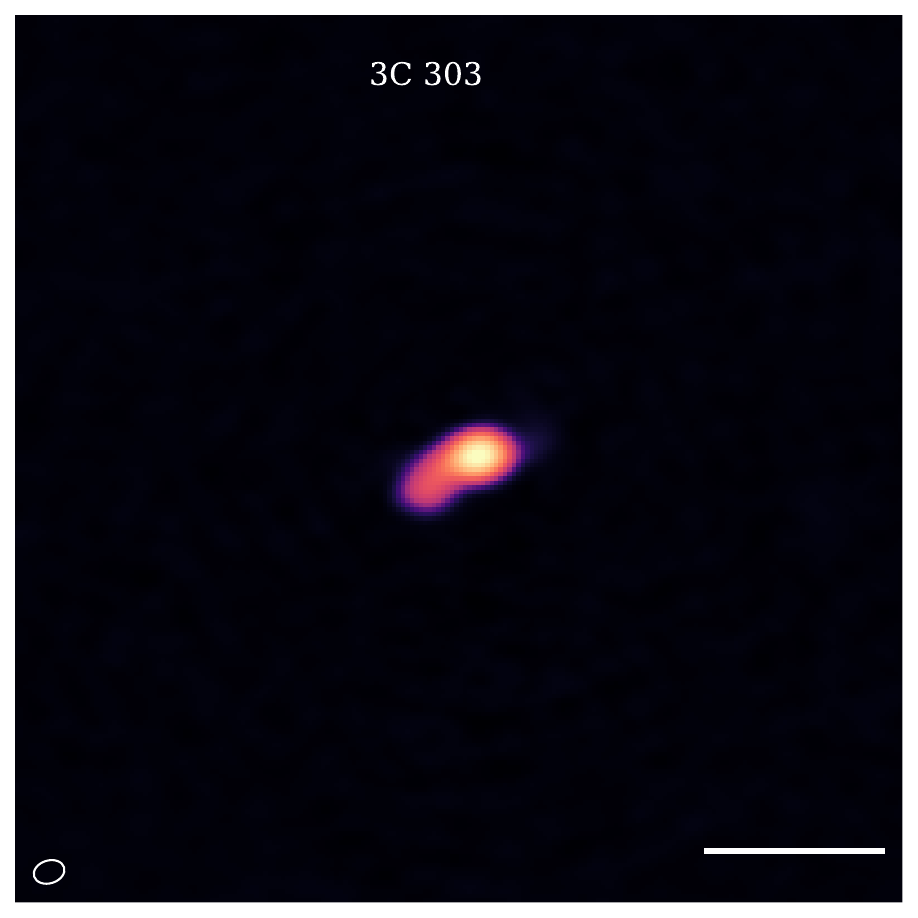}
\includegraphics[width=0.162\linewidth, trim={0.cm 0.cm 0.cm 0.cm},clip]{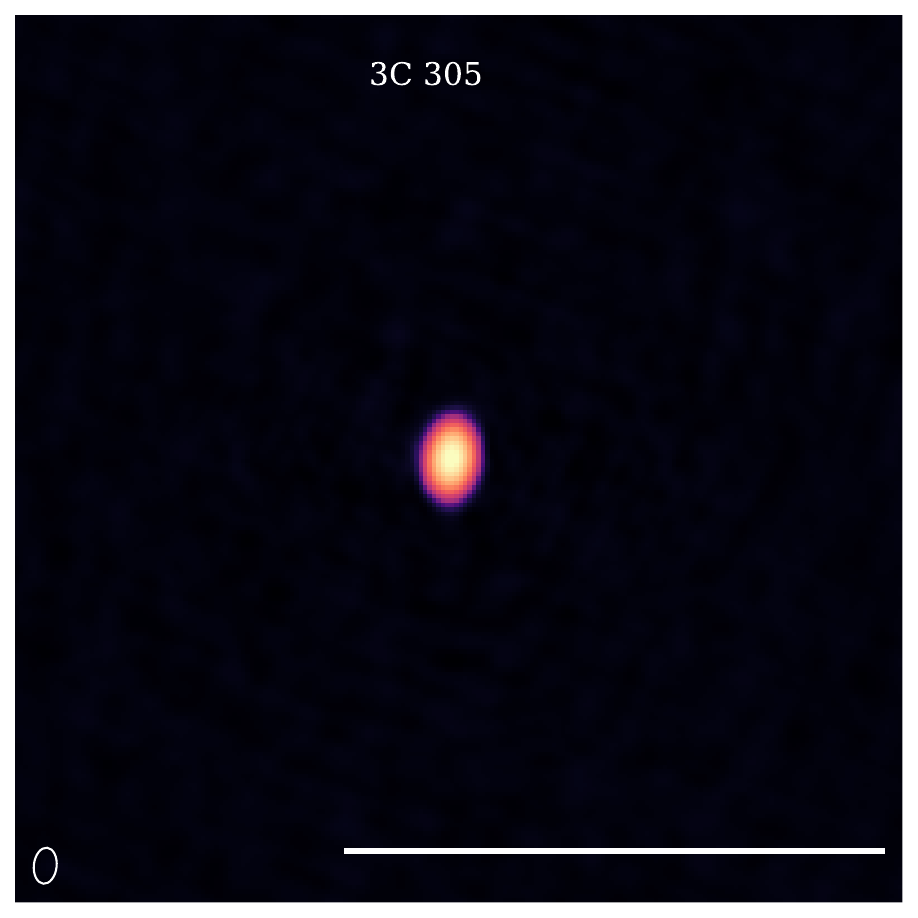}
\includegraphics[width=0.162\linewidth, trim={0.cm 0.cm 0.cm 0.cm},clip]{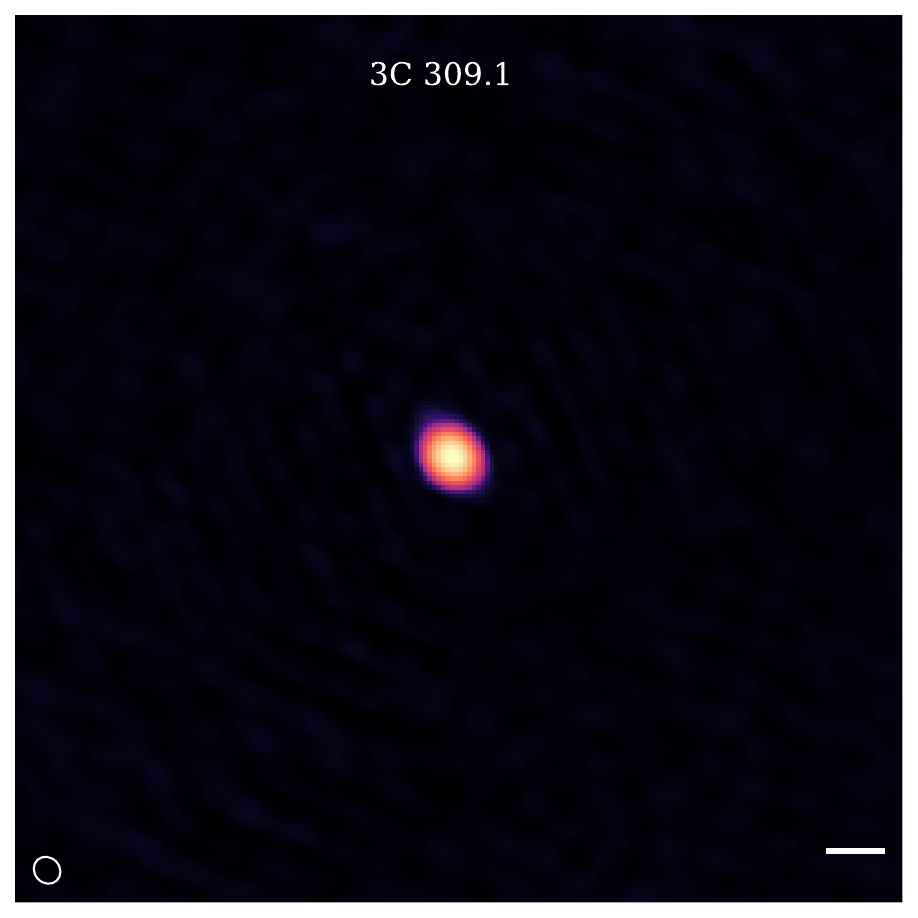}
\includegraphics[width=0.162\linewidth, trim={0.cm 0.cm 0.cm 0.cm},clip]{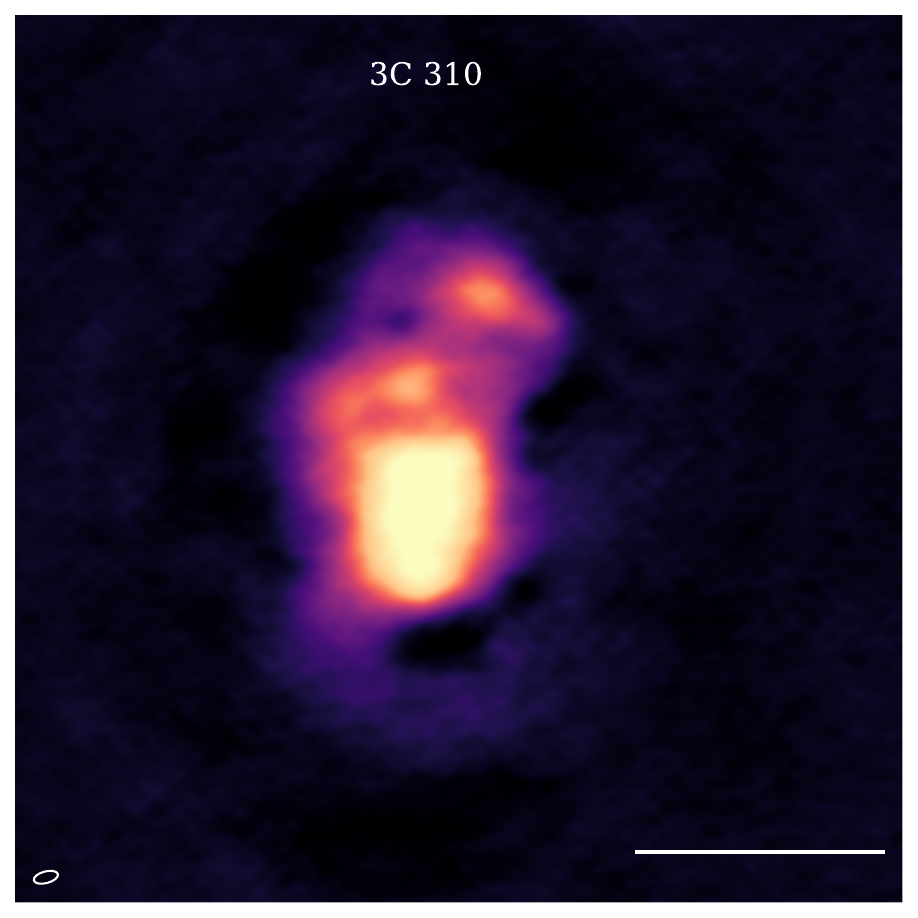}
\includegraphics[width=0.162\linewidth, trim={0.cm 0.cm 0.cm 0.cm},clip]{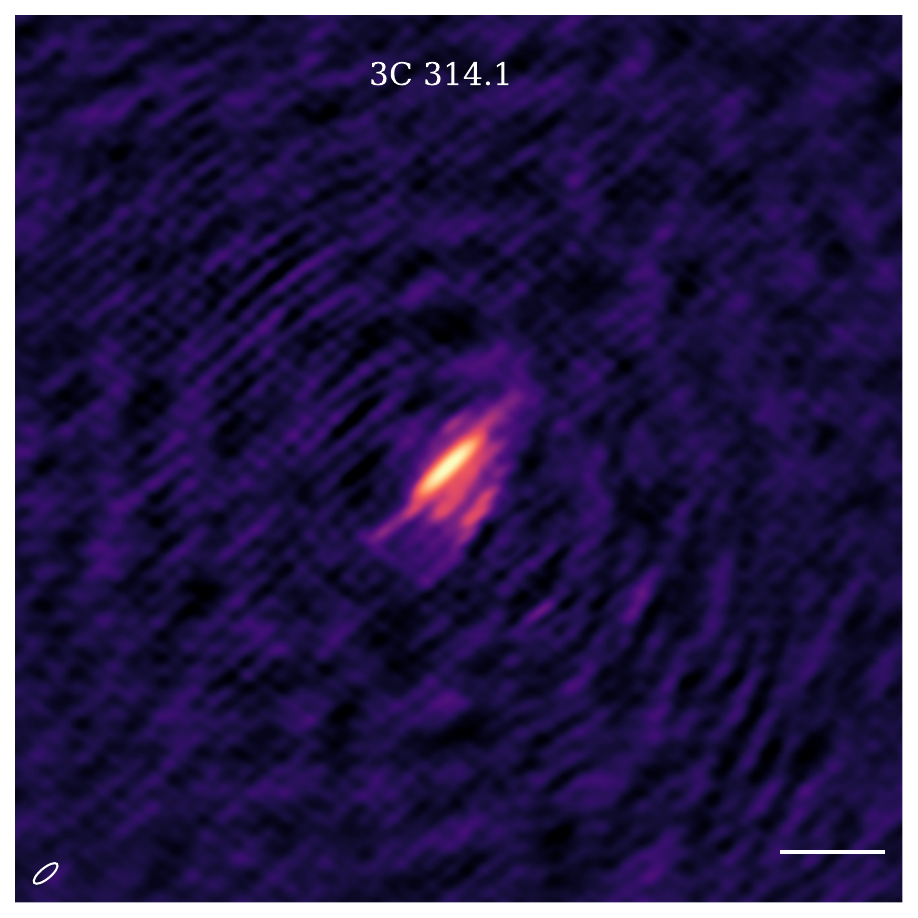}
\includegraphics[width=0.162\linewidth, trim={0.cm 0.cm 0.cm 0.cm},clip]{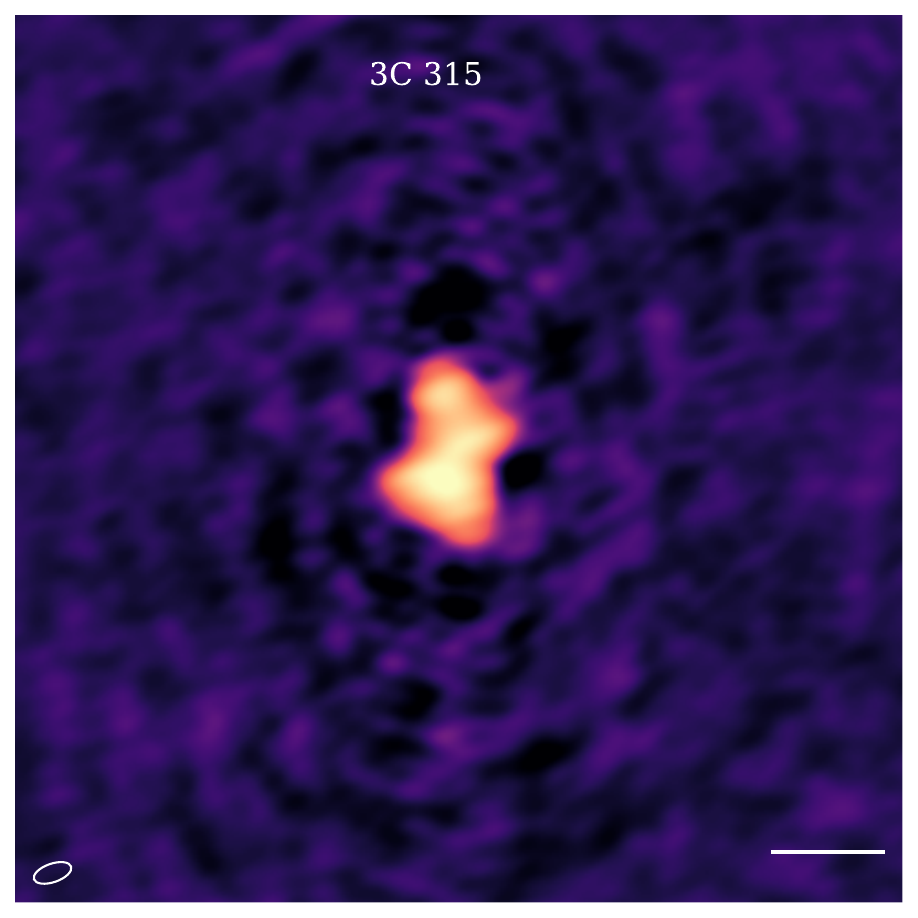}
\includegraphics[width=0.162\linewidth, trim={0.cm 0.cm 0.cm 0.cm},clip]{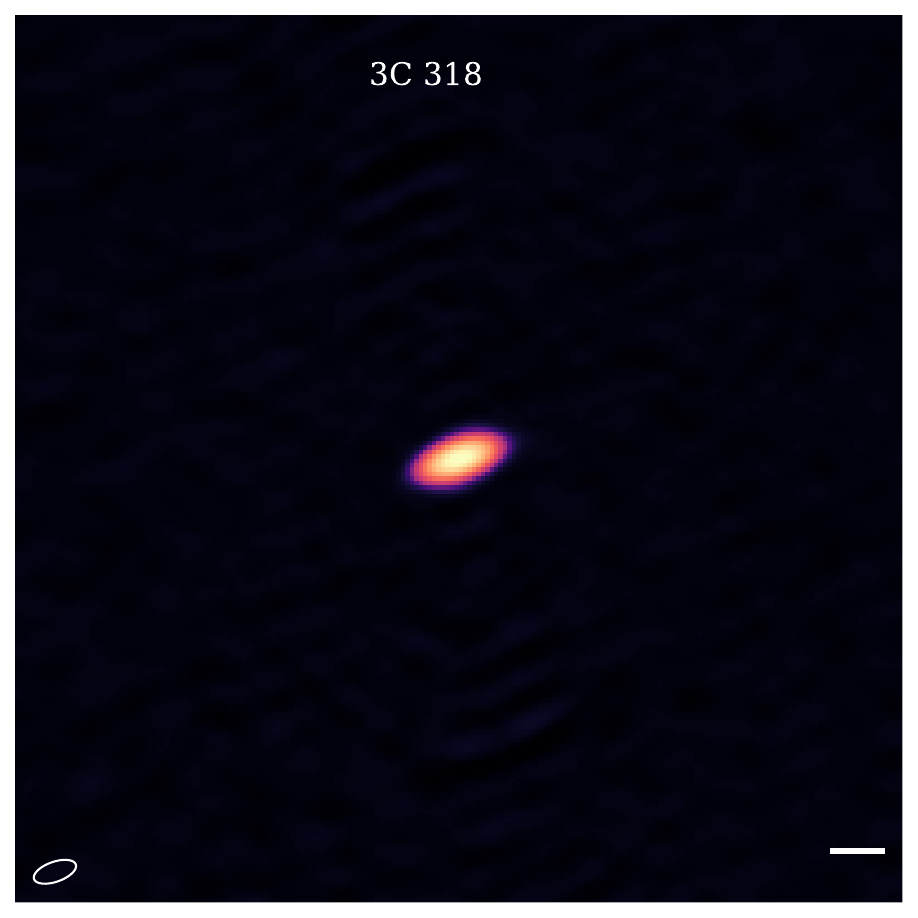}
\includegraphics[width=0.162\linewidth, trim={0.cm 0.cm 0.cm 0.cm},clip]{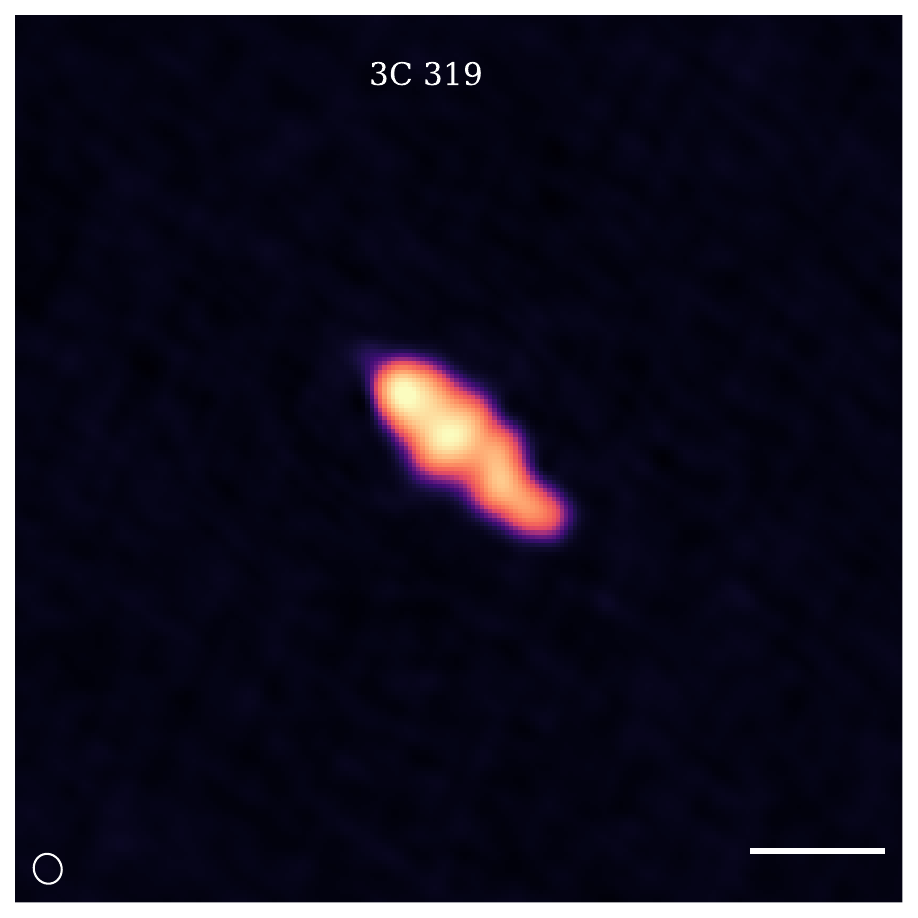}
\includegraphics[width=0.162\linewidth, trim={0.cm 0.cm 0.cm 0.cm},clip]{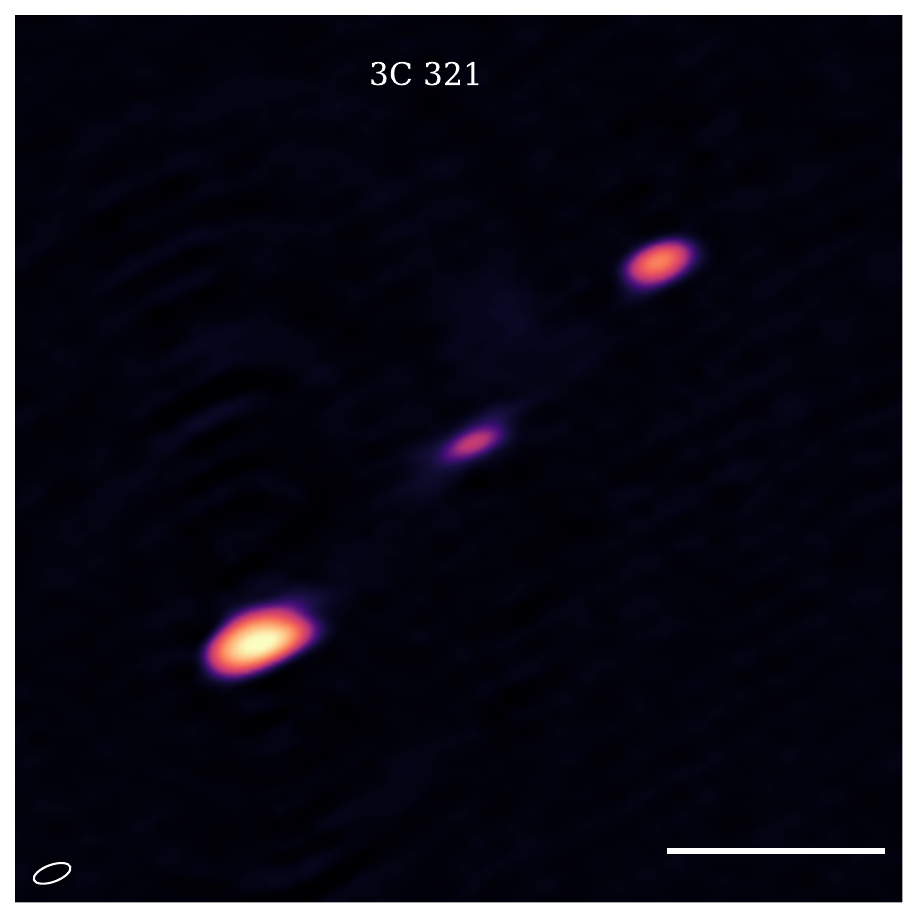}
\includegraphics[width=0.162\linewidth, trim={0.cm 0.cm 0.cm 0.cm},clip]{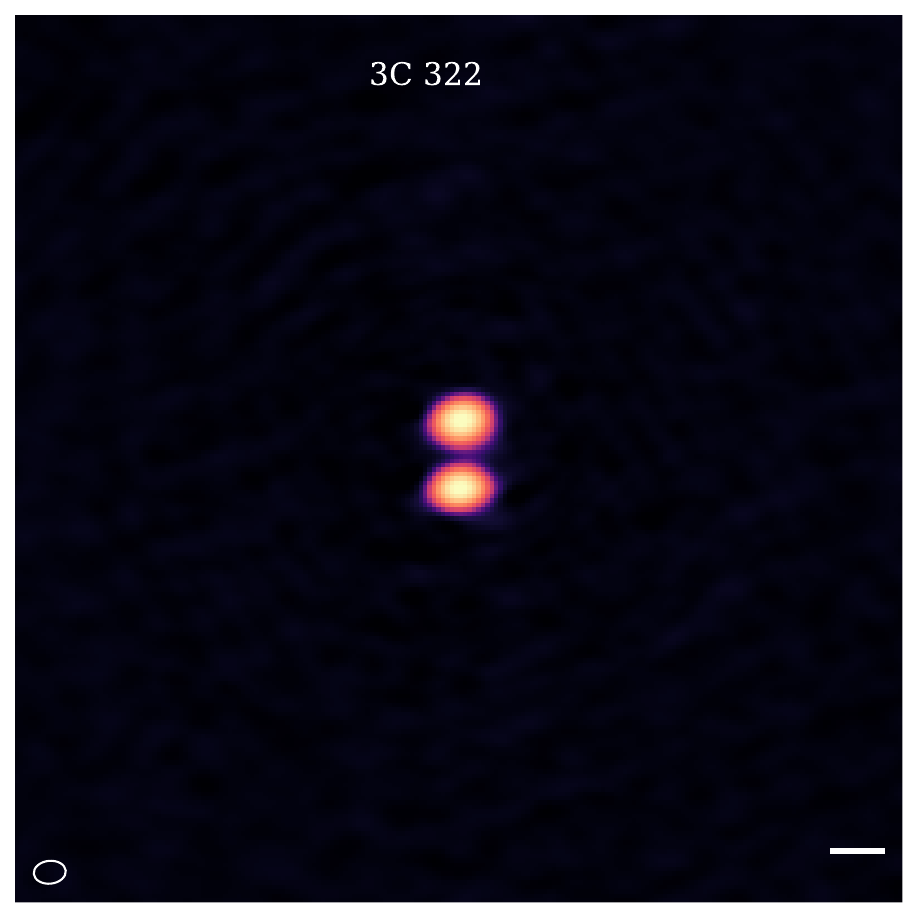}
\includegraphics[width=0.162\linewidth, trim={0.cm 0.cm 0.cm 0.cm},clip]{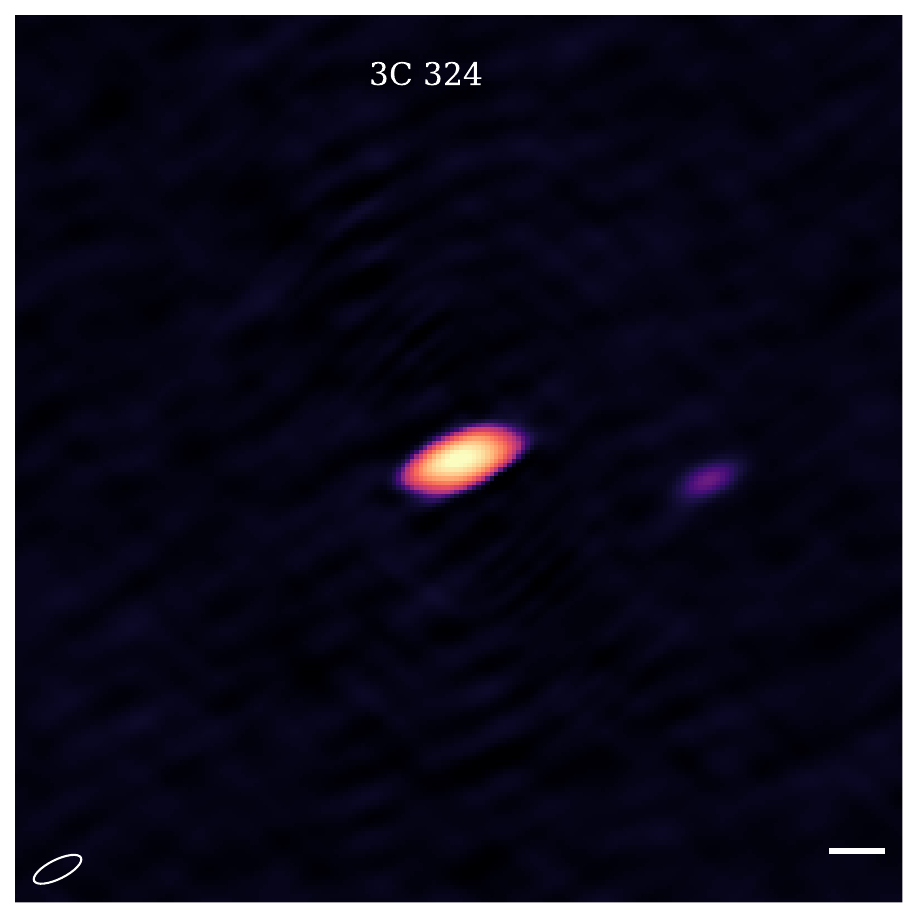}
\includegraphics[width=0.162\linewidth, trim={0.cm 0.cm 0.cm 0.cm},clip]{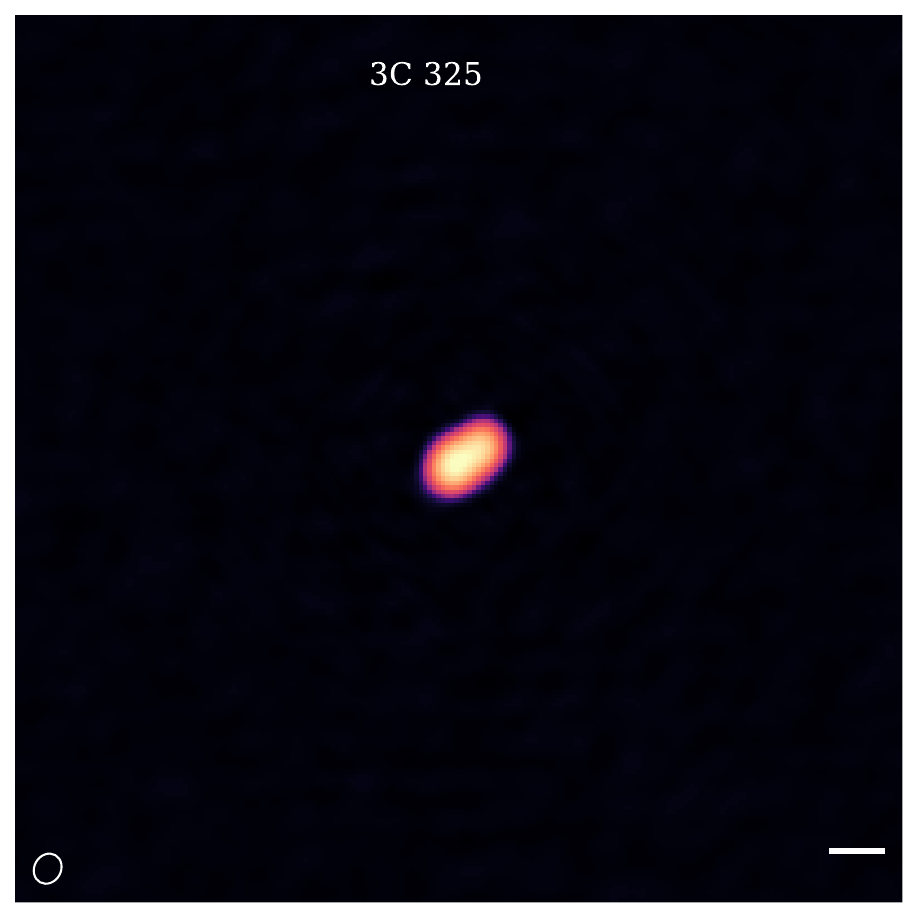}
\includegraphics[width=0.162\linewidth, trim={0.cm 0.cm 0.cm 0.cm},clip]{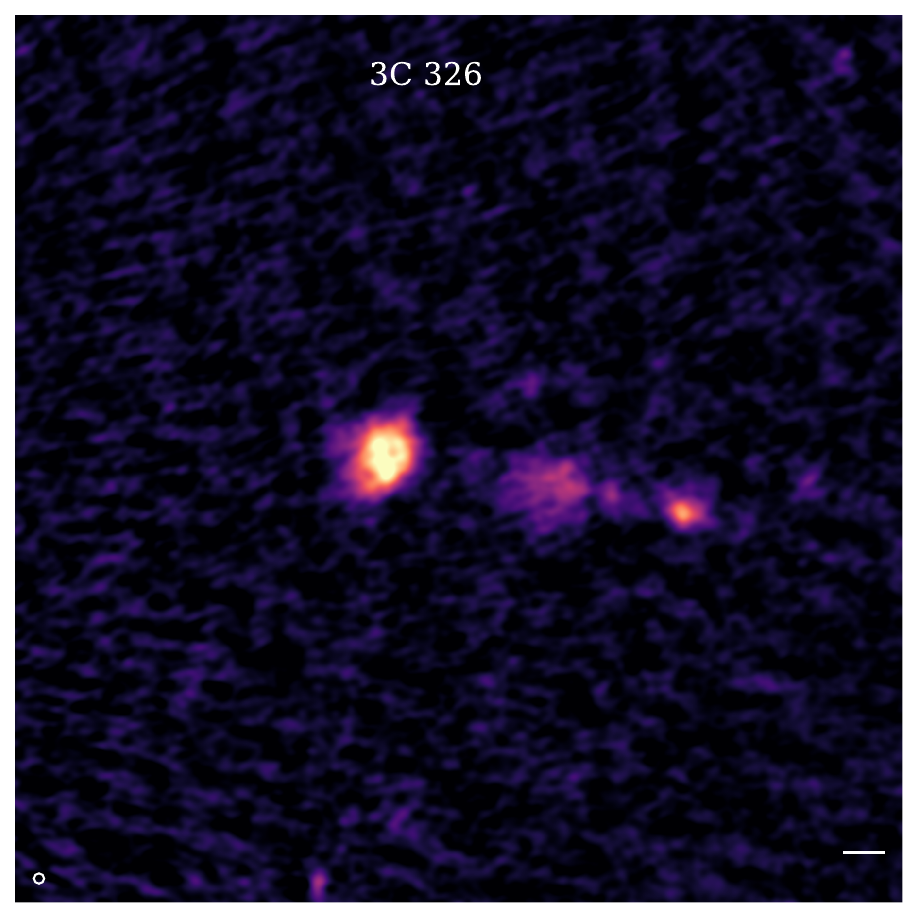}
\includegraphics[width=0.162\linewidth, trim={0.cm 0.cm 0.cm 0.cm},clip]{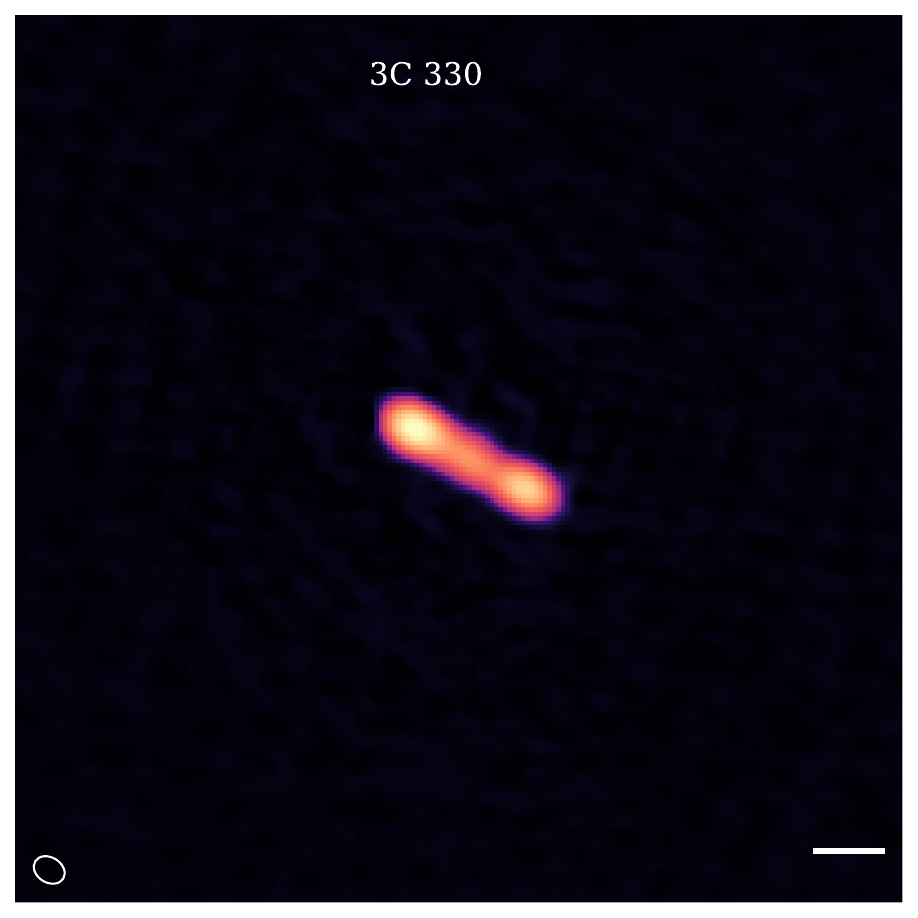}
\includegraphics[width=0.162\linewidth, trim={0.cm 0.cm 0.cm 0.cm},clip]{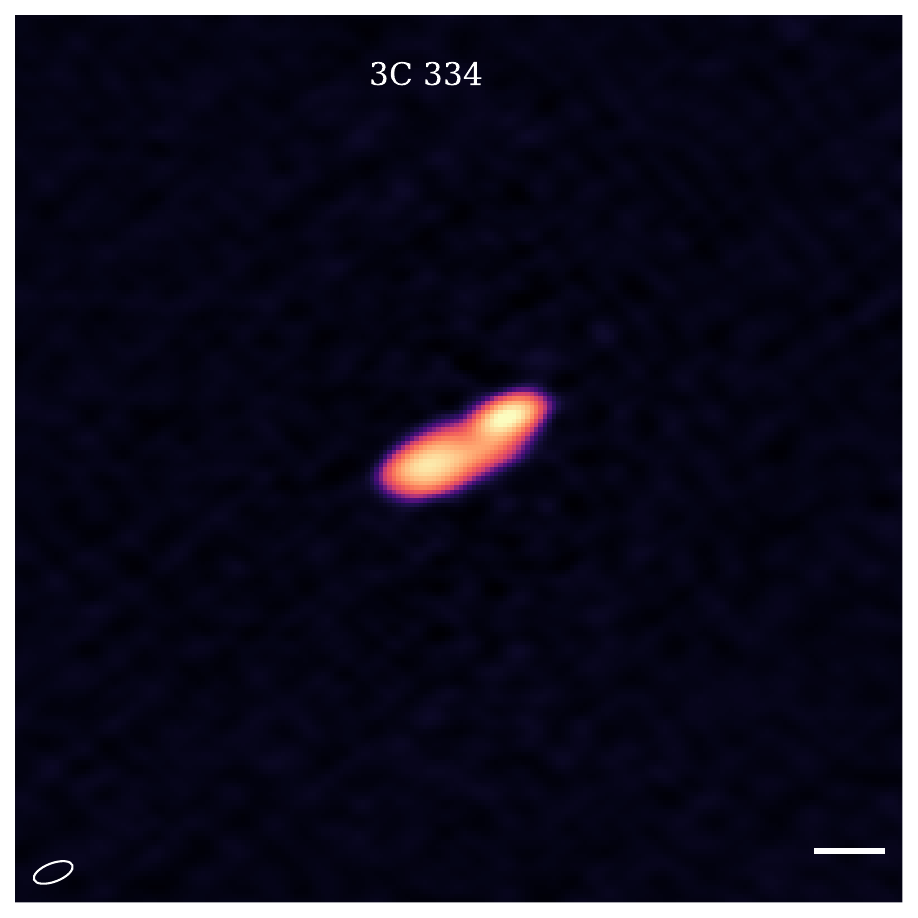}
\includegraphics[width=0.162\linewidth, trim={0.cm 0.cm 0.cm 0.cm},clip]{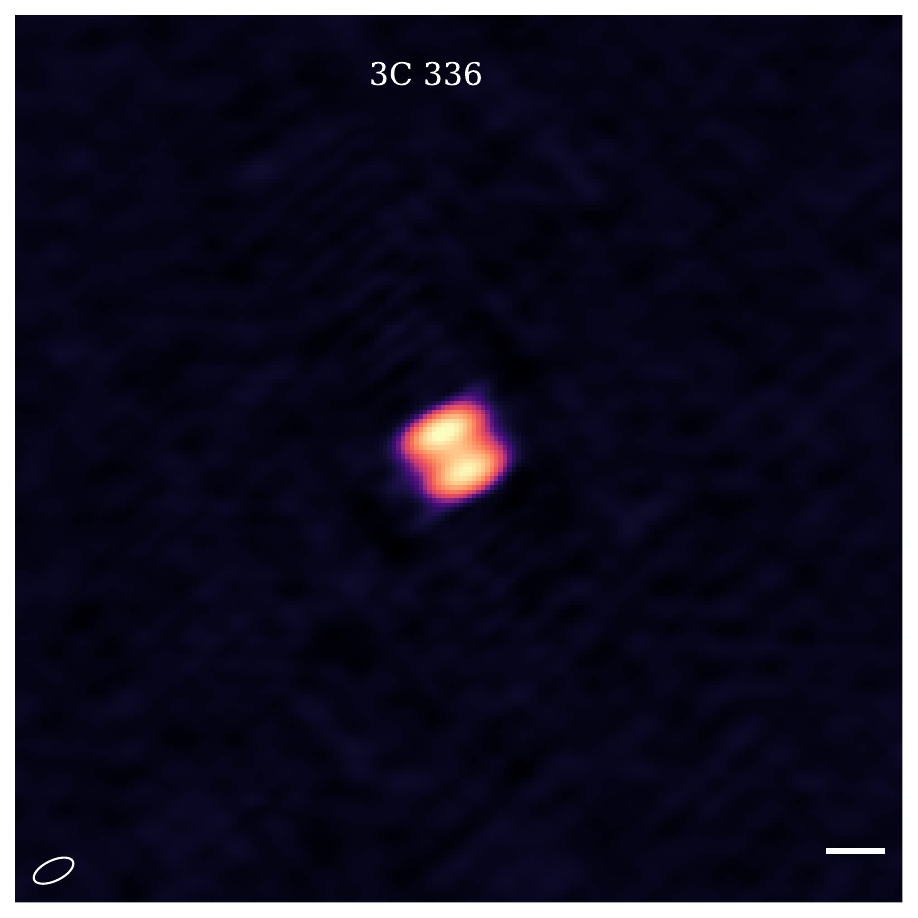}
\includegraphics[width=0.162\linewidth, trim={0.cm 0.cm 0.cm 0.cm},clip]{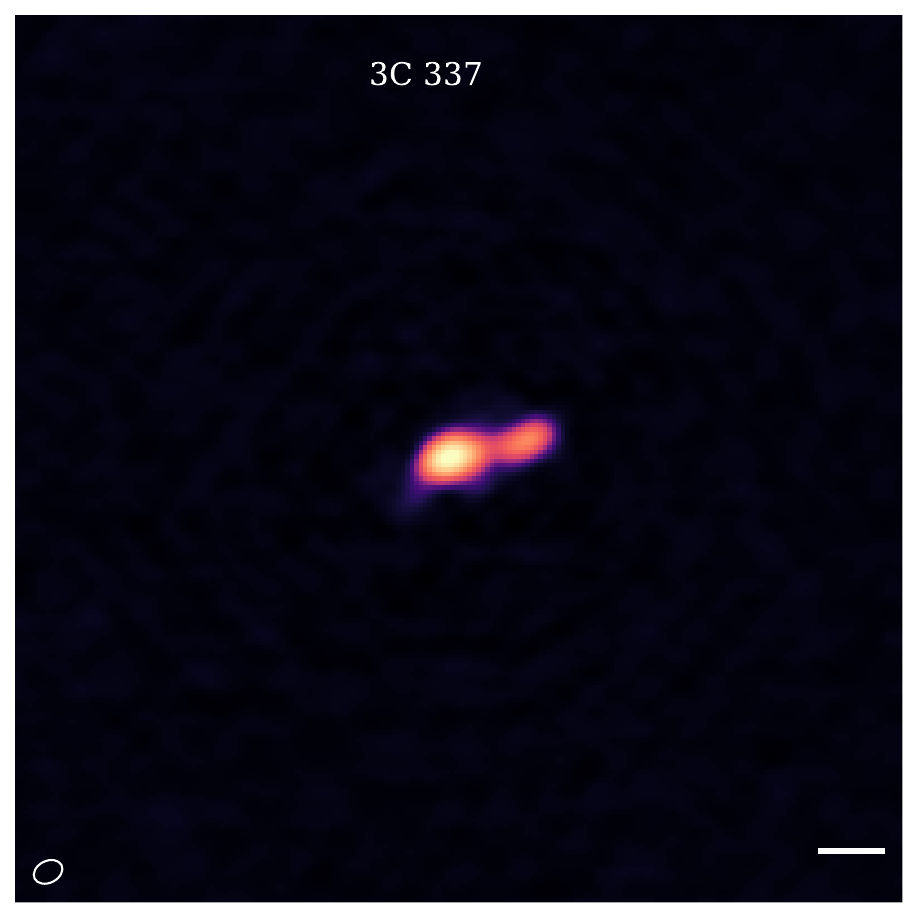}
\includegraphics[width=0.162\linewidth, trim={0.cm 0.cm 0.cm 0.cm},clip]{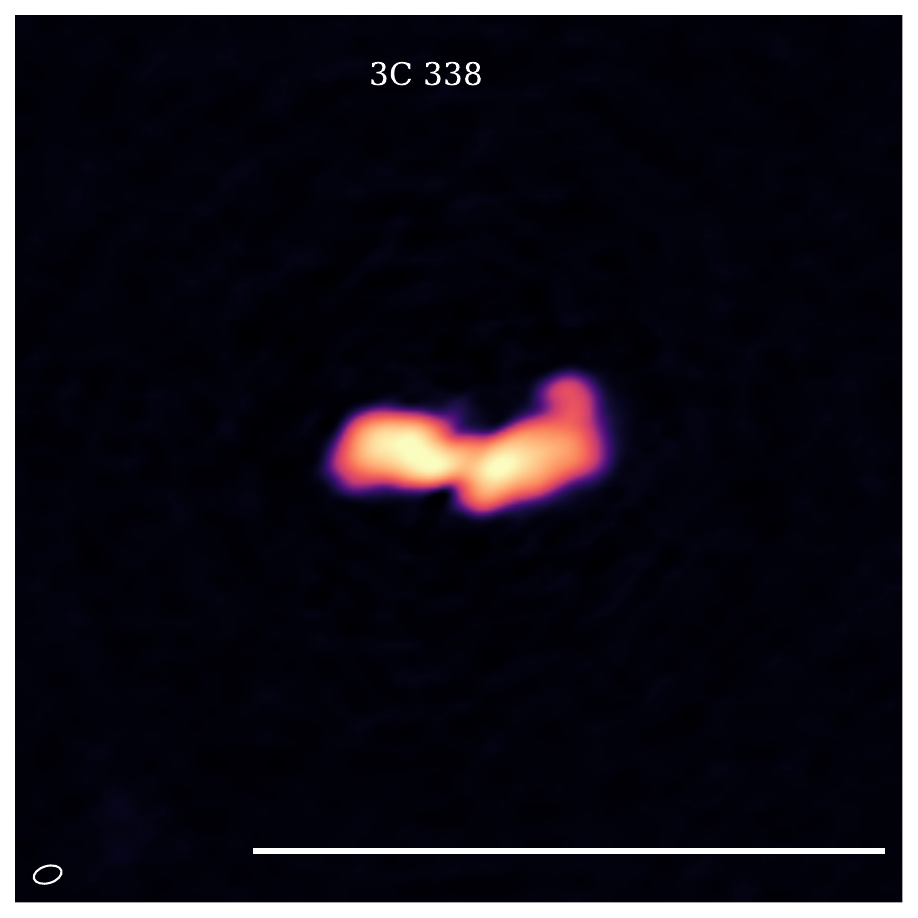}
\includegraphics[width=0.162\linewidth, trim={0.cm 0.cm 0.cm 0.cm},clip]{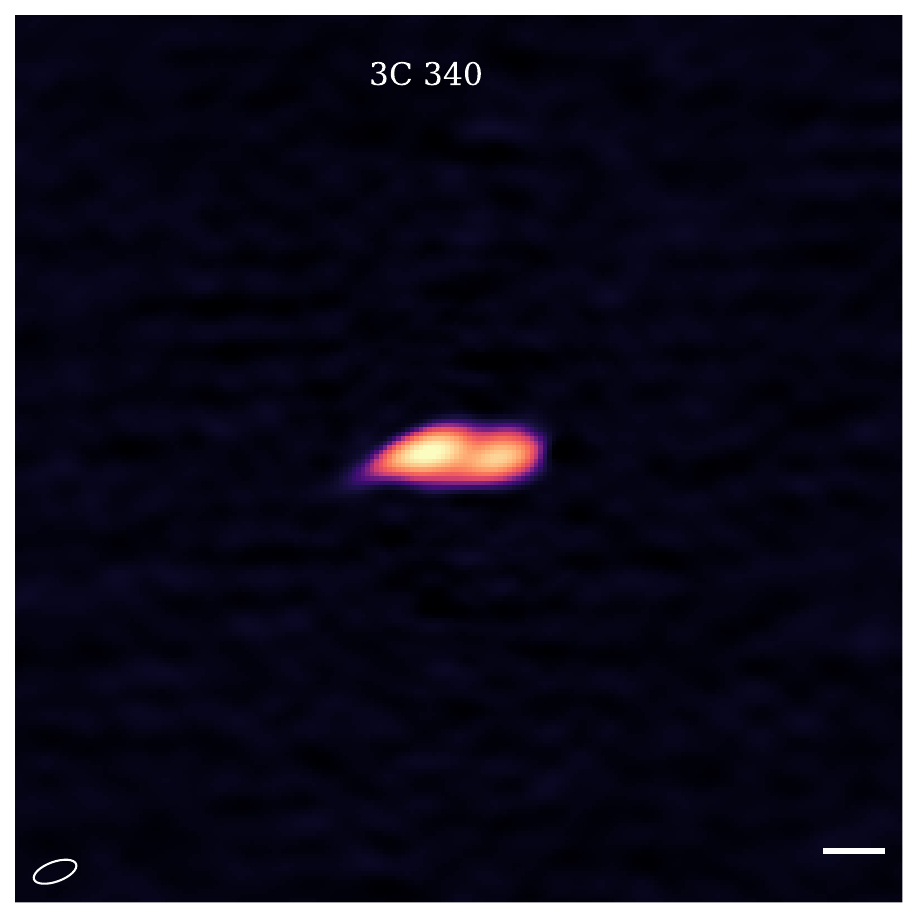}
\includegraphics[width=0.162\linewidth, trim={0.cm 0.cm 0.cm 0.cm},clip]{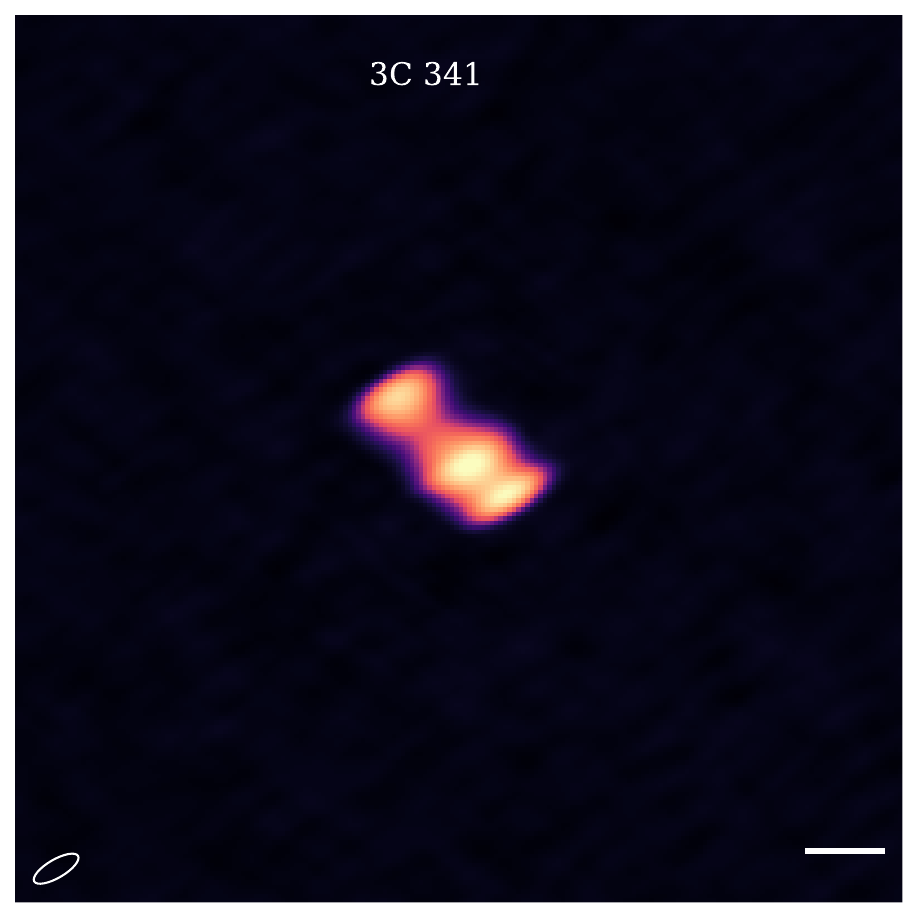}
\includegraphics[width=0.162\linewidth, trim={0.cm 0.cm 0.cm 0.cm},clip]{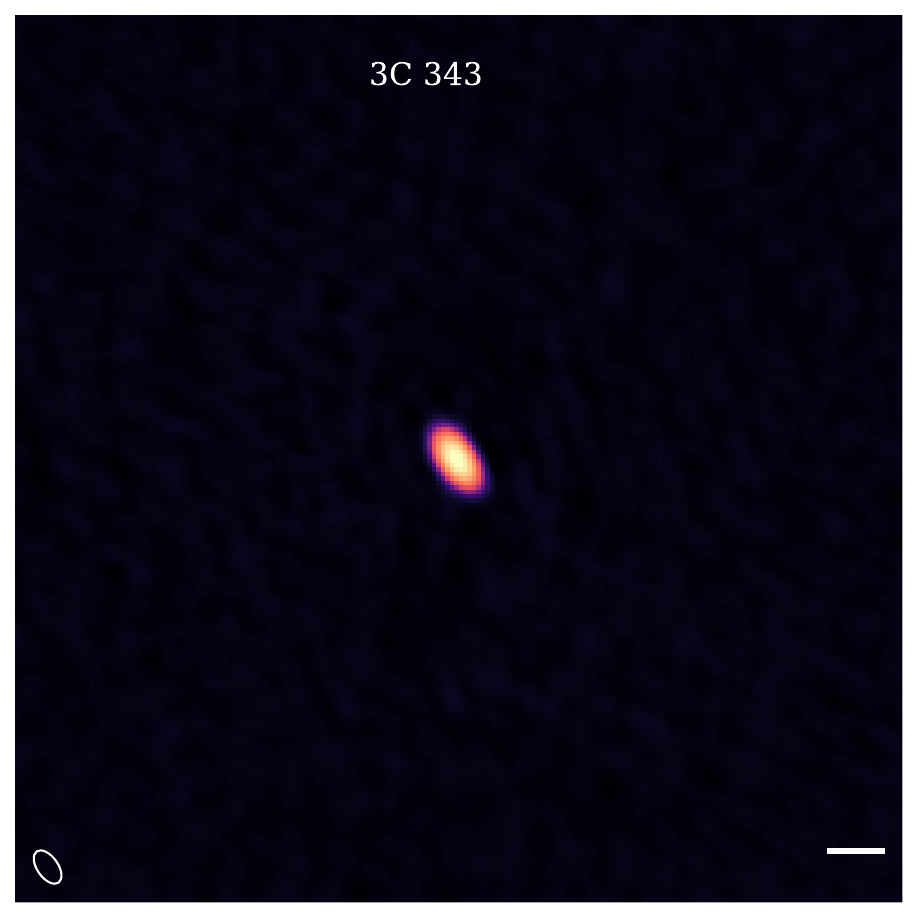}
\includegraphics[width=0.162\linewidth, trim={0.cm 0.cm 0.cm 0.cm},clip]{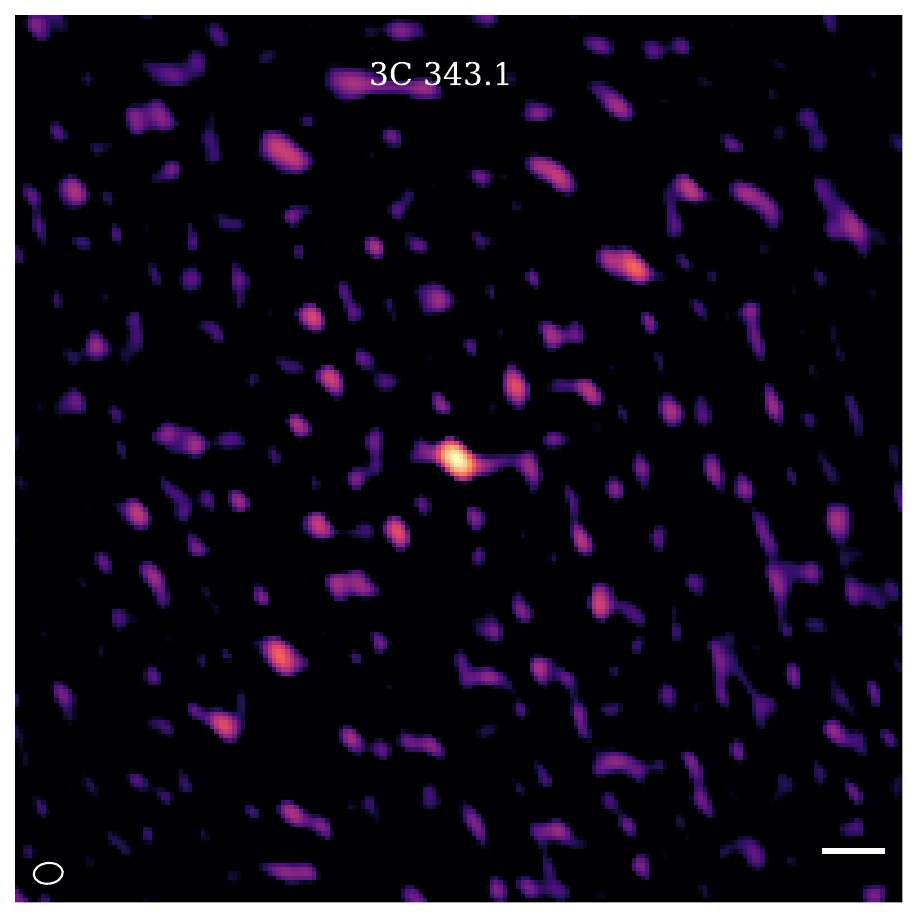}
\includegraphics[width=0.162\linewidth, trim={0.cm 0.cm 0.cm 0.cm},clip]{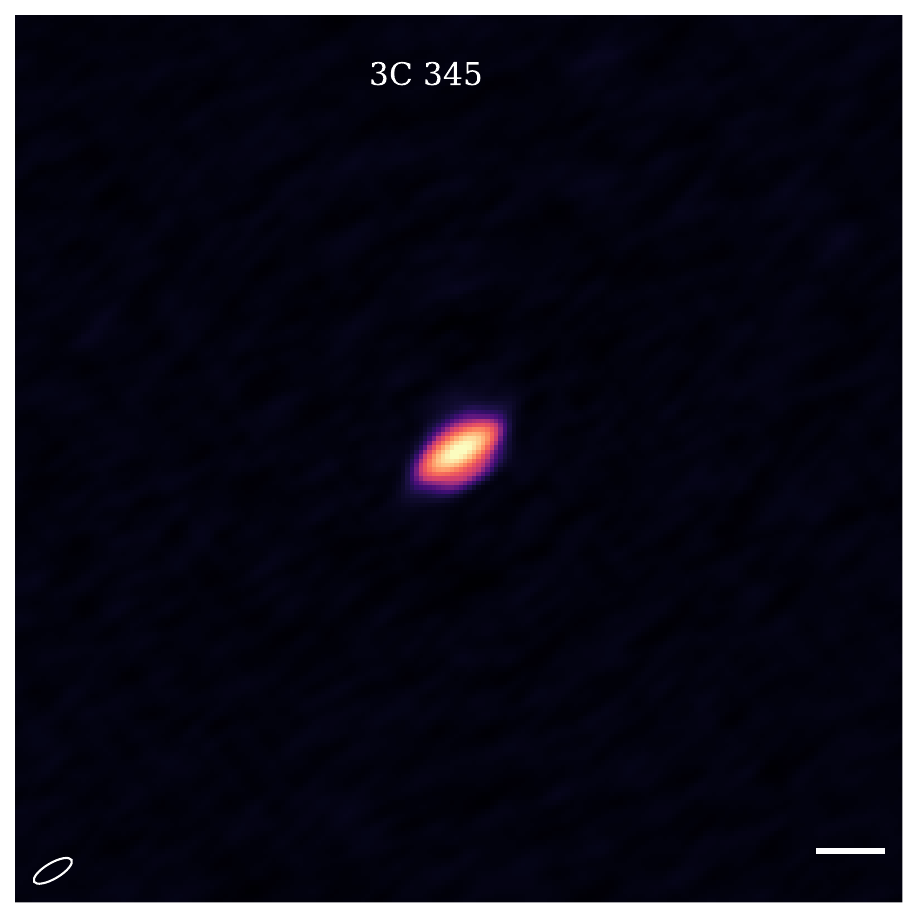}
\includegraphics[width=0.162\linewidth, trim={0.cm 0.cm 0.cm 0.cm},clip]{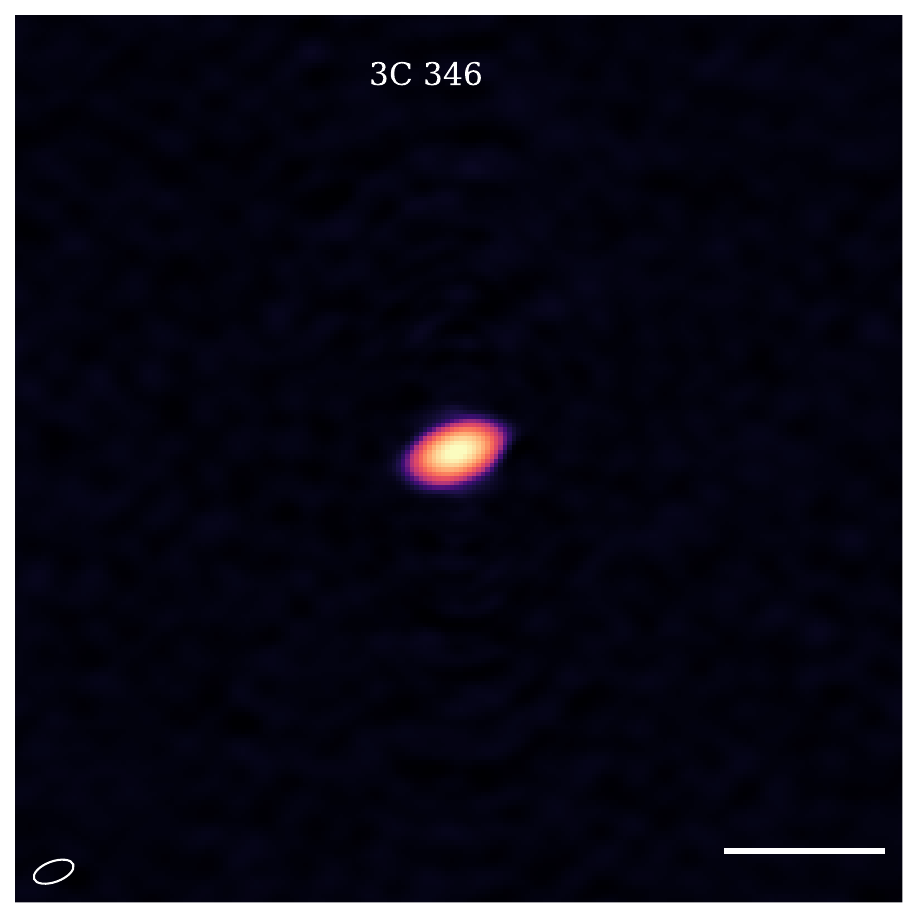}
\includegraphics[width=0.162\linewidth, trim={0.cm 0.cm 0.cm 0.cm},clip]{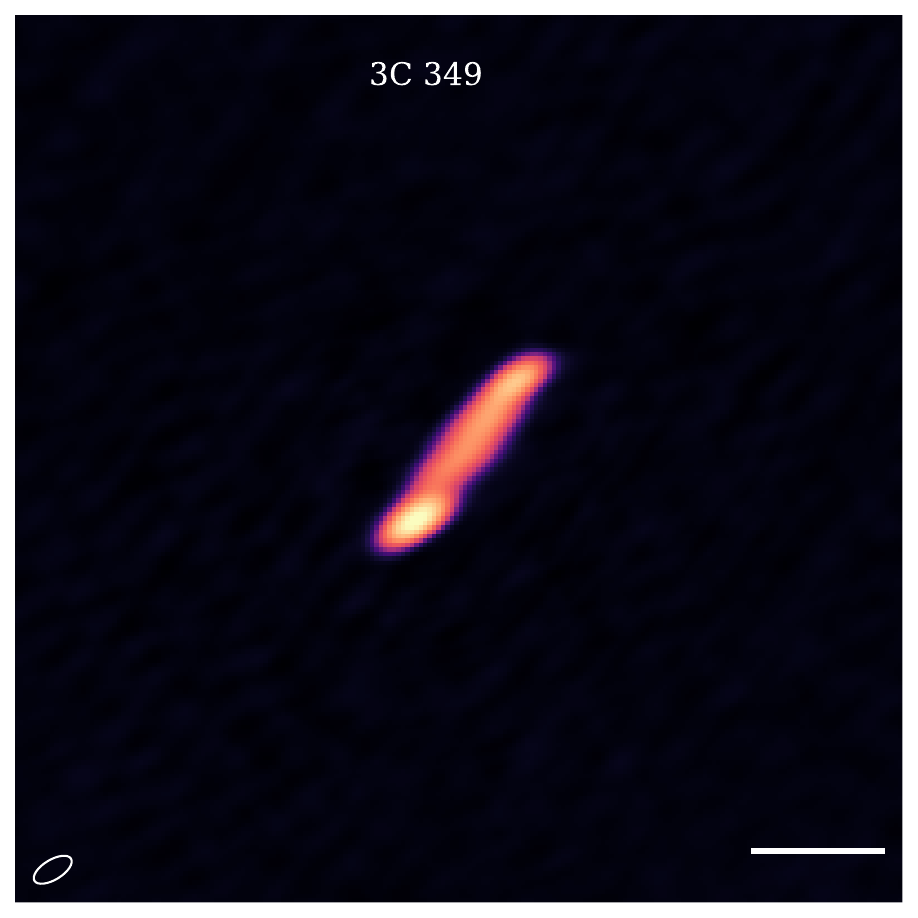}
\includegraphics[width=0.162\linewidth, trim={0.cm 0.cm 0.cm 0.cm},clip]{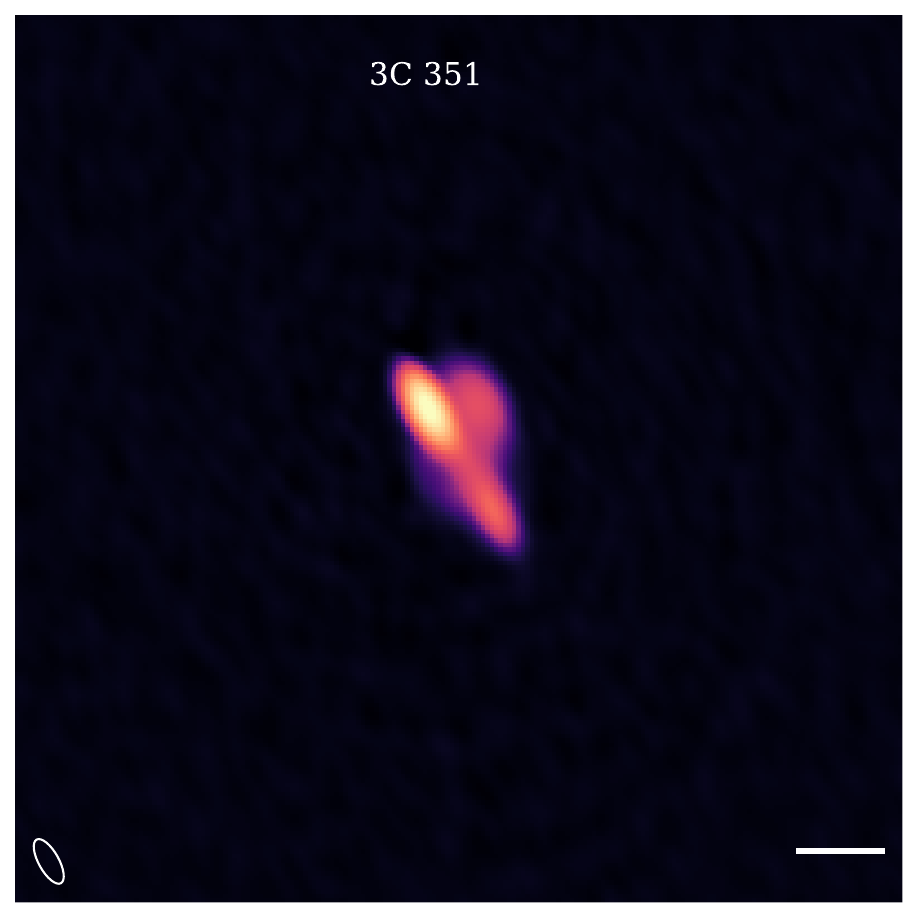}
\includegraphics[width=0.162\linewidth, trim={0.cm 0.cm 0.cm 0.cm},clip]{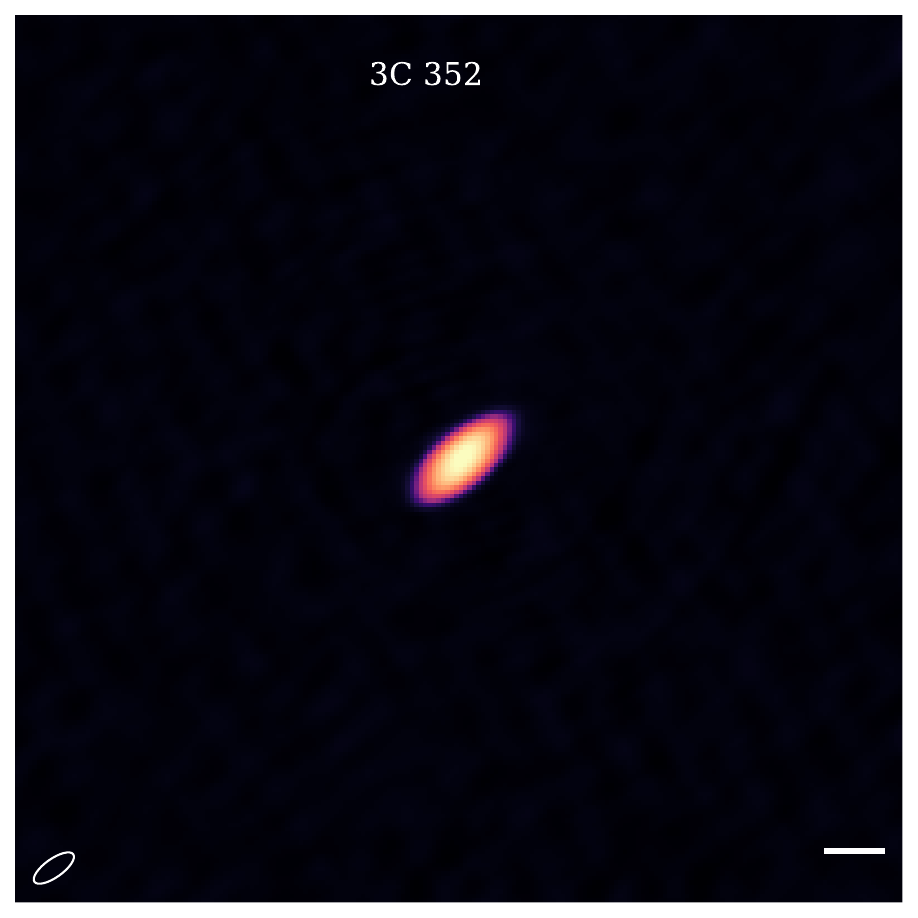}
\includegraphics[width=0.162\linewidth, trim={0.cm 0.cm 0.cm 0.cm},clip]{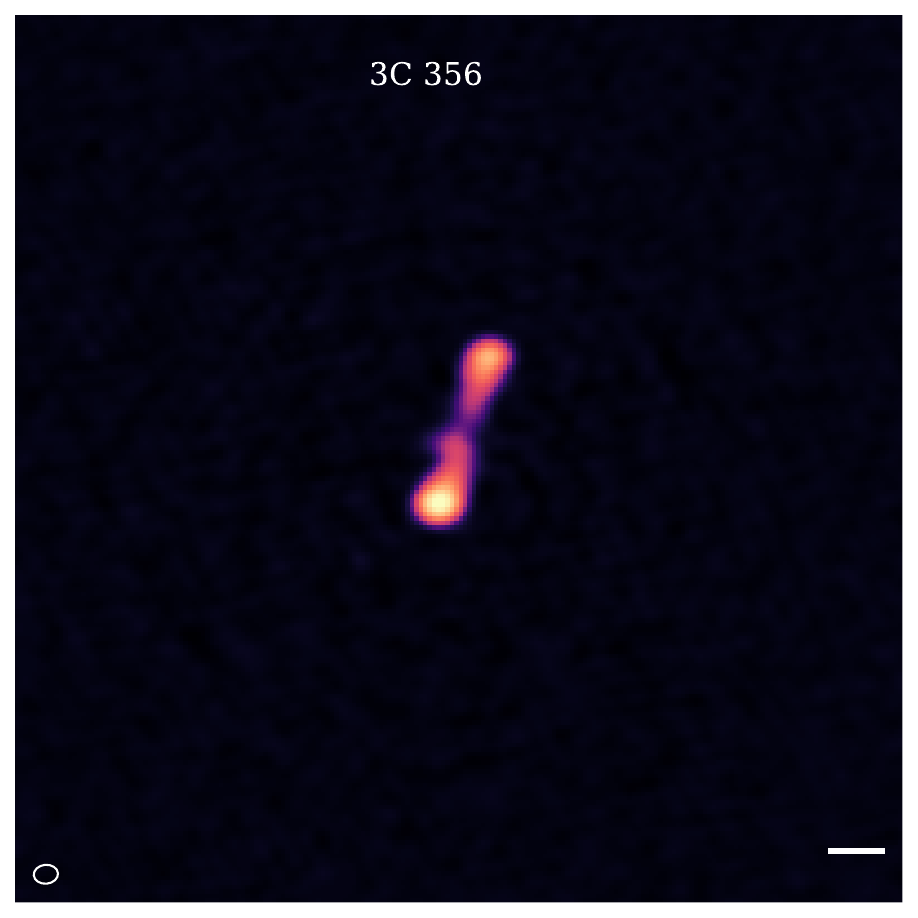}
\includegraphics[width=0.162\linewidth, trim={0.cm 0.cm 0.cm 0.cm},clip]{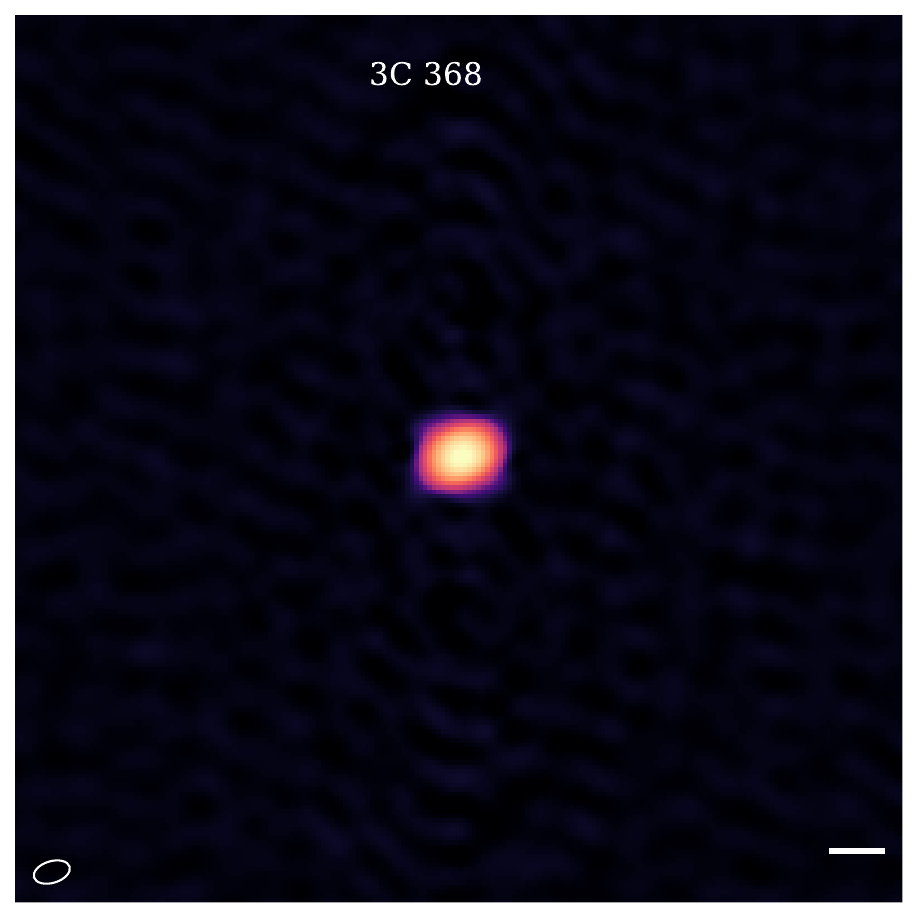}
\includegraphics[width=0.162\linewidth, trim={0.cm 0.cm 0.cm 0.cm},clip]{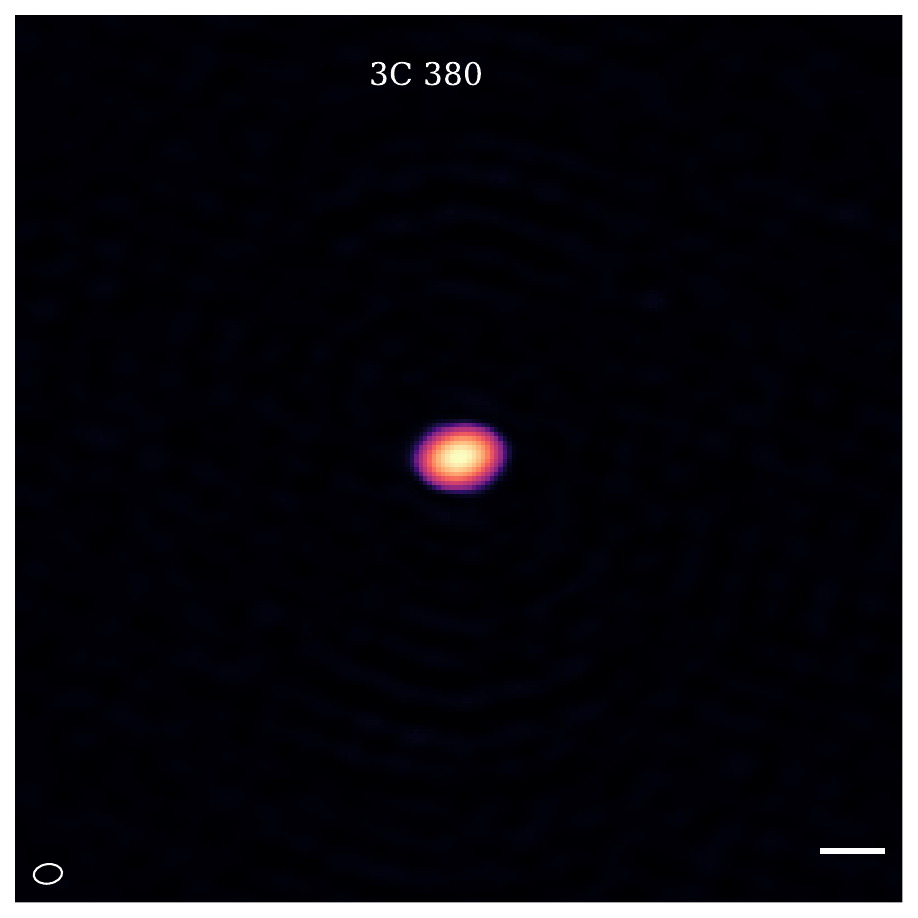}
\caption{continued.}
\end{figure}
\setcounter{figure}{0} 

\begin{figure}[H]
\includegraphics[width=0.162\linewidth, trim={0.cm 0.cm 0.cm 0.cm},clip]{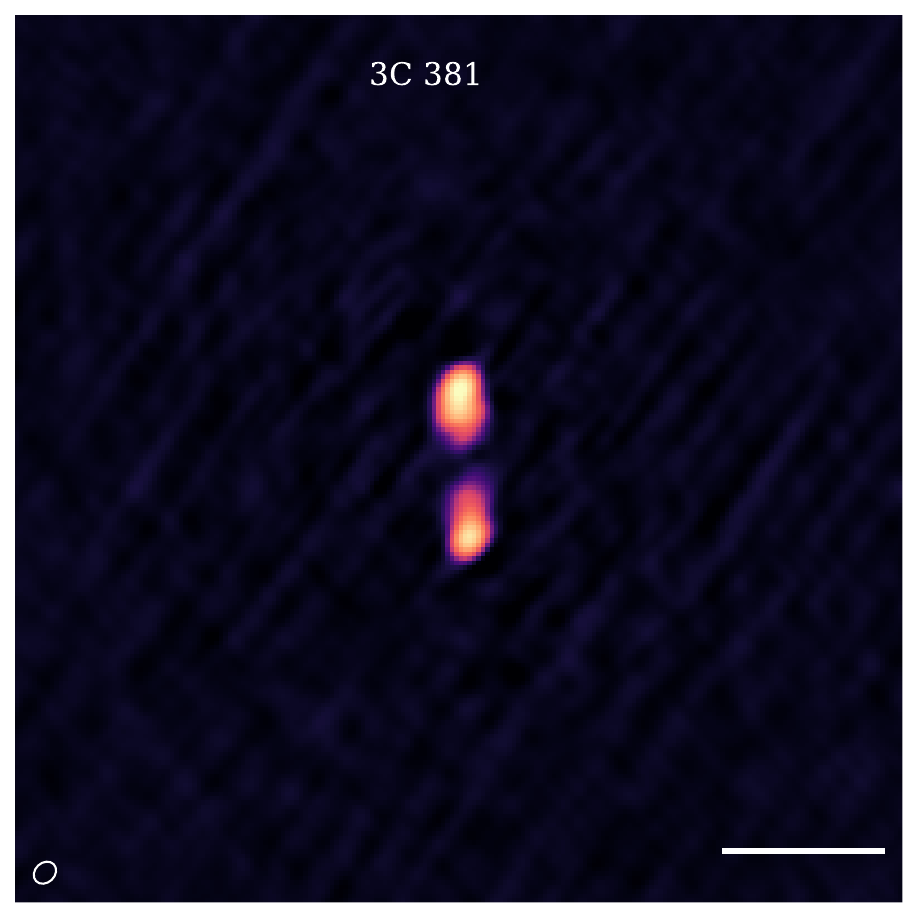}
\includegraphics[width=0.162\linewidth, trim={0.cm 0.cm 0.cm 0.cm},clip]{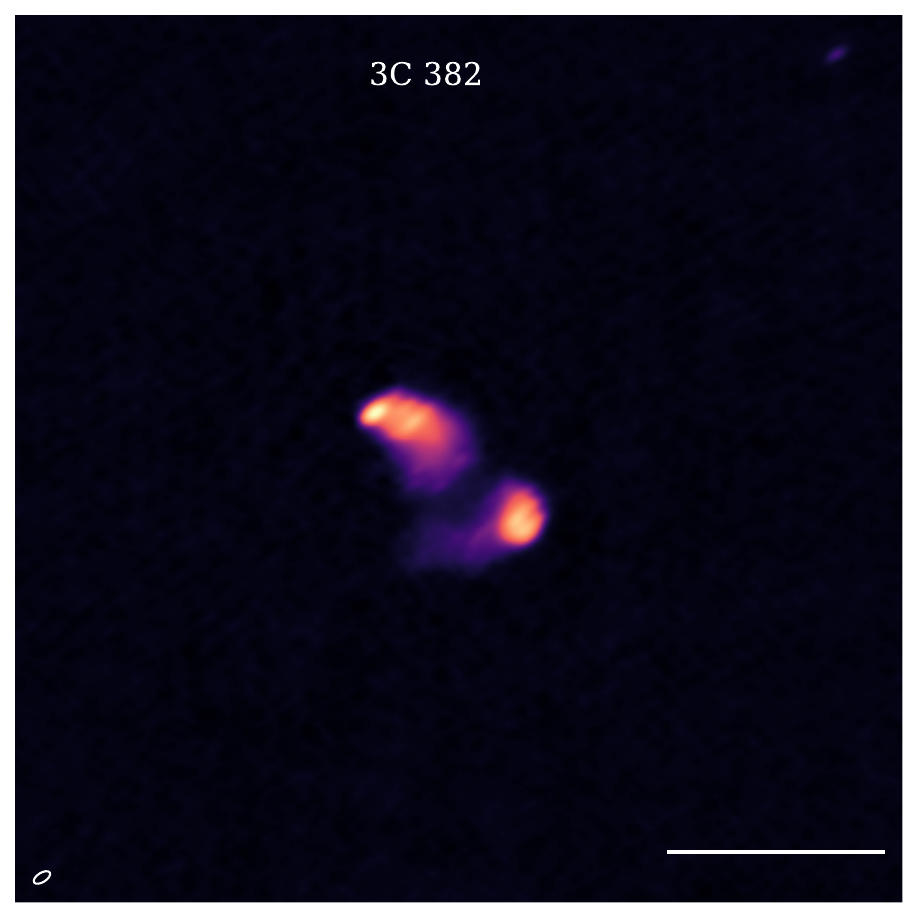}
\includegraphics[width=0.162\linewidth, trim={0.cm 0.cm 0.cm 0.cm},clip]{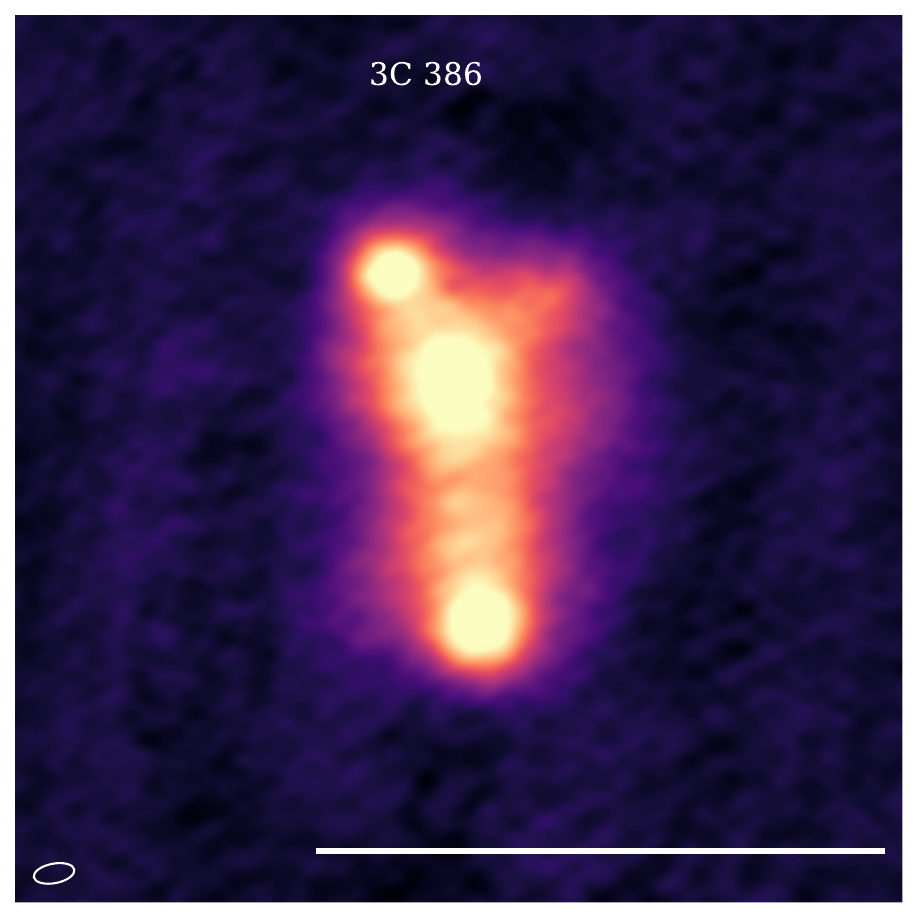}
\includegraphics[width=0.162\linewidth, trim={0.cm 0.cm 0.cm 0.cm},clip]{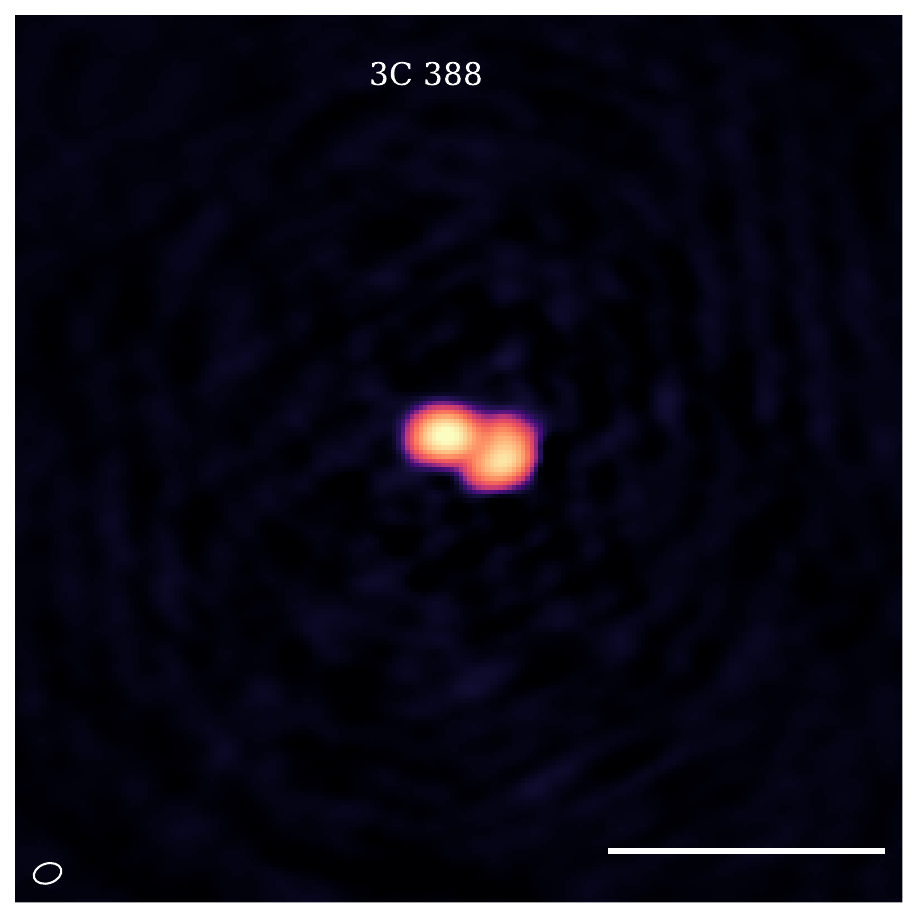}
\includegraphics[width=0.162\linewidth, trim={0.cm 0.cm 0.cm 0.cm},clip]{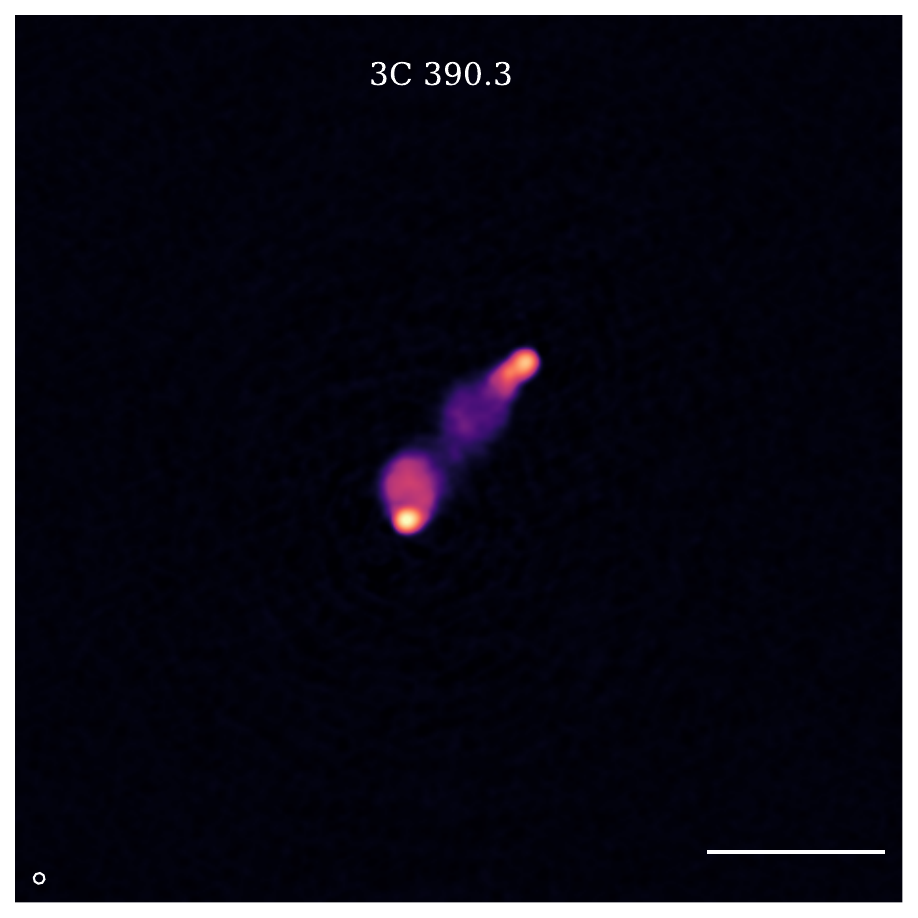}
\includegraphics[width=0.162\linewidth, trim={0.cm 0.cm 0.cm 0.cm},clip]{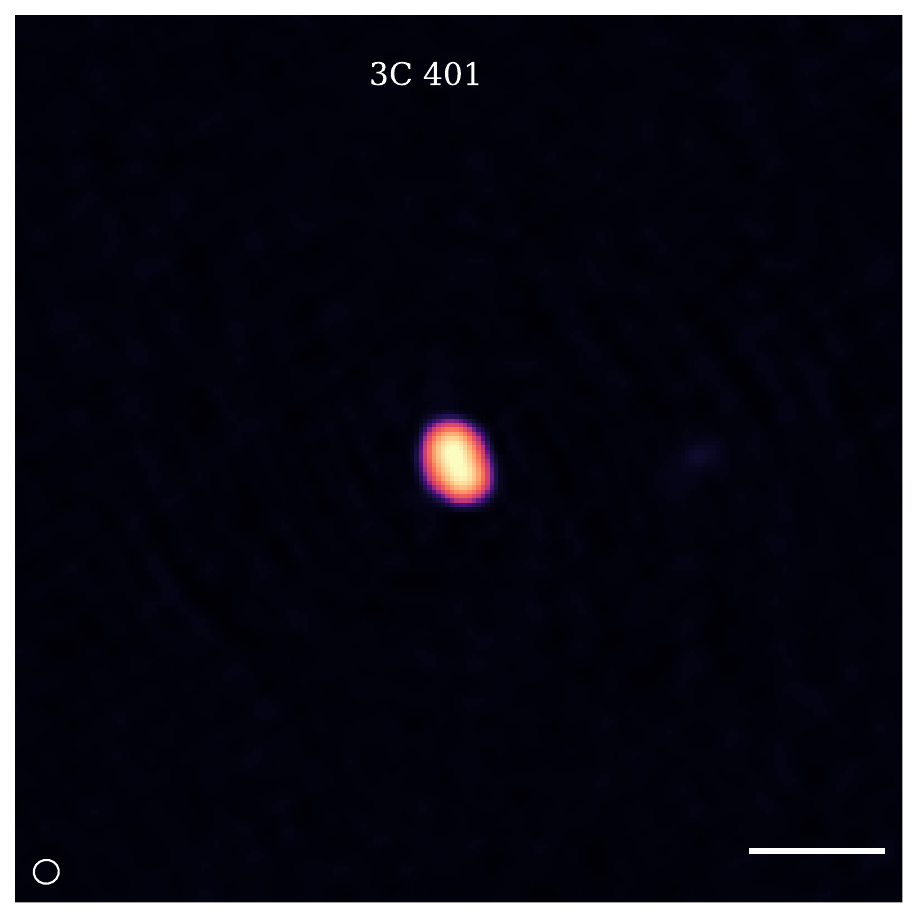}
\includegraphics[width=0.162\linewidth, trim={0.cm 0.cm 0.cm 0.cm},clip]{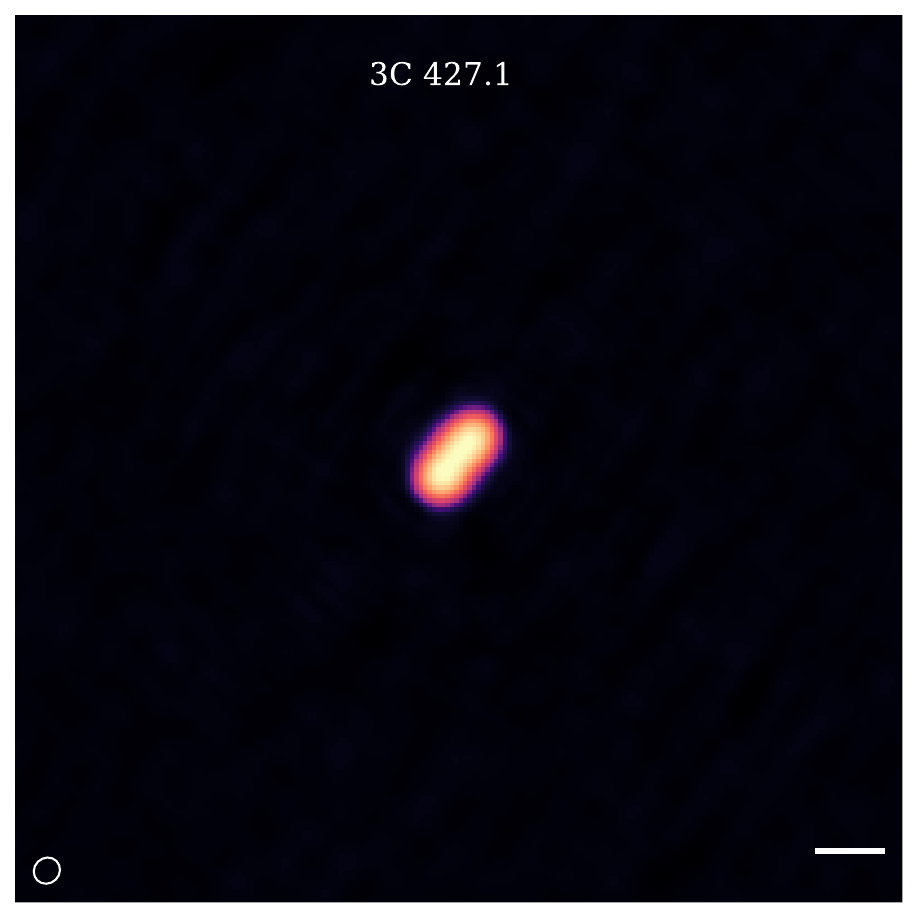}
\includegraphics[width=0.162\linewidth, trim={0.cm 0.cm 0.cm 0.cm},clip]{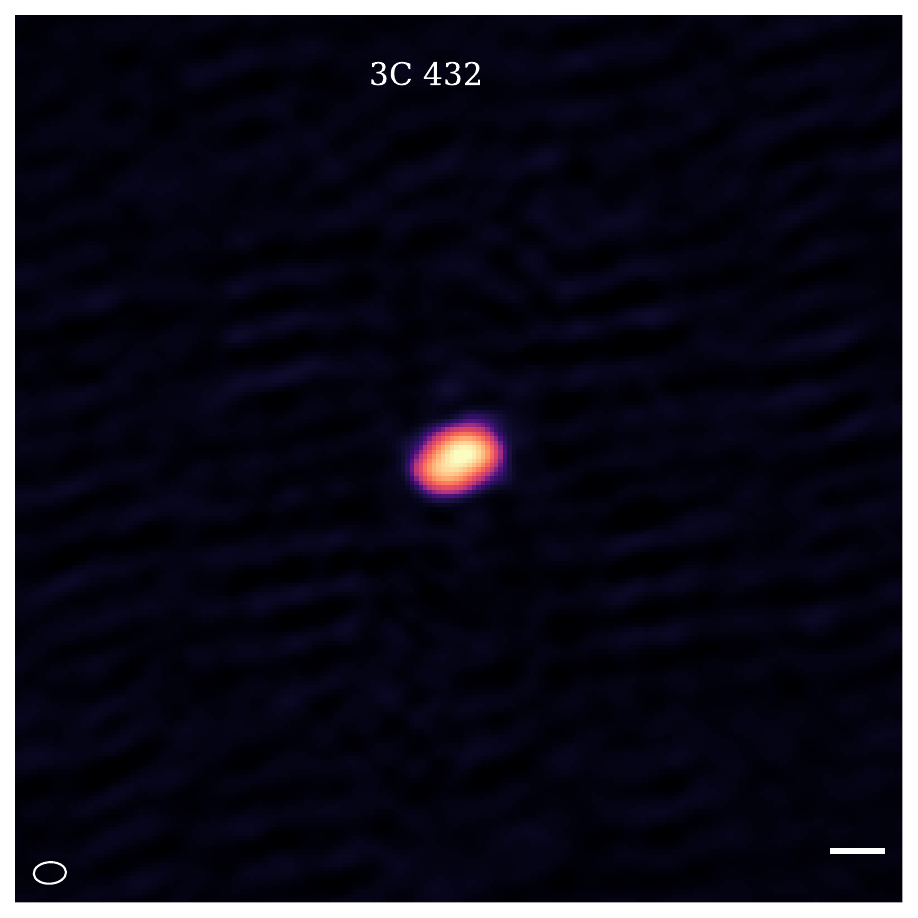}
\includegraphics[width=0.162\linewidth, trim={0.cm 0.cm 0.cm 0.cm},clip]{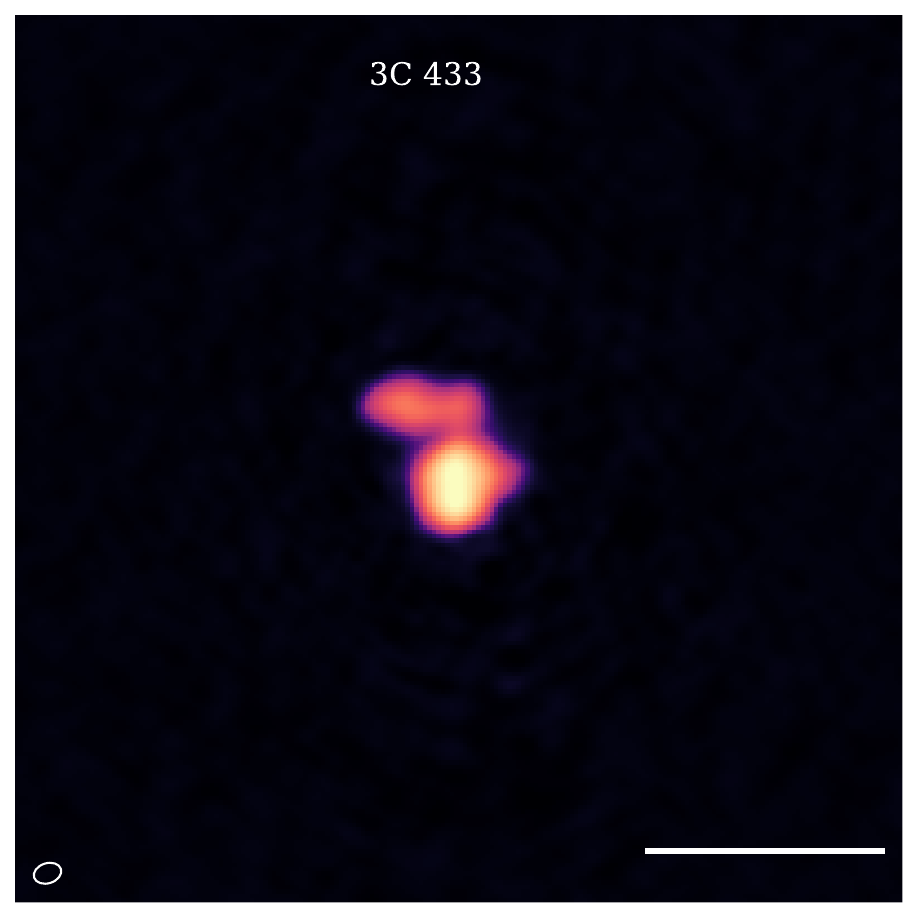}
\includegraphics[width=0.162\linewidth, trim={0.cm 0.cm 0.cm 0.cm},clip]{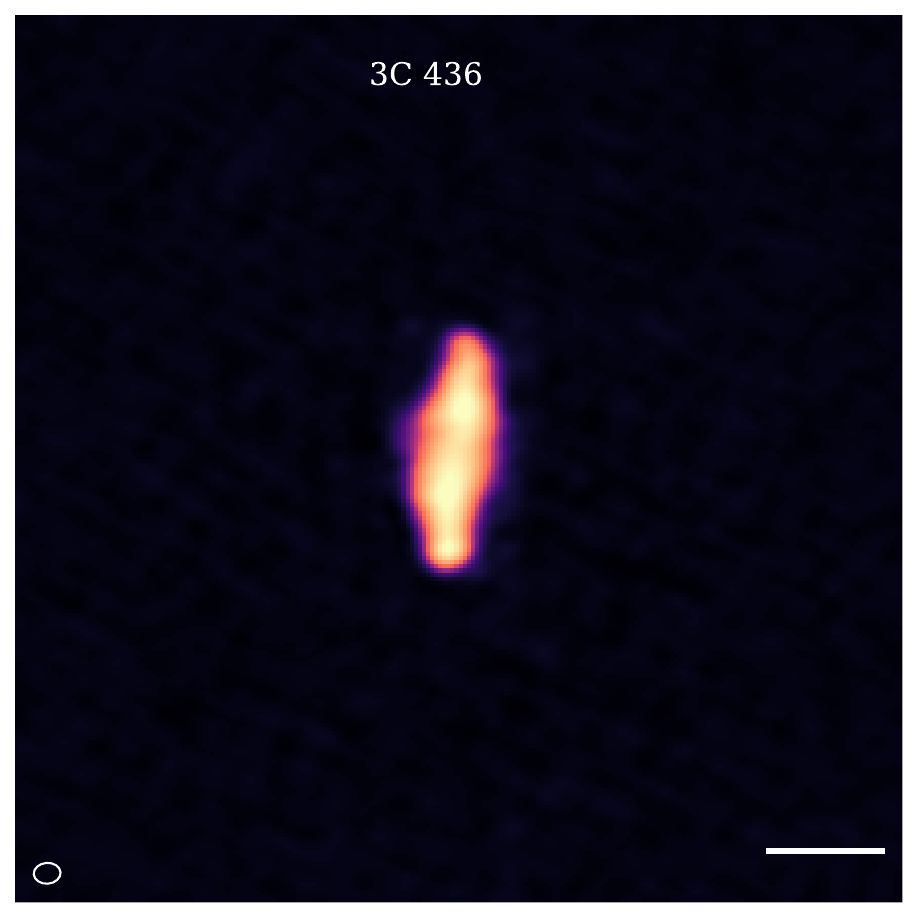}
\includegraphics[width=0.162\linewidth, trim={0.cm 0.cm 0.cm 0.cm},clip]{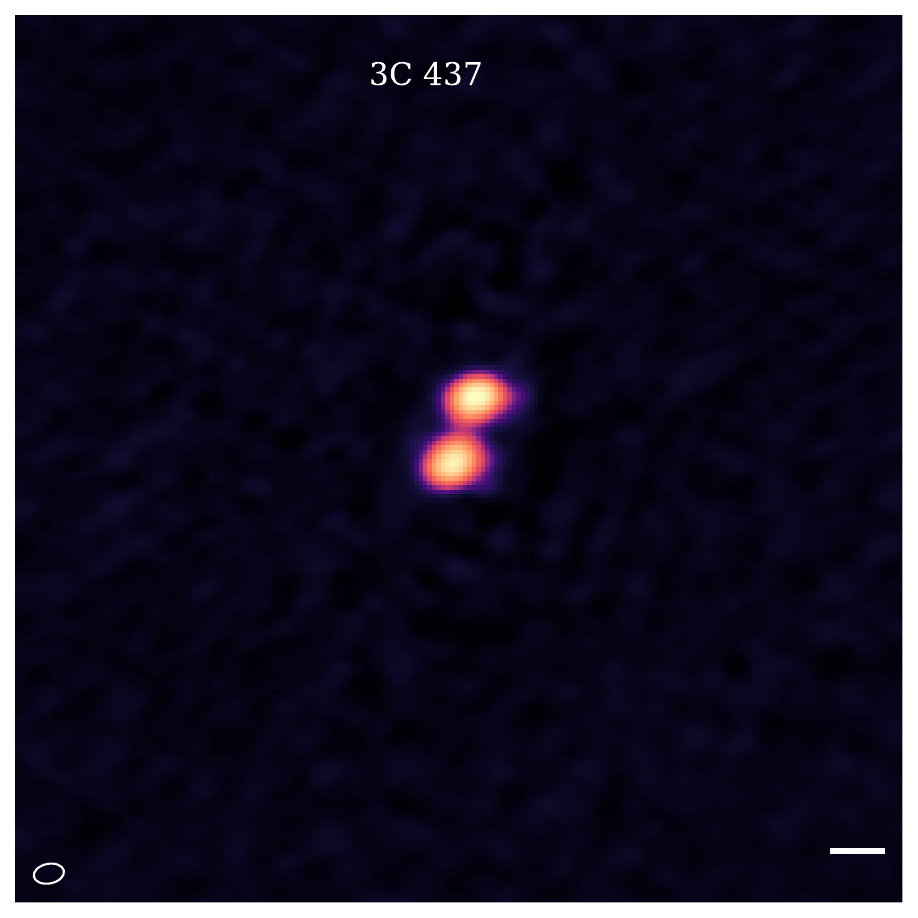}
\includegraphics[width=0.162\linewidth, trim={0.cm 0.cm 0.cm 0.cm},clip]{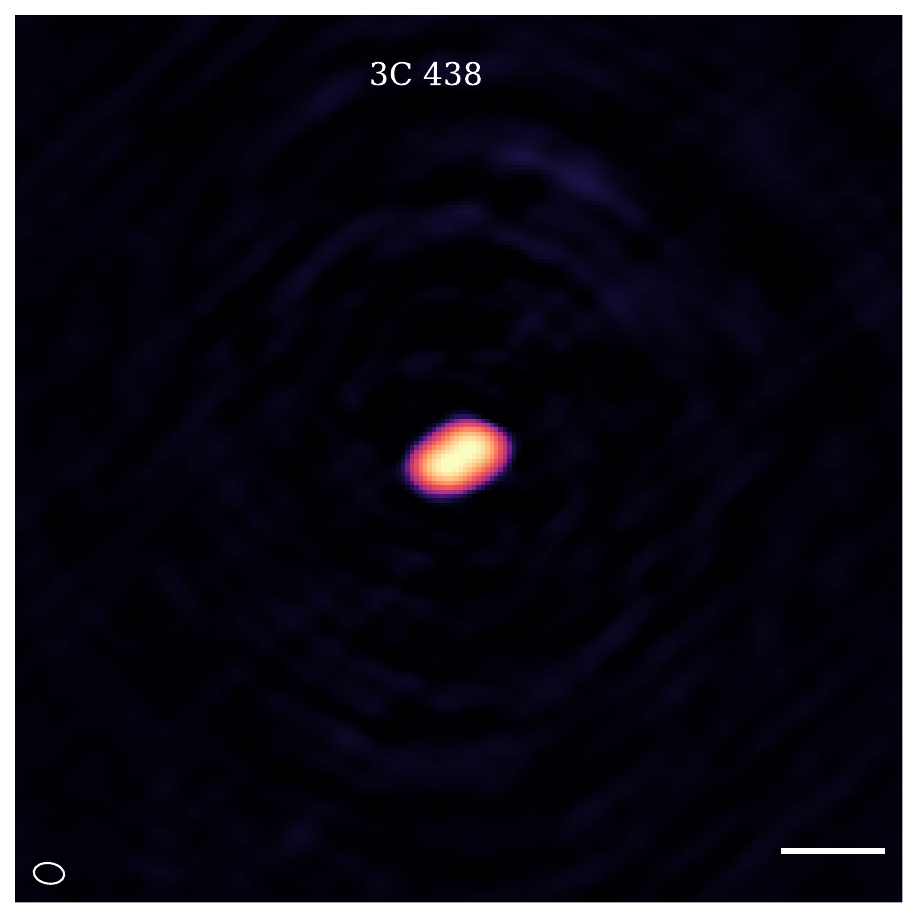}
\includegraphics[width=0.162\linewidth, trim={0.cm 0.cm 0.cm 0.cm},clip]{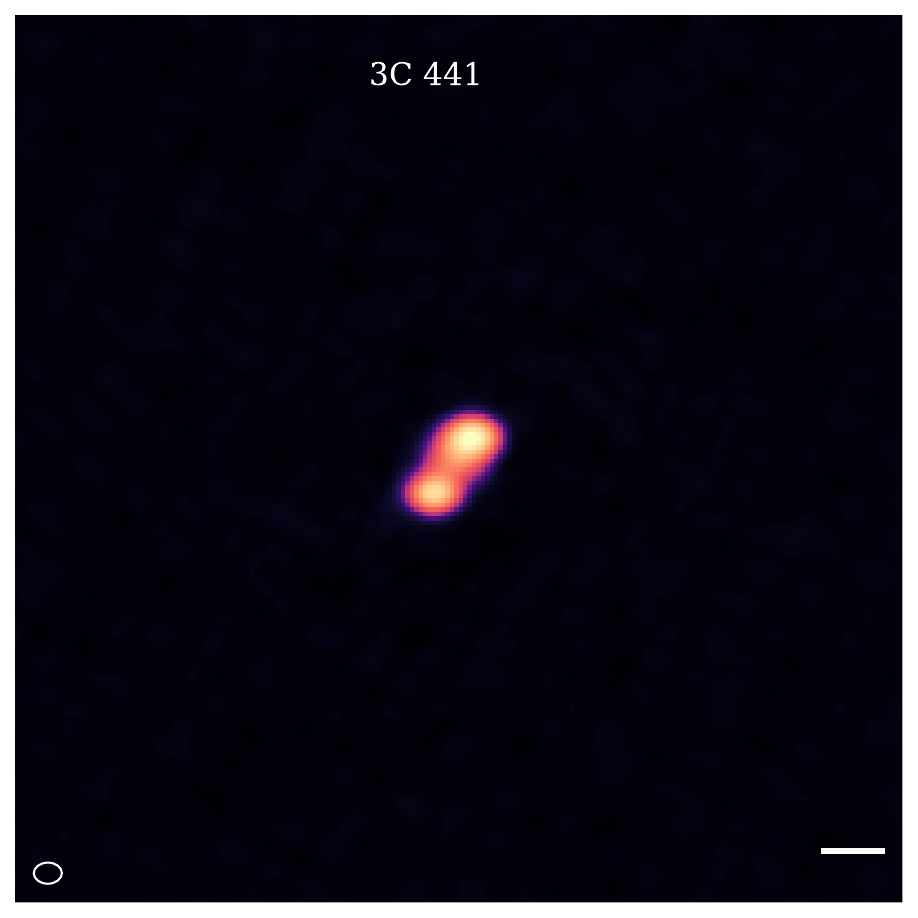}
\includegraphics[width=0.162\linewidth, trim={0.cm 0.cm 0.cm 0.cm},clip]{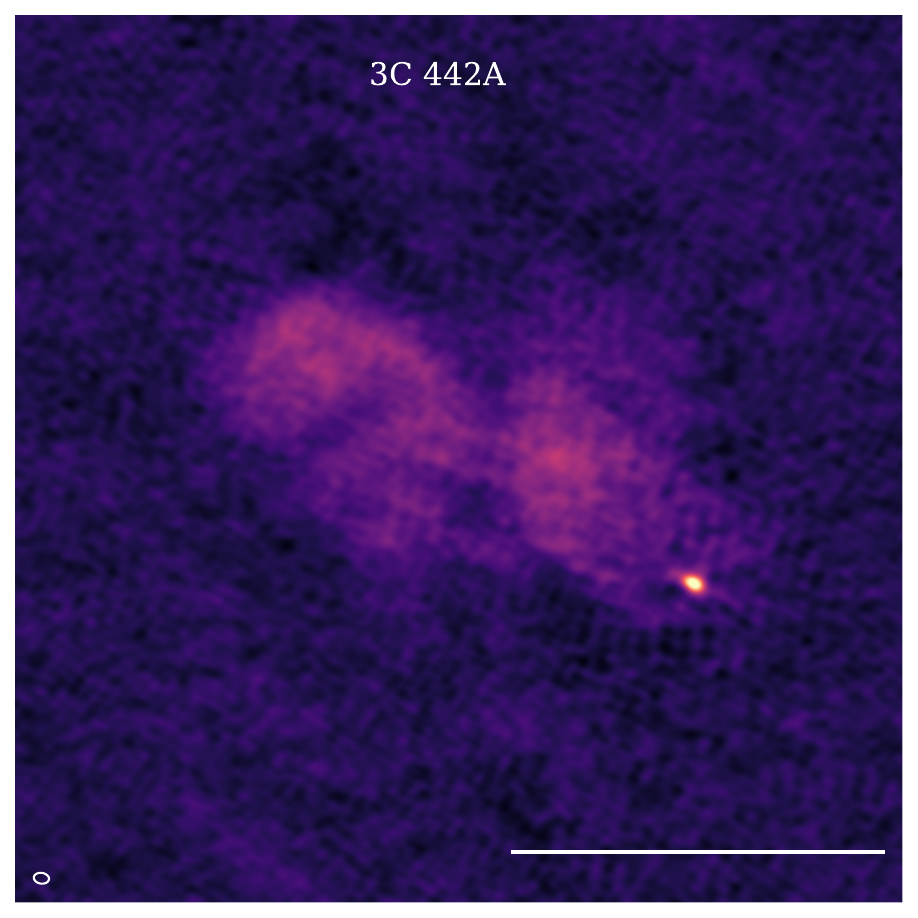}
\includegraphics[width=0.162\linewidth, trim={0.cm 0.cm 0.cm 0.cm},clip]{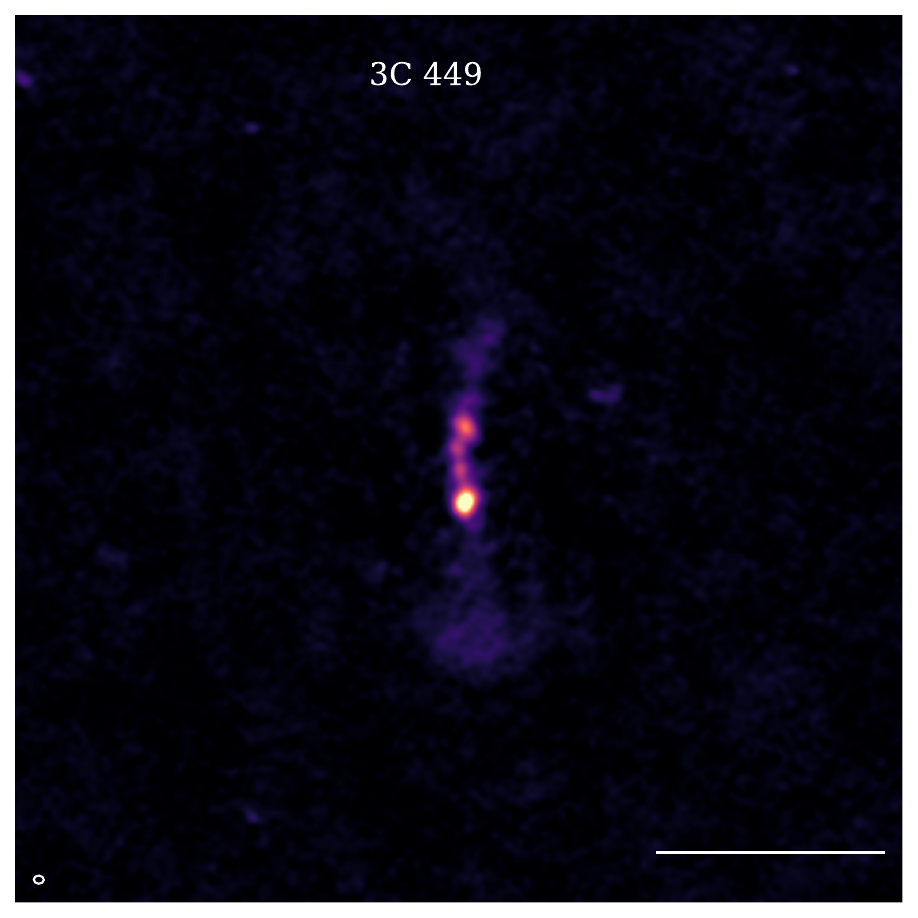}
\includegraphics[width=0.162\linewidth, trim={0.cm 0.cm 0.cm 0.cm},clip]{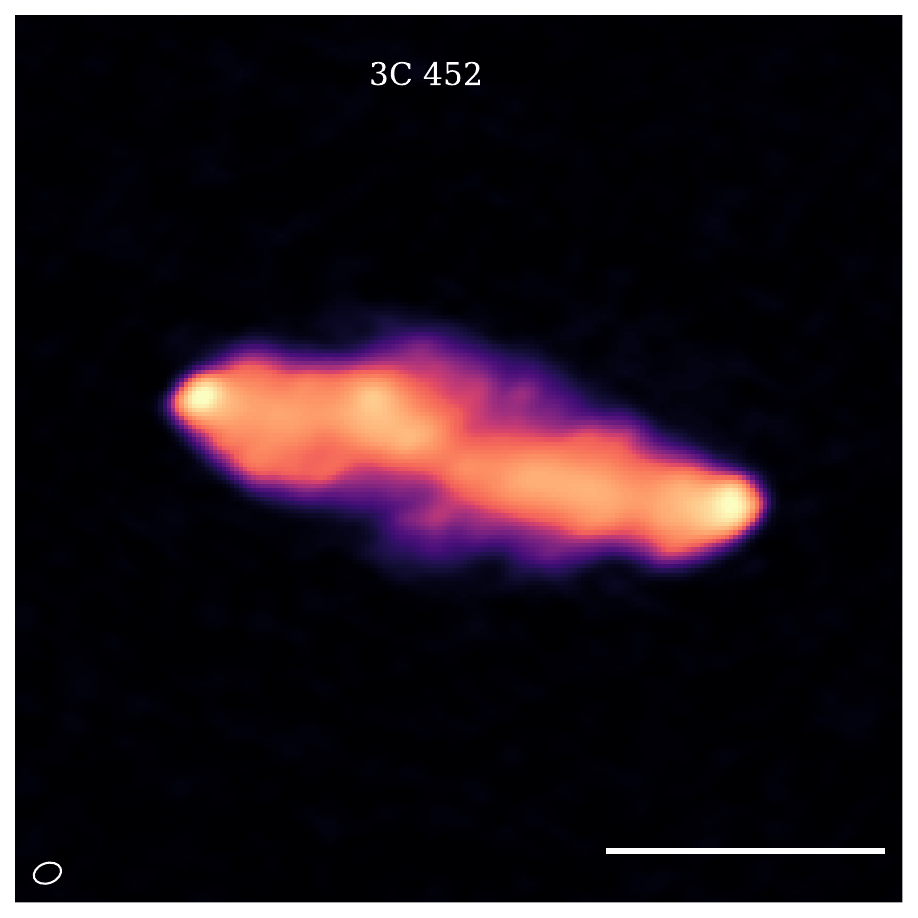}
\includegraphics[width=0.162\linewidth, trim={0.cm 0.cm 0.cm 0.cm},clip]{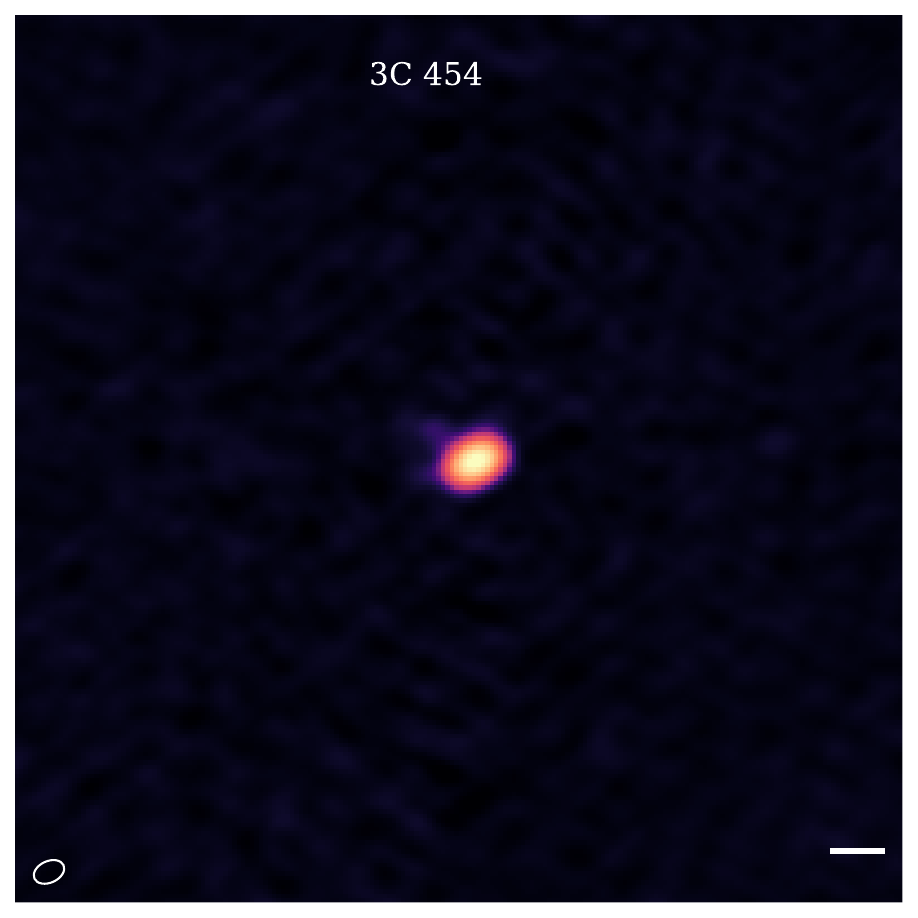}
\includegraphics[width=0.162\linewidth, trim={0.cm 0.cm 0.cm 0.cm},clip]{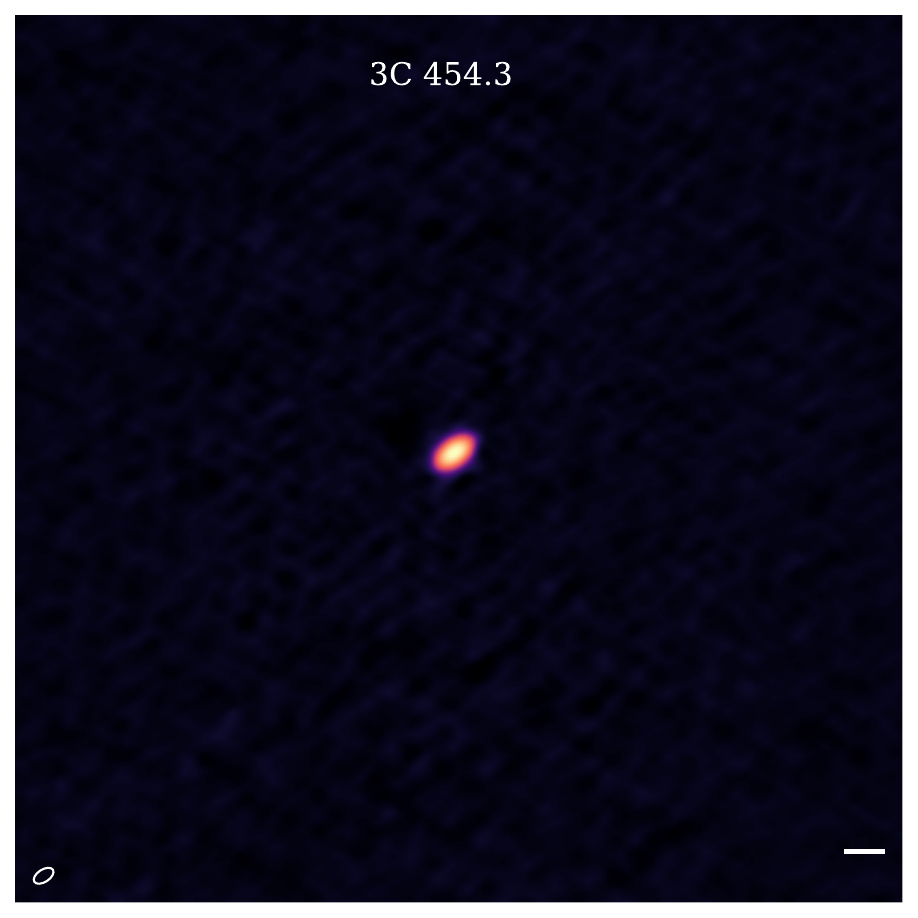}
\includegraphics[width=0.162\linewidth, trim={0.cm 0.cm 0.cm 0.cm},clip]{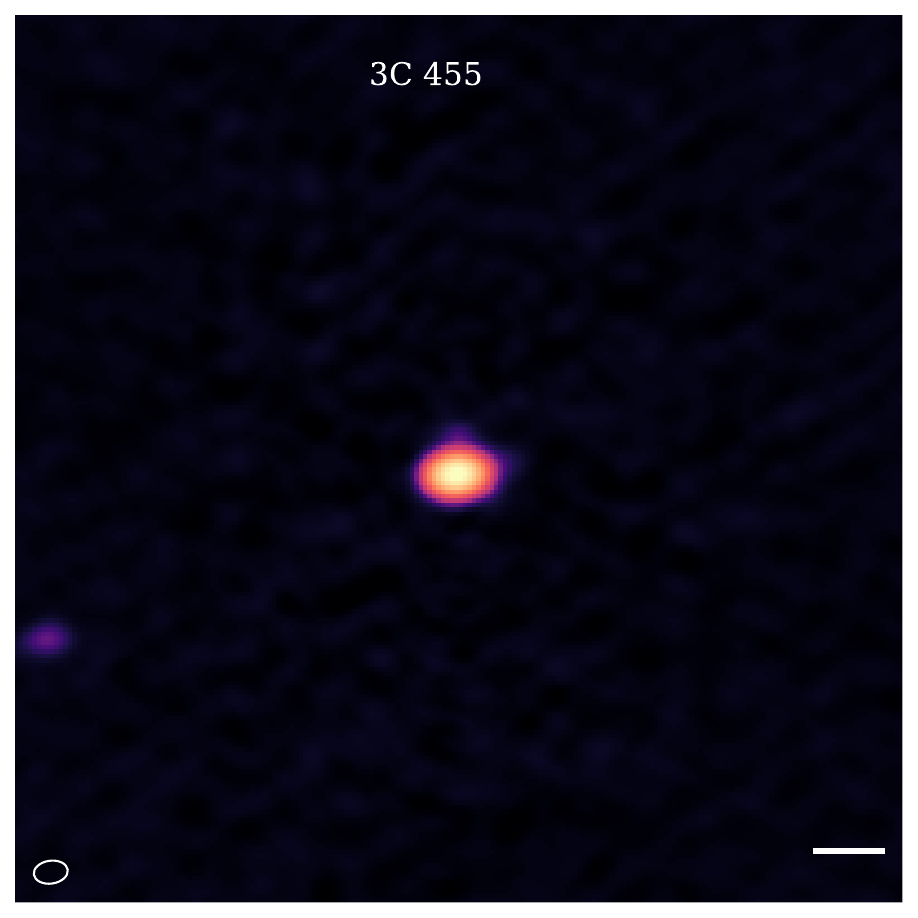}
\includegraphics[width=0.162\linewidth, trim={0.cm 0.cm 0.cm 0.cm},clip]{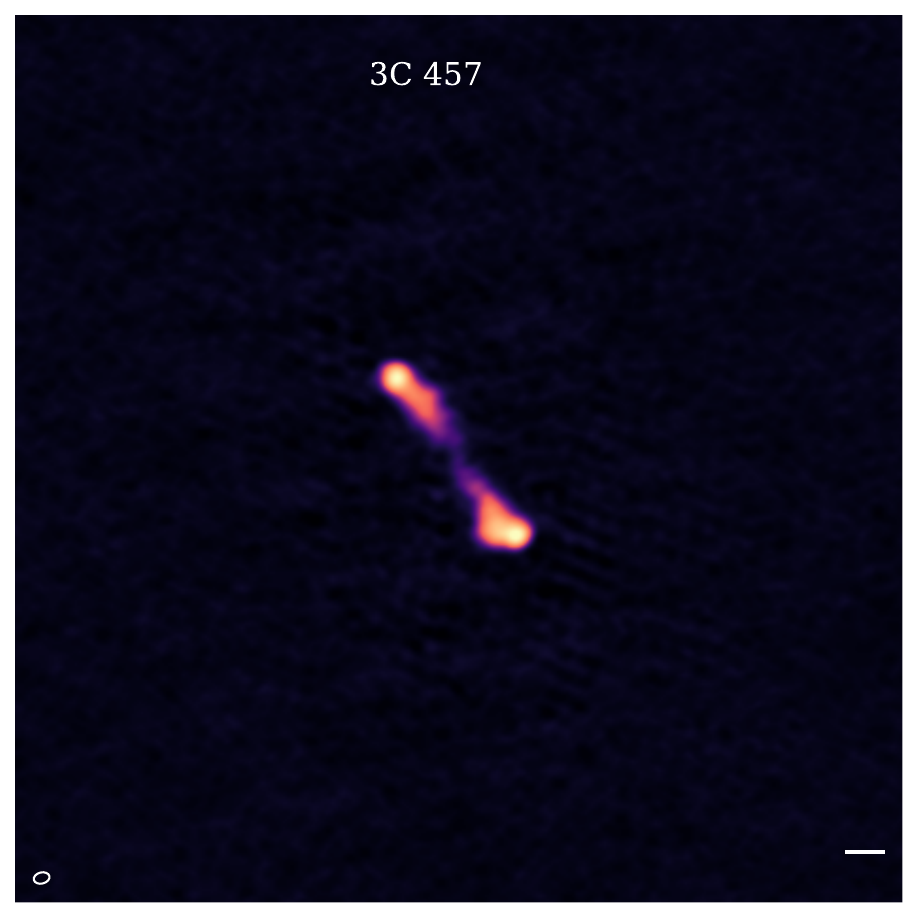}
\includegraphics[width=0.162\linewidth, trim={0.cm 0.cm 0.cm 0.cm},clip]{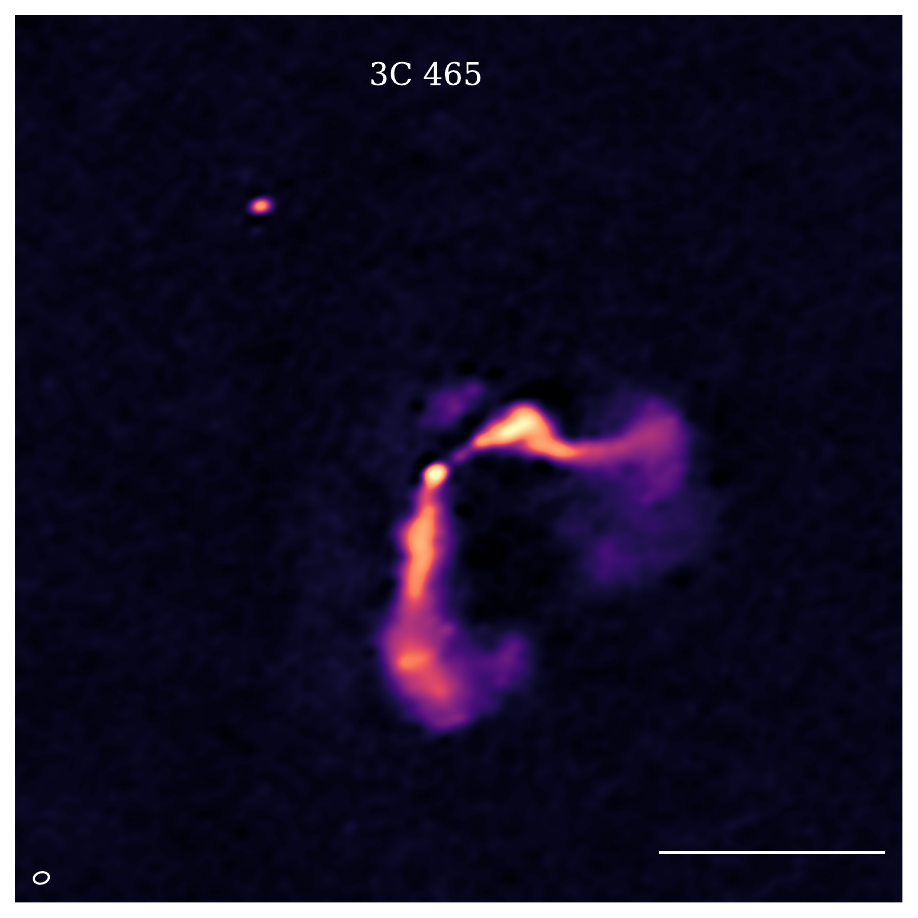}
\includegraphics[width=0.162\linewidth, trim={0.cm 0.cm 0.cm 0.cm},clip]{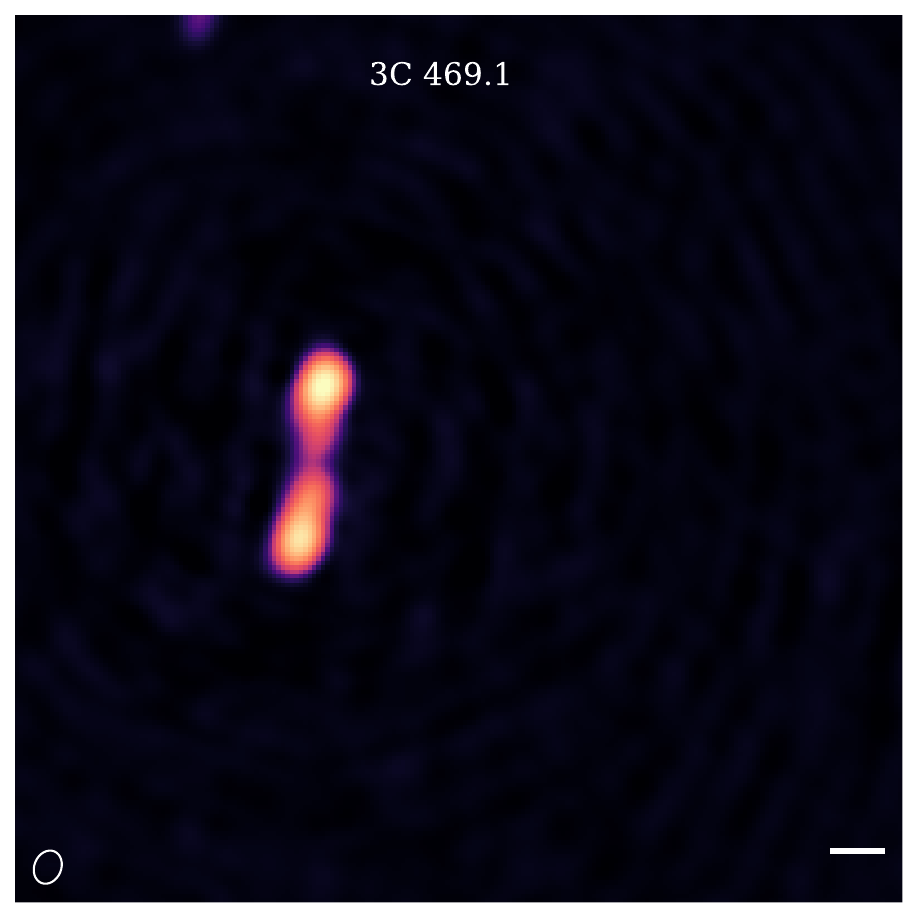}
\includegraphics[width=0.162\linewidth, trim={0.cm 0.cm 0.cm 0.cm},clip]{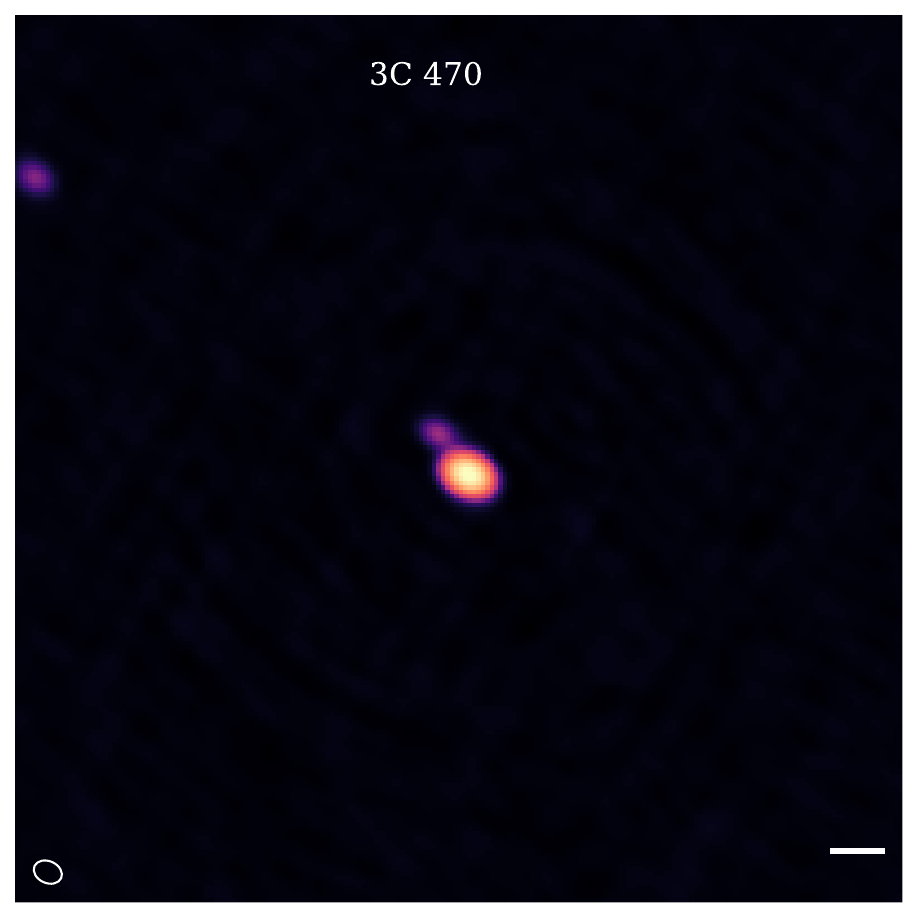}
\includegraphics[width=0.162\linewidth, trim={0.cm 0.cm 0.cm 0.cm},clip]{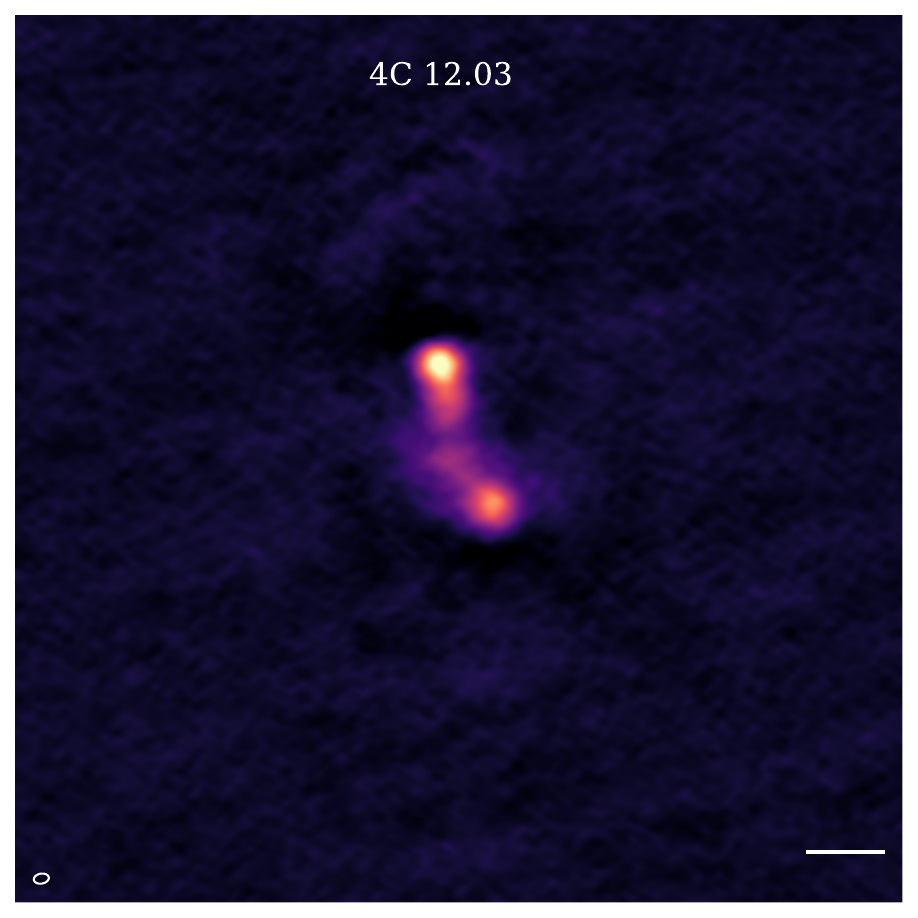}
\includegraphics[width=0.162\linewidth, trim={0.cm 0.cm 0.cm 0.cm},clip]{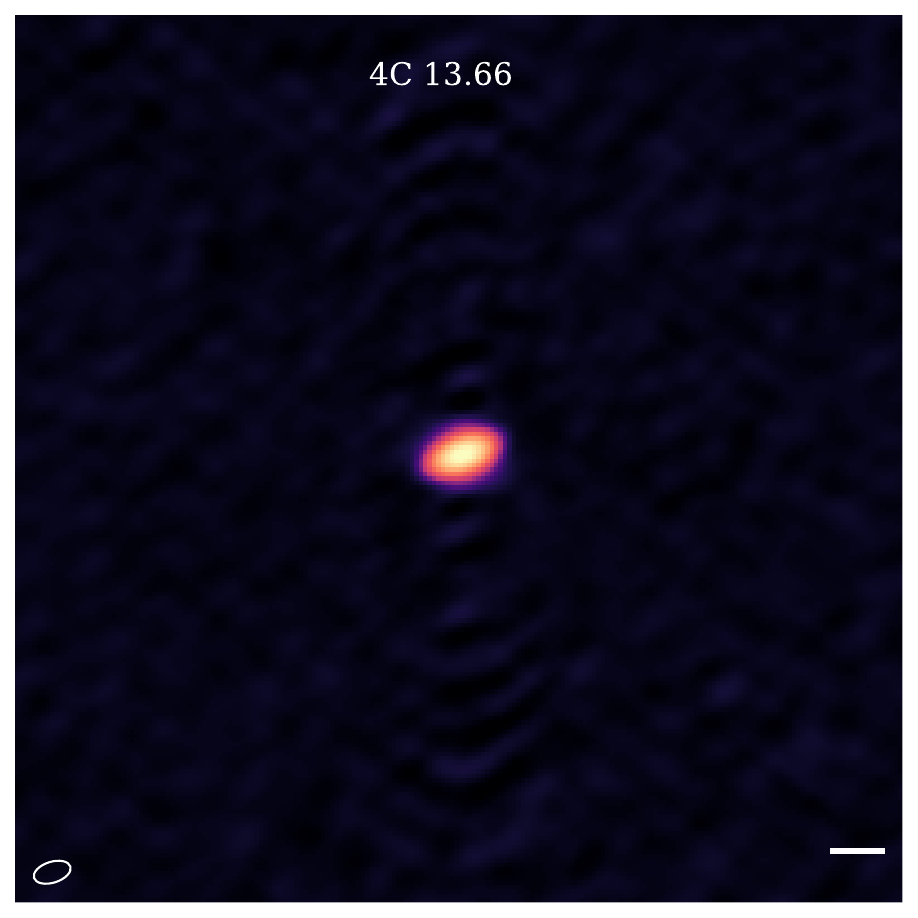}
\includegraphics[width=0.162\linewidth, trim={0.cm 0.cm 0.cm 0.cm},clip]{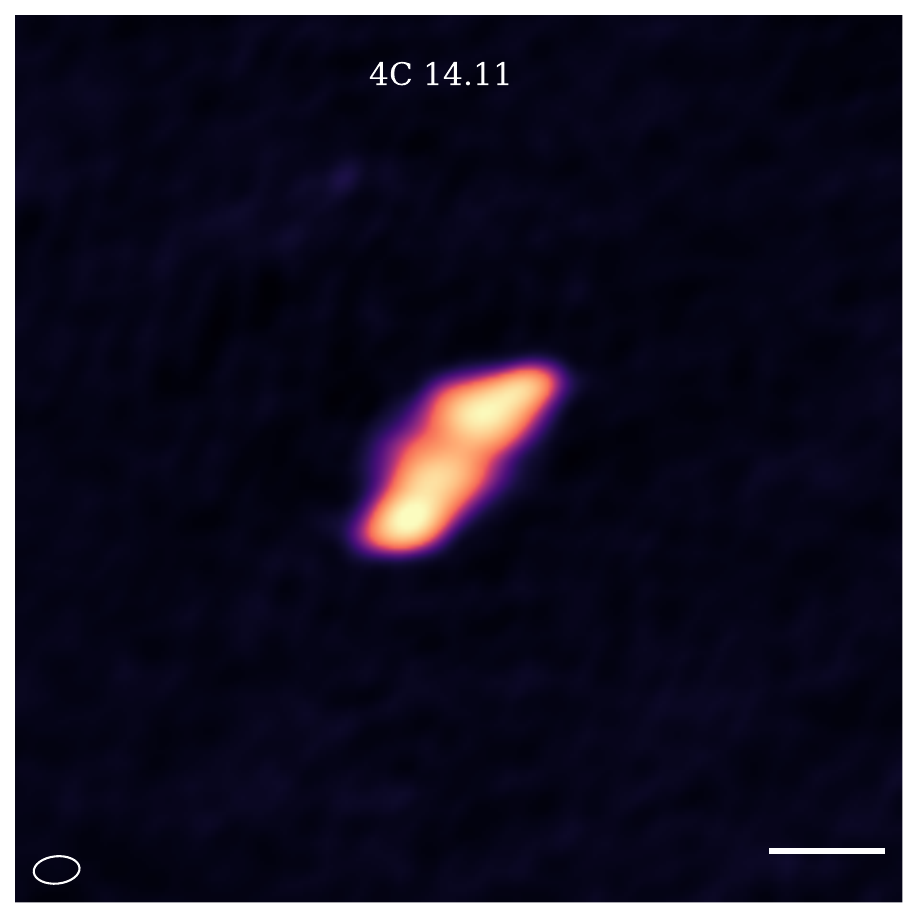}
\includegraphics[width=0.162\linewidth, trim={0.cm 0.cm 0.cm 0.cm},clip]{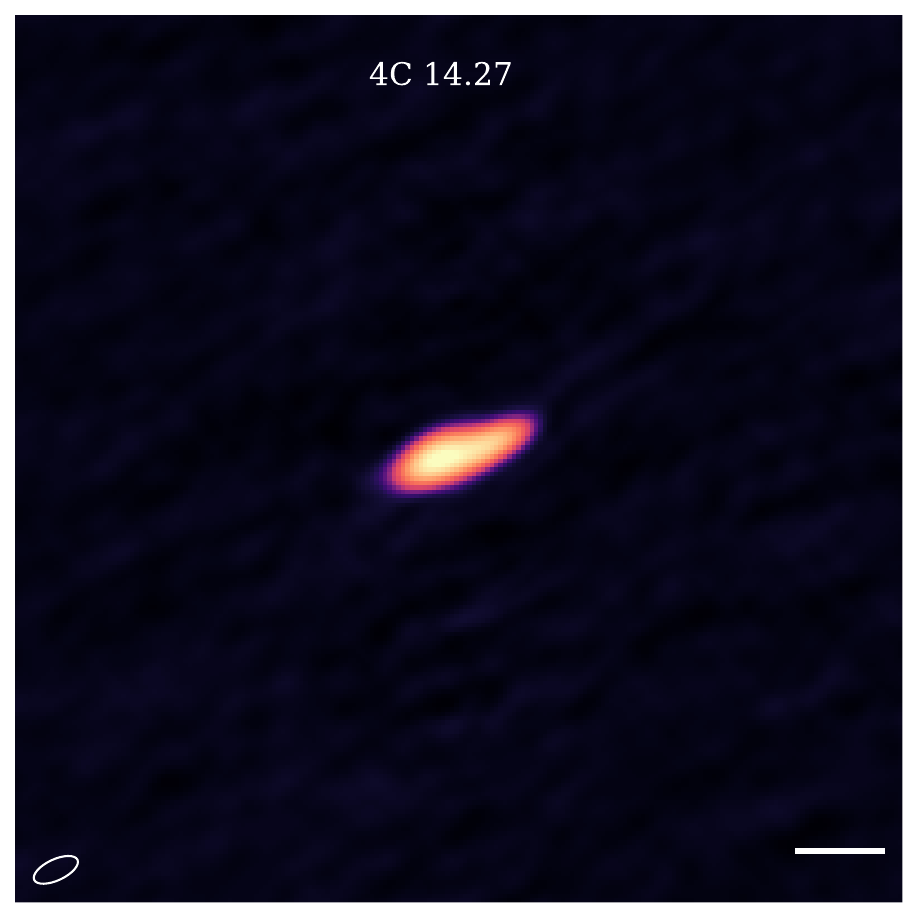}
\includegraphics[width=0.162\linewidth, trim={0.cm 0.cm 0.cm 0.cm},clip]{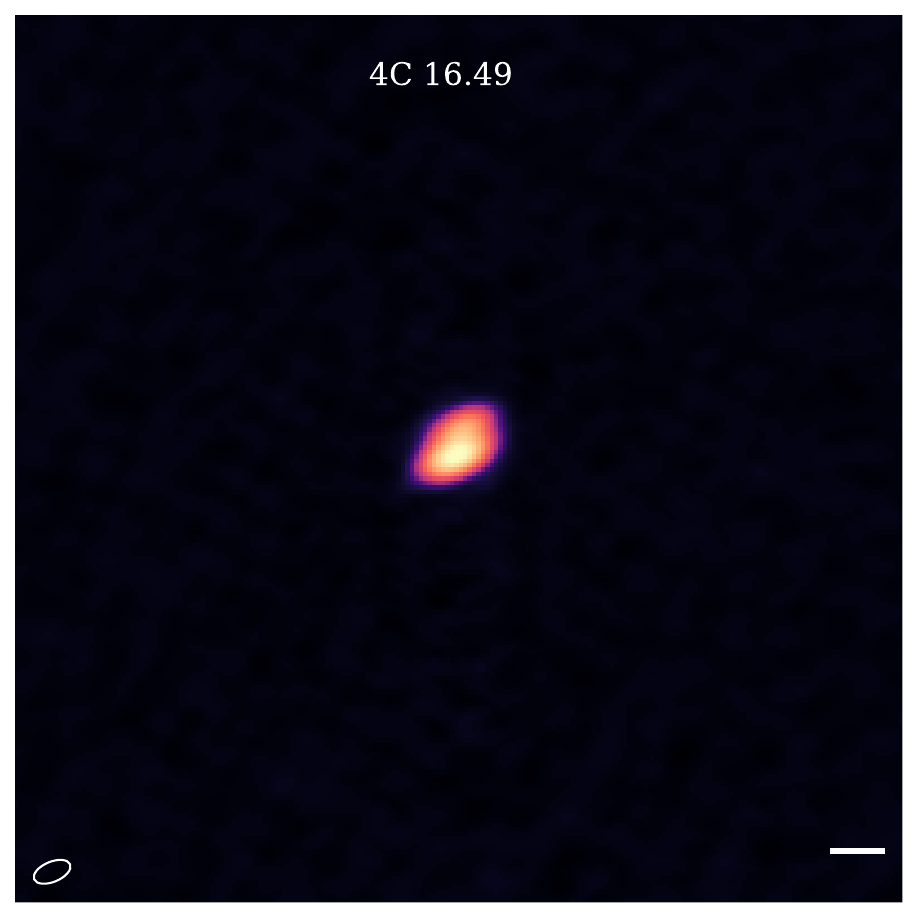}
\includegraphics[width=0.162\linewidth, trim={0.cm 0.cm 0.cm 0.cm},clip]{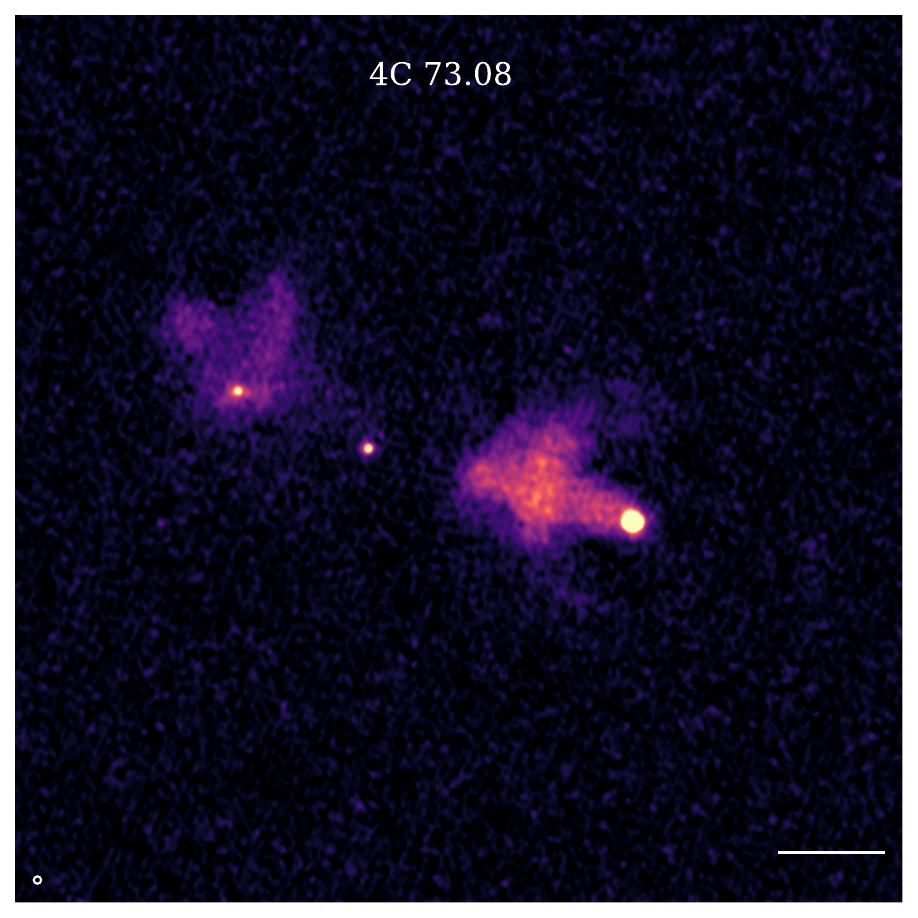}
\includegraphics[width=0.162\linewidth, trim={0.cm 0.cm 0.cm 0.cm},clip]{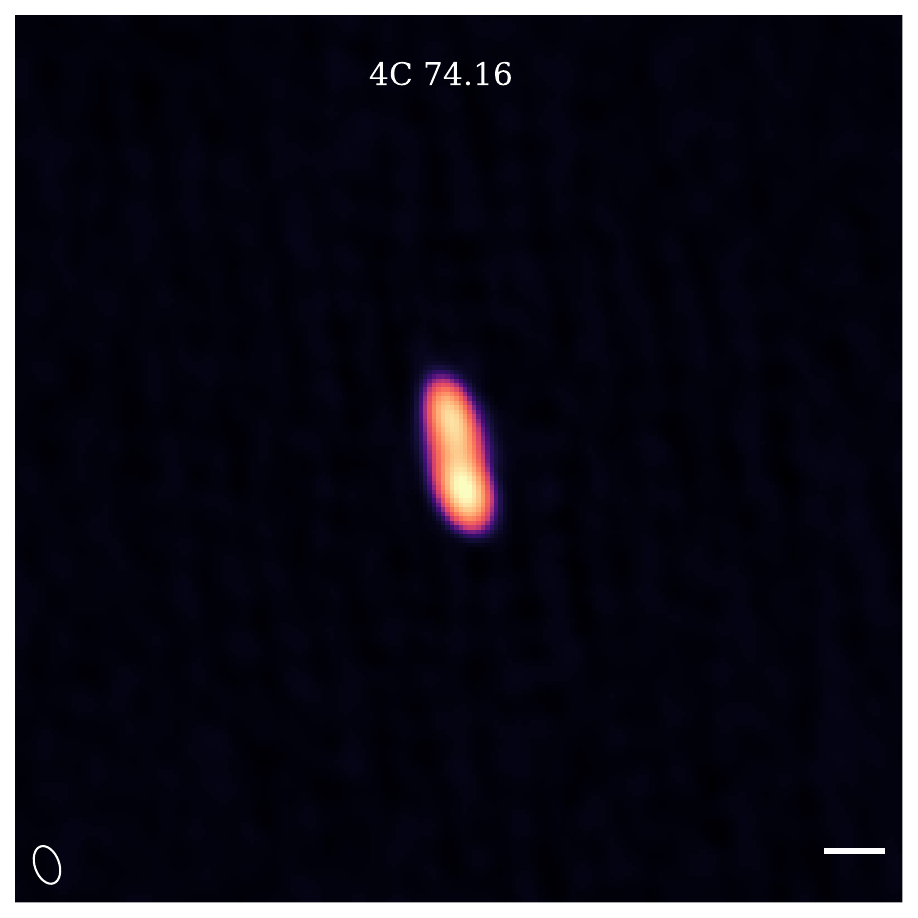}
\includegraphics[width=0.162\linewidth, trim={0.cm 0.cm 0.cm 0.cm},clip]{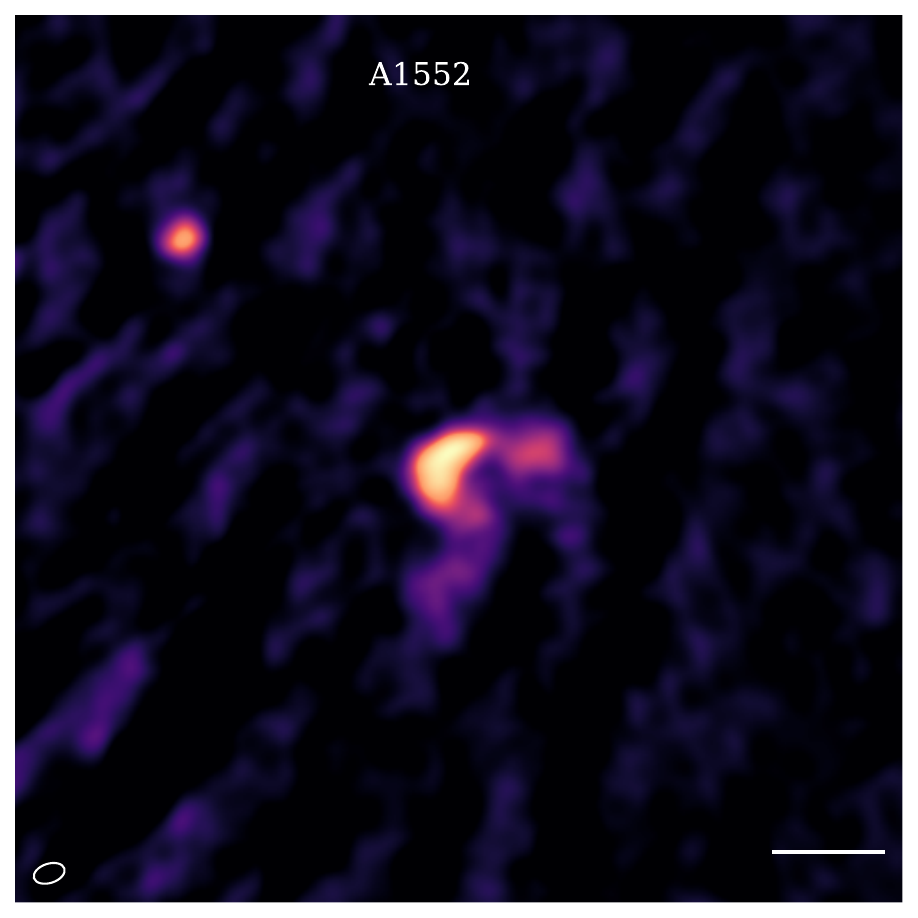}
\includegraphics[width=0.162\linewidth, trim={0.cm 0.cm 0.cm 0.cm},clip]{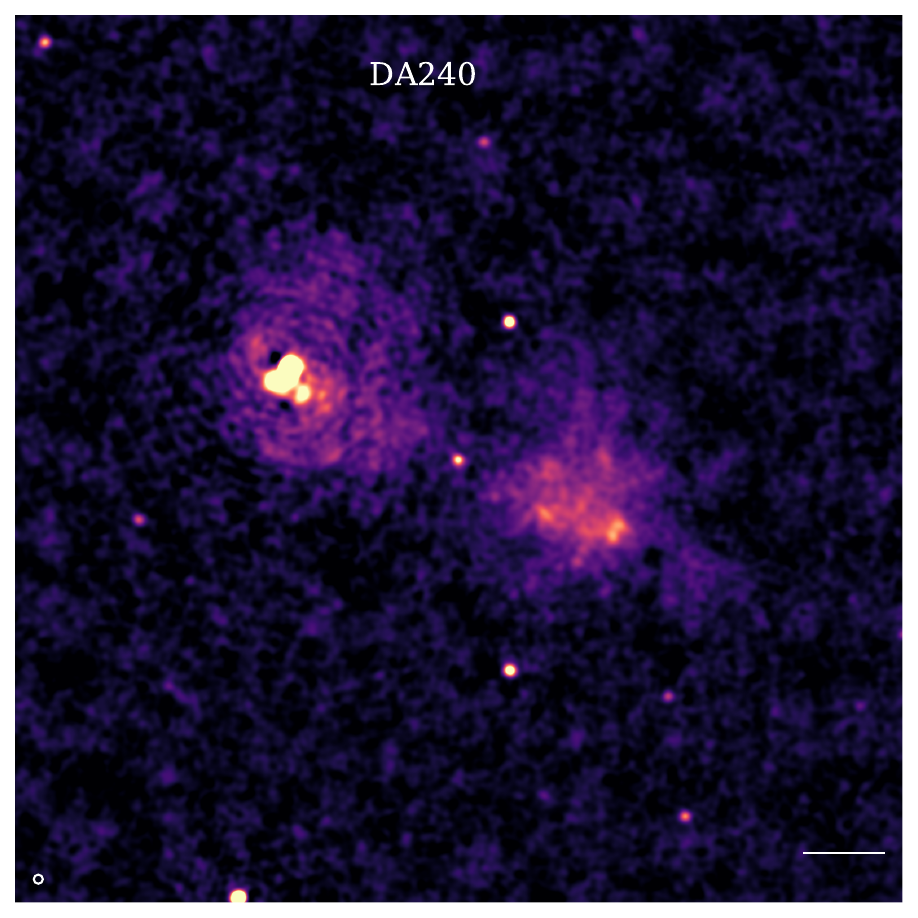}
\includegraphics[width=0.162\linewidth, trim={0.cm 0.cm 0.cm 0.cm},clip]{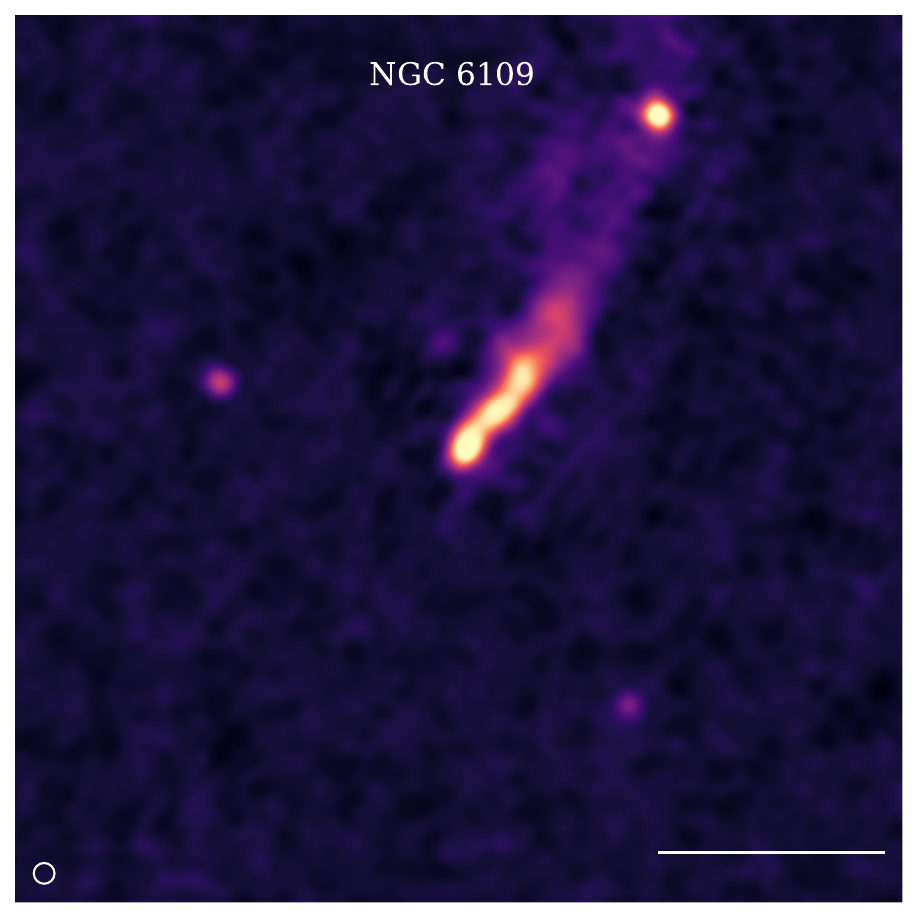}
\includegraphics[width=0.162\linewidth, trim={0.cm 0.cm 0.cm 0.cm},clip]{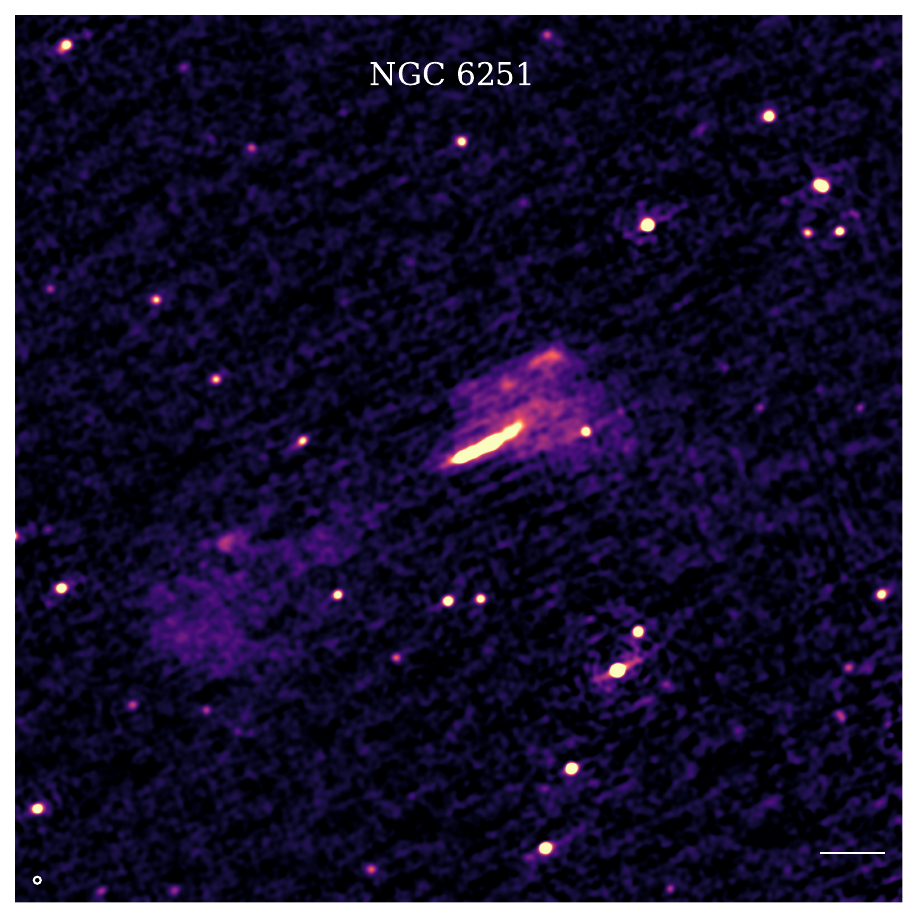}
\includegraphics[width=0.162\linewidth, trim={0.cm 0.cm 0.cm 0.cm},clip]{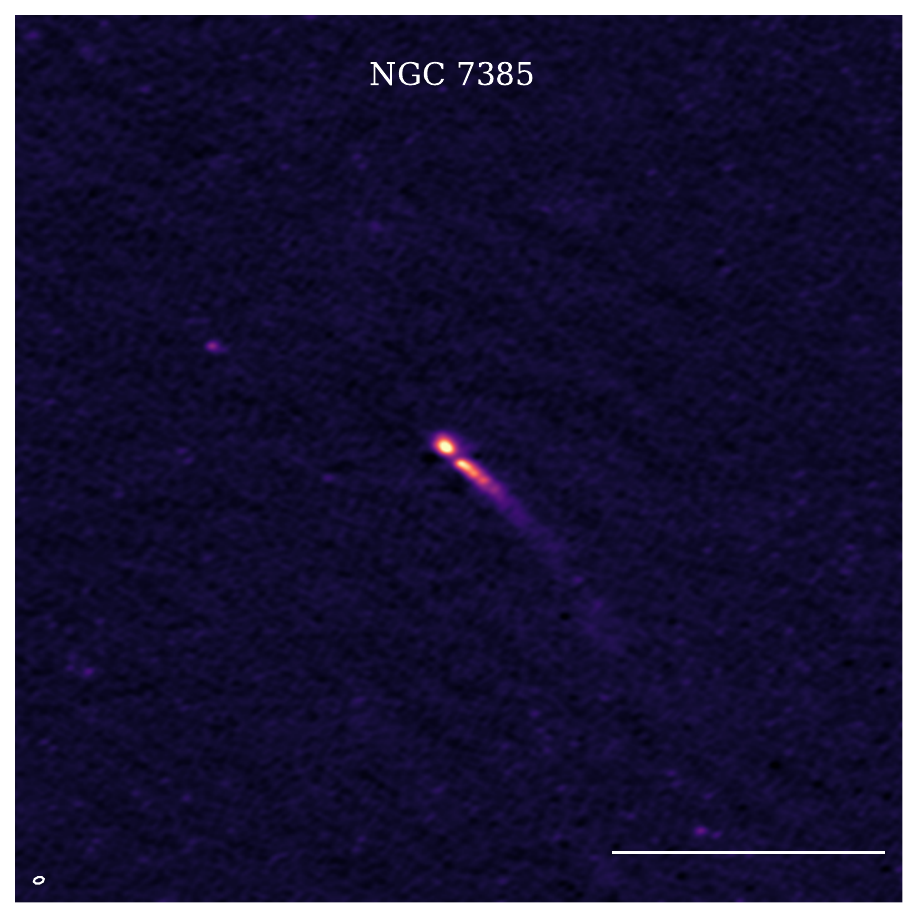}
\caption{continued.}
\label{fig:all-maps}
\end{figure}

\newpage
\section{SEDs of objects in the catalogue}
\label{app:seds}

\begin{figure}[H]
\includegraphics[width=0.162\linewidth, trim={0.cm .0cm 1.5cm 1.5cm},clip]{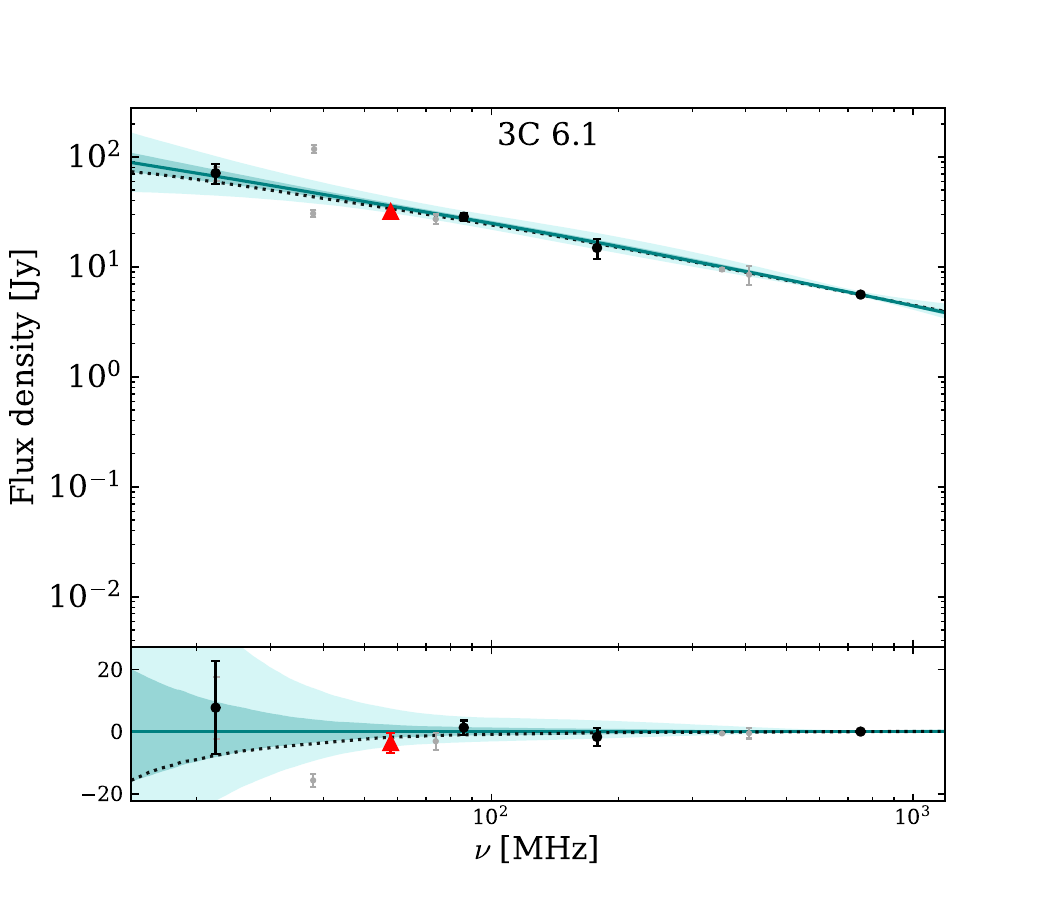}
\includegraphics[width=0.162\linewidth, trim={0.cm .0cm 1.5cm 1.5cm},clip]{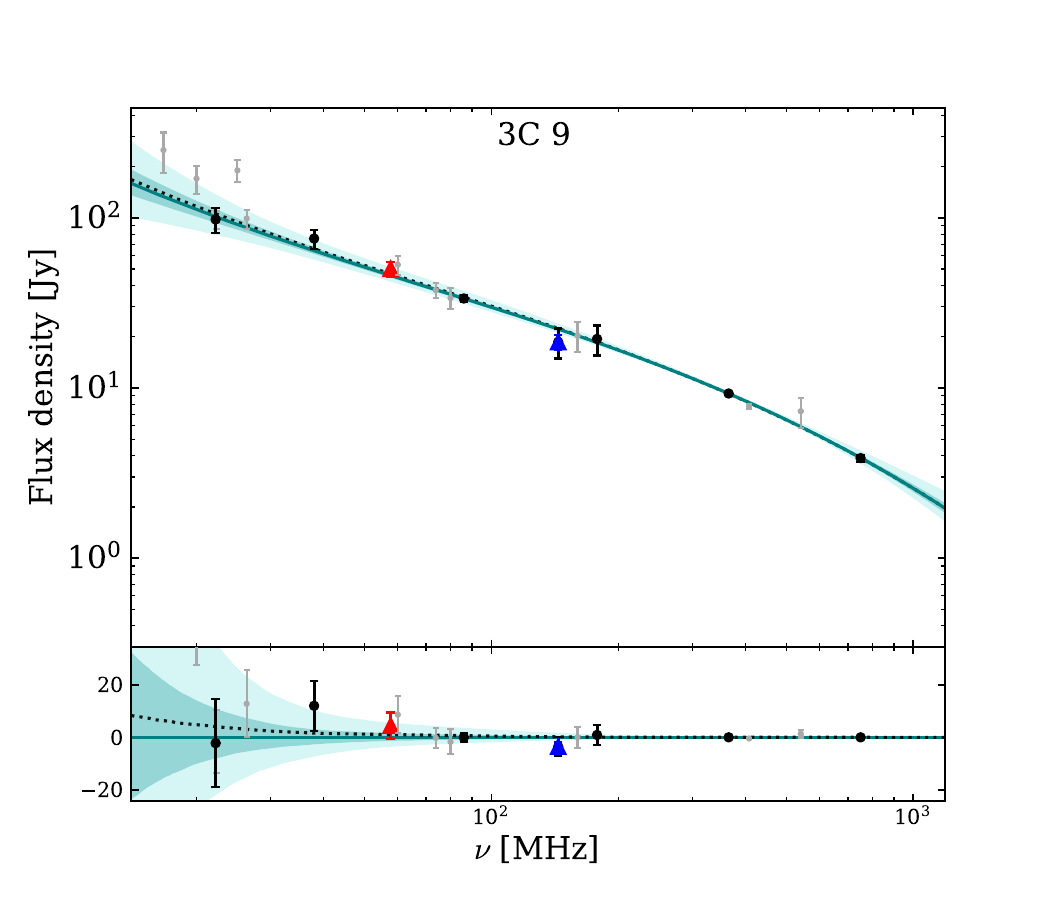}
\includegraphics[width=0.162\linewidth, trim={0.cm .0cm 1.5cm 1.5cm},clip]{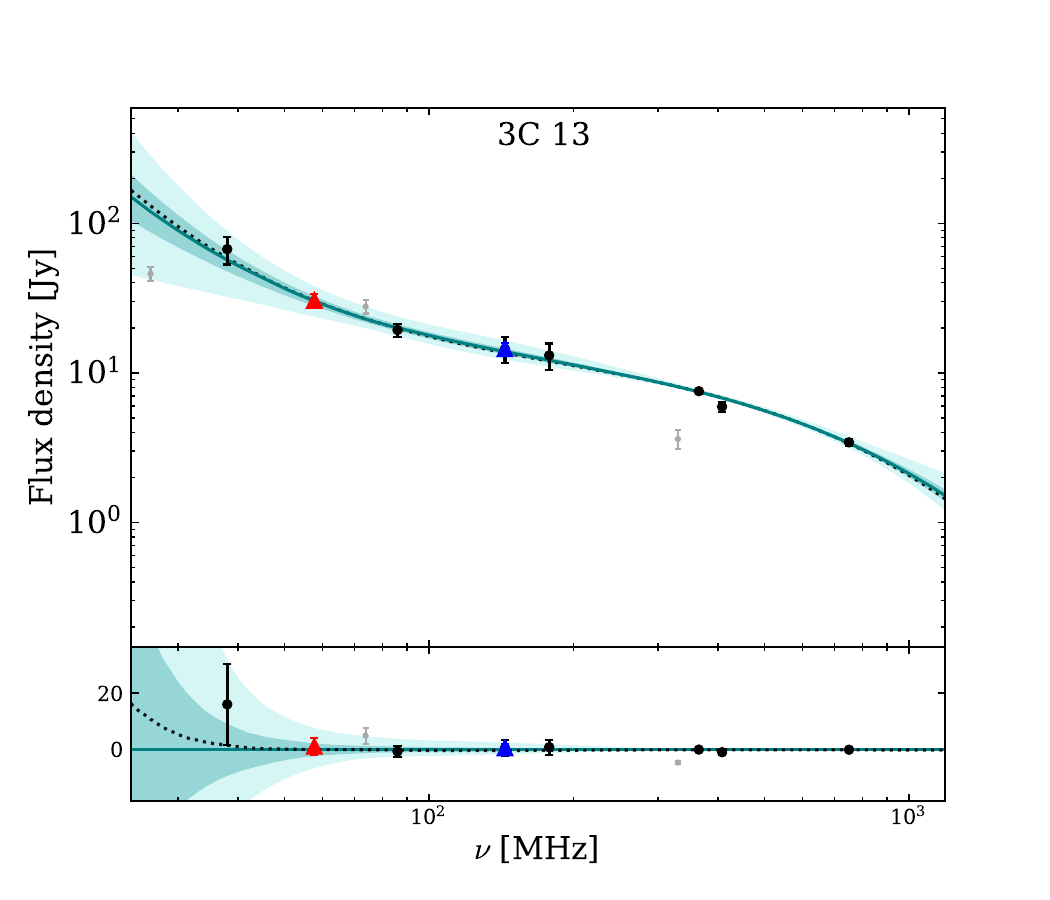}
\includegraphics[width=0.162\linewidth, trim={0.cm .0cm 1.5cm 1.5cm},clip]{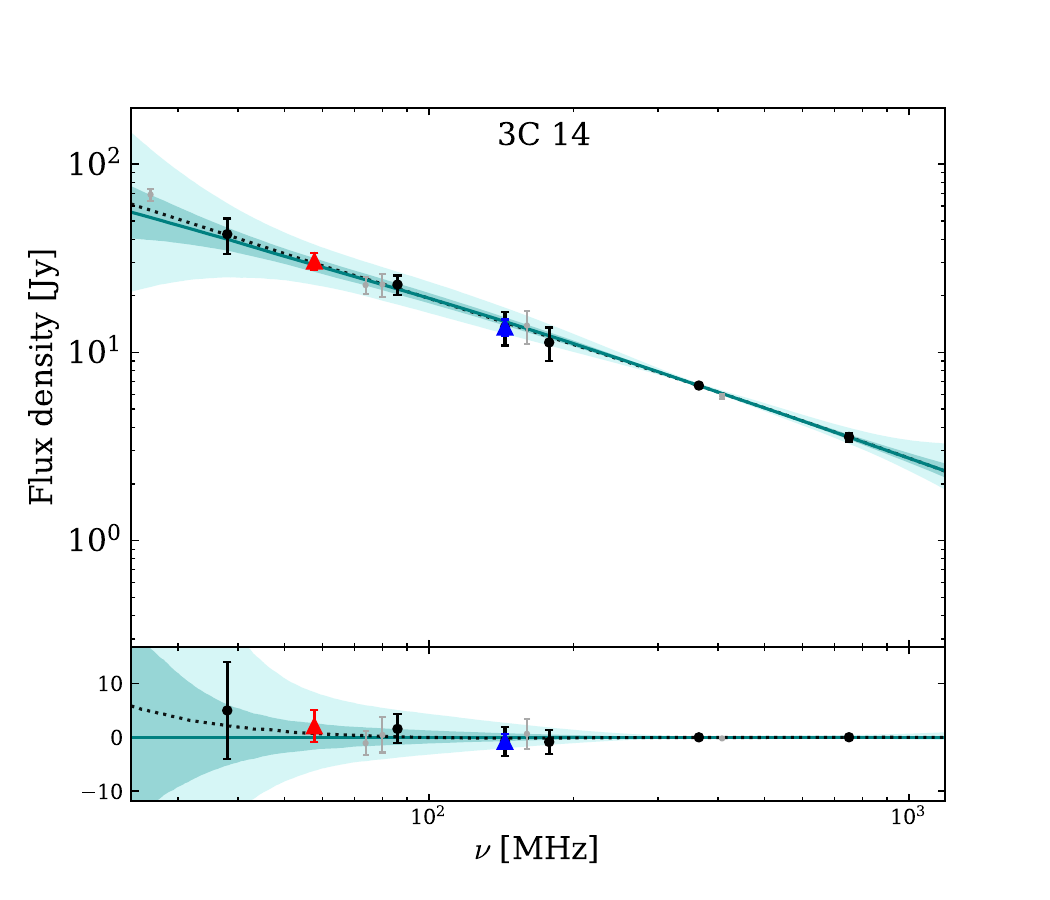}
\includegraphics[width=0.162\linewidth, trim={0.cm .0cm 1.5cm 1.5cm},clip]{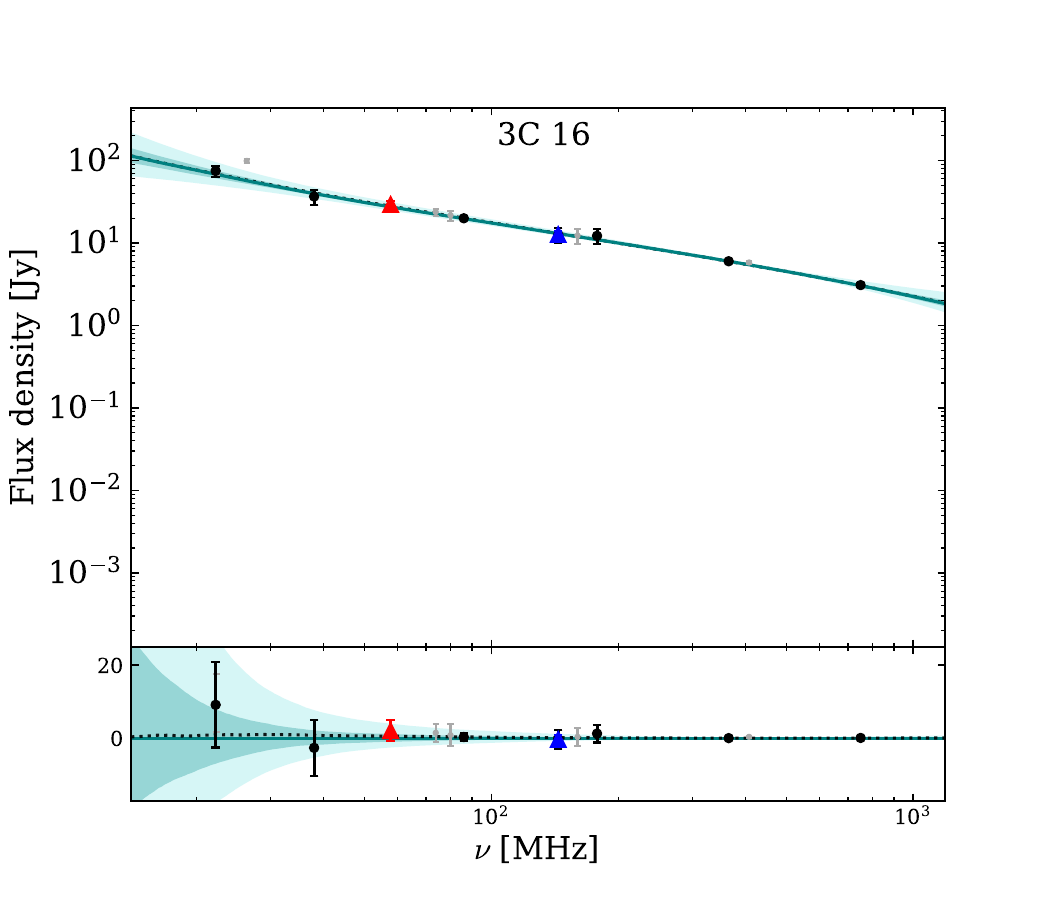}
\includegraphics[width=0.162\linewidth, trim={0.cm .0cm 1.5cm 1.5cm},clip]{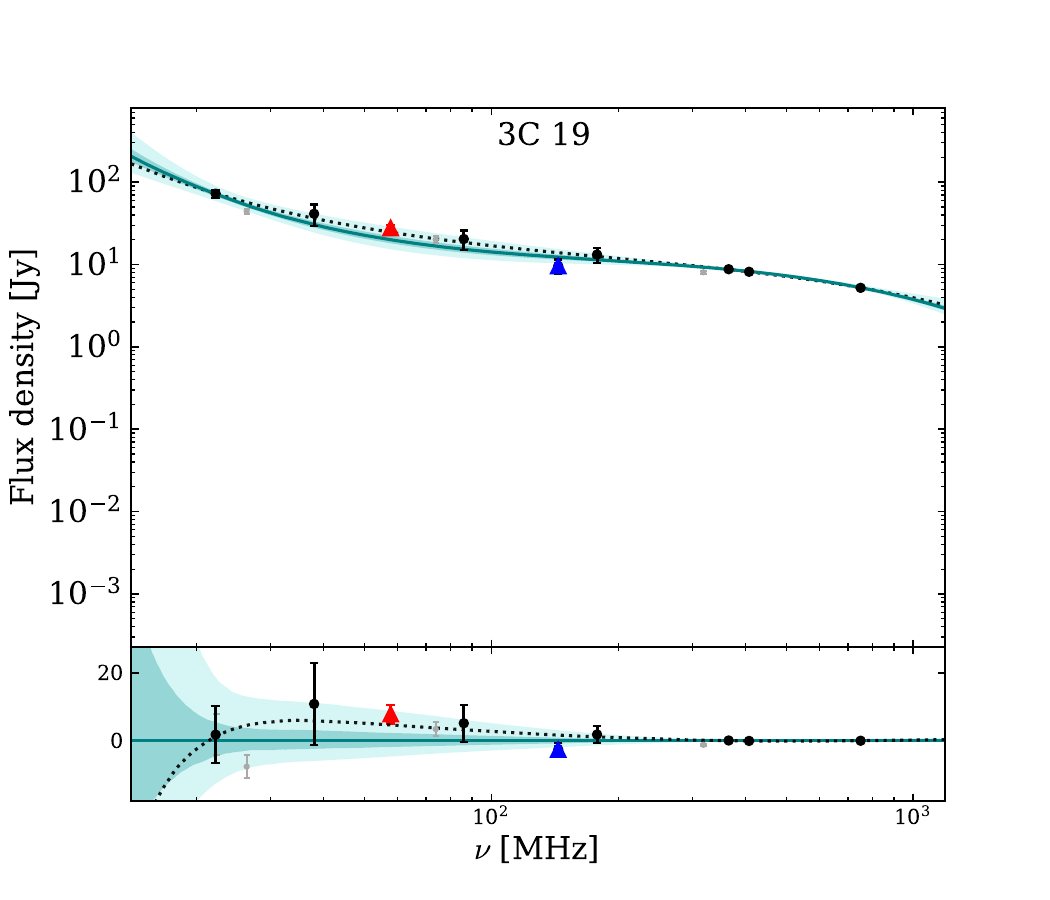}
\includegraphics[width=0.162\linewidth, trim={0.cm .0cm 1.5cm 1.5cm},clip]{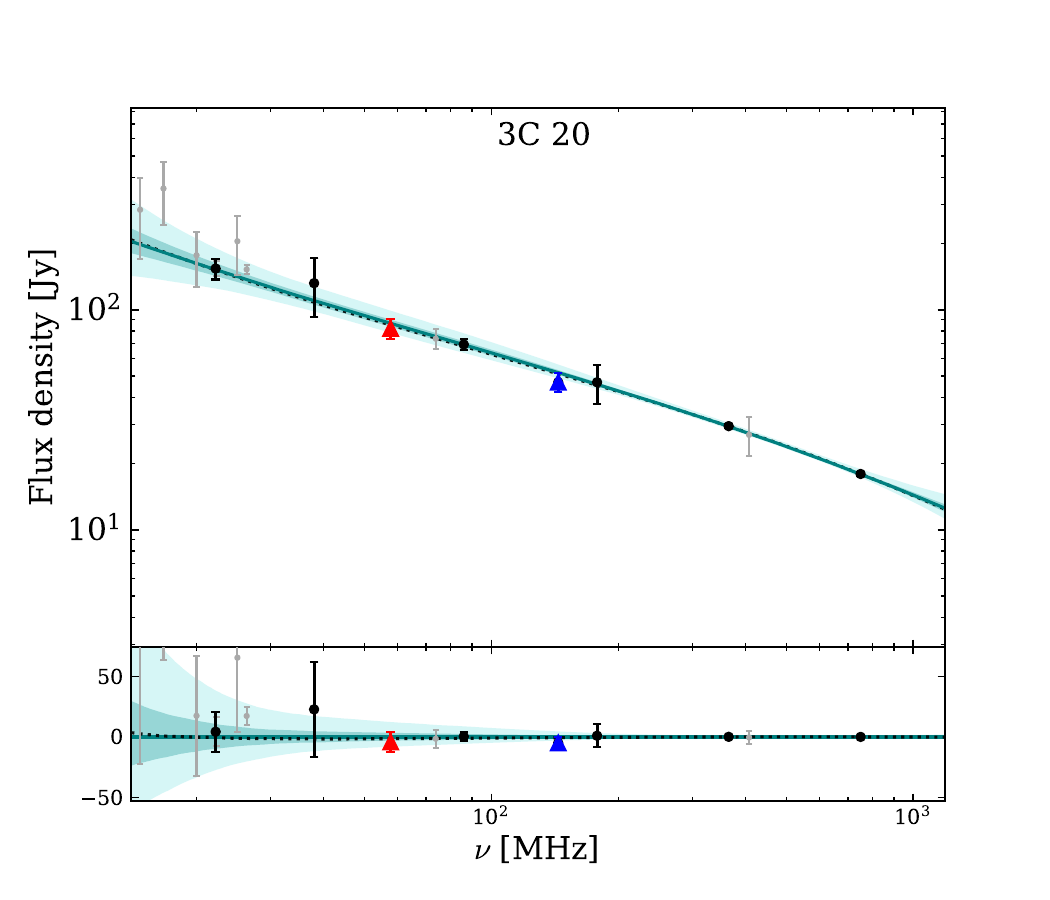}
\includegraphics[width=0.162\linewidth, trim={0.cm .0cm 1.5cm 1.5cm},clip]{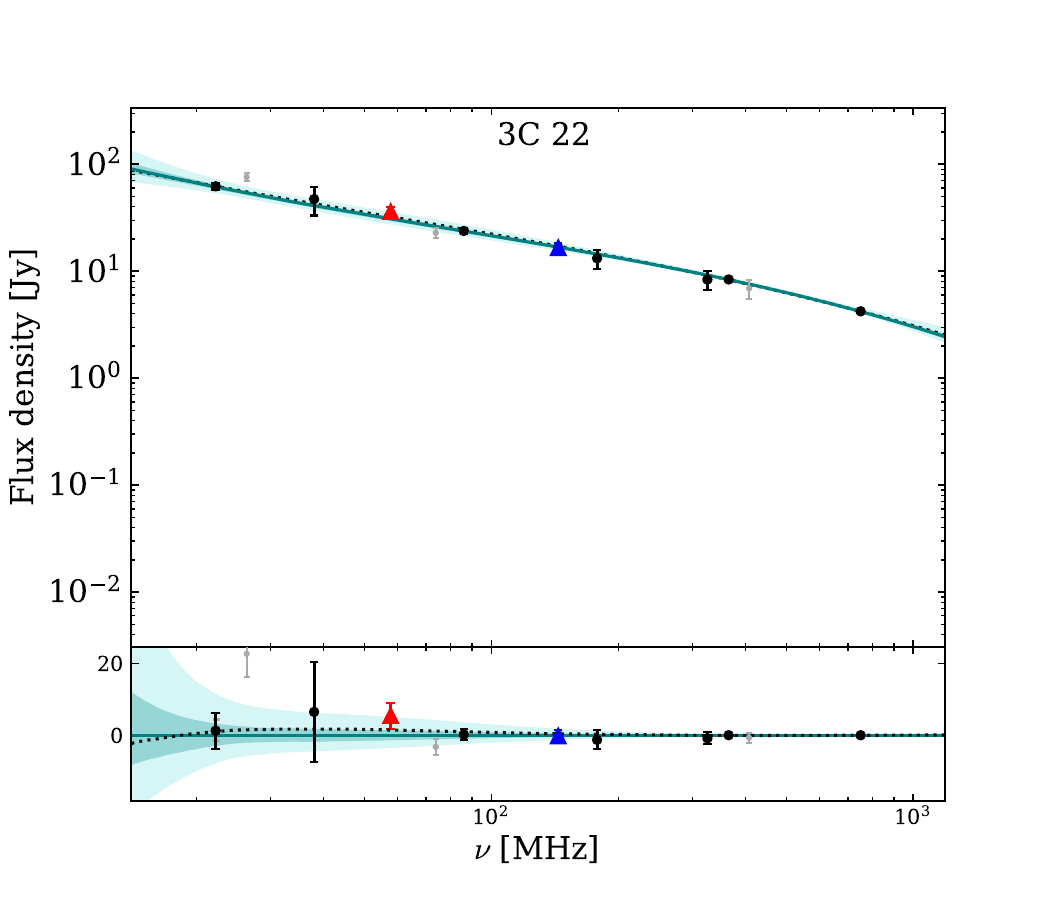}
\includegraphics[width=0.162\linewidth, trim={0.cm .0cm 1.5cm 1.5cm},clip]{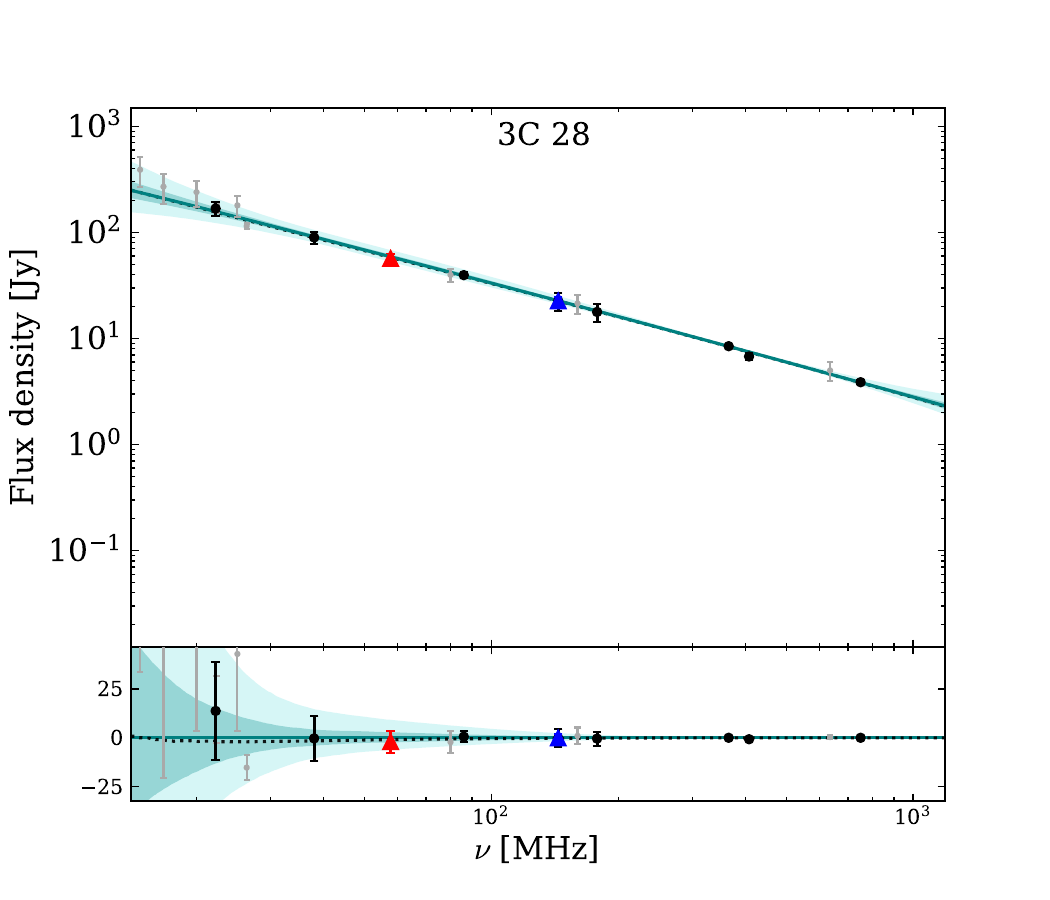}
\includegraphics[width=0.162\linewidth, trim={0.cm .0cm 1.5cm 1.5cm},clip]{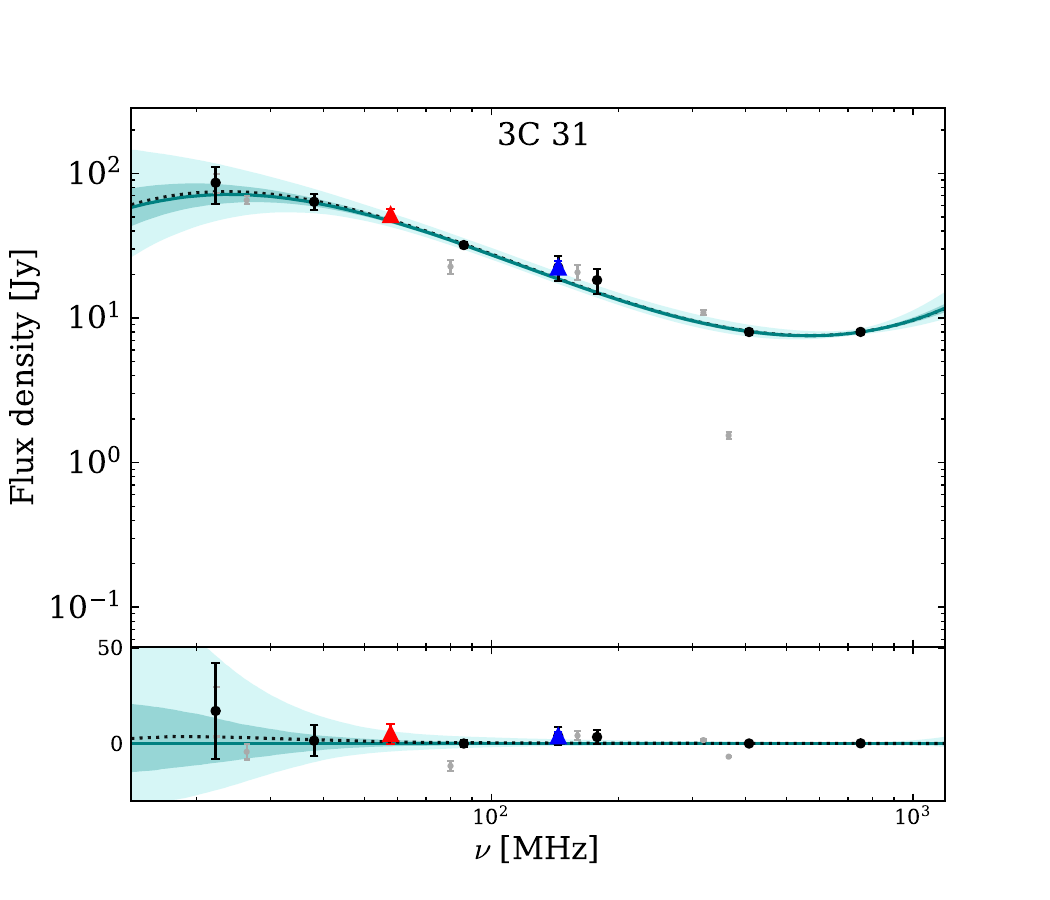}
\includegraphics[width=0.162\linewidth, trim={0.cm .0cm 1.5cm 1.5cm},clip]{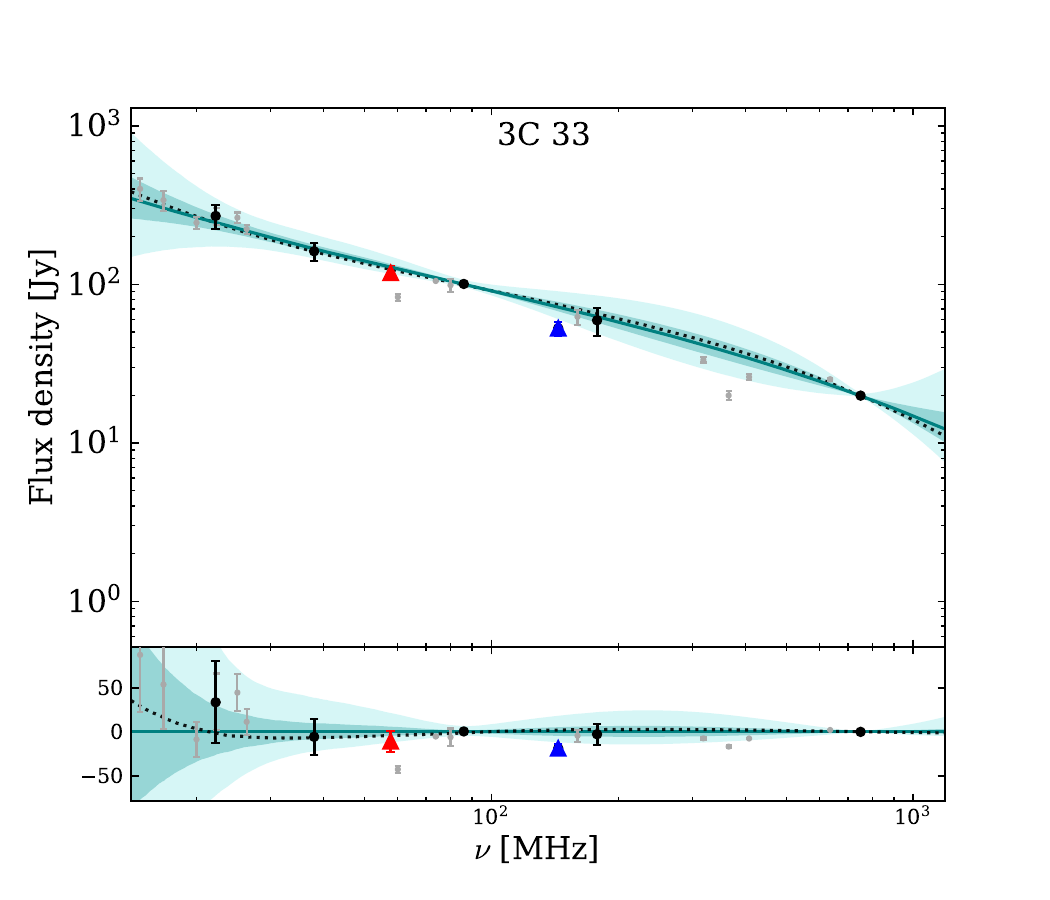}
\includegraphics[width=0.162\linewidth, trim={0.cm .0cm 1.5cm 1.5cm},clip]{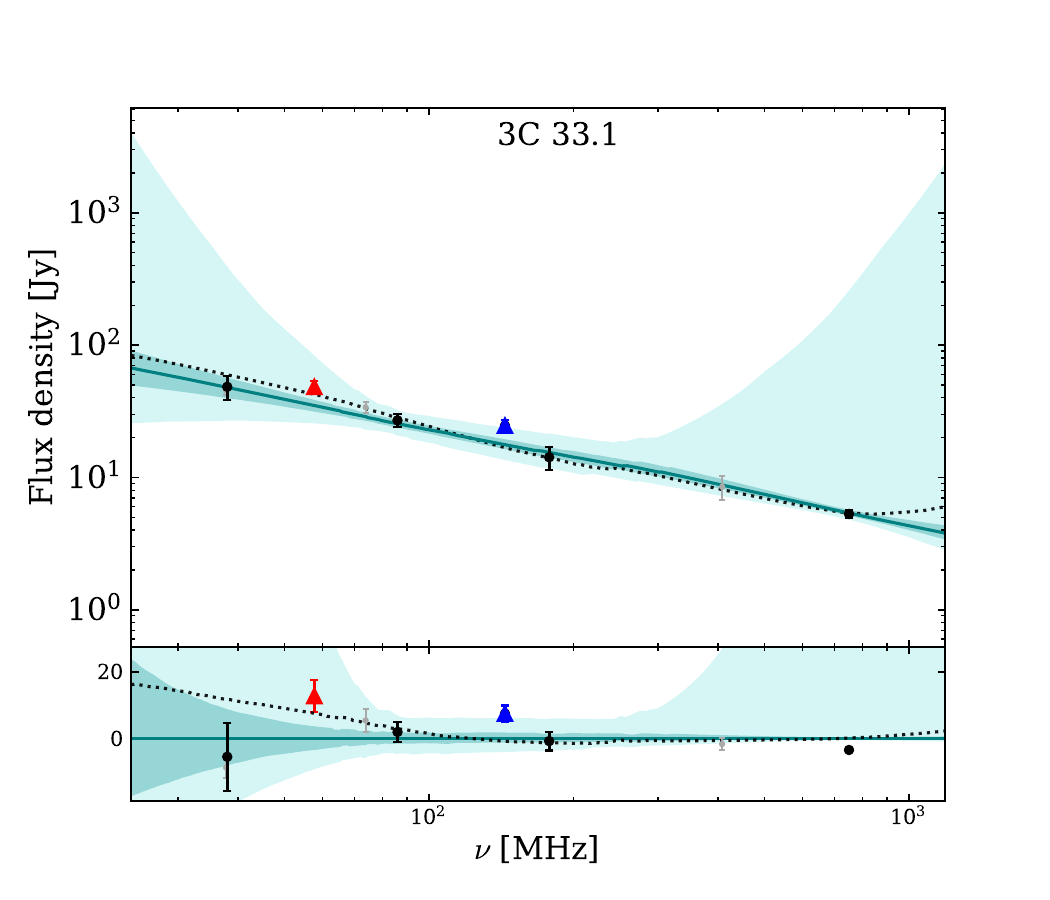}
\includegraphics[width=0.162\linewidth, trim={0.cm .0cm 1.5cm 1.5cm},clip]{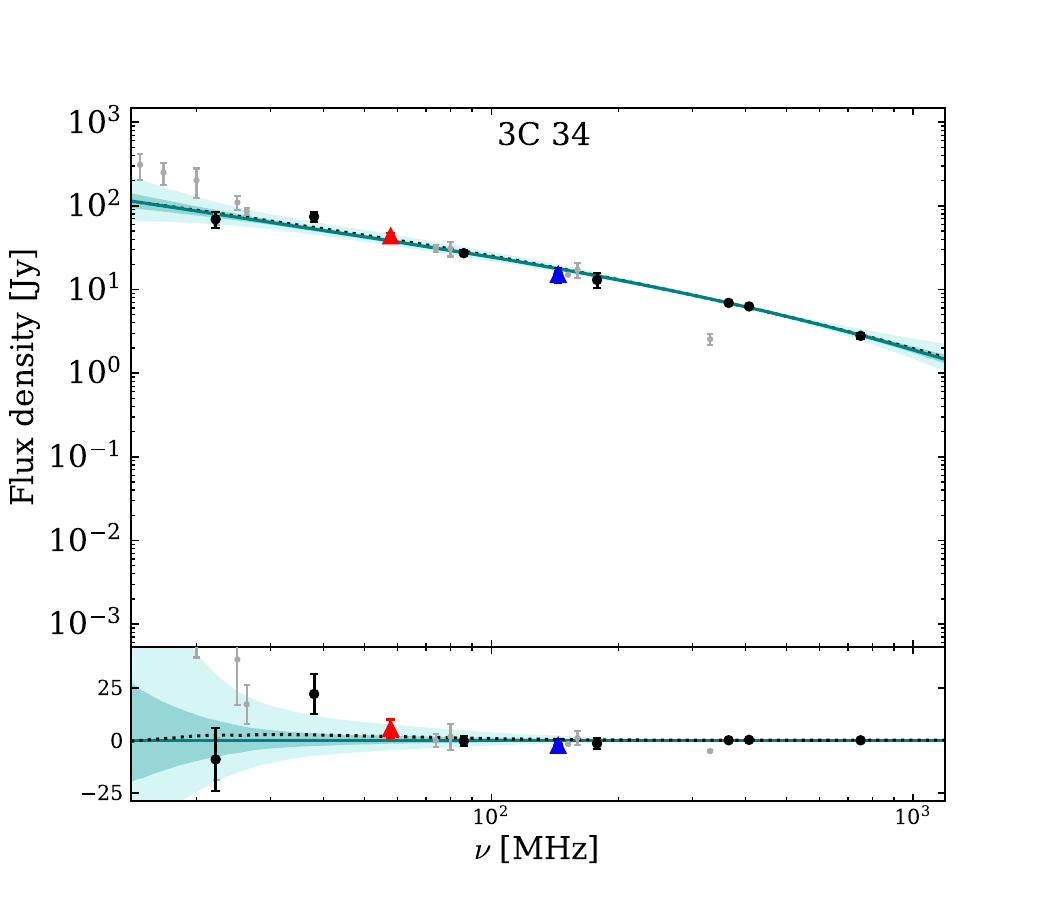}
\includegraphics[width=0.162\linewidth, trim={0.cm .0cm 1.5cm 1.5cm},clip]{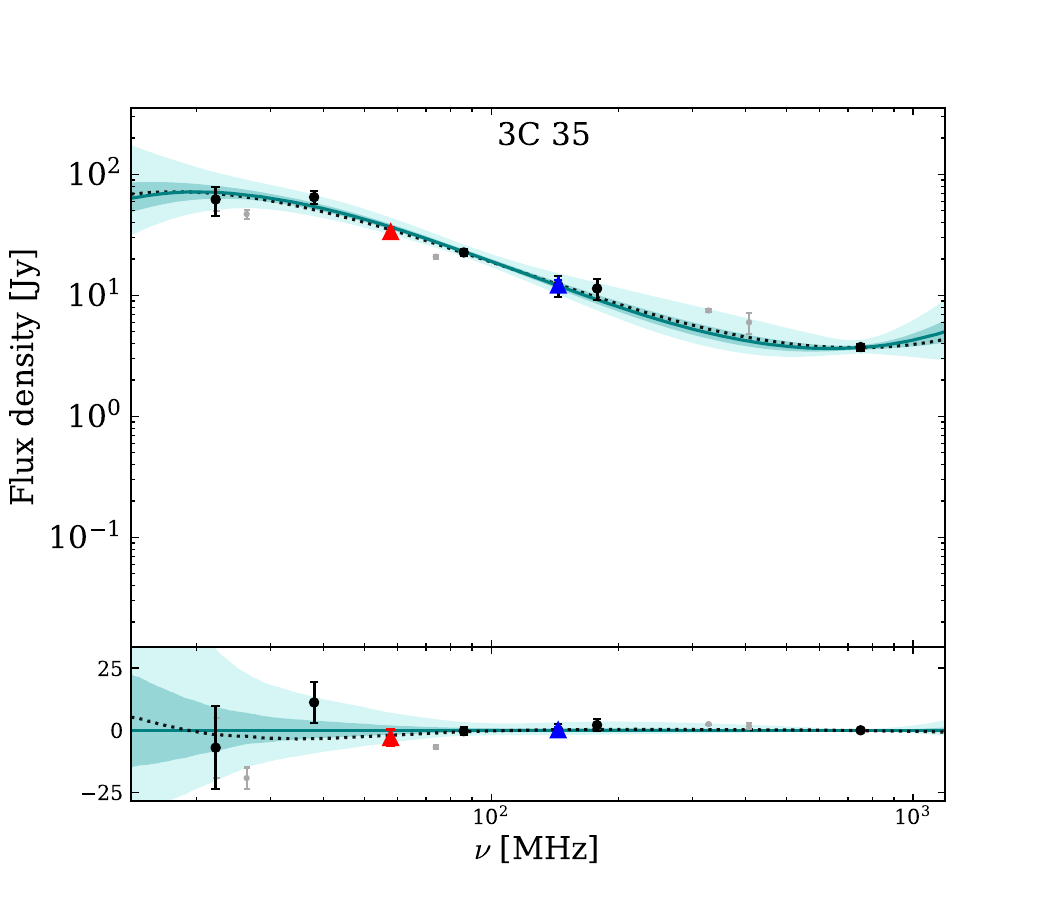}
\includegraphics[width=0.162\linewidth, trim={0.cm .0cm 1.5cm 1.5cm},clip]{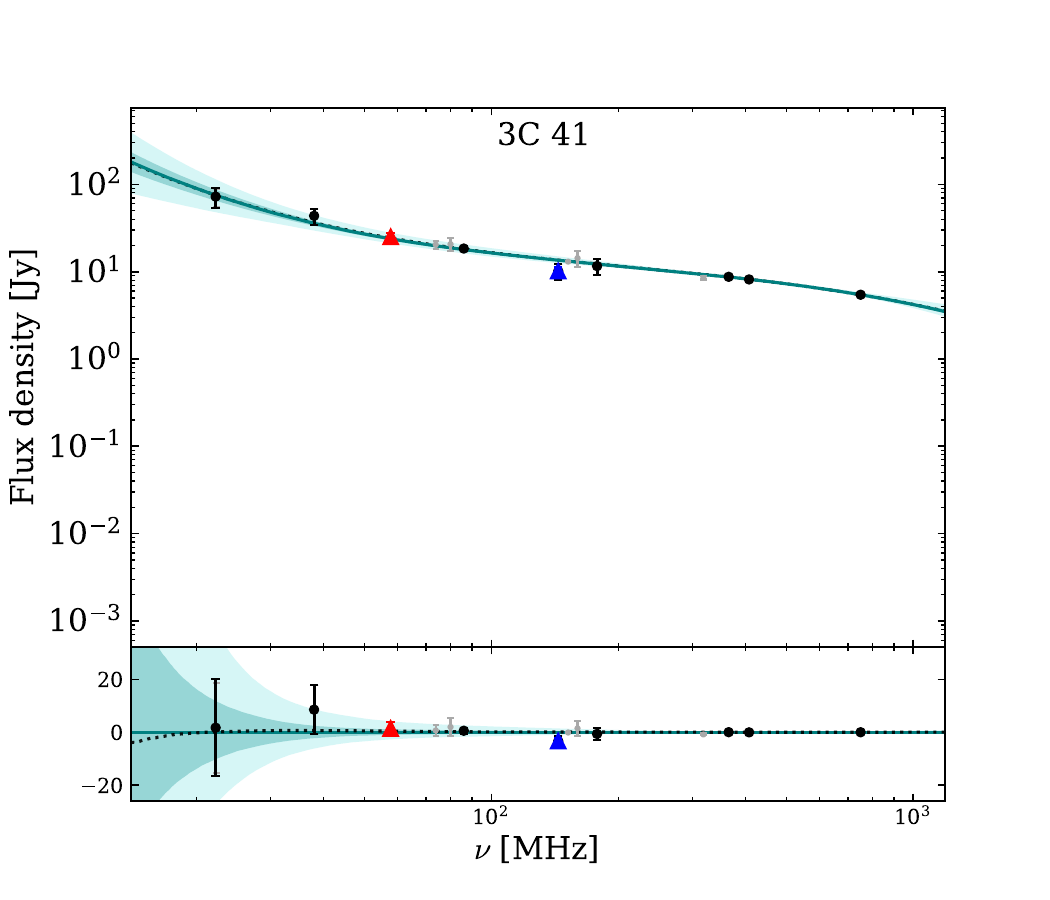}
\includegraphics[width=0.162\linewidth, trim={0.cm .0cm 1.5cm 1.5cm},clip]{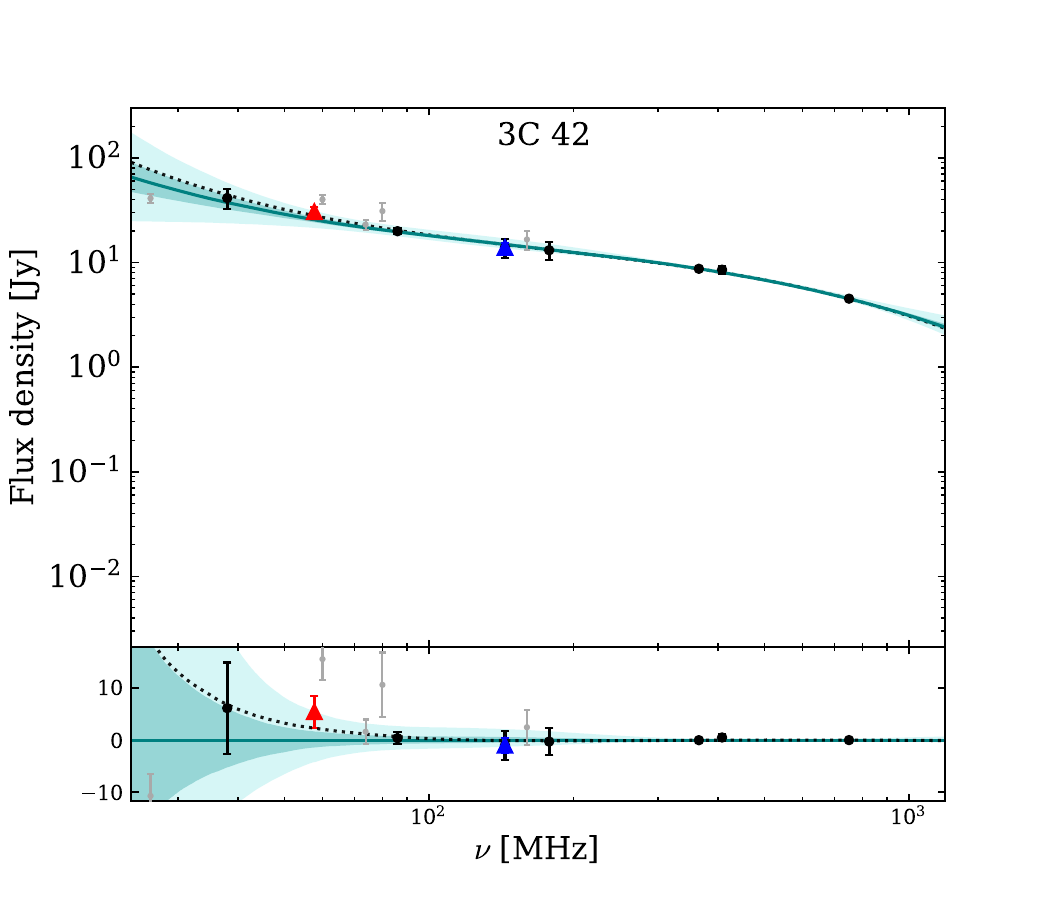}
\includegraphics[width=0.162\linewidth, trim={0.cm .0cm 1.5cm 1.5cm},clip]{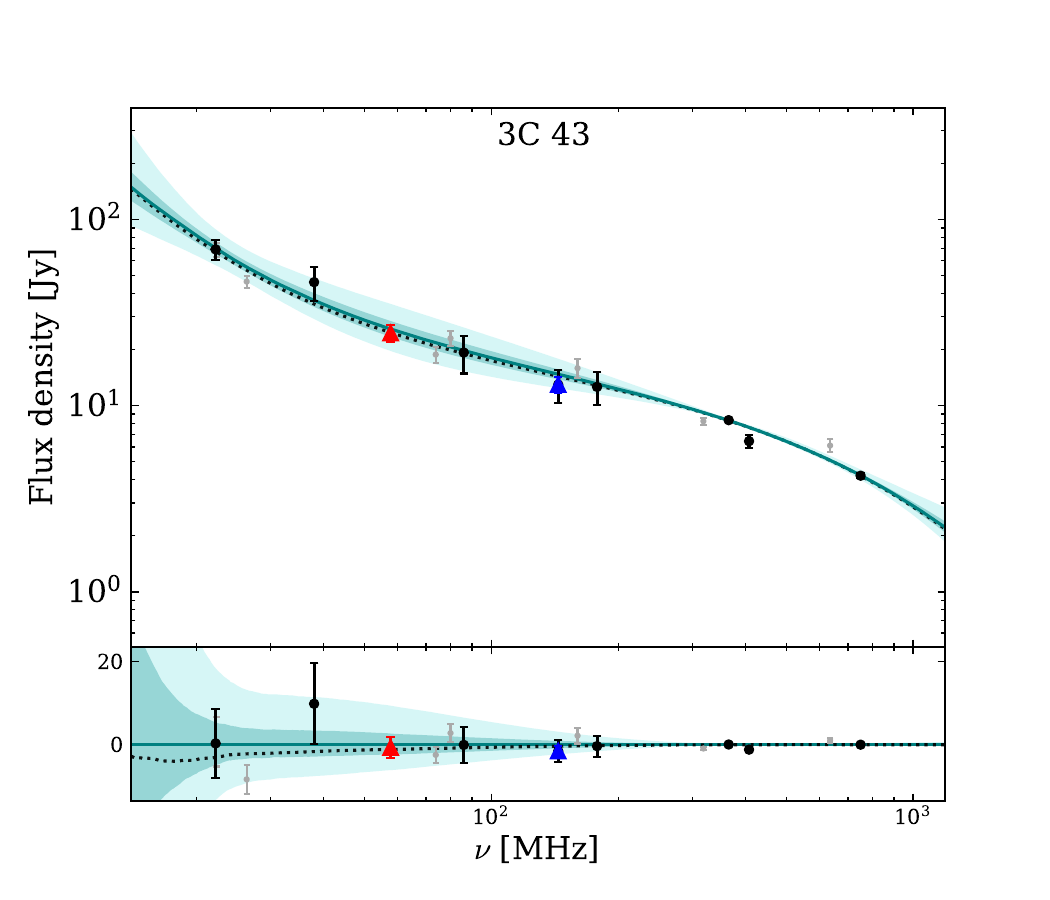}
\includegraphics[width=0.162\linewidth, trim={0.cm .0cm 1.5cm 1.5cm},clip]{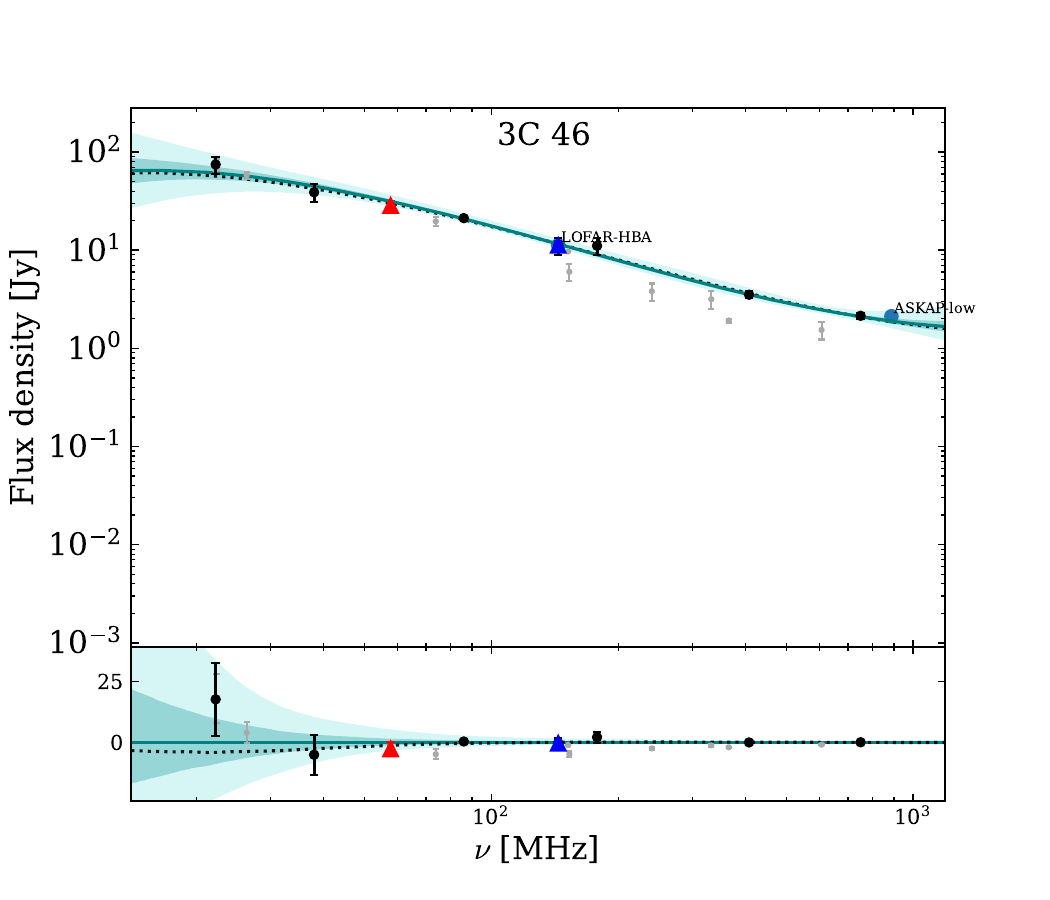}
\includegraphics[width=0.162\linewidth, trim={0.cm .0cm 1.5cm 1.5cm},clip]{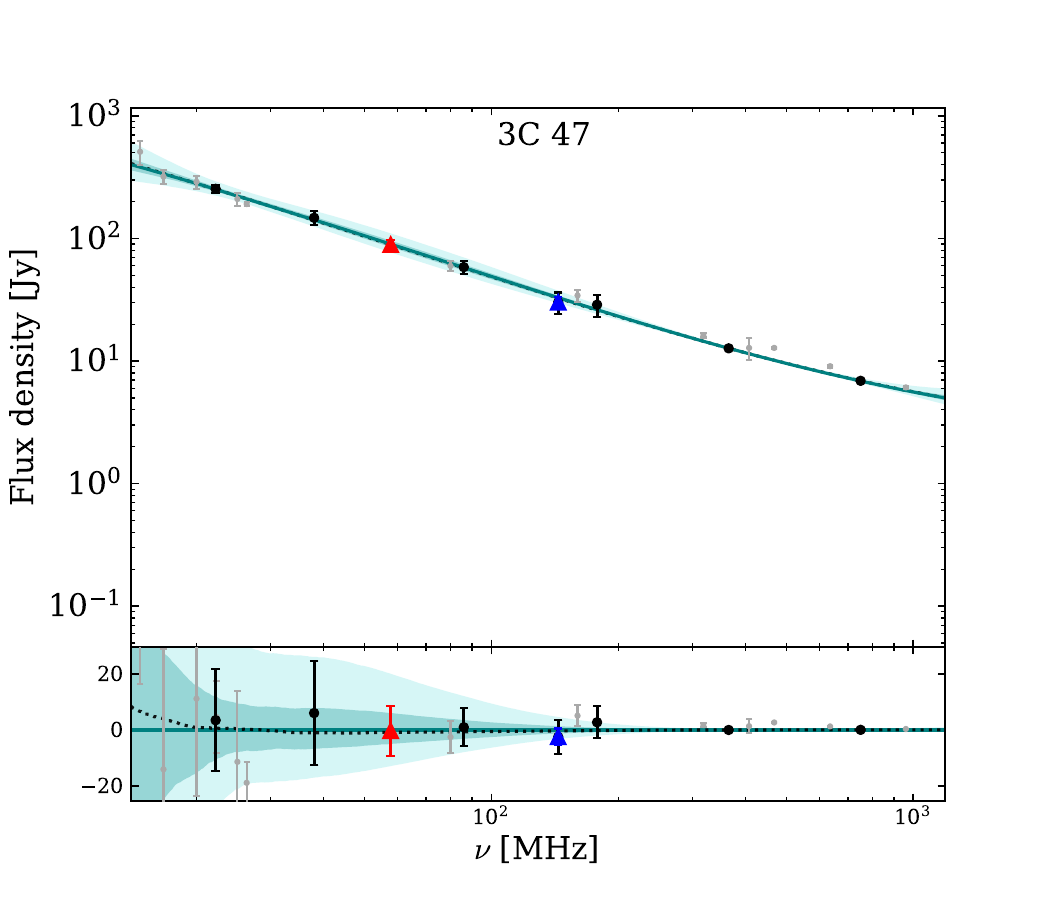}
\includegraphics[width=0.162\linewidth, trim={0.cm .0cm 1.5cm 1.5cm},clip]{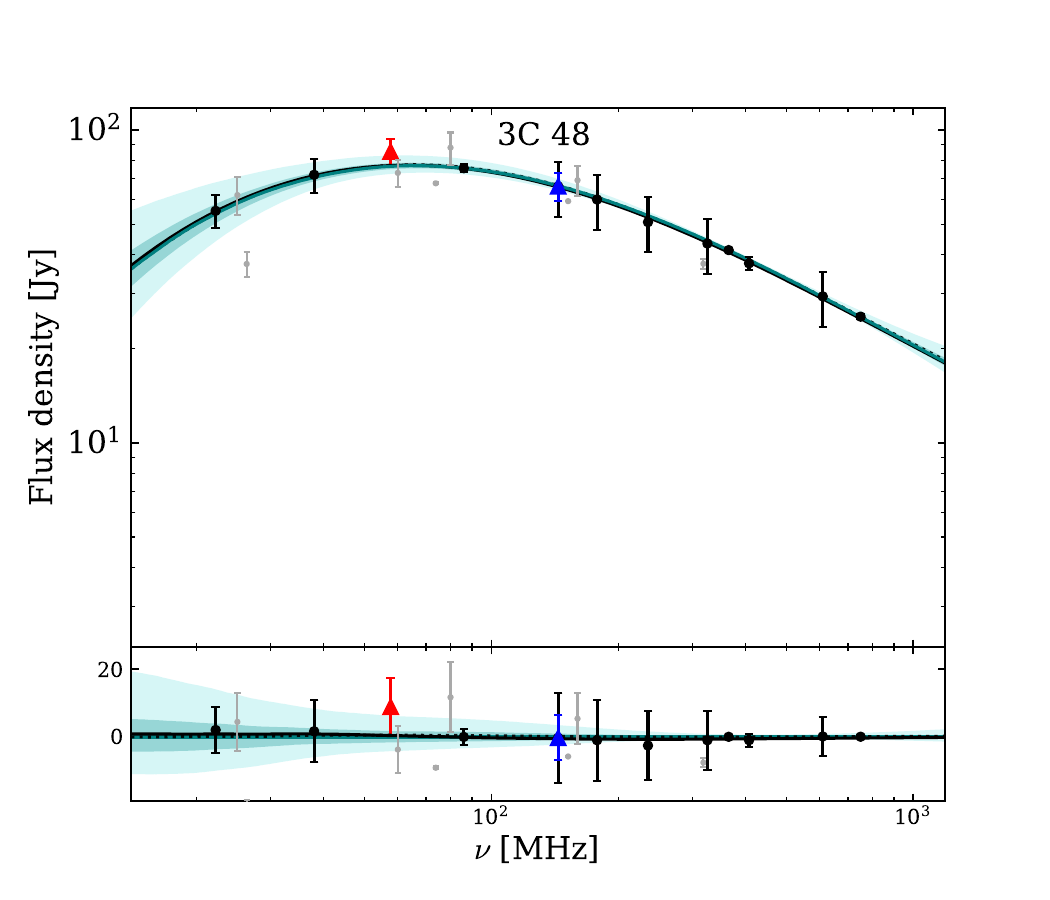}
\includegraphics[width=0.162\linewidth, trim={0.cm .0cm 1.5cm 1.5cm},clip]{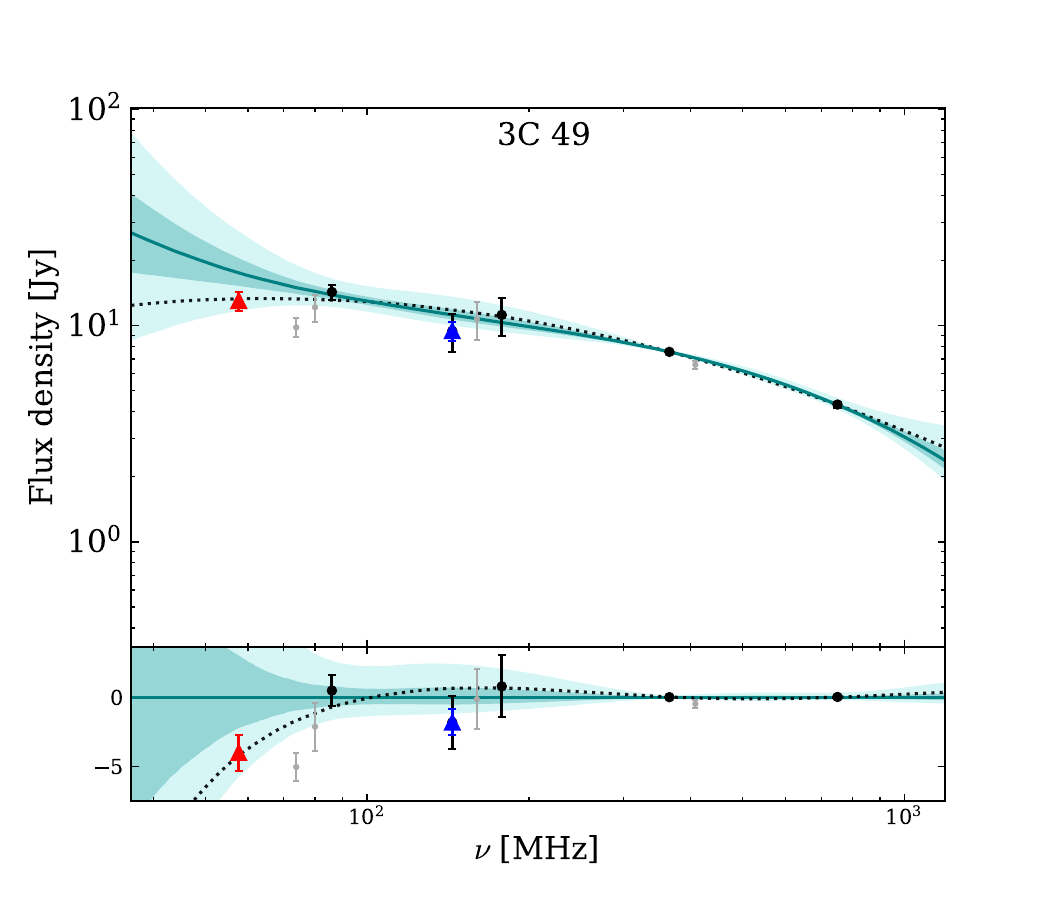}
\includegraphics[width=0.162\linewidth, trim={0.cm .0cm 1.5cm 1.5cm},clip]{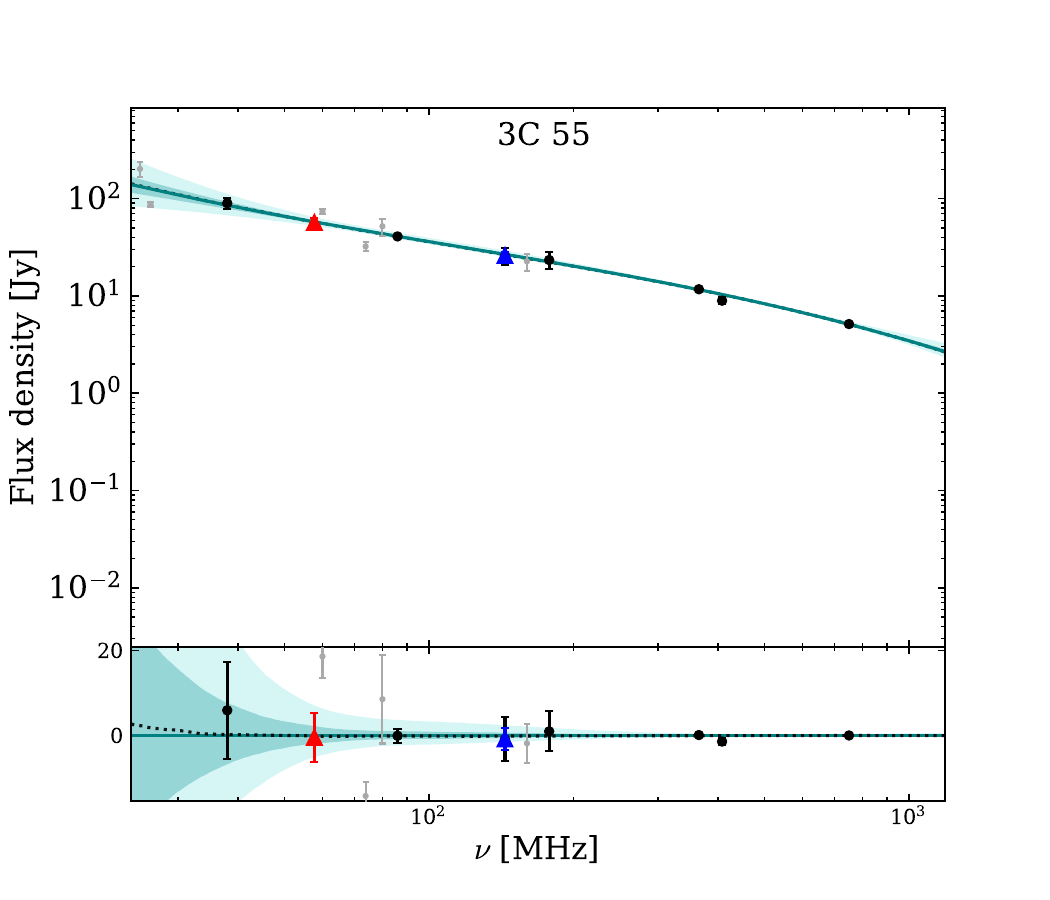}
\includegraphics[width=0.162\linewidth, trim={0.cm .0cm 1.5cm 1.5cm},clip]{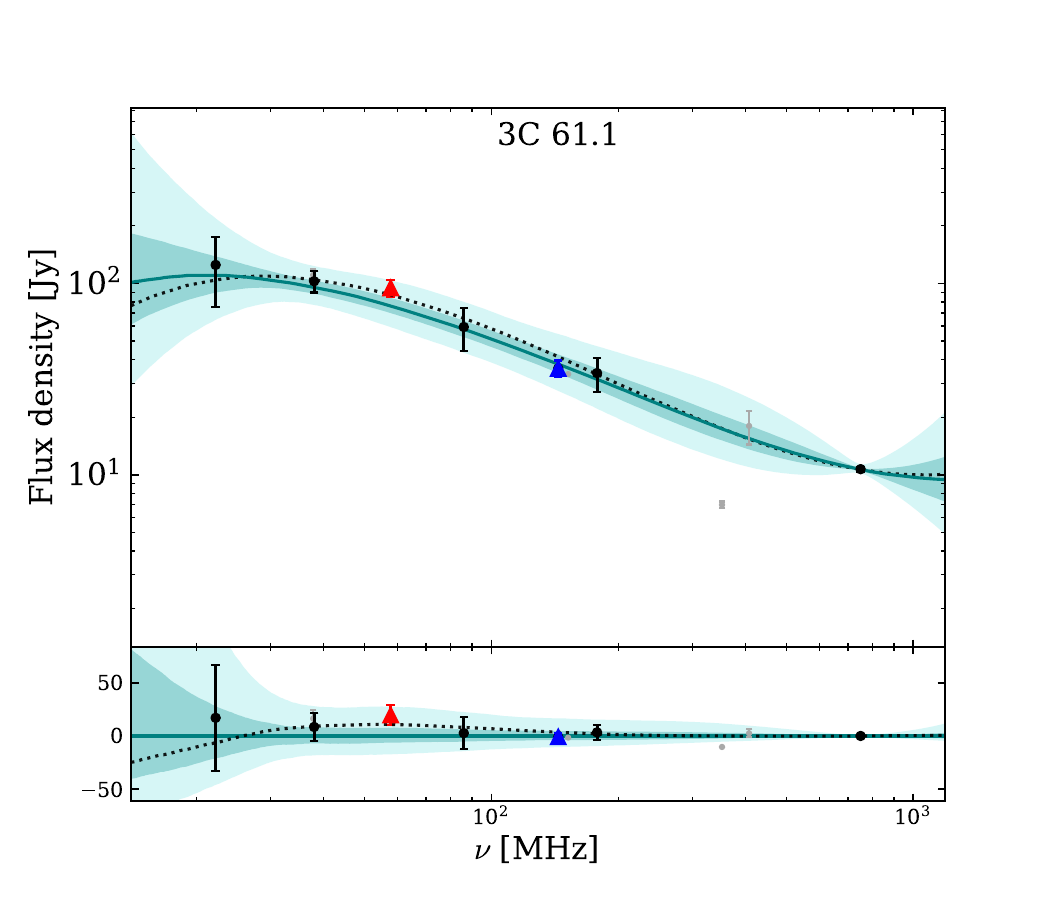}
\includegraphics[width=0.162\linewidth, trim={0.cm .0cm 1.5cm 1.5cm},clip]{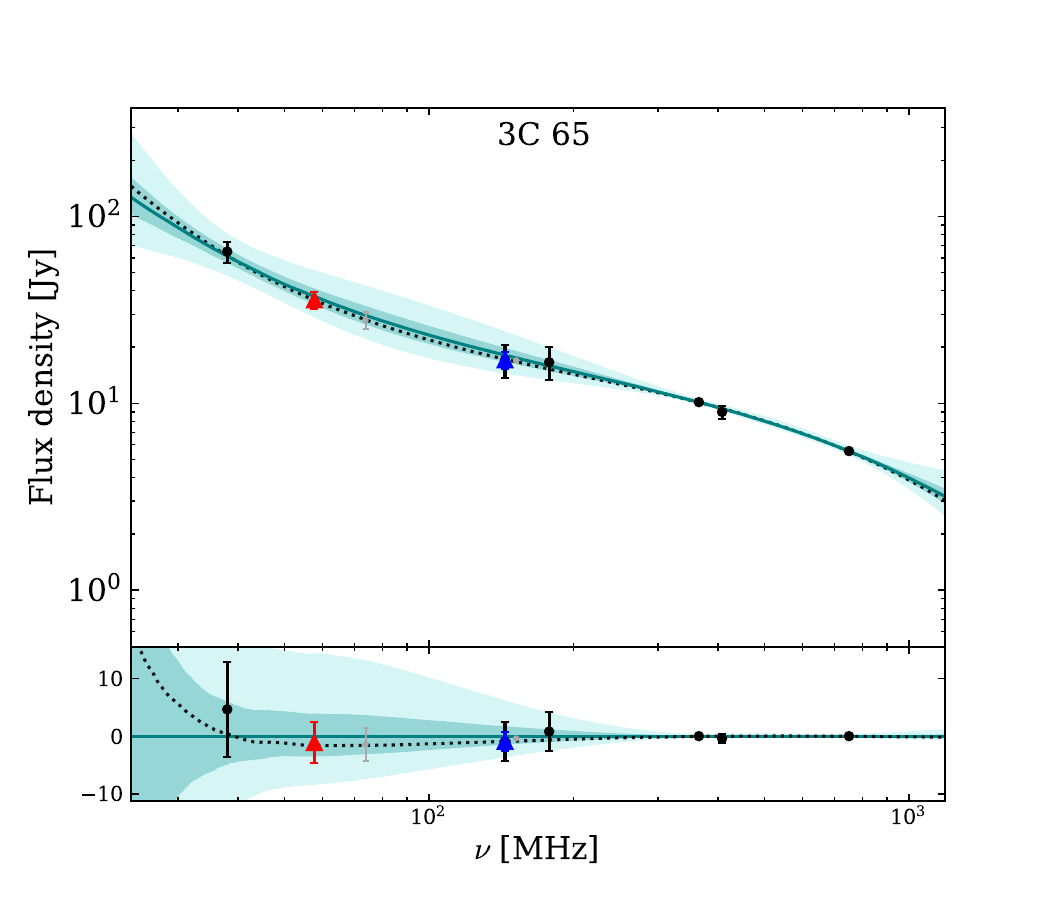}
\includegraphics[width=0.162\linewidth, trim={0.cm .0cm 1.5cm 1.5cm},clip]{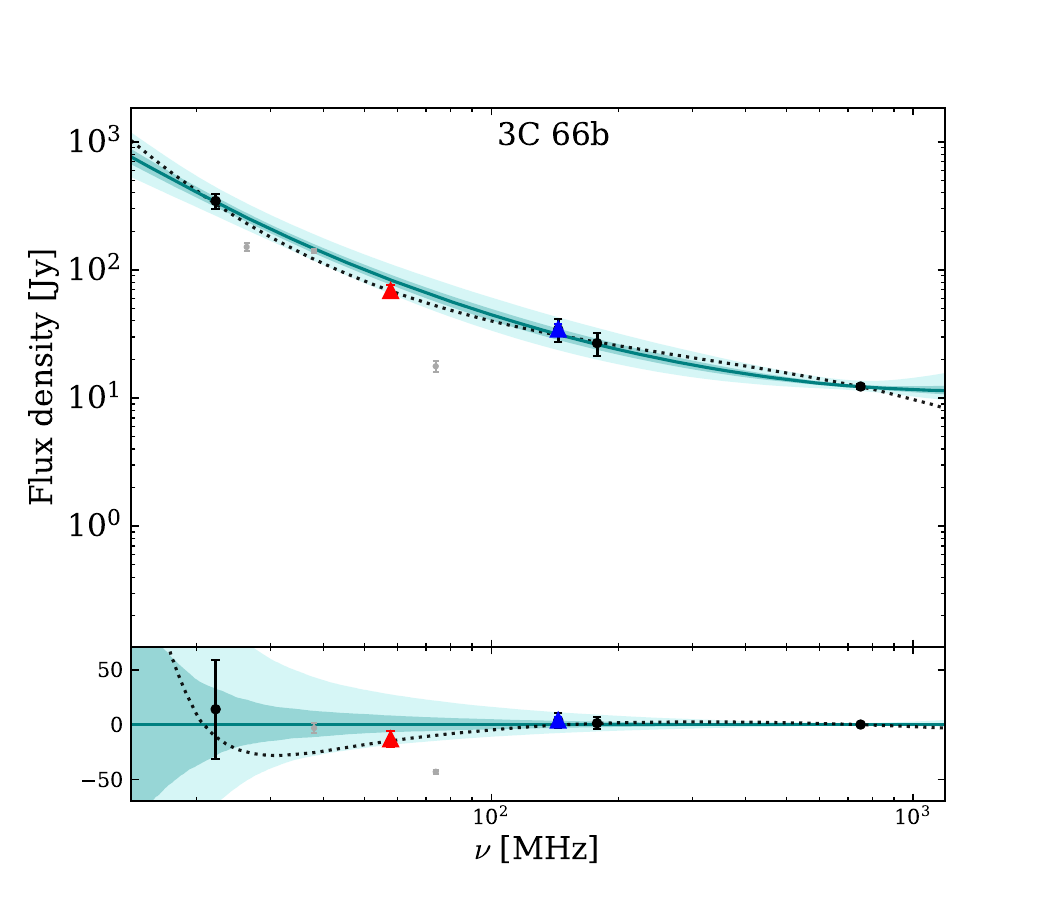}
\includegraphics[width=0.162\linewidth, trim={0.cm .0cm 1.5cm 1.5cm},clip]{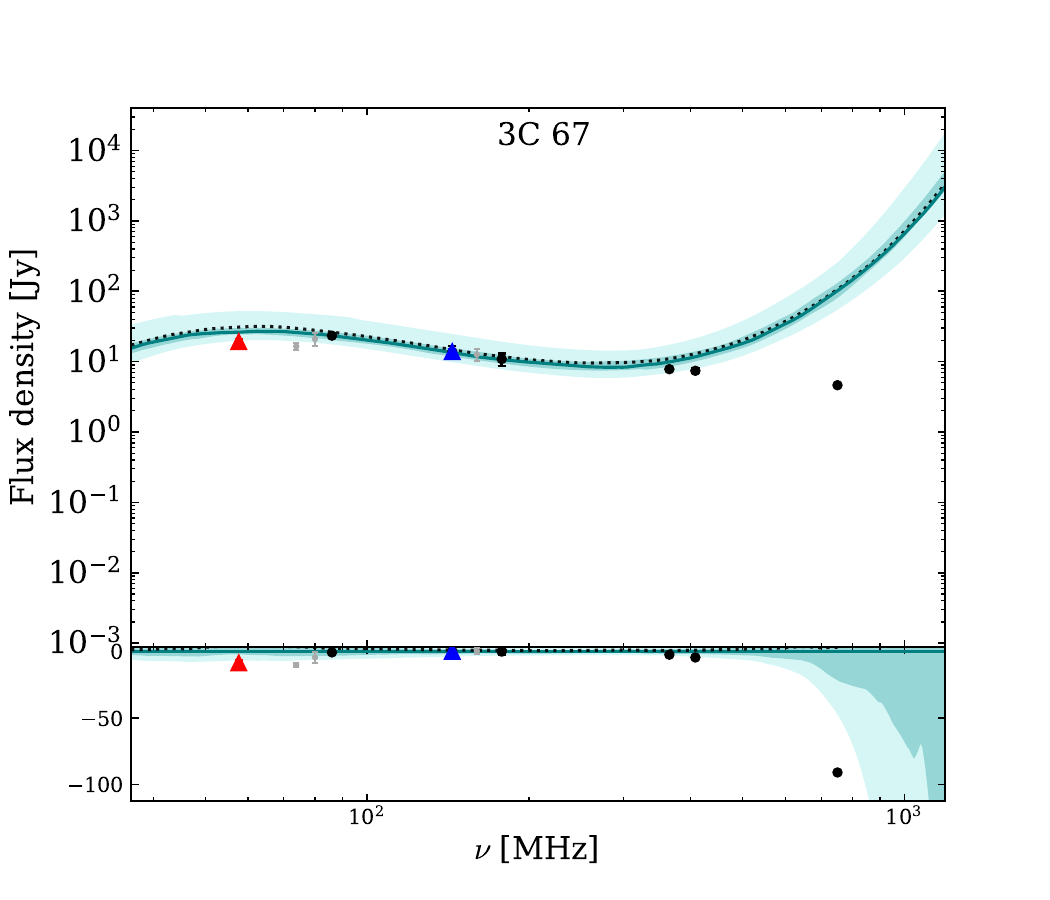}
\includegraphics[width=0.162\linewidth, trim={0.cm .0cm 1.5cm 1.5cm},clip]{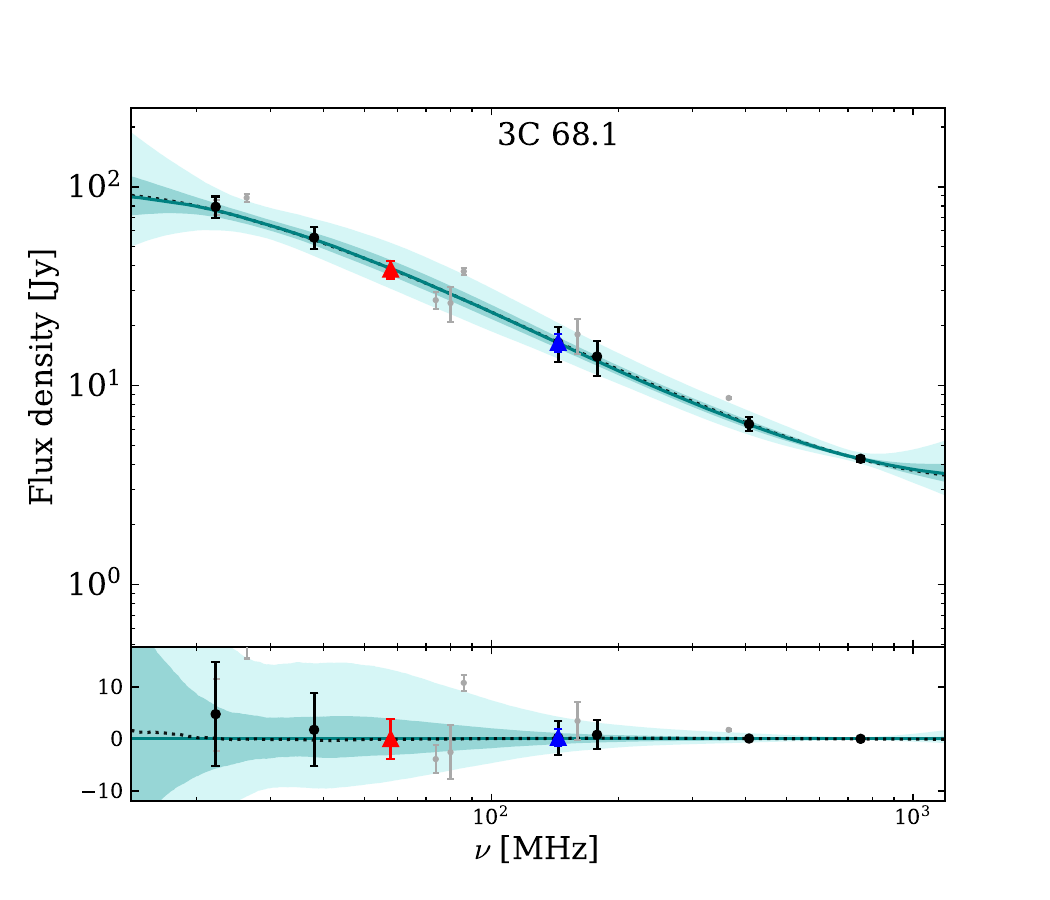}
\includegraphics[width=0.162\linewidth, trim={0.cm .0cm 1.5cm 1.5cm},clip]{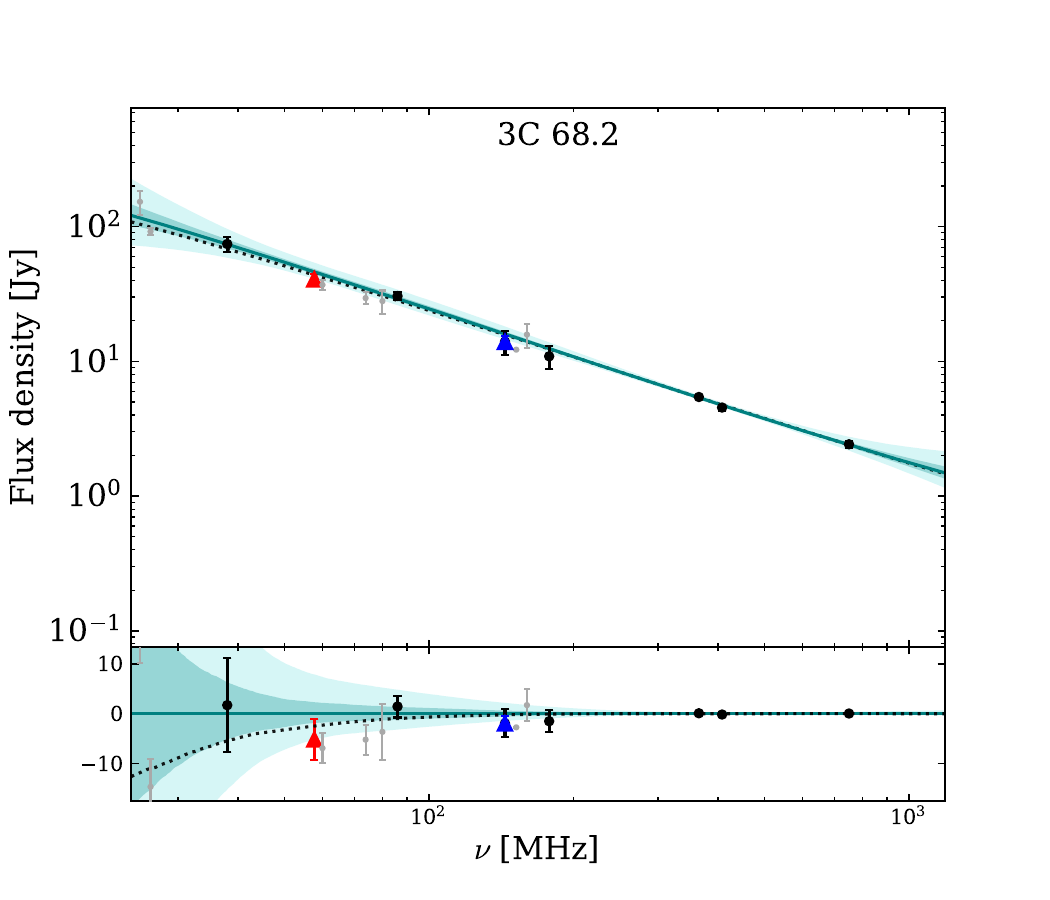}
\includegraphics[width=0.162\linewidth, trim={0.cm .0cm 1.5cm 1.5cm},clip]{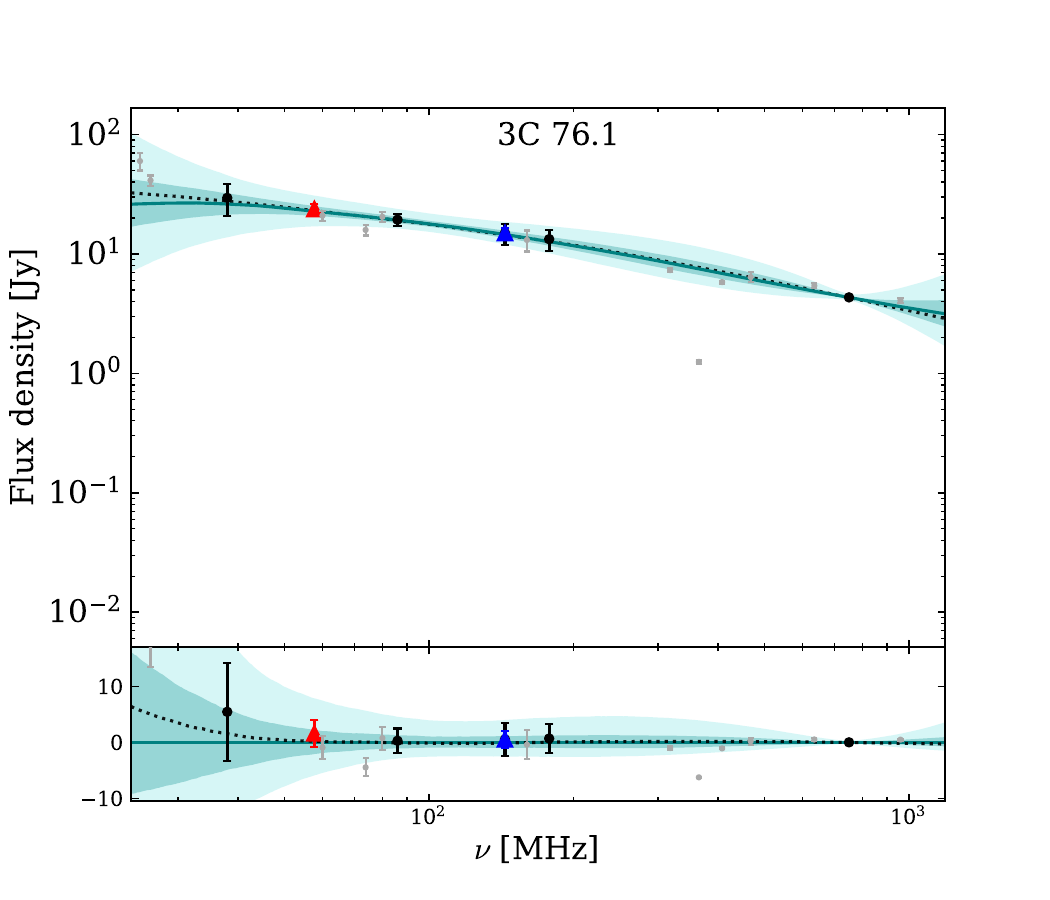}
\includegraphics[width=0.162\linewidth, trim={0.cm .0cm 1.5cm 1.5cm},clip]{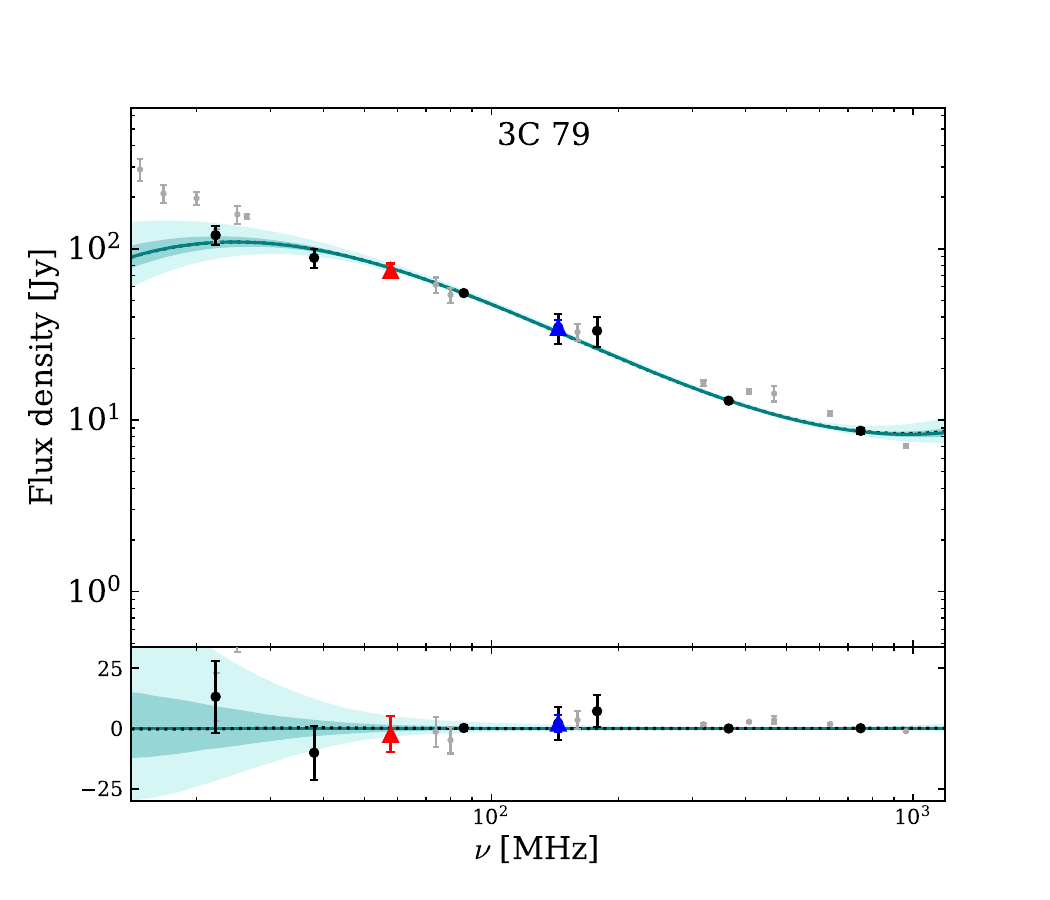}
\includegraphics[width=0.162\linewidth, trim={0.cm .0cm 1.5cm 1.5cm},clip]{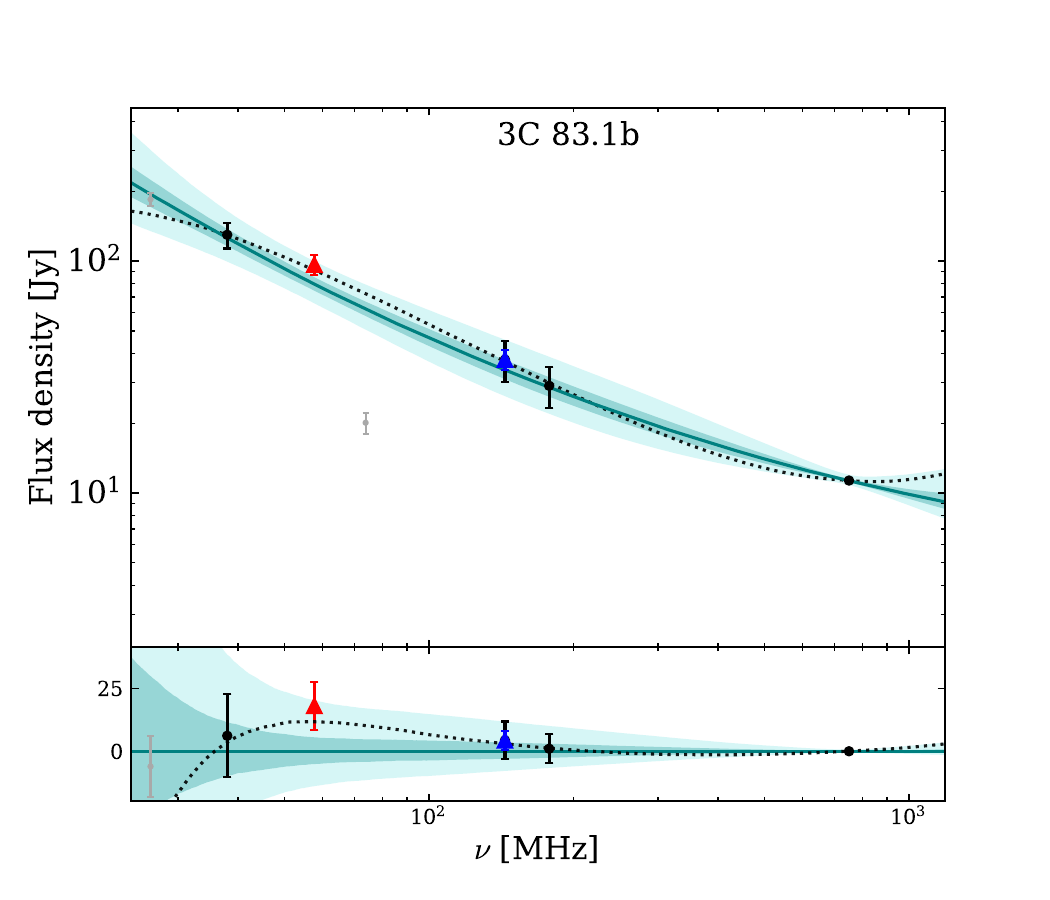}
\includegraphics[width=0.162\linewidth, trim={0.cm .0cm 1.5cm 1.5cm},clip]{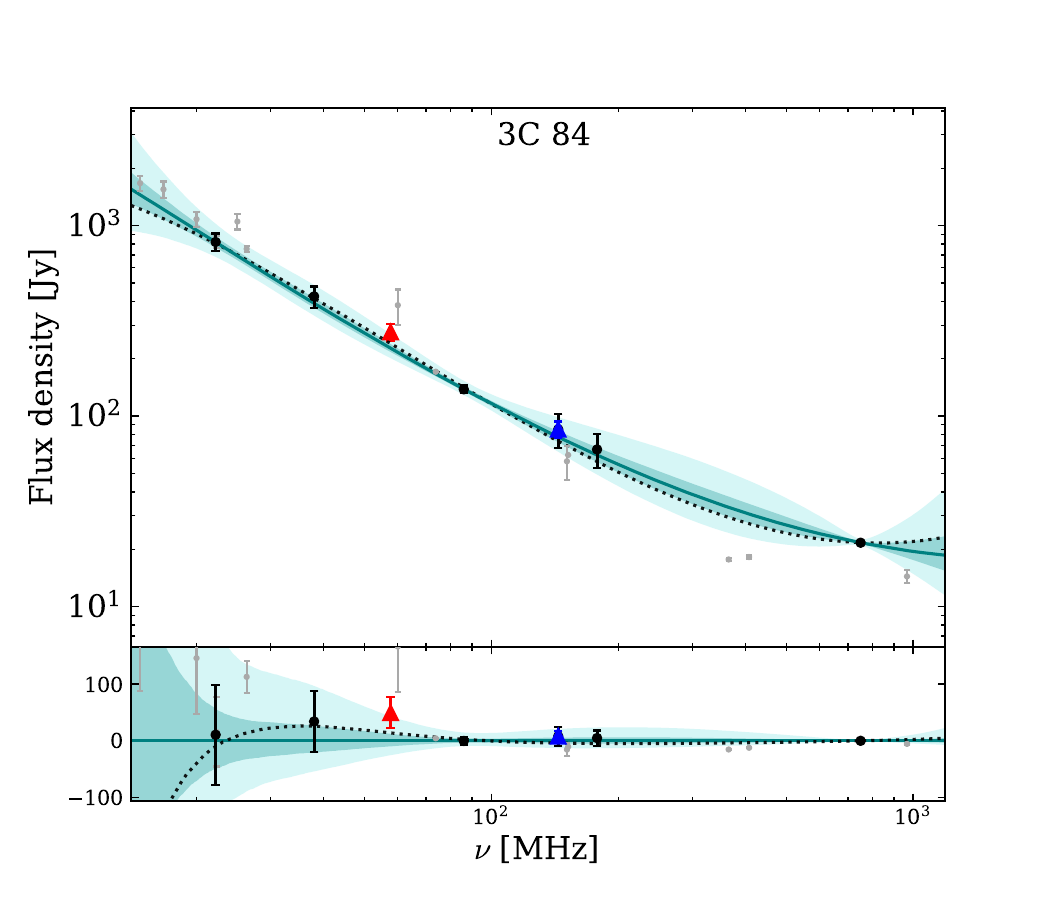}
\includegraphics[width=0.162\linewidth, trim={0.cm .0cm 1.5cm 1.5cm},clip]{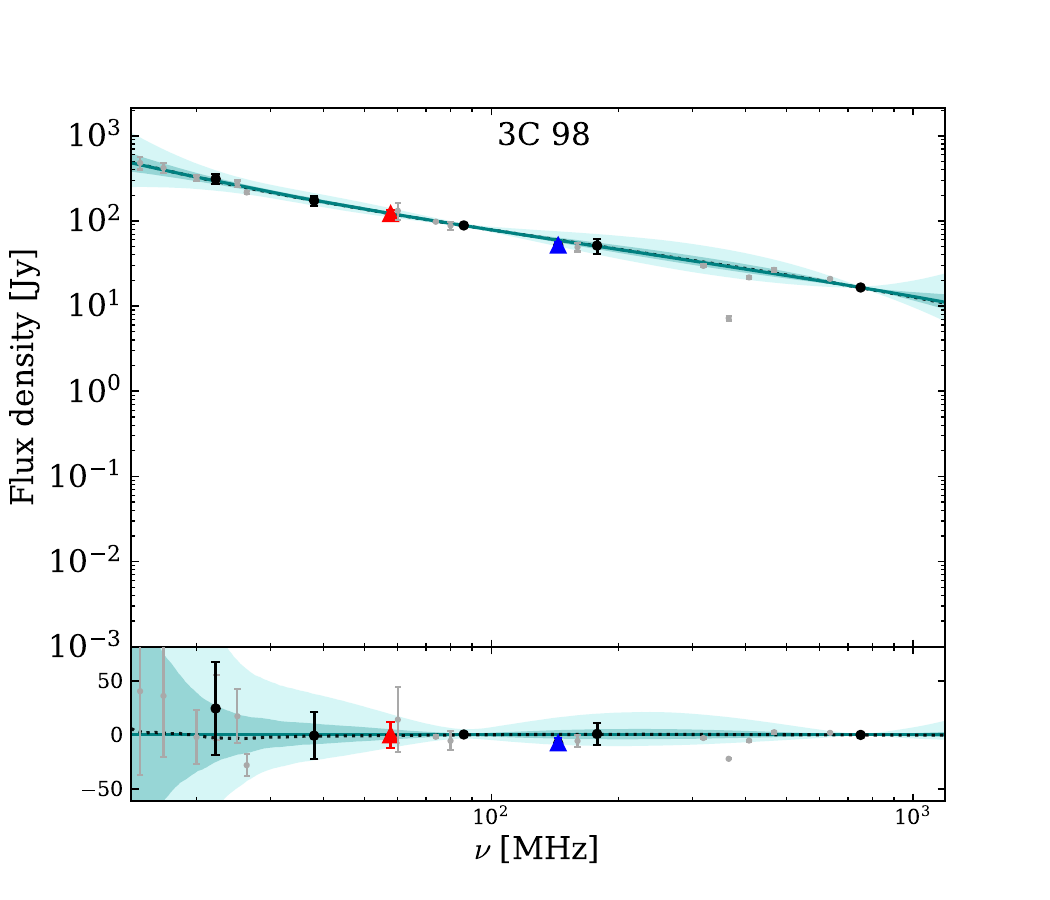}
\includegraphics[width=0.162\linewidth, trim={0.cm .0cm 1.5cm 1.5cm},clip]{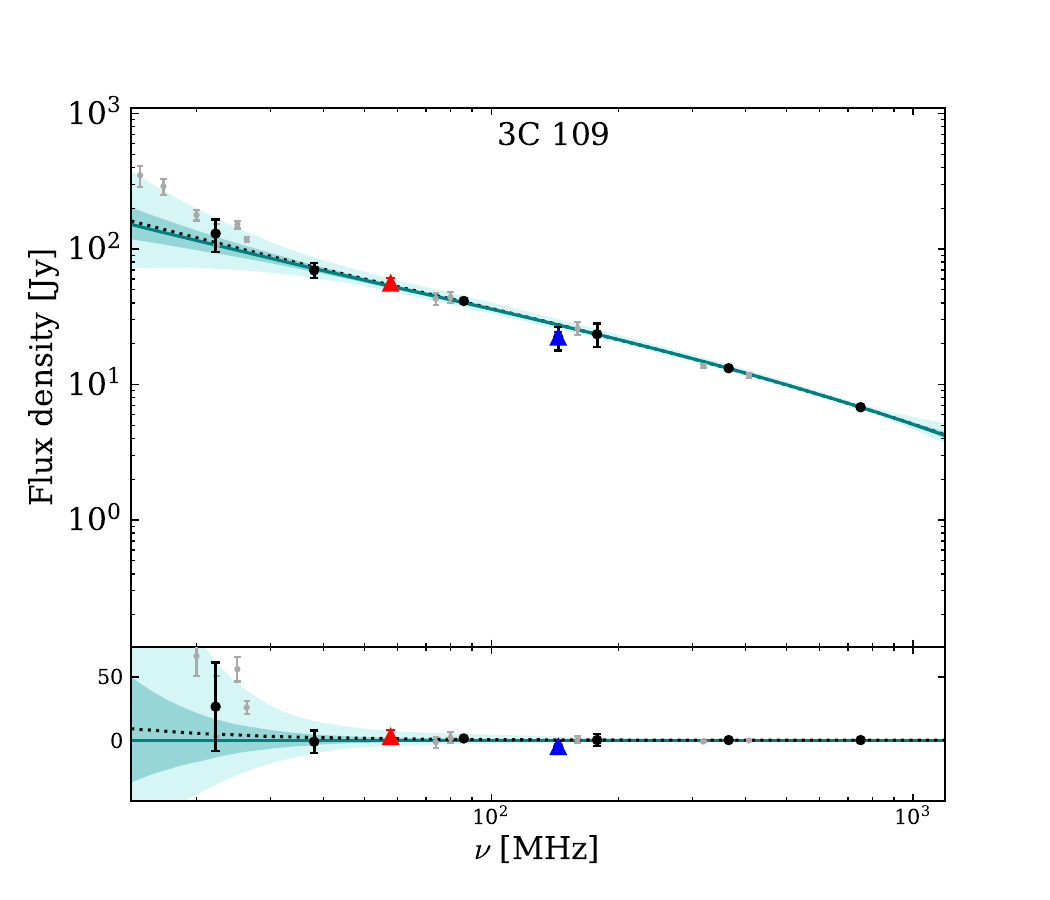}
\includegraphics[width=0.162\linewidth, trim={0.cm .0cm 1.5cm 1.5cm},clip]{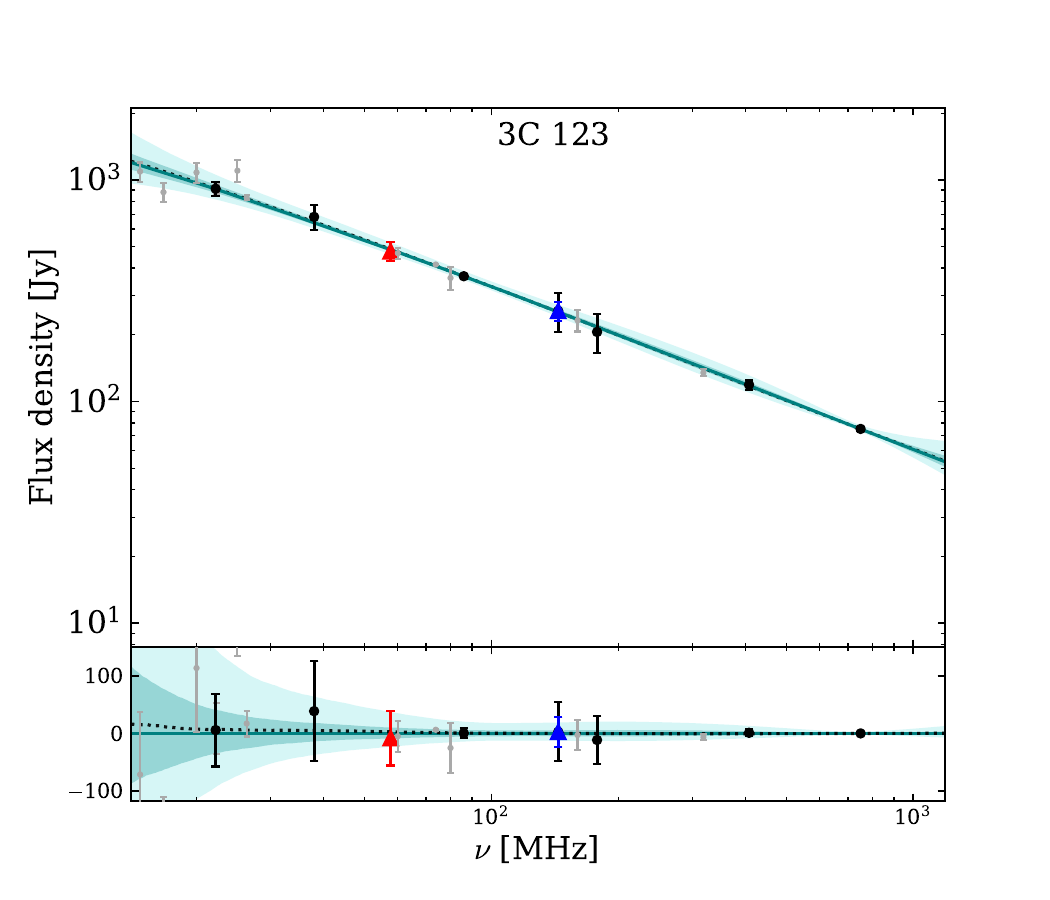}
\includegraphics[width=0.162\linewidth, trim={0.cm .0cm 1.5cm 1.5cm},clip]{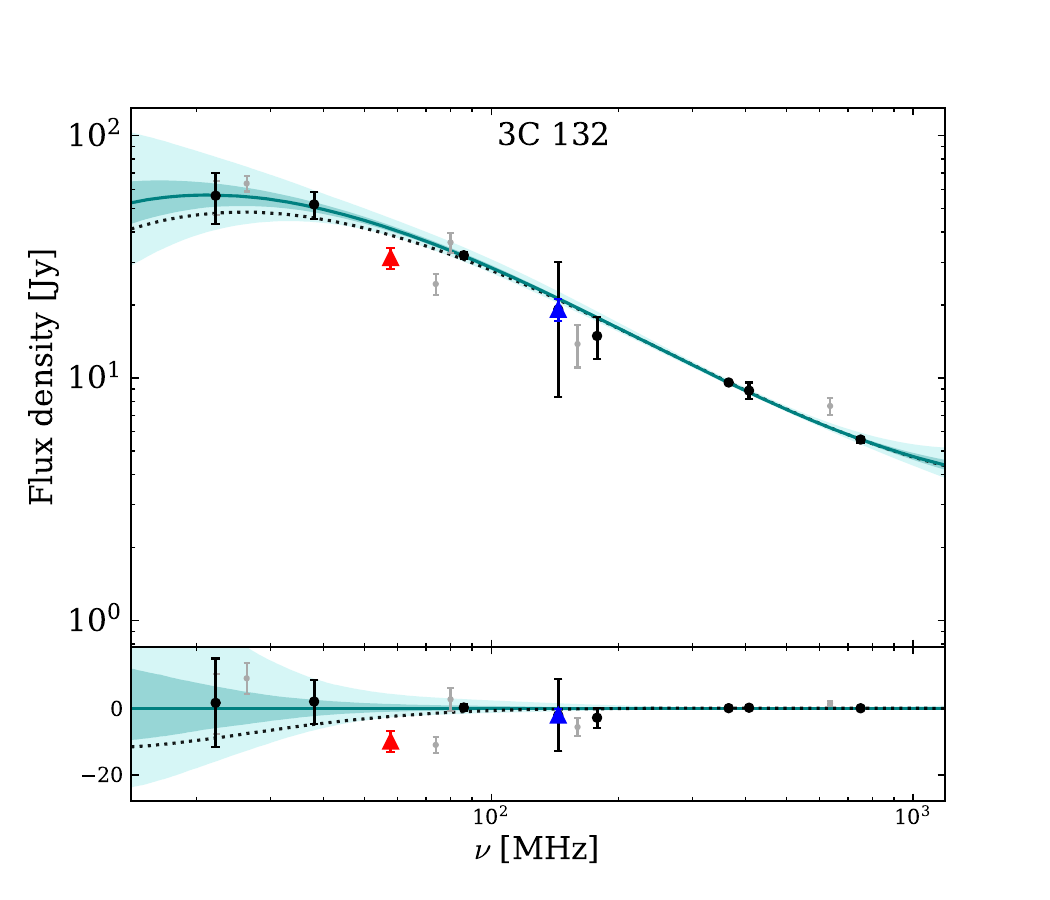}
\includegraphics[width=0.162\linewidth, trim={0.cm .0cm 1.5cm 1.5cm},clip]{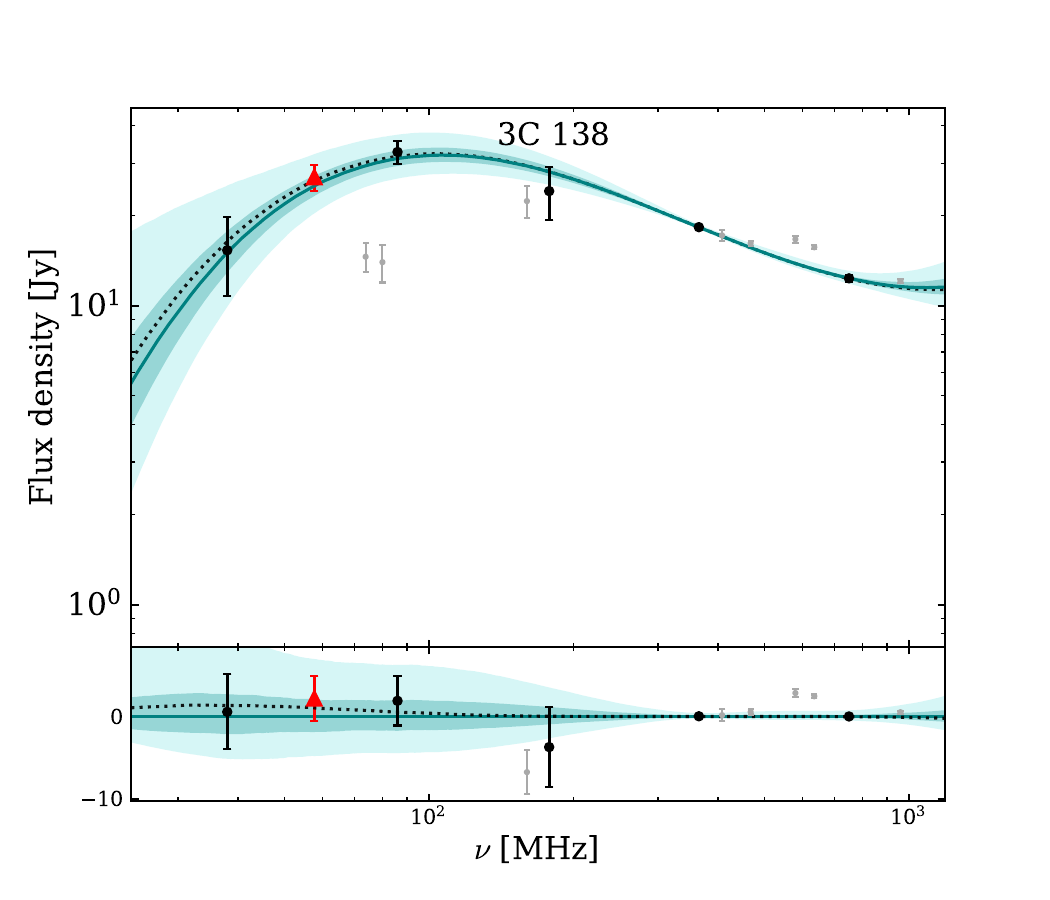}
\includegraphics[width=0.162\linewidth, trim={0.cm .0cm 1.5cm 1.5cm},clip]{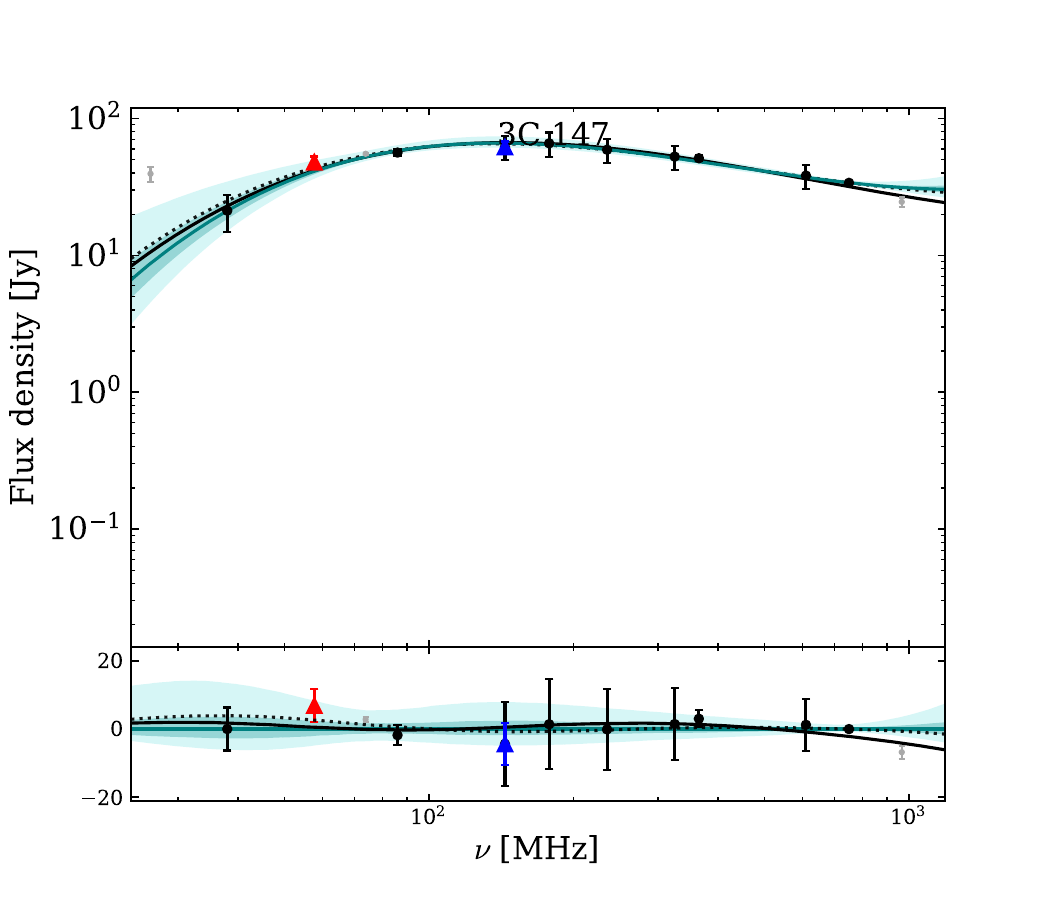}
\includegraphics[width=0.162\linewidth, trim={0.cm .0cm 1.5cm 1.5cm},clip]{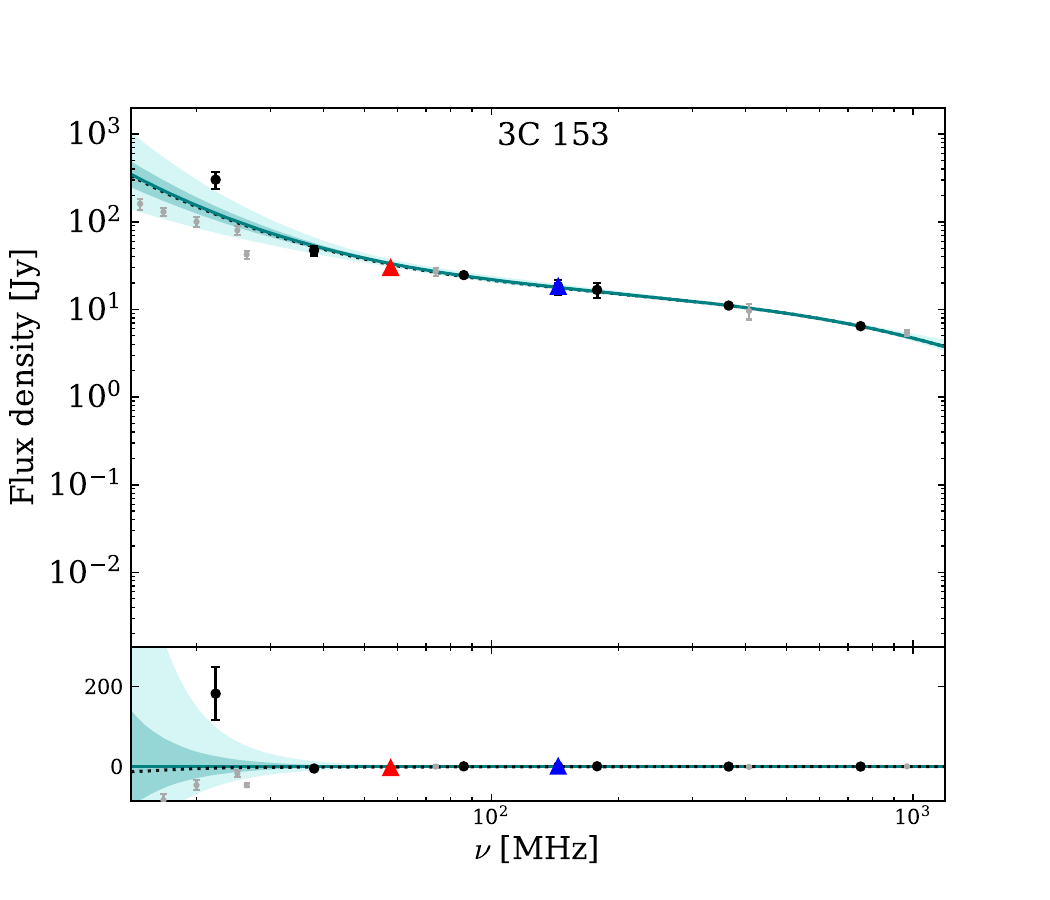}
\includegraphics[width=0.162\linewidth, trim={0.cm .0cm 1.5cm 1.5cm},clip]{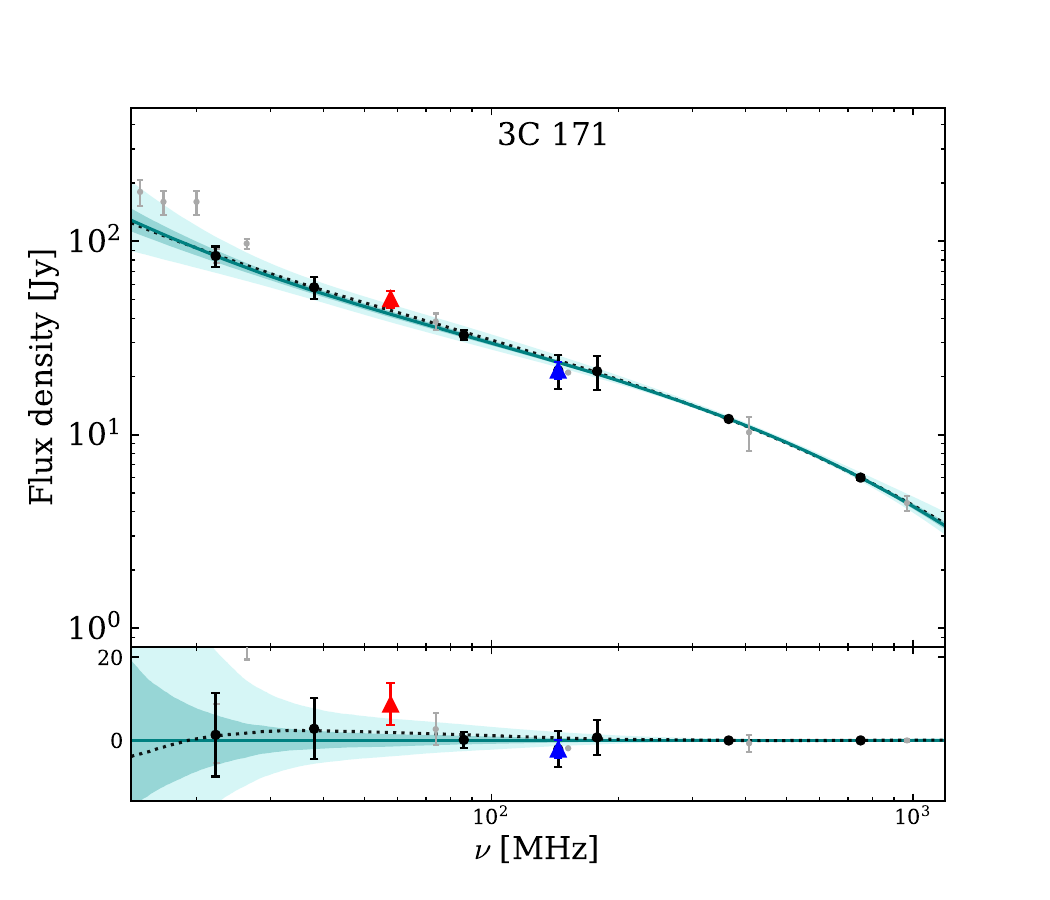}
\includegraphics[width=0.162\linewidth, trim={0.cm .0cm 1.5cm 1.5cm},clip]{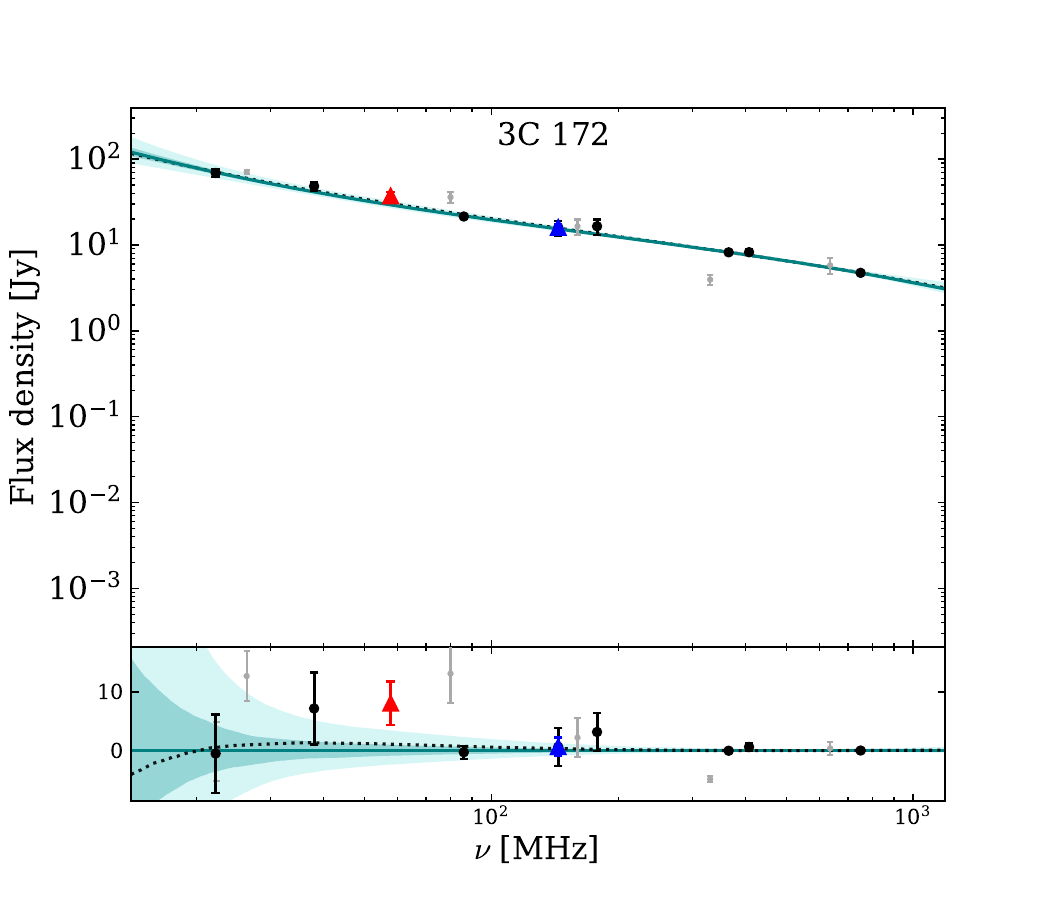}
\includegraphics[width=0.162\linewidth, trim={0.cm .0cm 1.5cm 1.5cm},clip]{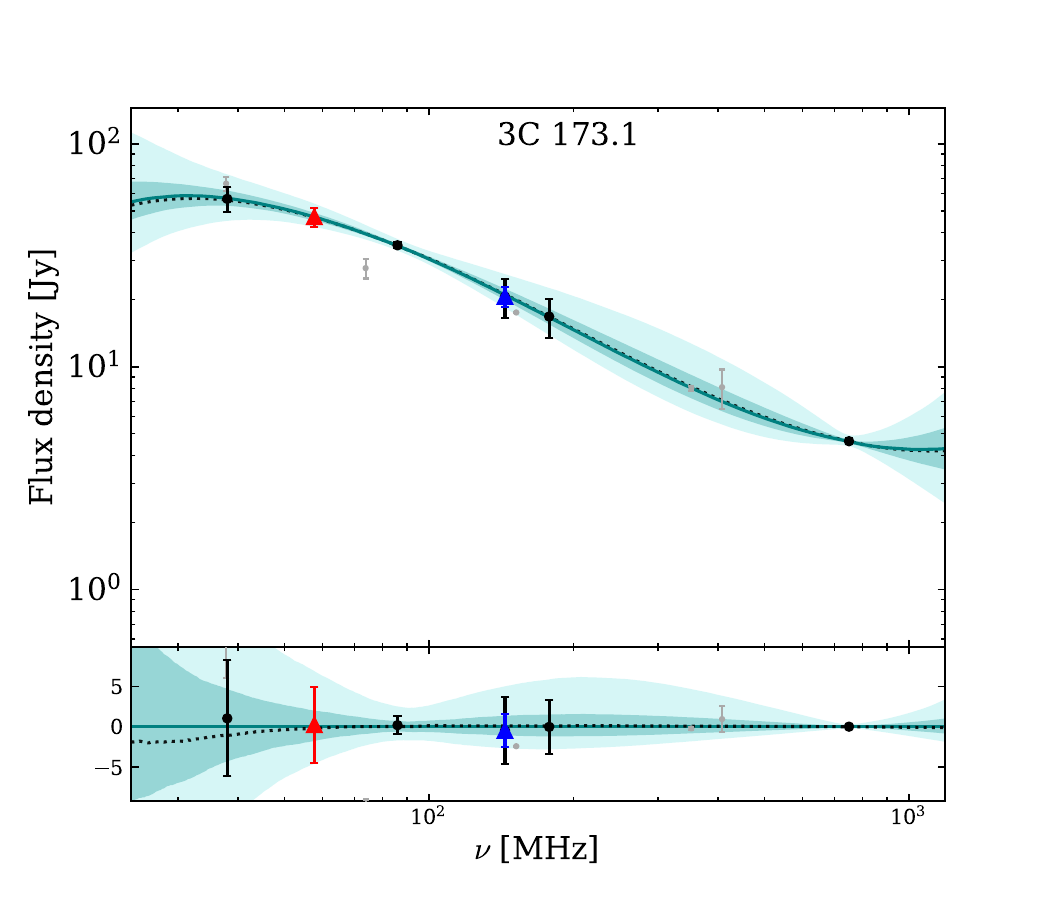}
\includegraphics[width=0.162\linewidth, trim={0.cm .0cm 1.5cm 1.5cm},clip]{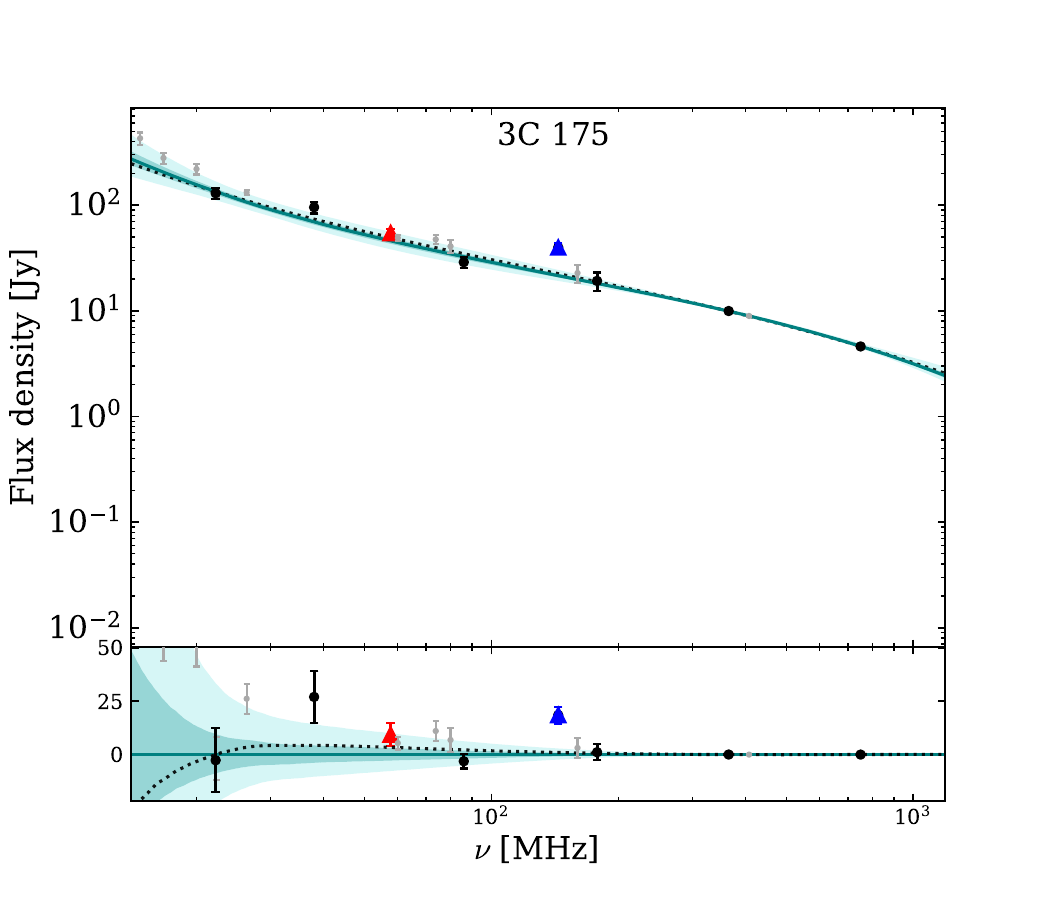}
\includegraphics[width=0.162\linewidth, trim={0.cm .0cm 1.5cm 1.5cm},clip]{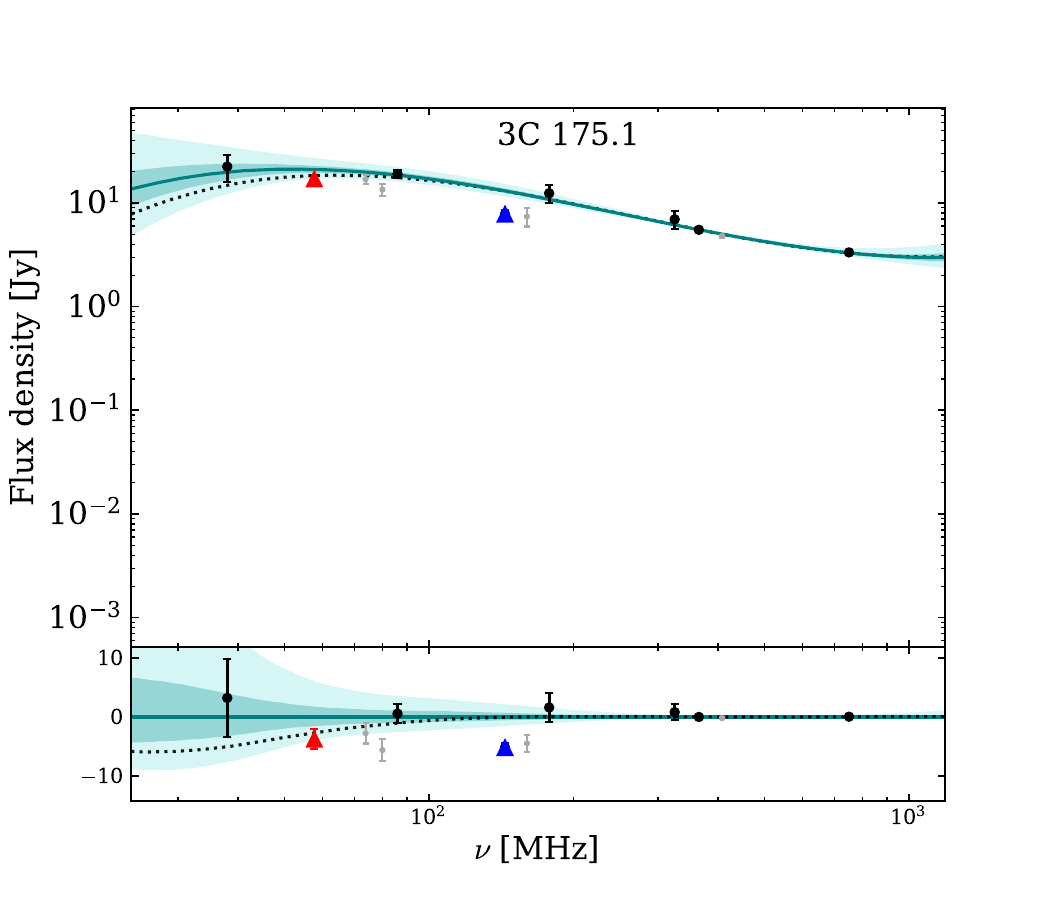}
\includegraphics[width=0.162\linewidth, trim={0.cm .0cm 1.5cm 1.5cm},clip]{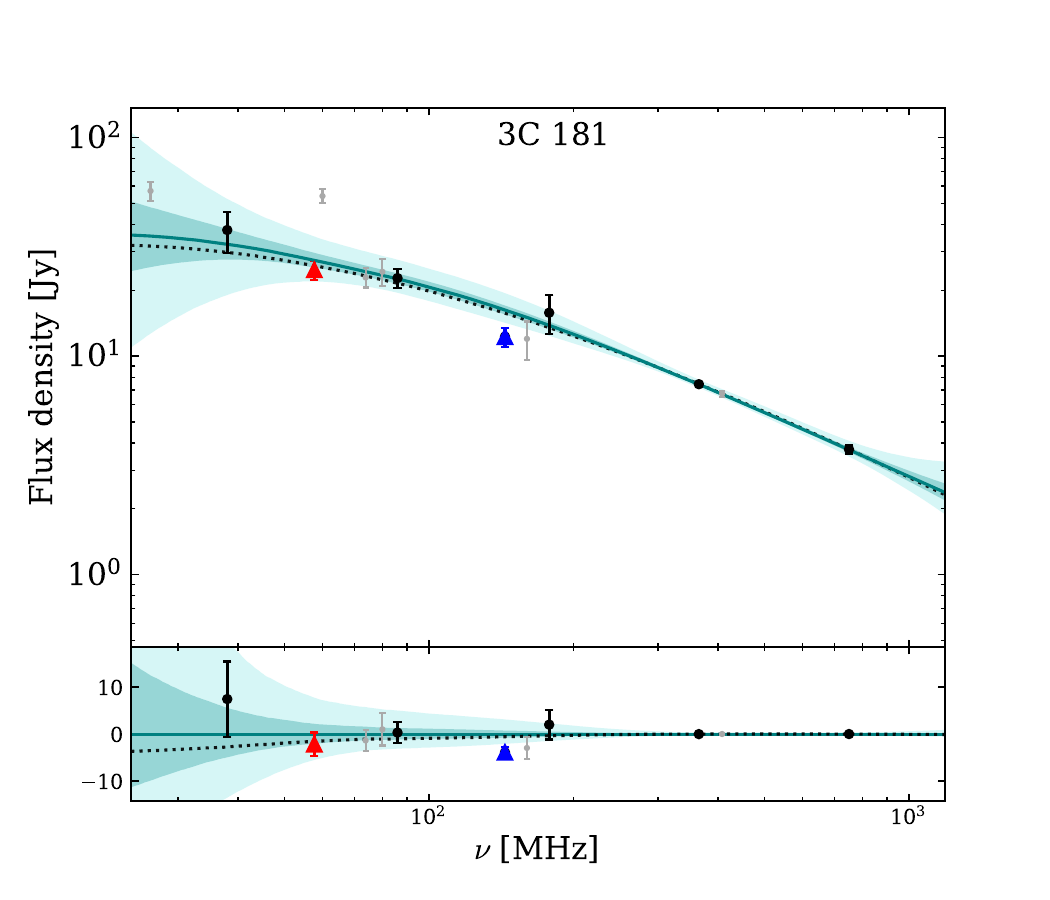}
\includegraphics[width=0.162\linewidth, trim={0.cm .0cm 1.5cm 1.5cm},clip]{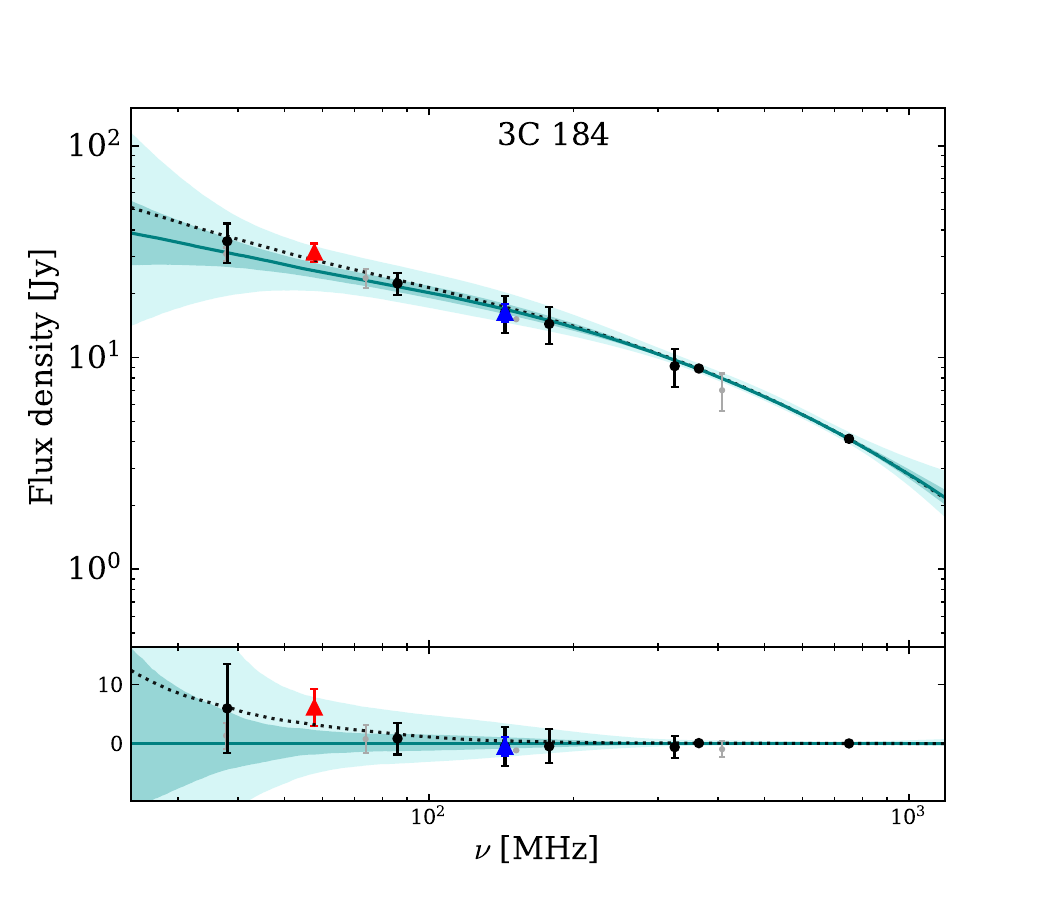}
\includegraphics[width=0.162\linewidth, trim={0.cm .0cm 1.5cm 1.5cm},clip]{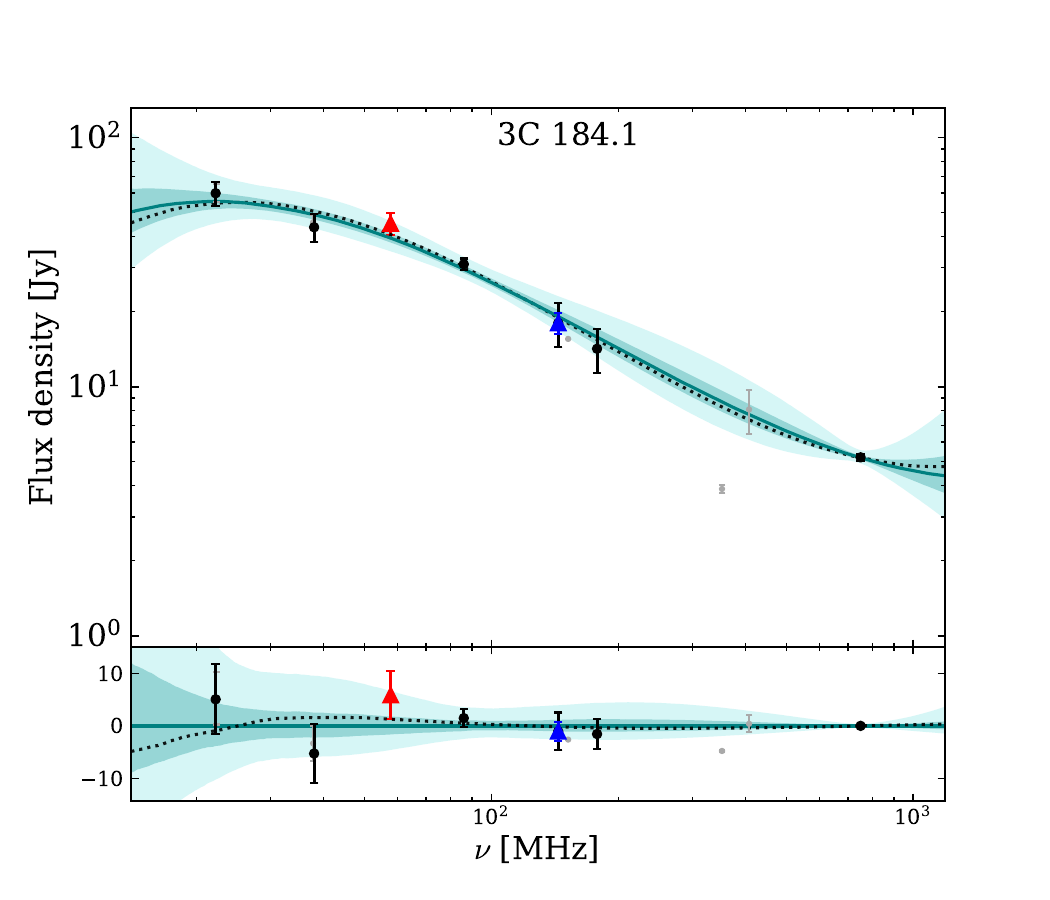}
\includegraphics[width=0.162\linewidth, trim={0.cm .0cm 1.5cm 1.5cm},clip]{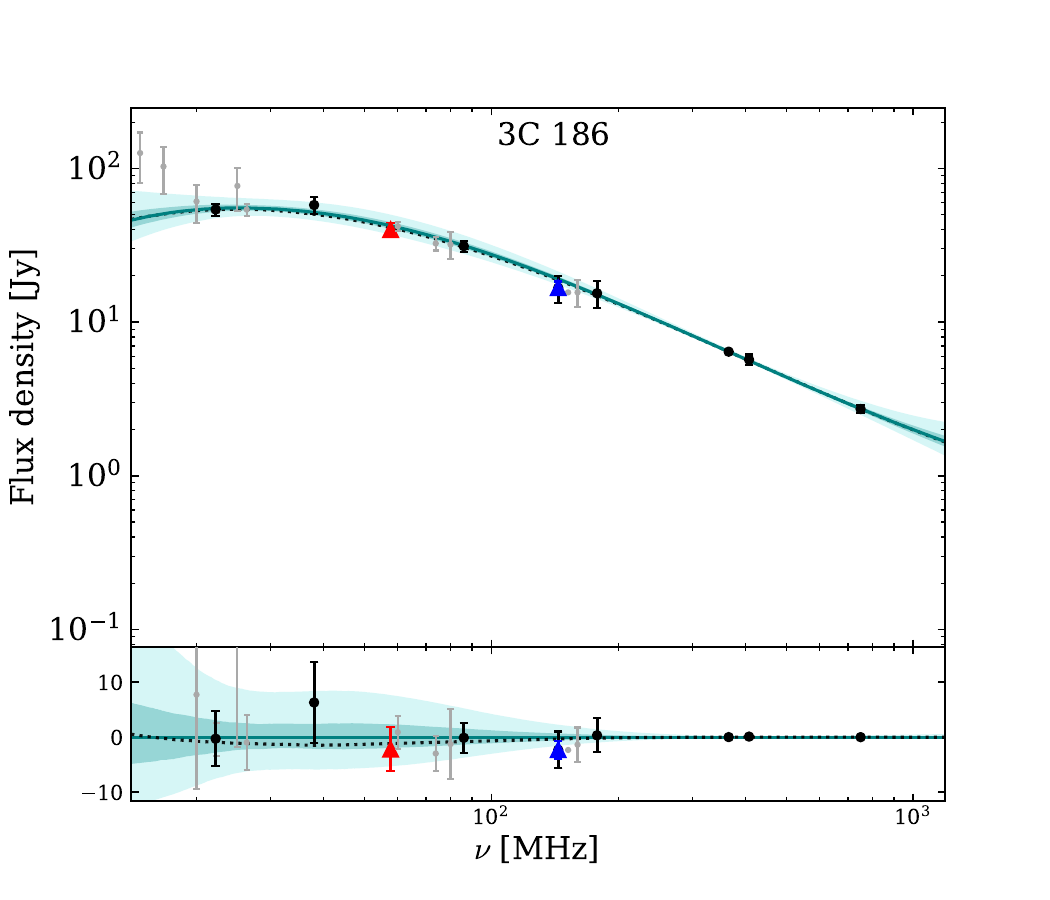}
\caption{Synchrotron spectrum for all sources in the catalog. Legend is the same as Figure~\ref{fig:calibrator_sed}.}
\end{figure}
\setcounter{figure}{0} 

\begin{figure}[H]
\centering
\includegraphics[width=0.162\linewidth, trim={0.cm .0cm 1.5cm 1.5cm},clip]{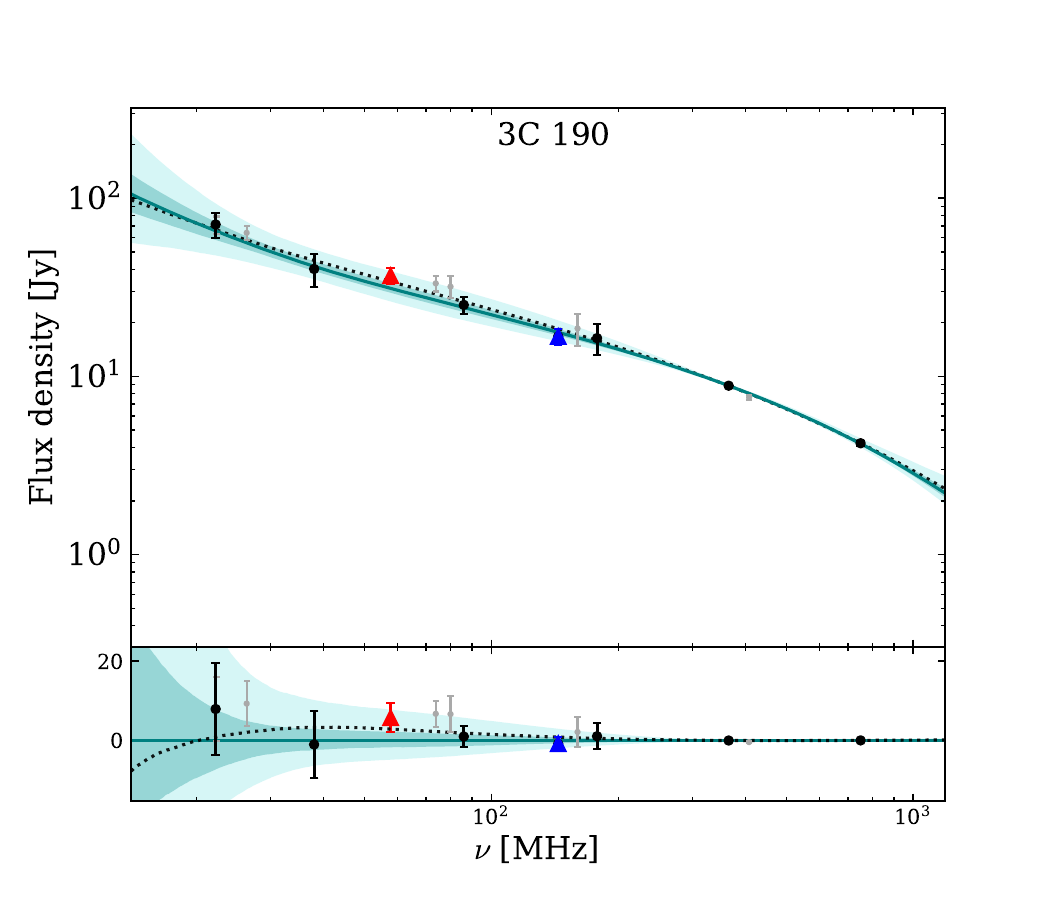}
\includegraphics[width=0.162\linewidth, trim={0.cm .0cm 1.5cm 1.5cm},clip]{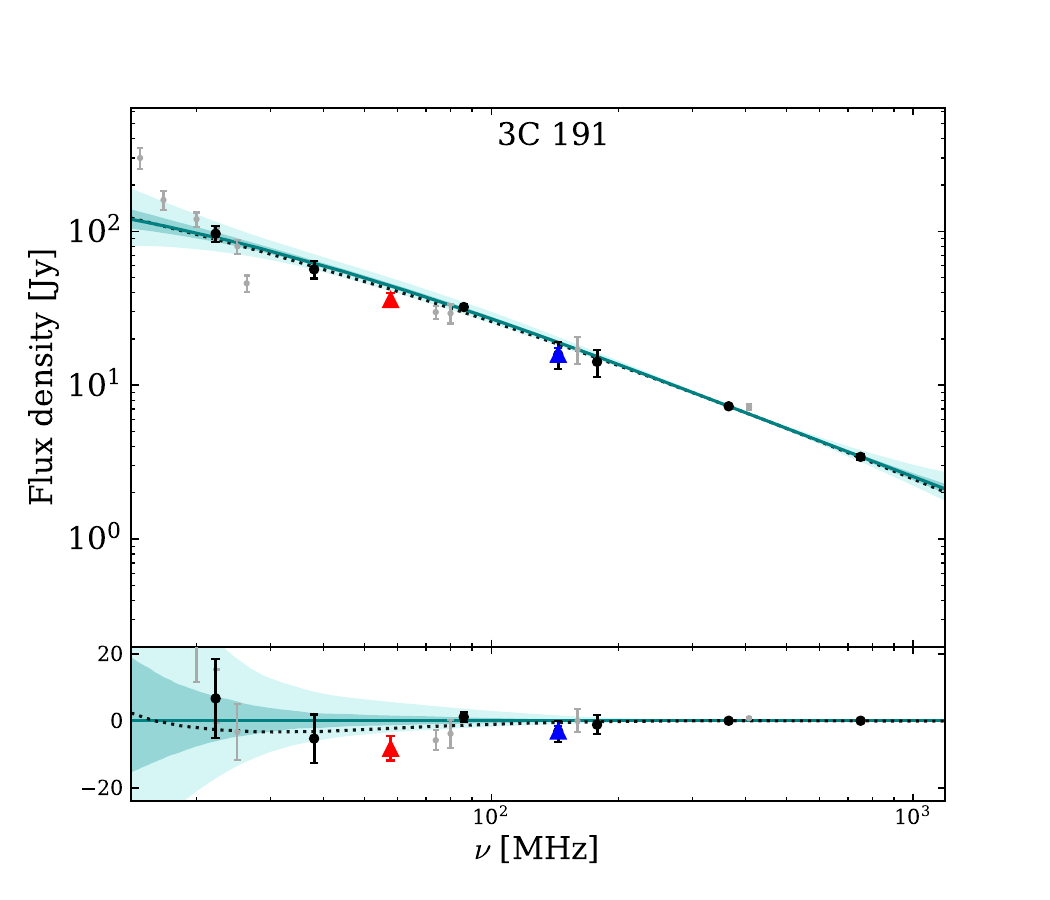}
\includegraphics[width=0.162\linewidth, trim={0.cm .0cm 1.5cm 1.5cm},clip]{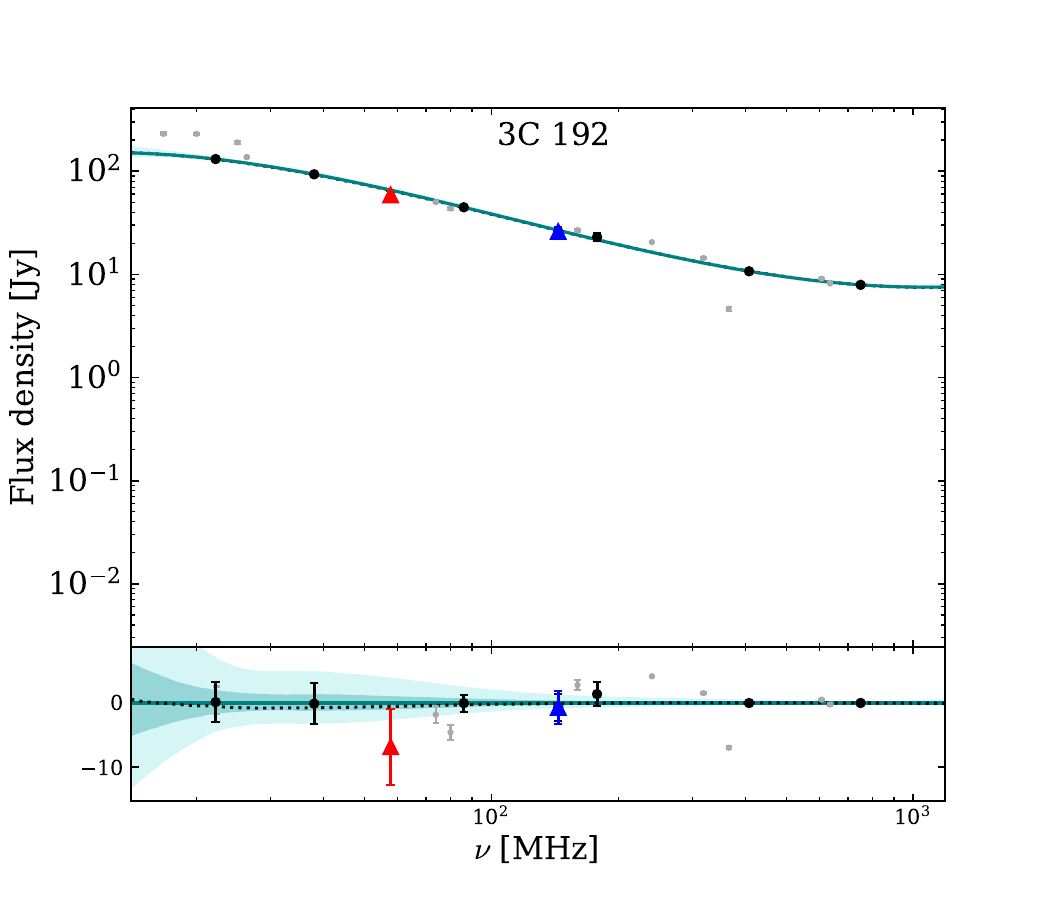}
\includegraphics[width=0.162\linewidth, trim={0.cm .0cm 1.5cm 1.5cm},clip]{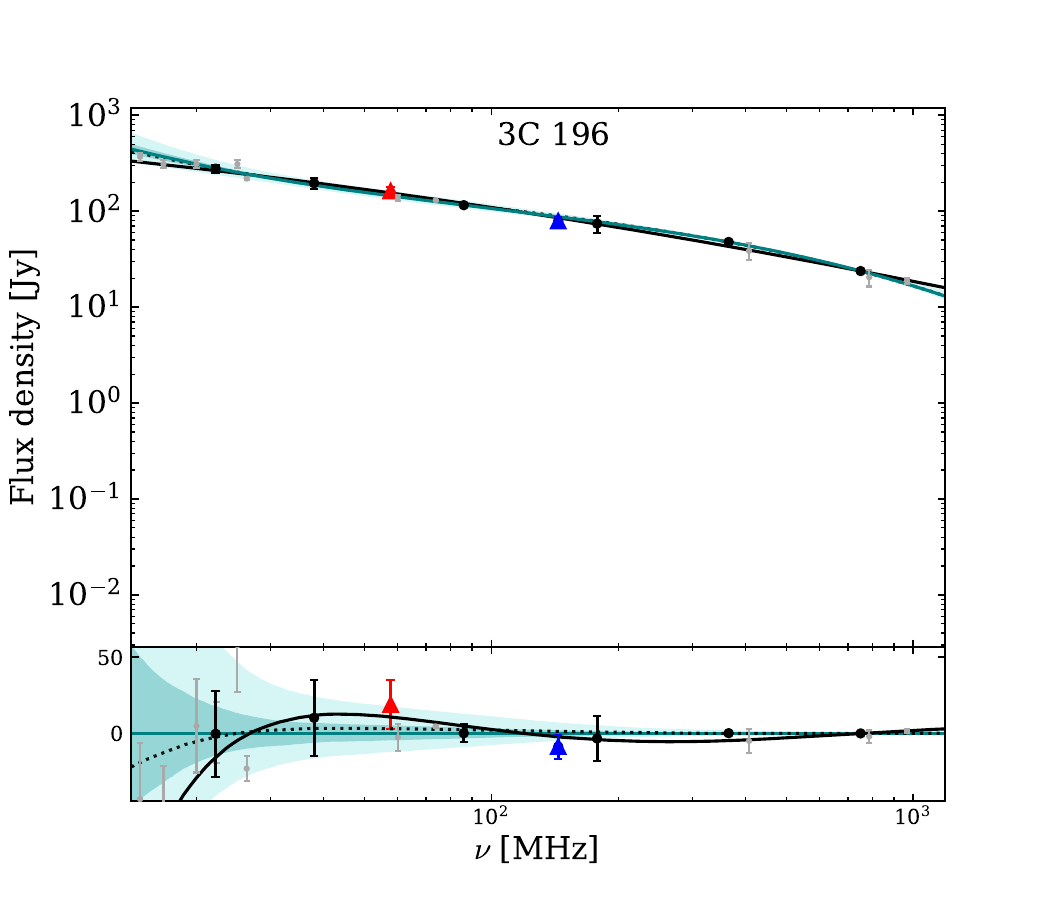}
\includegraphics[width=0.162\linewidth, trim={0.cm .0cm 1.5cm 1.5cm},clip]{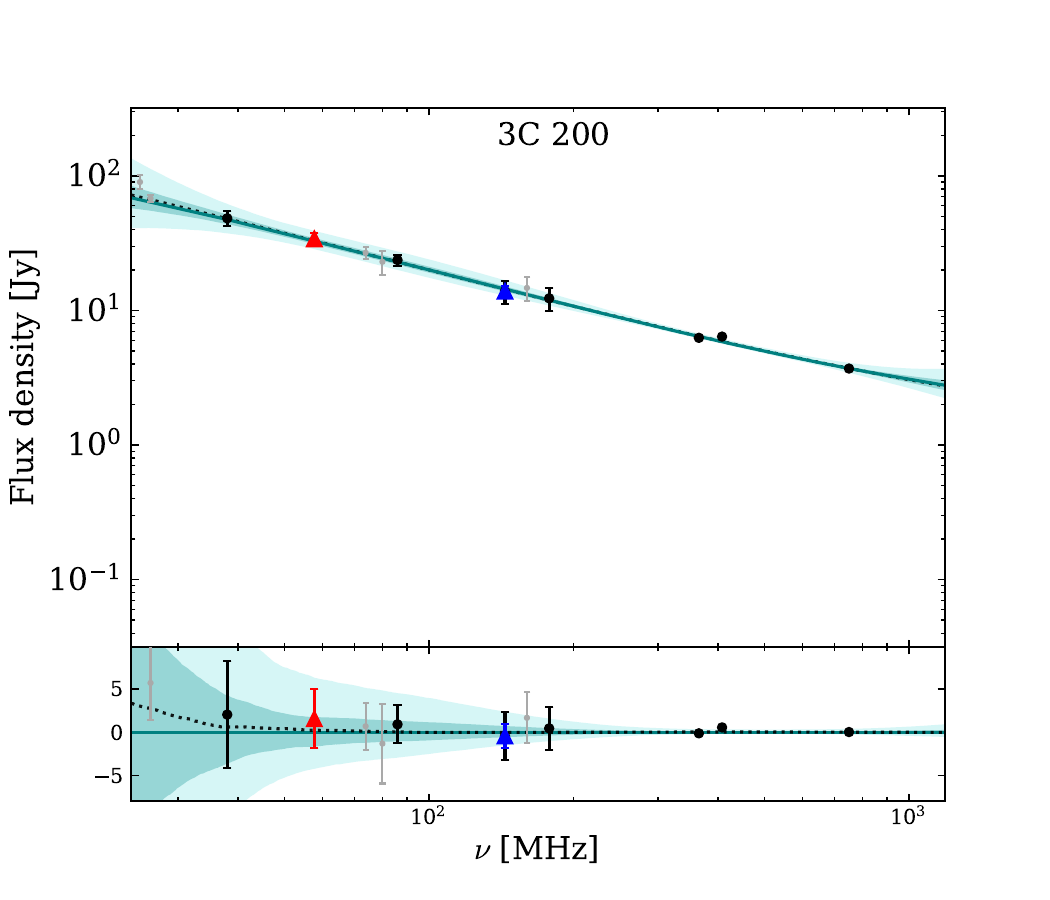}
\includegraphics[width=0.162\linewidth, trim={0.cm .0cm 1.5cm 1.5cm},clip]{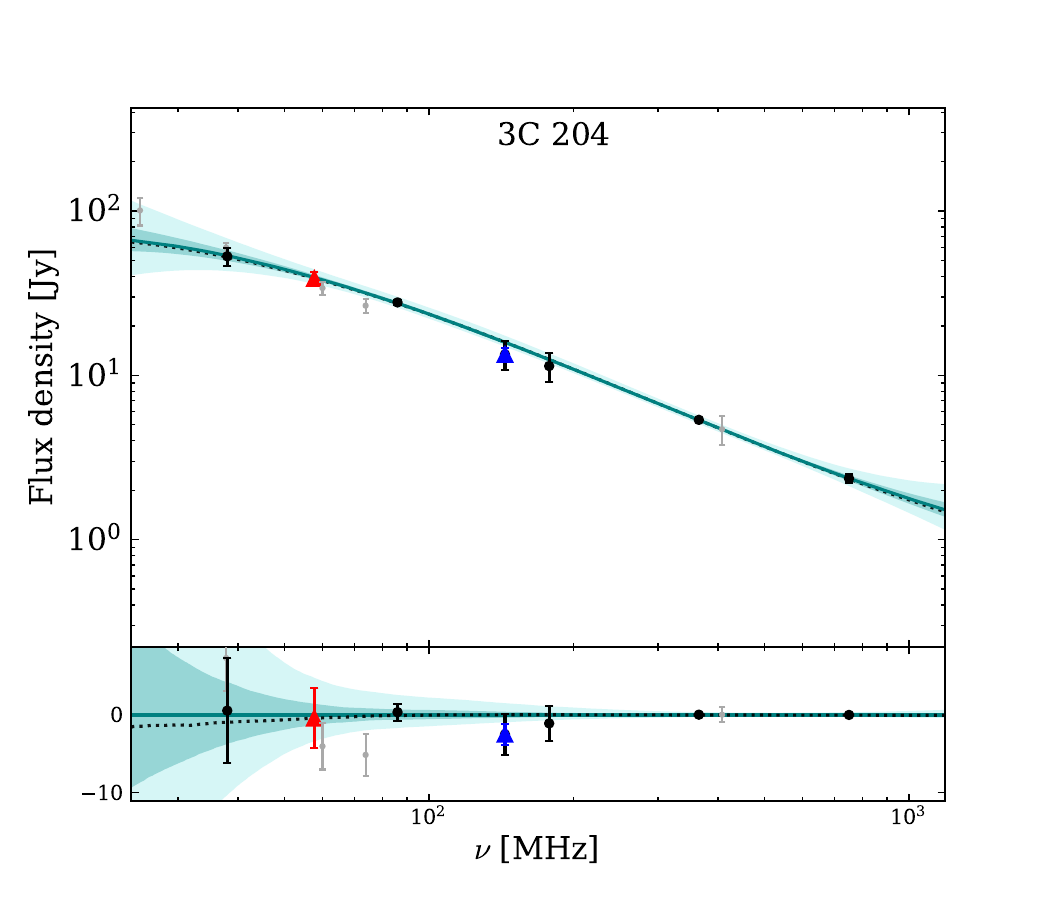}
\includegraphics[width=0.162\linewidth, trim={0.cm .0cm 1.5cm 1.5cm},clip]{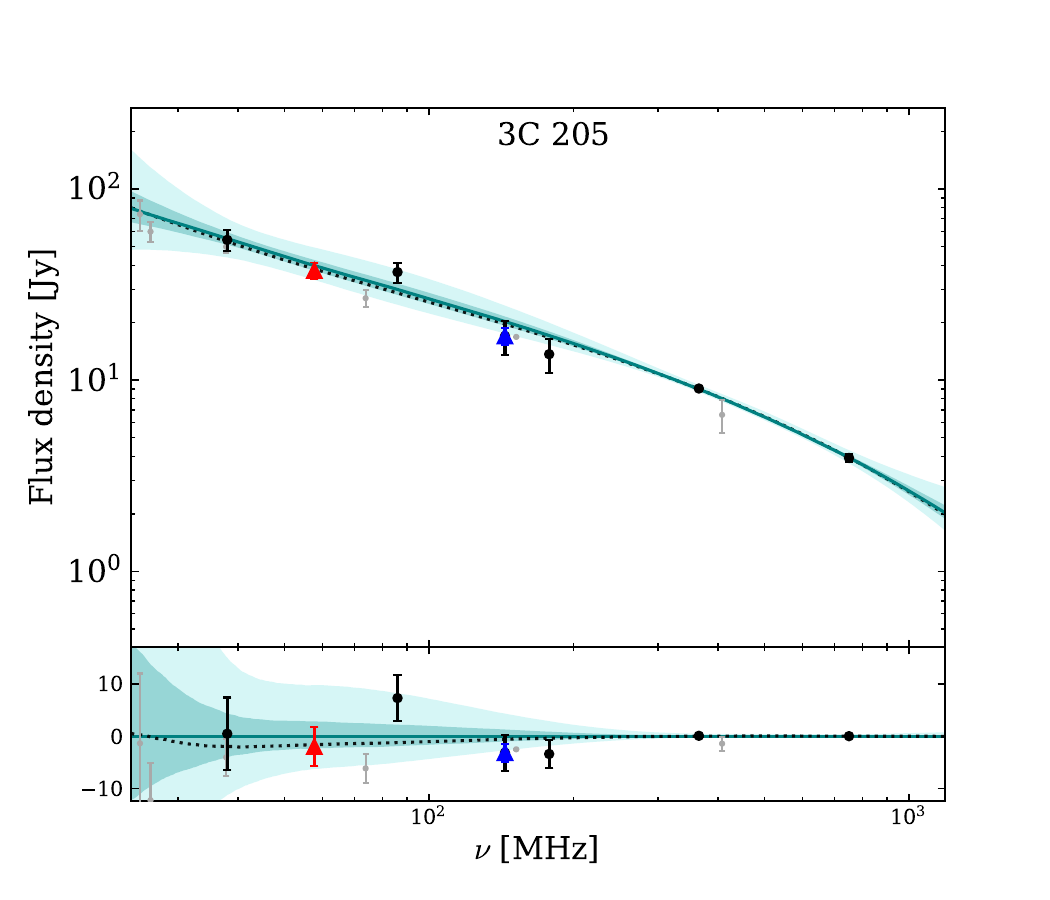}
\includegraphics[width=0.162\linewidth, trim={0.cm .0cm 1.5cm 1.5cm},clip]{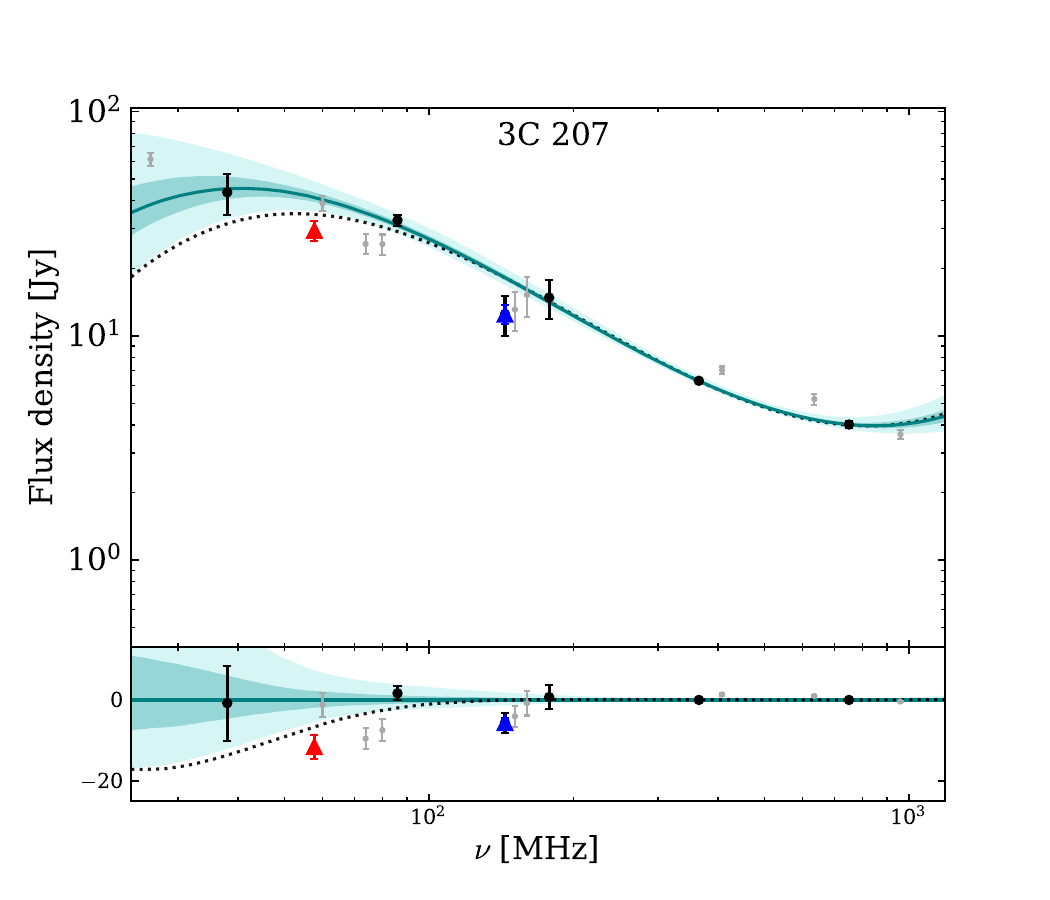}
\includegraphics[width=0.162\linewidth, trim={0.cm .0cm 1.5cm 1.5cm},clip]{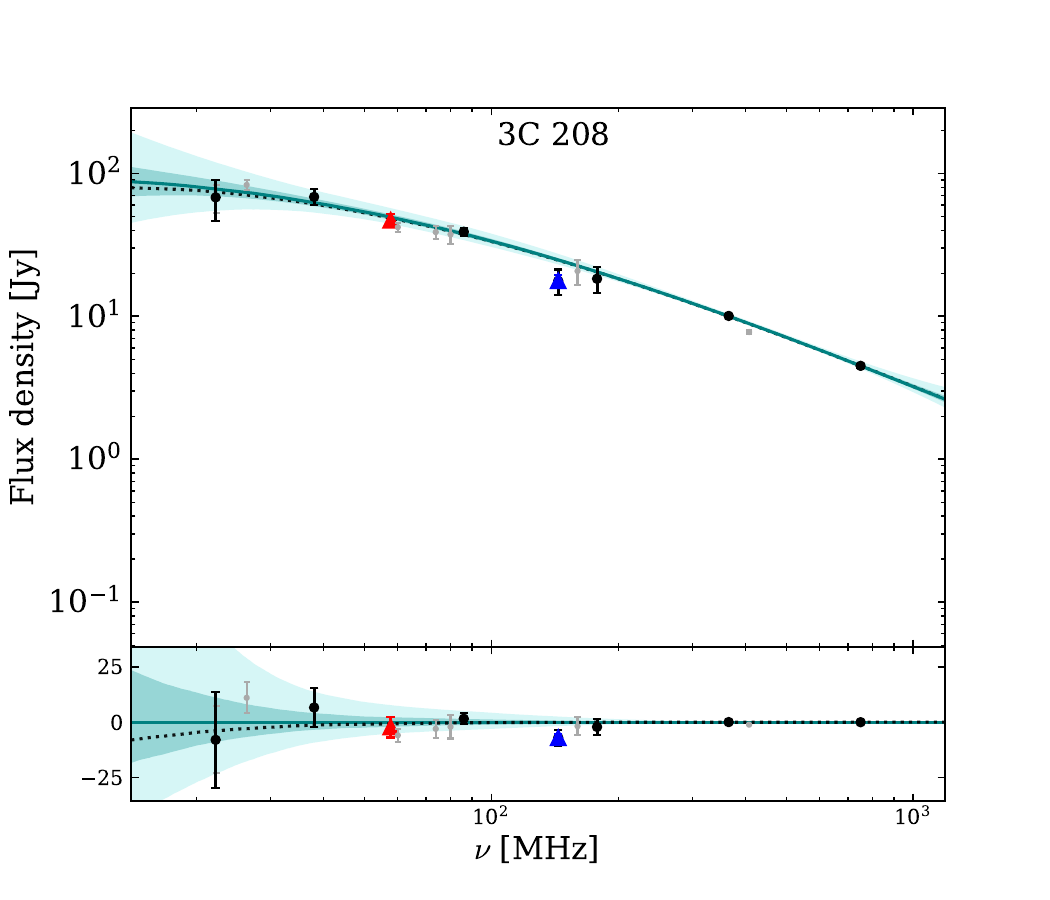}
\includegraphics[width=0.162\linewidth, trim={0.cm .0cm 1.5cm 1.5cm},clip]{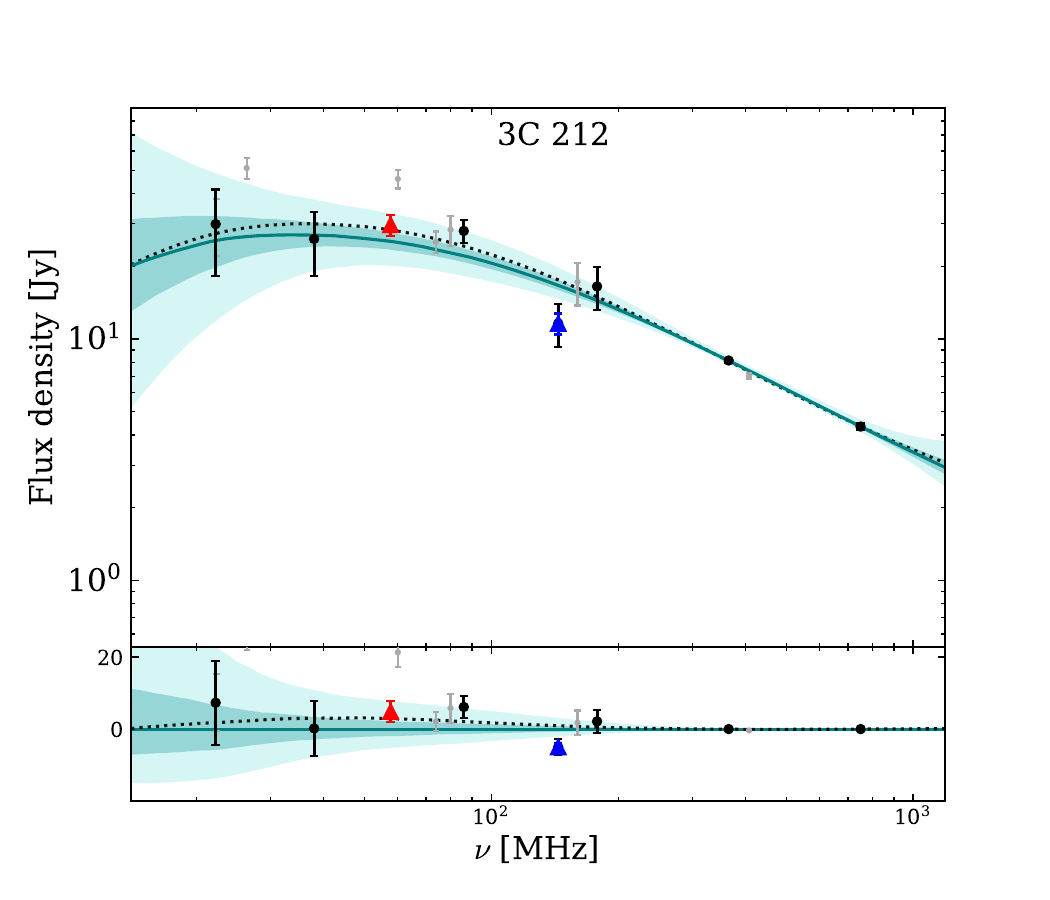}
\includegraphics[width=0.162\linewidth, trim={0.cm .0cm 1.5cm 1.5cm},clip]{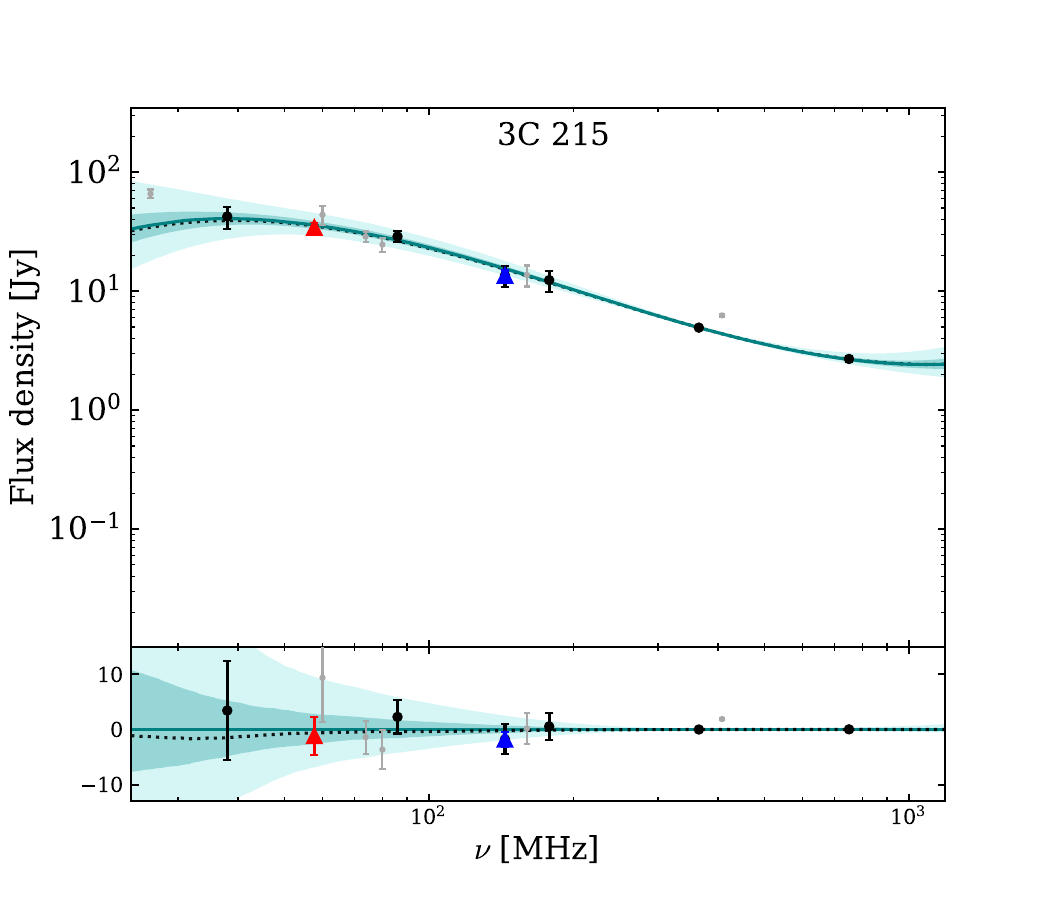}
\includegraphics[width=0.162\linewidth, trim={0.cm .0cm 1.5cm 1.5cm},clip]{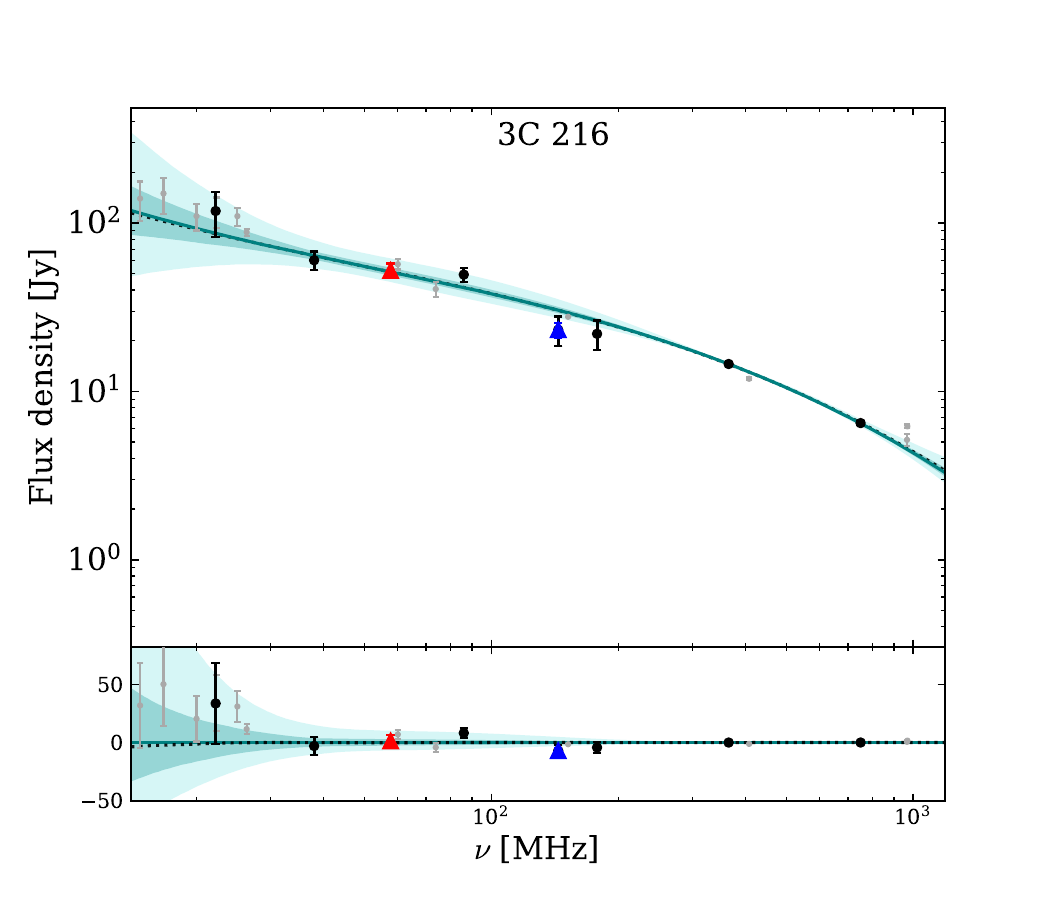}
\includegraphics[width=0.162\linewidth, trim={0.cm .0cm 1.5cm 1.5cm},clip]{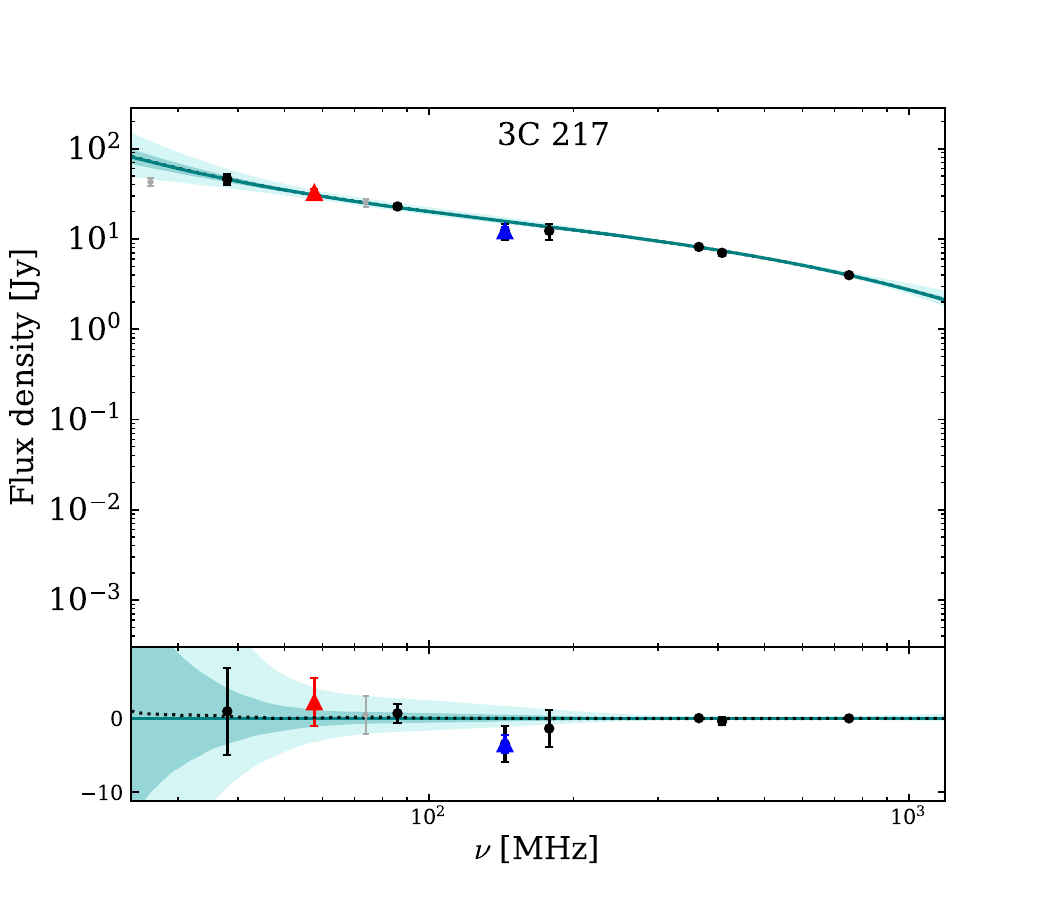}
\includegraphics[width=0.162\linewidth, trim={0.cm .0cm 1.5cm 1.5cm},clip]{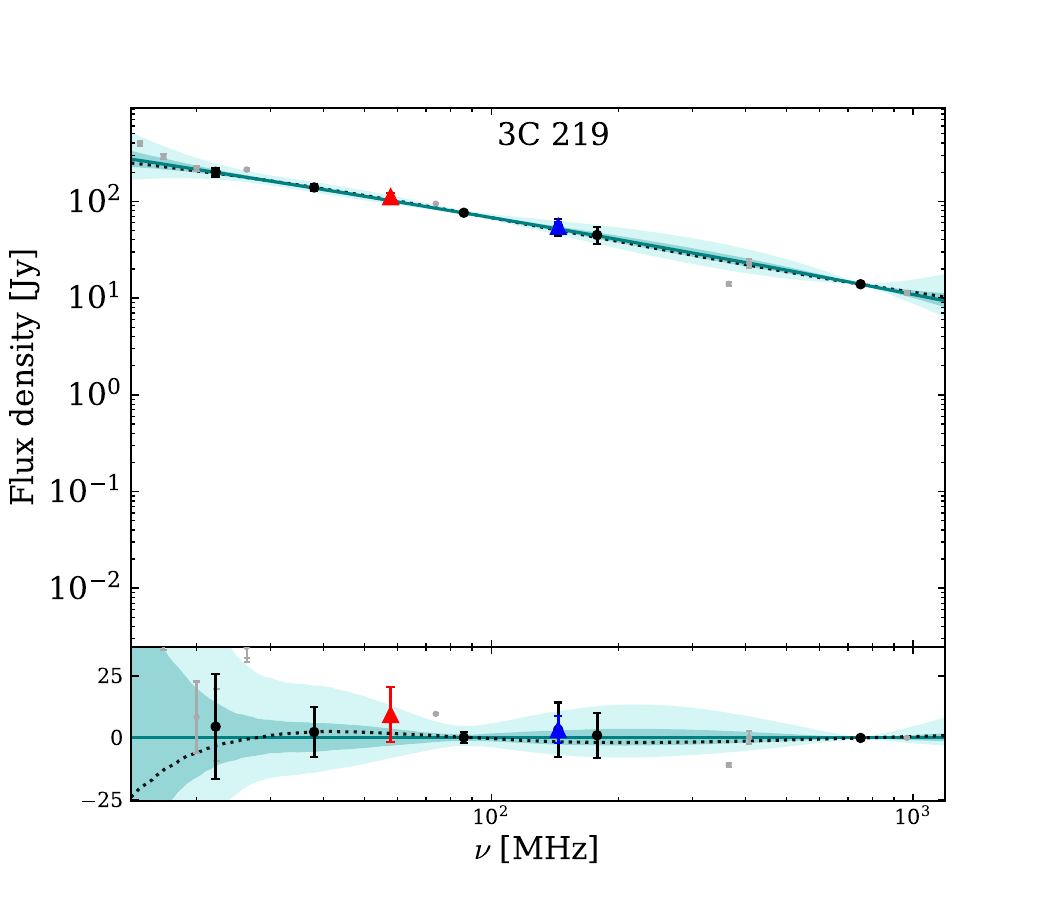}
\includegraphics[width=0.162\linewidth, trim={0.cm .0cm 1.5cm 1.5cm},clip]{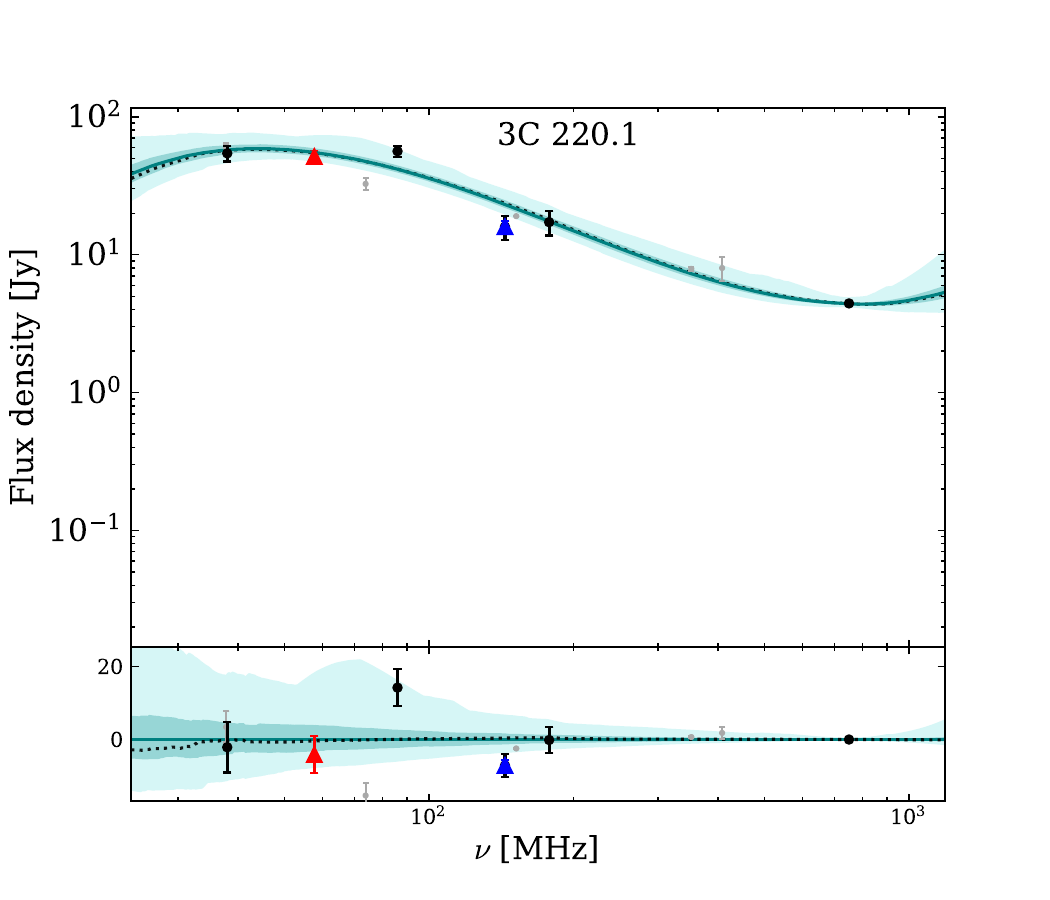}
\includegraphics[width=0.162\linewidth, trim={0.cm .0cm 1.5cm 1.5cm},clip]{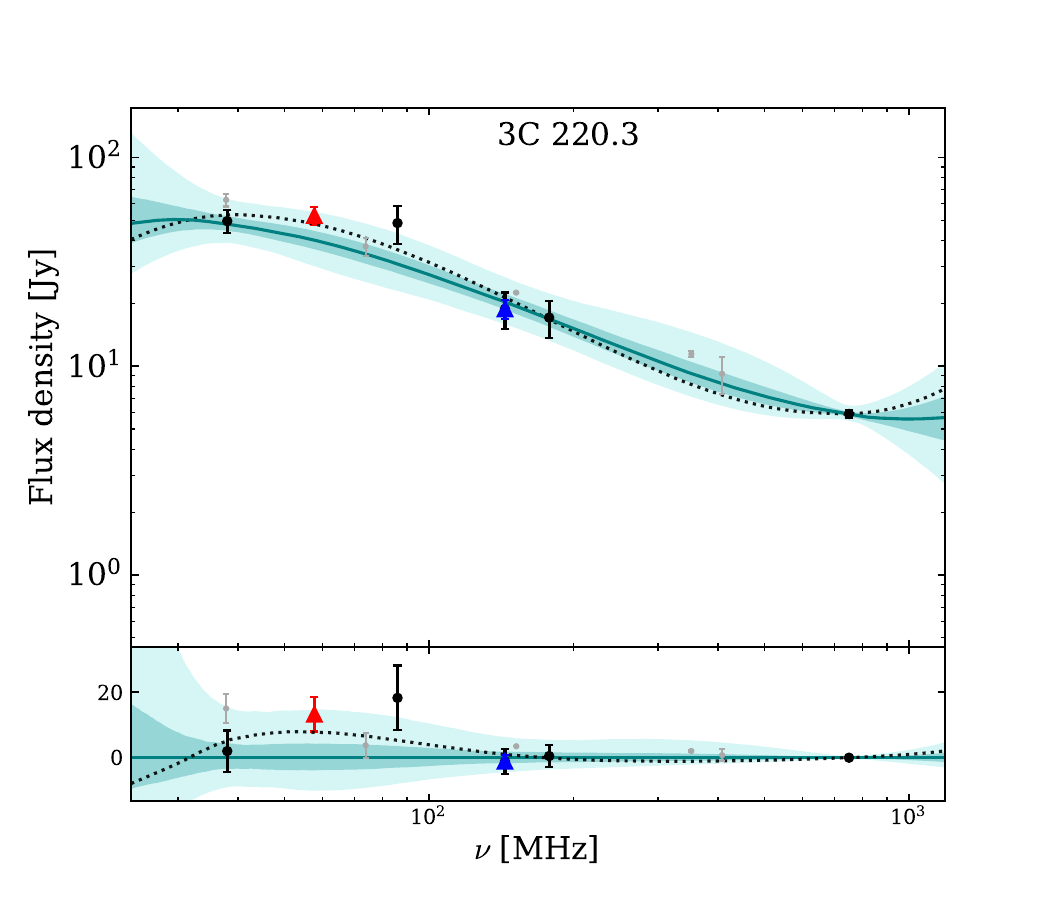}
\includegraphics[width=0.162\linewidth, trim={0.cm .0cm 1.5cm 1.5cm},clip]{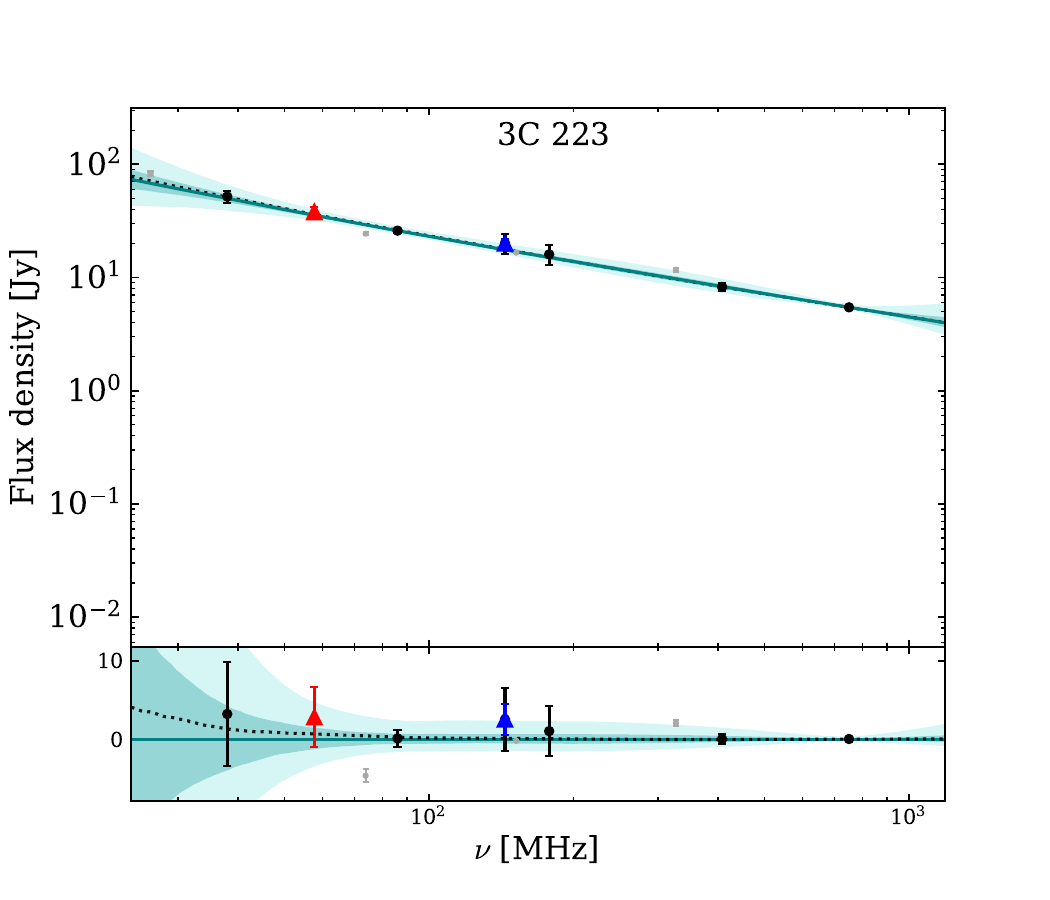}
\includegraphics[width=0.162\linewidth, trim={0.cm .0cm 1.5cm 1.5cm},clip]{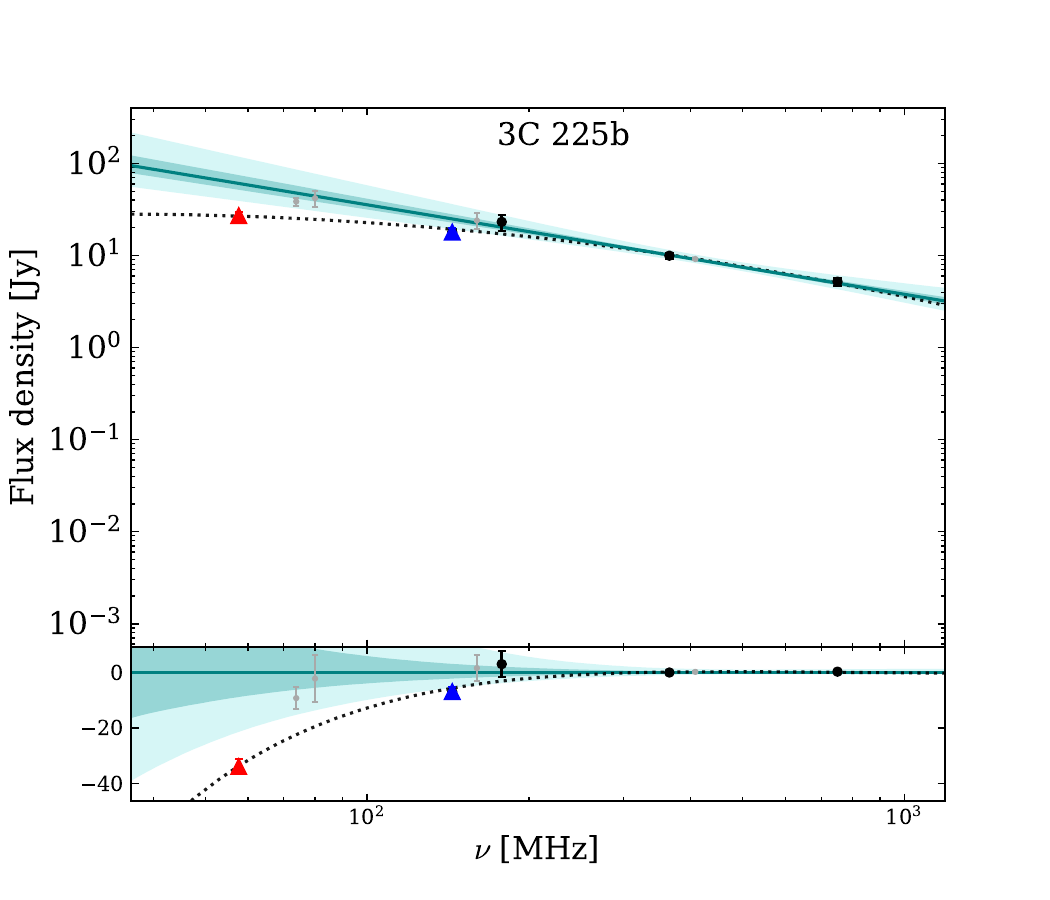}
\includegraphics[width=0.162\linewidth, trim={0.cm .0cm 1.5cm 1.5cm},clip]{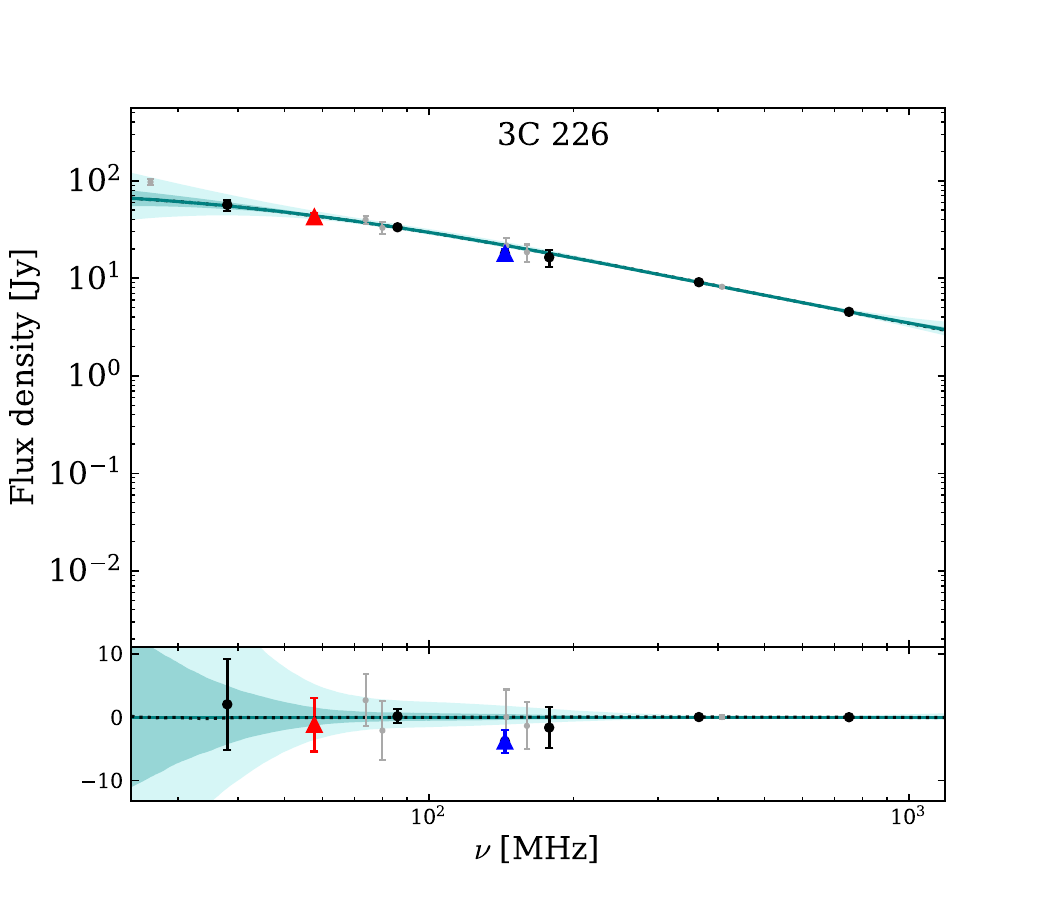}
\includegraphics[width=0.162\linewidth, trim={0.cm .0cm 1.5cm 1.5cm},clip]{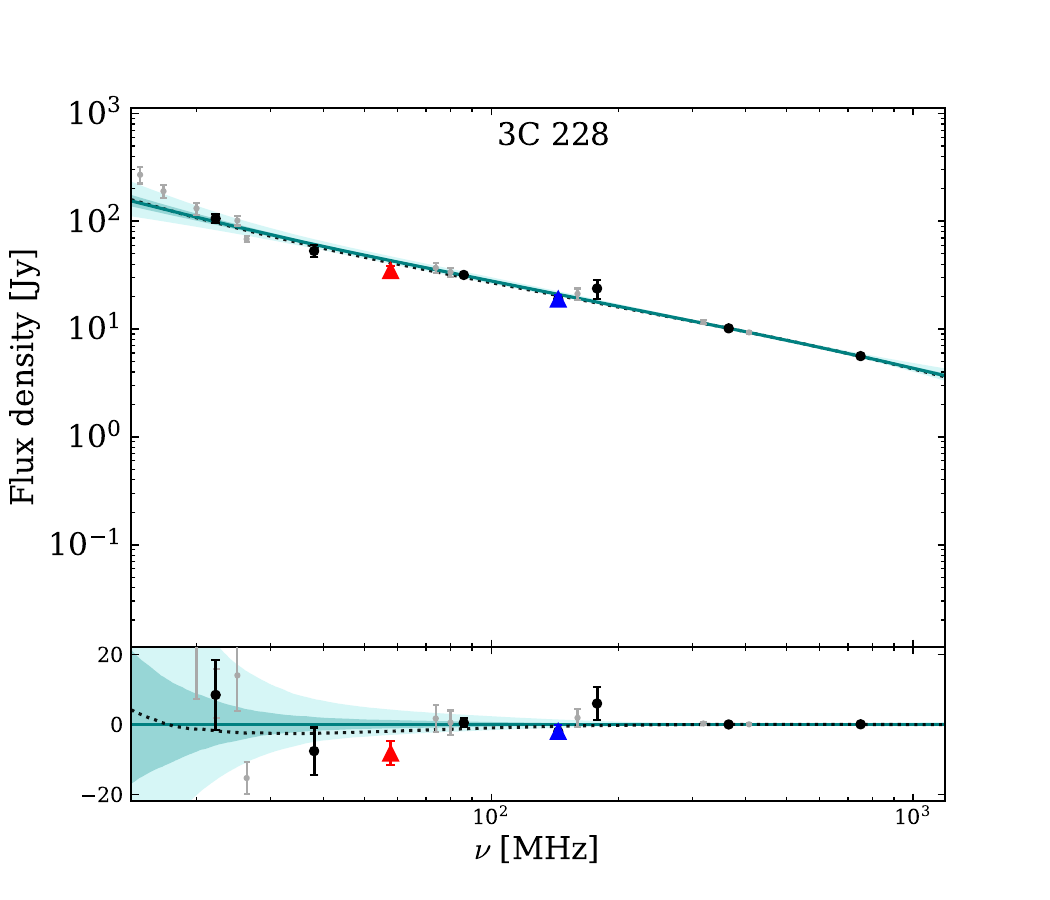}
\includegraphics[width=0.162\linewidth, trim={0.cm .0cm 1.5cm 1.5cm},clip]{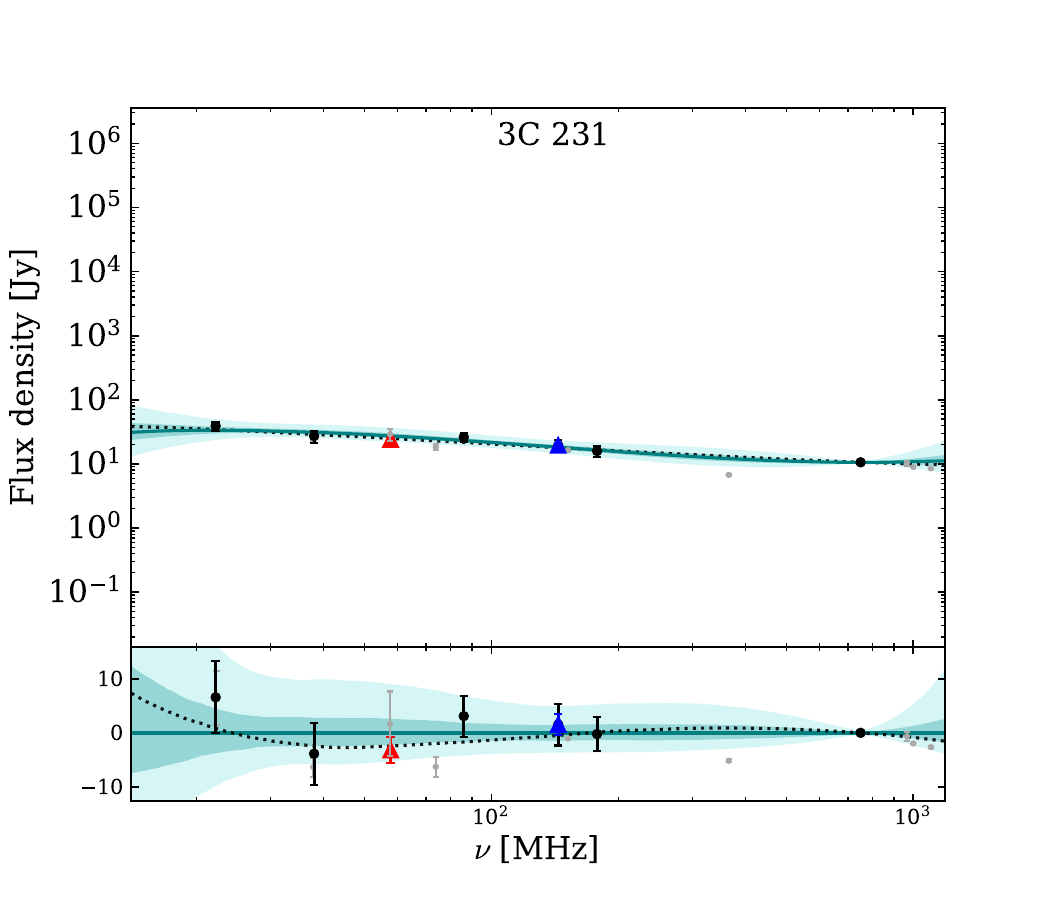}
\includegraphics[width=0.162\linewidth, trim={0.cm .0cm 1.5cm 1.5cm},clip]{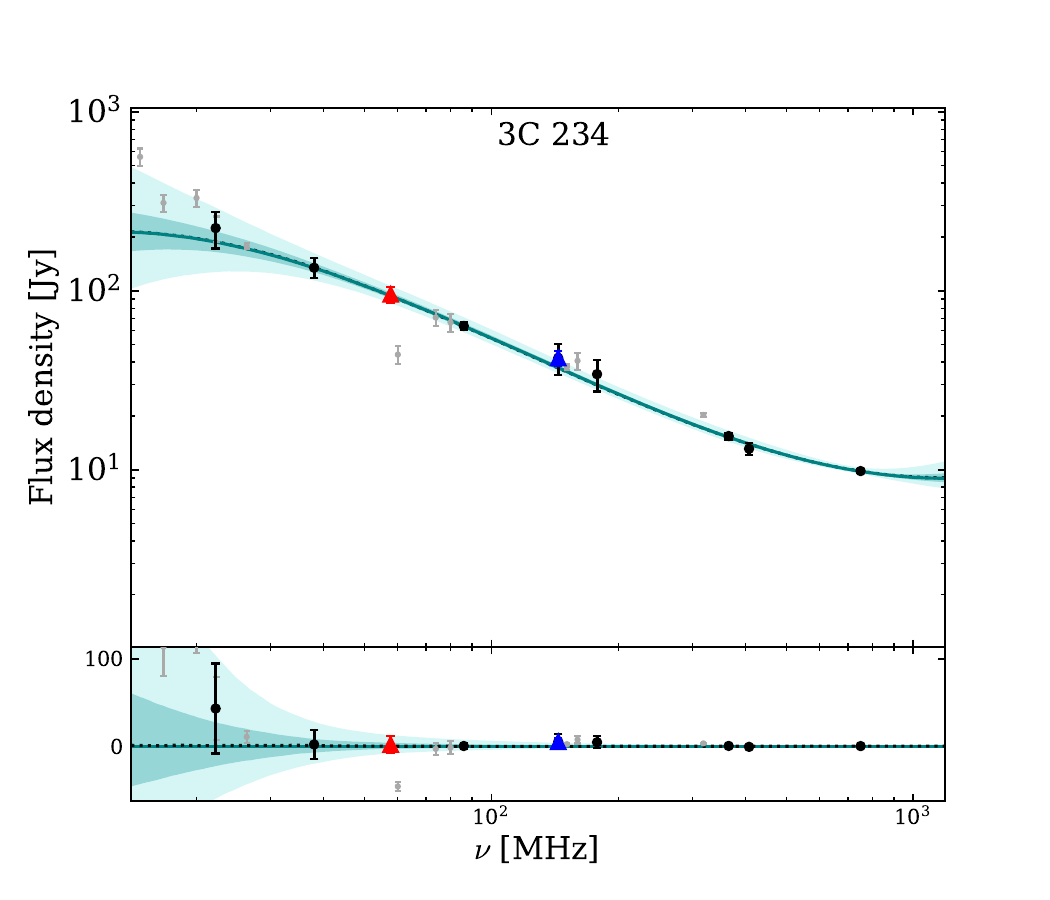}
\includegraphics[width=0.162\linewidth, trim={0.cm .0cm 1.5cm 1.5cm},clip]{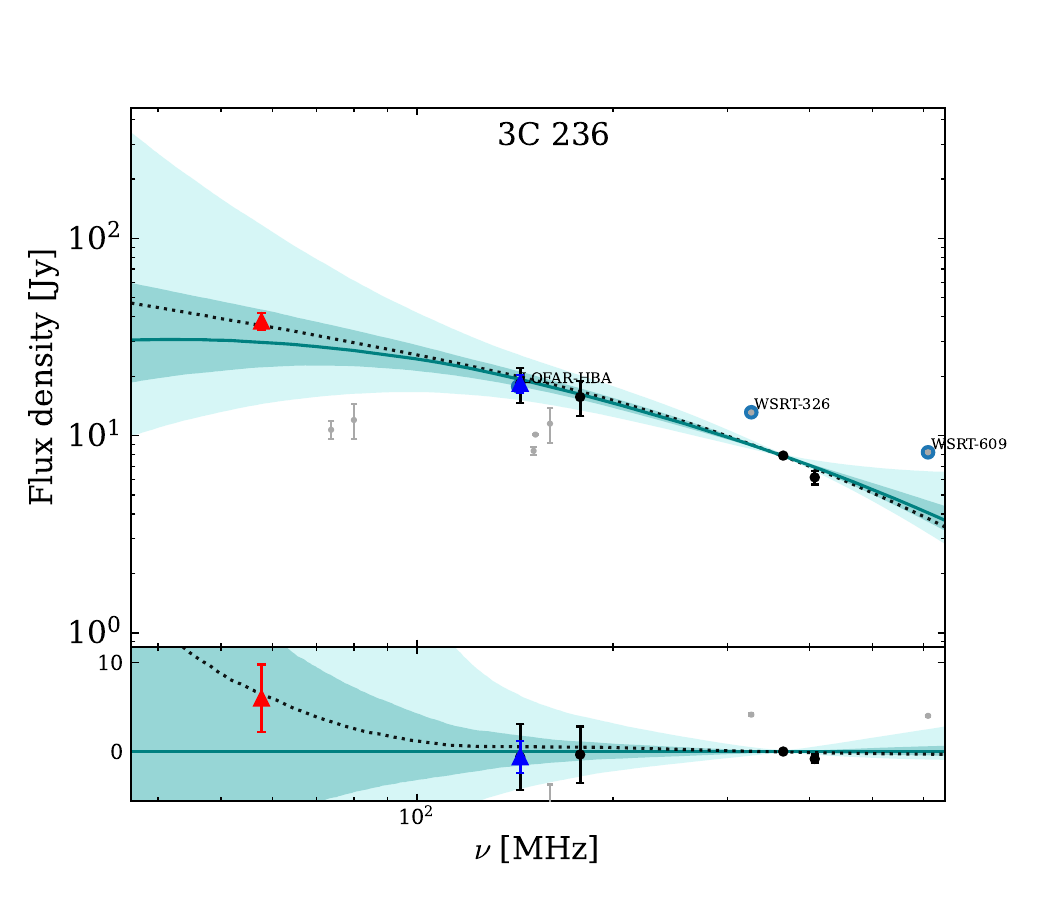}
\includegraphics[width=0.162\linewidth, trim={0.cm .0cm 1.5cm 1.5cm},clip]{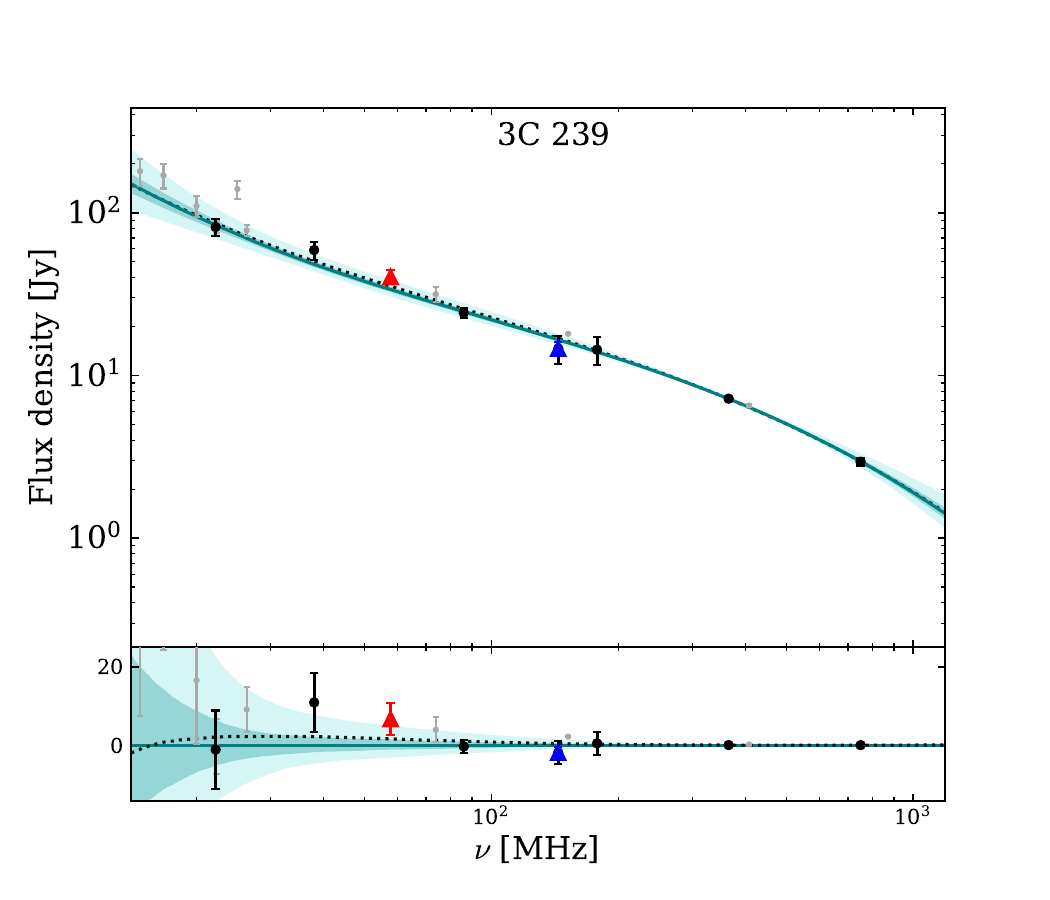}
\includegraphics[width=0.162\linewidth, trim={0.cm .0cm 1.5cm 1.5cm},clip]{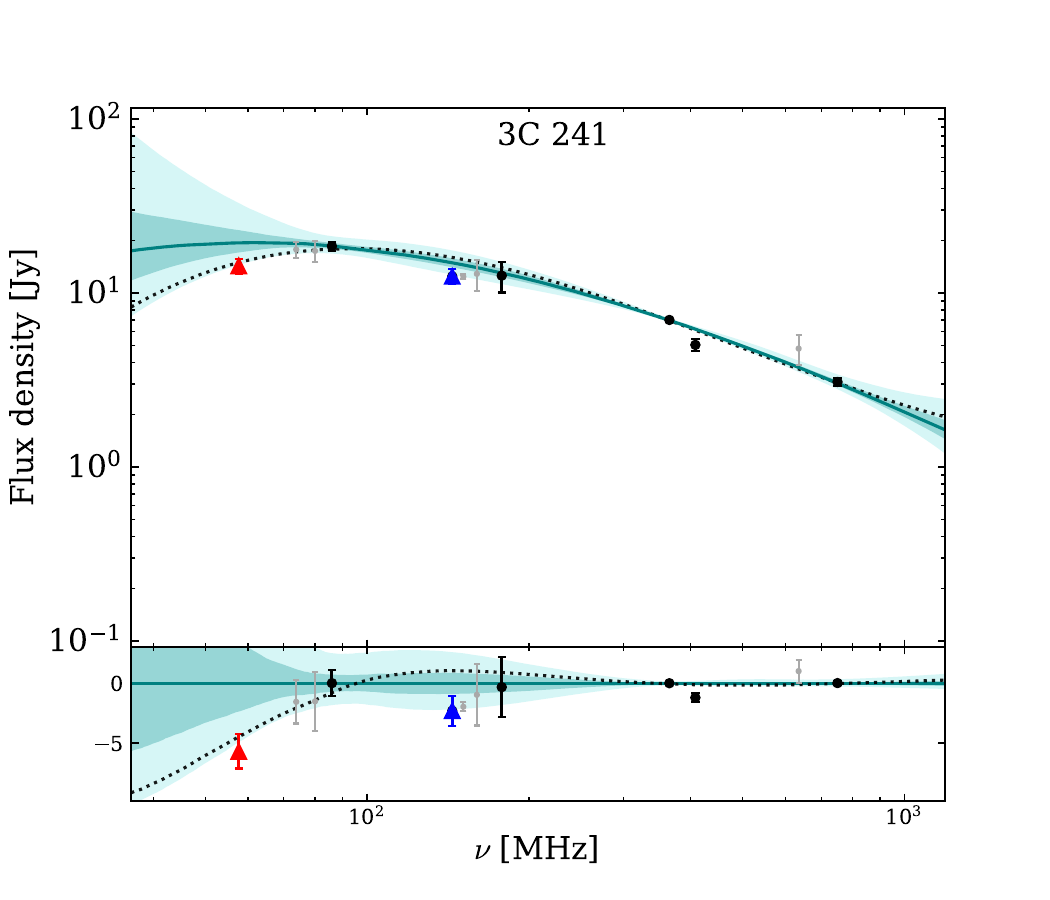}
\includegraphics[width=0.162\linewidth, trim={0.cm .0cm 1.5cm 1.5cm},clip]{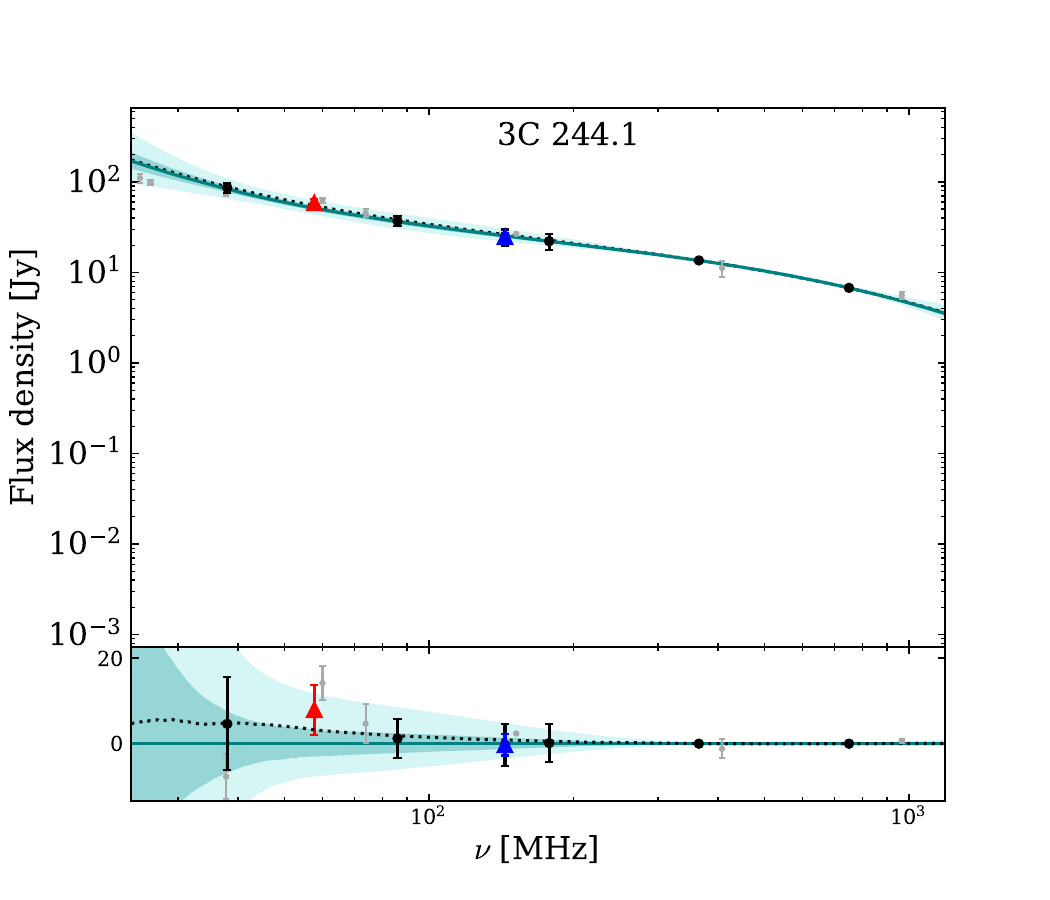}
\includegraphics[width=0.162\linewidth, trim={0.cm .0cm 1.5cm 1.5cm},clip]{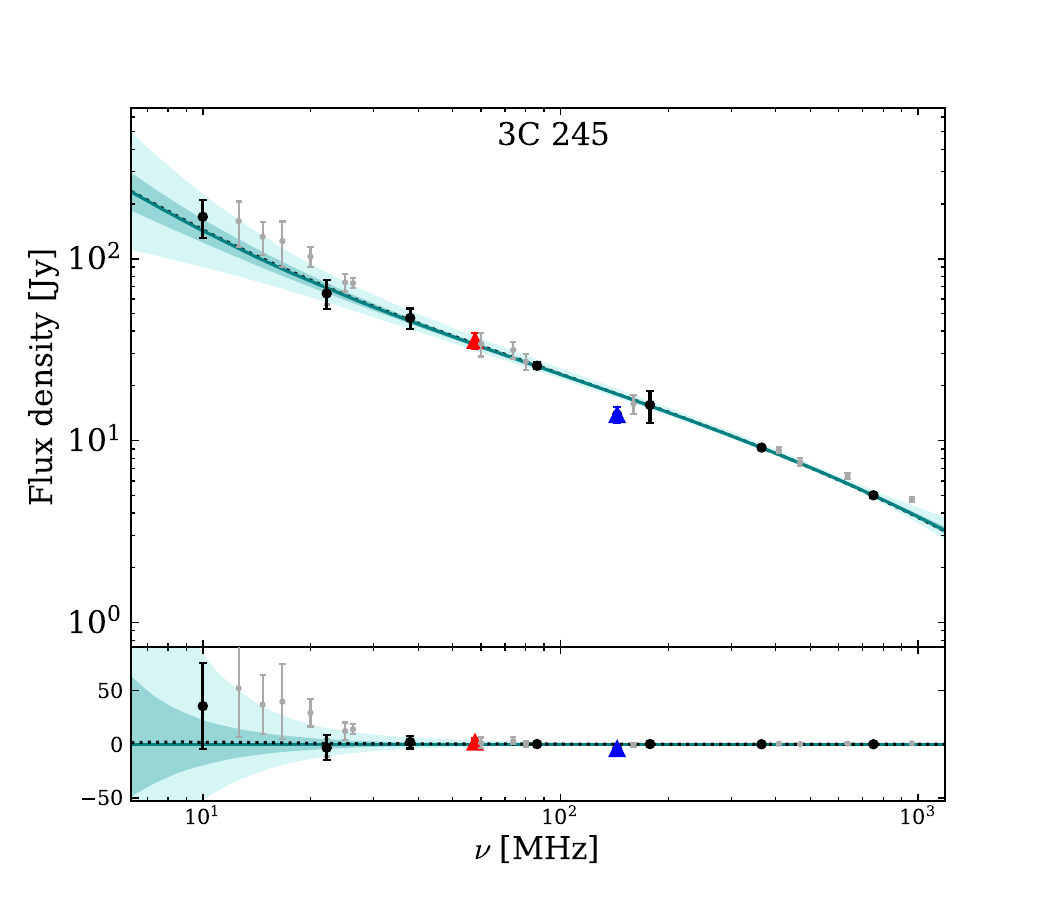}
\includegraphics[width=0.162\linewidth, trim={0.cm .0cm 1.5cm 1.5cm},clip]{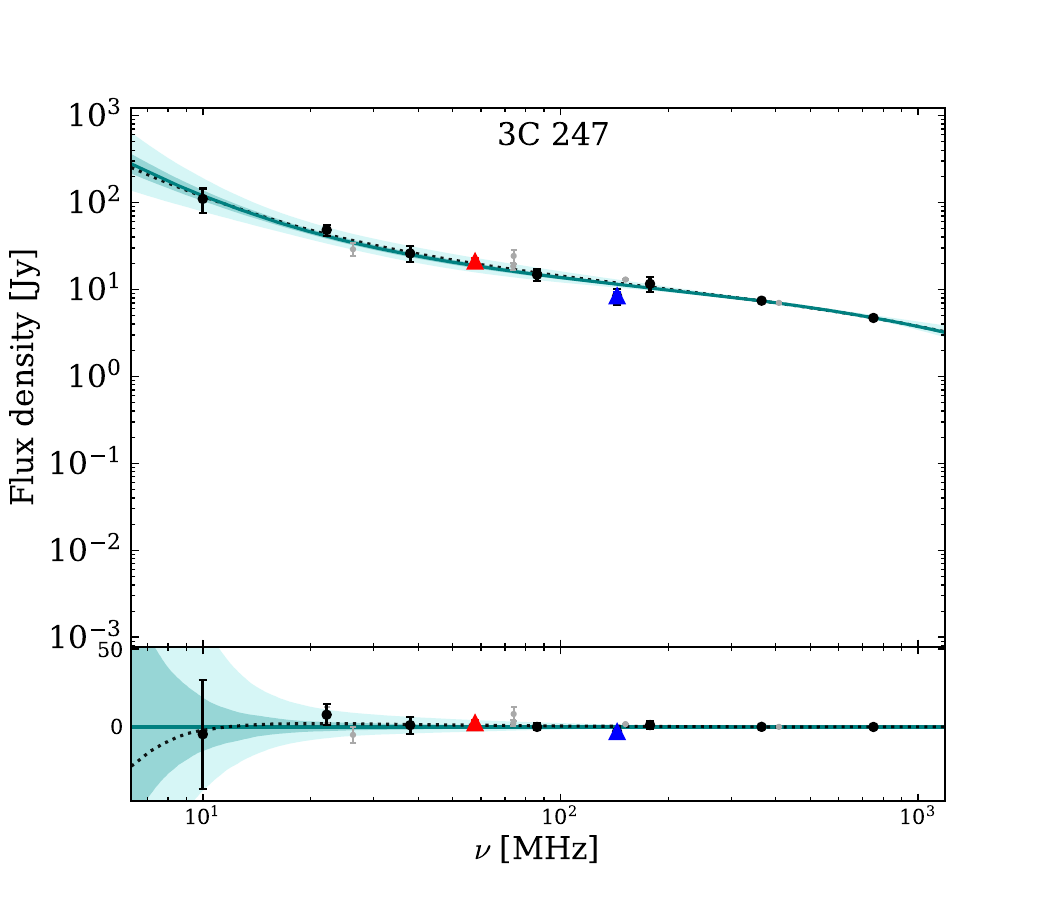}
\includegraphics[width=0.162\linewidth, trim={0.cm .0cm 1.5cm 1.5cm},clip]{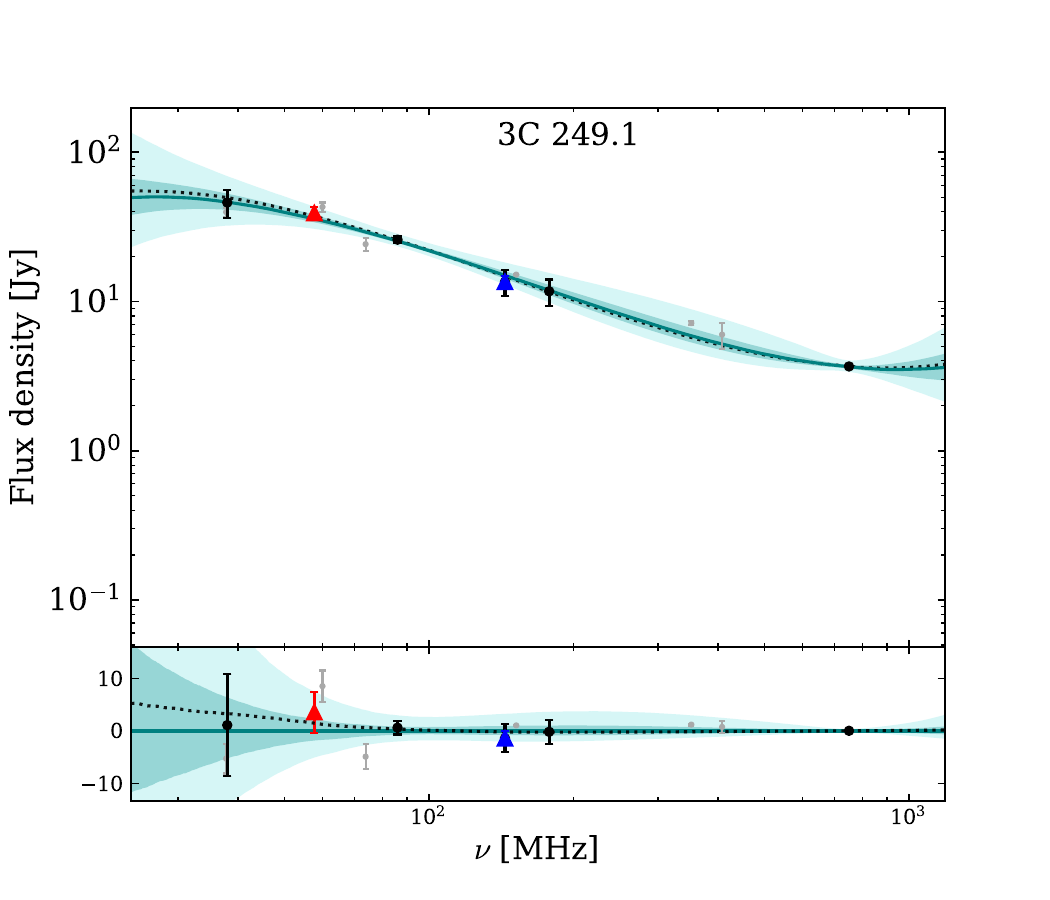}
\includegraphics[width=0.162\linewidth, trim={0.cm .0cm 1.5cm 1.5cm},clip]{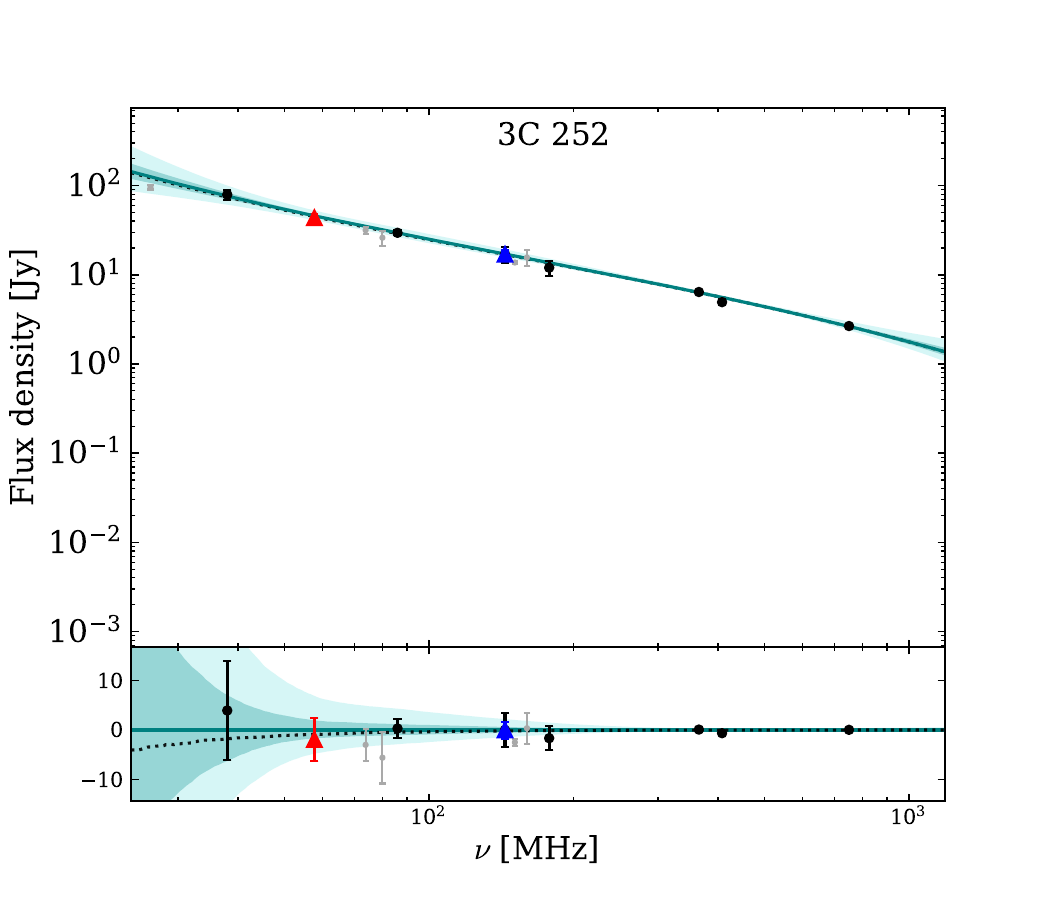}
\includegraphics[width=0.162\linewidth, trim={0.cm .0cm 1.5cm 1.5cm},clip]{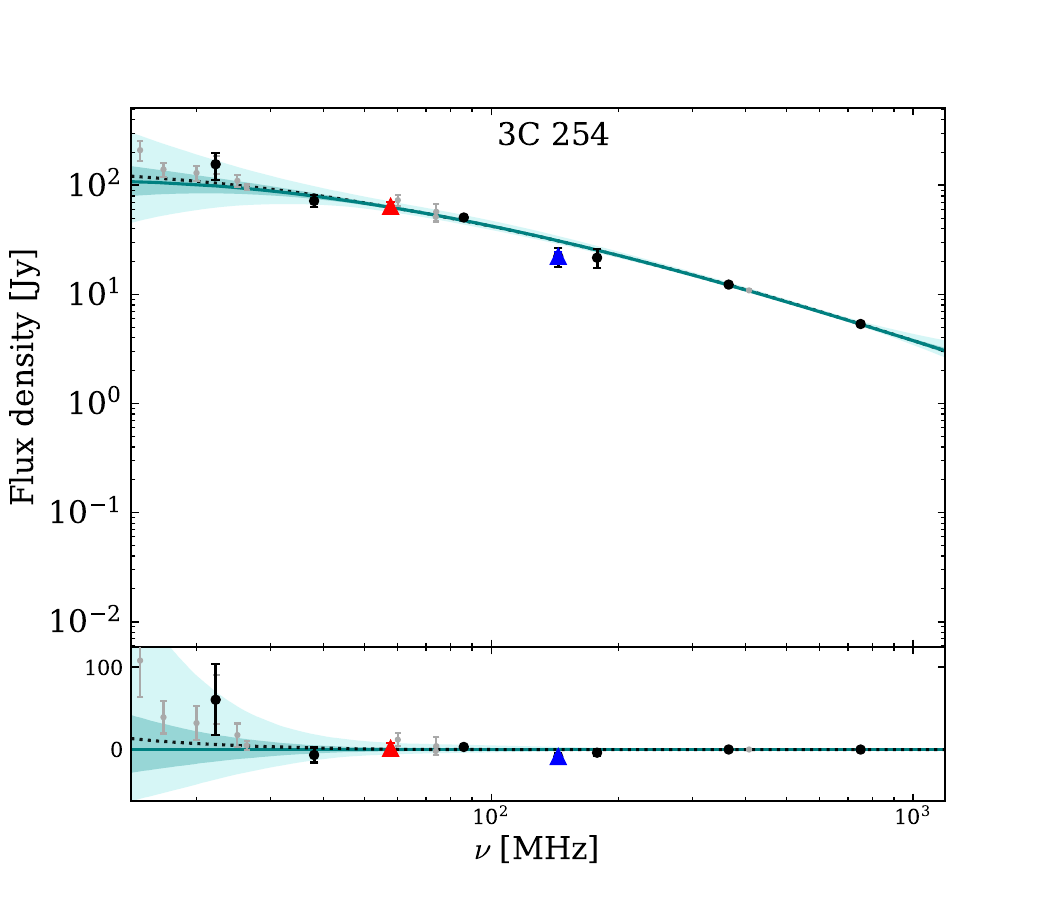}
\includegraphics[width=0.162\linewidth, trim={0.cm .0cm 1.5cm 1.5cm},clip]{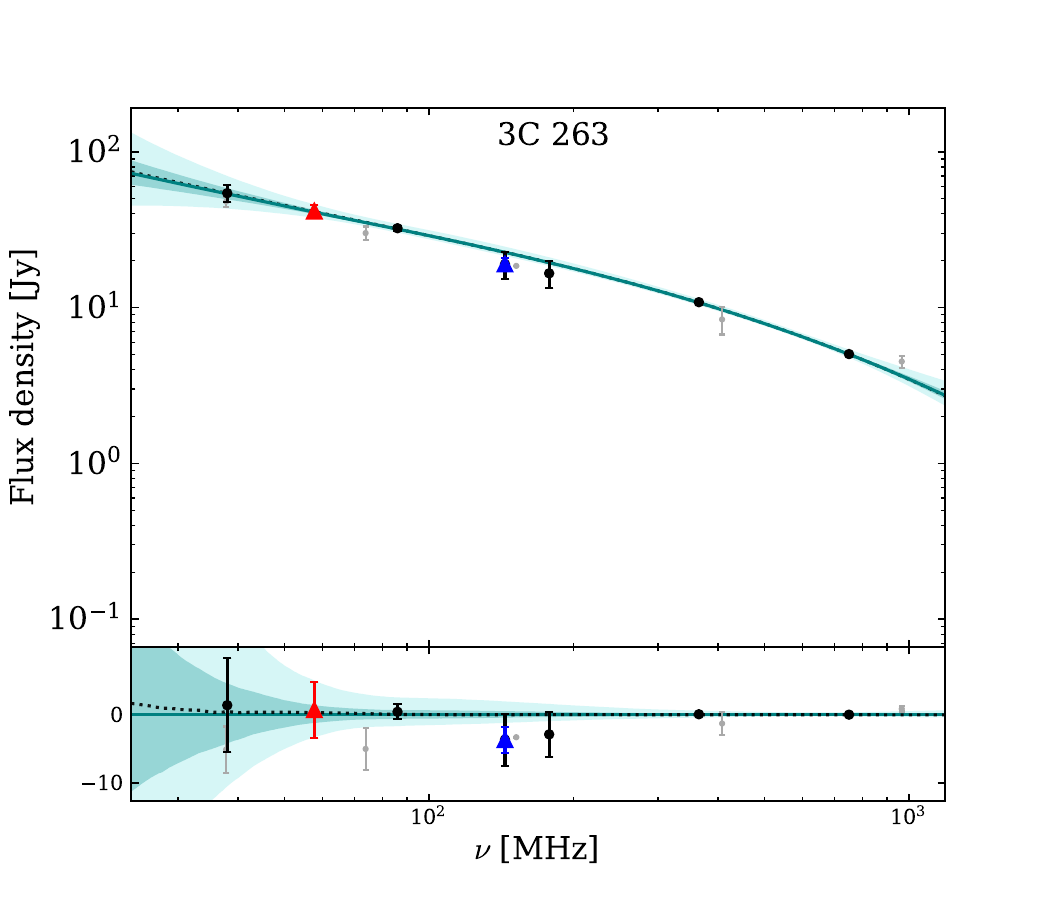}
\includegraphics[width=0.162\linewidth, trim={0.cm .0cm 1.5cm 1.5cm},clip]{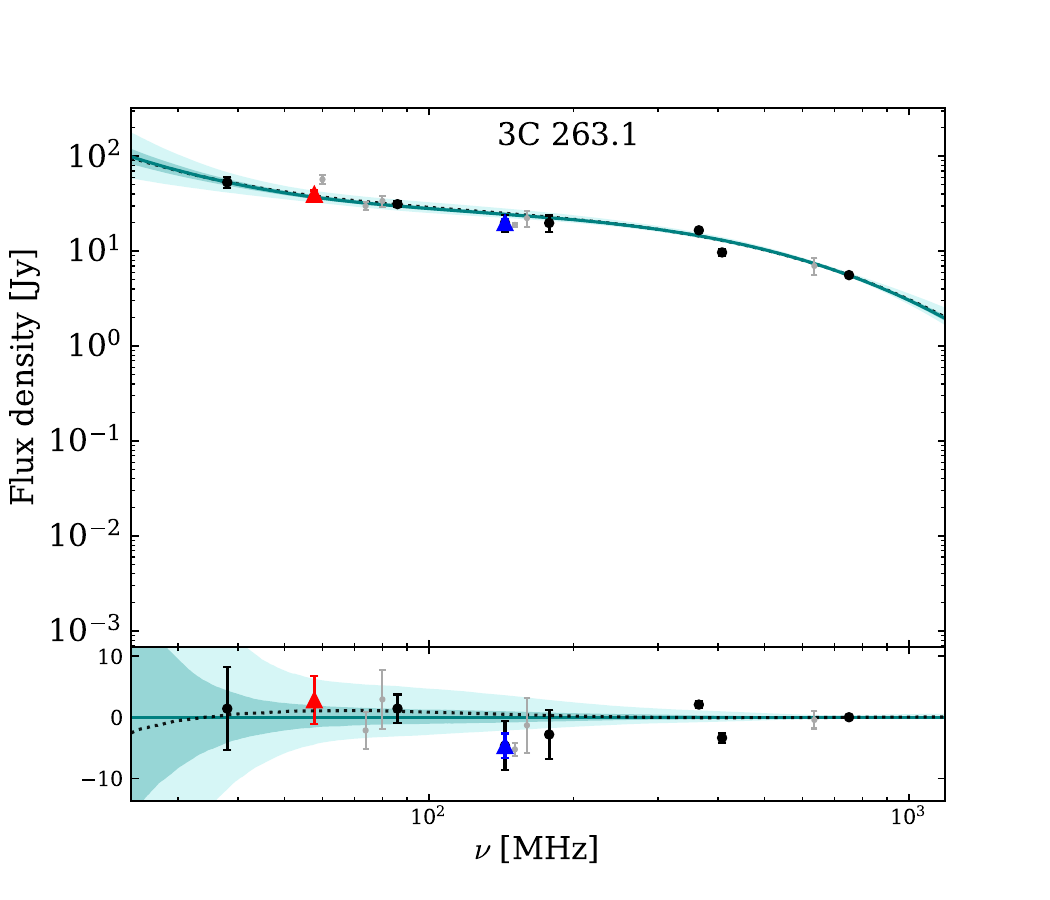}
\includegraphics[width=0.162\linewidth, trim={0.cm .0cm 1.5cm 1.5cm},clip]{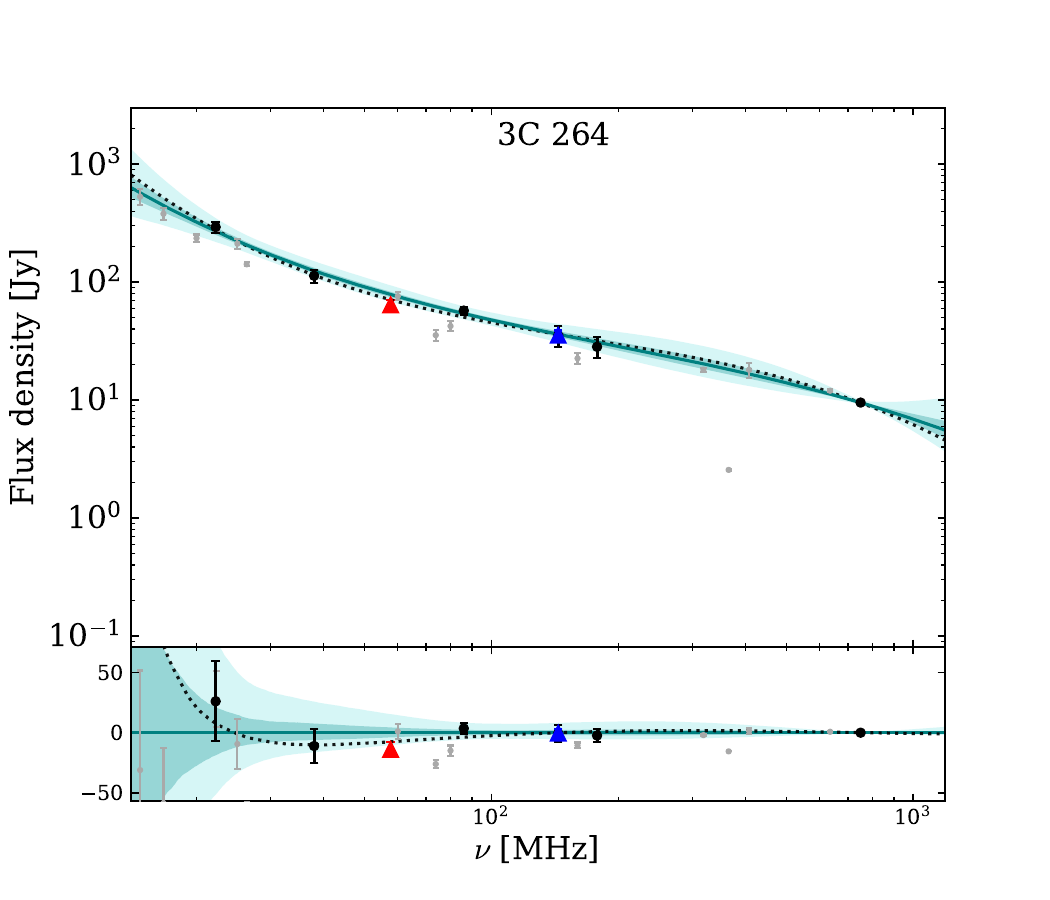}
\includegraphics[width=0.162\linewidth, trim={0.cm .0cm 1.5cm 1.5cm},clip]{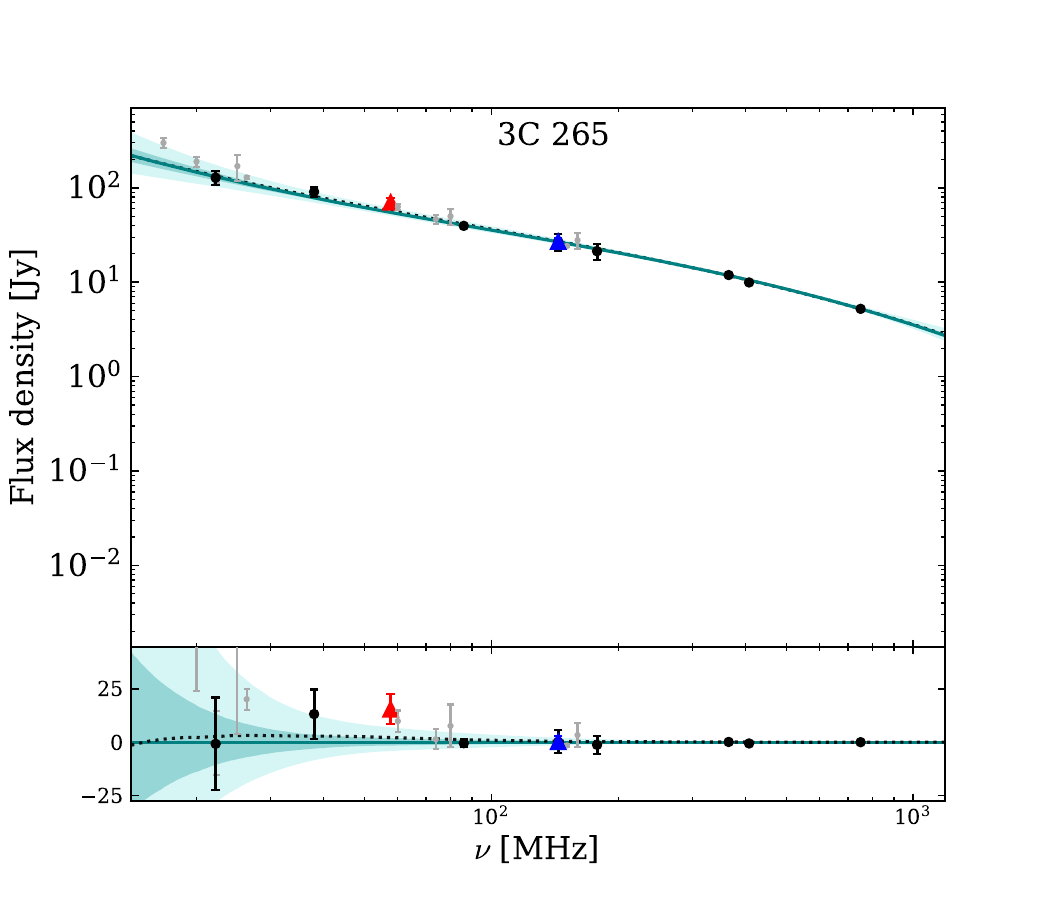}
\includegraphics[width=0.162\linewidth, trim={0.cm .0cm 1.5cm 1.5cm},clip]{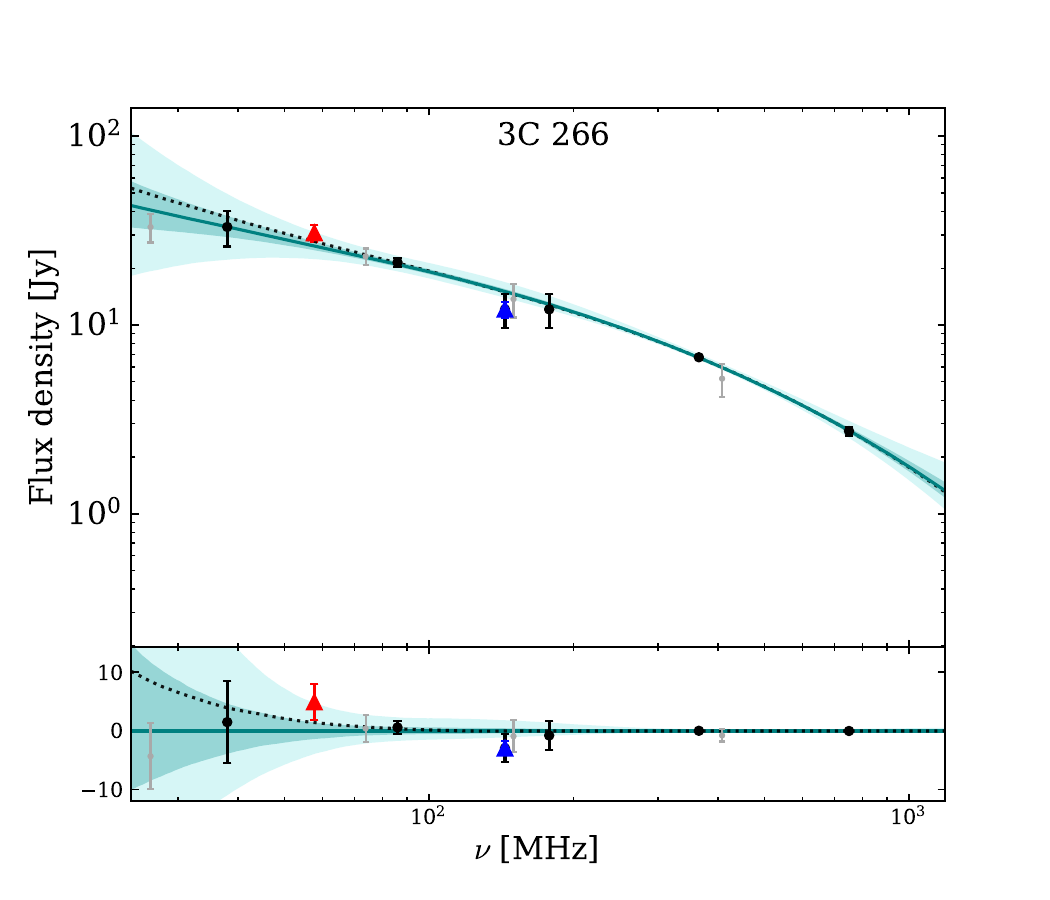}
\includegraphics[width=0.162\linewidth, trim={0.cm .0cm 1.5cm 1.5cm},clip]{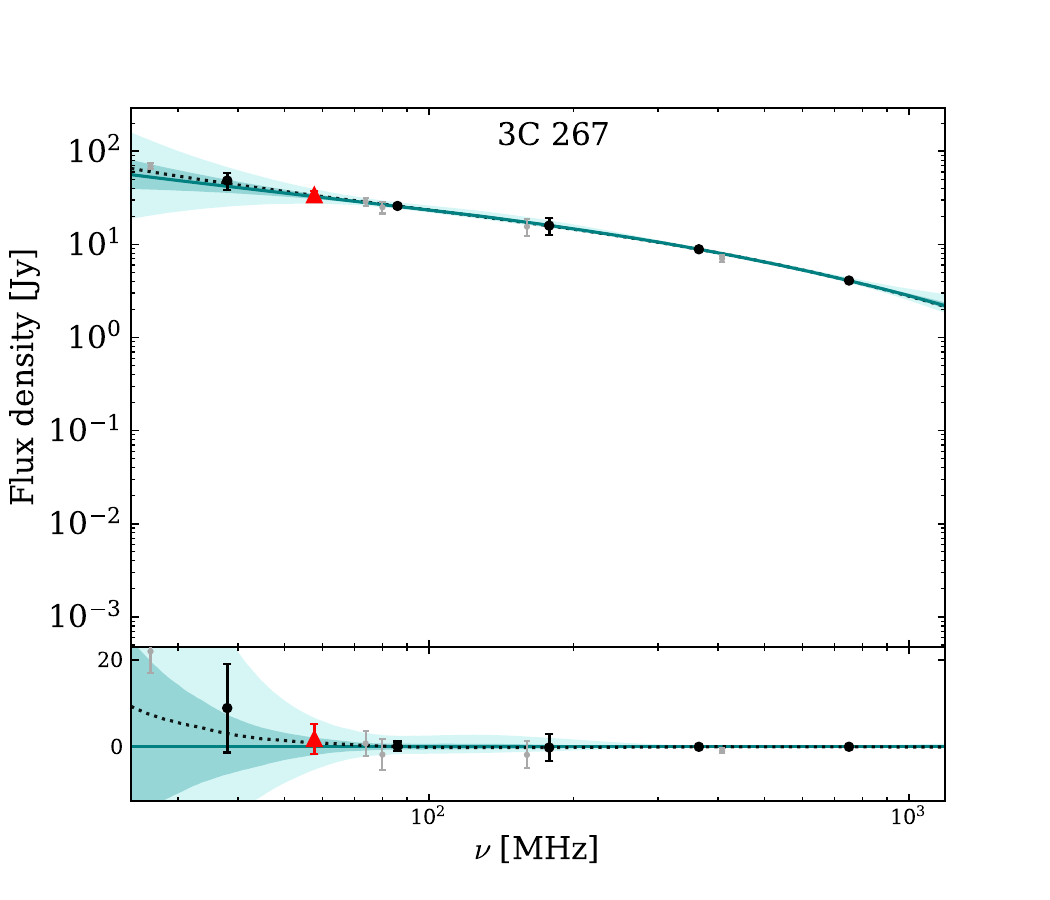}
\includegraphics[width=0.162\linewidth, trim={0.cm .0cm 1.5cm 1.5cm},clip]{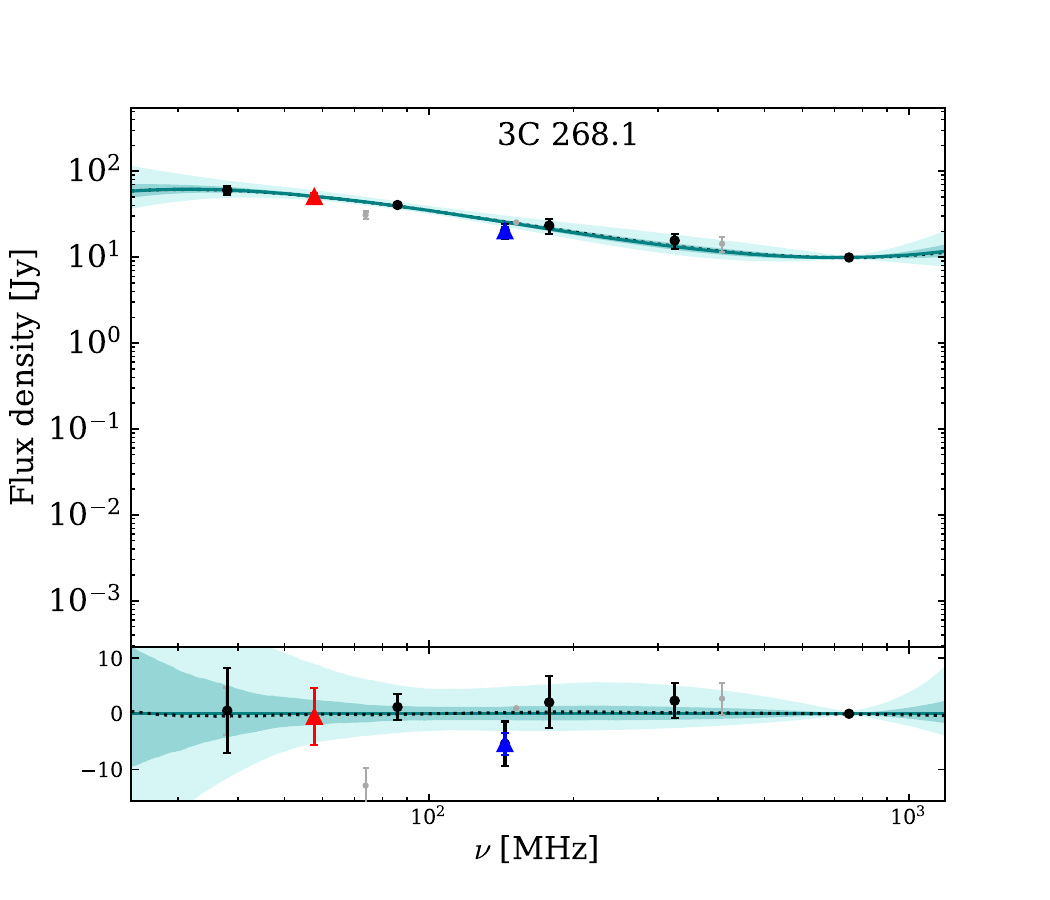}
\includegraphics[width=0.162\linewidth, trim={0.cm .0cm 1.5cm 1.5cm},clip]{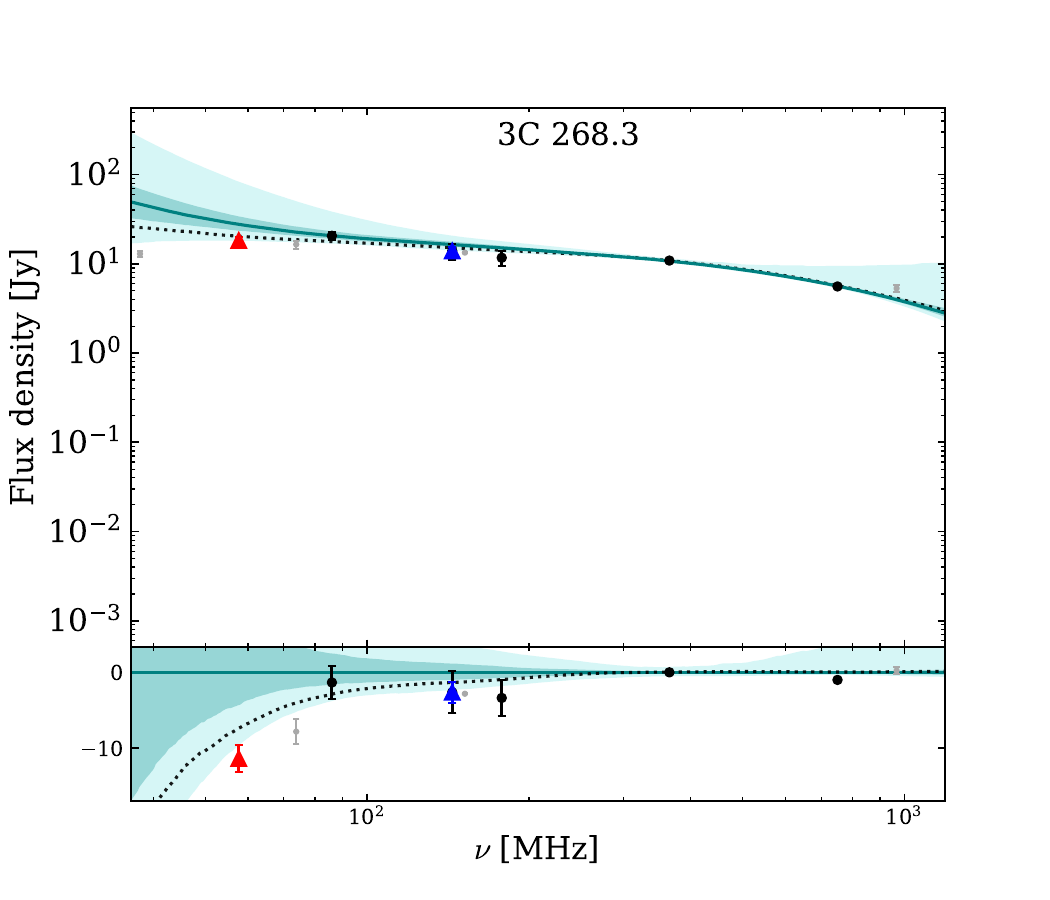}
\includegraphics[width=0.162\linewidth, trim={0.cm .0cm 1.5cm 1.5cm},clip]{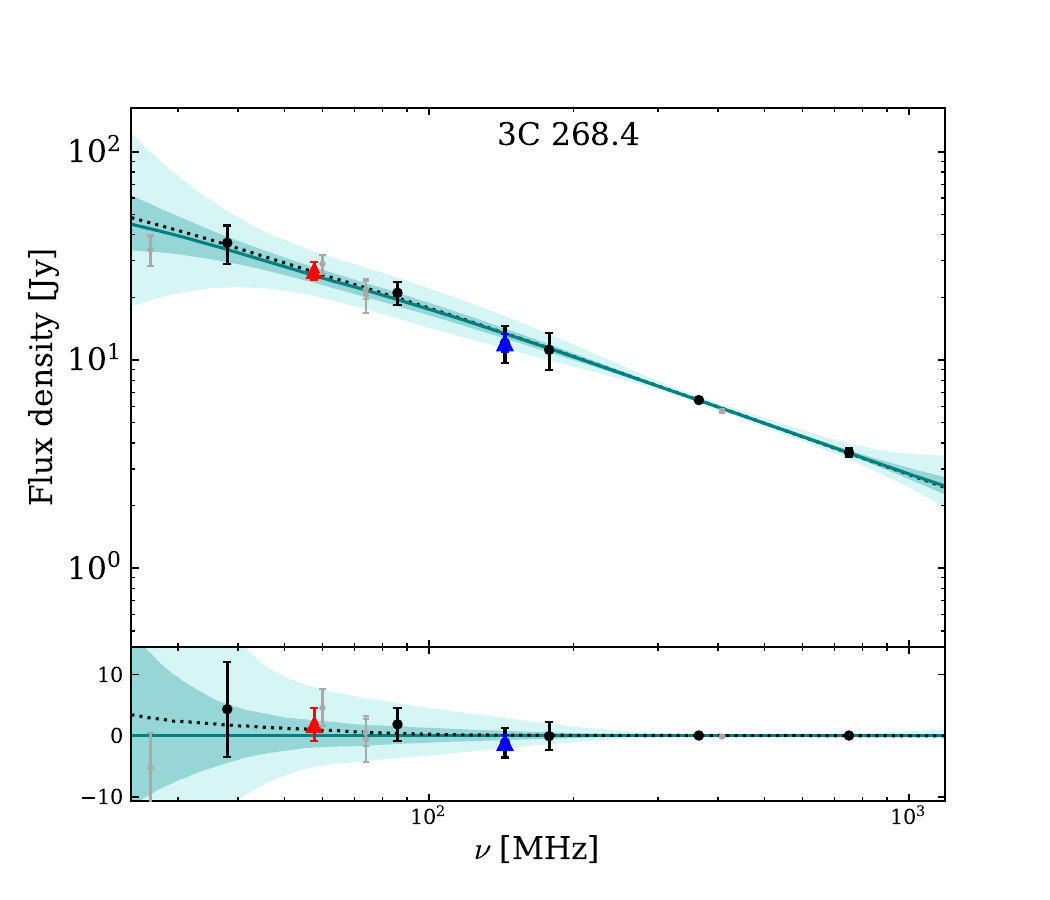}
\includegraphics[width=0.162\linewidth, trim={0.cm .0cm 1.5cm 1.5cm},clip]{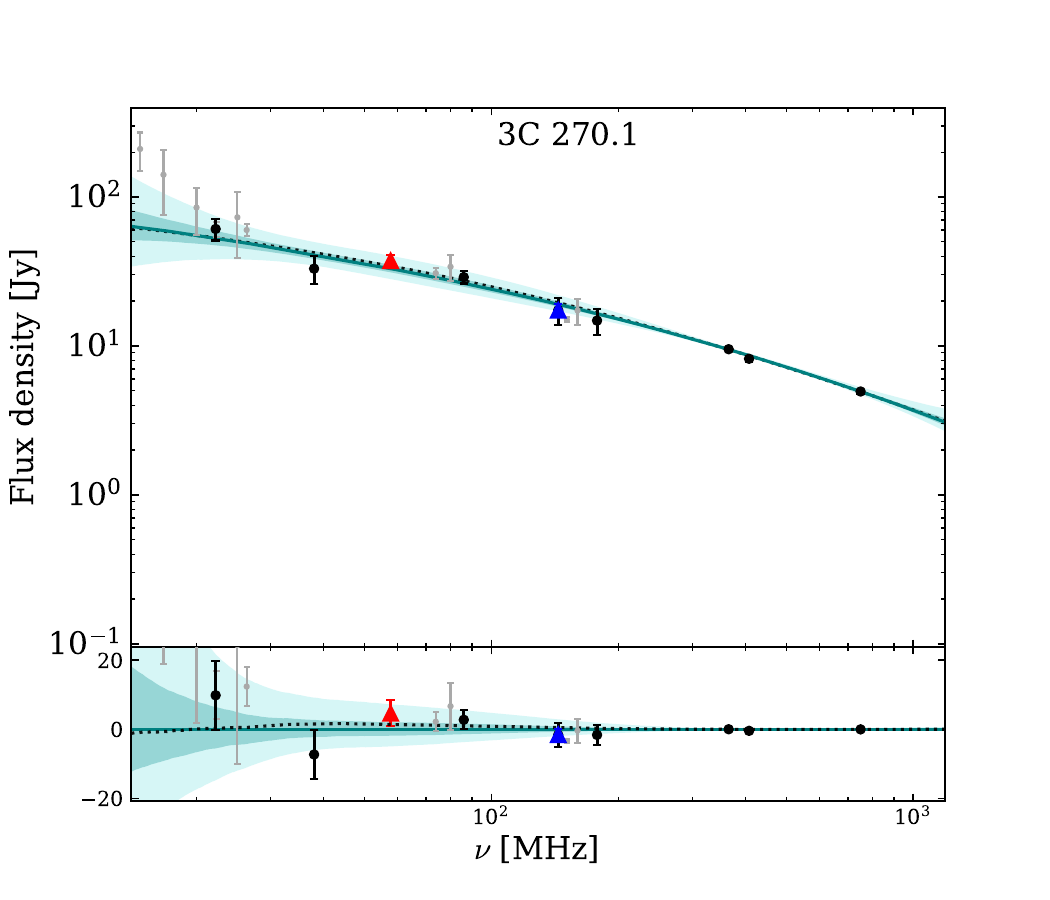}
\includegraphics[width=0.162\linewidth, trim={0.cm .0cm 1.5cm 1.5cm},clip]{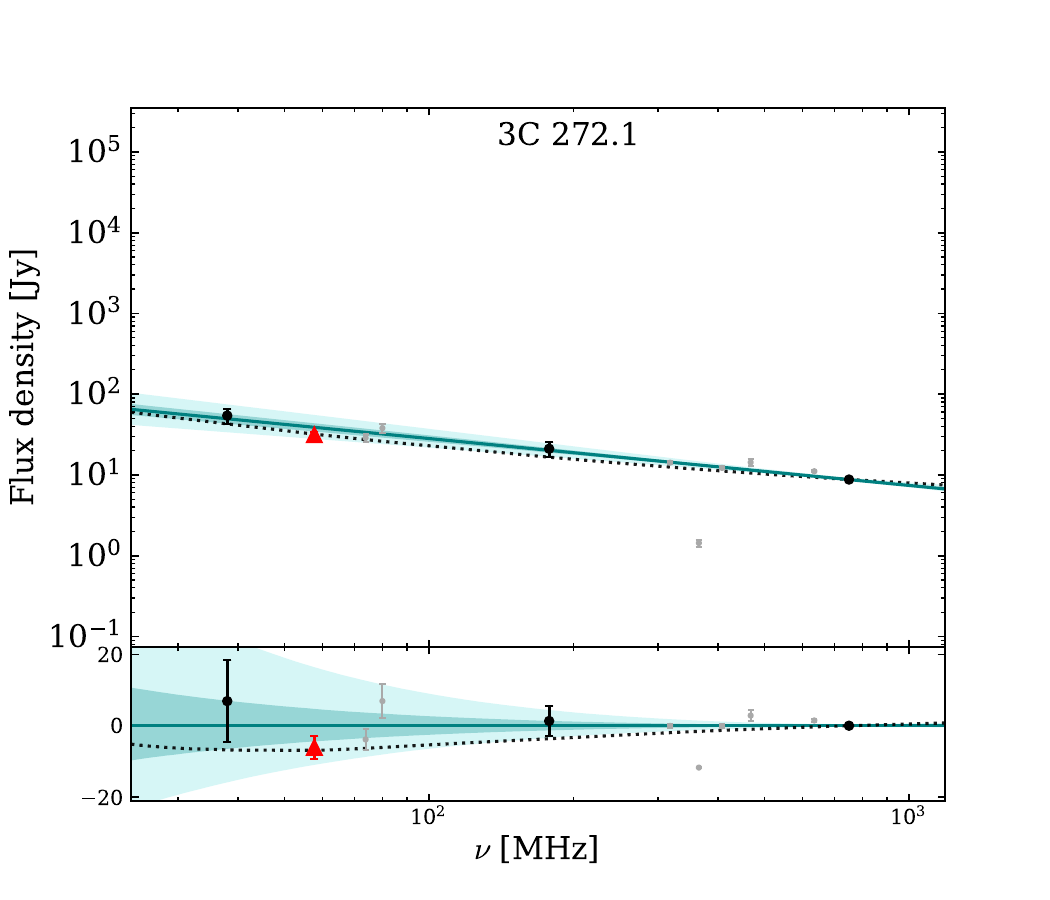}
\includegraphics[width=0.162\linewidth, trim={0.cm .0cm 1.5cm 1.5cm},clip]{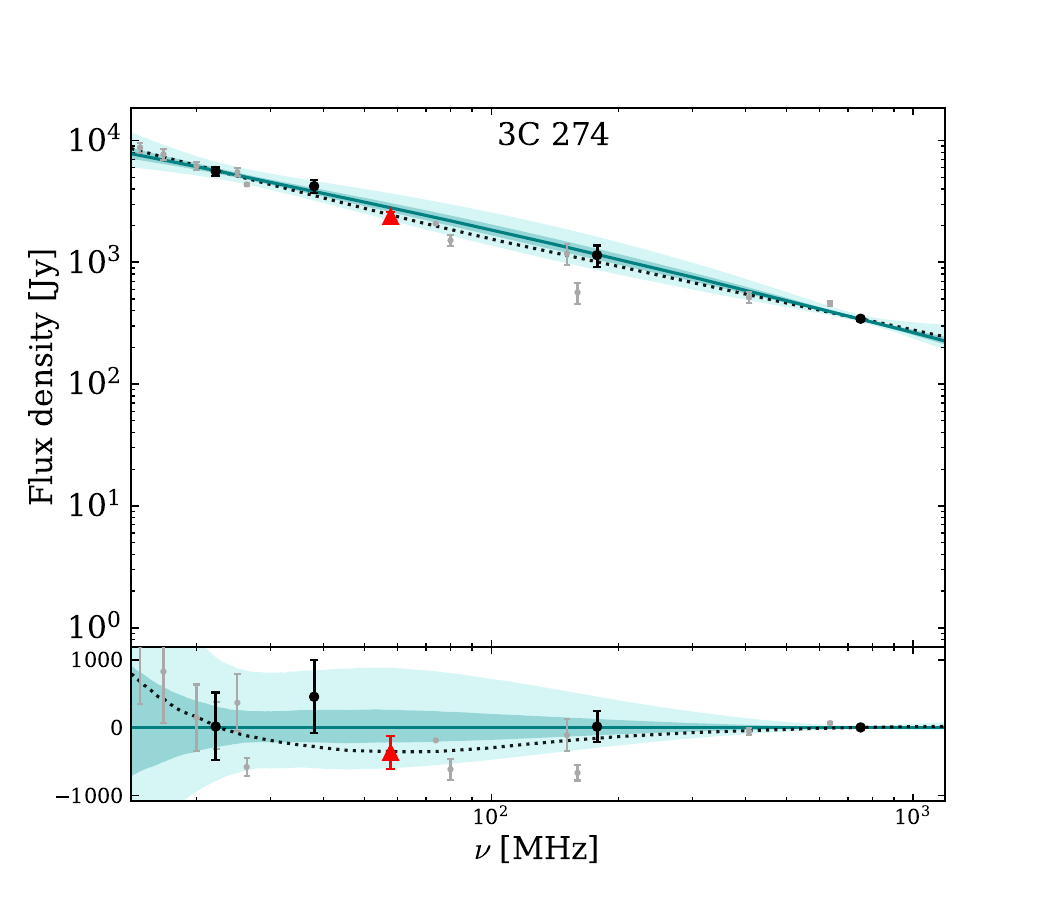}
\includegraphics[width=0.162\linewidth, trim={0.cm .0cm 1.5cm 1.5cm},clip]{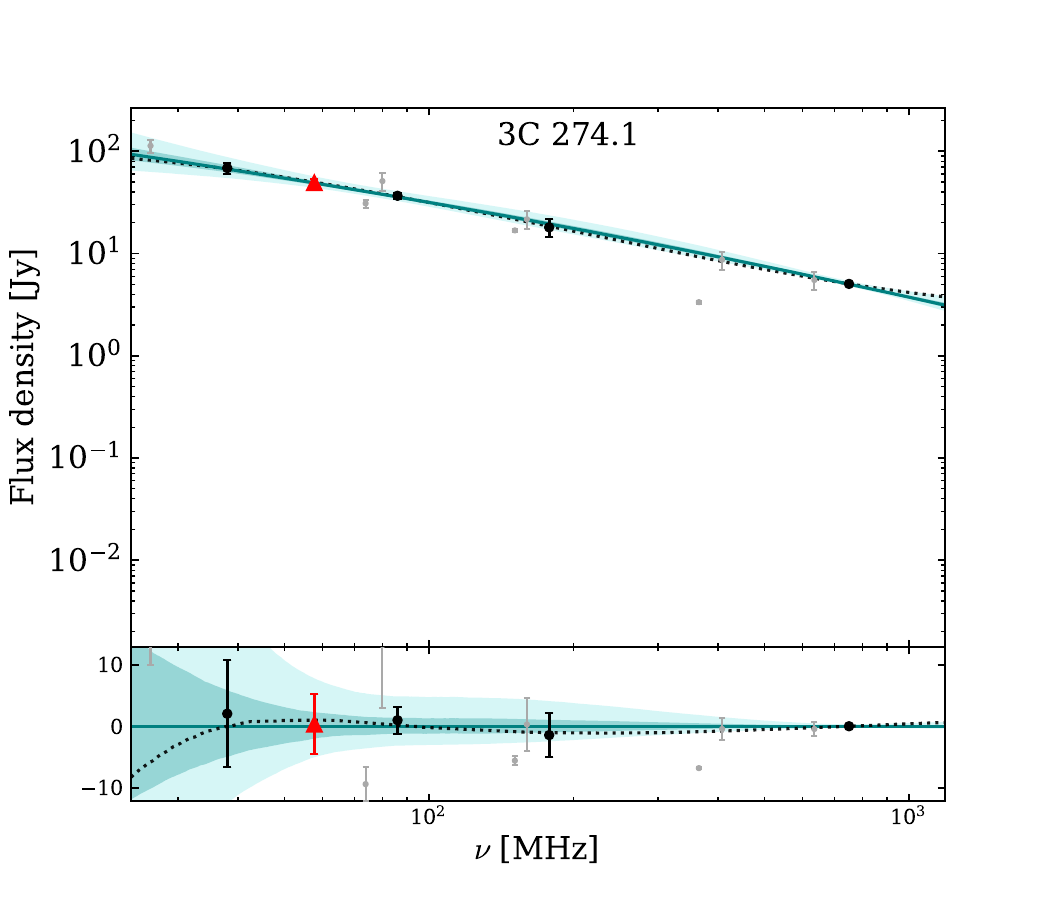}
\includegraphics[width=0.162\linewidth, trim={0.cm .0cm 1.5cm 1.5cm},clip]{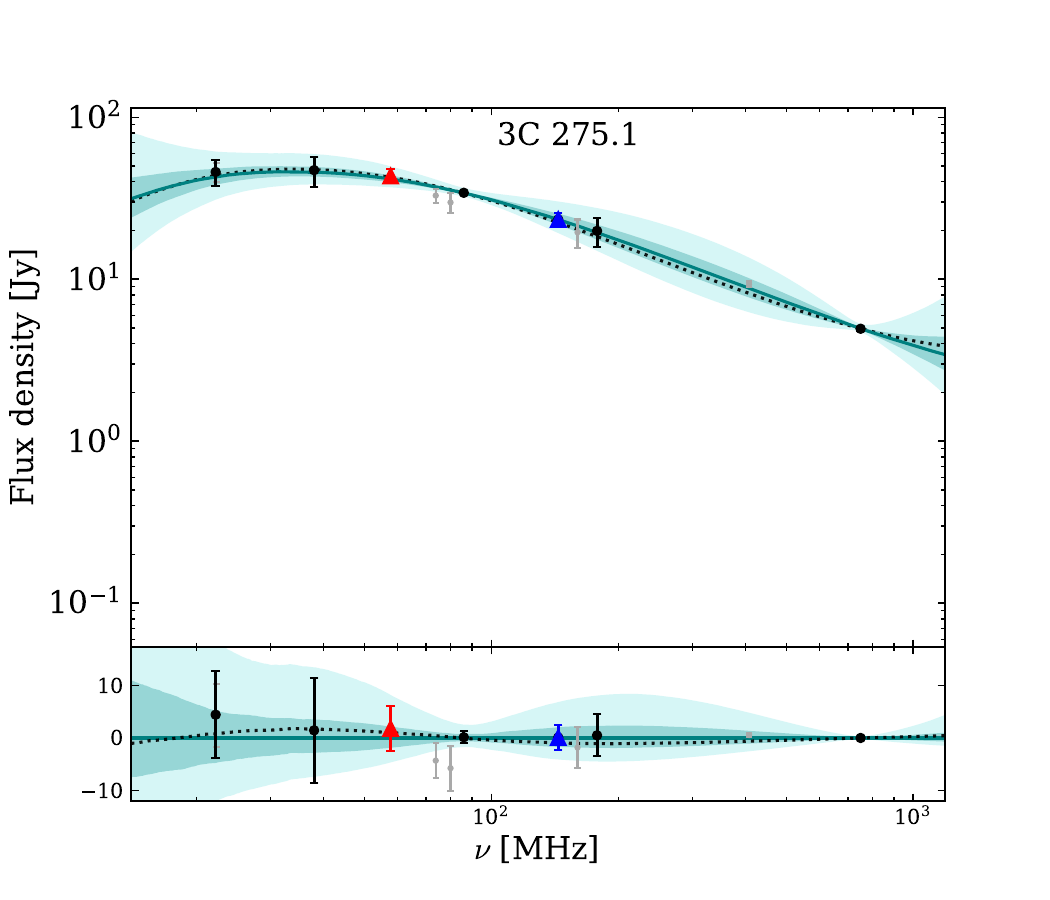}
\includegraphics[width=0.162\linewidth, trim={0.cm .0cm 1.5cm 1.5cm},clip]{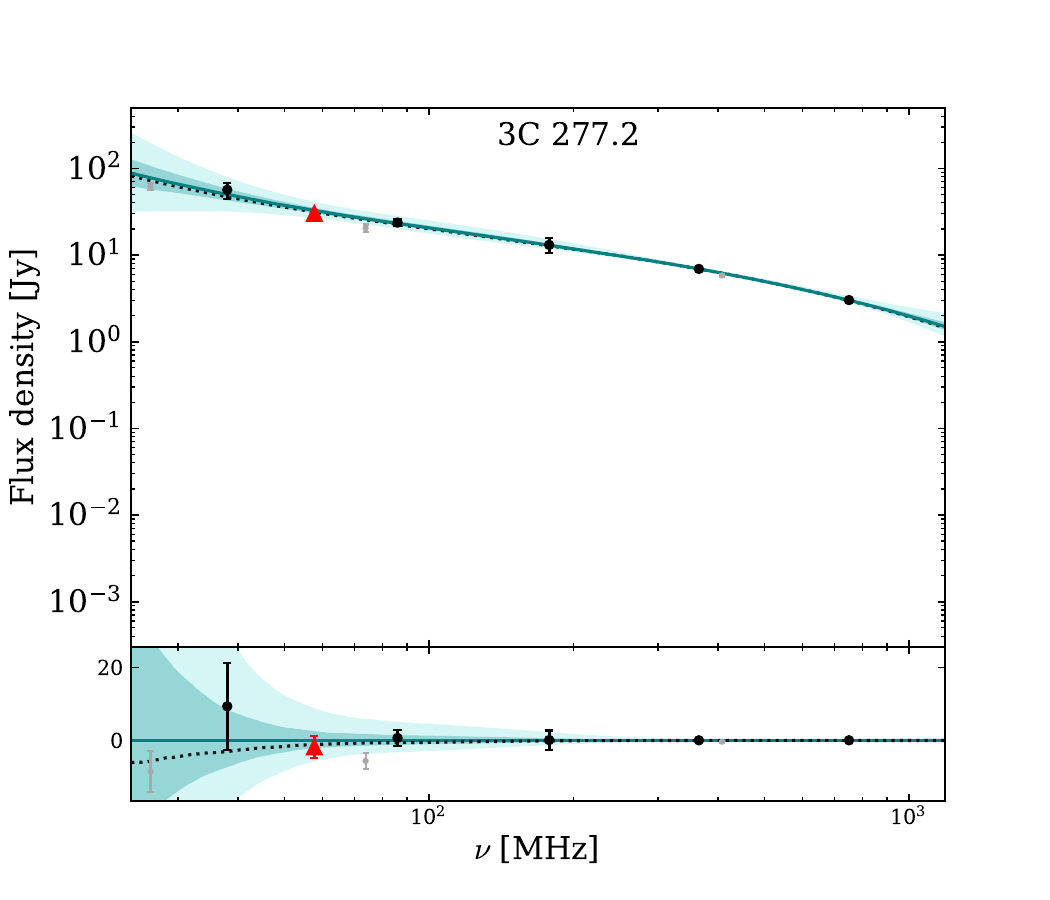}
\includegraphics[width=0.162\linewidth, trim={0.cm .0cm 1.5cm 1.5cm},clip]{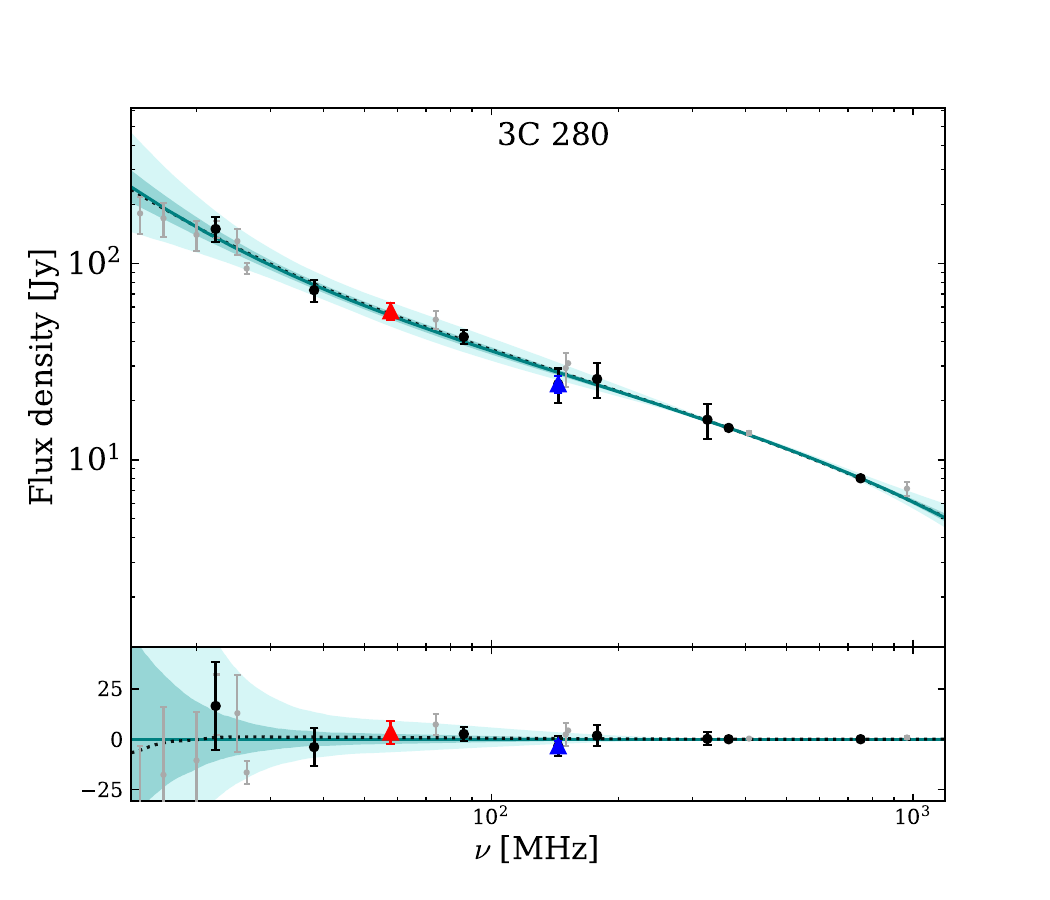}
\includegraphics[width=0.162\linewidth, trim={0.cm .0cm 1.5cm 1.5cm},clip]{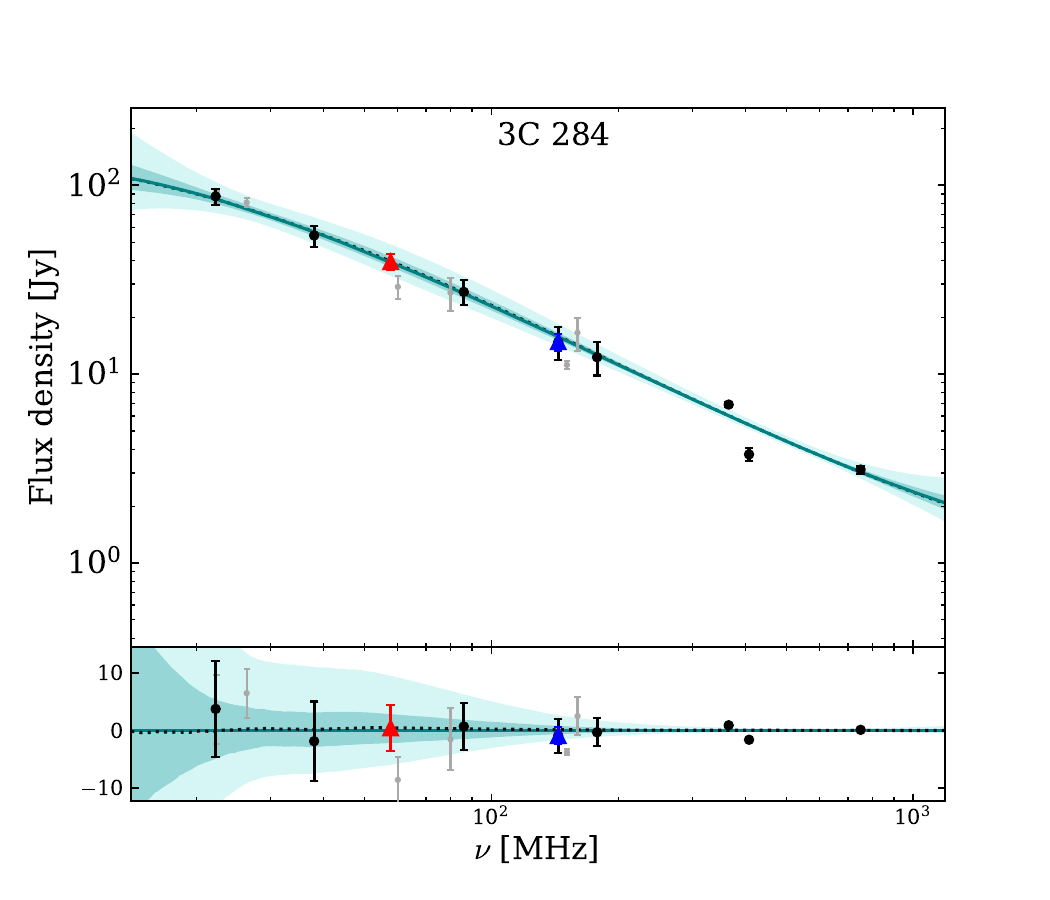}
\includegraphics[width=0.162\linewidth, trim={0.cm .0cm 1.5cm 1.5cm},clip]{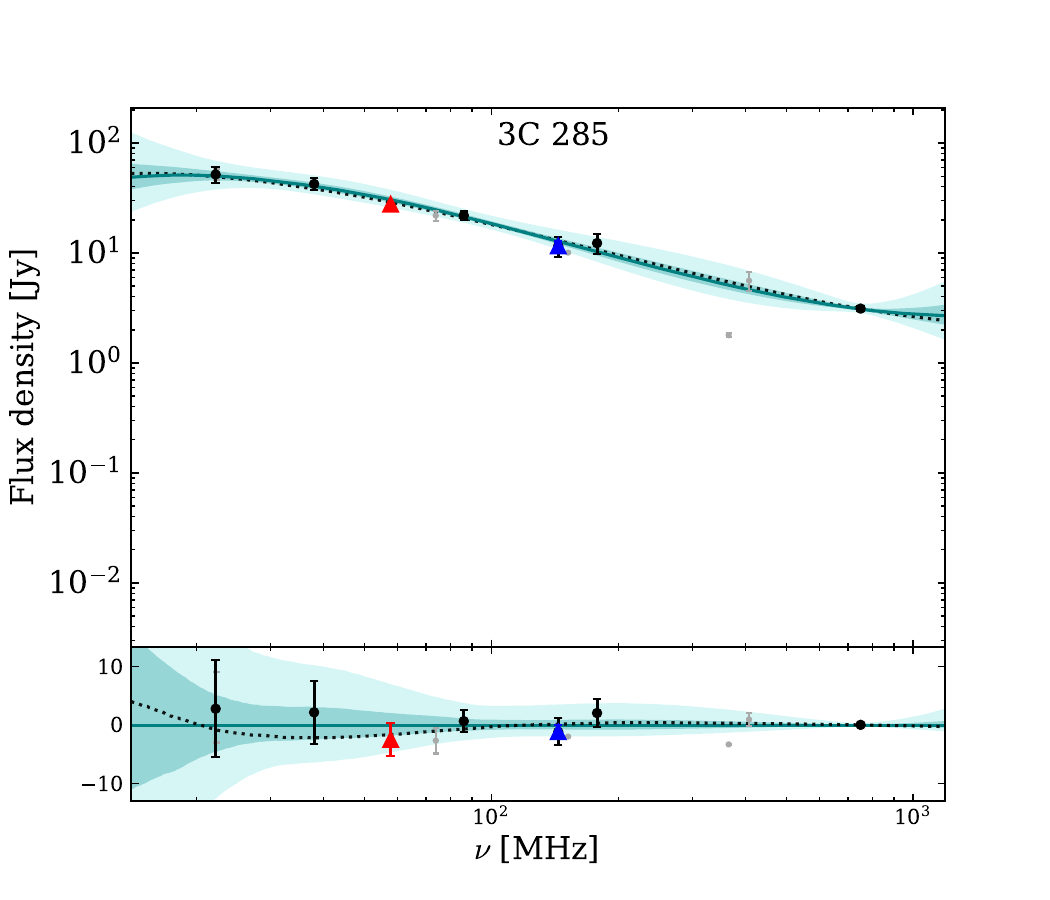}
\includegraphics[width=0.162\linewidth, trim={0.cm .0cm 1.5cm 1.5cm},clip]{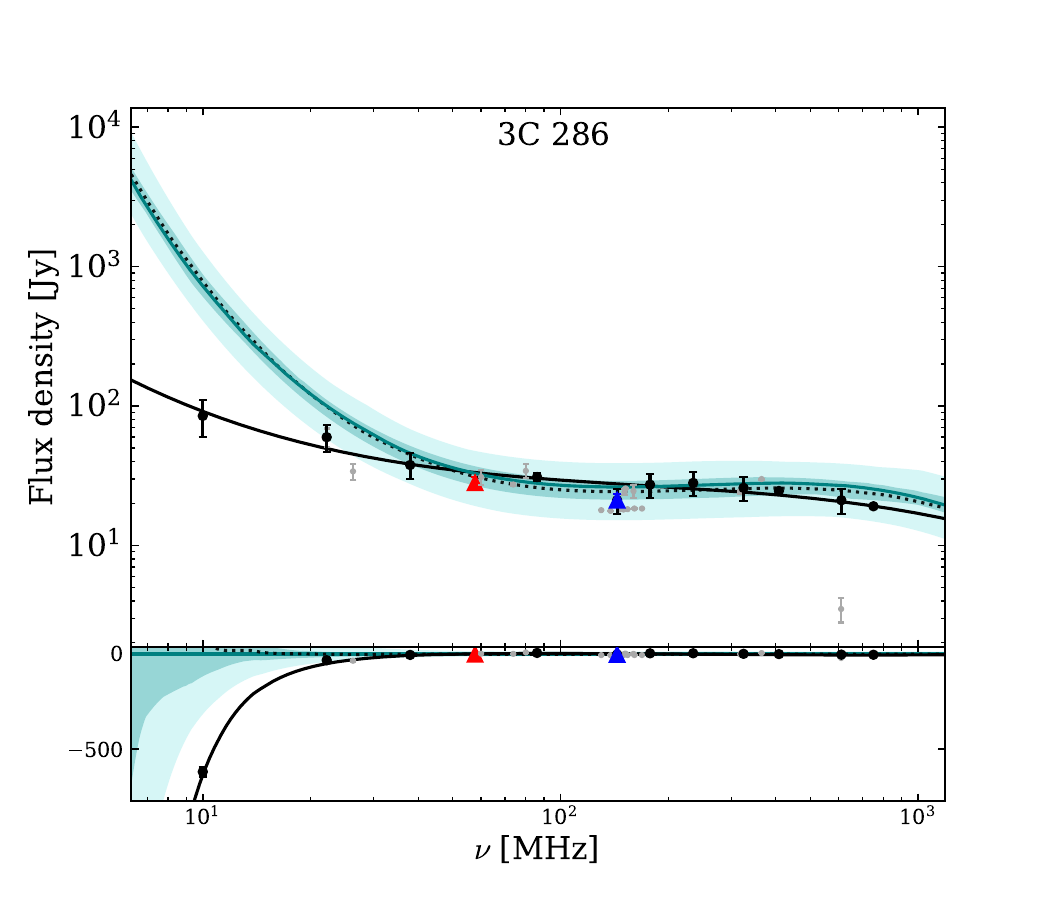}
\includegraphics[width=0.162\linewidth, trim={0.cm .0cm 1.5cm 1.5cm},clip]{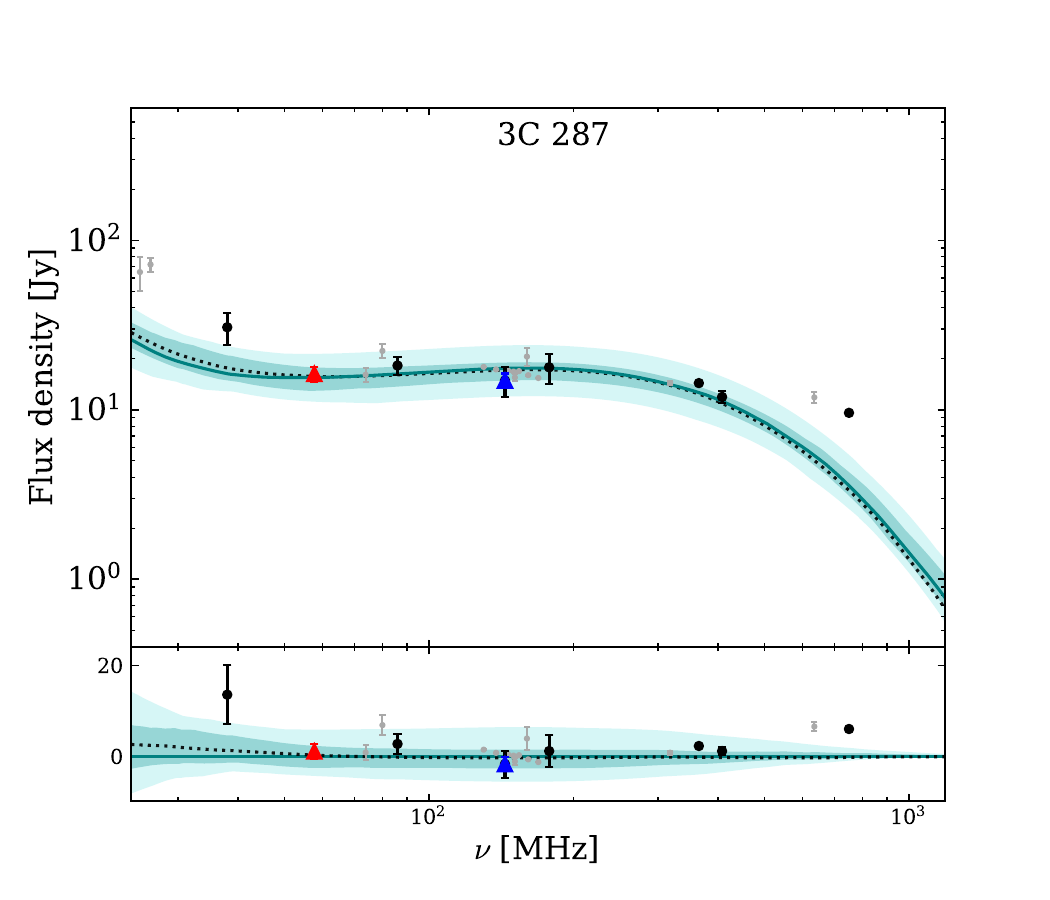}
\includegraphics[width=0.162\linewidth, trim={0.cm .0cm 1.5cm 1.5cm},clip]{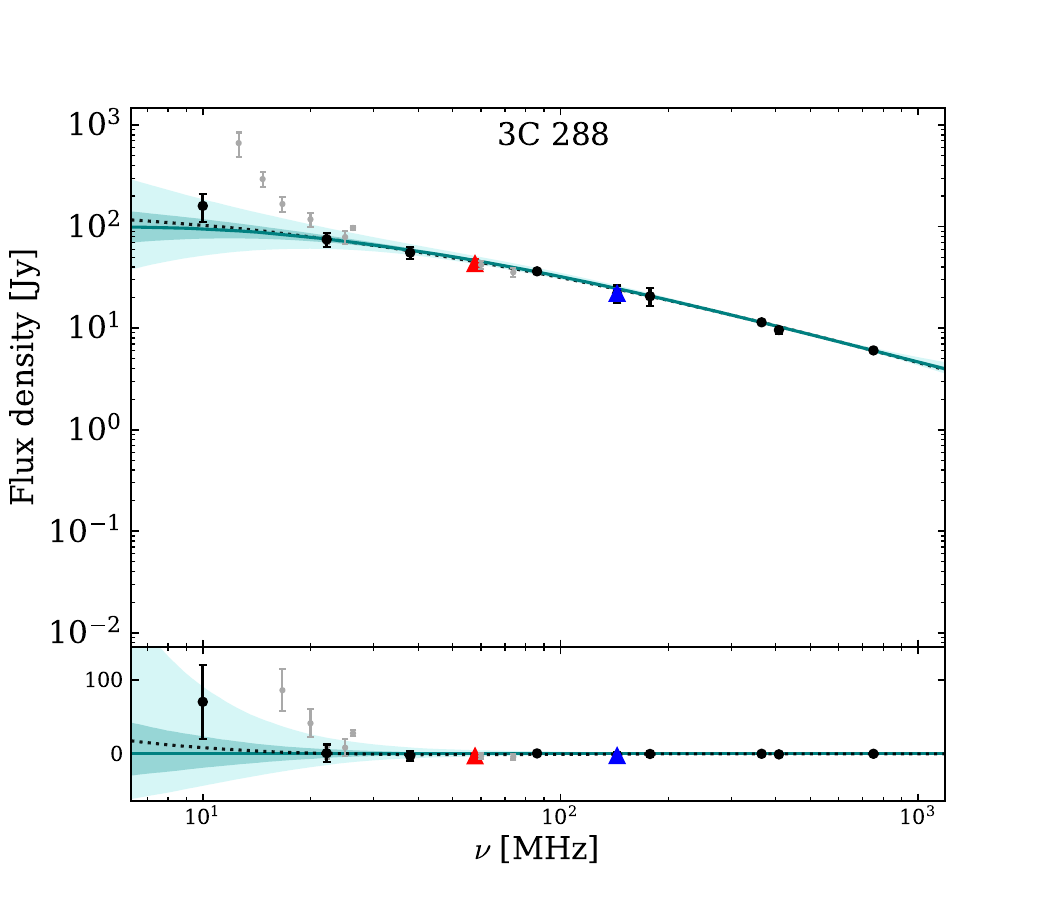}
\includegraphics[width=0.162\linewidth, trim={0.cm .0cm 1.5cm 1.5cm},clip]{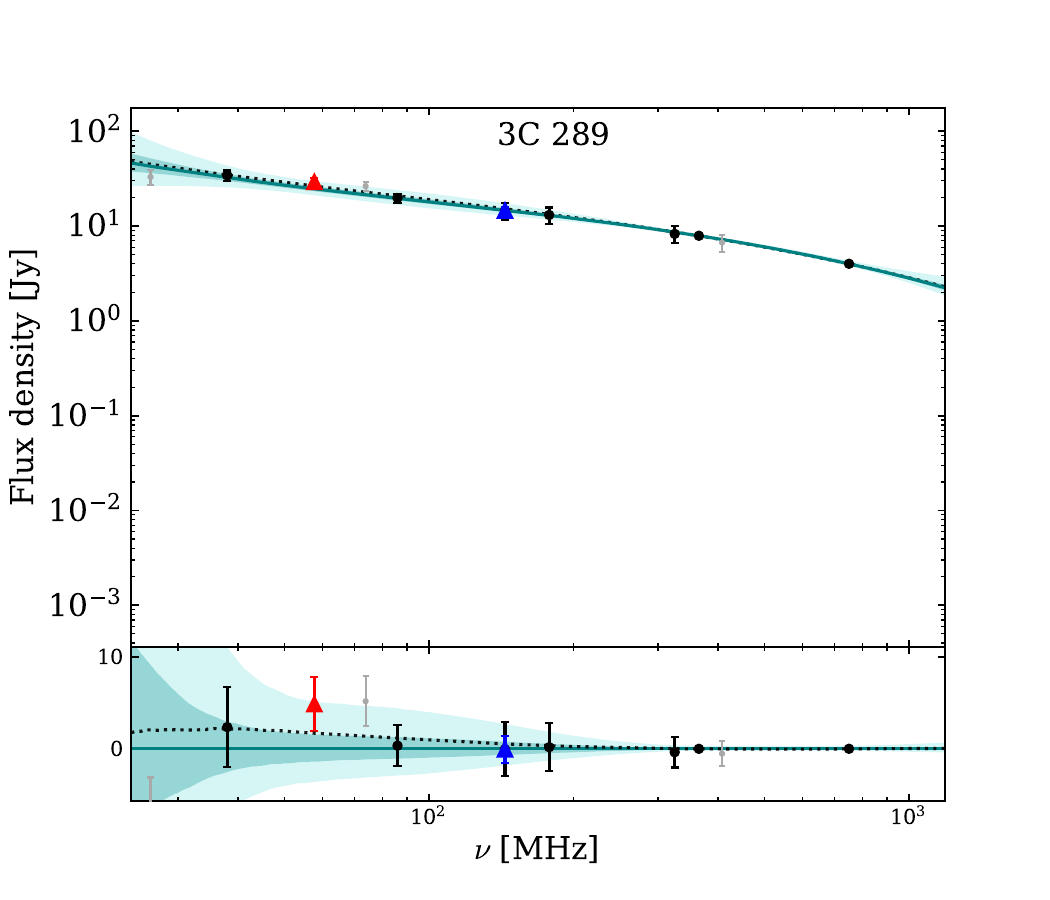}
\includegraphics[width=0.162\linewidth, trim={0.cm .0cm 1.5cm 1.5cm},clip]{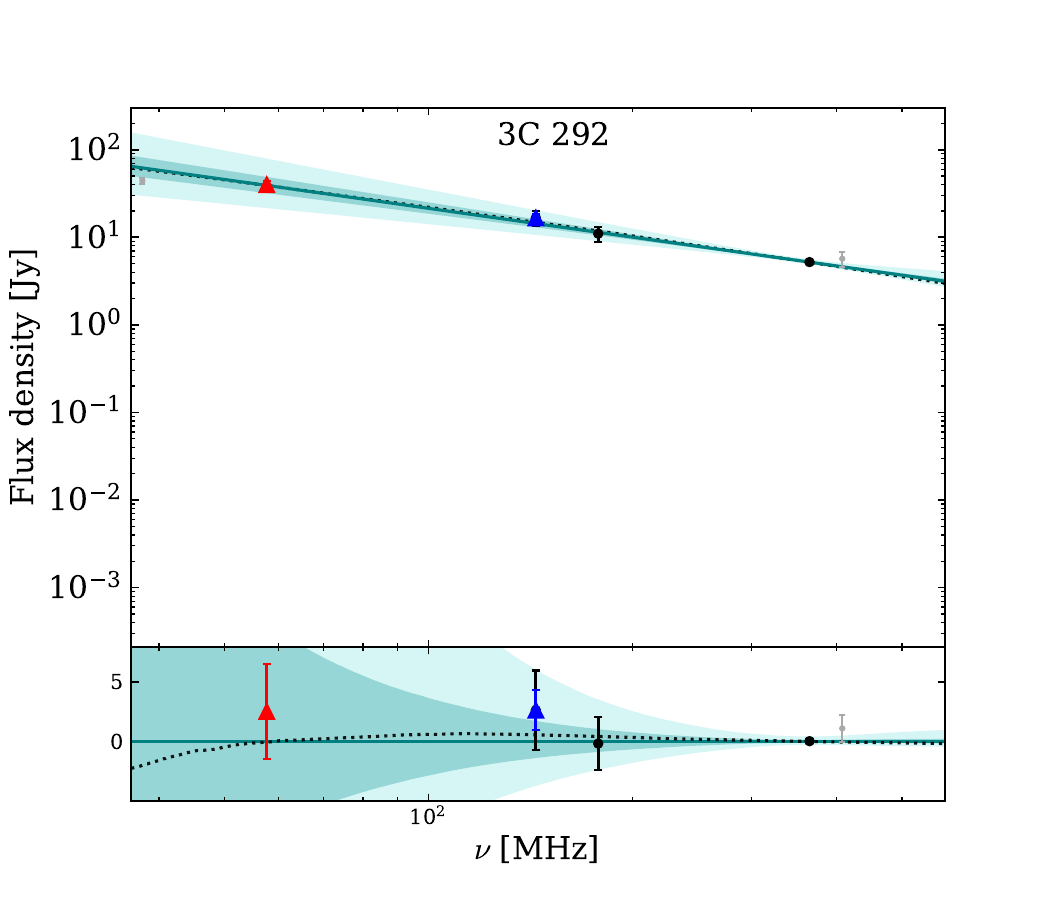}
\caption{continued.}
\end{figure}
\setcounter{figure}{0}

\begin{figure}[H]
\centering
\includegraphics[width=0.162\linewidth, trim={0.cm .0cm 1.5cm 1.5cm},clip]{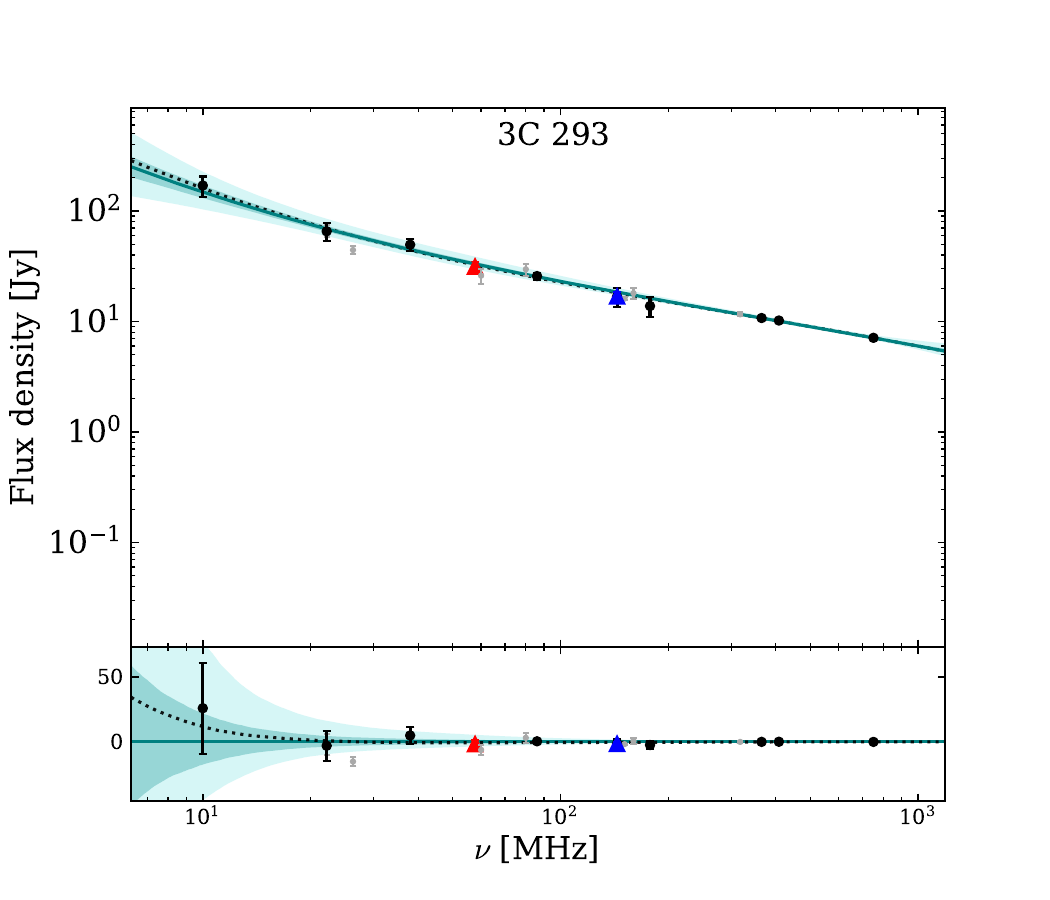}
\includegraphics[width=0.162\linewidth, trim={0.cm .0cm 1.5cm 1.5cm},clip]{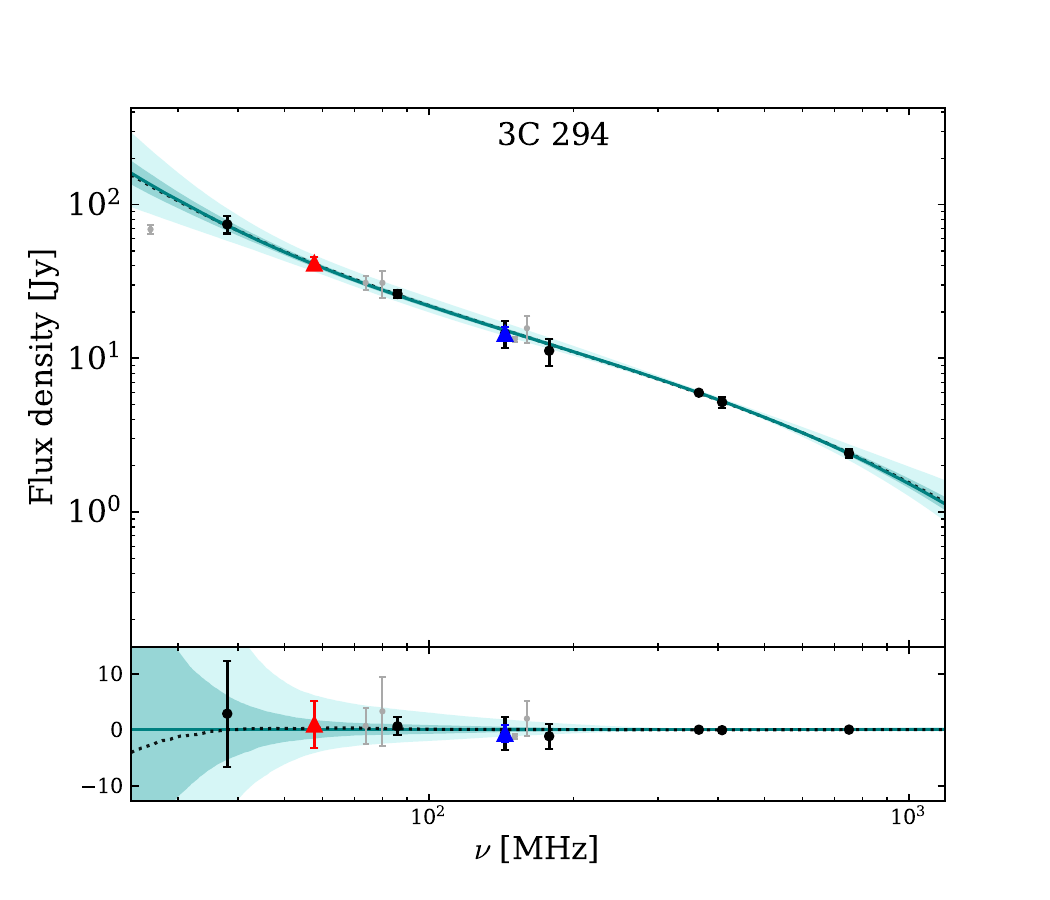}
\includegraphics[width=0.162\linewidth, trim={0.cm .0cm 1.5cm 1.5cm},clip]{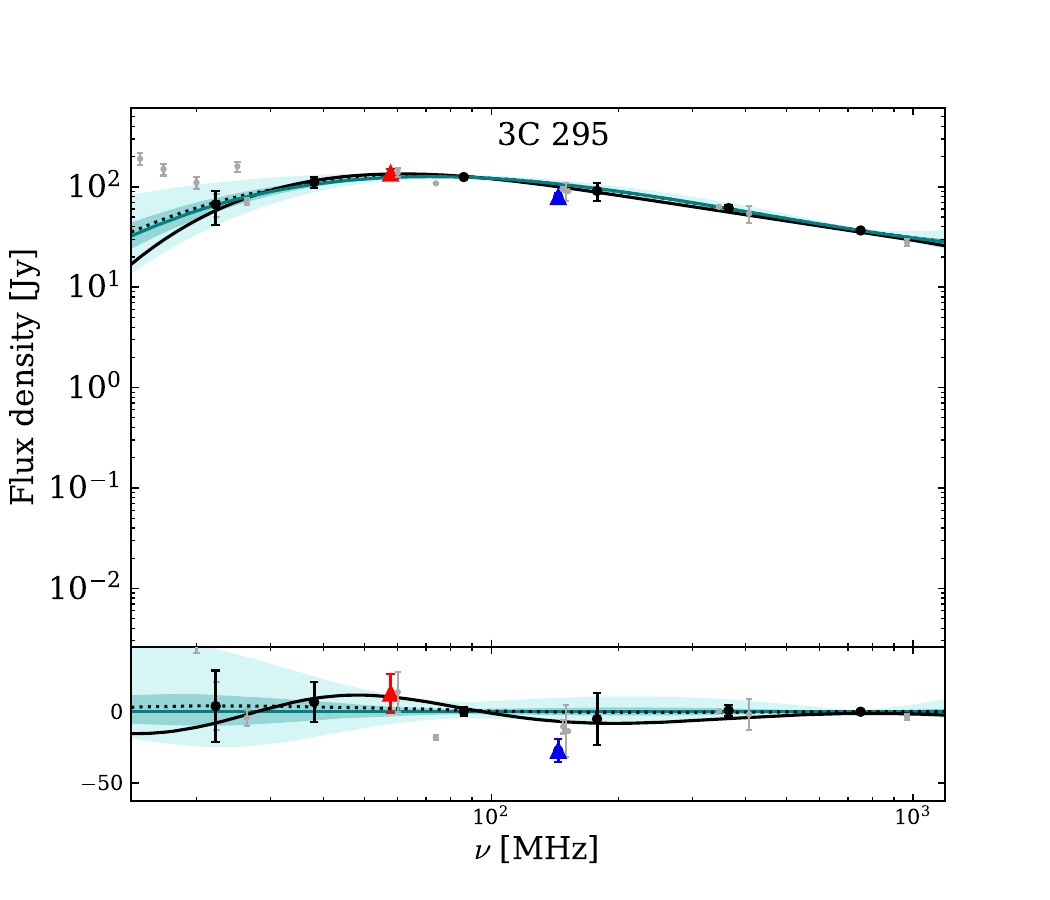}
\includegraphics[width=0.162\linewidth, trim={0.cm .0cm 1.5cm 1.5cm},clip]{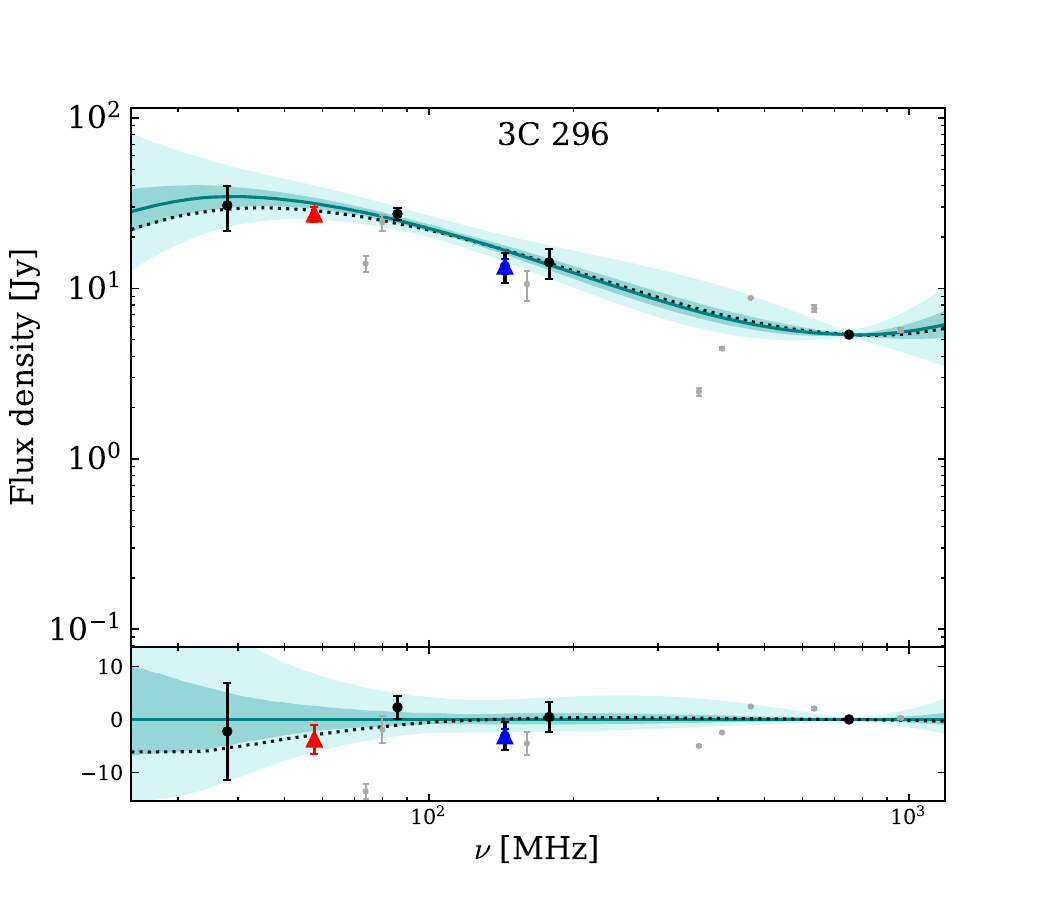}
\includegraphics[width=0.162\linewidth, trim={0.cm .0cm 1.5cm 1.5cm},clip]{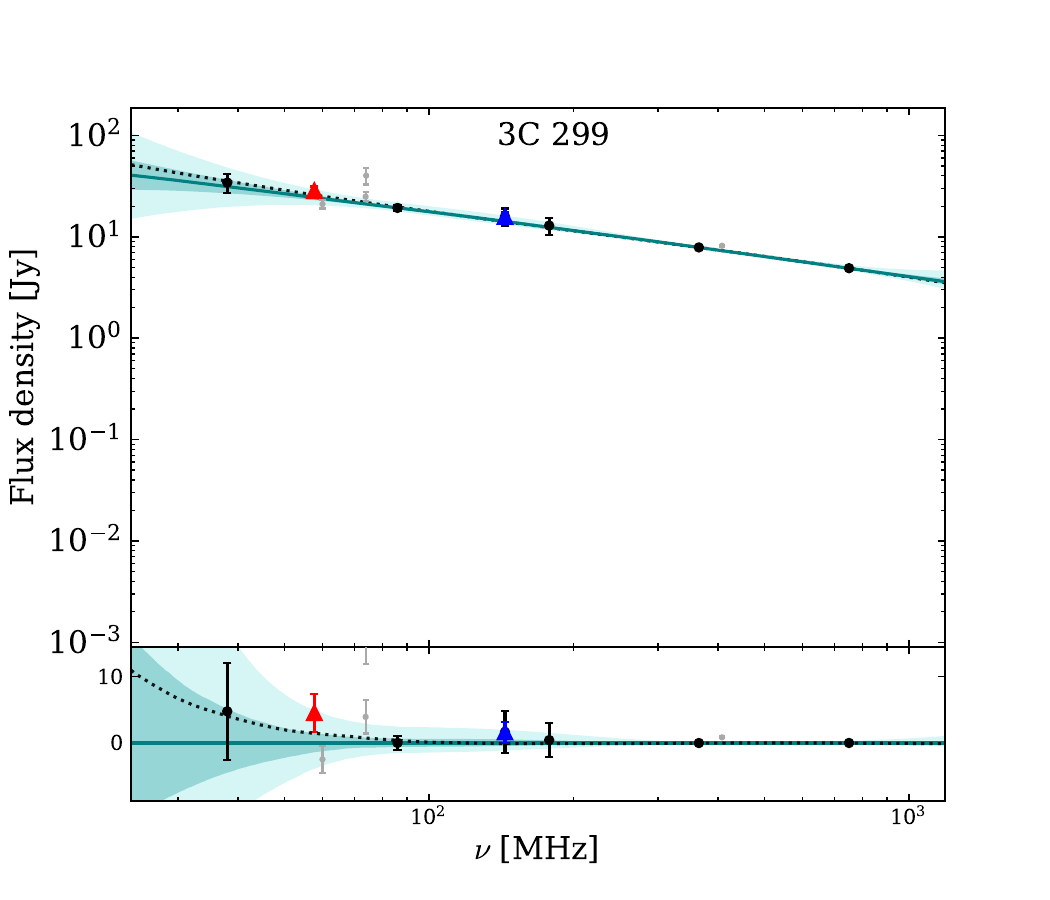}
\includegraphics[width=0.162\linewidth, trim={0.cm .0cm 1.5cm 1.5cm},clip]{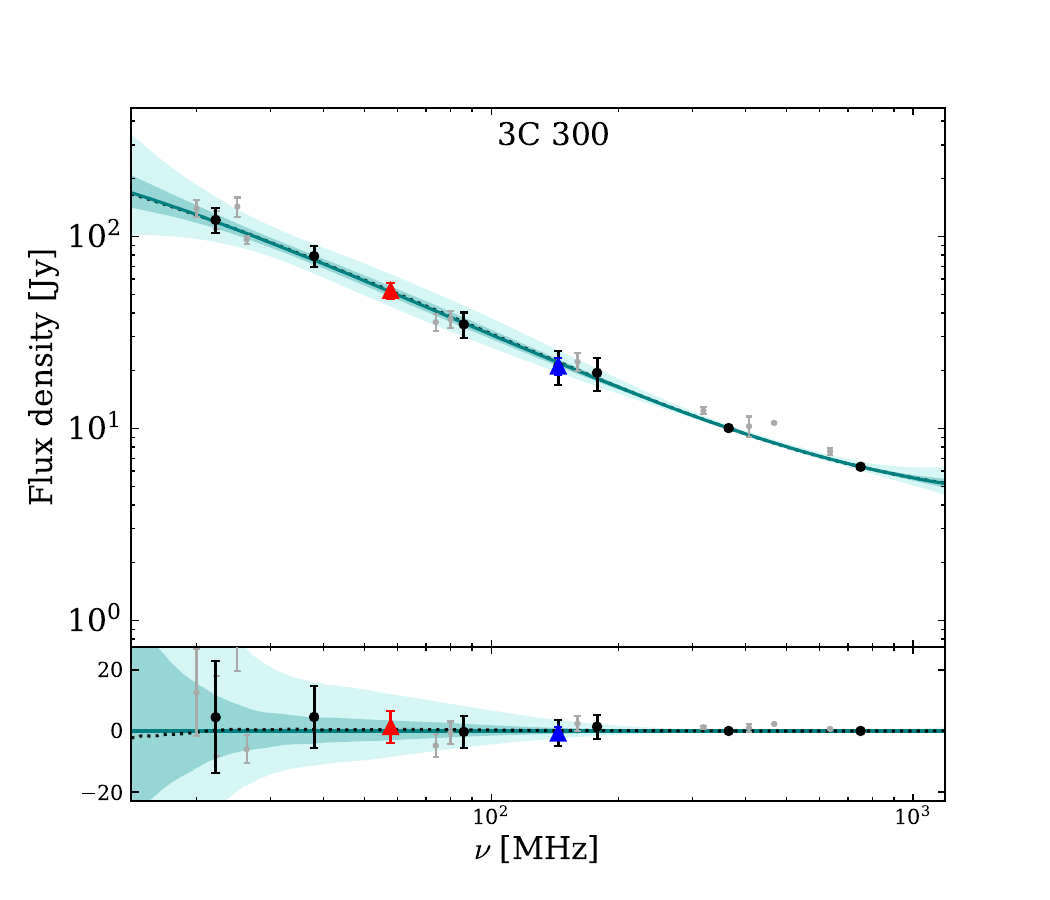}
\includegraphics[width=0.162\linewidth, trim={0.cm .0cm 1.5cm 1.5cm},clip]{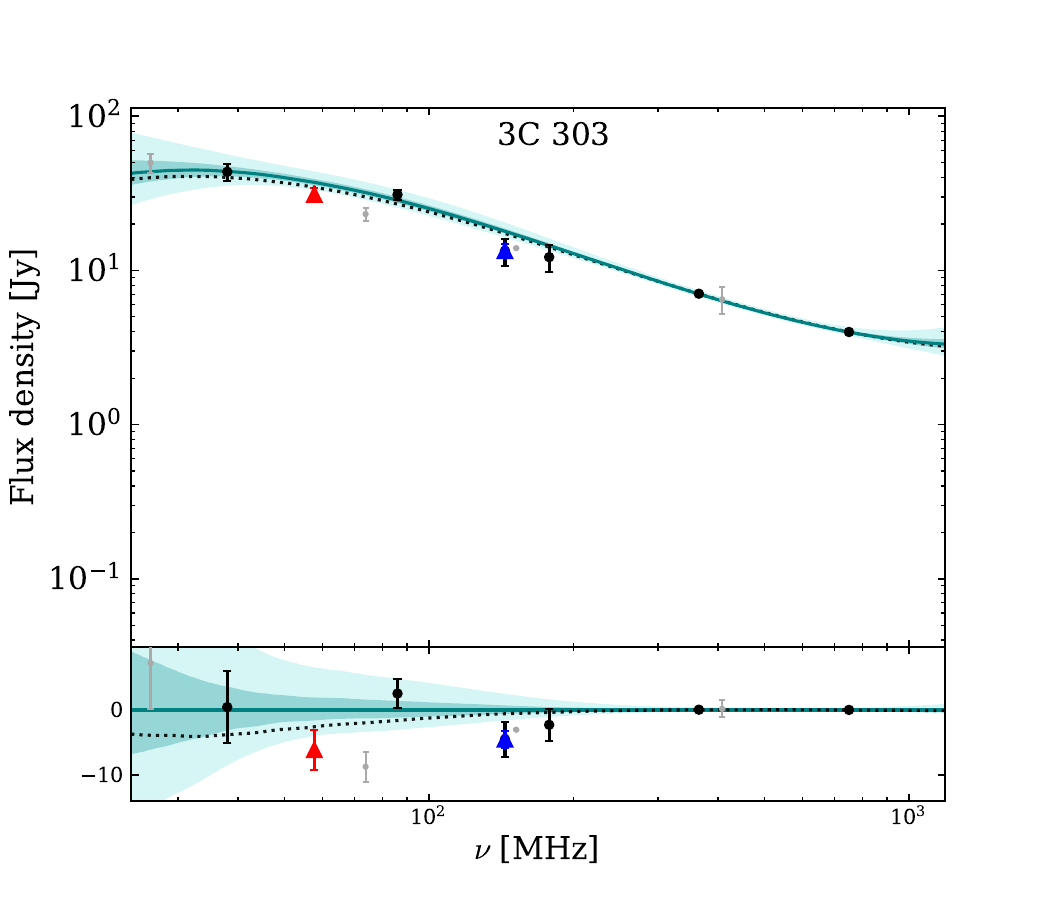}
\includegraphics[width=0.162\linewidth, trim={0.cm .0cm 1.5cm 1.5cm},clip]{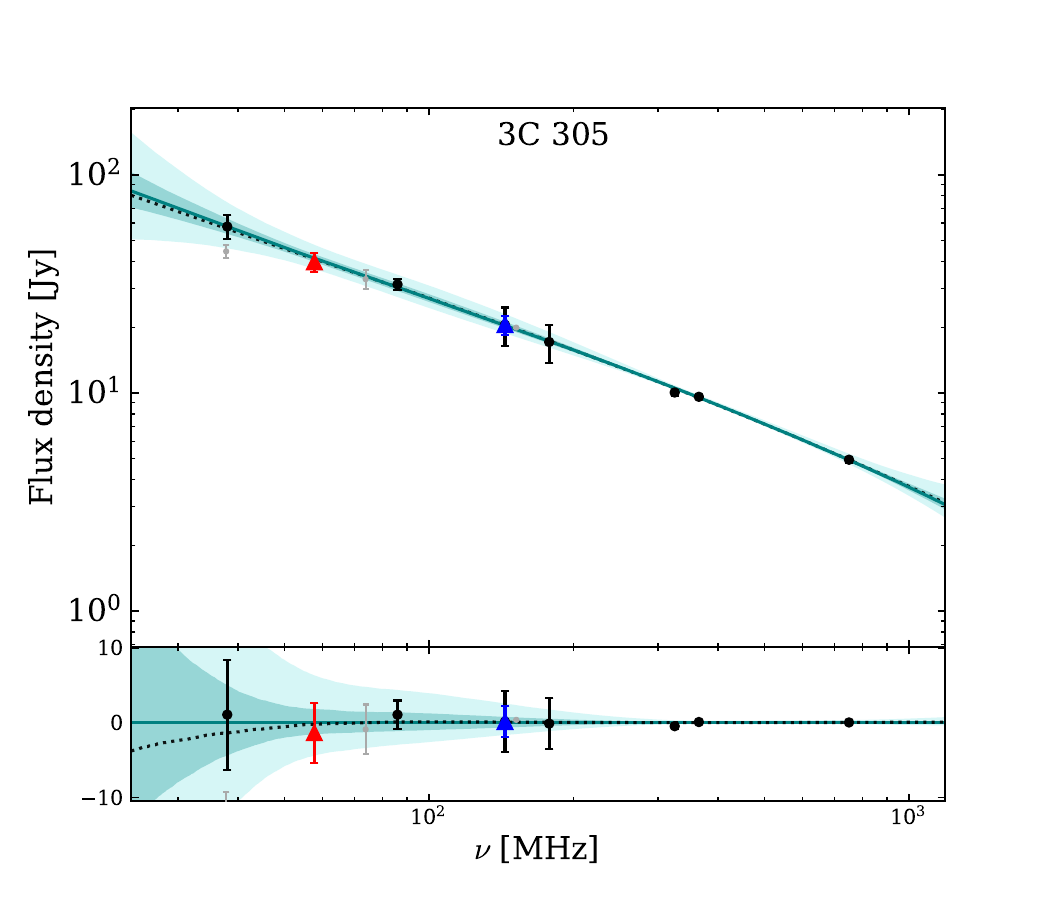}
\includegraphics[width=0.162\linewidth, trim={0.cm .0cm 1.5cm 1.5cm},clip]{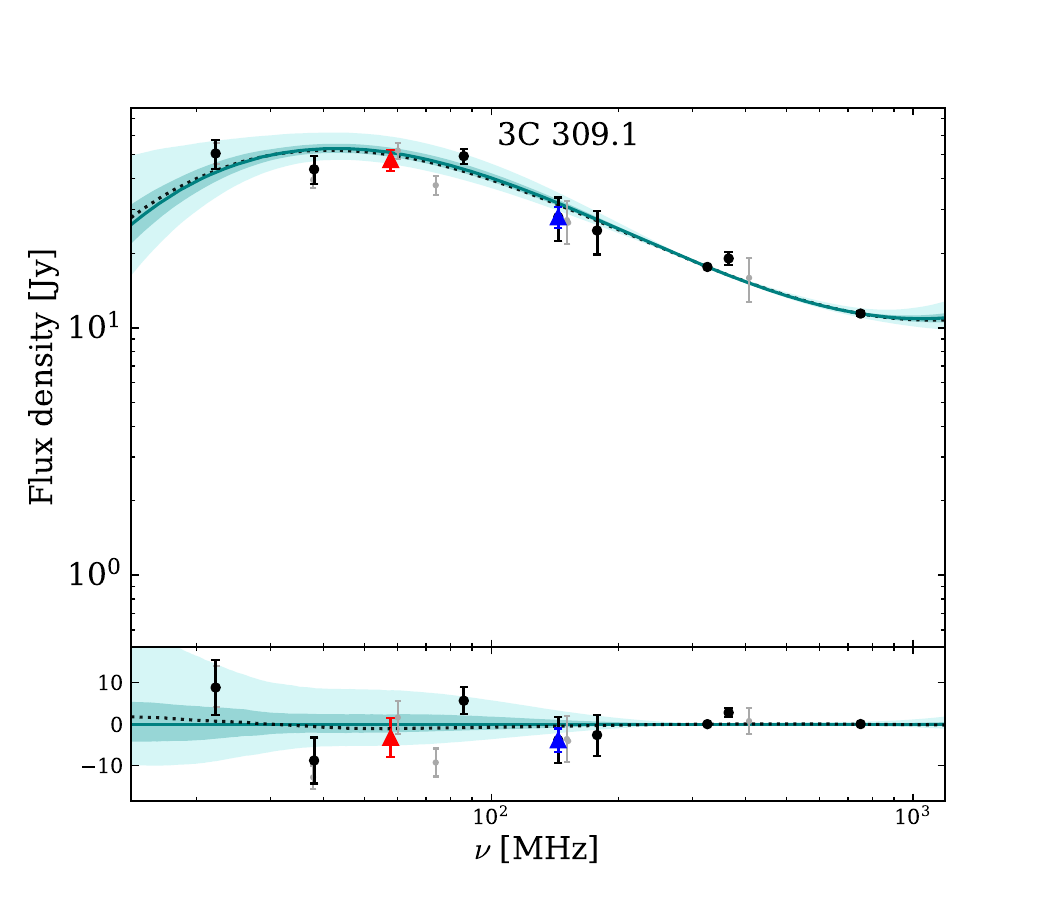}
\includegraphics[width=0.162\linewidth, trim={0.cm .0cm 1.5cm 1.5cm},clip]{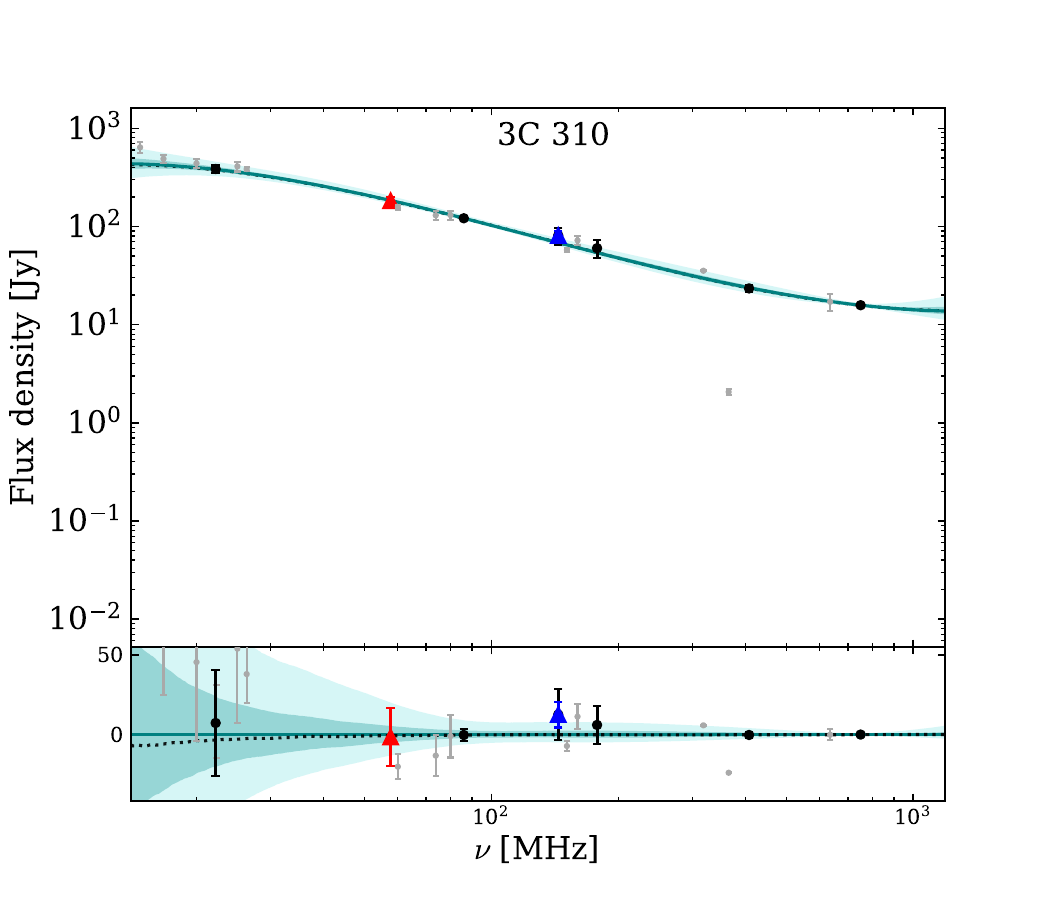}
\includegraphics[width=0.162\linewidth, trim={0.cm .0cm 1.5cm 1.5cm},clip]{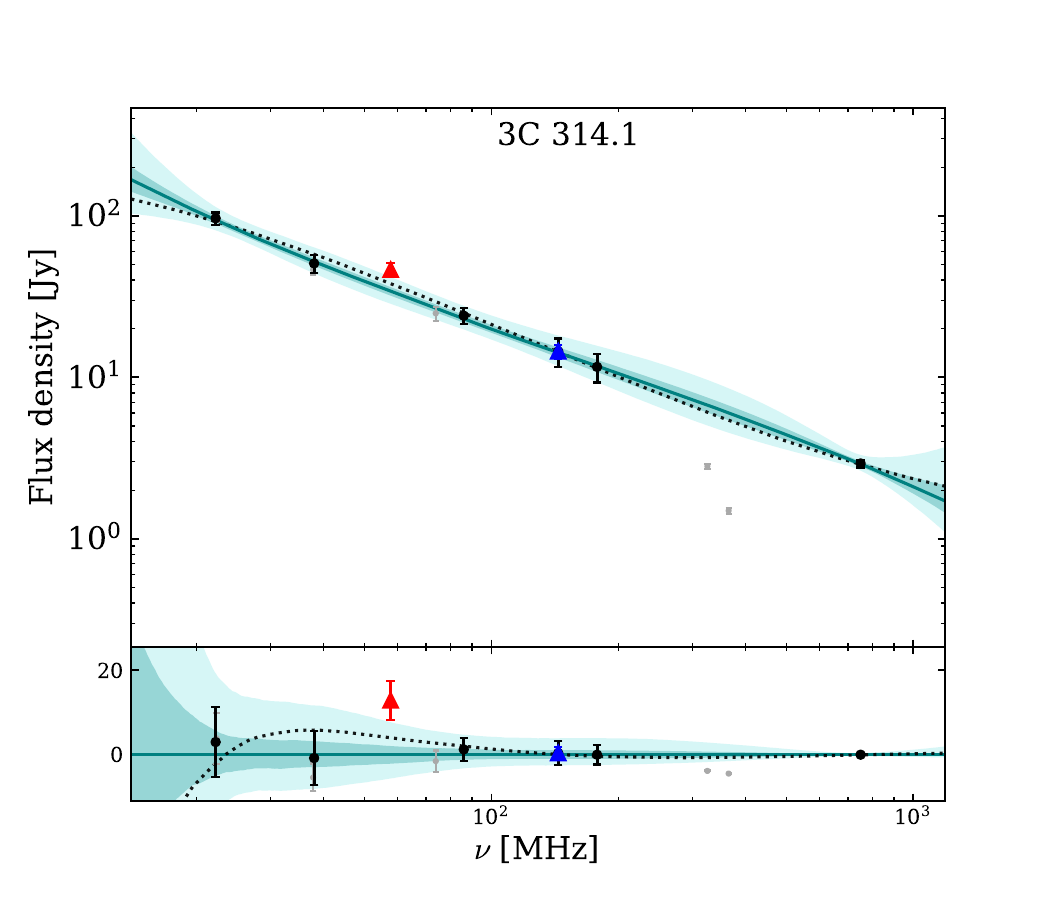}
\includegraphics[width=0.162\linewidth, trim={0.cm .0cm 1.5cm 1.5cm},clip]{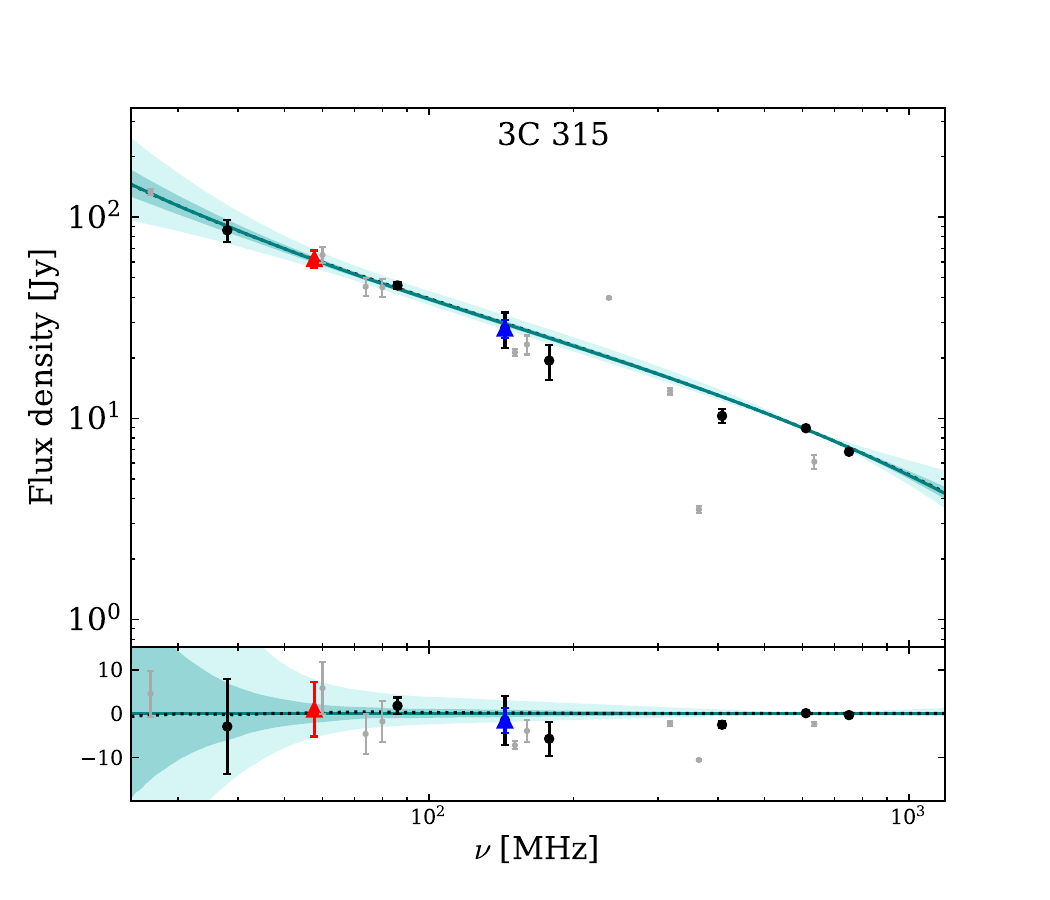}
\includegraphics[width=0.162\linewidth, trim={0.cm .0cm 1.5cm 1.5cm},clip]{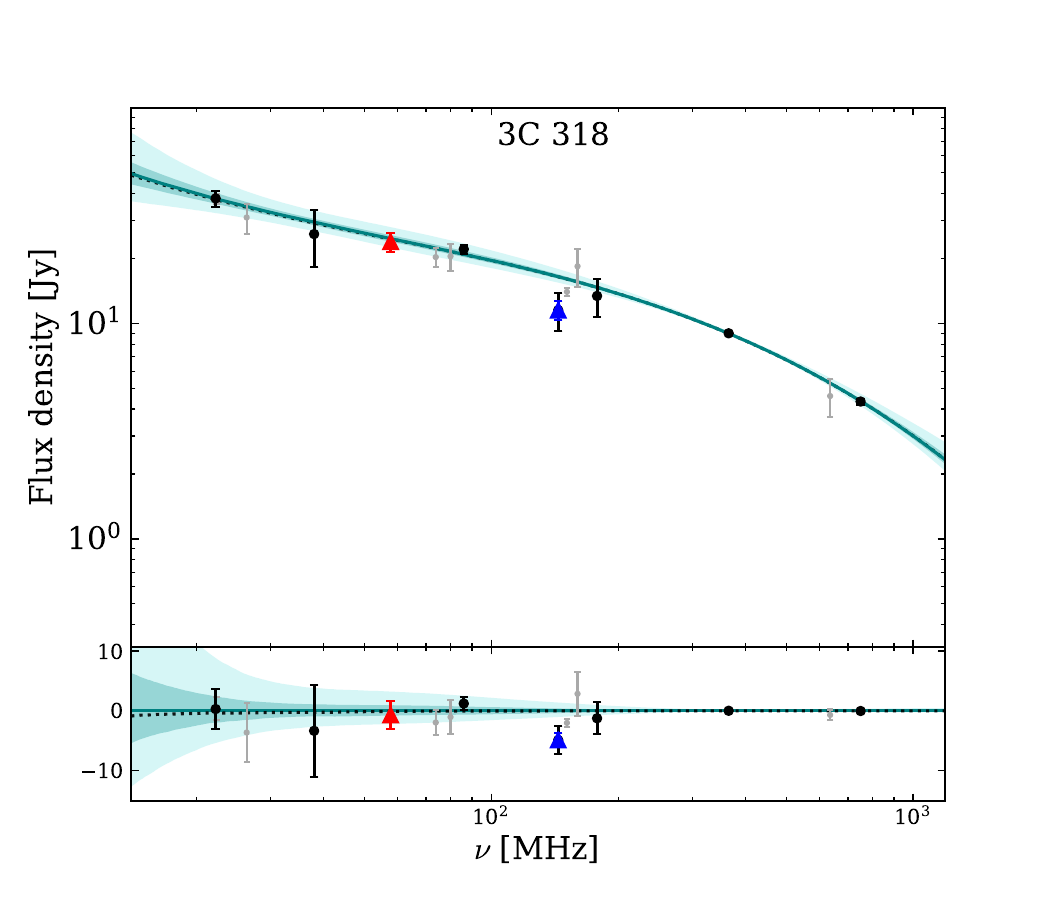}
\includegraphics[width=0.162\linewidth, trim={0.cm .0cm 1.5cm 1.5cm},clip]{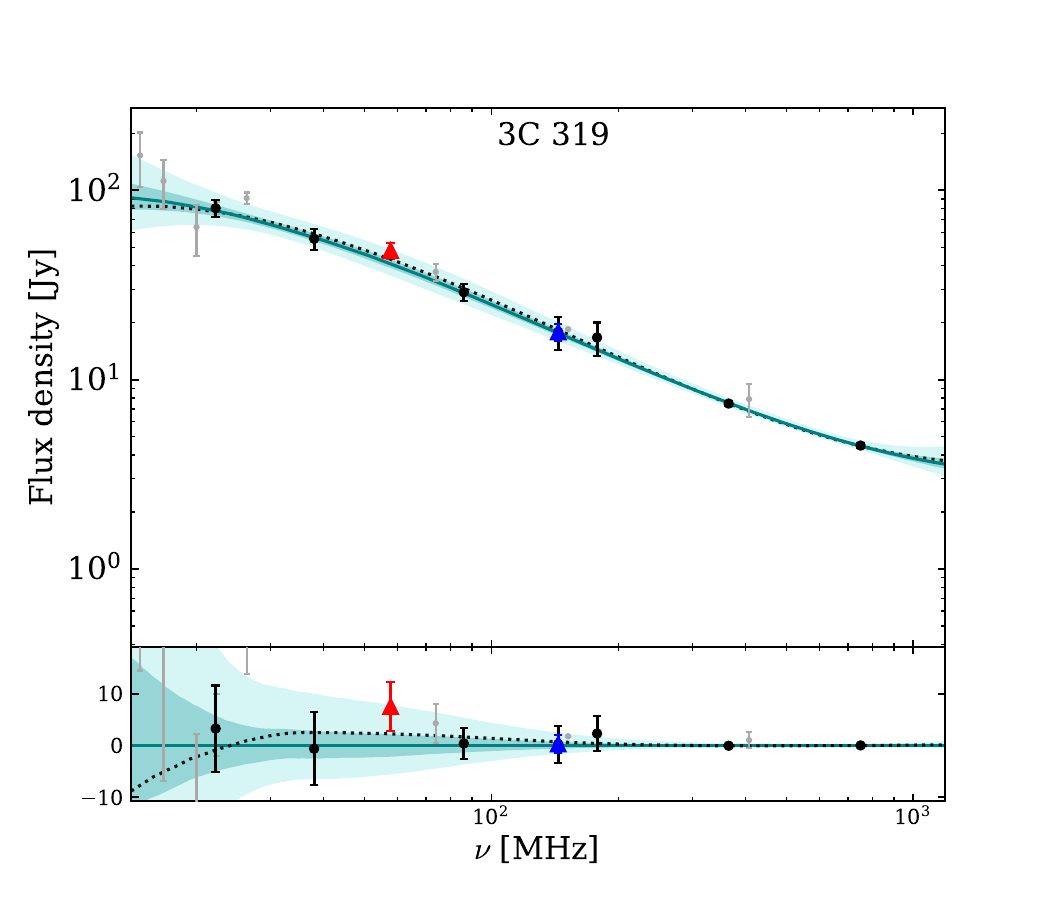}
\includegraphics[width=0.162\linewidth, trim={0.cm .0cm 1.5cm 1.5cm},clip]{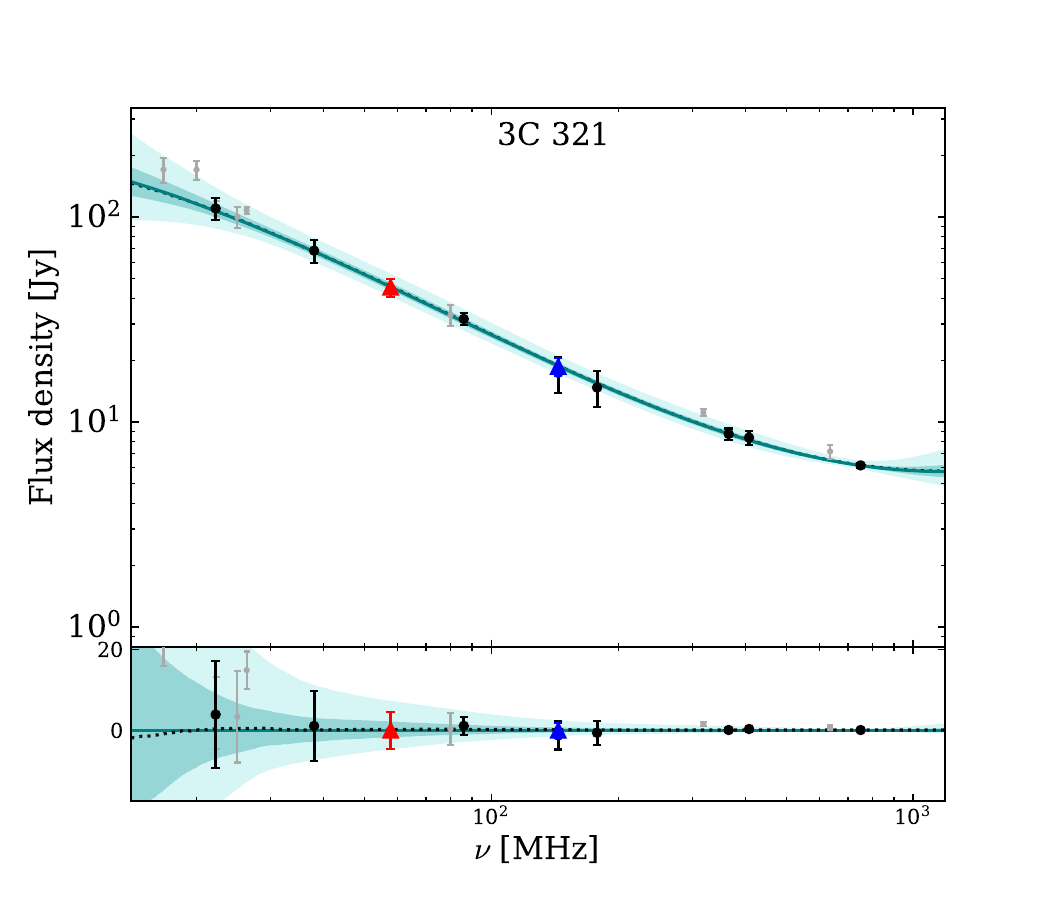}
\includegraphics[width=0.162\linewidth, trim={0.cm .0cm 1.5cm 1.5cm},clip]{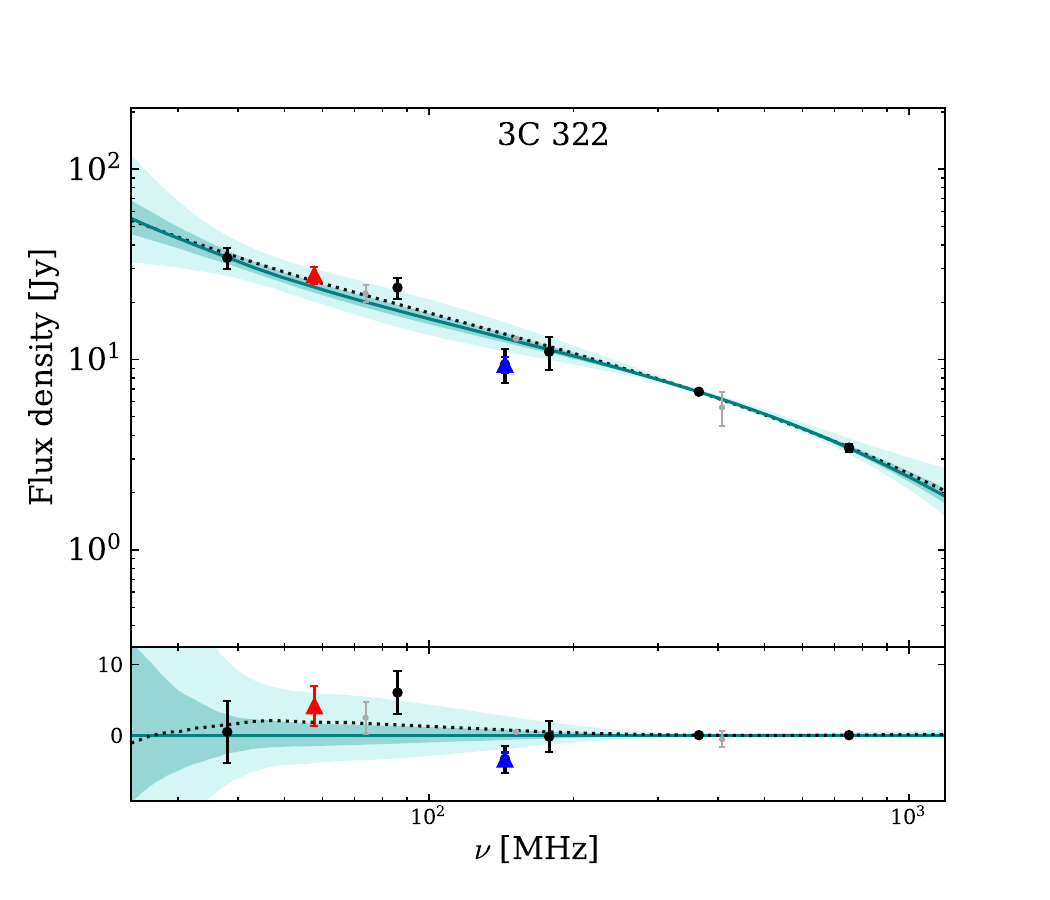}
\includegraphics[width=0.162\linewidth, trim={0.cm .0cm 1.5cm 1.5cm},clip]{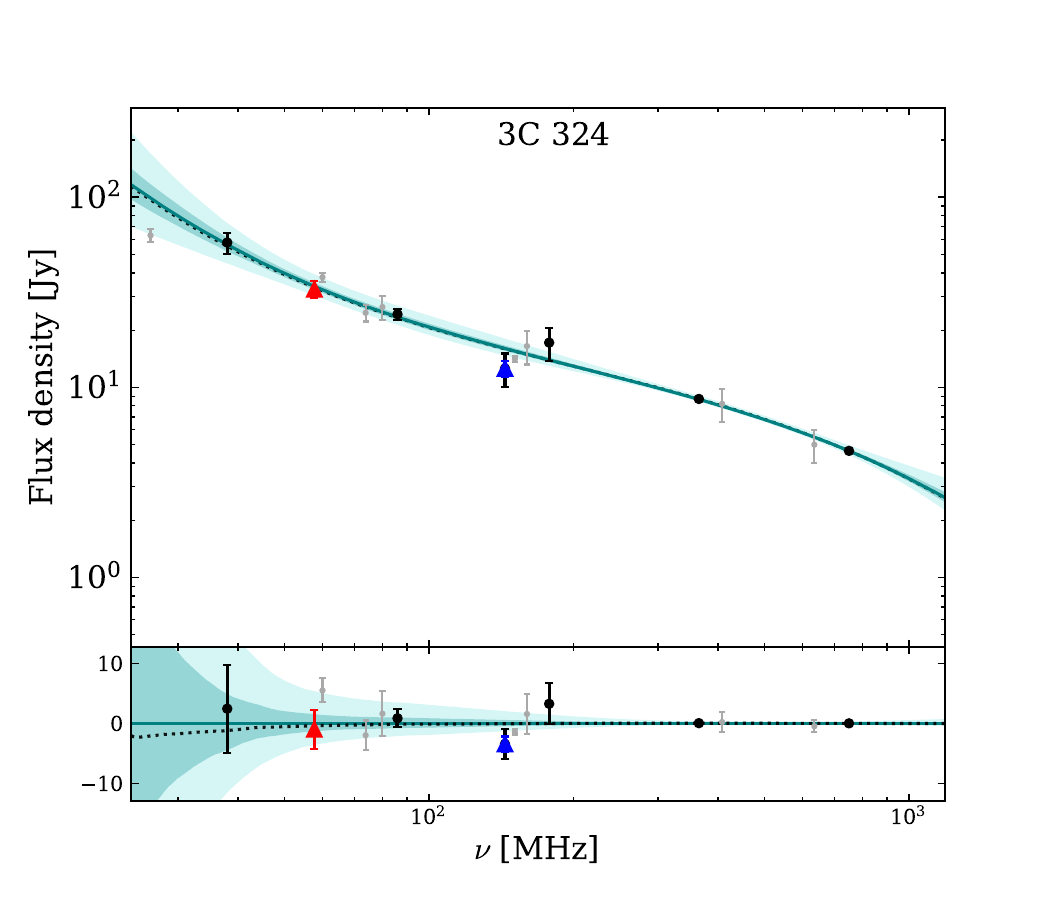}
\includegraphics[width=0.162\linewidth, trim={0.cm .0cm 1.5cm 1.5cm},clip]{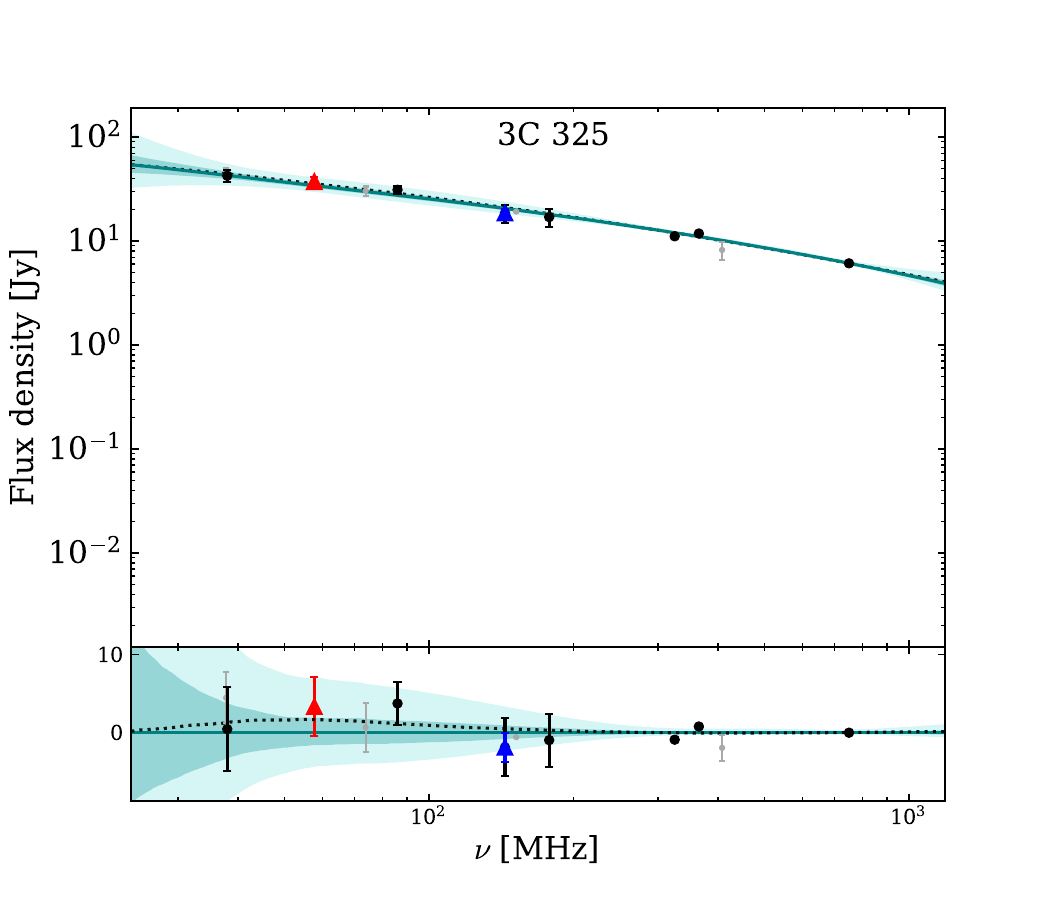}
\includegraphics[width=0.162\linewidth, trim={0.cm .0cm 1.5cm 1.5cm},clip]{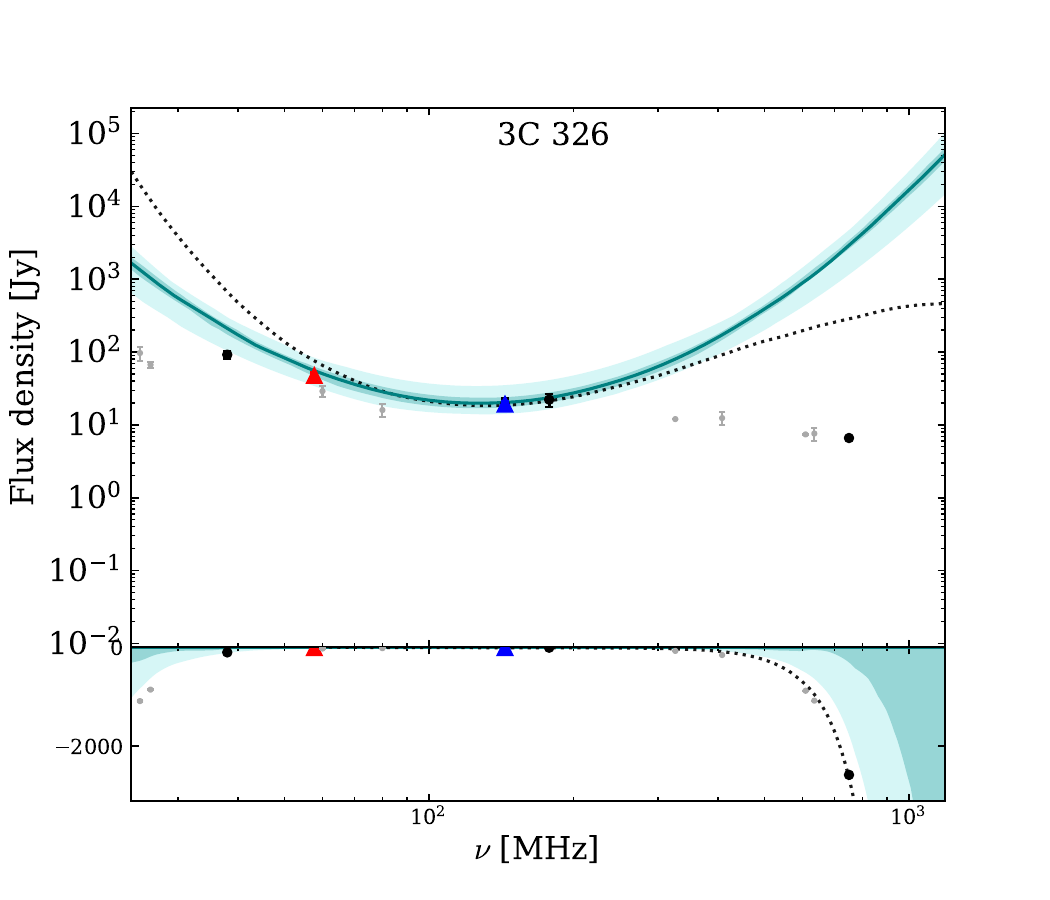}
\includegraphics[width=0.162\linewidth, trim={0.cm .0cm 1.5cm 1.5cm},clip]{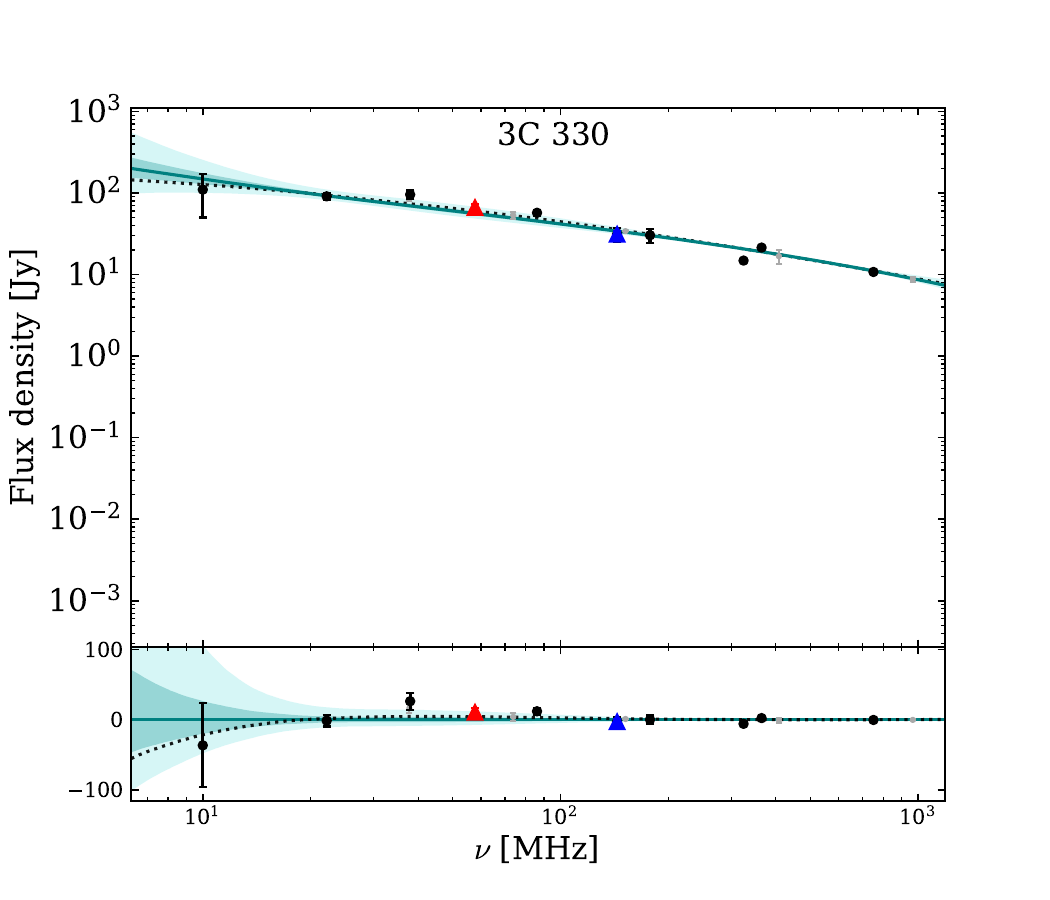}
\includegraphics[width=0.162\linewidth, trim={0.cm .0cm 1.5cm 1.5cm},clip]{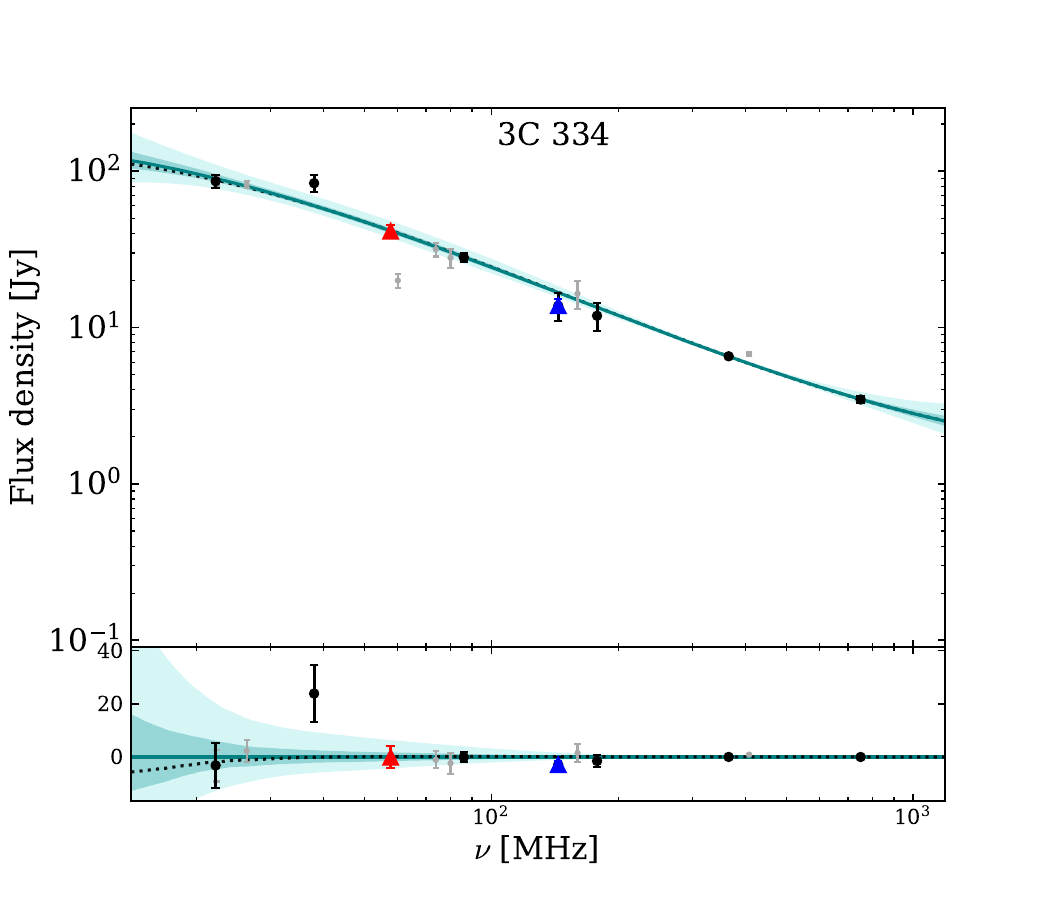}
\includegraphics[width=0.162\linewidth, trim={0.cm .0cm 1.5cm 1.5cm},clip]{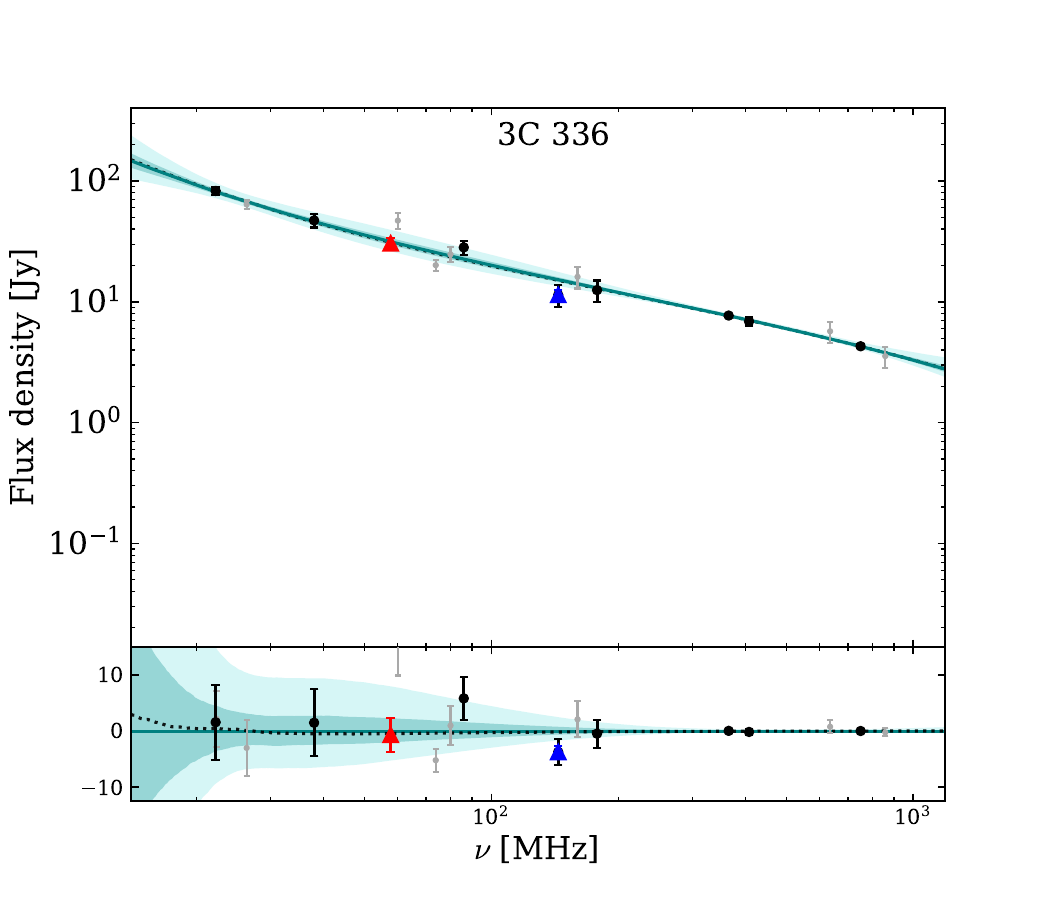}
\includegraphics[width=0.162\linewidth, trim={0.cm .0cm 1.5cm 1.5cm},clip]{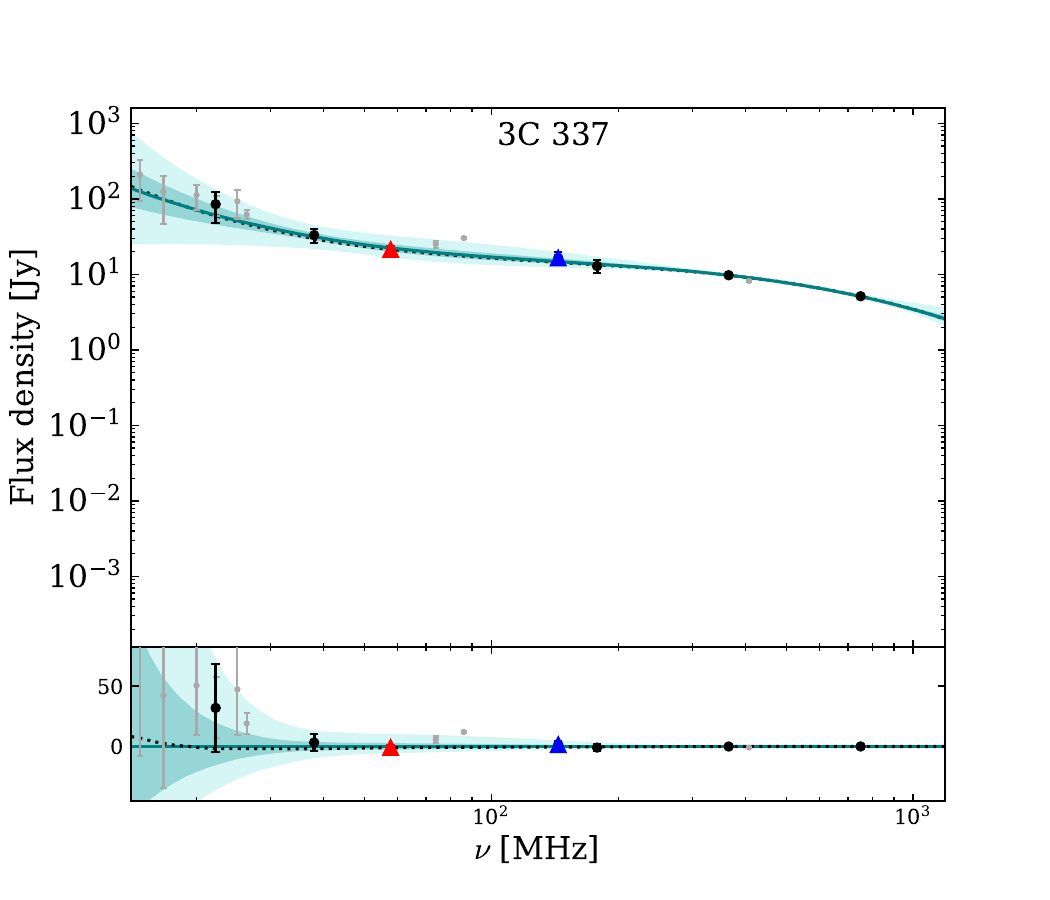}
\includegraphics[width=0.162\linewidth, trim={0.cm .0cm 1.5cm 1.5cm},clip]{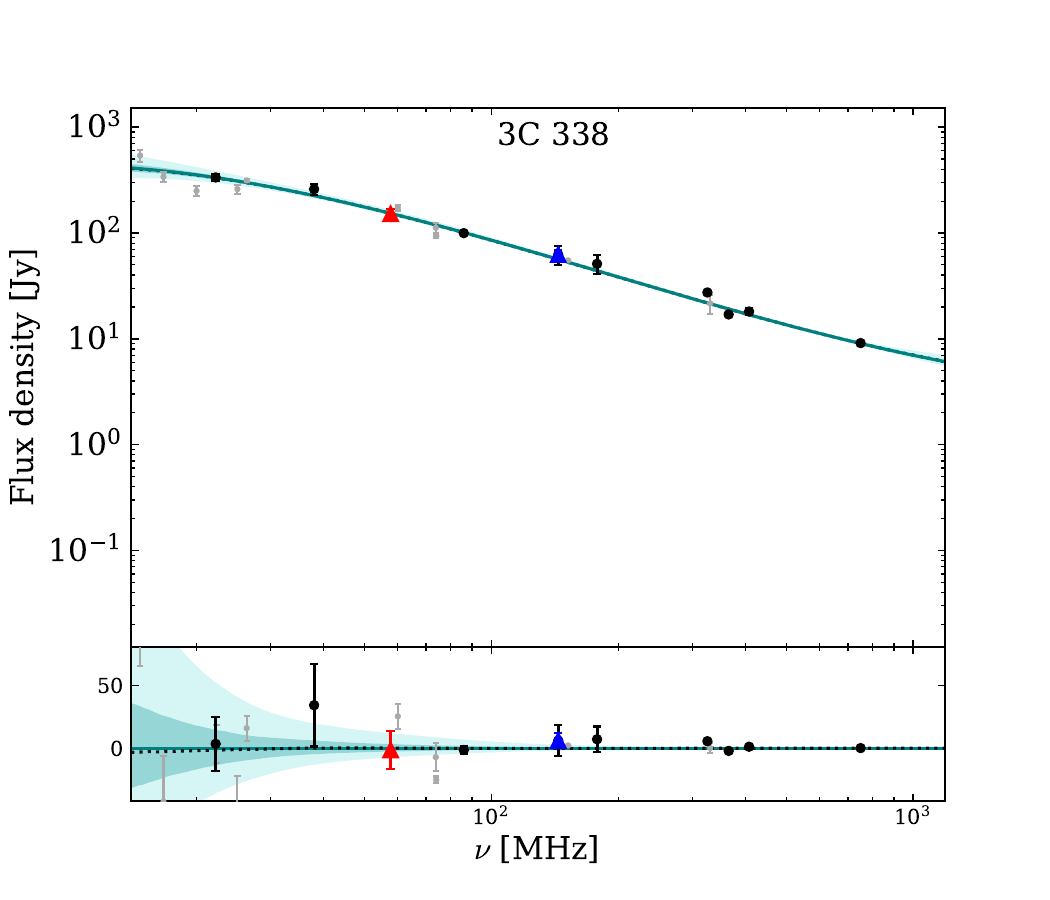}
\includegraphics[width=0.162\linewidth, trim={0.cm .0cm 1.5cm 1.5cm},clip]{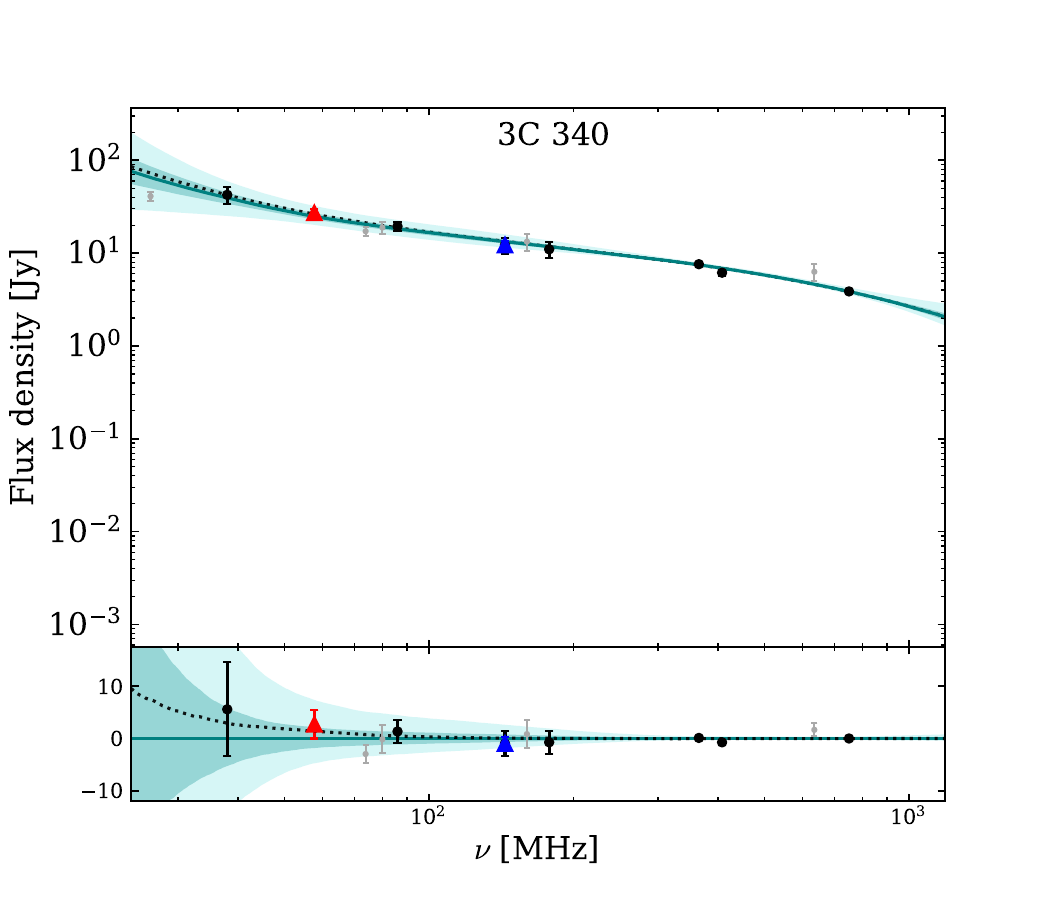}
\includegraphics[width=0.162\linewidth, trim={0.cm .0cm 1.5cm 1.5cm},clip]{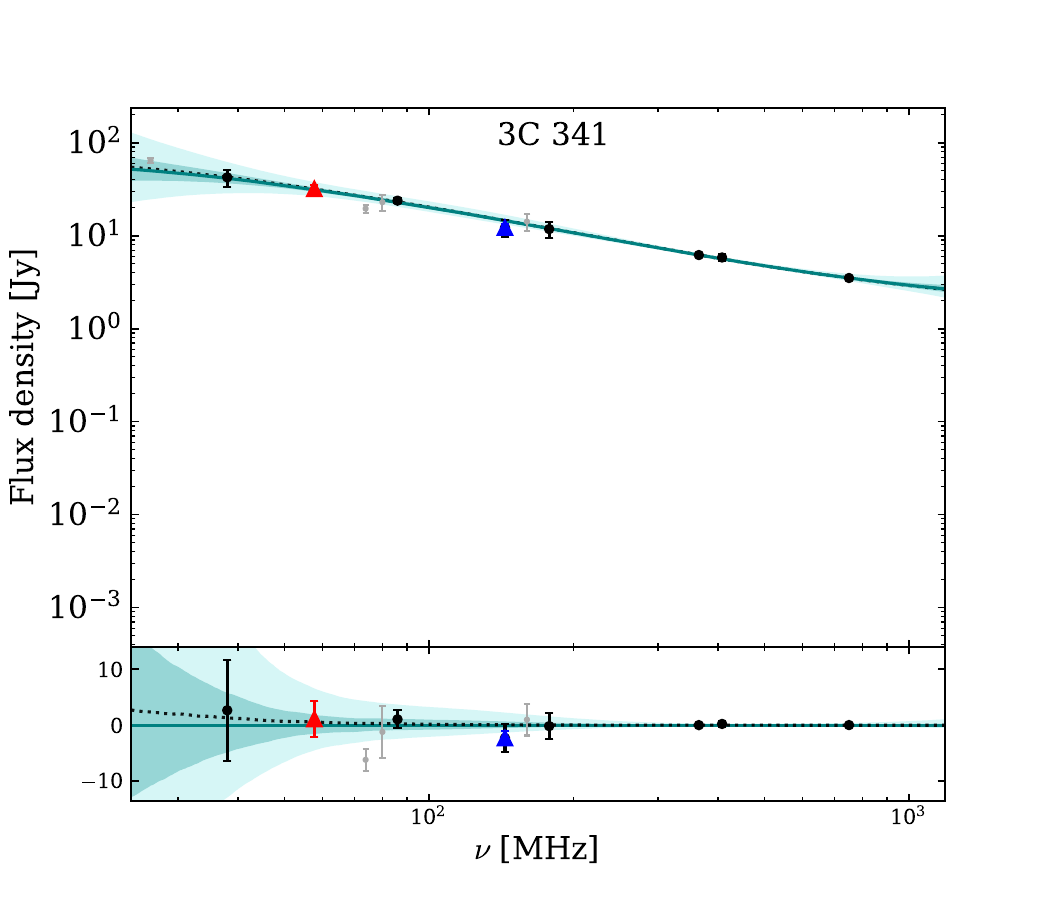}
\includegraphics[width=0.162\linewidth, trim={0.cm .0cm 1.5cm 1.5cm},clip]{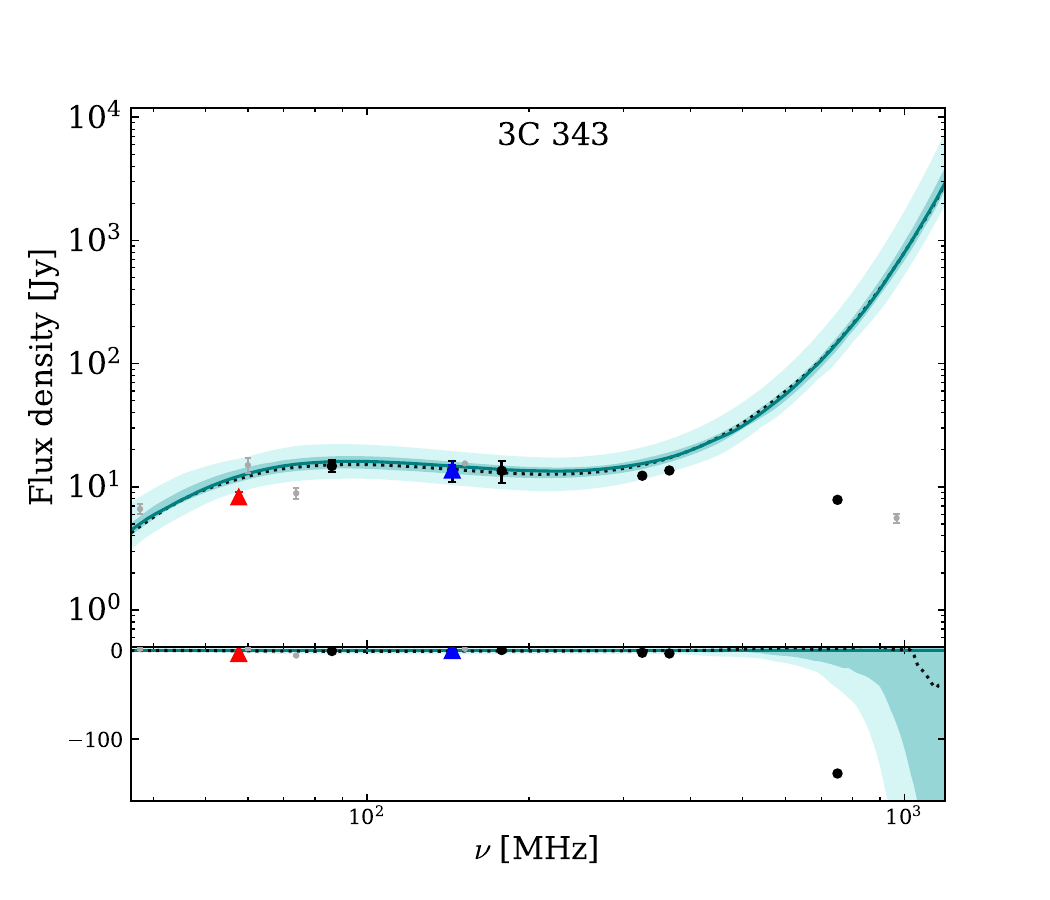}
\includegraphics[width=0.162\linewidth, trim={0.cm .0cm 1.5cm 1.5cm},clip]{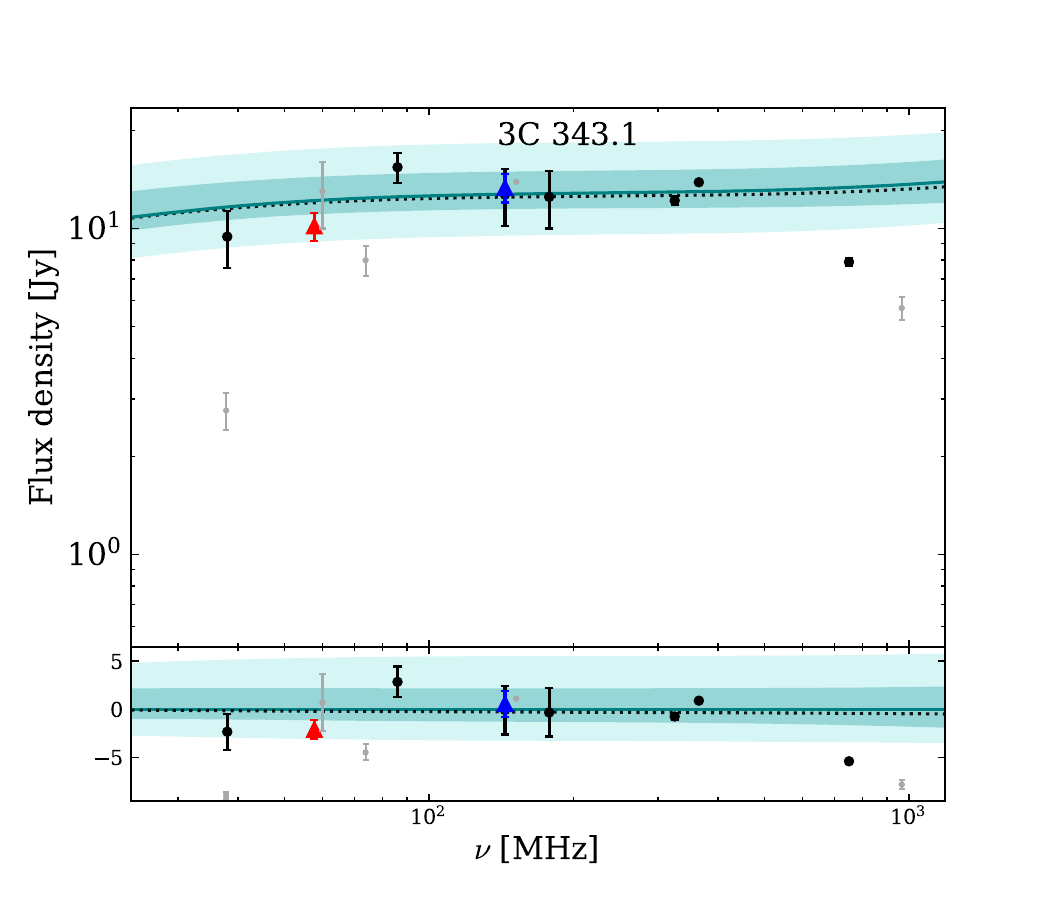}
\includegraphics[width=0.162\linewidth, trim={0.cm .0cm 1.5cm 1.5cm},clip]{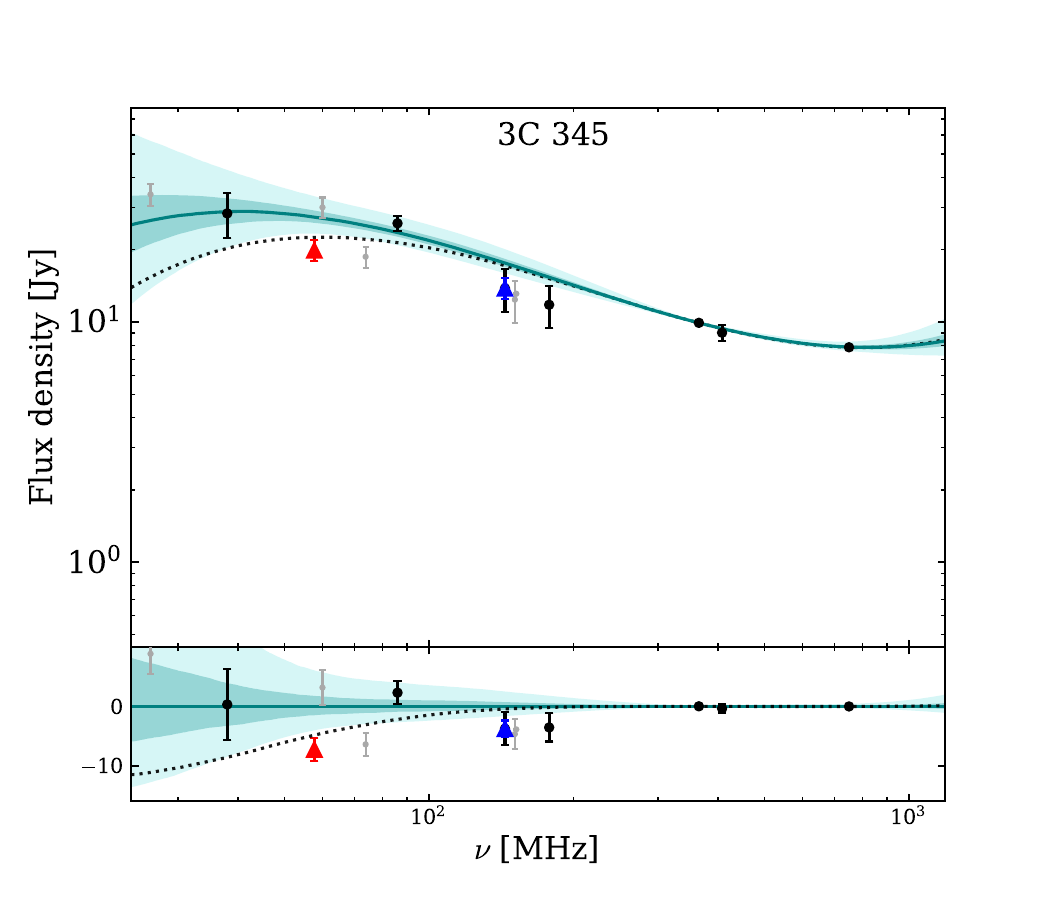}
\includegraphics[width=0.162\linewidth, trim={0.cm .0cm 1.5cm 1.5cm},clip]{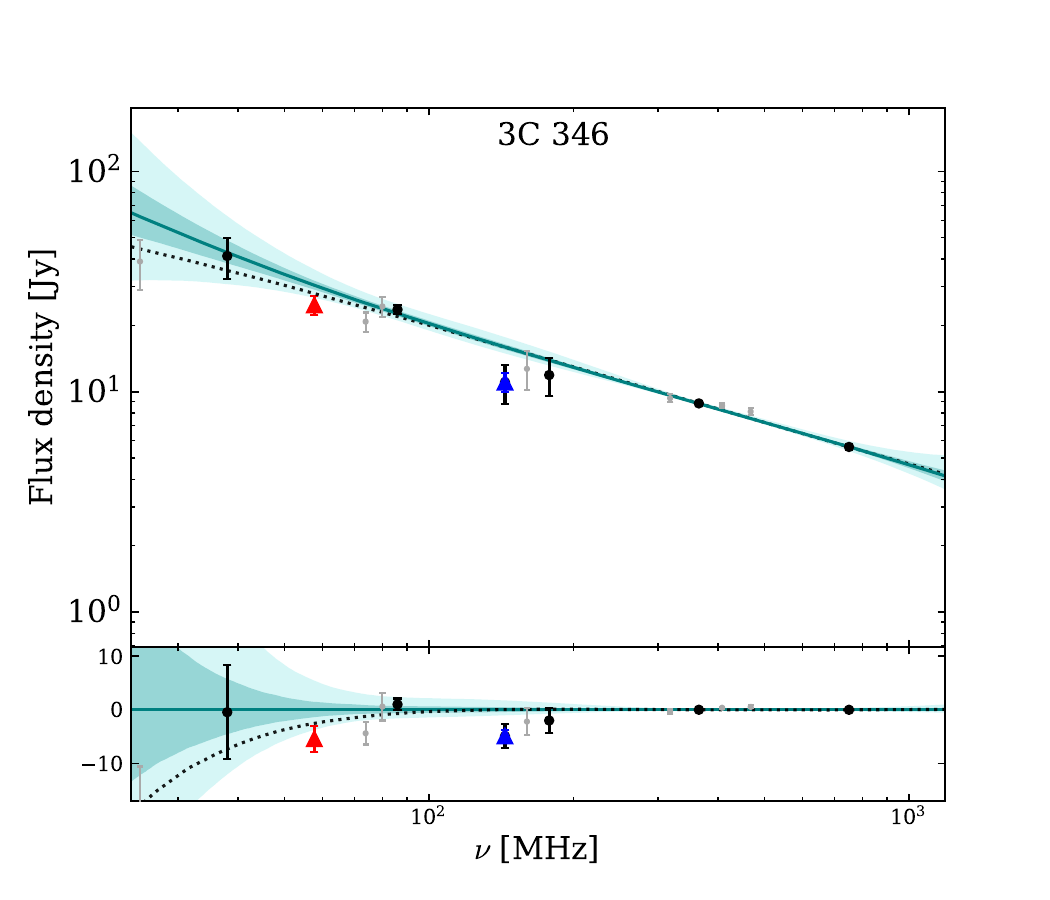}
\includegraphics[width=0.162\linewidth, trim={0.cm .0cm 1.5cm 1.5cm},clip]{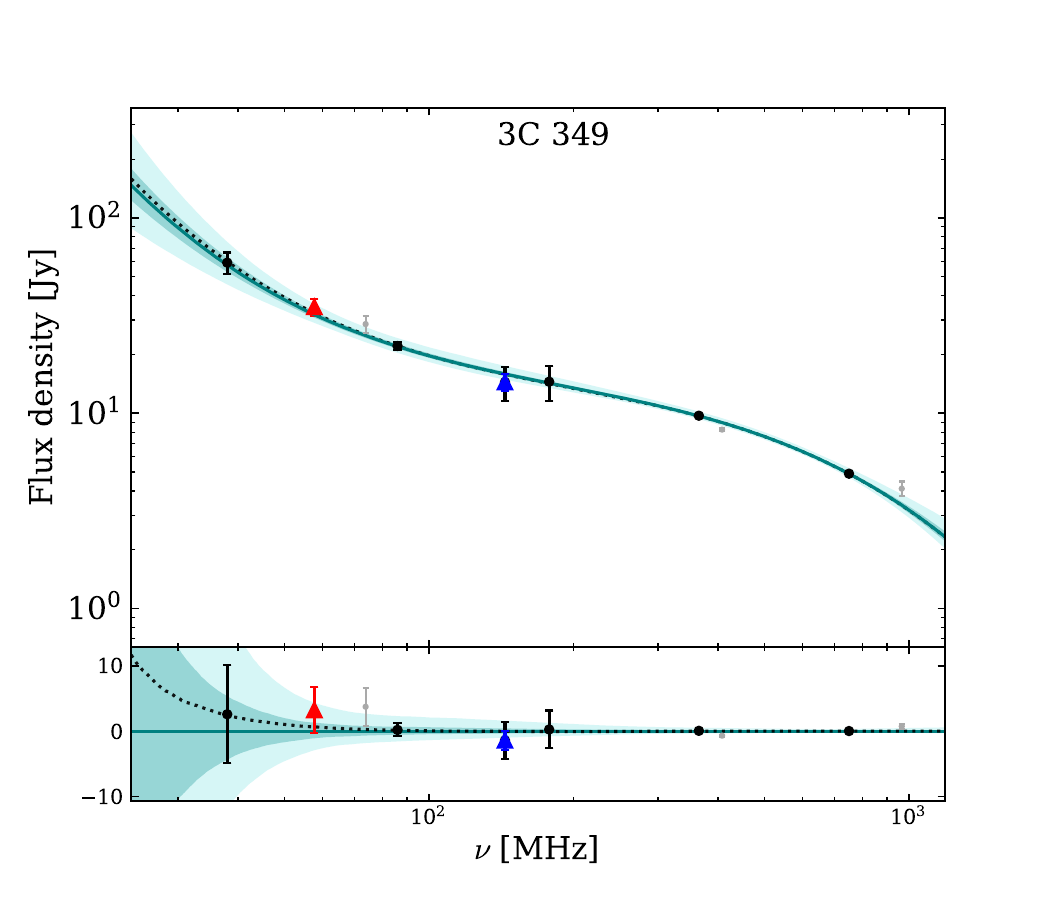}
\includegraphics[width=0.162\linewidth, trim={0.cm .0cm 1.5cm 1.5cm},clip]{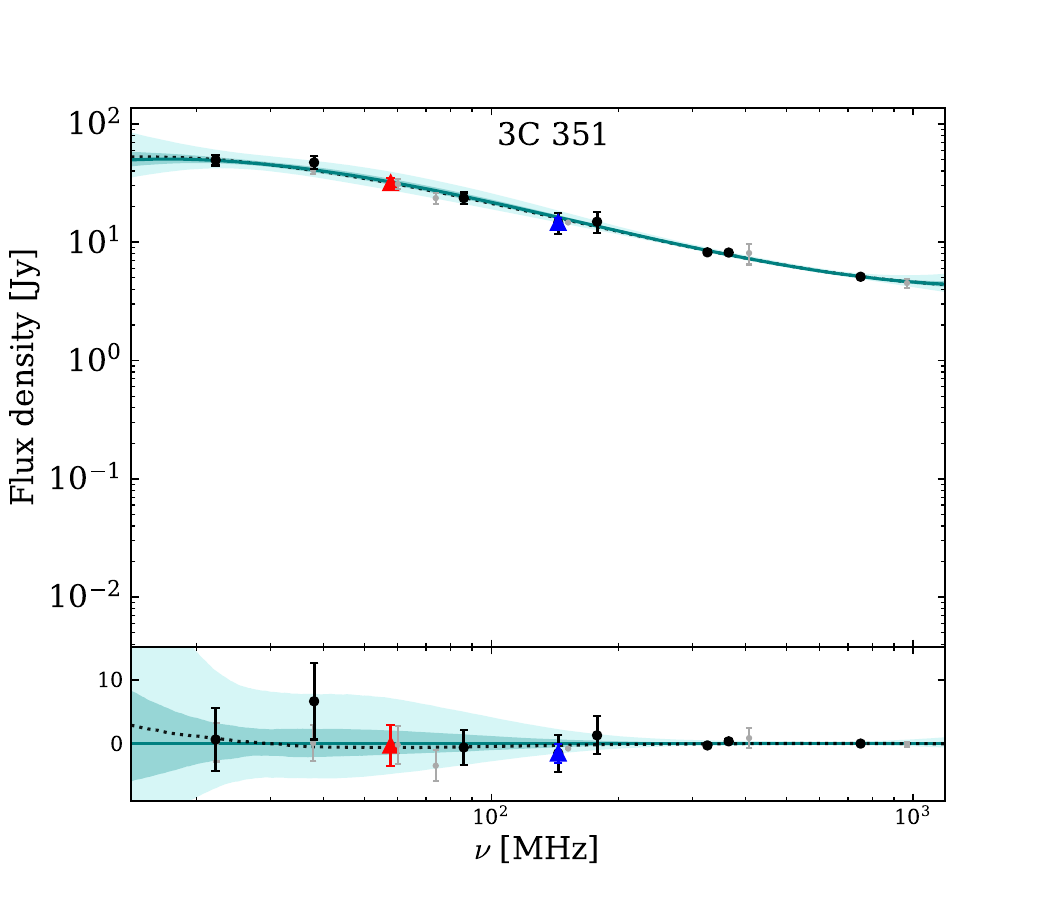}
\includegraphics[width=0.162\linewidth, trim={0.cm .0cm 1.5cm 1.5cm},clip]{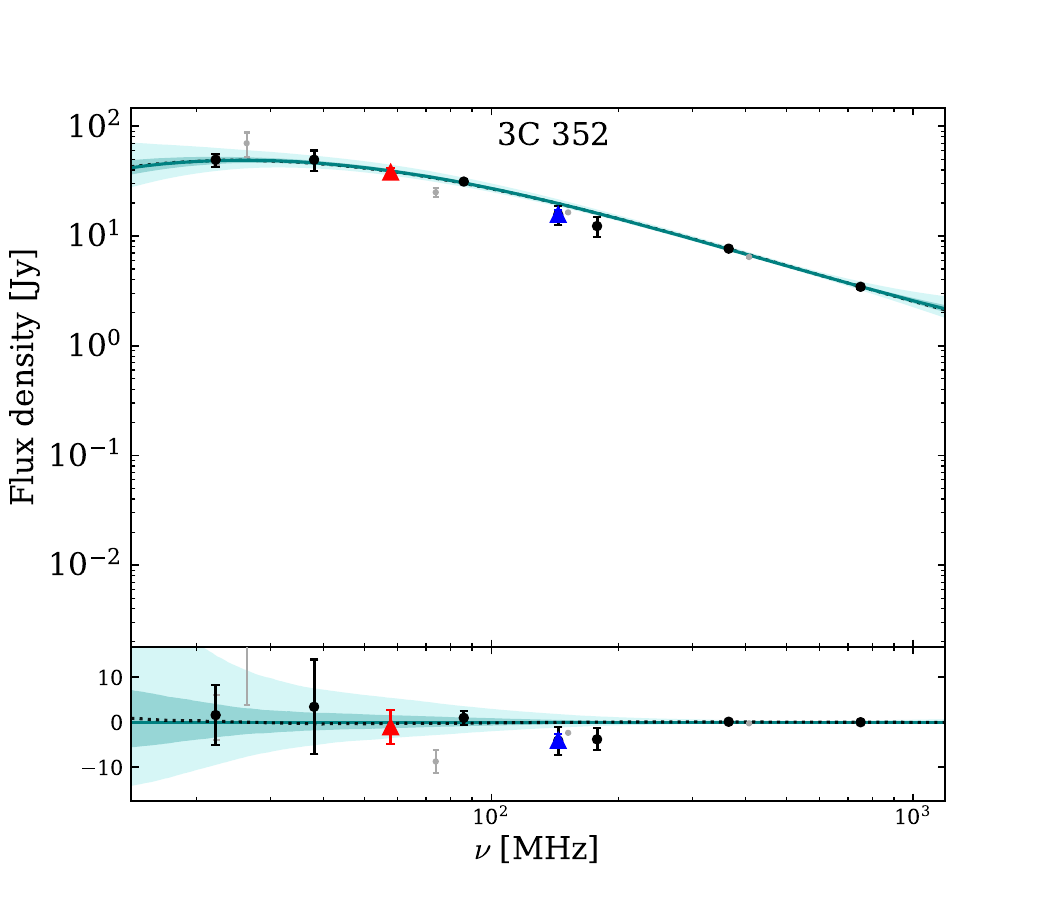}
\includegraphics[width=0.162\linewidth, trim={0.cm .0cm 1.5cm 1.5cm},clip]{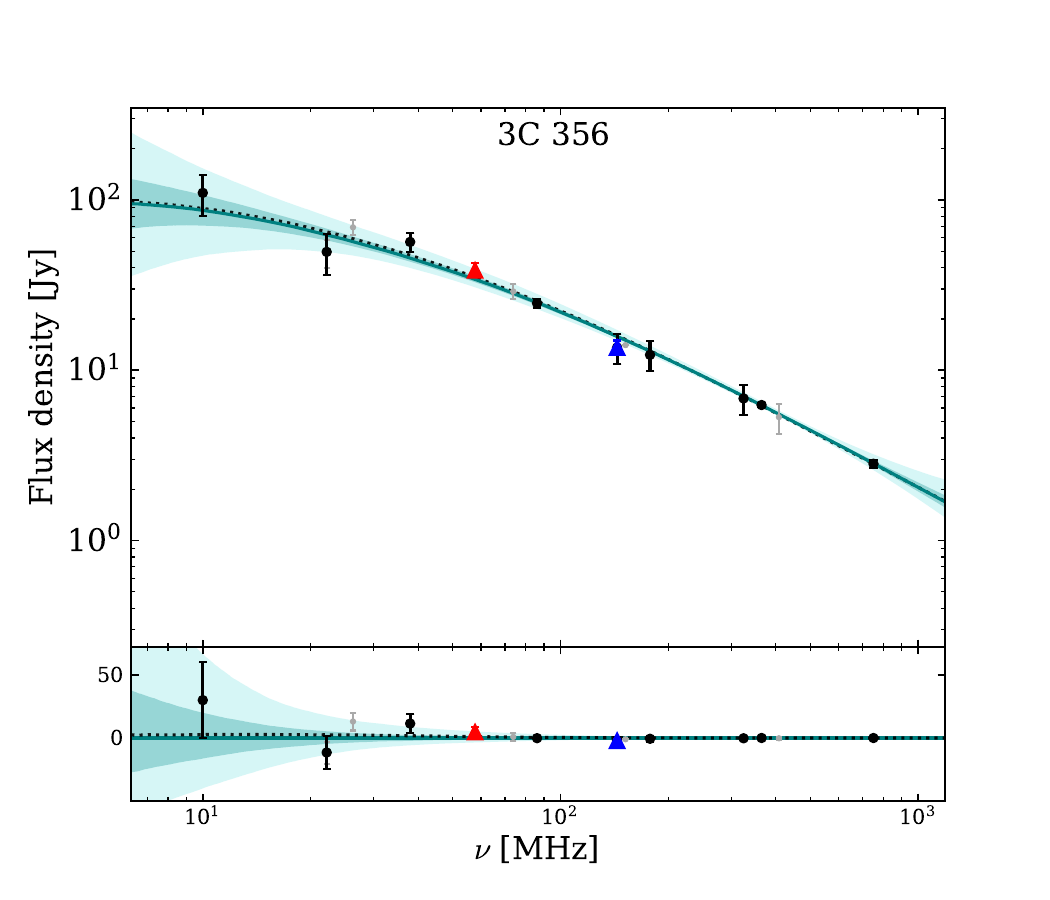}
\includegraphics[width=0.162\linewidth, trim={0.cm .0cm 1.5cm 1.5cm},clip]{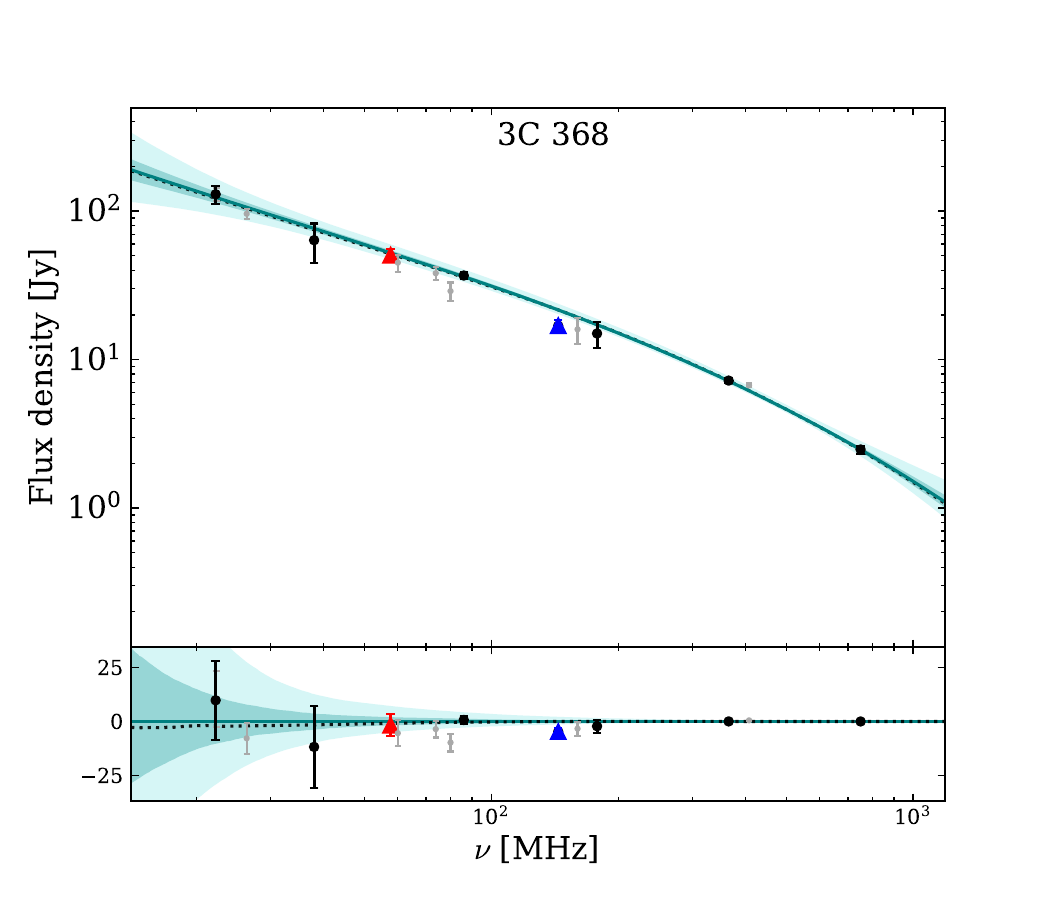}
\includegraphics[width=0.162\linewidth, trim={0.cm .0cm 1.5cm 1.5cm},clip]{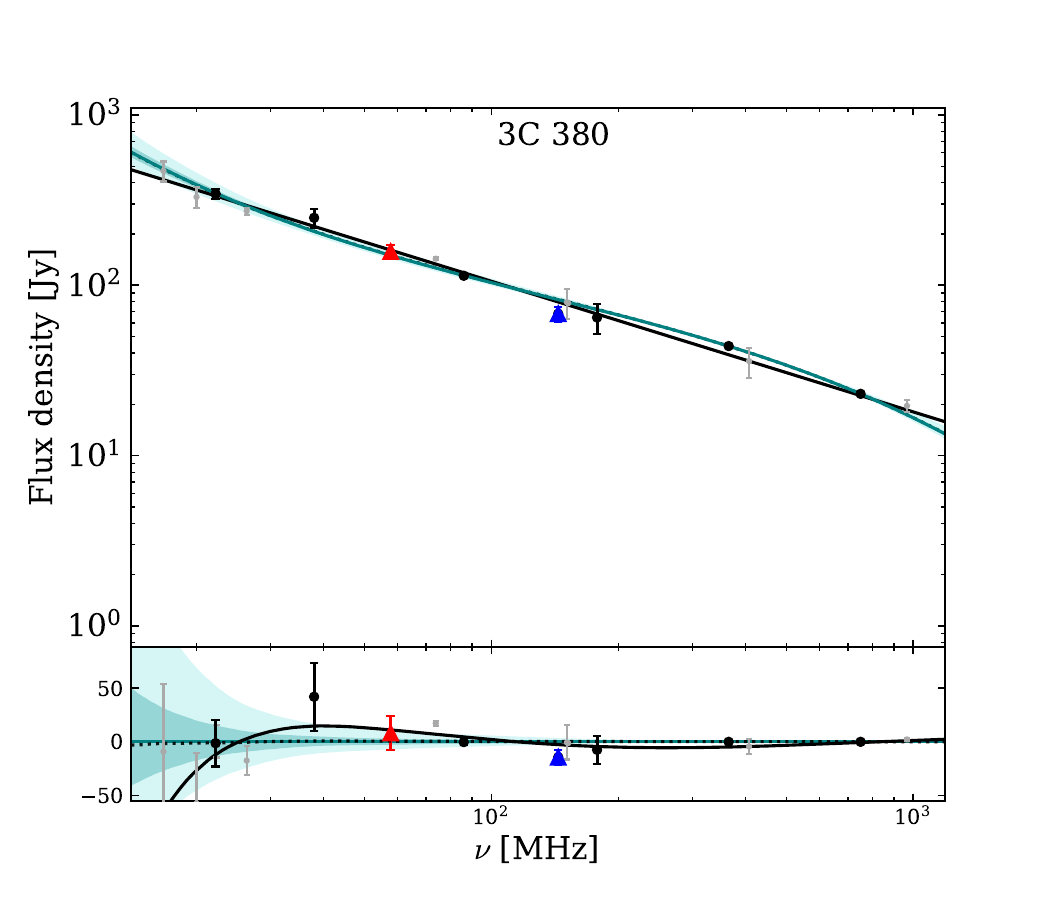}
\includegraphics[width=0.162\linewidth, trim={0.cm .0cm 1.5cm 1.5cm},clip]{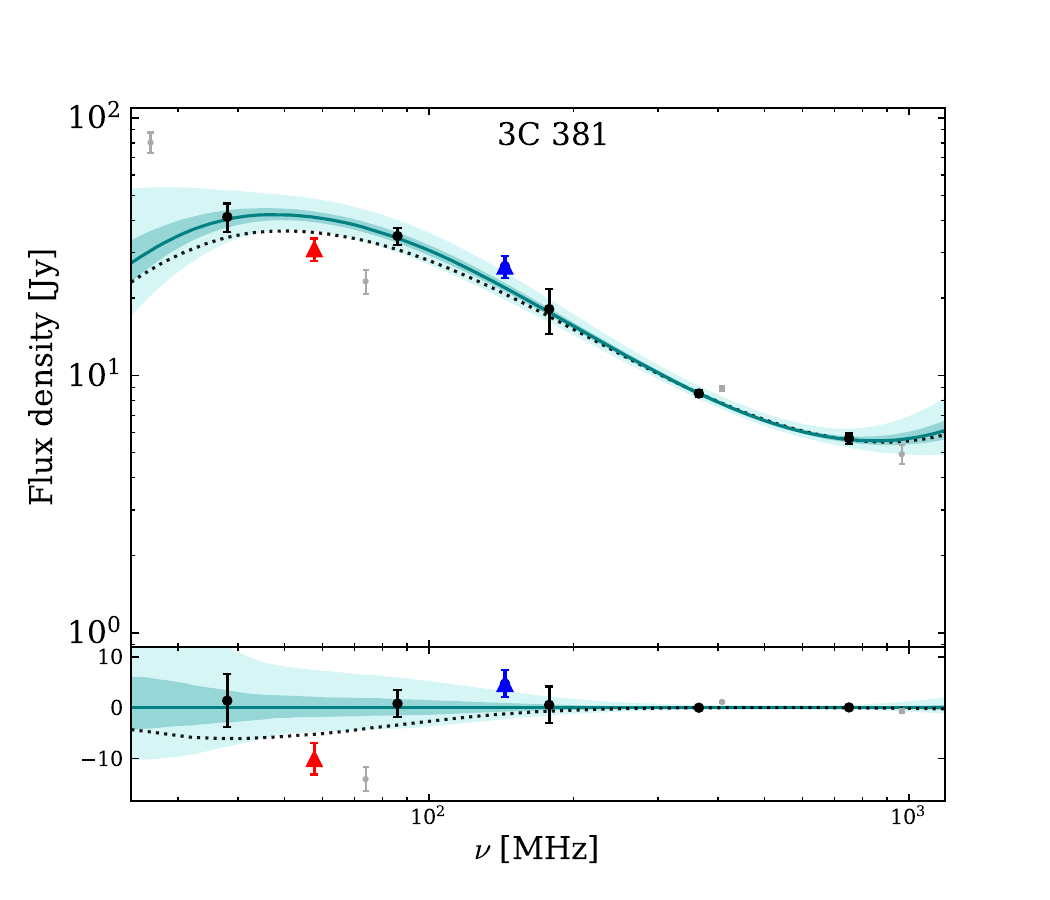}
\includegraphics[width=0.162\linewidth, trim={0.cm .0cm 1.5cm 1.5cm},clip]{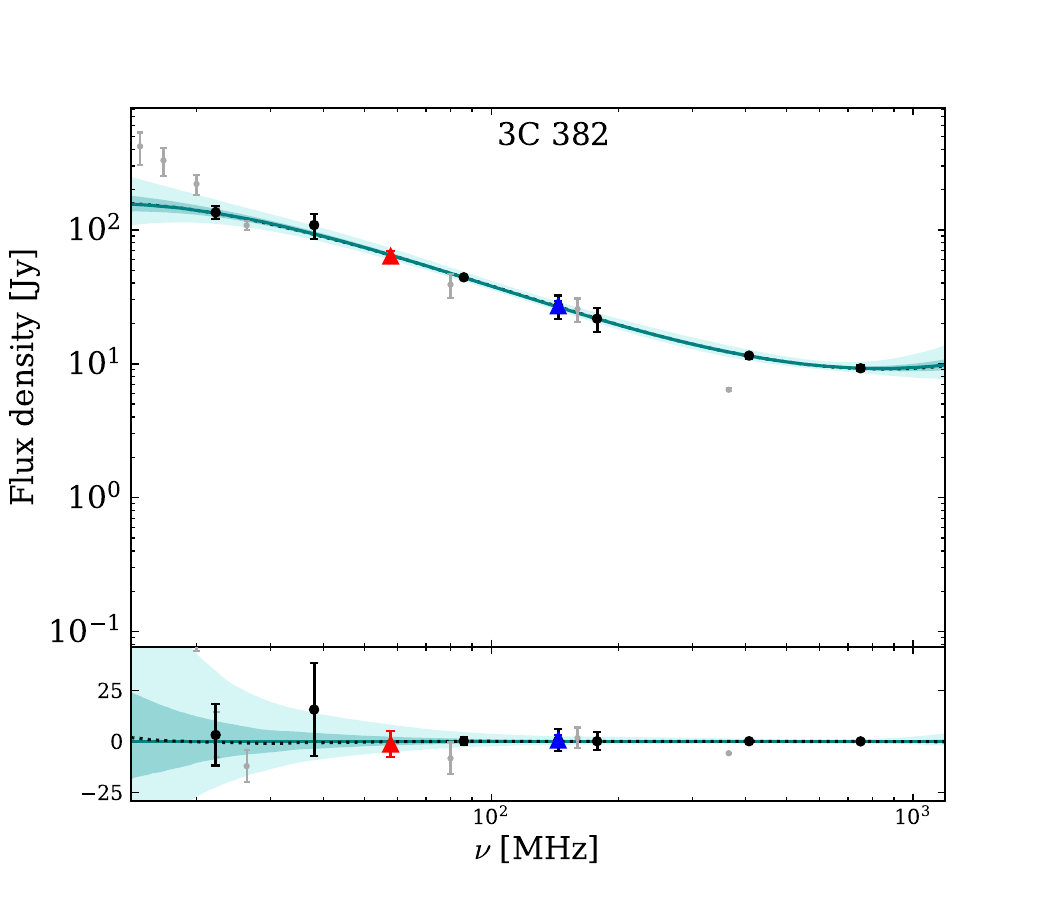}
\includegraphics[width=0.162\linewidth, trim={0.cm .0cm 1.5cm 1.5cm},clip]{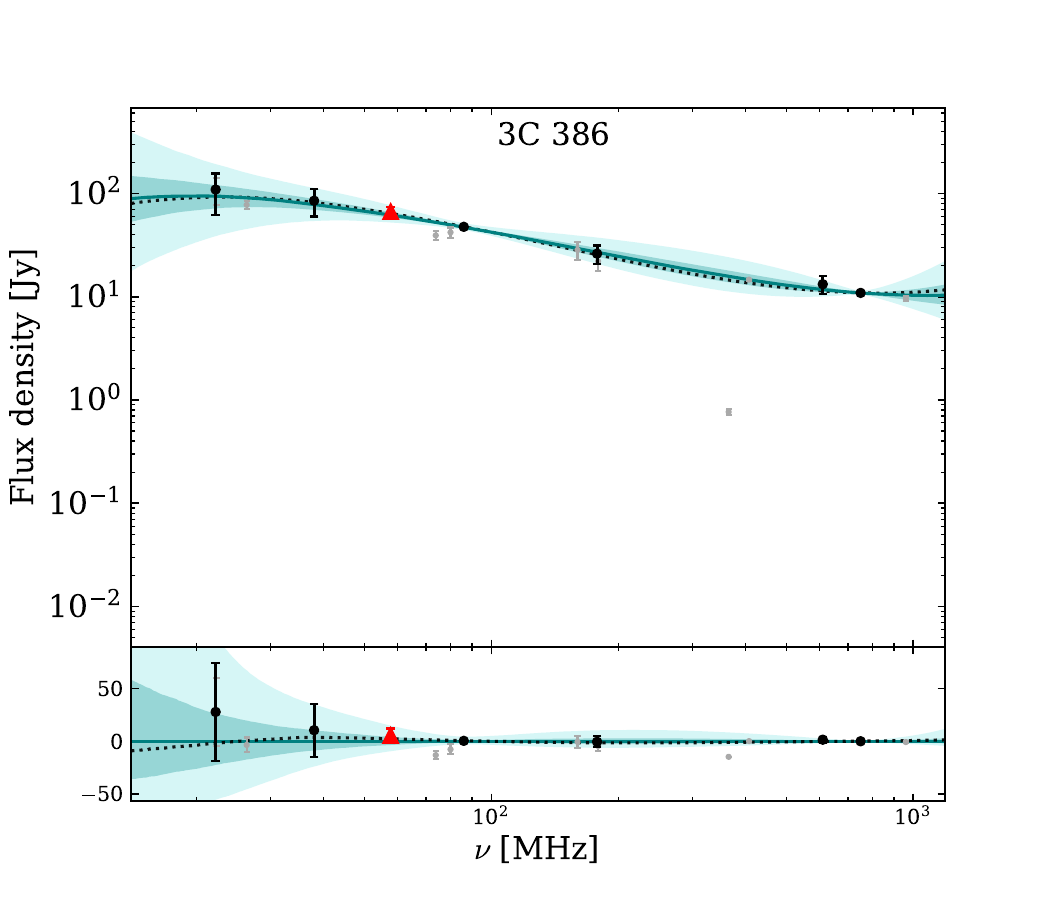}
\includegraphics[width=0.162\linewidth, trim={0.cm .0cm 1.5cm 1.5cm},clip]{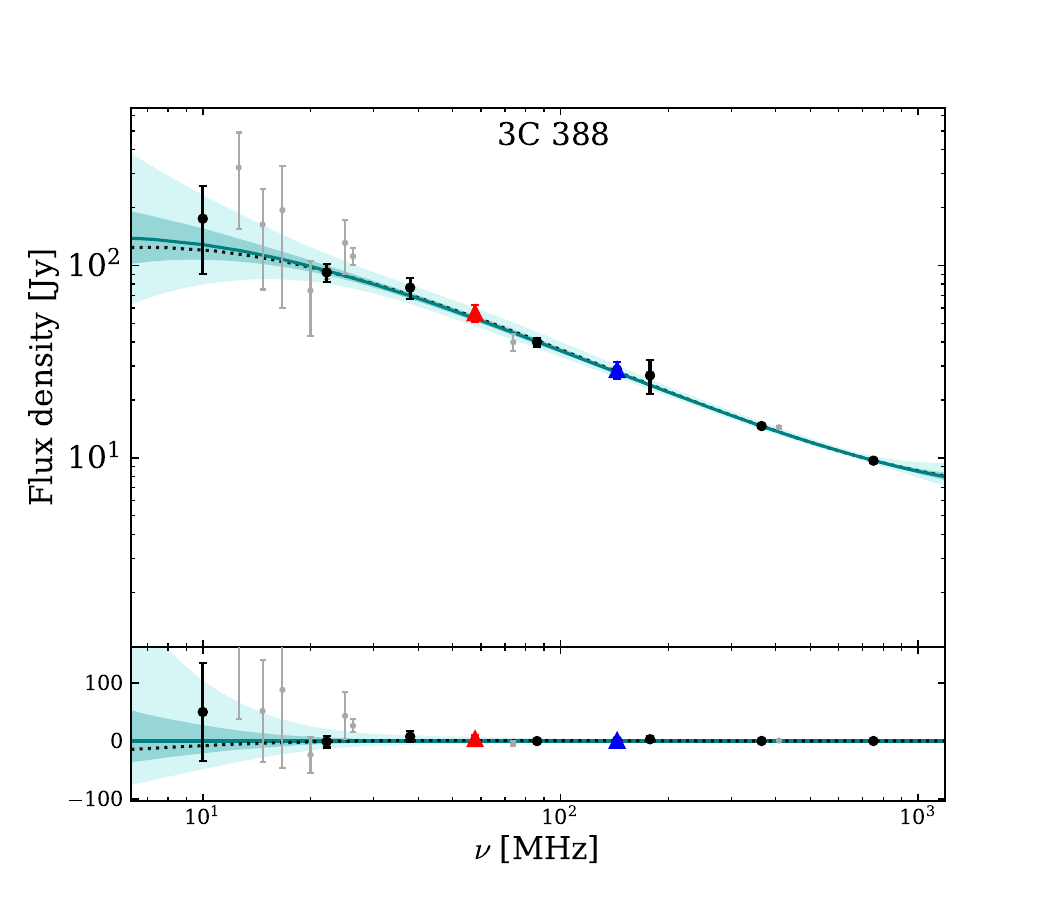}
\includegraphics[width=0.162\linewidth, trim={0.cm .0cm 1.5cm 1.5cm},clip]{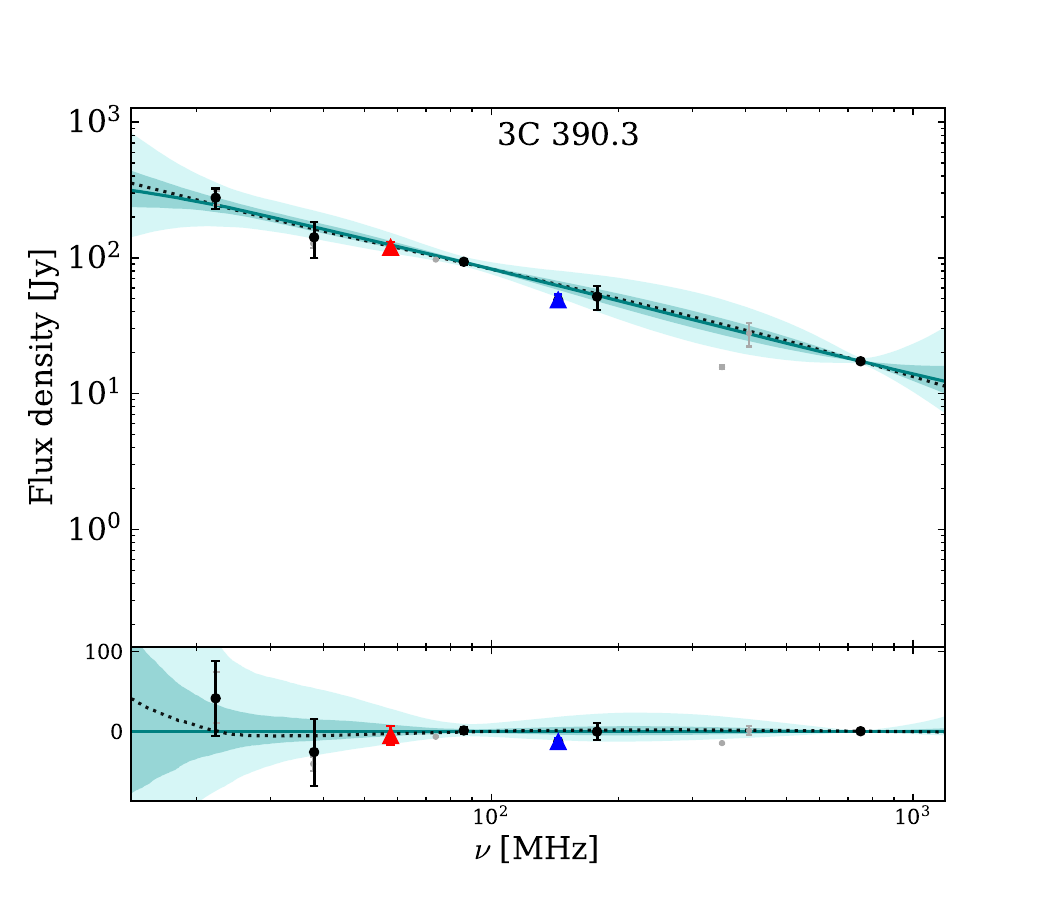}
\includegraphics[width=0.162\linewidth, trim={0.cm .0cm 1.5cm 1.5cm},clip]{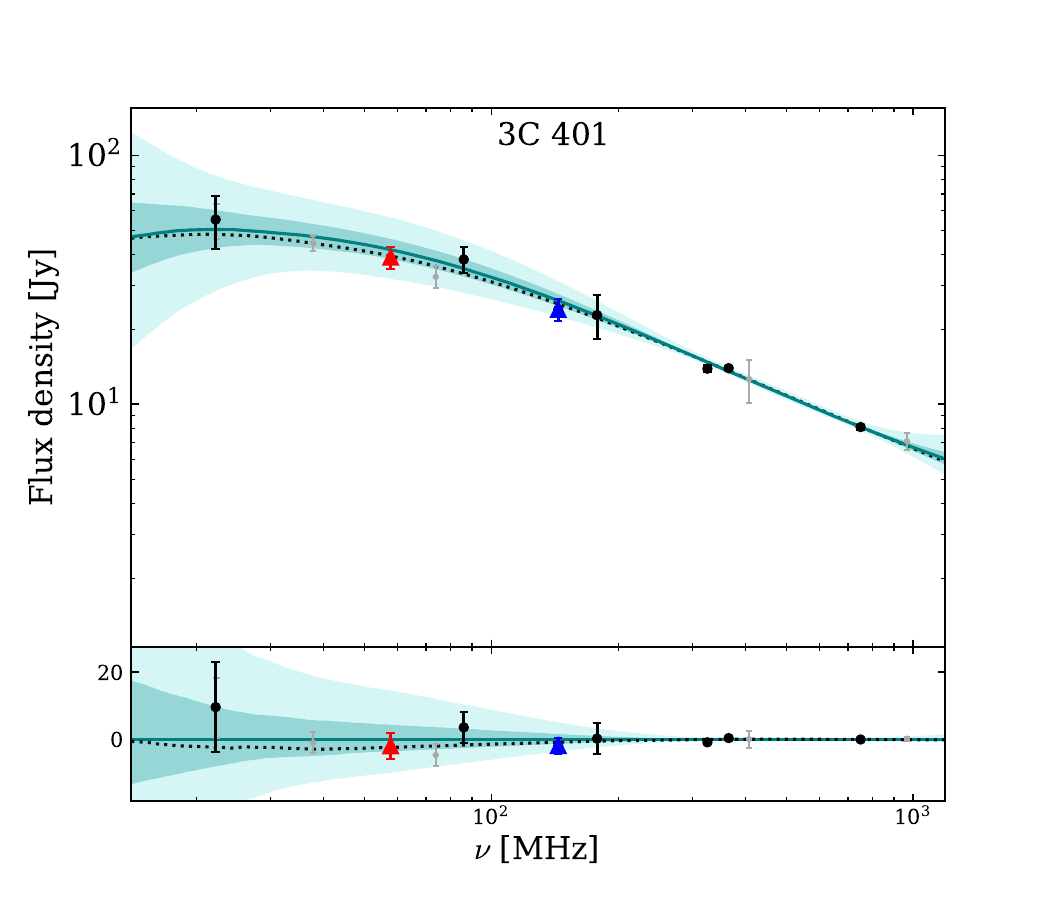}
\includegraphics[width=0.162\linewidth, trim={0.cm .0cm 1.5cm 1.5cm},clip]{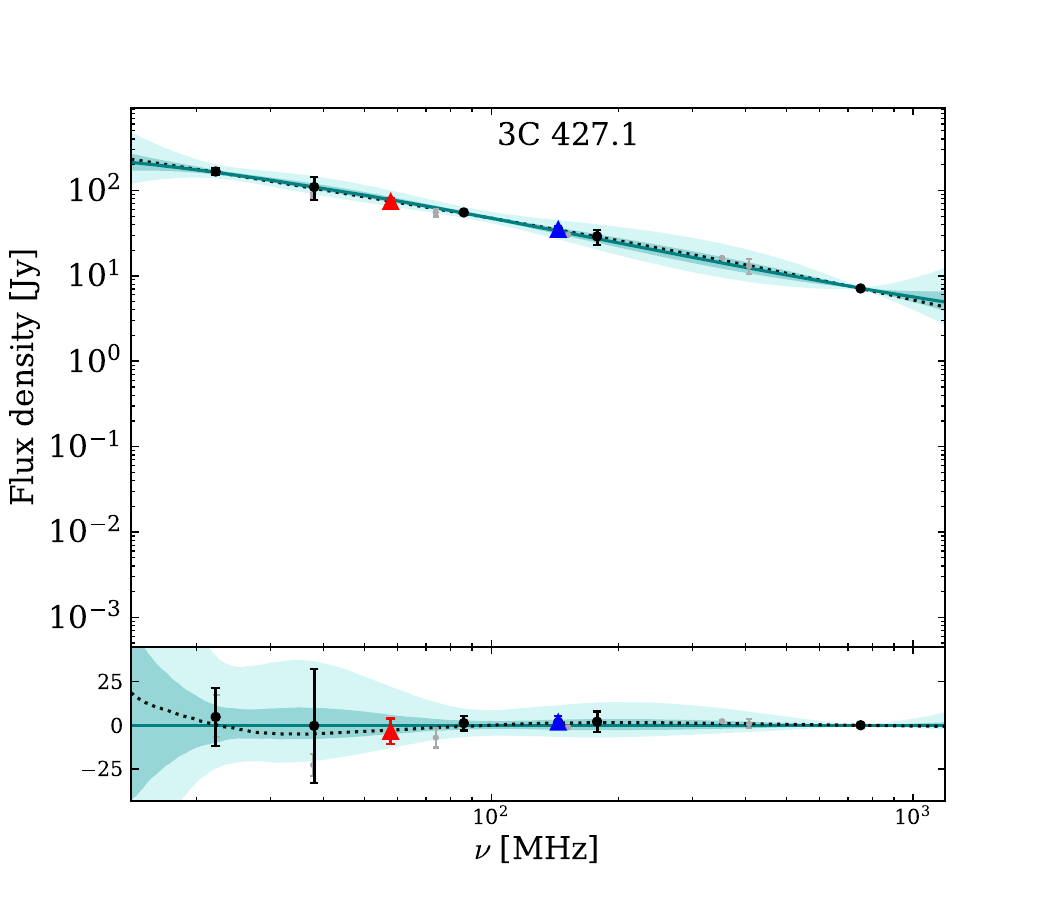}
\includegraphics[width=0.162\linewidth, trim={0.cm .0cm 1.5cm 1.5cm},clip]{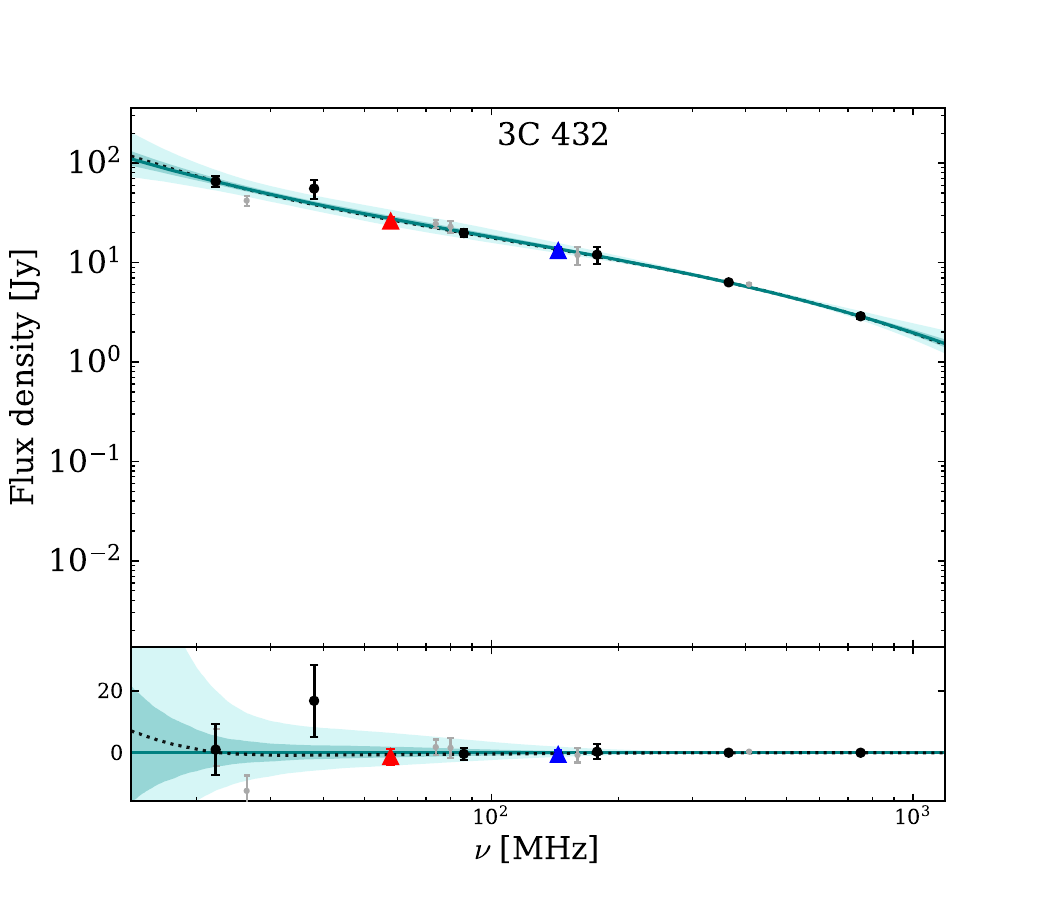}
\includegraphics[width=0.162\linewidth, trim={0.cm .0cm 1.5cm 1.5cm},clip]{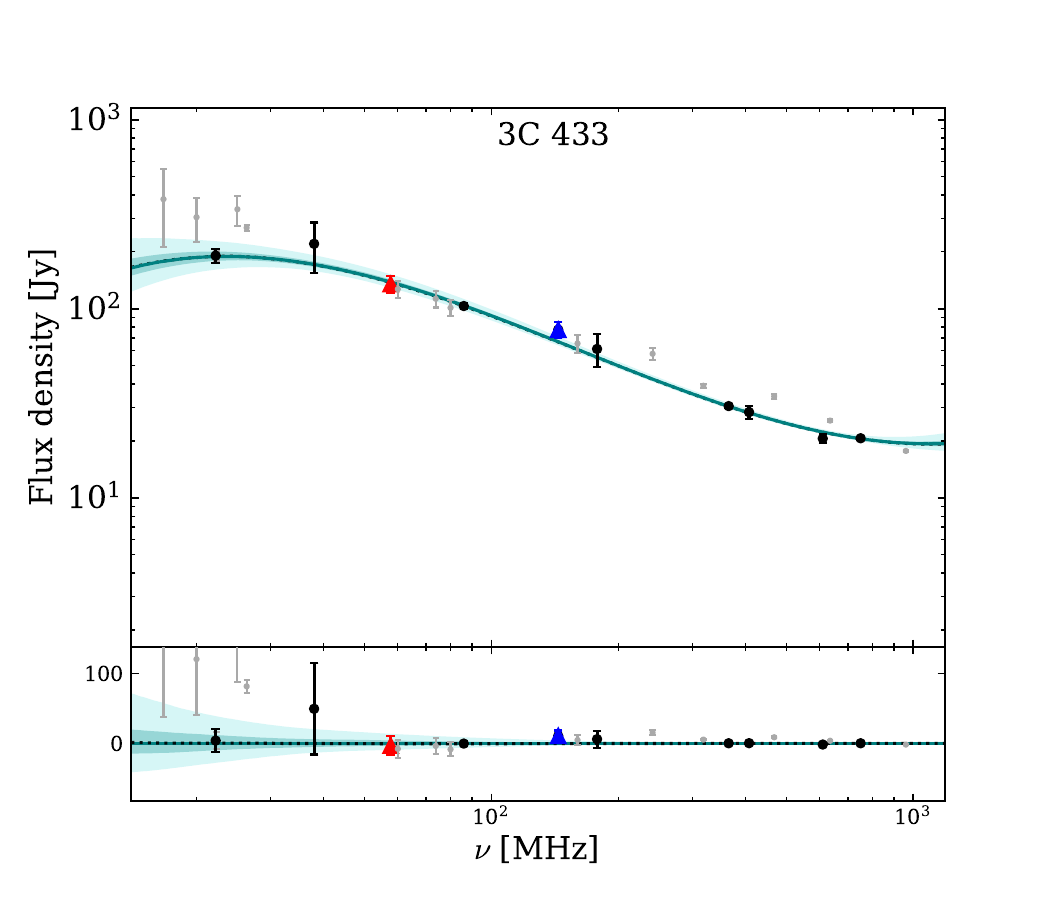}
\includegraphics[width=0.162\linewidth, trim={0.cm .0cm 1.5cm 1.5cm},clip]{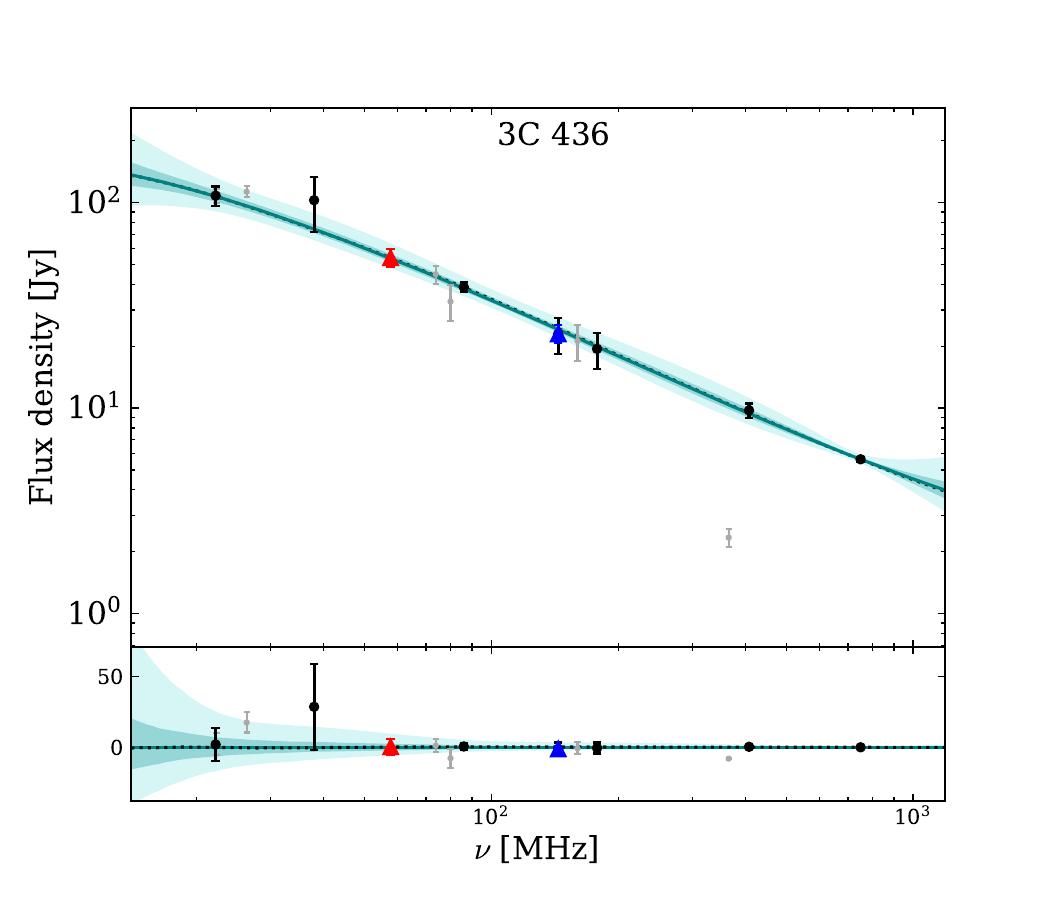}
\includegraphics[width=0.162\linewidth, trim={0.cm .0cm 1.5cm 1.5cm},clip]{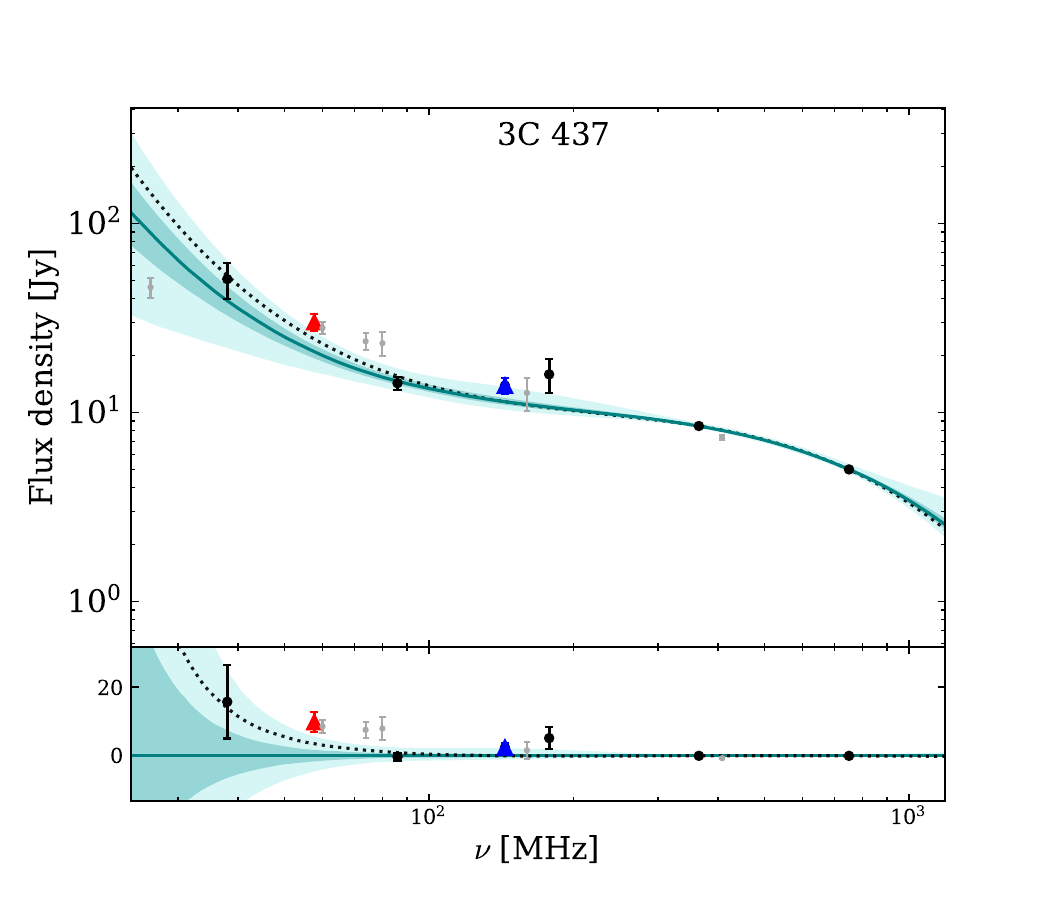}
\includegraphics[width=0.162\linewidth, trim={0.cm .0cm 1.5cm 1.5cm},clip]{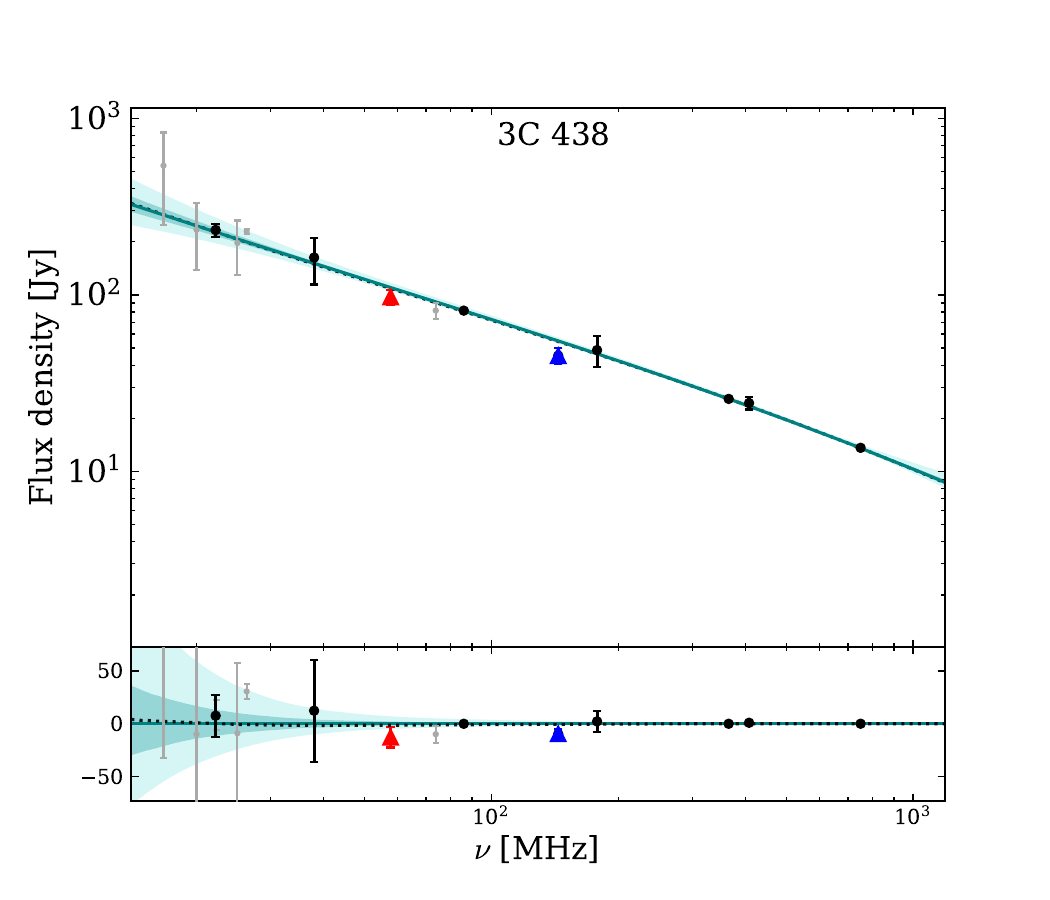}
\includegraphics[width=0.162\linewidth, trim={0.cm .0cm 1.5cm 1.5cm},clip]{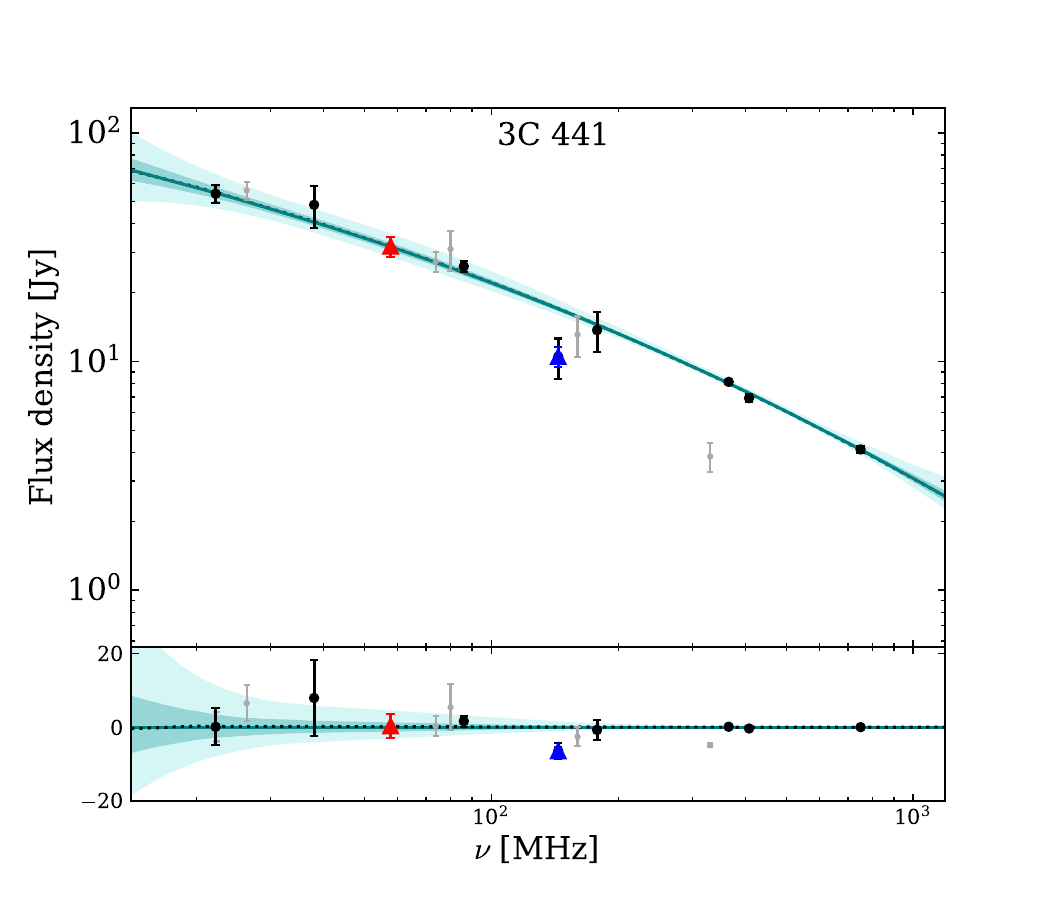}
\includegraphics[width=0.162\linewidth, trim={0.cm .0cm 1.5cm 1.5cm},clip]{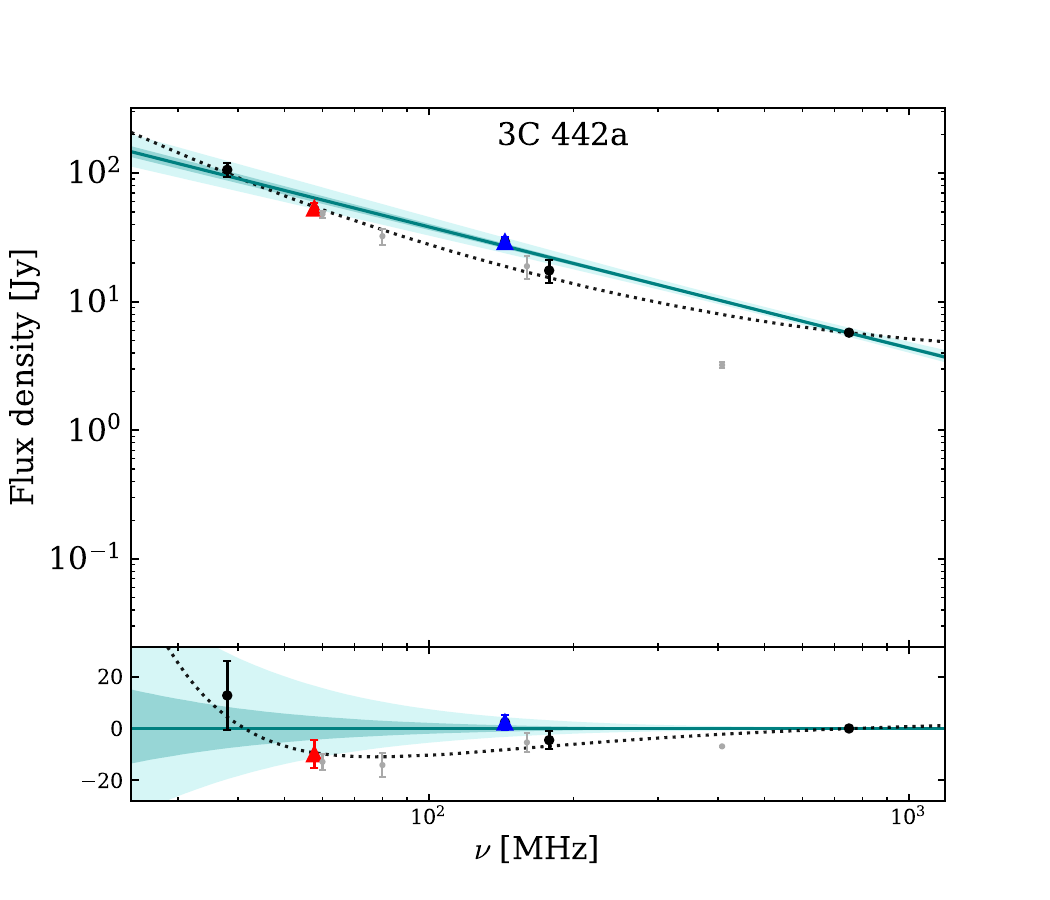}
\includegraphics[width=0.162\linewidth, trim={0.cm .0cm 1.5cm 1.5cm},clip]{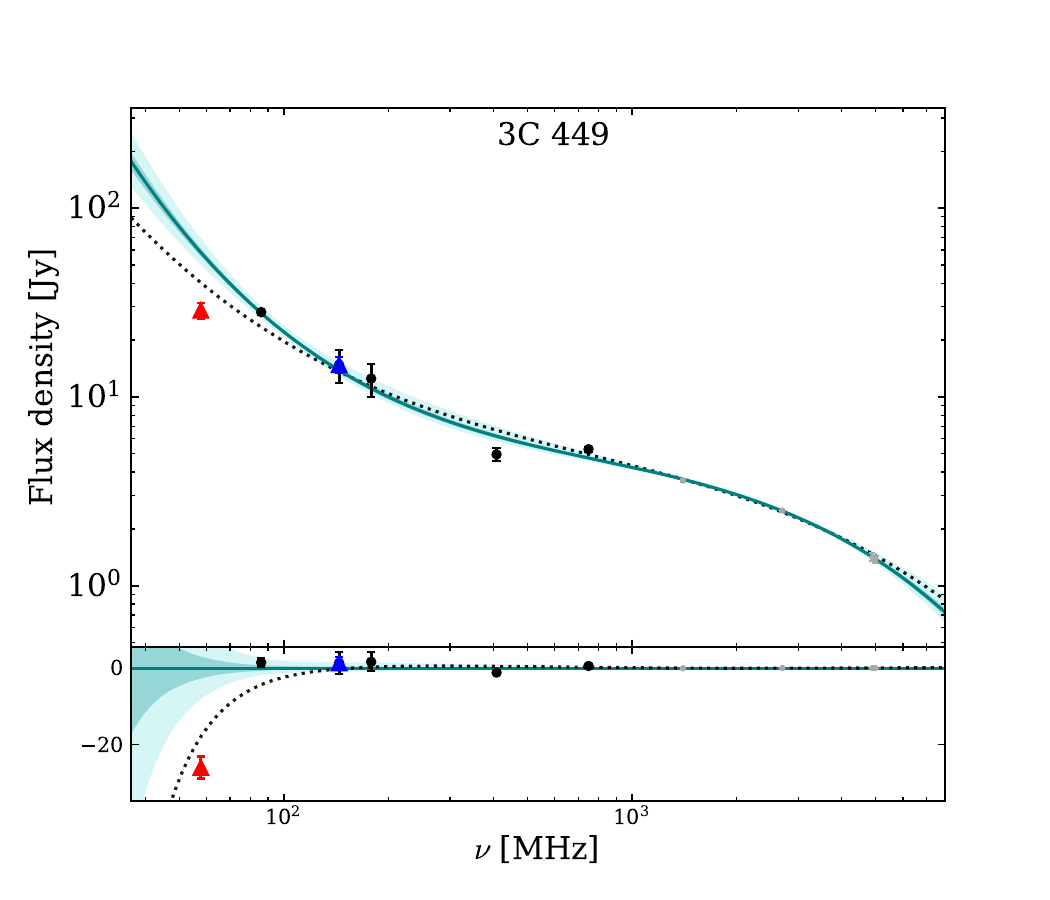}
\includegraphics[width=0.162\linewidth, trim={0.cm .0cm 1.5cm 1.5cm},clip]{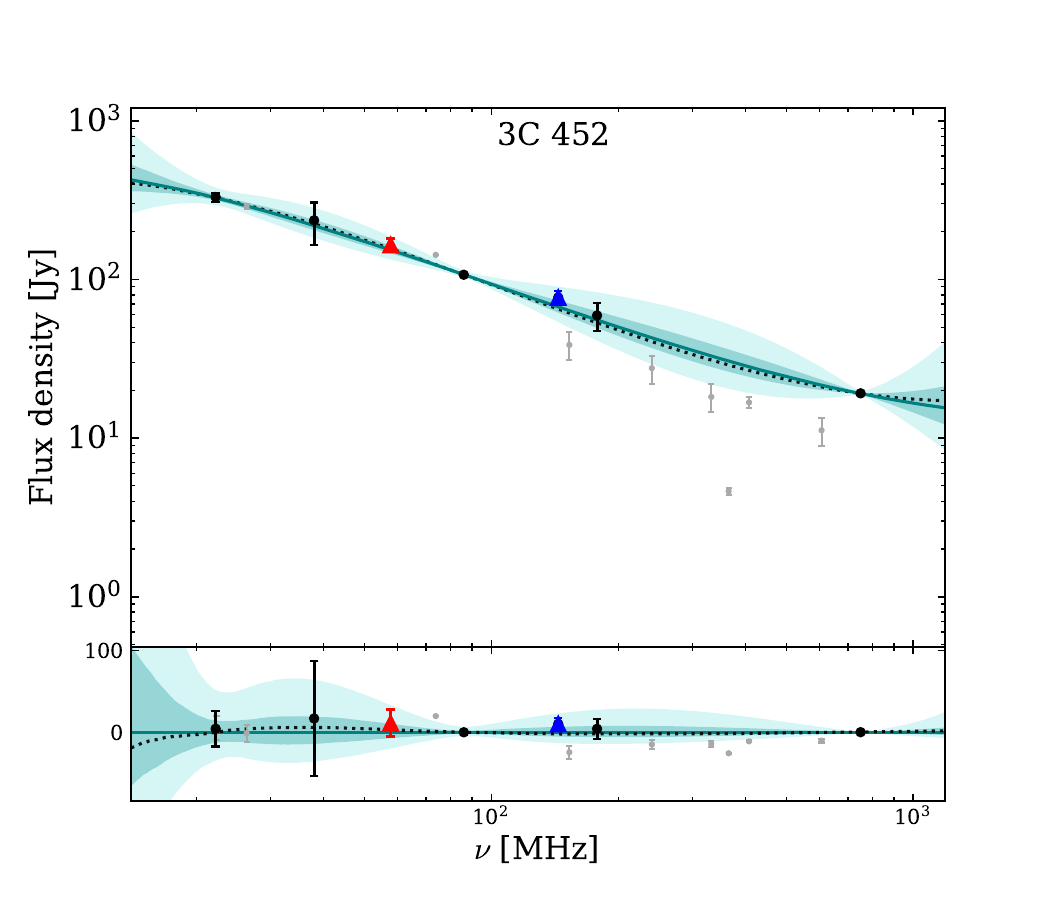}
\includegraphics[width=0.162\linewidth, trim={0.cm .0cm 1.5cm 1.5cm},clip]{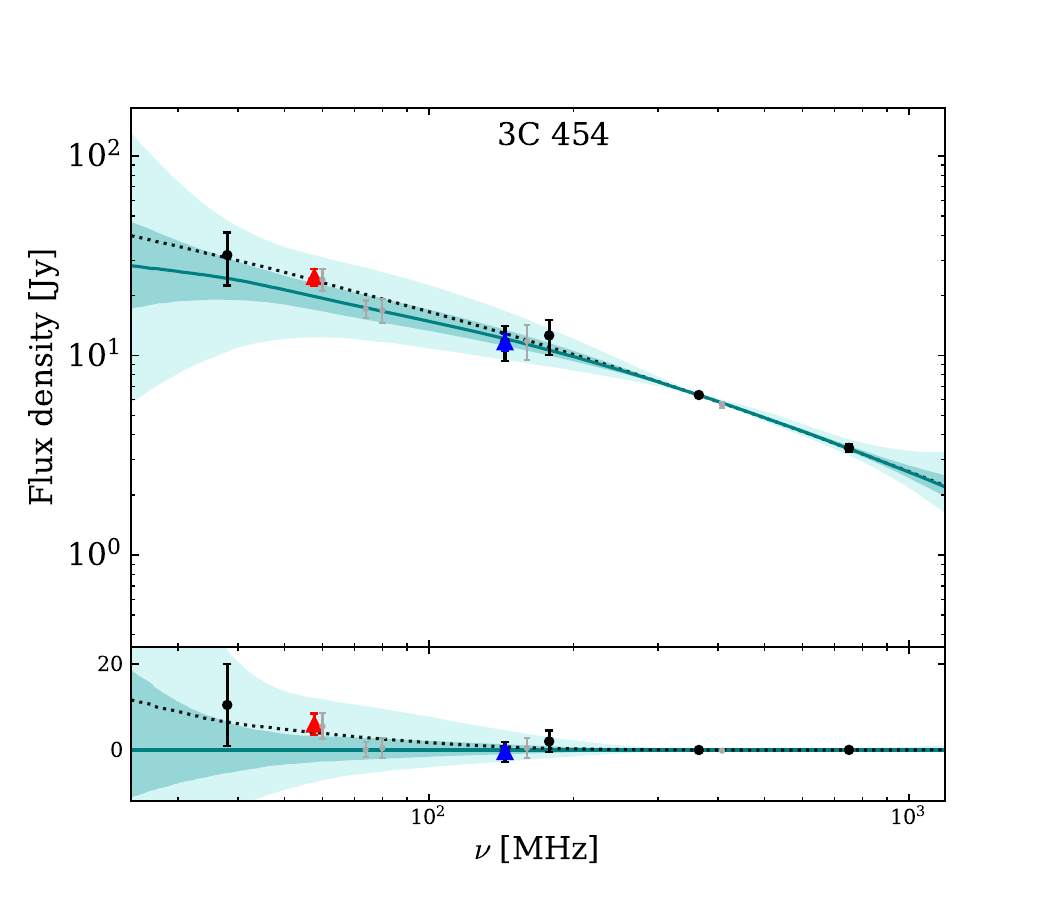}
\includegraphics[width=0.162\linewidth, trim={0.cm .0cm 1.5cm 1.5cm},clip]{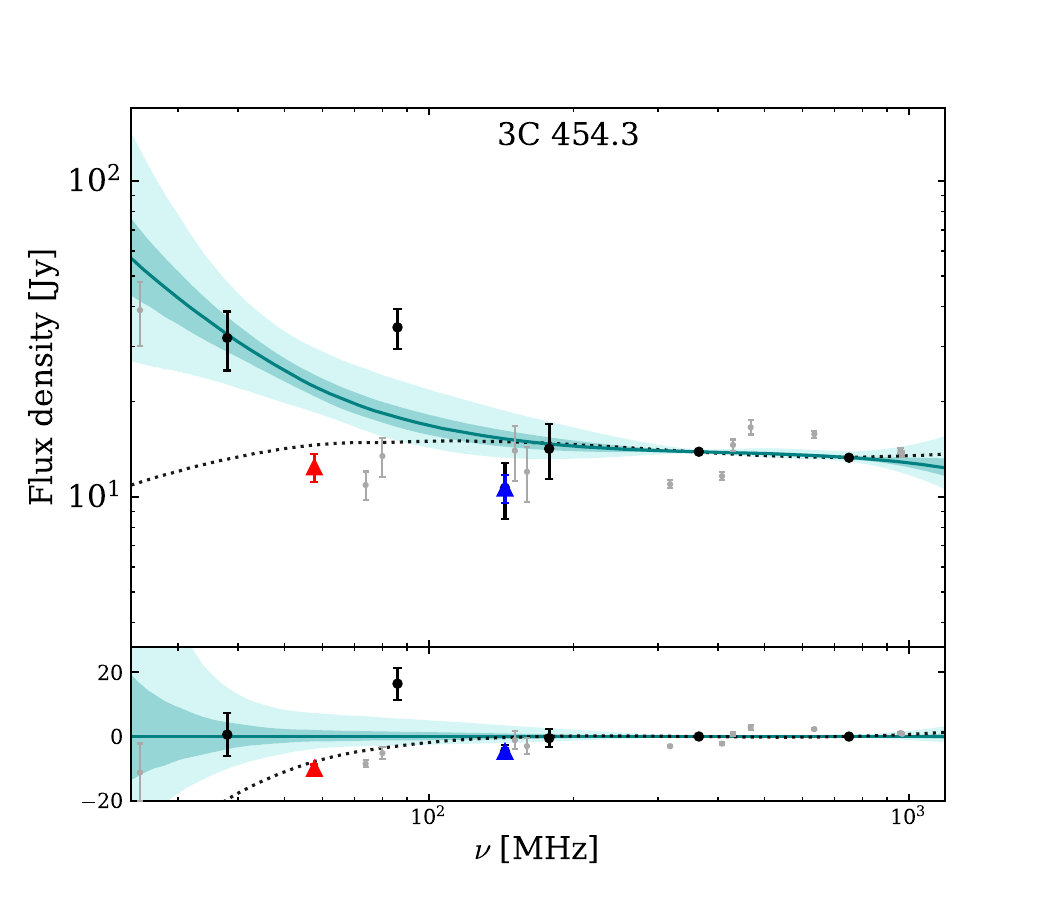}
\caption{continued.}
\end{figure}
\setcounter{figure}{0}

\begin{figure}[H]
\includegraphics[width=0.162\linewidth, trim={0.cm .0cm 1.5cm 1.5cm},clip]{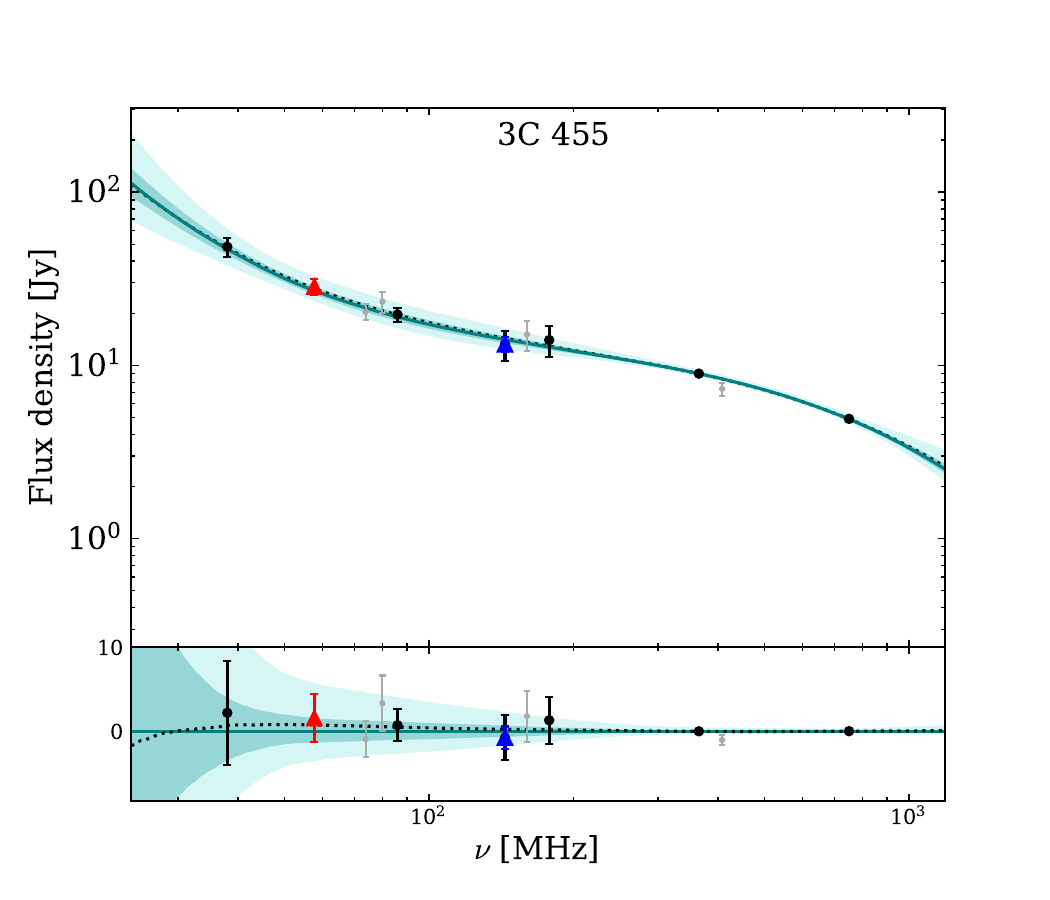}
\includegraphics[width=0.162\linewidth, trim={0.cm .0cm 1.5cm 1.5cm},clip]{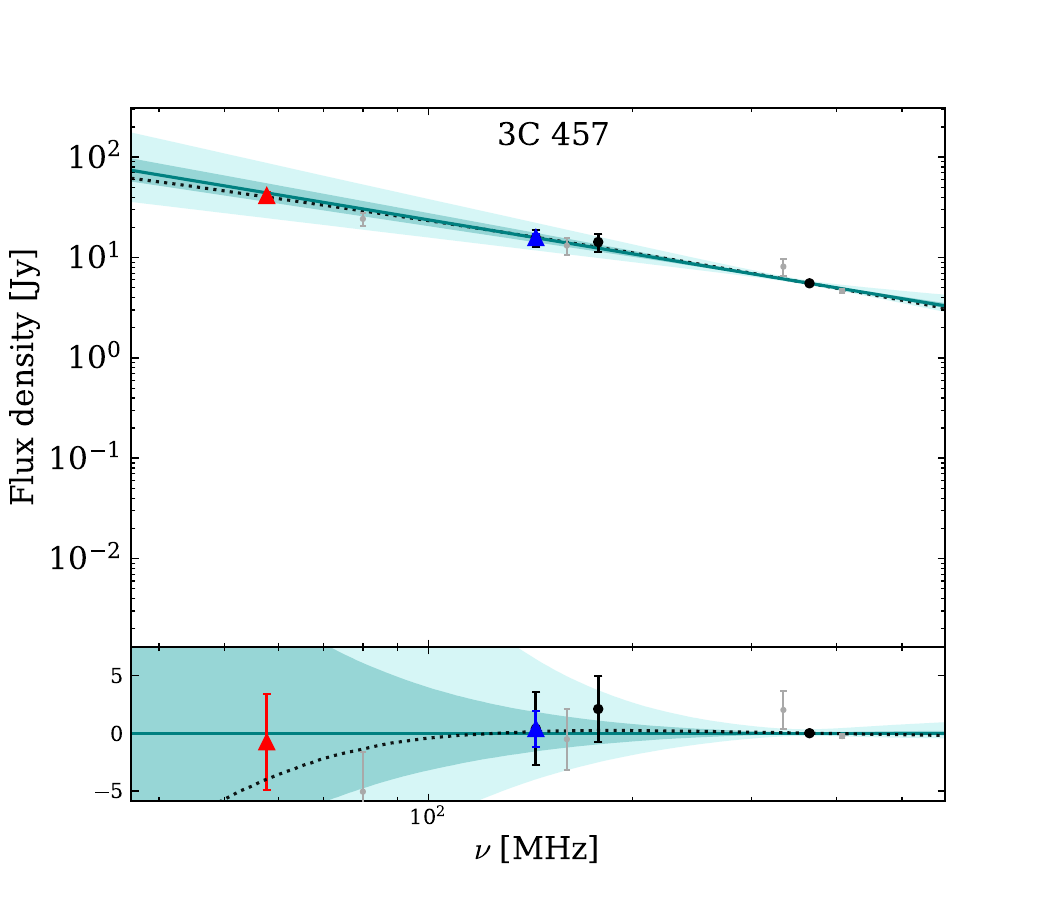}
\includegraphics[width=0.162\linewidth, trim={0.cm .0cm 1.5cm 1.5cm},clip]{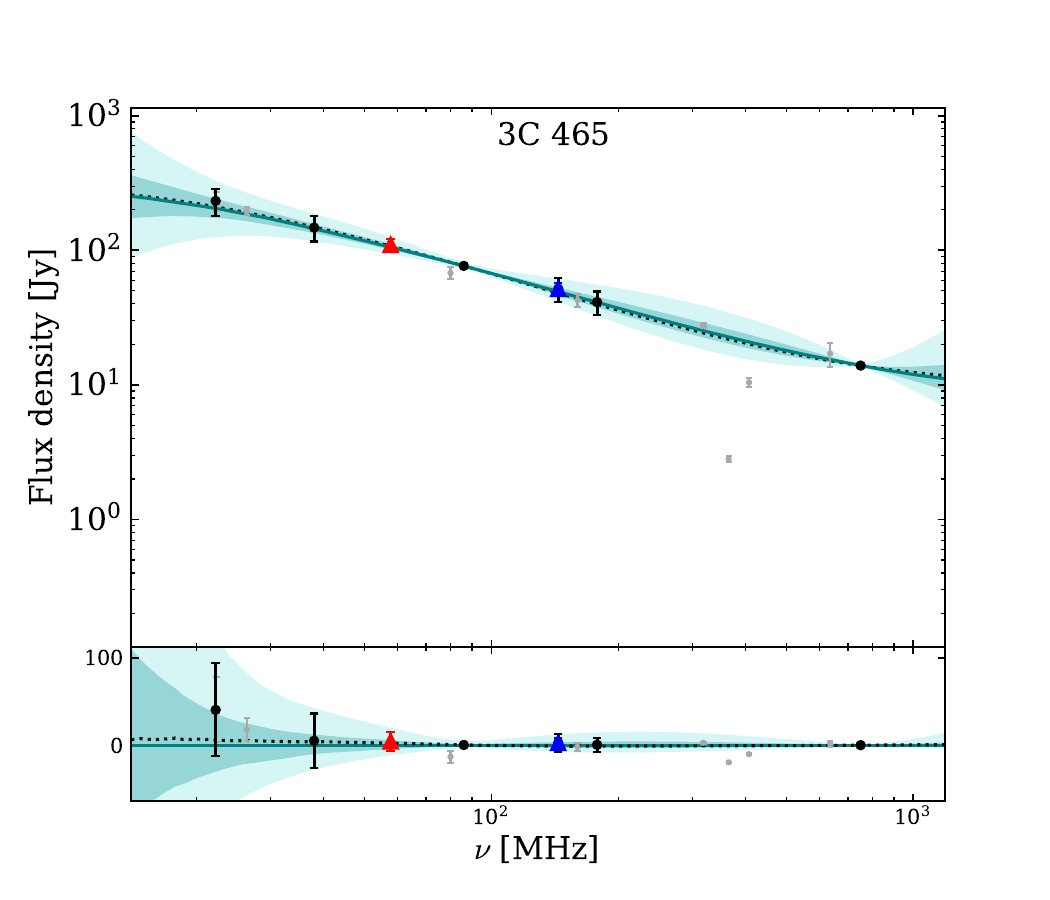}
\includegraphics[width=0.162\linewidth, trim={0.cm .0cm 1.5cm 1.5cm},clip]{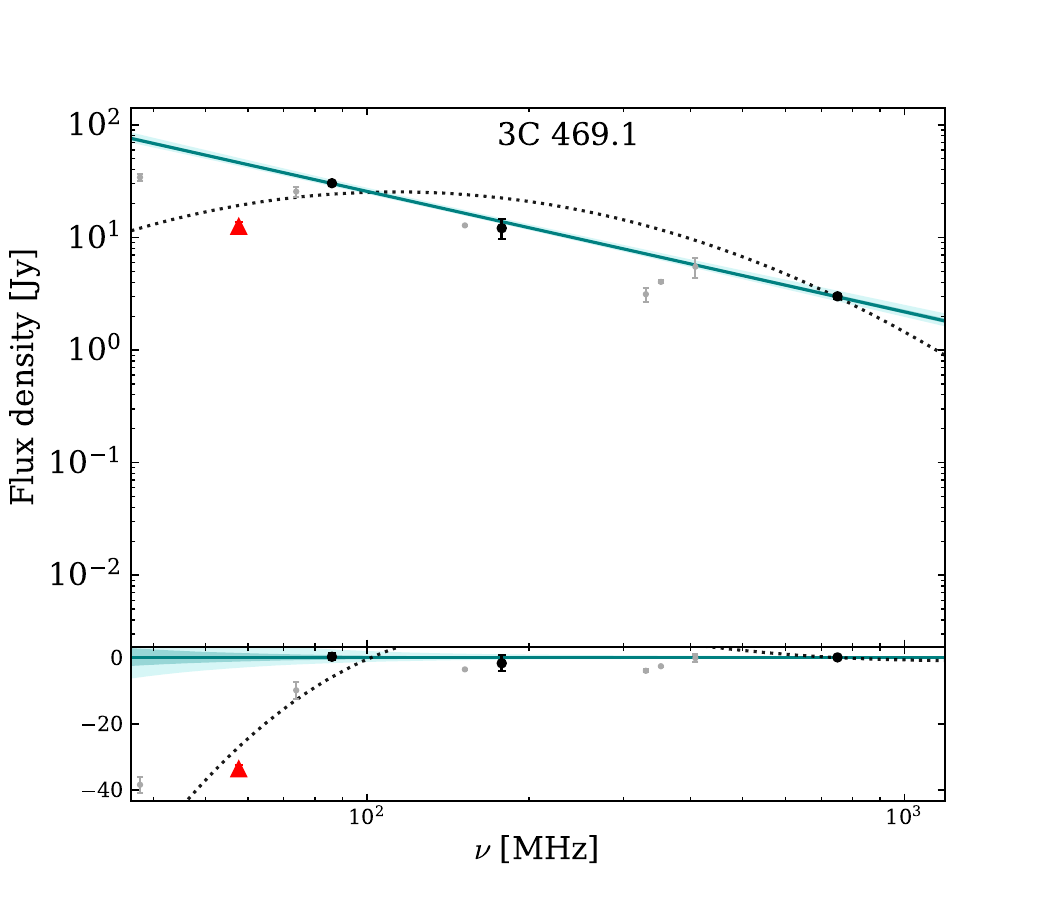}
\includegraphics[width=0.162\linewidth, trim={0.cm .0cm 1.5cm 1.5cm},clip]{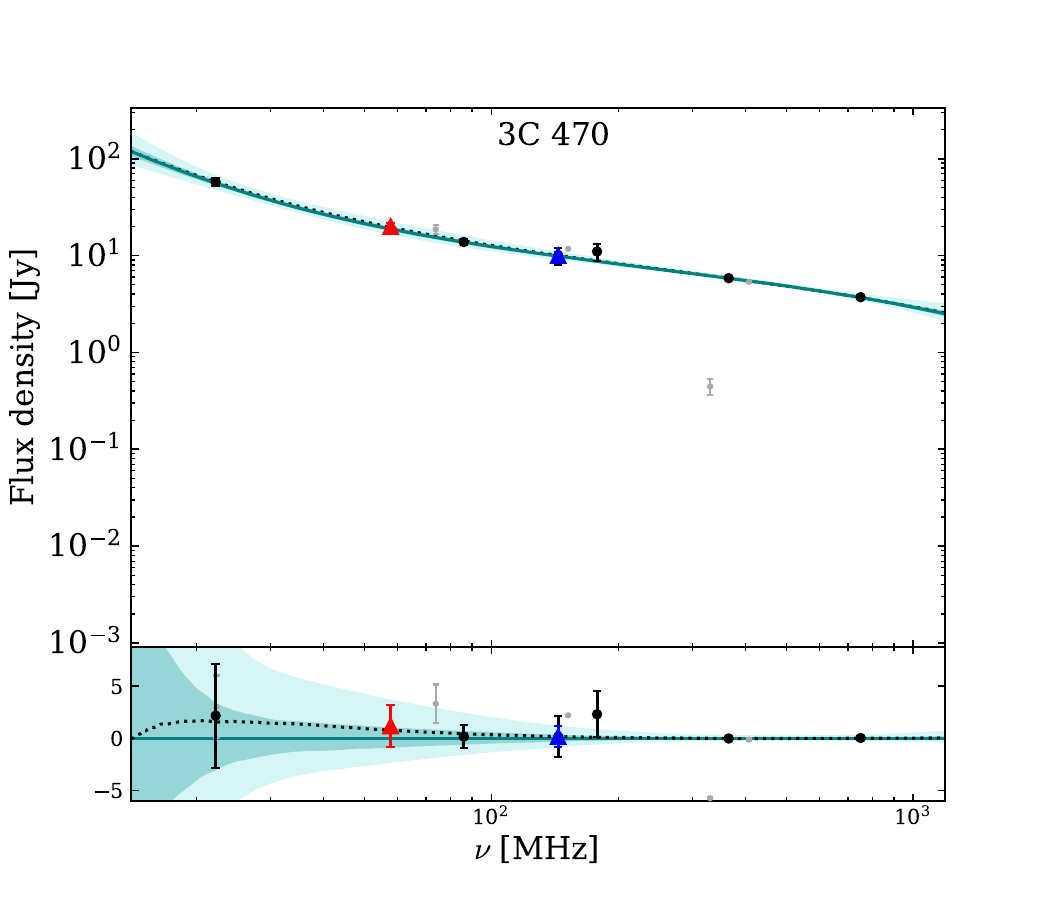}
\includegraphics[width=0.162\linewidth, trim={0.cm .0cm 1.5cm 1.5cm},clip]{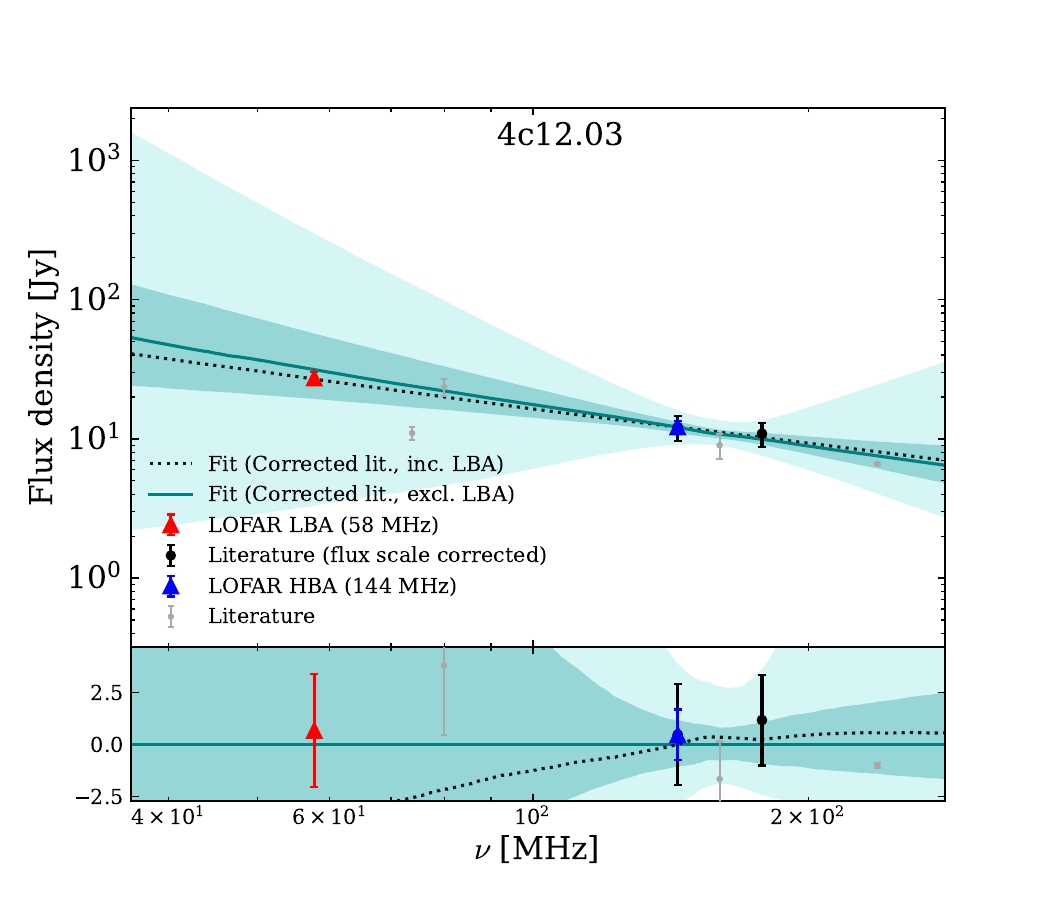}
\includegraphics[width=0.162\linewidth, trim={0.cm .0cm 1.5cm 1.5cm},clip]{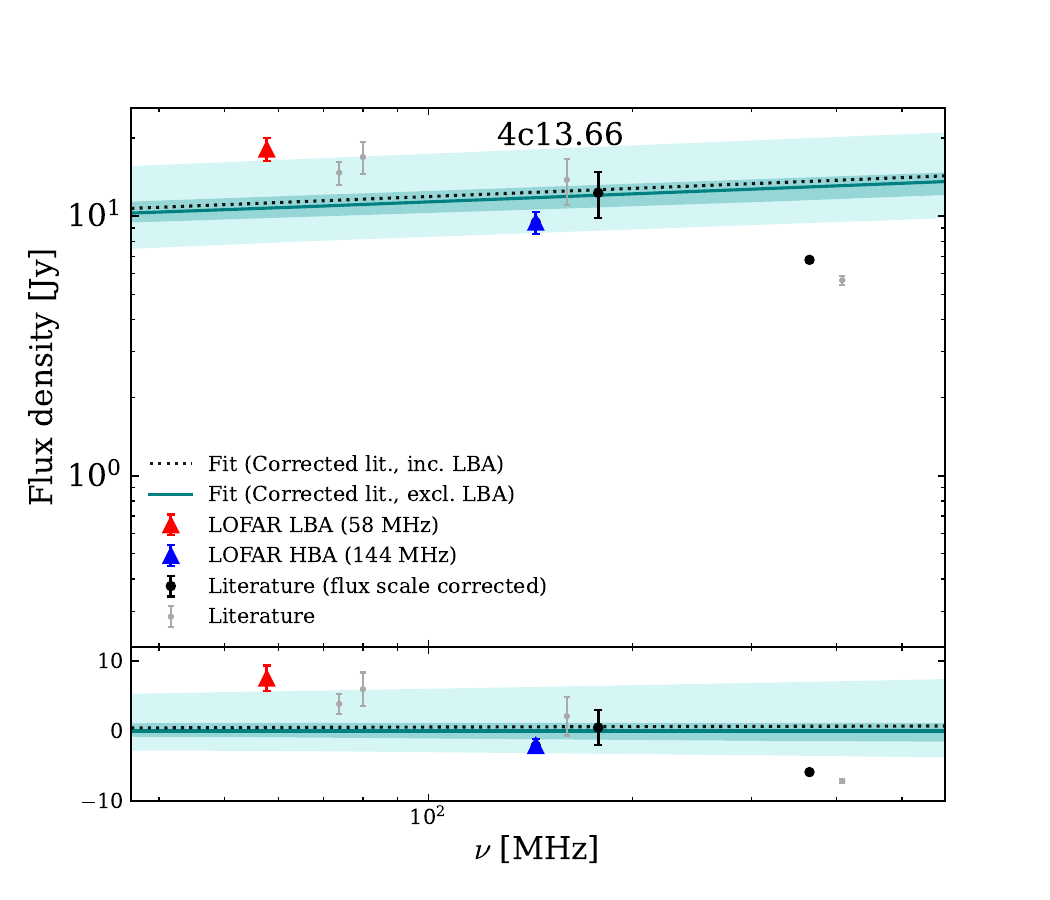}
\includegraphics[width=0.162\linewidth, trim={0.cm .0cm 1.5cm 1.5cm},clip]{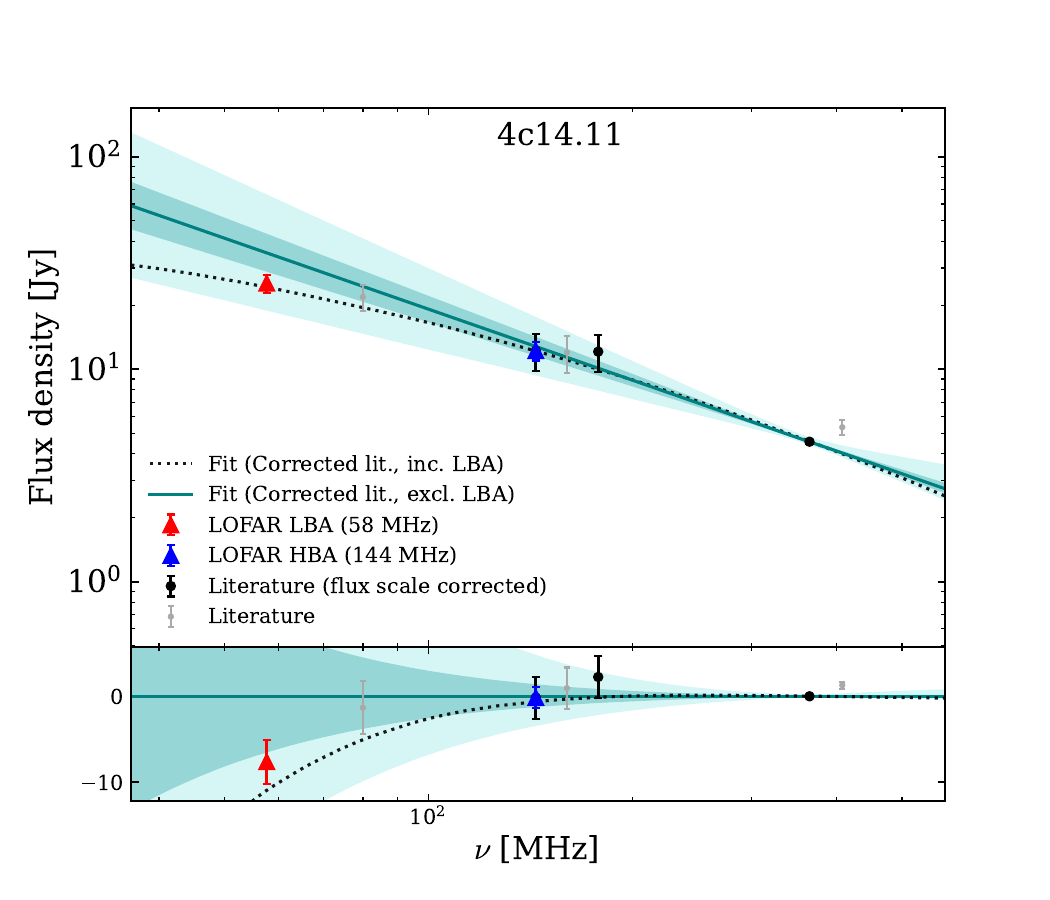}
\includegraphics[width=0.162\linewidth, trim={0.cm .0cm 1.5cm 1.5cm},clip]{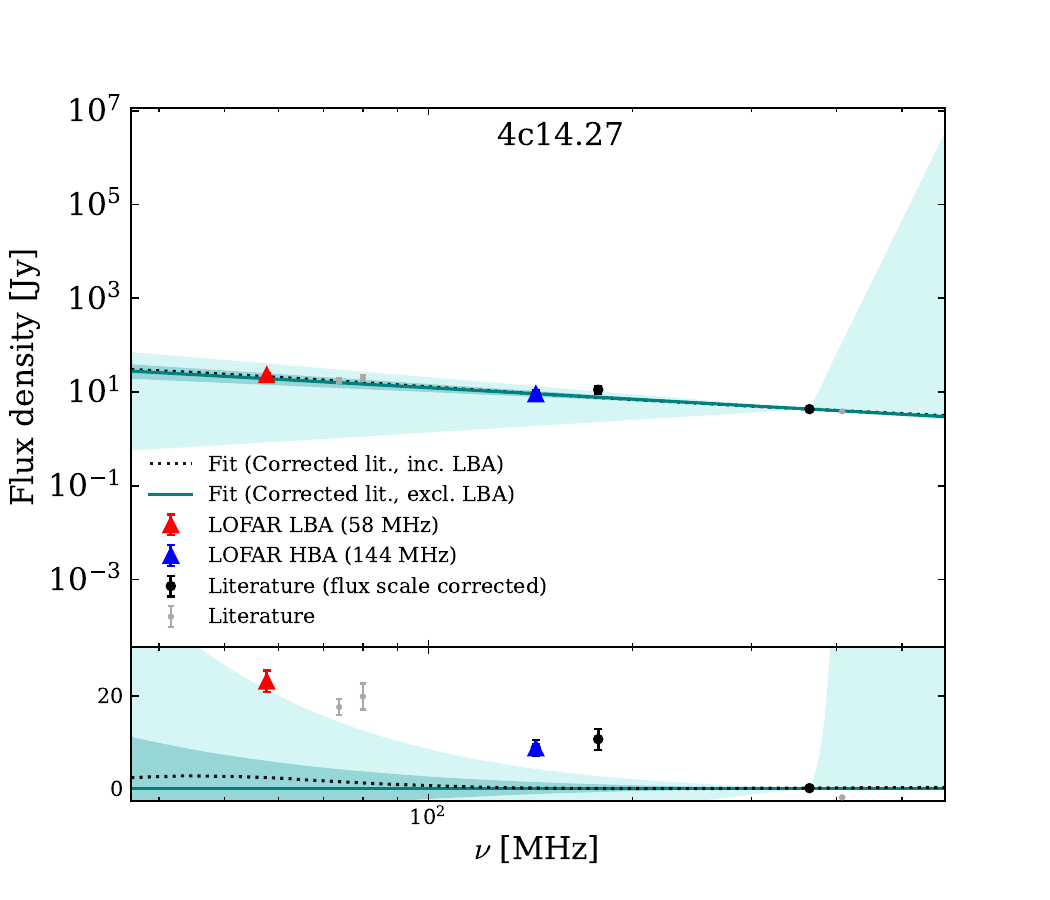}
\includegraphics[width=0.162\linewidth, trim={0.cm .0cm 1.5cm 1.5cm},clip]{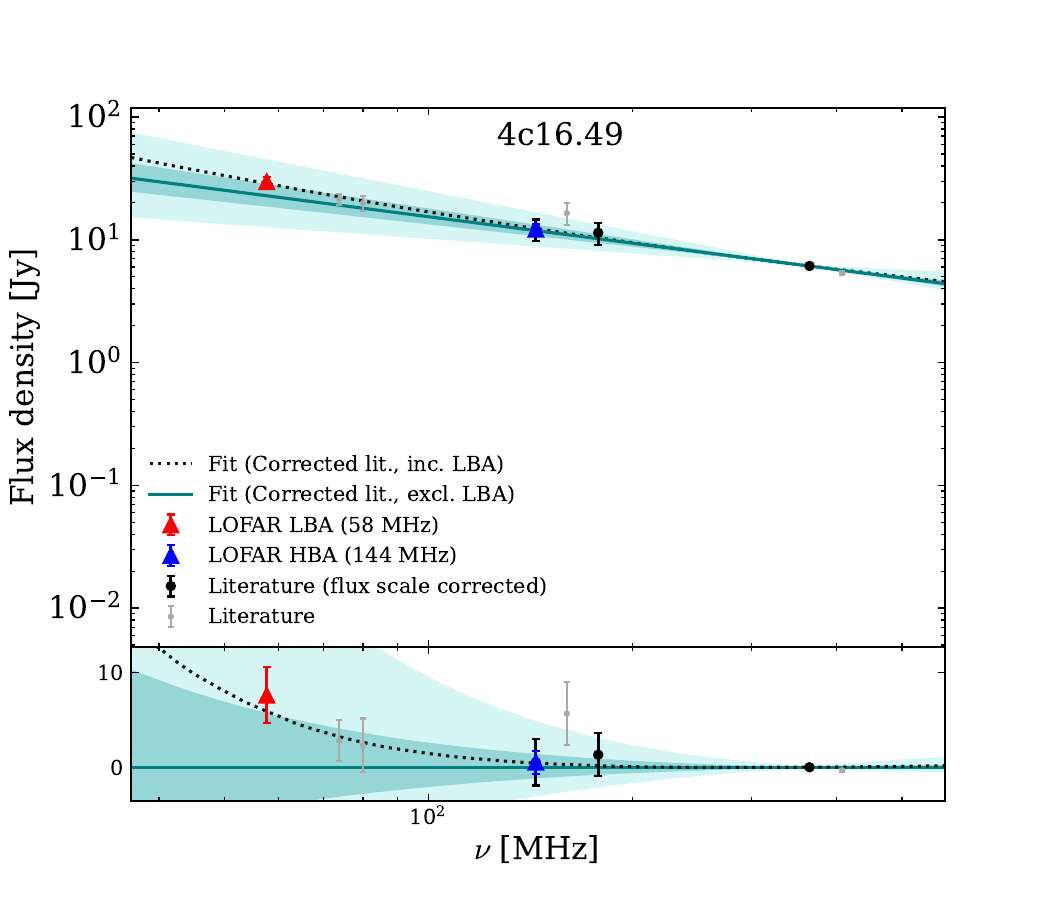}
\includegraphics[width=0.162\linewidth, trim={0.cm .0cm 1.5cm 1.5cm},clip]{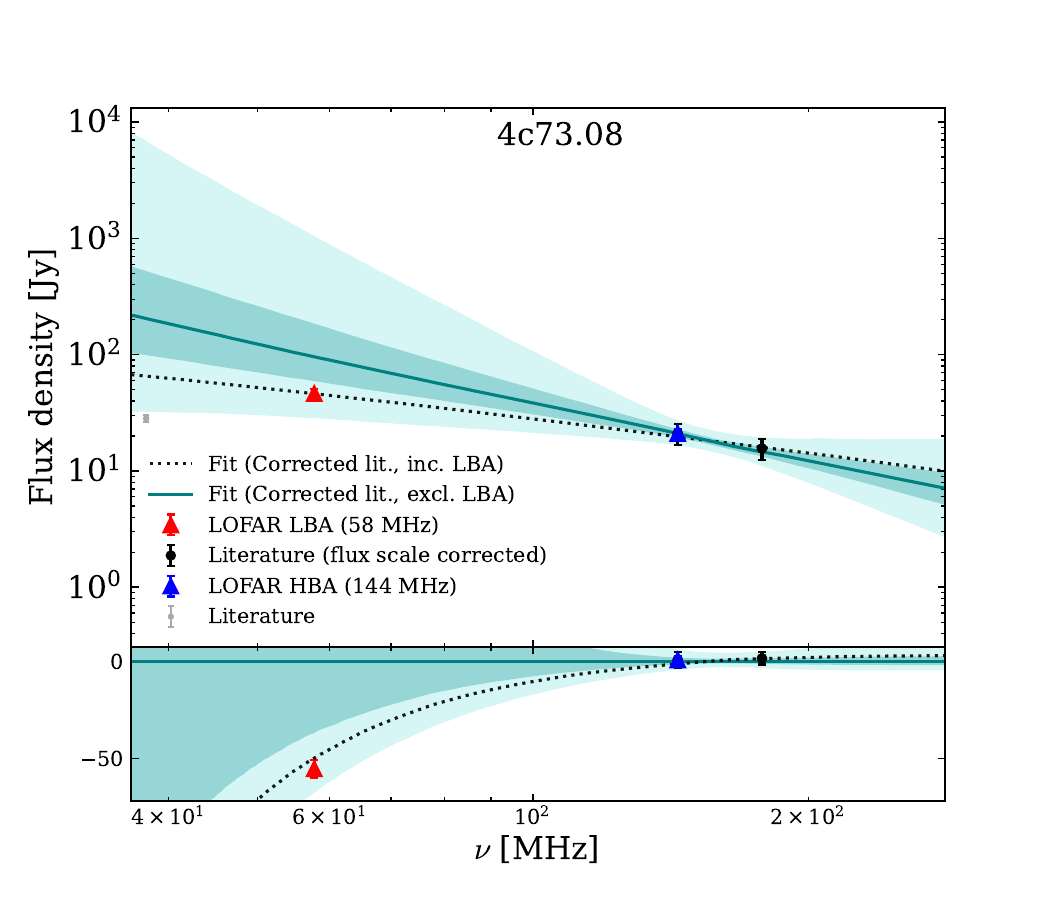}
\includegraphics[width=0.162\linewidth, trim={0.cm .0cm 1.5cm 1.5cm},clip]{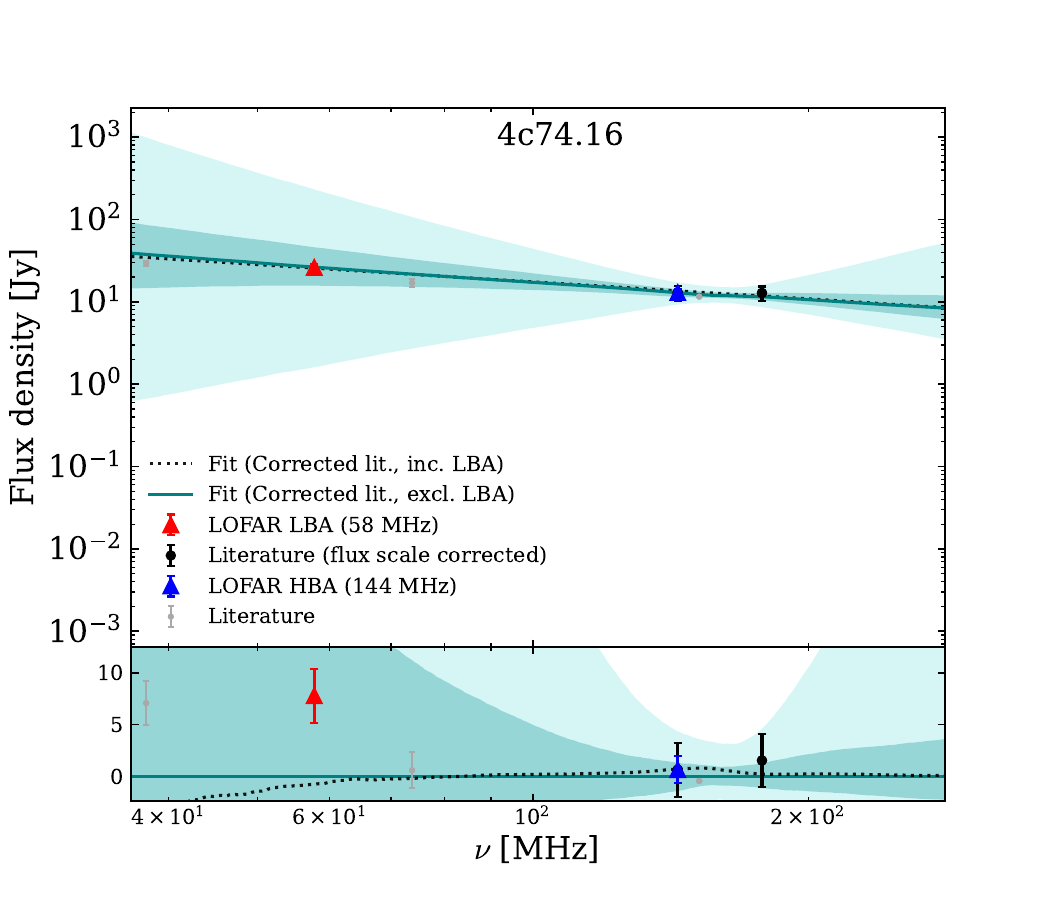}
\includegraphics[width=0.162\linewidth, trim={0.cm .0cm 1.5cm 1.5cm},clip]{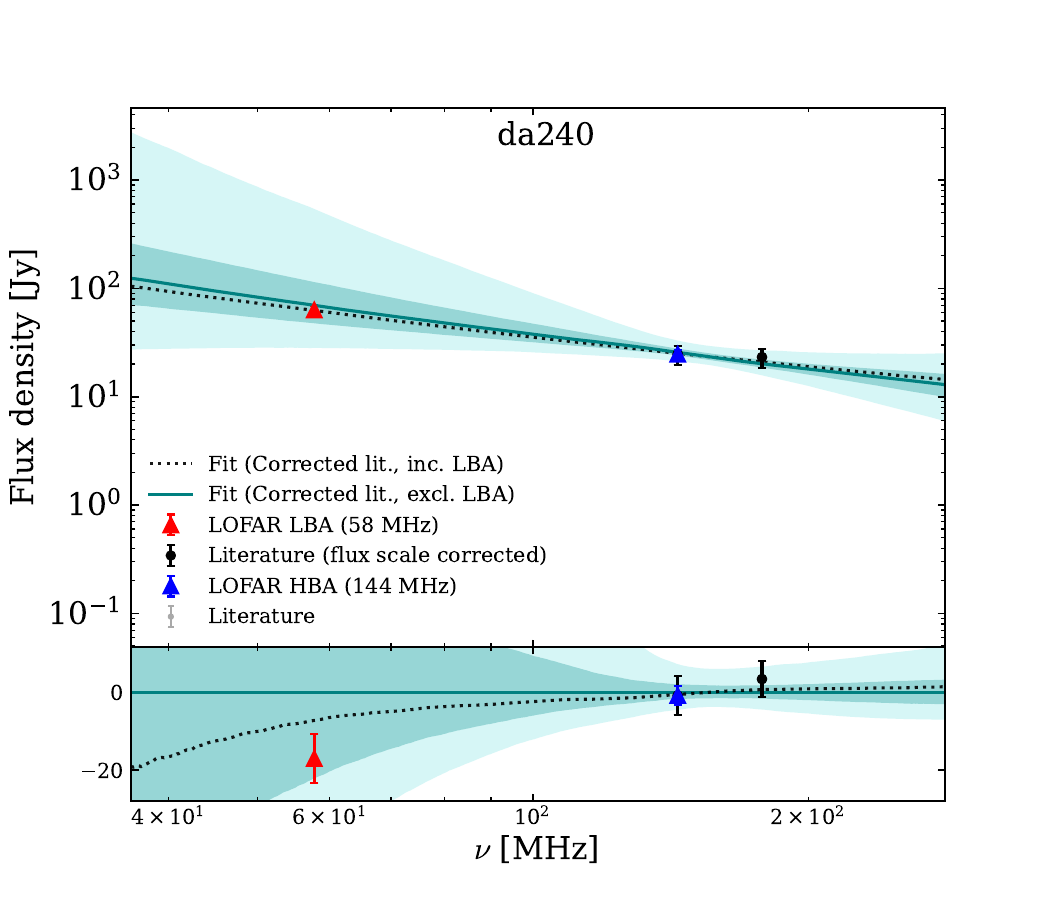}
\includegraphics[width=0.162\linewidth, trim={0.cm .0cm 1.5cm 1.5cm},clip]{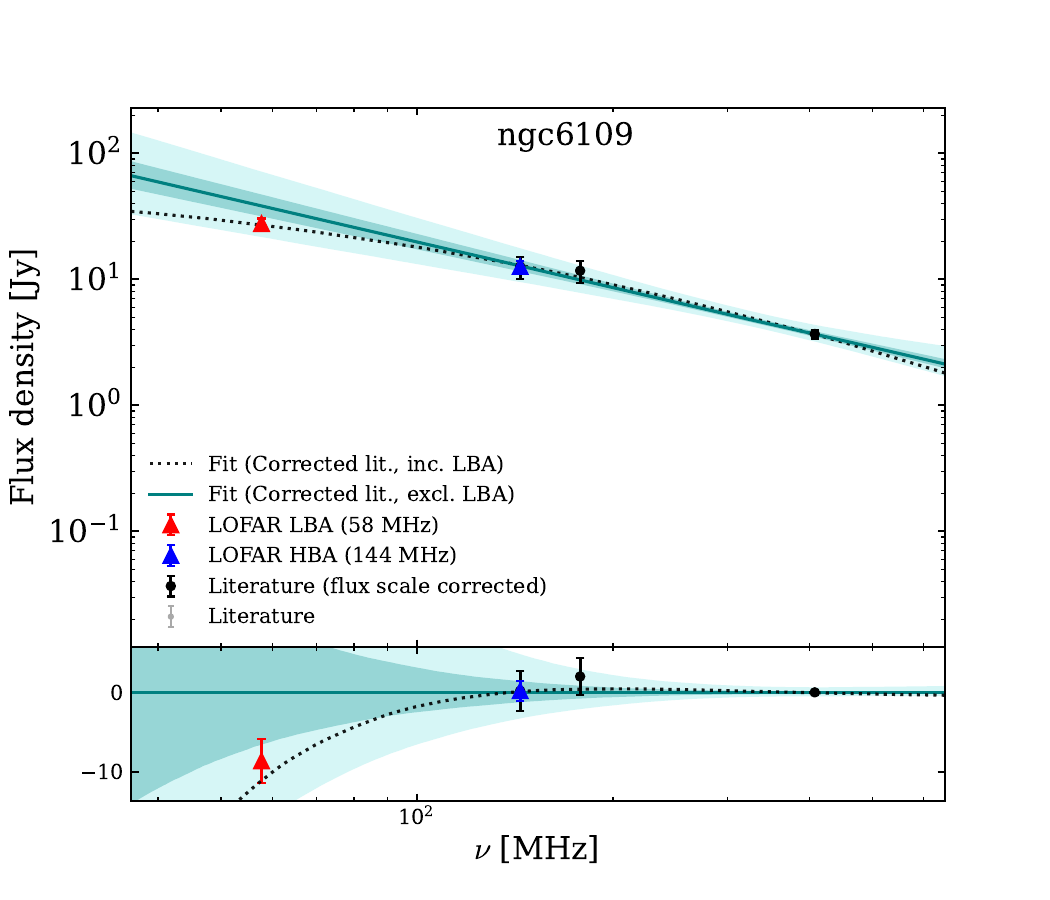}
\includegraphics[width=0.162\linewidth, trim={0.cm .0cm 1.5cm 1.5cm},clip]{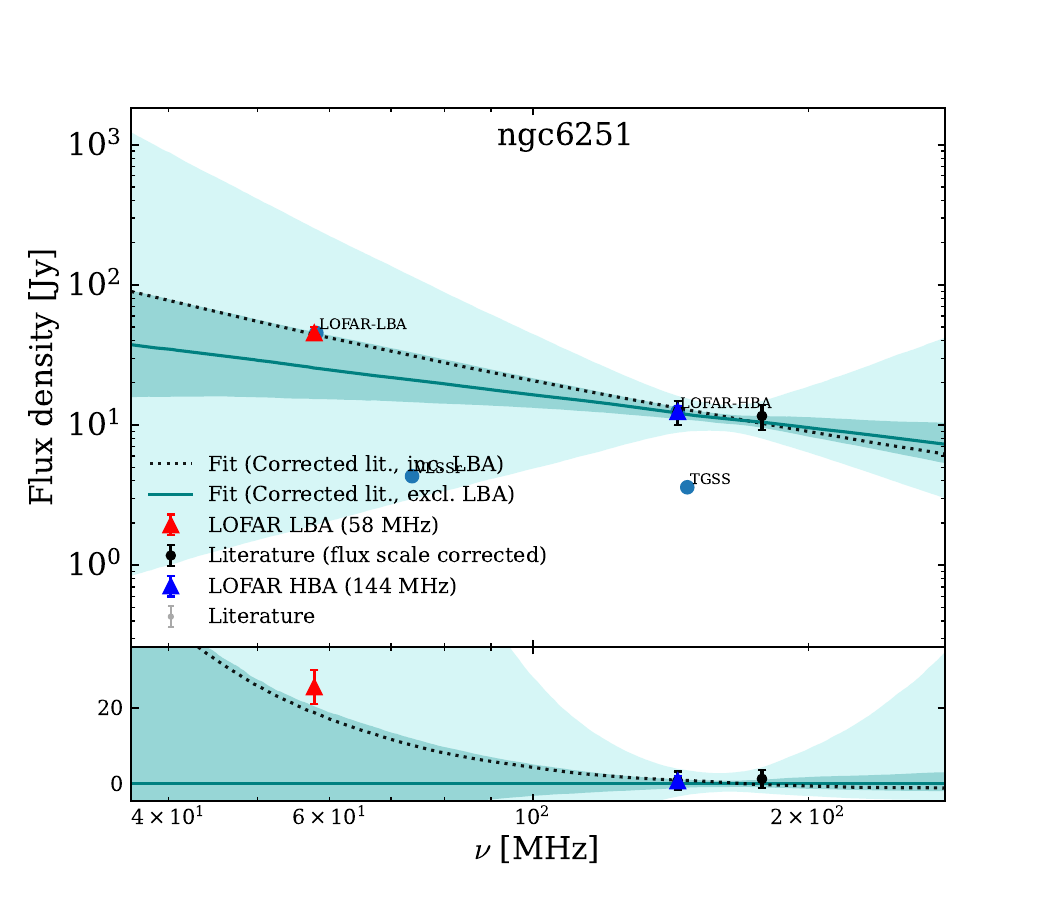}
\includegraphics[width=0.162\linewidth, trim={0.cm .0cm 1.5cm 1.5cm},clip]{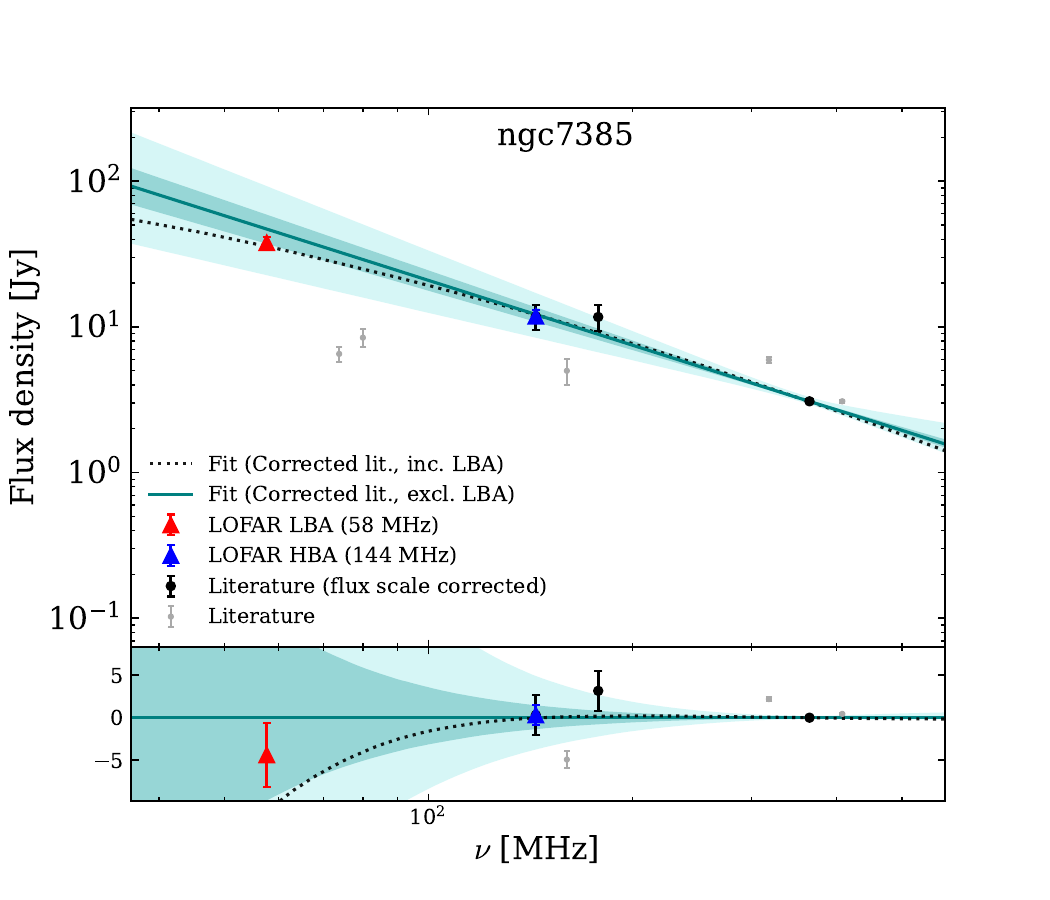}
\caption{continued.}
\label{fig:SEDs}
\end{figure}

\section{Summary of the 3CRR catalogue observed at 58~MHz.}
\label{app:all_sources}
\begin{longtable}{lcccccccccc}
\hline\hline
Object & RA & DEC  & z & $S_{58\;\mathrm{MHz}}$ & $\sigma_S$ & Resolution &$\sigma_{\text{rms}}$ & $S_{144\;\mathrm{MHz}}$ & $\alpha_{57}^{144}$\\ 
name & (J2000) & (J2000) &  &  &  &  & & & \\ 
& [h:m:s] & [d:m:s] &  & [Jy] & [Jy] & [$\arcsec\times\arcsec$] & [mJy~beam$^{-1}$] & [Jy] &\\ 
(1)&(2)&(3)&(4)&(5)&(6)&(7)&(8)&(9)&(10)\\ 
\hline
\endfirsthead
\caption{continued.}\\
\hline\hline
(1)&(2)&(3)&(4)&(5)&(6)&(7)&(8)&(9)&(10)&\\
\hline
\endhead
\hline
\endfoot

3C 6.1 & 0:16:31.23 & +79:16:50.38 & 0.840& 31.0 & 3.1 & $13.8\times10.5$ & 9.2 &  &  \\
3C 9 & 0:20:25.22 & +15:40:54.80 & 2.020& 50.2 & 5.0 & $15.6\times9.7$ & 10.3 & 18.6 & 1.08 \\
3C 13 & 0:34:14.56 & +39:24:16.67 & 1.351& 33.1 & 3.3 & $14.1\times8.3$ & 10.4 & 14.5 & 0.91 \\
3C 14 & 0:36:06.44 & +18:37:59.12 & 1.469& 31.2 & 3.1 & $16.0\times9.6$ & 8.7 & 13.6 & 0.91 \\
3C 16 & 0:37:44.57 & +13:19:54.98 & 0.406& 31.2 & 3.1 & $20.6\times9.6$ & 15.3 & 12.6 & 0.99 \\
3C 19 & 0:40:55.02 & +33:10:07.32 & 0.482& 27.5 & 2.8 & $11.1\times6.0$ & 13.7 & 9.6 & 1.15 \\
3C 20 & 0:43:09.18 & +52:03:36.11 & 0.174& 86.5 & 8.6 & $13.9\times9.4$ & 9.0 & 46.8 & 0.67 \\
3C 22 & 0:50:56.22 & +51:12:03.42 & 0.936& 59.5 & 6.0 & $12.6\times8.4$ & 15.0 & 16.6 & 1.40 \\
3C 28 & 0:55:50.61 & +26:24:37.48 & 0.195& 64.3 & 6.4 & $13.6\times8.7$ & 8.8 & 22.6 & 1.14 \\
3C 31 & 1:07:24.97 & +32:24:45.11 & 0.017& 43.7 & 4.4 & $26.1\times16.5$ & 10.5 & 22.5 & 0.73 \\
3C 33 & 1:08:52.87 & +13:20:14.35 & 0.061& 127.1 & 12.7 & $15.2\times9.9$ & 13.4 & 52.8 & 0.96 \\
3C 33.1 & 1:09:44.27 & +73:11:57.34 & 0.181& 44.4 & 4.4 & $14.1\times12.1$ & 10.0 & 24.6 & 0.64 \\
3C 34 & 1:10:18.67 & +31:47:20.47 & 0.690& 42.8 & 4.3 & $15.3\times8.8$ & 8.5 & 15.2 & 1.14 \\
3C 223 & 9:39:52.76 & +35:53:58.96 & 0.137& 41.0 & 4.1 & $14.5\times14.5$ & 10.5 & 20.1 & 0.78 \\
3C 225b & 9:42:15.35 & +13:45:49.57 & 0.583& 55.6 & 5.6 & $31.6\times8.2$ & 14.1 & 17.9 & 1.24 \\
3C 274\tablefootmark{a} & 12:30:49.42 & +12:23:28.03 & 0.0043& 2443.0 & 244.3 & $20.2\times8.0$ & 27.5 &  &  \\
 &  &  &  & &\vdots   & &  &  &  \\
  &  &  &  & &   & &  &  &  \\

\end{longtable}
\tablefoot{Illustrative selection of the summary of the entire catalog from this work. (4) The redshift of the object taken from the NASA Extragalactic Database, (5) Measured integrated flux density in Jy on the SH flux scale, (6) 10\% total uncertainty on the flux estimate, (7) Resolution of the produced total brightness maps, (8) RMS noise of those maps, (9) Dynamic range of final image defined as the ratio of the peak flux to the lowest pixel value, (10) estimated angular size of the sources in our data \& (11) integrated spectral index of the source fitted from a power law.}
\tablefoottext{a}{Also known as Virgo A or M87}

\end{appendix}

\end{document}